\documentclass[12pt]{book}

\usepackage[utf8]{inputenc} % set input encoding (not needed with XeLaTeX)

\usepackage[top=2.5cm,left=2.5cm,right=2.5cm,bottom=3cm]{geometry} % to change the page dimensions
\usepackage{graphicx} % support the \includegraphics command and options

 \usepackage[parfill]{parskip} % Activate to begin paragraphs with an empty line rather than an indent

\usepackage{booktabs} % for much better looking tables
\usepackage{array} % for better arrays (eg matrices) in maths
\usepackage{paralist} % very flexible & customisable lists (eg. enumerate/itemize, etc.)
\usepackage{verbatim} % adds environment for commenting out blocks of text & for better verbatim
\usepackage{subfig} % make it possible to include more than one captioned figure/table in a single float
\usepackage{hyperref}
\hypersetup{colorlinks,linkcolor=black,urlcolor=blue,pdftex}
\usepackage[english]{babel}

\usepackage{fancyhdr} % This should be set AFTER setting up the page geometry
\usepackage{sectsty}
\usepackage[nottoc,notlof,notlot]{tocbibind} % Put the bibliography in the ToC
\usepackage[titles,subfigure]{tocloft} % Alter the style of the Table of Contents

\usepackage{fontenc}
\usepackage{amsfonts}
\usepackage{wrapfig}
\usepackage{multirow}
\usepackage{amssymb}
\usepackage{quoting}
\usepackage{rotating}
\quotingsetup{font=normalsize}
\usepackage{amsthm}
\theoremstyle{plain}
\newtheorem*{teo}{Theorem}
\usepackage{bm}
\usepackage{amsmath}
\usepackage[mathscr]{eucal}

\newcommand{\bsig}{\boldsymbol{\sigma}}
\newcommand{\beps}{\boldsymbol{\varepsilon}}
\newcommand{\bk}{\boldsymbol{\kappa}}
\newcommand{\bom}{\boldsymbol{\omega}}
\newcommand{\bnu}{\boldsymbol{\nu}}
\newcommand{\btau}{\boldsymbol{\tau}}
\newcommand{\bb}{\boldsymbol{\beta}}
\newcommand{\bt}{\boldsymbol{\theta}}
\newcommand{\bp}{\boldsymbol{\phi}}
\newcommand{\bdelta}{\boldsymbol{\delta}}
\newcommand{\bvp}{\boldsymbol{\varphi}}
\newcommand{\bgamma}{\boldsymbol{\gamma}}
\newcommand{\bg}{\boldsymbol{\gamma}}
\newcommand{\bl}{\boldsymbol{\lambda}}
\newcommand{\bkappa}{\boldsymbol{\kappa}}
\newcommand{\bmu}{\boldsymbol{\mu}}
\newcommand{\au}{\alpha_1}
\newcommand{\ad}{\alpha_2}
\newcommand{\ai}{\alpha_i}
\newcommand{\bepsi}{\boldsymbol{\epsilon}}

\renewcommand{\Phi}{\varPhi}
\newcommand{\gr}{\mathbf}
\newcommand{\eps}{\varepsilon}

\newcommand{\R}{\mathbb{R}}
\newcommand{\tr}{\mathrm{tr}}
\renewcommand{\div}{\mathrm{div}}

\newcommand{\eu}{\gr{e}_1}
\newcommand{\ed}{\gr{e}_2}
\newcommand{\et}{\gr{e}_3}
\newcommand{\ex}{\gr{e}_\mathit{x}}
\newcommand{\ey}{\gr{e}_\mathit{y}}
\newcommand{\ez}{\gr{e}_\mathit{z}}
\newcommand{\q}{\mathbf{q}}
\newcommand{\m}{\mathbf{m}}
\newcommand{\Eu}{\mathcal{E}}

\newcommand{\Rep}{\mathcal{R}}
\renewcommand{\S}{\mathcal{S}}
\newcommand{\bu}{\mathbf{u}}
\newcommand{\bv}{\mathbf{v}}
\newcommand{\bw}{\mathbf{w}}
\newcommand{\bo}{\mathbf{o}}
\newcommand{\e}{\mathbf{e}}
\newcommand{\Q}{\mathbf{Q}}
\renewcommand{\O}{\mathbf{O}}
\newcommand{\I}{\mathbf{I}}
\newcommand{\B}{\mathbf{B}}

\renewcommand{\b}{\mathbf{b}}
\renewcommand{\t}{\mathbf{t}}
\renewcommand{\r}{\mathbf{r}}

\renewcommand{\d}{\mathbf{d}}
\newcommand{\M}{\mathbf{M}}
\newcommand{\N}{\mathbf{N}}
\newcommand{\mn}{\mathcal{N}}
\newcommand{\mpe}{\mathscr{P}}
\newcommand{\W}{\mathbf{W}}
\newcommand{\n}{\mathbf{n}}
\newcommand{\T}{\mathbf{T}}
\renewcommand{\L}{\mathbf{L}}
\newcommand{\F}{\mathbf{F}}

\newcommand{\f}{\mathbf{f}}
\newcommand{\g}{\mathbf{g}}
\newcommand{\X}{\mathbf{X}}

\newcommand{\bi}{\begin{itemize}}
\newcommand{\ei}{\end{itemize}}
\renewcommand{\i}{\item}
\renewcommand{\(}{\begin{columns}}
\renewcommand{\)}{\end{columns}}
\newcommand{\<}[1]{\begin{column}{#1}}
\renewcommand{\>}{\end{column}}
\newcommand{\be}{\begin{equation}}
\newcommand{\ee}{\end{equation}}
\newcommand{\bes}{\begin{equation*}}
\newcommand{\ees}{\end{equation*}}
\newcommand{\besp}{\begin{split}}
\newcommand{\esp}{\end{split}}

\begin{document}
    \frontmatter
\title{\begin{figure}[t]
	\begin{center}
         \includegraphics[scale=1]{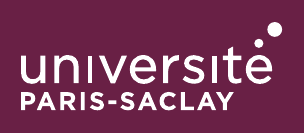}\vspace{-2cm}\\
%	\caption{}
%	\label{fig1}
	\end{center}
	\end{figure}
	\footnotesize{Class notes of the course}\vspace{3mm}\\\huge{\textbf{Modeling of Solids}}\vspace{3mm} \\}

\author{%{Class notes}%Paolo Vannucci
\\%vspace{.5cm}
%Master MMM:\\ {\it Mathematical Methods for Mechanics}
}

\date{\vspace{.5cm}
\includegraphics[scale=.3]{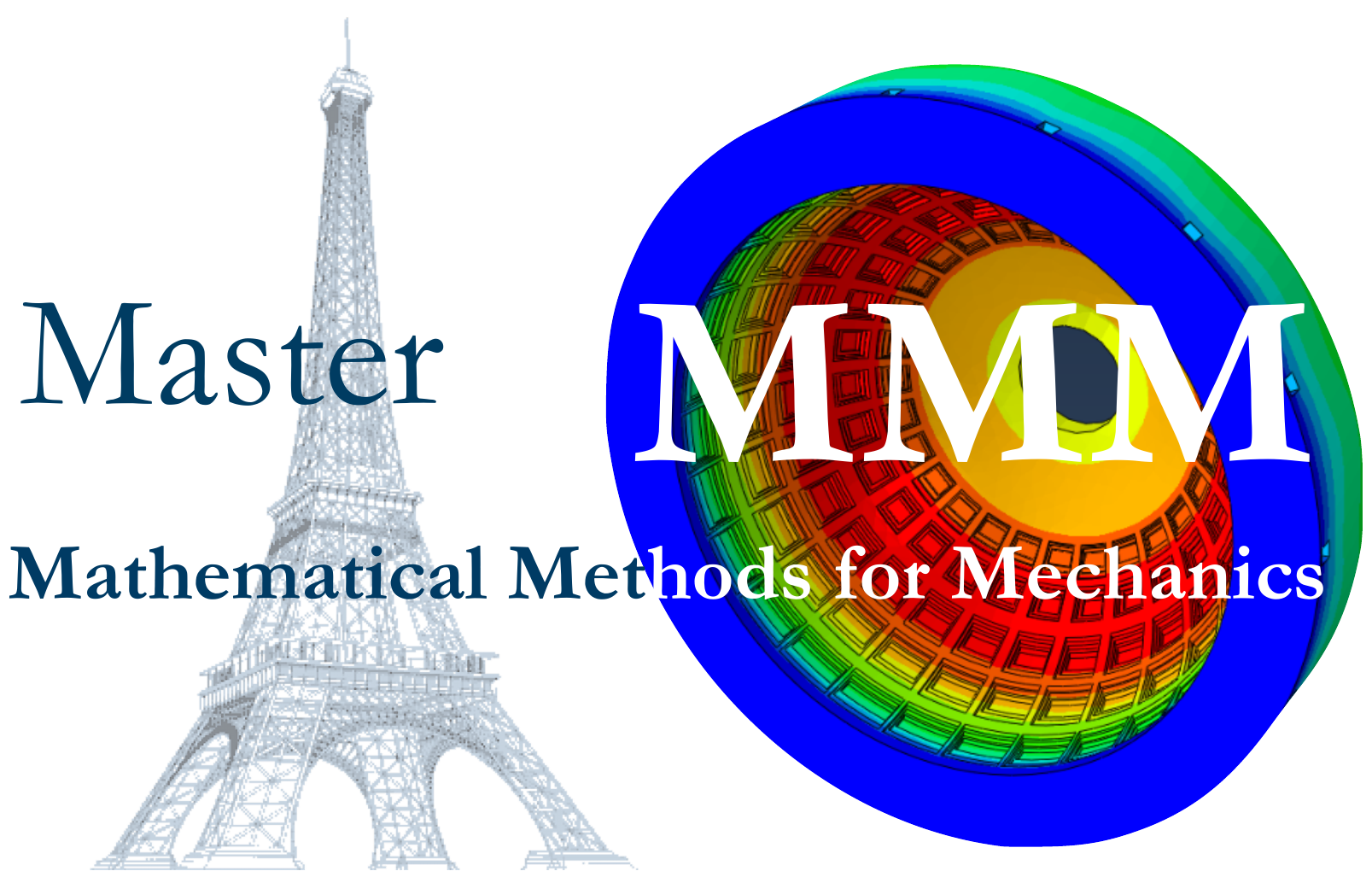}\vspace{2.cm}\\
Paolo Vannucci\\
\vspace{.2cm}
{\href{mailto:paolo.vannucci@uvsq.fr}{paolo.vannucci@uvsq.fr}}\vspace{1.5cm}\\
\bigskip\bigskip\bigskip{\footnotesize Last update: February 7, 2024}}

\maketitle

    \chapter{Preface}
    
    This text is the support for the course of Modeling of Solids,  of the Master of Mechanics of the University Paris-Saclay - Curriculum MMM: Mathematical Methods for Mechanics, held at Versailles.
    
    The course is the continuation of the course Continuum Mechanics - Solids, and as such it is an introduction, for graduate students, to some typical topics of the theory of solid bodies. 
    
     The different arguments are dealt with in a simple, succinct way, the objective being to give to students the fundamentals of each argument. Only static problems are considered, being the dynamic of structures dealt with in other courses.
     %The topics are:  cables, arches, slender rods, membranes, plates and shells.
     \\\\\\
{\it Versailles, December 7, 2015}   
%    \chapter{Remerciements}
%    \chapter{Introduction}
\tableofcontents

    % Corps du livre
    \mainmatter

\chapter{Some elements of differential geometry}
\label{ch:1}
\section{Curves of points, vectors and tensors}
Be $\mathcal{R}=\{o; \gr{e}_1,\gr{e}_2,\gr{e}_2\}$ a reference frame of the euclidean space $\mathcal{E}$\footnote{
The Euclidean space $\mathcal{E}$ is the ordinary three-dimensional space, whose elements are geometric points $p$. A reference frame $\mathcal{R}=\{o; \gr{e}_1,\gr{e}_2,\gr{e}_2\}$  in $\mathcal{E}$ is composed by a point $o$, called the {\it origin} of the frame and by a basis $\{\gr{e}_1,\gr{e}_2,\gr{e}_2\}$ of the space vector $\mathcal{V}$ associated to $\mathcal{E}$, called the {\it space of translations}: any vector $\gr{v}\in\mathcal{V}$ is a function that operates on the points $p\in\mathcal{E}:\ \forall \gr{v}\in\mathcal{V}, \ \gr{v}(p)=q$. Concerning the basis associated to the reference frame $\mathcal{R}$, we assume it to be {\it orthonormal}: $\gr{e}_i\cdot\gr{e}_j=\delta_{ij}\ \forall i,j=1,2,3.$ A {\it second-rank tensor} is any linear mapping $\gr{L}:\mathcal{V}\rightarrow\mathcal{V}$. The space of all the tensors forms a manifold, indicated by Lin($\mathcal{V}$). Any tensor can be expressed, in $\mathcal{R}$, as a sum of dyads: $\gr{L}=L_{ij}\gr{e}_i\otimes\gr{e}_j$. A {\it dyad} $\gr{a}\otimes\gr{b}$ is defined as $(\gr{a}\otimes\gr{b})\gr{v}=\gr{b}\cdot\gr{v}\ \gr{a}\ \forall\gr{v}\in\mathcal{V}$.
See ref. \ref{ref:geodiff} for more details on tensors.}
 and let us consider a point $p=(p_1,p_2,p_3)$. If the three coordinates are three continuous functions $p_i(t)$ over the interval $[t_1,t_2]\in\R$, then the mapping $p(t):[t_1,t_2]\rightarrow\mathcal{E}$ is a {\it curve} in $\mathcal{E}$.

\begin{figure}
\begin{center}
\includegraphics[scale=1]{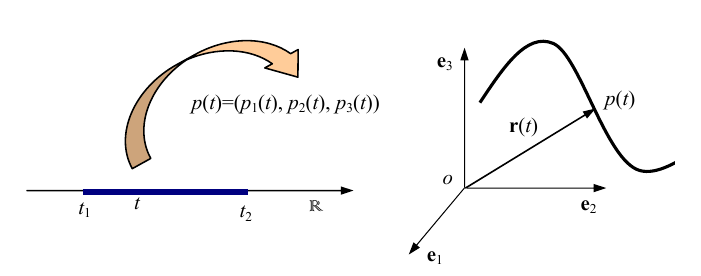}
\caption{Mapping of a curve of points.}
\label{fig:f1_1}
\end{center}
\end{figure}

The independent variable $t$ is the {\it parameter} and the equation
\begin{equation}
p(t)=(p_1(t),p_2(t),p_3(t))\ \rightarrow\ \left\{
\begin{array}{l}
p_1=p_1(t)\\
p_2=p_2(t)\\
p_3=p_3(t)
\end{array}
\right.
\end{equation}
is the {\it parametric point equation of the curve}: to each value of $t\in[t_1,t_2]$ it corresponds a point of the curve in $\mathcal{E}$, see Fig. \ref{fig:f1_1}.
The curve is {\it smooth} whenever the functions $p_i(t)$ are of class C$^1$.

The vector function $\gr{r}(t)=p(t)-o$ is the {\it position vector} of point $p$ in $\mathcal{R}$; the equation
\begin{equation}
\gr{r}(t)=r_1(t)\gr{e}_1+r_2(t)\gr{e}_2+r_3(t)\gr{e}_3\ \rightarrow\ \left\{
\begin{array}{l}
r_1=r_1(t)\\
r_2=r_2(t)\\
r_3=r_3(t)
\end{array}
\right.
\end{equation}
is the {\it parametric vector equation} of the curve: to each value of $t\in[t_1,t_2]$ it corresponds a vector of $\mathcal{V}$ that determines a point of the curve in $\mathcal{E}$ through the operation $p(t)=o+\gr{r}(t)$.

The expression $p=p(t)$ is a {\it curve of points} while $\gr{r}=\gr{r}(t)$ is a {\it curve of vectors}. In the same way, we can introduce a {\it curve of tensors}: if the components $L_{ij}(t)$ are continuous functions of a parameter $t$, the mapping $\gr{L}(t):[t_1,t_2]\rightarrow Lin(\mathcal{V})$ defined as
\begin{equation}
\gr{L}(t)=L_{ij}(t)\gr{e}_i\otimes\gr{e}_j,\ \ i,j=1,2,3,
\end{equation}
is a {\it curve of tensors}.

To be noticed that the choice of the parameter is {\it not unique}: the equation $p=p[\tau(t)]$ still represents the same curve $p=p(t)$, through the {\it change of parameter} $\tau=\tau(t)$.

\section{Differentiation of a curve}
\begin{figure}[h]
\begin{center}
\includegraphics[scale=1]{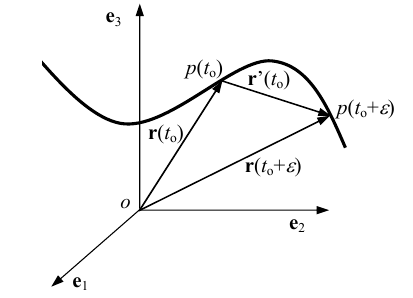}
\caption{Derivative of a curve.}
\label{fig:f1_2}
\end{center}
\end{figure}

We define {\it derivative} of a curve of points $p=p(t)$ in $t=t_0$ the limit
\begin{equation}
\frac{dp(t)}{dt}=\lim_{\eps\rightarrow0}\frac{p(t_0+\eps)-p(t_0)}{\eps};
\end{equation}
being defined as a difference of points, $\dfrac{dp(t)}{dt}$ is a vector, see Fig. \ref{fig:f1_2}

In a similar way we define the derivative of a curve of vectors,
\begin{equation}
\frac{d\gr{r}(t)}{dt}=\lim_{\eps\rightarrow0}\frac{\gr{r}(t_0+\eps)-\gr{r}(t_0)}{\eps};
\end{equation}
and of a curve of tensors
\begin{equation}
\frac{d\gr{L}(t)}{dt}=\lim_{\eps\rightarrow0}\frac{\gr{L}(t_0+\eps)-\gr{L}(t_0)}{\eps};
\end{equation}
being defined as differences of vectors or of tensors, the derivative of a curve of vectors is a vector and that of a curve of tensors is a tensor.

We indicate also $\dfrac{dp}{dt}$ as $p'$ and, if $t$ is the time, as $\dot{p}$; the same will be do also for the derivatives of vector and tensor curves.

The following rules for the differentiation of products or sums of curves can be easily shown using the above definitions:
\begin{equation}
\begin{split}
&(\gr{u}+\gr{v})'=\gr{u}'+\gr{v}',\\
&(\alpha \gr{v})'=\alpha'\gr{v}+\alpha\gr{v}',\ \ \alpha(t):\R\rightarrow\R,\\
&(\gr{u}\cdot\gr{v})'=\gr{u}'\cdot\gr{v}+\gr{u}\cdot\gr{v}',\\
&(\gr{u}\times\gr{v})'=\gr{u}'\times\gr{v}+\gr{u}\times\gr{v}',\\
&[\gr{v}(\alpha(t))]'=\gr{v}'(\alpha(t))\ \alpha'(t),\ \ \alpha(t):\R\rightarrow\R,\\
&(\gr{u}\otimes\gr{v})'=\gr{u}'\otimes\gr{v}+\gr{u}\otimes\gr{v}',\\
&(\gr{Lu})'=\gr{L}'\gr{u}+\gr{L}\gr{u}',\\
&(\gr{LM})'=\gr{L}'\gr{M}+\gr{L}\gr{M}',\\
&(\gr{L}^{-1})'=-\gr{L}^{-1}\gr{L}'\gr{L}^{-1},\\
&(\det\gr{L})'=\det\gr{L}\ \gr{L}'^\top\cdot\gr{L}^{-1}.
\end{split}
\end{equation}

An important case is that of a vector $\gr{v}(t)$ whose norm $v(t)$ is constant $\forall t$:
\begin{equation}
\label{eq:constvect}
(v^2)'=(\gr{v}\cdot\gr{v})'=\gr{v}'\cdot\gr{v}+\gr{v}\cdot\gr{v}'=2\gr{v}'\cdot\gr{v}=0:
\end{equation}
the derivative of such a vector is  orthogonal to it $\forall t$. The contrary is also true, as immediately apparent.

The {\it second derivative} of a curve is simply defined as the derivative of the derivative of a curve. In such a way, derivatives of any order can be defined and calculated.

Finally, using the above rules and assuming that the reference frame $\mathcal{R}$ is independent from $t$, we get easily that
\begin{equation}
\begin{split}
&p'(t)=p'_i(t)\ \gr{e}_i,\\
&\gr{v}'(t)=v'_i(t)\ \gr{e}_i,\\
&\gr{L}'(t)=L'_{ij}(t)\ \gr{e}_i\otimes\gr{e}_j.
\end{split}
\end{equation}

\section{Integral of a curve of vectors, length of a curve}
\begin{figure}[h]
\begin{center}
\includegraphics[scale=.9]{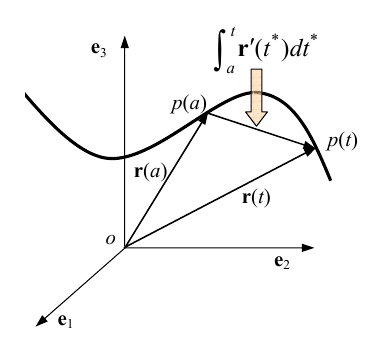}
\caption{Integral of a vector curve.}
\label{fig:f1_3}
\end{center}
\end{figure}

We define {\it integral of a curve of vectors} $\r(t)$ between $a$ and $b \in[t_1,t_2]$ the curve that is obtained integrating each component of the curve:
\be
\begin{split}
&\int_a^b\gr{r}(t)\ dt=\int_a^br_i(t)\ dt\ \gr{e}_i.
\end{split}
\ee
If the curve is regular, we can generalize the second fundamental theorem of the integral calculus
\be
\gr{r}(t)=\gr{r}(a)+\int_a^t\gr{r}'(t^*)\ dt^*.
\ee
Because
\be
\gr{r}(t)=p(t)-o,\ \ \ \gr{r}'(t)=(p(t)-o)'=p'(t),
\ee
we get also
\be
p(t)=p(a)+\int_a^tp'(t^*)\ dt^*.
\ee
The integral of a vector function is the generalization of the vector sum, see Fig. \ref{fig:f1_3}.

Be $\r(t):[a,b]\rightarrow\Eu$ a regular curve, $\sigma$ a partition of the type $a=t_0<t_1<...<t_n=b$, and 
\be
\sigma_{max}={\max_{i=1,...,n}}|t_i-t_{i-1}|.
\ee
The length $\ell_\sigma$ of the polygonal line whose vertices are the points $\r(t_i)$  is hence:
\be
\ell_\sigma=\sum_{i=1}^n{|\r(t_i)-\r(t_{i-1})|}.
\ee
We define {\it length of the curve} $\r(t)$ the (positive) number
\be
\ell:=\sup_{\sigma}\ell_\sigma.
\ee
\begin{teo} Be $\gr{r}(t):[a,b]\Rightarrow\Eu$ a regular curve; then
\be
\ell=\int_a^b|\gr{r}'(t)|dt.
\ee
\begin{proof}
By the fundamental theorem of calculus,
\be
\r(t_i)-\r(t_{i-1})=\int_{t_{i-1}}^{t_i}\r'(t)dt\ \rightarrow\ |\r(t_i)-\r(t_{i-1})|=\left|\int_{t_{i-1}}^{t_i}\r'(t)dt\right|\leq\int_{t_{i-1}}^{t_i}|\r'(t)|dt,
\ee
whence
\be
\label{eq:longueur1}
\ell\leq\int_a^b|\r'(t)|dt.
\ee
Because $\r'(t)$ is continuous on $[a,b]$, $\forall \eps>0\ \exists\delta>0$ such that $|t-\overline{t}|<\delta\Rightarrow|\r'({t})-\r'(\overline{t})|<\eps$. Be $t\in[t_{i-1},t_i]$ and $\sigma_{max}<\delta$, which is always possible by the choice of the partition $\sigma$; by the triangular inequality,
\be
|\r'(t)|\leq|\r'(t)-\r'(t_i)|+|\r'(t_i)|<\eps+|\r'(t_i)|,
\ee
whence
\be
\besp
\int_{t_{i-1}}^{t_i}|\r'(t)|dt&<\int_{t_{i-1}}^{t_i}|\r'(t_i)|dt+\eps(t_i-t_{i-1})=\left|\int_{t_{i-1}}^{t_i}\r'(t_i)dt\right|+\eps(t_i-t_{i-1})\\
&\leq\left|\int_{t_{i-1}}^{t_i}\r'(t)dt\right|+\left|\int_{t_{i-1}}^{t_i}(\r'(t_i)-\r'(t))dt\right|+\eps(t_i-t_{i-1})\\
&\leq|\r(t_i)-\r(t_{i-1})|+2\eps(t_i-t_{i-1}).
\end{split}
\ee
Summing up over all the intervals $[t_{i-1},t_i]$ we get
\be
\int_a^b|\r'(t)|dt\leq\ell_\sigma+2\eps(b-a)\leq\ell+2\eps(b-a),
\ee
and because $\eps$ is arbitrary, 
\be
\int_a^b|\r'(t)|dt\leq\ell,
\ee
which by eq. (\ref{eq:longueur1}) implies the thesis.
\end{proof}
\end{teo}

\begin{teo}
The length of a curve does not depend upon its parameterization.
\begin{proof}
Be $\r(t):[a,b]\rightarrow\Eu$ a regular curve and $t=t(\tau):[c,d]\rightarrow[a,b]$ a change of parameter; then
\be
\ell=\int_a^b|\r'(t)|dt=\int_c^d|\r'(t(\tau))t'(\tau)|d\tau=\int_c^d|\r'(\tau)|d\tau.
\ee
\end{proof}
\end{teo}
A simple way to determine a point $p(t)$ on a curve is to fix a point $p_0$ on the curve and to measure the length $s(t)$ of the arc of curve between $p_0=p(t=0)$ and $p(t)$. This length $s(t)$ is called  {\it curvilinear abscissa}\footnote{The curvilinear abscissa is also called {\it arc-length} or {\it natural parameter}.}:
\be
\label{eq:esse}
s(t)=\int_{0}^t|\gr{r}'(t^*)|dt^*=\int_{0}^t|(p(t^*)-o)'|dt^*.
\ee
From eq. (\ref{eq:esse}) we get
\be
\frac{ds}{dt}=|\gr{r}'(t)|>0,%=\sqrt{\left(\frac{dr_1}{dt}\right)^2+\left(\frac{dr_2}{dt}\right)^2+\left(\frac{dr_3}{dt}\right)^2}>0
\ee
so that $s(t)$ is an increasing function of $t$ and the length of an infinitesimal arc is
\be
ds=\sqrt{dr_1^2+dr_2^2+dr_3^2}.%=\sqrt{dx^2+dy^2+dz^2}.
\ee
For a plane curve $y=f(x)$, we can always put $t=x$, which gives the parametric equation
\be
p(t)=(t,f(t)),
\ee
or in vector form
\be
\gr{r}(t)=t\ \gr{e}_1+f(t)\ \gr{e}_2,
\ee
from which we obtain
\be
\label{eq:ds}
\frac{ds}{dt}=|\gr{r}'(t)|=|p'(t)|=\sqrt{1+f'^2(t)},
\ee
that gives the length of a plane curve between $t=x_0$ and $t=x$ as a function of the abscissa $x$:
\be
s(x)=\int_{x_0}^x\sqrt{1+f'^2(t)}dt.
\ee

\section{The Frenet-Serret basis}
We define the {\it tangent vector} $\btau(t)$ to a regular curve $p=p(t)$ the vector
\begin{equation}
\label{eq:tau}
\btau(t)=\dfrac{p'(t)}{|p'(t)|}.
\end{equation}

By the definition of derivative, this unit vector is always oriented as the increasing values of $t$; the straight line tangent to the curve in $p_0=p(t_0)$ has hence equation
\begin{equation}
q(\bar{t})=p(t_0)+\bar{t}\ \btau(t_0).
\end{equation}

If the curvilinear abscissa $s$ is chosen as parameter for the curve, through the change of parameter $s=s(t)$ we get
\begin{equation}
\btau(t)=\dfrac{p'(t)}{|p'(t)|}=\dfrac{p'[s(t)]}{|p'[s(t)]|}=\frac{1}{s'(t)}\frac{dp(s)}{ds}\frac{ds(t)}{dt}=\frac{dp(s)}{ds}\ \rightarrow \btau(s)=p'(s).
\end{equation}
So, if the parameter of the curve is $s$, the derivative of the curve is  $\btau$, i.e. it is automatically a unit vector. The above equation, in addition, shows that the change of parameter does not change the direction of the tangent, because just a scalar, the derivative of the parameter's change, multiplies the vector. Nevertheless, generally speaking, a change of parameter can change the orientation of the curve.

Because the norm of $\btau$ is constant, its derivative is a vector orthogonal to $\btau$, see eq. (\ref{eq:constvect}). That is why we call {\it principal normal vector} to a curve the unit vector
\begin{equation}
\label{eq:normal}
\bnu(t)=\frac{\btau'(t)}{|\btau'(t)|}.
\end{equation}

$\bnu$ is defined only on the points of the curve where $\btau'\neq\gr{o}$ which implies that $\bnu$ is not defined on the points of a straight line. This simply means that there is not, among the infinite unit normal vectors to a straight line, a normal with special properties, a {\it principal} one, uniquely linked to $\btau$.

Unlike $\btau$, whose orientation changes with the choice of the parameter, $\bnu$ is an {\it intrinsic} local characteristic of the curve: {\it it is not affected by the choice of the parameter}. In fact, by its same definition, $\bnu$ does not depend upon the reference frame; then, because the direction of $\btau$ is also independent upon the parameter's choice,  the only factor that could affect $\bnu$ is the orientation of the curve, that depends upon the parameter. But a change of the orientation affects, in (\ref{eq:normal}), both $\btau$ and the sign of the increment $dt$, so that $\btau'(t)=d\btau/dt$ does not change, neither $\bnu$, which is hence an intrinsic property of the curve.

\begin{figure}[h]
\begin{center}
\includegraphics[scale=1]{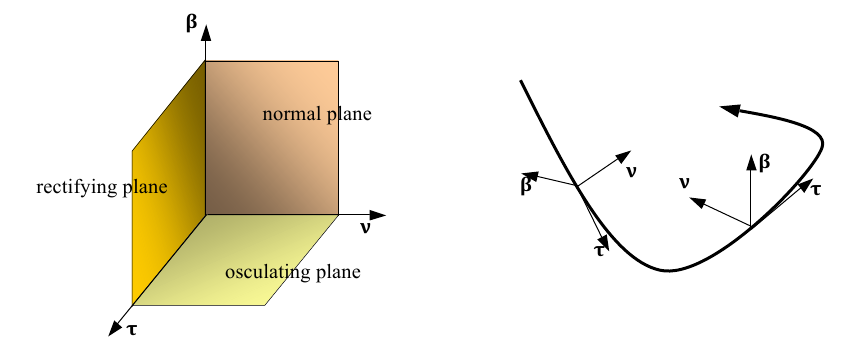}
\caption{The Frenet-Serret basis.}
\label{fig:f1_4}
\end{center}
\end{figure}

The vector 
\begin{equation}
\bb(t)=\btau(t)\times\bnu(t)
\end{equation}
is called the {\it binormal vector}; by construction, it is orthogonal to $\btau$ and $\bnu$ and it is a unit vector. In addition, it is evident that
\begin{equation}
\btau\times\bnu\cdot\bb=1,
\end{equation}
so the set $\{\btau,\bnu,\bb\}$ forms a positively oriented othonormal basis that can be defined at any regular point of a curve with $\btau'\neq\gr{o}$. Such a basis is called the {\it Frenet-Serret local basis}, local in the sense that it changes with the position along the curve. The plane $\btau-\bnu$ is the {\it osculating plane}, the plane $\bnu-\bb$ the {\it normal plane} and the plane $\bb-\btau$ the {\it rectifying plane}, see Fig. \ref{fig:f1_4}.
The osculating  plane is particularly important: if we consider a plane passing through three not aligned points of the curve, when these points become closer and closer, still remaining on the curve, the plane tends to the osculating plane: the osculating plane at a point of a curve is the plane that better approaches the curve near the point.
A plane curve is entirely contained in the osculating plane, which is fixed. 

The principal normal $\bnu$ is always oriented towards the part  of the space, with respect to the rectifying plane, where the curve is; in particular, for a plane curve, $\bnu$ is always directed towards the concavity of the curve.
To show it, it is sufficient to prove that the vector $p(t+\eps)-p(t)$ forms with $\bnu$ an angle $\psi\leq\pi/2$, i.e. that $(p(t+\eps)-p(t))\cdot\bnu\geq0$. In fact,
\begin{equation}
\begin{split}
&p(t+\eps)-p(t)=\eps\ p'(t)+\frac{1}{2}\eps^2p''(t)+o(\eps^2)\ \rightarrow\\
&(p(t+\eps)-p(t))\cdot\bnu=\frac{1}{2}\eps^2p''(t)\cdot\bnu+o(\eps^2),
\end{split}
\end{equation}
but
\begin{equation}
p''(t)\cdot\bnu=(\btau'|p'|+\btau|p'|')\cdot\bnu=(|\btau'||p'|\bnu+\btau|p'|')\cdot\bnu=|\btau'||p'|,
\end{equation}
so that, to within infinitesimal quantities of order $o(\eps^2)$, we obtain
\begin{equation}
(p(t+\eps)-p(t))\cdot\bnu=\frac{1}{2}\eps^2|\btau'||p'|\geq0.
\end{equation}

\section{Curvature of a curve}
\begin{figure}[h]
\begin{center}
\includegraphics[scale=1]{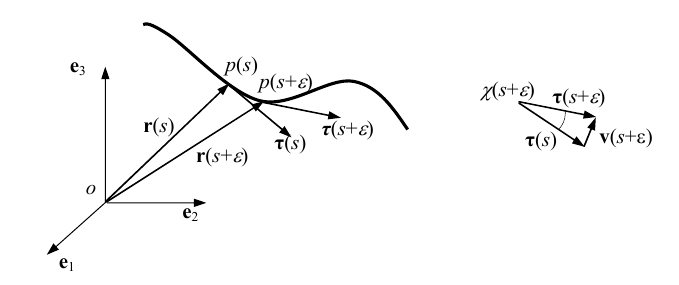}
\caption{Curvature of a curve.}
\label{fig:f1_5}
\end{center}
\end{figure}

It is important, in several situations, to evaluate how much a curve moves away from a straight line, in the neighborhood of a point. To do that, we calculate the angle formed by the tangents at two close points, determined by the curvilinear abscissae $s$ and $s+\eps$, and we measure the angle $\chi(s,\eps)$ that they form, see Fig. \ref{fig:f1_5}.

We then define {\it curvature} of the curve in $p=p(s)$ the limit
\begin{equation}
c(s)=\lim_{\eps\rightarrow0}\left|\frac{\chi(s,\eps)}{\eps}\right|.
\end{equation}
The curvature is hence a non-negative scalar that measures the rapidity of variation of the direction of the curve per unit length of the curve (that is why $c(s)$ is defined as a function of the curvilinear abscissa); by its same definition, the curvature is an {\it intrinsic property} of the curve, i.e. independent from the parameter's choice. For a straight line, the curvature is identically null everywhere.

The curvature is linked to the second derivative of the curve:
\begin{equation}
\begin{split}
c(s)=&\lim_{\eps\rightarrow0}\left|\frac{\chi(s,\eps)}{\eps}\right|=\lim_{\eps\rightarrow0}\left|\frac{\sin\chi(s,\eps)}{\eps}\right|=\lim_{\eps\rightarrow0}\left|\frac{2}{\eps}\sin\frac{\chi(s,\eps)}{2}\right|=\\
&\lim_{\eps\rightarrow0}\left|\frac{\gr{v}(s,\eps)}{\eps}\right|=\lim_{\eps\rightarrow0}\left|\frac{\btau(s+\eps)-\btau(s)}{\eps}\right|=|\btau'(s)|=|p''(s)|.
\end{split}
\end{equation}

Another formula for the calculation of $c(s)$ can be obtained if we consider that
\begin{equation}
\frac{d\btau[s(t)]}{dt}=\frac{d\btau}{ds}\frac{ds}{dt}=\frac{d\btau}{ds}|p'(t)|\ \rightarrow\ \frac{d\btau}{ds}=\frac{1}{|p'(t)|}\frac{d\btau}{dt},
\end{equation}
so that
\begin{equation}
\label{eq:curv2}
c(s)=|\btau'(s)|=\frac{1}{|p'(t)|}\left|\frac{d\btau}{dt}\right|=\frac{|\btau'(t)|}{|p'(t)|}.
\end{equation}

A better formula can be obtained as follows:
\begin{equation}
\begin{split}
\frac{d\btau}{ds}=&\frac{1}{|p'(t)|}\frac{d\btau}{dt}=\frac{1}{|p'(t)|}\frac{d}{dt}\frac{p'(t)}{|p'(t)|}=
\frac{1}{|p'|}\frac{p''|p'|-p'\dfrac{p''\cdot p'}{|p'|}}{|p'|^2}=\\
&\frac{p''-\btau\ p''\cdot\btau}{|p'|^2}=(\gr{I}-\btau\otimes\btau)\frac{p''}{|p'|^2}.
\end{split}
\end{equation}
By consequence,
\begin{equation}
c(s)=\left|\frac{d\btau(s)}{ds}\right|=\frac{1}{|p'|^2}|(\gr{I}-\btau\otimes\btau)p''|.
\end{equation}
Now, we use the following general formula expressing a skew tensor $\gr{W}$:
\begin{equation}
\gr{WW}=-\frac{1}{2}|\gr{W}|^2(\gr{I}-\gr{w}\otimes\gr{w});
\end{equation}
if we use this formula for $\btau$, so that $\gr{W}$ is the axial tensor of $\btau$, we get
\begin{equation}
\gr{I}-\btau\otimes\btau=-2\frac{\gr{WW}}{|\gr{W}|^2}=-\gr{WW},
\end{equation}
because if $\btau=(\tau_1,\tau_2,\tau_3)$, then
\begin{equation}
\begin{split}
|\gr{W}|^2=&\gr{W}\cdot\gr{W}=\left[\begin{array}{ccc}0 & -\tau_3 & \tau_2 \\\tau_3& 0 & -\tau_1 \\-\tau_2 & \tau_1 & 0\end{array}\right]\cdot\left[\begin{array}{ccc}0 & -\tau_3 & \tau_2 \\\tau_3& 0 & -\tau_1 \\-\tau_2 & \tau_1 & 0\end{array}\right]=\\
&2(\tau_1^2+\tau_2^2+\tau_3^2)=2.
\end{split}
\end{equation}
So, recalling that for any skew tensor $\gr{W}$,
\begin{equation}
\gr{W}\ \gr{u}=\gr{w}\times\gr{u}\ \ \forall \gr{u}\in\mathcal{V},
\end{equation}
with $\gr{w}$ the axial vector of $\gr{W}$, we get
\begin{equation}
\begin{split}
|(\gr{I}-\btau\otimes\btau)p''|=&|-\gr{WW}p''|=|-\gr{W}(\btau\times p'')|=|-\btau\times(\btau\times p'')|=\\
&|\btau\times(\btau\times p'')|=|\btau\times p''|=\frac{|p'\times p''|}{|p'|},
\end{split}
\end{equation}
so that finally
\begin{equation}
\label{eq:curv1}
c=\frac{|p'\times p''|}{|p'|^3}.
\end{equation}

Applying this last formula to a plane curve $p(t)=(x(t),y(t))$, we get
\begin{equation}
c=\frac{|x'y''-x''y'|}{(x'^2+y'^2)^\frac{3}{2}}
\end{equation}
and if the curve is given in the form $y=y(x)$, so that the parameter $t=x$, then we obtain
\begin{equation}
c=\frac{|y''|}{(1+y'^2)^\frac{3}{2}}.
\end{equation}

This last formula shows that if $|y'|\ll1$, like in the infinitesimal theory of strain, then
\begin{equation}
c\simeq|y''|.
\end{equation}

\section{The Frenet-Serret formulae}
From eqs. (\ref{eq:normal}) for $t=s$ and (\ref{eq:curv2}) we get
\begin{equation}
\label{eq:fs1}
\frac{d\btau}{ds}=c\ \bnu
\end{equation}
which is the {\it first Frenet-Serret Formula}, giving the variation of $\btau$ per unit length of the curve. Such a variation is a vector whose norm is the curvature and that has as direction that of $\bnu$.

Let us now consider the variation of $\bb$ per unit length of the curve; because $\bb$ is a unit vector, we have
\begin{equation}
\frac{d\bb}{ds}\cdot\bb=0,
\end{equation}
and 
\begin{equation}
\bb\cdot\btau=0\ \Rightarrow\ \frac{d(\bb\cdot\btau)}{ds}=\frac{d\bb}{ds}\cdot\btau+\bb\cdot\frac{d\btau}{ds}=0.
\end{equation}
Through eq. (\ref{eq:fs1}) and because $\bb\cdot\bnu=0$ we get
\begin{equation}
\frac{d\bb}{ds}\cdot\btau=-c\ \bb\cdot\bnu=0,
\end{equation}
so that $\dfrac{d\bb}{ds}$ is necessarily parallel to $\bnu$. We then put
\begin{equation}
\label{eq:fs2}
\frac{d\bb}{ds}=\vartheta\bnu,
\end{equation}
which is the {\it second Frenet-Serret formula}. The scalar $\vartheta(s)$ is called the {\it torsion of the curve} in $p=p(s)$. So, we see that the variation of $\bb$ per unit length is a vector parallel to $\bnu$ and proportional to the torsion of the curve.

We can now find the variation of $\bnu$ per unit length of the curve:
\begin{equation}
\frac{d\bnu}{ds}=\frac{d(\bb\times\btau)}{ds}=\frac{d\bb}{ds}\times\btau+\bb\times\frac{d\btau}{ds}=\vartheta\ \bnu\times\btau+c\ \bb\times\bnu,
\end{equation}
so finally
\begin{equation}
\label{eq:fs3}
\frac{d\bnu}{ds}=-c\ \btau-\vartheta\ \bb,
\end{equation}
which is the {\it third Frenet-Serret formula}: the variation of $\bnu$ per unit length of the curve is a vector of the rectifying plane.

The three formulae of Frenet-Serret (discovered independently by J. F. Frenet in 1847 and by J. A. Serret in 1851) can be condensed in the symbolic matrix product
\begin{equation}
\left\{\begin{array}{c}\btau' \\\bnu' \\\bb'\end{array}\right\}=
\left[\begin{array}{ccc}0 & c & 0 \\-c & 0 & -\vartheta \\0 & \vartheta & 0\end{array}\right]
\left\{\begin{array}{c}\btau \\\bnu \\\bb\end{array}\right\}.
\end{equation}

\section{The torsion of a curve}
We have introduced the torsion of a curve in the previous section, with the second formula of Frenet-Serret.
The torsion measures the deviation of a curve from flatness: if a curve is planar, it belongs to the osculating plane and $\bb$, which is perpendicular to the osculating pane, is hence a constant vector. So, its derivative is  null and by the Frenet-Serret second formula $\vartheta=0$.

Conversely, if $\vartheta=0$ everywhere, $\bb$ is a constant vector and hence the osculating plane does not change and the curve is planar. So we have that {\it a curve is planar if and only if the torsion is null $\forall p(s)$}.

Using the Frenet-Serret formulae in the expression of $p'''(s)$ we get a formula for the torsion:
\begin{equation}
\begin{split}
&p'(t)=|p'|\btau=\frac{dp}{ds}\frac{ds}{dt}=s'\btau\ \Rightarrow\ |p'|=s'\ \rightarrow\\
&p''(t)=s''\btau+s'\btau'=s''\btau+s'^2\frac{d\btau}{ds}=s''\btau+c\ s'^2\bnu\ \rightarrow\\
&p'''(t)=s'''\btau+s''\btau'+(c\ s'^2)'\bnu+c\ s'^2\bnu'=\\
&\hspace{15mm}s'''\btau+s''s'\frac{d\btau}{ds}+(c\ s'^2)'\bnu+c\ s'^3\frac{d\bnu}{ds}=\\
&\hspace{15mm}s'''\btau+s''s'c\bnu+(c\ s'^2)'\bnu-c\ s'^3(c\btau+\vartheta\bb)=\\
&\hspace{15mm}(s'''-c^2s'^3)\btau+(s''s'c+c's'^2+2c\ s's'')\bnu-c\ s'^3\vartheta\bb,
\end{split}
\end{equation}
so that, through eq. (\ref{eq:curv1}), we get
\begin{equation}
\begin{split}
p'\times p''\cdot p'''=&s'\btau\times(s''\btau+c\ s'^2\bnu)\cdot[(s'''-c^2s'^3)\btau+\\
&(s''s'c+c's'^2+2c\ s's'')\bnu-c\ s'^3\vartheta\bb]=\\
&-c^2s'^6\vartheta=-c^2|p'|^6\vartheta=-\frac{|p'\times p''|^2}{|p'|^6}|p'|^6\vartheta,
\end{split}
\end{equation}
so that, finally, 
\begin{equation}
\vartheta=-\frac{p'\times p''\cdot p'''}{|p'\times p''|^2}.
\end{equation}
To remark that while the curvature is linked to the second derivative of the curve, the torsion is a function also of the third derivative.

Unlike curvature, which is intrinsically positive, the torsion can be negative. %In particular, see Fig. \ref{fig:f1_6}, if once fixed the curvilinear abscissa $s$ and proceeding along the curve according to the increasing values of $s$, the curve crosses the osculating plane on the side of $\bb$, then $\vartheta<0$, while it is positive in the opposite case. 
In fact, still using the Frenet-Serret formulae,
\begin{equation}
\begin{split}
p(s+\eps)-p(s)=&\eps\ p'+\frac{1}{2}\eps^2p''+\frac{1}{6}\eps^3p'''+o(\eps^3)=\\
&\eps\btau+\frac{1}{2}\eps^2c\bnu+\frac{1}{6}\eps^3(c\bnu)'+o(\eps^3)=\\
&\eps\btau+\frac{1}{2}\eps^2c\bnu+\frac{1}{6}\eps^3(c'\bnu-c^2\btau-c\ \vartheta\bb)+o(\eps^3)\ \rightarrow\\
&(p(s+\eps)-p(s))\cdot\bb=-\frac{1}{6}\eps^3c\ \vartheta+o(\eps^3).
\end{split}
\end{equation}

\begin{figure}[h]
\begin{center}
\includegraphics[scale=1]{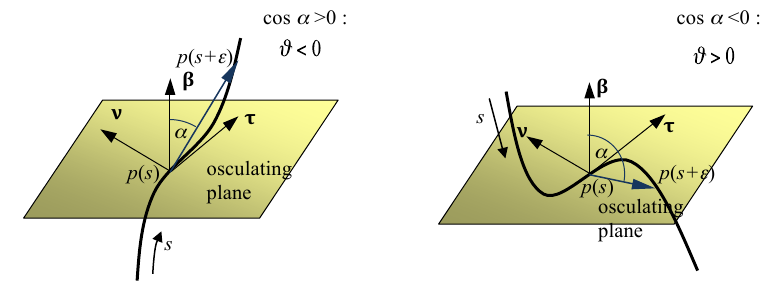}
\caption{Torsion of a curve.}
\label{fig:f1_6}
\end{center}
\end{figure}

The above dot product determines if the point $p(s+\eps)$ is located, with respect to the osculating plane, on the side of $\bb$ or on the opposite one, see Fig. \ref{fig:f1_6}: if following the curve for increasing values of $s,\ \eps>0$, the point passes into the semi-space of $\bb$ from the opposite one, because $1/6\ c\ \eps^3>0$, it will be $\vartheta<0$, while in the opposite case it will be $\vartheta>0$.

This result is intrinsic, i.e. it does not depend upon the choice of the parameter, hence of the positive orientation of the curve; in fact, $\bnu$ is intrinsic, but changing the orientation of the curve, $\btau$, and hence $\bb$, change of orientation.

 \section{Osculating sphere and circle}
 The {\it osculating sphere}\footnote{The word osculating comes from the latin word {\it osculo} that means to kiss; an osculating sphere or circle or plane is a geometric object that is very close to the curve, as close as two  lovers are in a kiss.} to a curve at a point $p$ is a sphere to which the curve tends to adhere in the neighborhood of $p$. Mathematically speaking, if $q_s$ is the center of the sphere relative to point $p(s)$, then
 \begin{equation}
 |p(s+\eps)-q_s|^2= |p(s)-q_s|^2+o(\eps^3).
 \end{equation}
 Using this definition, discarding the terms of order $o(\eps^3)$ and using the Frenet-Serret formulae, we get:
 \begin{equation}
 \begin{split}
  |p(s+\eps)-q_s|^2=&|p(s)-q_s+\eps p'+\frac{1}{2}\eps^2p''+\frac{1}{6}\eps^3p'''+o(\eps^3)|^2=\\
  &|p(s)-q_s+\eps \btau+\frac{1}{2}\eps^2c\ \bnu+\frac{1}{6}\eps^3(c\bnu)'+o(\eps^3)|^2=\\
  &|p(s)-q_s|^2+2\eps(p(s)-q_s)\cdot\btau+\eps^2+\eps^2c(p(s)-q_s)\cdot\bnu+\\
  &\frac{1}{3}\eps^3(p(s)-q_s)\cdot(c'\bnu-c^2\btau-c\ \vartheta\bb)+o(\eps^3),
 \end{split}
 \end{equation}
 which gives
 \begin{equation}
 \begin{split}
 &(p(s)-q_s)\cdot\btau=0,\\
  &(p(s)-q_s)\cdot\bnu=-\frac{1}{c}=-\rho,\\
   &(p(s)-q_s)\cdot\bb=-\frac{c'}{c^2\vartheta}=\frac{\rho'}{\vartheta},
 \end{split}
 \end{equation}
 and finally
 \begin{equation}
 \label{eq:spherecenter}
 q_s=p+\rho\ \bnu-\frac{\rho'}{\vartheta}\bb,
 \end{equation}
so the center of the sphere belongs to the normal plane; the sphere is not defined for a plane curve. $\rho$ is the {\it radius of curvature} of the curve, defined as
\begin{equation}
\rho=\frac{1}{c}.
\end{equation}

The radius of the osculating sphere is
\begin{equation}
\rho_s=|p-q_s|=\sqrt{\rho^2+\left(\frac{\rho'}{\vartheta}\right)^2}.
\end{equation}

\begin{figure}[h]
\begin{center}
\includegraphics[scale=.5]{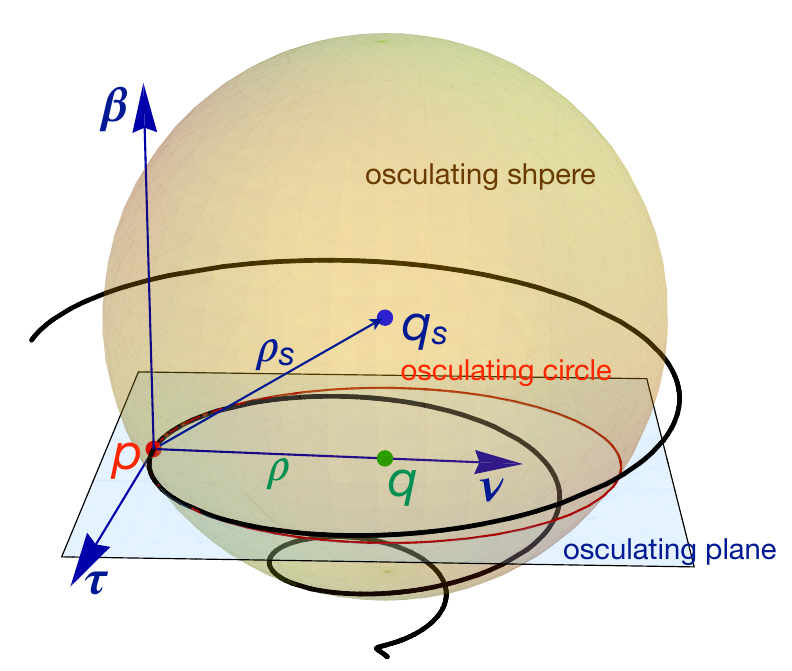}
\caption{Osculating plane, circle and sphere for a point $p$ of a conical helix.}
\label{fig:f1_7}
\end{center}
\end{figure}

The intersection between the osculating sphere and the osculating plane at a same point $p$ is the {\it osculating circle}. This circle has the property to share the same tangent in $p$ with the curve and its radius is the radius of curvature, $\rho$. From eq. (\ref{eq:spherecenter}) we get the position of the osculating circle center $q$:
\begin{equation}
q=p+\rho\ \bnu.
\end{equation}
An example can be seen in Fig. \ref{fig:f1_7}, where the osculating plane, circle and sphere are shown for a point $p$ of a conical helix. 

The osculating circle is a diametral circle of the osculating sphere only when $q=q_s$, so if and only if
\begin{equation}
\frac{\rho'}{\vartheta}=-\frac{c'}{c^2\vartheta}=0,
\end{equation}
i.e. when the curvature is constant.

\section{Surfaces in $\Eu$, coordinate lines, tangent planes}
\label{sec:surf1}
\begin{figure}[ht]
	\begin{center}
         \includegraphics[scale=.7]{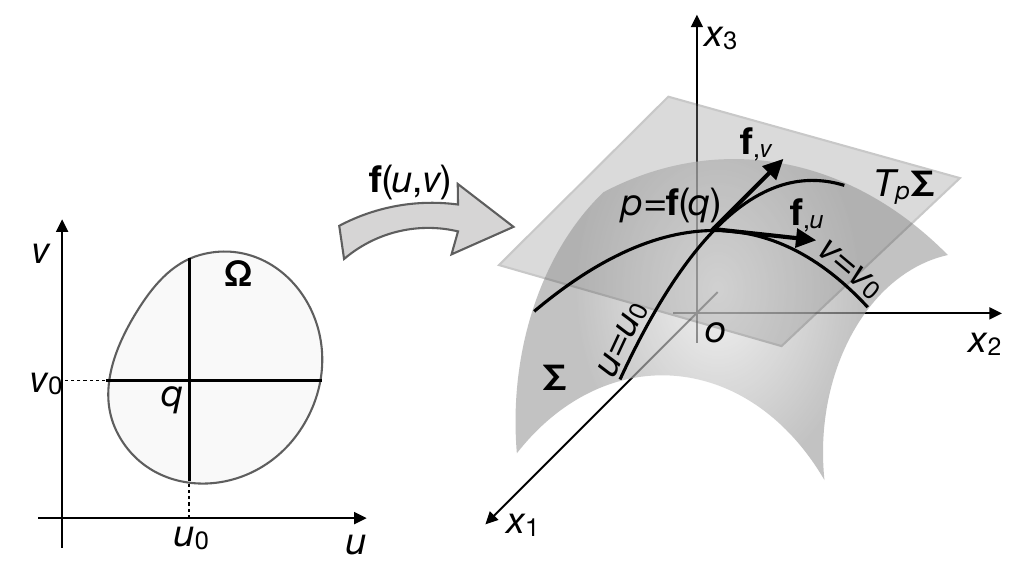}
	\caption{General scheme of a surface  and of the tangent space at a point $p$.}
	\label{fig:28}
	\end{center}
\end{figure}

A function $\f(u,v):\Omega\subset\R^2\rightarrow\Eu$ of class $\geq$C$^1$ and such that  its Jacobian 
\be
J=\left[\begin{array}{cc}
\dfrac{\partial f_1}{\partial u}&\dfrac{\partial f_1}{\partial v}\medskip\\
\dfrac{\partial f_2}{\partial u}&\dfrac{\partial f_2}{\partial v}\medskip\\
\dfrac{\partial f_3}{\partial u}&\dfrac{\partial f_3}{\partial v}
\end{array}
\right]
\ee
has maximum rank (rank[J]=2) defines a {\it surface in $\Eu$}, see Fig. \ref{fig:28}. We say also that $\f$ is an {\it immersion} of $\Omega$ into $\Eu$ and that  the subset  $\Sigma\subset\Eu$ image of $\f$ is the {\it support} or {\it trace} of the surface $\f$. 

 We will also indicate derivatives with respect to the variables $u,v$ by, e.g., $\dfrac{\partial \f}{\partial u}=\f_{,u}$ etc. The condition on the rank of $J$ is equivalent to impose that
 \be
 \label{eq:regulsurf}
 \f_{,u}(u,v)\times\f_{,v}(u,v)\neq\bo\ \ \forall (u,v)\in\Omega.
 \ee
 The {\it normal to the surface $\f$} is the vector $\N\in\S$ defined as
 \be
 \label{eq:normsurfreg}
 \N:=\frac{\f_{,u}\times\f_{,v}}{|{\f_{,u}\times\f_{,v}}|}.
 \ee
 A {\it regular point of $\Sigma$} is a point where $\N$ is defined; if $\N$ is defined $\forall p\in\Sigma$ then the surface is said to be {\it regular}.
 
 A function $\bg(t):G\subset\R\rightarrow\Omega$ whose parametric equation is $\bg(t)=(u(t),v(t))$ describes a curve in $\Omega$ whose image, through $\f$, is a curve 
 \be
 \widehat{\bg}(t)=\f(u(t),v(t)):G\subset\R\rightarrow\Sigma\subset\Eu.
 \ee
 As a special case of curve in $\Omega$, let us consider the curves of the type $v=v_0$ or $u=u_0$, with $u_0,v_0$ some constants. Then, their image through $\f$ are two curves $\f(u,v_0),\f(u_0,v)$ on $\Sigma$ called {\it coordinate lines}, see Fig. \ref{fig:28}. The {\it tangent vectors to the coordinate lines} are respectively the vectors $\f_{,u}(u,v_0)$ and $\f_{,v}(u_0,v)$, while the tangent to a curve $\widehat{\bg}(t)=\f(u(t),v(t))$ is the vector 
 \be
 \label{eq:tangonsigma}
\widehat{\bg}'(t)=\f_{,u}\frac{du}{dt}+\f_{,v}\frac{dv}{dt},
 \ee
 i.e. the tangent vector to any curve on $\Sigma$ is a linear combination of the tangent vectors to the coordinate lines. To remark that the tangent vectors $\f_{,u}(u,v_0)$ and $\f_{,v}(u_0,v)$ are necessarily non null and  linear independent as consequence of the assumption on the rank of $J$, and hence of the existence of $\N$, i.e. of the regularity of $\Sigma$. They determine a plane that contains the tangents to all the curves on $\Sigma$ passing by $p=\f(u_0,v_0)$ and form a basis on this plane, called the {\it natural basis}. Such a plane is the {\it tangent plane to $\Sigma$ in $p$} and is indicated by $T_p\Sigma$; this plane is actually the space spanned by $\f_{,u}(u,v_0)$ and $\f_{,v}(u_0,v)$ and  is also called the {\it tangent vector space}.
 
 Let us consider two open subsets $\Omega_1,\Omega_2\subset\R^2$; a {\it diffeomorphism\footnote{The definition of diffeomorphism, of course, can be given for subsets of $\R^n,n\geq1$; here, we bound the definition to the case of interest.} of class} C$^k$ between $\Omega_1$ and $\Omega_2$ is a bijective map $\vartheta:\Omega_1\rightarrow\Omega_2$ of class C$^k$ with also its inverse of class C$^k$; the diffeomorphism is {\it smooth} if $k=\infty$.
 
 Let $\Omega_1,\Omega_2$  be two open subsets of $\R^2$, $\f:\Omega_2\rightarrow\Eu$ a surface and  $\vartheta:\Omega_1\rightarrow\Omega_2$ a smooth diffeomorphism. Then the surface $\F=\f\circ\vartheta:\Omega_1\rightarrow\Eu$ is a {\it change of parameterization} for $\f$. In practice, the function defining the surface changes, but not $\Sigma$, its trace in $\Eu$. Let $(U,V)$ be the coordinates in $\Omega_1$ and $(u,v)$ in $\Omega_2$; then, by the chain rule,
 \be
 \besp
& \F_{,U}=\f_{,u}\frac{\partial u}{\partial U}+\f_{,v}\frac{\partial v}{\partial U},\\
&  \F_{,V}=\f_{,u}\frac{\partial u}{\partial V}+\f_{,v}\frac{\partial v}{\partial V},
  \end{split}
 \ee
 or, denoting by $J_\vartheta$ the Jacobian of $\vartheta$, 
 \be
 \left\{\begin{array}{c}\F_{,U}\\\F_{,V}\end{array}\right\}=\left[J_\vartheta\right]^\top\left\{\begin{array}{c}\f_{,u}\\\f_{,v}\end{array}\right\},
 \ee
 whence
 \be
 \F_{,U}\times\F_{,V}=\det [J_\vartheta]\ \f_{,u}\times\f_{,v}.
 \ee
 This result shows that the regularity of the surface, condition (\ref{eq:regulsurf}), the tangent plane and the tangent space vector {\it do not depend upon the parameterization of $\Sigma$}. 
 From the last equation, we get also
 \be
 \N(U,V)=\mathrm{sgn}(\det [J_\vartheta])\ \N(u,v);
 \ee
 we say that the change of parameterization {\it preserves the orientation if $\det [J_\vartheta]>0$}, and that it {\it inverses the parameterization} in the opposite case.

 \section{Surfaces of revolution}
 A {\it surface of revolution} is a surface whose trace  is obtained letting rotate a plane curve, say $\bg$, around an axis, say $x_3$. To be more specific, and without loss of generality, let $\bg:G\subset\R\rightarrow\R^2$ be a regular curve of the plane $x_2=0$, whose parametric equation is
 \be
 \label{eq:curvesurfrevol}
 \bg(u):\ \left\{\begin{array}{l}x_1=\phi(u),\\x_3=\psi(u),\end{array}\right.\ \ \ \phi(u)>0\ \forall u\in G.
 \ee
 Then, the subset $\Sigma_\gamma\subset\Eu$ defined by
 \be
 \Sigma_\gamma:=\left\{(x_1,x_2,x_3)\in\Eu|x_1^2+x_2^2=\phi^2(u),x_3=\psi(u),u\in G\right\}
 \ee
 is  the trace of a surface of revolution of the curve $\bg(u)$ around the axis $x_3$. A general parameterization of such a surface is 
 \be
 \label{eq:revolsurf}
 \f(u,v):G\times(-\pi,\pi]\rightarrow\Eu|\ \ \ 
 \left\{
 \begin{array}{l}
 x_1=\phi(u)\cos v,\\
 x_2=\phi(u)\sin v,\\
 x_3=\psi(u). 
 \end{array}
 \right.
 \ee
 It is readily checked that this parameterization actually defines a regular surface:
 \be
 \f_{,u}=\left\{\begin{array}{c}\phi'(u)\cos v\\\phi'(u)\sin v\\\psi'(u)\end{array}\right\},\ \ \f_{,v}=\left\{\begin{array}{c}-\phi(u)\sin v\\\phi(u)\cos v\\0\end{array}\right\}\ \rightarrow\ \f_{,u}\times\f_{,v}=\left\{\begin{array}{c}-\phi(u)\psi'(u)\cos v\\-\phi(u)\psi'(u)\sin v\\\phi(u)\phi'(u)\end{array}\right\}
 \ee
 so that 
 \be
 |\f_{,u}\times\f_{,v}|=\phi^2(u)(\phi'^2(u)+\psi'^2(u))\neq0\ \forall u\in G
 \ee
 for being $\bg(u)$ a regular curve, i.e. with $\bg'(u)\neq\bo\ \forall u\in G$.
 A {\it meridian} is a curve in $\Eu$ intersection of the trace of $\f$, $\Sigma_\gamma$, with a plane containing the axis $x_3$; the equation of a meridian is obtained fixing the value of $v$, say $v=v_0$:
 \be
 \left\{\begin{array}{l}x_1=\phi(u)\cos v_0,\\x_2=\phi(u)\sin v_0,\\x_3=\psi(u).\end{array}\right.
 \ee
 A {\it parallel} is a curve in $\Eu$  intersection  of $\Sigma_\gamma$ with a plane orthogonal to $x_3$; the equation of a parallel, which is a circle with center on the axis $x_3$, is obtained fixing the value of $u$, say $u=u_0$:
 \be
 \left\{\begin{array}{l}x_1=\phi(u_0)\cos v,\\x_2=\phi(u_0)\sin v,\\x_3=\psi(u_0),\end{array}\right.
 \ee
 or also
 \be
  \left\{\begin{array}{l}x_1^2+x_2^2=\phi(u_0)^2,\\x_3=\psi(u_0);\end{array}\right.
 \ee
 the radius of the circle is $\phi(u_0)$.
 
 A {\it loxodrome} or {\it rhumb line} is a curve on $\Sigma_\gamma$ crossing all the meridians  at the same angle. 
 
 Some important examples of surfaces of revolution are:
 \begin{itemize}
 \item the {\it sphere}: 
 \be
 \f(u,v):\left[-\dfrac{\pi}{2},\dfrac{\pi}{2}\right]\times(-\pi,\pi]\rightarrow\Eu|\ \ \left\{\begin{array}{l}x_1=\cos u\cos v,\\x_2=\cos u\sin v,\\x_3=\sin v;\end{array}\right.
 \ee
 \item the {\it catenoid}:
 \be
 \f(u,v):\left[-a,a\right]\times(-\pi,\pi]\rightarrow\Eu|\ \ \left\{\begin{array}{l}x_1=\cosh u\cos v,\\x_2=\cosh u\sin v,\\x_3=u;\end{array}\right.
 \ee
 \item the {\it pseudo-sphere}:
  \be
  \label{eq:pseudosphere}
 \f(u,v):\left[0,a\right]\times(-\pi,\pi]\rightarrow\Eu|\ \ \left\{\begin{array}{l}x_1=\sin u\cos v,\\x_2=\sin u\sin v,\\x_3=\cos u+\ln\left(\tan\dfrac{u}{2}\right);\end{array}\right.
 \ee
 \item the {\it hyperbolic hyperboloid}:
  \be
 \f(u,v):\left[-a,a\right]\times(-\pi,\pi]\rightarrow\Eu|\ \ \left\{\begin{array}{l}x_1=\cos u-v\sin u,\\x_2=\sin u+v\cos u,\\x_3= v.\end{array}\right.
 \ee
 \end{itemize}
 \begin{figure}[ht]
	\begin{center}
         \includegraphics[height=.13\textheight]{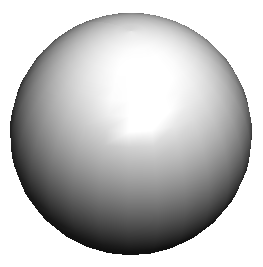}
         \includegraphics[height=.13\textheight]{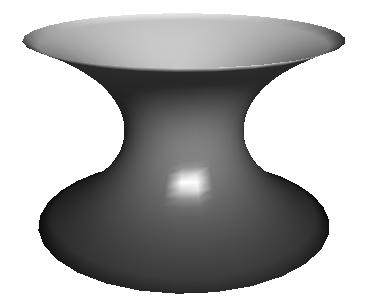}
         \includegraphics[height=.13\textheight]{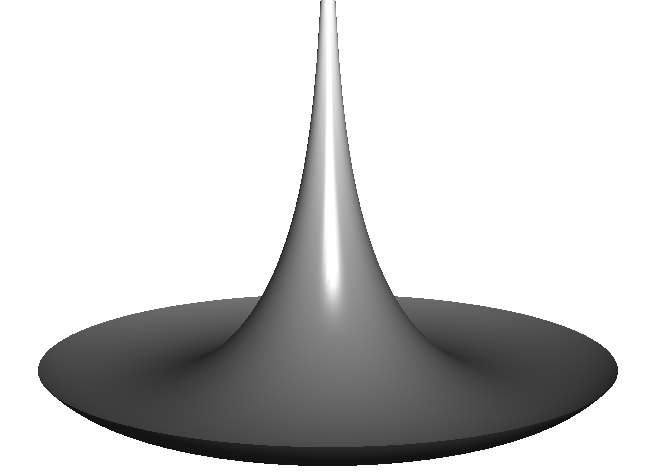}
         \includegraphics[height=.13\textheight]{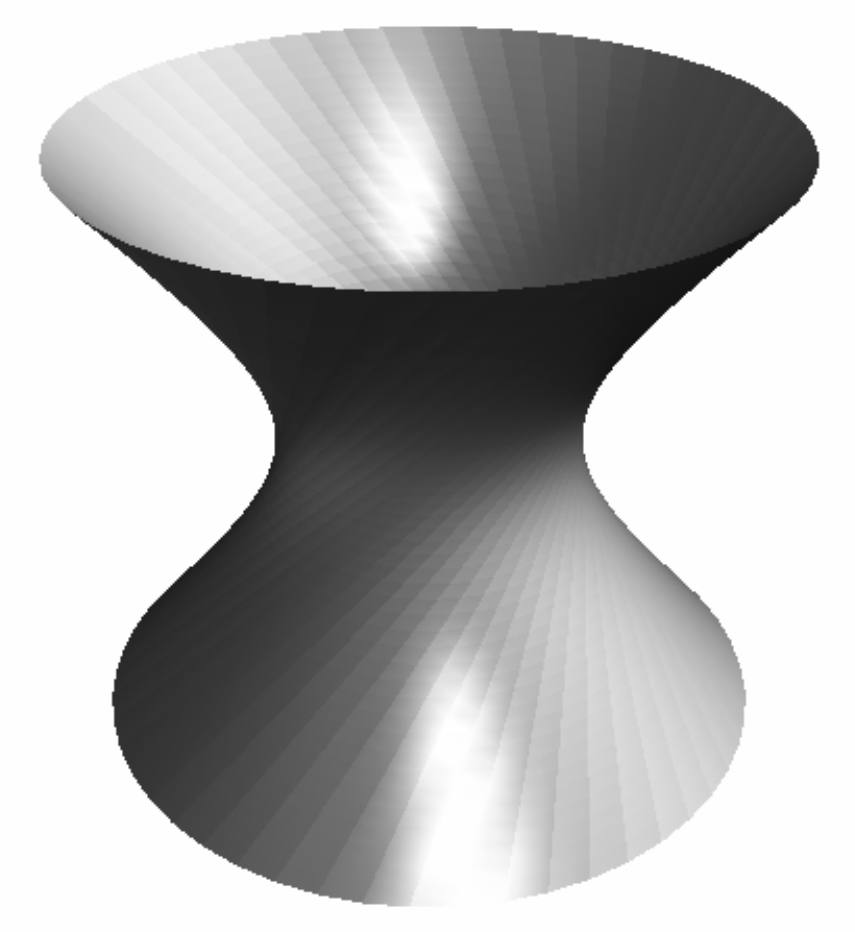}
	\caption{Surfaces of revolution. From the left: sphere, catenoid, pseudo-sphere, hyperbolic hyperboloid.}
	\label{fig:29}
	\end{center}
\end{figure}
  
  \section{Ruled surfaces}
 A {\it ruled surface} (also named a {\it scroll}) is a surface with the property that through every one of its points there is a straight line that lies on the  surface. 
 A ruled surface can be seen as the set of points swept by a moving straight line.  We say that a surface is {\it doubly ruled} if through every one of its points there are two distinct straight lines that lie on the  surface.
 
 Any ruled surface can be represented by a parameterization of the form
 \be
 \label{eq:ruledsurf}
 \f(u,v)=\bg(u)+v\bl(u),
 \ee
 where $\bg(u)$ is a regular smooth curve, the {\it directrix},  and $\bl(u)$ is a smooth curve. Fixing $u=u_0$ gives a {\it generator line} $\f(u_0,v)$ of the surface; the vectors $\bl(u)\neq\bo$ describe the directions of the generators. Some important examples of ruled surfaces are:
 \begin{itemize}
 \item {\it cones}: for these surfaces, all the straight lines pass through a point, the {\it apex} of the cone, choosing the apex as the origin, then it must be $\bl(u)=k\bg(u),\ k\in\R\rightarrow$
 \be
 \f(u,v)=v\bg(u);
 \ee 
\item {\it cylinders}:  a ruled surface is a cylinder $\iff\bl(u)=const$. In this case, it is always possible to choose $\bl(u)\in\S$ and  $\bg(u)$  a planar curve lying in a plane orthogonal to $\bl(u)$ (it is sufficient to choose the curve $\bg^*(u)=(\I-\bl(u)\otimes\bl(u))\bg(u)$);
 \item {\it helicoids}:  a surface  generated by rotating and simultaneously displacing a curve, the {\it profile curve}, along an axis is a helicoid. Any point of the profile curve is the starting point of a circular helix. Generally speaking, we get a helicoid if 
 \be
 \bg(u)=(0,0,\phi(u)),\ \ \bl(u)=(\cos u,\sin u,0),\ \ \ \phi(u):\R\rightarrow\R.
 \ee
 \item {\it Möbius strip}: it is a ruled surface with
 \be
 \bg(u)=(\cos 2u,\sin 2u,0),\ \ \ \bl(u)=(\cos u\cos 2u,\cos u\sin 2u,\sin u).
 \ee
  \end{itemize}
 \begin{figure}[ht]
	\begin{center}
         \includegraphics[height=.13\textheight]{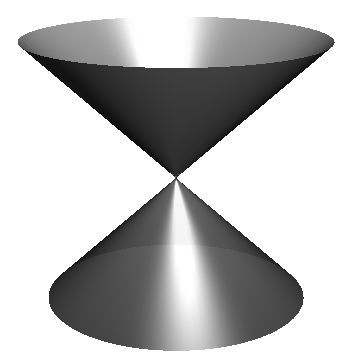}
         \includegraphics[height=.13\textheight]{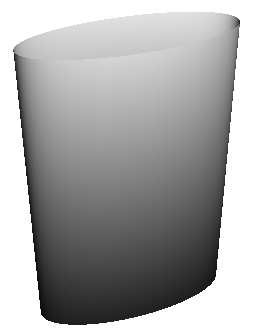}
         \includegraphics[height=.13\textheight]{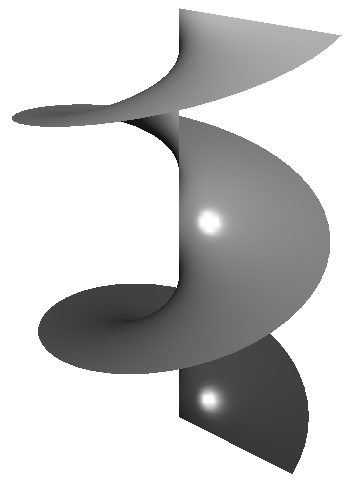}
         \includegraphics[height=.13\textheight]{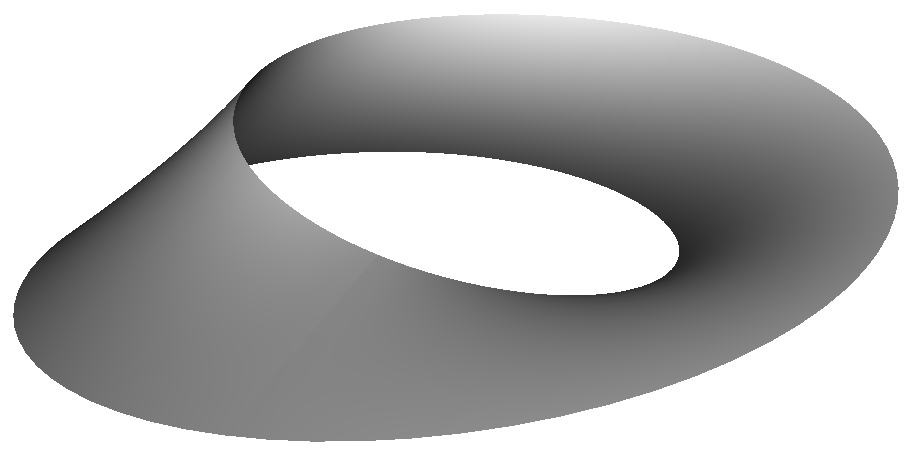}
	\caption{Ruled surfaces. From the left: elliptical cone, elliptical cylinder, helicoid, Möbius strip.}
	\label{fig:30}
	\end{center}
\end{figure}

 \section{First fundamental form of a surface}
 We call {\it first fundamental form of a surface}, denoted by $I(\cdot,\cdot)$, the restriction of the scalar product to the tangent vector space $T_p\Sigma$. We recall that a scalar product is a positive definite symmetric form. Let us consider two vectors $\bw_1=a_1\f_{,u}+b_1\f_{,v},\bw_2=a_2\f_{,u}+b_2\f_{,v}\in T_p\Sigma$; then
 \be
 I(\bw_1,\bw_2)=\bw_1\cdot\bw_2=a_1a_2\f^2_{,u}+(a_1b_2+a_2b_1)\f_{,u}\cdot\f_{,v}+b_1b_2\f^2_{,v}
 \ee
 is the first form of $\f(u,v)$; if $\bw_1=\bw_2=\bw=a\f_{,u}+b\f_{,v}$, then
  \be
 I(\bw)=\bw^2=a^2\f^2_{,u}+2ab\f_{,u}\cdot\f_{,v}+b^2\f^2_{,v}
 \ee
is a positive form $\forall \bw\in T_p\Sigma$. We can rewrite $I(\cdot,\cdot)$ in the form
\be
I(\bw_1,\bw_2)=\bw_1\cdot\g\ \bw_2,
\ee
 where\footnote{\label{note:firstform} Actually, $\g$ is  the metric tensor of $\Sigma$, see ref. \ref{ref:geodiff}. In fact, $\f_{,u}$ and $\f_{,v}$ are the tangent vectors to the coordinate lines on $\Sigma$, i.e. they coincide with the vectors $\g_k$s. Often, in texts on differential geometry, tensor $\g$ is indicated as \be\g=\left[\begin{array}{cc}E&F\\F&G\end{array}\right],\ee where $E:=\f_{,u}\cdot\f_{,u},F:=\f_{,u}\cdot\f_{,v},G:=\f_{,v}\cdot\f_{,v}$.}
 \be
 \g=\left[\begin{array}{cc}\f_{,u}\cdot\f_{,u}&\f_{,u}\cdot\f_{,v}\\\f_{,v}\cdot\f_{,u}&\f_{,v}\cdot\f_{,v}\end{array}\right].
 \ee

 Through $I(\cdot,\cdot)$ we can calculate some important  quantities regarding the geometry of $\Sigma$:
 \begin{itemize}
 \item metric on $\Sigma$: $\forall ds\in\Sigma$,
 \be
 ds^2=ds\cdot ds=I(ds);
 \ee
 so, if 
 \be
 ds=\f_{,u}du+\f_{,v}dv
 \ee
 then
 \be
 \label{eq:metricsigma}
 ds^2=\f_{,u}^2du^2+2\f_{,u}\cdot\f_{,v}du\ dv+\f_{,v}^2dv^2;
 \ee
 \item length $\ell$ of a curve $\bg:[t_1,t_2]\subset\R\rightarrow\Sigma$: we know, eq. (\ref{eq:esse}), that the length of a curve is the integral of the tangent vector:
 \be
 \ell=\int_{t_1}^{t_2}|\bg'(t)|dt=\int_{t_1}^{t_2}\sqrt{\bg'(t)\cdot\bg'(t)}dt
 \ee
and hence, see  eq. (\ref{eq:tangonsigma}), if we call $\bw=(u',v')$ the tangent vector to $\bg$, expressed by its components in the natural basis,
\be
\label{eq:lengthcurve}
\besp
\ell&=\int_{t_1}^{t_2}\sqrt{u'^2\f^2_{,u}+2u'v'\f_{,u}\cdot\f_{,v}+v'^2\f^2_{,v}}dt=\int_{t_1}^{t_2}\sqrt{(u',v')\cdot\g\ (u',v')}dt\\
&=\int_{t_1}^{t_2}\sqrt{I(\bw)}dt;
\end{split}
\ee
 \item angle $\theta$ formed by two vectors $\bw_1,\bw_2\in T_p\Sigma$:
 \be
 \cos\theta=\frac{\bw_1\cdot\bw_2}{|\bw_1||\bw_2|}=\frac{I(\bw_1,\bw_2)}{\sqrt{I(\bw_1)}\sqrt{I(\bw_2)}};
 \ee
 \item area of a small surface on $\Sigma$: be $\f_{,u}du$ and $\f_{,v}dv$ two small vectors on $\Sigma$, forming together the angle $\theta$, that are the transformed, through\footnote{For the sake of conciseness, from now on we will indicate a surface as the function $\f:\Omega\rightarrow\Sigma$, with $\f=\f(u,v), (u,v)\in\Omega\subset\R^2$ and $\Sigma\subset\Eu$.} $\f:\Omega\rightarrow\Sigma$, of two small orthogonal vectors $du,dv\in\Omega$; then the area $d\mathcal{A}$ of the parallelogram determined by them is
 \be
 \besp
 d\mathcal{A}&=|\f_{,u}du\times\f_{,v}dv|=|\f_{,u}\times\f_{,v}|du\ dv=\sqrt{\f^2_{,u}\f^2_{,v}\sin^2\theta}du\ dv\\
 &=\sqrt{\f^2_{,u}\f^2_{,v}(1-\cos^2\theta)}du\ dv=\sqrt{\f^2_{,u}\f^2_{,v}-\f^2_{,u}\f^2_{,v}\cos^2\theta}du\ dv\\
 &=\sqrt{\f^2_{,u}\f^2_{,v}-(\f_{,u}\cdot \f_{,v})^2}du\ dv=\sqrt{\det\g}du\ dv;
 \end{split}
 \ee
 the term $\sqrt{\det\g}$ is hence the {\it dilatation factor of the areas}; because $\det\g>0$, we see that the previous expression has a sense $\forall\f(u,v)$, i.e. for any parameterization of the surface.
  \end{itemize}

 \section{Second fundamental form of a surface}
\begin{figure}[ht]
	\begin{center}
         \includegraphics[width=.8\textwidth]{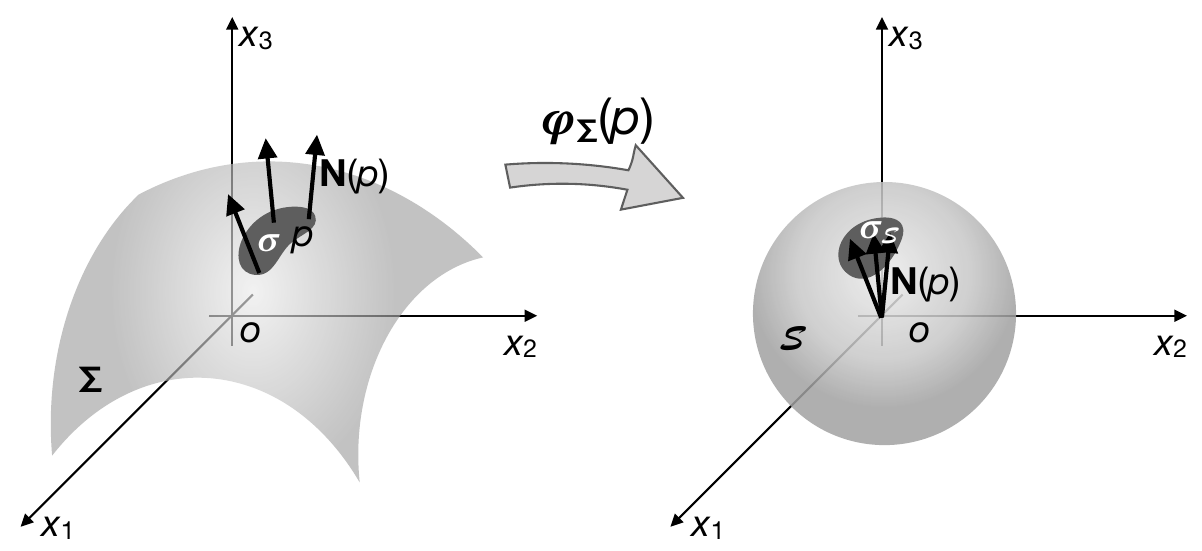}
	\caption{The map of Gauss.}
	\label{fig:37}
	\end{center}
\end{figure}

Be $\f:\Omega\rightarrow\Sigma$ a regular surface, $\{\f_{,u},\f_{,v}\}$ the natural basis for $T_p\Sigma$ and $\N\in\S$ the normal to $\Sigma$ defined as in (\ref{eq:normsurfreg}). We call {\it map of Gauss} of $\Sigma$ the map $\phi_\Sigma:\Sigma\rightarrow\S$ that associates to each $p\in\Sigma$ its $\N:\ \phi_\Sigma(p)=\N(p)$. To each subset $\sigma\subset\Sigma$ the map of Gauss associates hence a subset $\sigma_\S\subset\S$, Fig. \ref{fig:37} (e.g. the Gauss map of a plane is just a point of $\S$).

 We want to study how $\N(p)$ varies at the varying of $p$ on $\Sigma$. %The underlying idea is that the change of $\N(p)$ on $\Sigma$ is related to the curvature of the surface.
 To this purpose, we calculate the change of $\N$ per unit length of a curve $\bg(s)\in\Sigma$, i.e., we study how $\N$ varies along any curve of $\Sigma$ per unit of length of the curve itself; that is why we parameterize the curve with its arc-length $s$ \footnote{Actually, it is possible to introduce the following concepts also more generally, for any parameterization of the curve; anyway, for the sake of simplicity, we will just use the parameter $s$ in the following.}. Be $\N=N_i(u,v)\e_i$; then
 \be
 \besp
 \frac{d\N}{ds}&=\frac{dN_i(u(s),v(s))}{ds}\e_i=\left(\frac{\partial N_i}{\partial u}\frac{du}{ds}+\frac{\partial N_i}{\partial v}\frac{dv}{ds}\right)\e_i\\
 &= \nabla N_i\cdot\btau\e_i=(\e_i\otimes\nabla N_i)\btau=(\nabla \N)\ \btau=\frac{d\N}{d\btau}.
 \end{split}
 \ee
The change of $\N$ is hence related to the directional derivative of $\N$ along the tangent $\btau$ to $\bg(s)$, which is a linear operator on $T_p\Sigma$.  Moreover, as $\N\in\S$, then, cf. eq. (\ref{eq:constvect}),
\be
%\label{eq:orthderivN}
\N\cdot\N_{,u}=\N\cdot\N_{,v}=0\ \Rightarrow\  \N_{,u},\N_{,v}\in T_p\Sigma.
\ee
We then call {\it Weingarten operator} $\mathcal{L}_W:T_p\Sigma\rightarrow T_p\Sigma$ the  opposite of the directional derivative of $\N$:
\be
\mathcal{L}_W(\btau):=-\frac{d\N}{d\btau}.
\ee
 Hence,
 \be
 \label{eq:Wo1}
 \mathcal{L}_W(\f_{,u})=-\N_{,u},\ \ \mathcal{L}_W(\f_{,v})=-\N_{,v}.
 \ee
 Because $\mathcal{L}_W$ is linear, then it exists a tensor $\X$ on $T_p\Sigma$ such that
 \be
 \label{eq:Wo2}
 \mathcal{L}_W(\bv)=\X\bv\ \ \ \forall\bv\in T_p\Sigma.
 \ee
 For any two vectors $\bw_1,\bw_2\in T_p\Sigma$, we define {\it second fundamental form of a surface}, denoted by $II(\bw_1,\bw_2)$ the bilinear form
 \be
 II(\bw_1,\bw_2):=I(\mathcal{L}_W(\bw_1),\bw_2).
 \ee
 \begin{teo}(Symmetry of the second fundamental form): $\forall \bw_1,\bw_2\in T_p\Sigma,\ II(\bw_1,\bw_2)=II(\bw_2,\bw_1)$.
 \begin{proof}
 Because $I$ and $\mathcal{L}_W$ are linear, it is sufficient to prove the thesis for the natural basis $\{\f_{,u},\f_{,v}\}$ of $T_p\Sigma$, and, by the symmetry of $I$, it is sufficient to prove that 
 \be
 I(\mathcal{L}_W(\f_{,u}),\f_{,v})= I(\f_{,u},\mathcal{L}_W(\f_{,v})),
 \ee
 i.e. that
 \be
 I(-\N_{,u},\f_{,v})=I(\f_{,u},-\N_{,v})
 \ee
 and in the end that
 \be
\N_{,u}\cdot\f_{,v}=\f_{,u}\cdot\N_{,v}.
 \ee
To this end, we recall that
 \be
 \N\cdot\f_{,u}=0=\N\cdot\f_{,v},
 \ee
 so, differentiating the first equation by $v$ and the second one by $u$, we get
 \be
 \label{eq:orthsymL}
 \N_{,v}\cdot\f_{,u}=-\N\cdot\f_{,uv}=\N_{,u}\cdot\f_{,v}.
 \ee
 \end{proof}
 \end{teo}
 
The second fundamental form defines a quadratic, bilinear symmetric form:
\be
 II(\bw_1,\bw_2)=I(\mathcal{L}_W(\bw_1),\bw_2)= I(\bw_1,\mathcal{L}_W(\bw_2))=I(\bw_1,\X\bw_2)=\bw_1\cdot\g\X\bw_2=\bw_1\cdot\B\bw_2,
\ee
 where
 \be
 \label{eq:defineB}
 \B:=\g\X.
 \ee
In the natural basis $\{\f_{,u},\f_{,v}\}$ of $T_p\Sigma$, by eq. (\ref{eq:orthsymL}), it is\footnote{\label{note:secondform}In many texts on differential geometry, the following symbols are used: 
\be
\besp
&L=\f_{,uu}\cdot\N=-\f_{,u}\cdot\N_{,u},\\
&M=\f_{,uv}\cdot\N=-\f_{,u}\cdot\N_{,v},\\
&N=\f_{,vv}\cdot\N=-\f_{,v}\cdot\N_{,v}.
\end{split}
\ee
}
\be
\label{eq:composB}
B_{ij}=II(\f_{,i},\f_{,j})=I(\mathcal{L}_W(\f_{,i}),\f_{,j})=-\N_{,i}\cdot\f_{,j}=\N\cdot\f_{,ij};
\ee
tensor $\X$ can then be calculated by eq. (\ref{eq:defineB}):
\be
\label{eq:XgB}
\X=\g^{-1}\B.
\ee
By eq. (\ref{eq:composB}), because $\f_{,ij}=\f_{,ji}$ or simply because $II(\cdot,\cdot)$ is symmetric, we get that 
\be
\B=\B^\top.
\ee

 \section{Curvatures of a surface}
 Be $\f:\Omega\rightarrow\Sigma$ a regular surface and $\bg(s):G\subset\R\rightarrow\Sigma$  a regular curve on $\Sigma$ parameterized with the arc length $s$. We call {\it curvature vector of $\bg(s)$} the vector $\bkappa(s)$ defined as
 \be
 \bkappa(s):=c(s)\bnu(s)=\bg''(s),
 \ee
 where $\bnu(s)$ is the principal normal to $\bg(s)$. Then, we call {\it normal curvature $\kappa_N(s)$ of $\bg(s)$} the projection of $\bkappa(s)$ onto $\N(s)$, the normal to $\Sigma$:
 \be
 \kappa_N(s):=\bkappa(s)\cdot\N(s)=c(s)\ \bnu(s)\cdot\N(s)=\bg''(s)\cdot\N(s).
 \ee
 \begin{teo} The normal curvature $\kappa_N(s)$ of $\bg(s)\in\Sigma$ depends uniquely on $\btau(s)$:
 \be
 \label{eq:curvnormcalcul}
 \kappa_N(s)=\btau(s)\cdot\B\btau(s)=II(\btau(s),\btau(s)).
 \ee
 \begin{proof}
\be
 \bg(s)=\bg(u(s),v(s))\ \rightarrow\ \btau(s)=\bg'(s)=\f_{,u}u'+\f_{,v}v',
 \ee
 therefore $\btau=(u',v')$ in the natural basis and
 \be
 \bkappa(s)=\bg''(s)=\f_{,u}u''+\f_{,v}v''+\f_{,uu}u'^2+2\f_{,uv}u'v'+\f_{,vv}v'^2
 \ee
 and finally, by eqs. (\ref{eq:normsurfreg}) and (\ref{eq:composB}),
 \be
 \kappa_N(s)=\bg''(s)\cdot\N(s)=B_{11}u'^2+2B_{12}u'v'+B_{22}v'^2=\btau\cdot\B\btau=II(\btau,\btau).
 \ee
 \end{proof}
 \end{teo}
 If now $s=s(t)$ is a change of parameter for $\bg$, then
 \be
 \bg'(t)=|\bg'(t)|\btau(t),
 \ee
 so, by the linearity of $II(\cdot,\cdot)$ we get
 \be
 II(\bg'(t),\bg'(t))=|\bg'(t)|^2II(\btau(t),\btau(t))=|\bg'(t)|^2\kappa_N(t)
 \ee
 and finally
 \be
 \kappa_N(t)=\frac{II(\bg'(t),\bg'(t))}{I(\bg'(t),\bg'(t))}.
 \ee
 To each point $p\in\Sigma$ it corresponds  uniquely (in the assumption of regularity of the surface $\f:\Omega\rightarrow\Sigma$) a  tangent plane and a tangent space vector $T_p\Sigma$. In $p$, there are infinite tangent vectors to $\Sigma$, all of them belonging to $T_p\Sigma$. We can associate a curvature to each direction $\bt\in T_p\Sigma$, i.e. to each tangent direction, in the following way: let us consider the bundle  $\mathcal{H}$ of planes whose support is the straight line through $p$ and parallel to $\N$. Then any plane $H\in\mathcal{H}$ is a {\it normal plane to $\Sigma$ in $p$}; each normal plane is uniquely determined by a tangent direction $\bt$ and the (planar) curve $\bg_{N\bt}:=H\cap\Sigma$ is called a {\it normal section of $\Sigma$}. If $\bnu$ and $\N$ are respectively the principal normal to $\bg_{N\bt}$ and the normal to $\Sigma$ in $p$, then
\be
\bnu=\pm\N
\ee
 for each normal section. We have in this way defined a function that  to each tangent direction $\bt\in T_p\Sigma$ associates the normal curvature $\kappa_N$ of the normal section $\bg_{N\bt}$:
 \be
 \kappa_N:\S\cap T_p\Sigma\rightarrow\R|\ \ \ \kappa_N(\bt)=\frac{II(\bt,\bt)}{I(\bt,\bt)}.
 \ee
 By the bilinearity of the second fundamental form, $\kappa_N(\bt)=\kappa_N(-\bt)$.
 
 A point $p\in\Sigma$ is said to be a {\it umbilical point} if $\kappa_{N}(\bt)=const.\ \forall\bt$, it is a {\it planar point} if $\kappa_N(\bt)=0\ \forall\bt$. In all the other points, $\kappa_N$ takes a minimum and a maximum value on distinct directions $\bt\in T_p\Sigma$. %By eq. (\ref{eq:curvnormcalcul}), being  $\B=\B^\top$,  through the spectral theorem we get that the maximum and minimum  of $\kappa_N$ correspond to the two (real) eigenvalues of $\B$, and the directions $\bt$ whereupon the maximum and the minimum of $\kappa_N$ are get, to the respective eigenvectors, cf. Sect. \ref{sec:eigenvaluesvectors}. 

\begin{figure}[ht]
	\begin{center}
         \includegraphics[width=.45\textwidth]{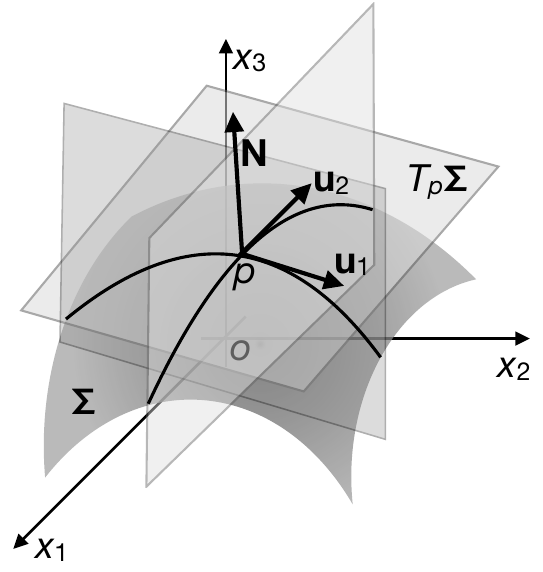}
	\caption{Principal curvatures.}
	\label{fig:38}
	\end{center}
\end{figure}

Because $\B=\B^\top$, by the spectral theorem it exists an orthonormal basis $\{\bu_1,\bu_2\}$ of $T_p\Sigma$ such that
\be
\B=\beta_i\bu_i\otimes\bu_i,
\ee
  with $\beta_i$ the eigenvalues of $\B$. In such a basis, by eq. (\ref{eq:defineB}) we get 
  \be
  \kappa_N(\bu_i)=\frac{II(\bu_i,\bu_i)}{I(\bu_i,\bu_i)}=\frac{\bu_i\cdot\B\bu_i}{\bu_i\cdot\g\bu_i}=\frac{\bu_i\cdot\g\X\bu_i}{\bu_i\cdot\g\bu_i}.
  \ee
  Then, because $\{\bu_1,\bu_2\}$ is an orthonormal basis, $\g=\I$ and
  \be
   \kappa_N(\bu_i)=\bu_i\cdot\X\bu_i=\beta_i,
  \ee
  i.e. $\X$ and $\B$ shares the same eigenvalues and eigenvectors. Moreover, we know that the two directions $\bu_1,\bu_2$ are the directions whereupon the quadratic form in the previous equation gets its maximum, $\kappa_1$, and minimum, $\kappa_2$, values, and in such a basis
  \be
  \X=\kappa_i\bu_i\otimes\bu_i.
  \ee
  We  call $\kappa_1$ and $\kappa_2$ the {\it principal curvatures of $\Sigma$ in $p$}  and $\bu_1,\bu_2$ the {\it principal directions of $\Sigma$ in $p$}, see Fig. \ref{fig:38}.

We call {\it Gaussian curvature $K$} the product of the principal curvatures:
 \be
 K:=\kappa_1\kappa_2=\det\X.
 \ee
By eq. (\ref{eq:XgB}) and the Theorem of Binet, it is also
\be
\label{eq:curvgaus2}
K=\frac{\det\B}{\det\g}.
\ee

 We define {\it mean curvature $H$ of a surface}\footnote{The concept of mean curvature of a surface was introduced for the first time by Sophie Germain, in her celebrated work on the elasticity of plates.} $\f:\Omega\rightarrow\Sigma$ at a point $p\in\Sigma$ the mean of the principal curvatures at $p$:
 \be
 H:=\frac{\kappa_1+\kappa_2}{2}=\frac{1}{2}\tr\X.
 \ee
 
   Of course, a change of parameterization of a surface can change the orientation, cf. Sect. \ref{sec:surf1}, which induces a change of $\N$ into its opposite one and by consequence of the sign of the second fundamental form and hence of the normal and principal curvatures. These last are hence defined  to less the sign, and  the mean curvature too, while the principal directions, umbilicality, flatness and Gaussian curvature are intrinsic to $\Sigma$, i.e. they do not depend on its parameterization. 
 
 \section{The theorem of Rodrigues}
 Principal directions of curvature have a property which is specified by the 
 \begin{teo}(Theoreom of Rodrigues): be $\f(u,v)$ a surface of class at least C$^2$ and $\bl=(\lambda_u,\lambda_v)\in T_p\Sigma$; then
 \be
 \frac{d\N(p)}{d\bl}=-\kappa_\lambda\bl 
 \ee
 if and only if $\bl$ is a principal direction; $\kappa_\lambda$ is the principal curvature relative to $\bl$.
 \begin{proof}
Let $\bl$ be a principal direction of $T_p\Sigma$. Because $|\N|=1$, then 
\be
\label{eq:rodrigues1}
\frac{d\N}{d\bl}\cdot\bl=0;
\ee
 moreover, 
 \be
 \label{eq:rodrigues2}
 \frac{d\N}{d\bl}=\nabla\N\ \bl=\left[\begin{array}{ccc}0&0&0\\0&0&0\\\N_{,u}&\N_{,v}&1\end{array}\right]\left(\begin{array}{c}\lambda_u\\\lambda_v\\0\end{array}\right)=\N_{,u}\lambda_u+\N_{,v}\lambda_v.
 \ee
 Be $\bmu=(\mu_u,\mu_v)$ the other principal direction of $T_p\Sigma$; then
 \be
 \bl\cdot\bmu=0\ \rightarrow\ I(\bl,\bmu)=II(\bl,\bmu)=0.
 \ee
 Moreover
 \be
 \frac{d\N}{d\bl}\cdot\bmu=-II(\bl,\bmu)=0
 \ee
 which implies, together with eq. (\ref{eq:rodrigues1}),
 \be
 \label{eq:rodrigues3}
 \frac{d\N}{d\bl}=\alpha \bl.
\ee
 Therefore
 \be
 \frac{d\N}{d\bl}\cdot\bl=-II(\bl)=\alpha\bl\cdot\bl=\alpha I(\bl)
 \ee
 and finally
 \be
 \alpha=-\frac{II(\bl)}{I(\bl)}=-\kappa_\lambda.
 \ee
 Contrarily, if we assume eq. (\ref{eq:rodrigues3}), like before we get $\alpha=-\kappa_\lambda$ and to end we just need to prove that $\bl$ is a principal direction. From eqs. (\ref{eq:rodrigues2}) and (\ref{eq:rodrigues3}) we get
 \be
 \lambda_u\N_{,u}+\lambda_v\N_{,v}=-\kappa_\lambda(\lambda_u\f_{,u}+\lambda_v\f_{,v}).
 \ee
 Projecting this equation onto $\f_{,u}$ and $\f_{,v}$ gives the two equations
 \be
 \label{eq:rodrigues4}
 \besp
& L\lambda_u+M\lambda_v=\kappa_\lambda(E\lambda_u+F\lambda_v),\\
& M\lambda_u+N\lambda_v=\kappa_\lambda(E\lambda_u+G\lambda_v),
 \end{split}
  \ee
 with the symbols $E,F,G,L,M$ and $N$ defined in Notes \ref{note:firstform} and \ref{note:secondform}.
Let $\bw=(w_u,w_v)\in T_p\Sigma$ and consider the function 
\be
\zeta(\bw,\kappa_\lambda)=II(\bw)-\kappa_\lambda I(\bw);
\ee
it is easy to check that $\zeta,\dfrac{\partial\zeta}{\partial w_u}$ and $\dfrac{\partial\zeta}{\partial w_v}$ take zero value for $\bw=\bl_0$, with $\bl_0$ the eigenvector of the principal direction relative to $\kappa_\lambda$, which gives the system of equations
\be
\left\{\besp
&II(\bl_0)-\kappa_\lambda I(\bl_0)=0,\\
&\frac{\partial II(\bl_0)}{\partial w_u}-\kappa_\lambda\frac{\partial II(\bl_0)}{\partial w_u}=0,\\
&\frac{\partial II(\bl_0)}{\partial w_v}-\kappa_\lambda\frac{\partial II(\bl_0)}{\partial w_v}=0.
\end{split}
\right.
\ee
Developing the derivatives and making some standard passages, eq.  (\ref{eq:rodrigues4}) is found again, which proves that $\bl$ is necessarily the principal direction relative to $\kappa_\lambda$.
\end{proof}
 \end{teo}
 This theorems states hence that the derivative of $\N$ along a given direction is a vector parallel to such a direction only when this is a principal direction of curvature.

 \section{Classification of the points of a surface}
 Be $\f:\Omega\rightarrow\Sigma$ a regular surface  and  $p\in \Sigma$ a non-planar point. Then, we say that
 \begin{itemize}
 \item $p$ is an {\it elliptic point} if $K(p)>0$;
 \item $p$ is a {\it hyperbolic point} if $K(p)<0$;
 \item $p$ is a {\it parabolic point} if $K(p)=0$. 
 \end{itemize}
 To remark that, by eq. (\ref{eq:curvgaus2}), because $\det\g>0$, the value of $\det\B$ is sufficient to determine the type of a point on $\Sigma$.
 \begin{teo}
 If $p$ is an elliptical point of $\sigma$, then it exists a neighbourhood $U\in\Sigma$ of $p$ such that all the  points $q\in U$ belong to the same  half-space into which $\Eu$ is divided by the tangent plane $T_p\Sigma$. 
 \begin{proof}
For the sake of simplicity and without loss of generality, we can always chose a parameterization $\f(u,v)$  of the surface such that $p=\f(0,0)$. Expanding $\f(u,v)$ into a Taylor's series around $(0,0)$ we get the position of a point $q=\f(u,v)\in\Sigma$ in the nighbourhood of $p$ (though not indicated for the sake of shortness, all the derivatives are intended to be calculated at $(0,0)$):
\be
\f(u,v)=\f_{,u}u+\f_{,v}v+\frac{1}{2}(\f_{,uu}u^2+2\f_{,uv}uv+\f_{,vv}v^2)+o(u^2+v^2).
\ee
The distance with sign $d(q)$ of $q\in\Sigma$ from the tangent plane $T_p\Sigma$ is the projection onto $\N$, i.e.:
\be
\besp
d(q)&=\frac{1}{2}(\f_{,uu}u^2+2\f_{,uv}uv+\f_{,vv}v^2)\cdot\N+o(u^2+v^2)\\
&=\frac{1}{2}(B_{11}u^2+2B_{12}uv+B_{22}v^2)+o(u^2+v^2),
\end{split}
\ee
or, which is the same, once put $\bw=u\f_{,u}+v\f_{,v}$,
\be
\label{eq:distanceq}
d(q)=II(\bw,\bw)+o(u^2+v^2).
\ee
If $p$ is an elliptic point, the principal curvatures have the same sign because $K=\kappa_1\kappa_2>0\Rightarrow$ the sign of $II(\bw,\bw)$ does not depend upon $\bw$, i.e. upon the tangent vector. As a consequence the sign of $d(q)$ does not change with $\bw\Rightarrow\forall q\in U,\ \Sigma$ is on the same side of the tangent plane $T_p\Sigma$.
 \end{proof}
 \end{teo}
   \begin{figure}[ht]
	\begin{center}
        \includegraphics[height=.12\textheight]{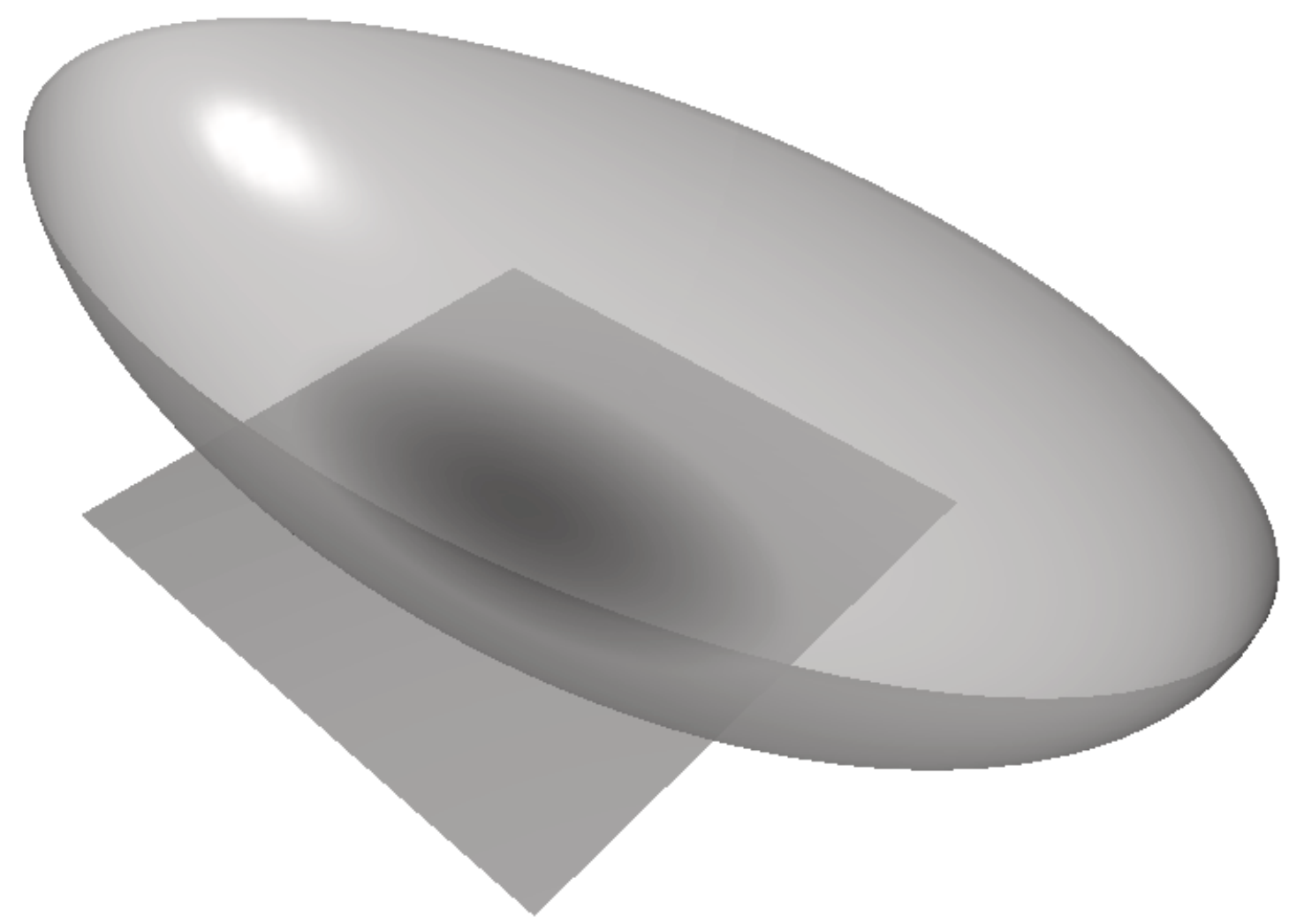}\ 
        \includegraphics[height=.12\textheight]{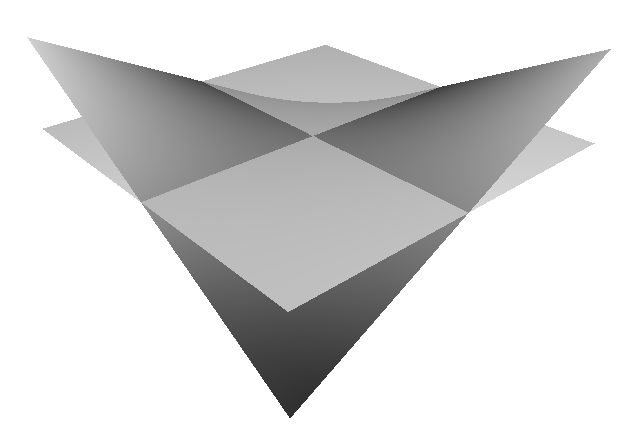}\ 
        \includegraphics[height=.12\textheight]{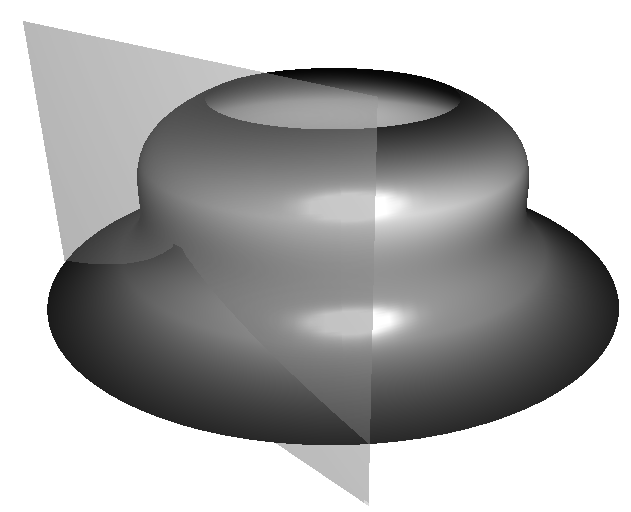}\ 
        \includegraphics[height=.12\textheight]{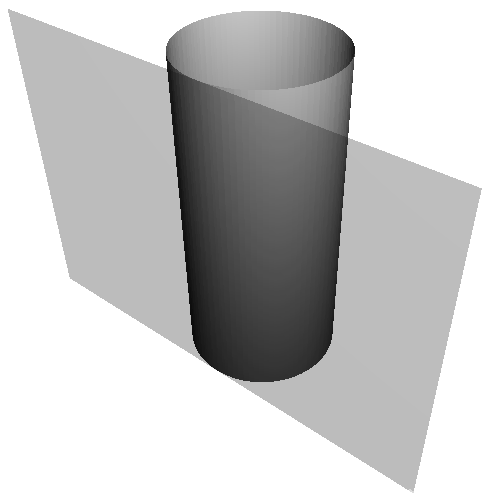}
	\caption{Elliptic, left,  hyperbolic, center, and parabolic, the two last on the right, points.}
	\label{fig:39}
	\end{center}
\end{figure}

\begin{teo}
 If $p$ is a hyperbolic point of $\Sigma$, then for each neighbourhood $U\in\Sigma$ of $p$ there are  points $q\in U$  that are in  half-spaces on the opposite sides with respect to the tangent plane $T_p\Sigma$. 
  \begin{proof}
The proof is identical to that of the previous theorem, until eq. (\ref{eq:distanceq}); if now $p$ is a hyperbolic point, the principal curvatures have opposite  sign and by consequence $d(q)$ changes of sign at least two times in any neighbourhood $U$ of $p\Rightarrow$ there are points $q\in U$ lying in  half-spaces on the opposite sides with respect to the tangent plane $T_p\Sigma$.
 \end{proof}
 \end{teo}
% The typical situation of an elliptic and a hyperbolic point is depicted in Fig. \ref{fig:39}, right.

 In a parabolic point, there are different possibilities: $\Sigma$ is on one side of the space with respect to $T_p\Sigma$, like for the case of a cylinder, or not, like, e.g., for the points $(0,v)$ of the surface, see Fig. \ref{fig:39},
 \be
 \left\{\begin{array}{l}x=(u^3+2)\cos v,\\y=(u^3+2)\sin v,\\z=-u.\end{array}\right.
 \ee

\begin{figure}[ht]
	\begin{center}
        \includegraphics[height=.15\textheight]{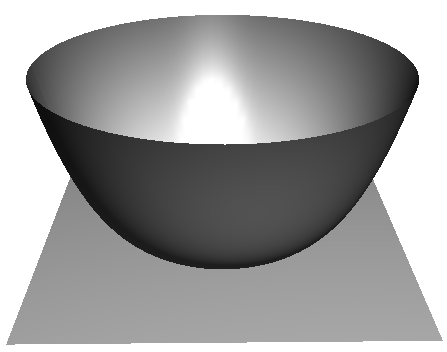}\ \ \
        \includegraphics[height=.15\textheight]{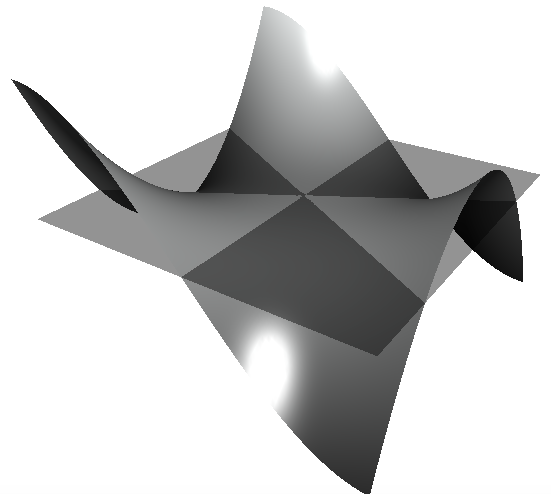}
	\caption{Two different planar points.}
	\label{fig:43}
	\end{center}
\end{figure}

This is the case also for planar points: e.g., the point $(0,0,0)$ is a planar point for both the surfaces
\be
z=x^4+y^4,\ \ \ z=x^3-3xy^2,
\ee
 but in the first case, all the surface is on one side from the tangent plane, while it is on both sides for the second case (the so-called {\it monkey's saddle}), see Fig.\ref{fig:43}.

 \section{Lines of curvature, conjugated directions, asymptotic directions}
 \label{sec:linescurvat}
 A {\it line of curvature} is a curve on a surface with the property to be tangent, at each point, to a principal direction.  
 \begin{teo}
 The lines of curvature of a surface are the solutions to the differential equation
 \be
 X_{21}u'^2+(X_{22}-X_{11})u'v'-X_{12}v'^2=0.
 \ee
 \begin{proof}
 A curve $\bg(t):G\subset\R\rightarrow\Sigma\subset\Eu$ is a line of curvature $\iff$
 \be
 \bg'(t)=\f_{,u}u'+\f_{,v}v'
 \ee
 is an eigenvector of $\X(t)\ \forall t$, i.e. $\iff$ it exists a function $\mu(t)$ such that
 \be
 \X(t)\bg'(t)=\mu(t)\bg'(t)\ \ \forall t.
 \ee
 In the natural basis of $T_p\Sigma$, this condition reads like (we omit the dependence upon $t$ for the sake of simplicity)
 \be
 \left[\begin{array}{cc}X_{11}&X_{12}\\X_{21}&X_{22}\end{array}\right]\left\{\begin{array}{c}u'\\v'\end{array}\right\}=\mu\left\{\begin{array}{c}u'\\v'\end{array}\right\},
 \ee
 which is satisfied $\iff$ the two vectors at the left and right sides are proportional, i.e. if
 \be
 \det\left[\begin{array}{cc}X_{11}u'+X_{12}v'&u'\\X_{21}u'+X_{22}v'&v'\end{array}\right]=0\ \rightarrow\  X_{21}u'^2+(X_{22}-X_{11})u'v'-X_{12}v'^2=0.
 \ee
 \end{proof}
 \end{teo}
 As a corollary, if $\X$ is diagonal, then the coordinate lines are at the same time  principal directions and lines of curvature.

 Be $p$ a non-planar point of a surface $\f:\Omega\rightarrow\Sigma$ and $\bv_1,\bv_2$ two vectors of $T_p\Sigma$. We say that $\bv_1$ and $\bv_2$ are {\it conjugated} if $II(\bv_1,\bv_2)=0$. The directions corresponding to $\bv_1$ and $\bv_2$ are called {\it conjugated directions}. Hence, the principal directions at a point $p$ are conjugated; if $p$ is an umbilical point, any two orthogonal directions are conjugated.
 
 The direction of a vector $\bv\in T_p\Sigma$ is said to be {\it asymptotic} if it is {\it autoconjugated}, i.e. if $II(\bv,\bv)=0$. An asymptotic direction is hence a direction where the normal curvature is null. In a hyperbolic point there are two asymptotic directions, in a parabolic point only one and in an elliptic point there are not asymptotic directions. An {\it asymptotic line} is a curve on a surface with the property of being tangent at every point to an asymptotic direction.
 The asymptotic lines are the solution of the differential equation
 \be
 II(\bg',\bg')=0\ \rightarrow\ B_{11}u'^2+2B_{12}u'v'+B_{22}v'^2=0;
 \ee
 in particular, if $B_{11}=B_{22}=0$ and $\B\neq\O$, then the coordinate lines are asymptotic lines. Asymptotic lines exist only in the regions where $K\leq0$.

 \section{The Dupin's conical curves}
\begin{figure}[ht]
	\begin{center}
        \includegraphics[width=.9\textwidth]{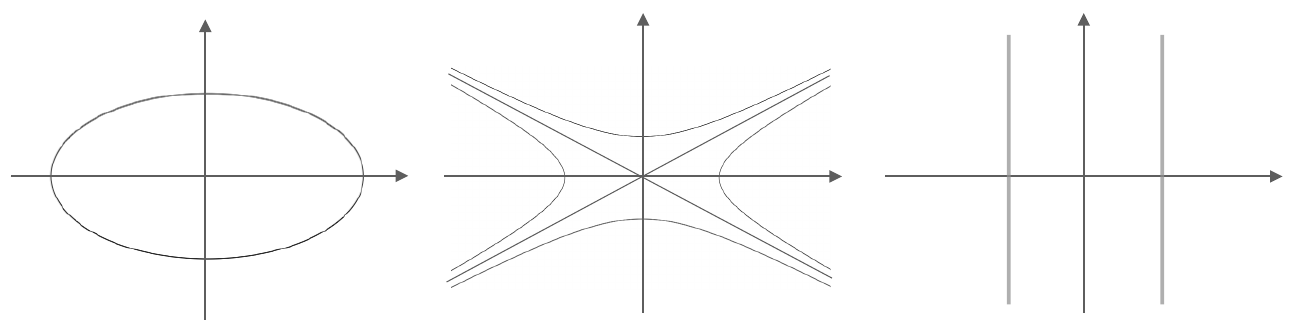}
	\caption{The conical curves of Dupin; from the left: elliptic, hyperbolic and parabolic points.}
	\label{fig:46}
	\end{center}
\end{figure}

 The {\it conical curves of Dupin} are the real curves in $T_p\Sigma$ whose equations are
 \be
 II(\bv,\bv)=\pm1,\ \ \ \bv\in\S.
 \ee
 Be $\{\bu_1,\bu_2\}$ the basis of the principal directions. Using polar coordinates, we can write
 \be
 \bv=\rho\e_\rho,\ \ \ \e_\rho=\cos\theta\bu_1+\sin\theta\bu_2.
 \ee
 Therefore,
 \be
 II(\bv,\bv)=\rho^2II(\e_\rho,\e_\rho)=\rho^2\kappa_N(\e_\rho),
 \ee
 and the conicals' equations are
 \be
 \rho^2(\kappa_1\cos^2\theta+\kappa_2\sin^2\theta)=\pm1.
 \ee
 With the Cartesian coordinates $\xi=\rho\cos\theta,\eta=\rho\sin\theta$, we get
 \be
 \kappa_1\xi^2+\kappa_2\eta^2=\pm1.
 \ee
 The type of conical curves depend upon the kind of point on $\Sigma$:
 \begin{itemize}
 \item elliptical points: the principal curvatures have the same sign $\rightarrow$ one of the conical curves is an ellipse, the other one the null set (actually, it is not a real curve);
 \item hyperbolic points: the principal curves have opposite signs $\rightarrow$ the conical curves are conjugated hyperbolae whose asymptotes coincide with the asymptotic directions;
 \item parabolic points: at least one of the principal curvatures is null $\rightarrow$ one of the conical curves degenerates into a couple of parallel straight lines, corresponding to the asymptotic direction, the other one is the null set. 
 \end{itemize}
 The three possible cases are depicted in Fig. \ref{fig:46}
 
  \section{The Gauss-Codazzi compatibility conditions}
  \label{sec:gausscodazzi}
 Let us consider a surface $\Sigma$ whose points are determined by the vector function $\r:\Omega\subset\R^2\rightarrow\Sigma\subset\Eu,\r(\au,\ad)=x_i(\au,\ad)\bepsi_i$, with $\bepsi_i, i=1,2,3,$ the vectors of the orthonormal basis of the reference frame $\Rep=\{o;\bepsi_1,\bepsi_2,\bepsi_3\}$ and the parameters $\au,\ad$ chosen in such a way that the lines $\au=const.,\ad=const.$ are lines of  curvature, i.e. tangent at each point to the principal directions of curvature and hence mutually orthogonal\footnote{The symbol $\r$ is here preferred to $\f$, like $\au$ to $u$ and $\ad$ to $v$, to recall that we have made the particular choice of coordinate lines that are lines of curvature. All the developments could be done in a more general case, but this choice is made to obtain simpler relations, that preserves anyway the generality because the lines of curvature exist everywhere.}. 
 With such a choice, cf. eq. (\ref{eq:metricsigma}),
 \be
 ds^2=A_1^2d\au^2+A_2^2d\ad^2,
 \ee
 with
 \be
 \label{eq:paramlamé}
 \besp
& A_1=\sqrt{\r_{,\au}^2}=\sqrt{\frac{dx_i}{d\au}\frac{dx_i}{d\au}},\\
& A_2=\sqrt{\r_{,\ad}^2}=\sqrt{\frac{dx_i}{d\ad}\frac{dx_i}{d\ad}}
 \end{split}
 \ee
  the so-called {\it Lamé's parameters}. We remark that along the lines of curvatures, i.e. the lines $\alpha_i=const., i=1,2$, that in short from now on we will call the {\it lines $\alpha_i$}, it is
  \be
  \label{eq:distanzelinee}
  \besp
  &ds_1=A_1d\au,\\
 & ds_2=A_2d\ad,
  \end{split}
  \ee
  and hence, 
  \be
  \label{eq:vecttanglc}
  \besp
  \bl_1=\frac{ds_1}{d\au}=A_1\eu,\\
  \bl_2=\frac{ds_2}{d\ad}=A_2\ed
  \end{split}
  \ee
  are the vectors tangent to the lines of curvature.
  Be 
  \be
  \label{eq:triedre}
  \eu=\frac{1}{A_1}\r_{,\au},\ \ \ed=\frac{1}{A_2}\r_{,\ad},\ \ \et=\eu\times\ed(=\N);
  \ee
  these three vectors form the  orthonormal (local) natural basis $\texttt{e}=\{\eu,\ed,\et\}$. We always make the choice of $\au,\ad$ such that $\e_3$ is always directed to the convex side of $\Sigma$ if the point is elliptic or parabolic, or to the side of the centres of negative curvature, if the point is hyperbolic. 
  
  We consider a vector $\bv=\bv(p),\ p\in\Sigma$,
  \be
  \bv=v_1\eu+v_2\ed+v_3\et,
  \ee
  and we want to calculate how it transforms  when $p$ changes. To this end, we need to calculate how $\eu,\ed,\et$ change with $\au,\ad$. Be $q\in\Sigma$ a point in the neighborhood of $p$ on the line $\alpha_i$ and let us first consider the change of $\et$ in passing from $p$ to $q$. Because $p$ and $q$ belong to the same line $\alpha_i$, by the Theorem of Rodrigues we get (no summation on $i$ in the following equations)
  \be
  \frac{\partial\et}{\partial\bl_i}=-\kappa_i\bl_i,\ i=1,2,
  \ee
i.e., by  eq. (\ref{eq:vecttanglc}),
\be
\frac{\partial\et}{\partial\ai}=\frac{A_i}{R_i}\e_i,
\ee
 with 
 \be
 R_i=-\frac{1}{\kappa_i}
 \ee
 the (principal) radius of curvature along the line $\ai$. The sign minus in the previous equation is due to the choice done above for orienting $\et=\N$, that gives always $\N=-\bnu$, with $\bnu$ the principal normal to the line $\ai$. This result can be obtained also directly, see Fig. \ref{fig:52}:
      \begin{figure}[ht]
	\begin{center}
        \includegraphics[height=.2\textheight]{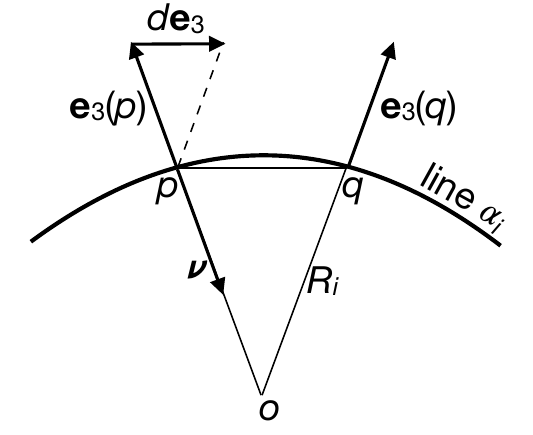}
	\caption{Variation of $\N=\et$ along a line of curvature.}
	\label{fig:52}
	\end{center}
\end{figure}
\be
\et(q)=\et(p)+d\et
\ee
 and  in the limit of $q\rightarrow p$, $d\et$ tends to be parallel to $q-p$ and
  \be
  \lim_{q\rightarrow p}(q-p)=\bl_i=A_i\e_i.
  \ee
 By the similitude of the triangles, it is evident that
 \be
 \frac{|d\et|}{|\et|}=\frac{|q-p|}{R_i};
 \ee
moreover,
\be
d\et=\frac{\partial\et}{\partial\ai}d\ai\e_i.
\ee
Finally, as $|\et|=1$, we get again
\be
\label{eq:codazzi0}
\frac{\partial\et}{\partial\ai}=\frac{A_i}{R_i}\e_i.
\ee
  Implicitly, in this last proof, we have used the Theorem of Rodrigues, because we have assumed that $d\et$ is parallel to $\bl_i$, as it is, because line $\ai$ is a line of curvature.  
  
  We pass now to determine the changes of $\eu$ and $\ed$; to this purpose, we remark that
  \be
  \frac{\partial\r_{,\au}}{\partial\ad}=\frac{\partial^2\r}{\partial\ad\partial\au}=\frac{\partial^2\r}{\partial\au\partial\ad}= \frac{\partial\r_{,\ad}}{\partial\au},
  \ee
  so by eq. (\ref{eq:triedre}) we get
  \be
  \label{eq:codazzi1}
  \frac{\partial (A_1\eu)}{\partial\ad}= \frac{\partial (A_2\ed)}{\partial\au}.
  \ee
Let us study now $\dfrac{\partial\e_j}{\partial\ai}$; as $|\e_j|=1, j=1,2$,
  \be
  \label{eq:codazzi2}
  \frac{\partial\e_j}{\partial\ai}\cdot\e_j=0\ \ \forall i,j=1,2.
  \ee
  Because $\eu\cdot\ed=0$,
  \be
   \frac{\partial\eu}{\partial\au}\cdot\ed=\frac{\partial(\eu\cdot\ed)}{\partial\au}-\eu\cdot\frac{\partial\ed}{\partial\au}=-\eu\cdot\frac{\partial\ed}{\partial\au}.
  \ee
 By eq. (\ref{eq:codazzi1}) we get 
 \be
 \frac{\partial\ed}{\partial\au}=\frac{1}{A_2}\frac{\partial( A_1\eu)}{\partial\ad}-\frac{1}{A_2}\frac{\partial A_2}{\partial\au}\ed,
 \ee 
  that injected in the previous equation gives, by eq. (\ref{eq:codazzi2}),
  \be
   \frac{\partial\eu}{\partial\au}\cdot\ed=-\frac{1}{A_2}\frac{\partial (A_1\eu)}{\partial\ad}\cdot\eu+\frac{1}{A_2}\frac{\partial A_2}{\partial\au}\ed\cdot\eu=-\frac{A_1}{A_2}\frac{\partial\eu}{\partial\ad}\cdot\eu-\frac{1}{A_2}\frac{\partial A_1}{\partial\ad}\eu\cdot\eu=-\frac{1}{A_2}\frac{\partial A_1}{\partial\ad}.
  \ee
  Then, because $\eu\cdot\et=0$,
  \be
  \frac{\partial\eu}{\partial\au}\cdot\et=
  \frac{\partial(\eu\cdot\et)}{\partial\au}-\eu\cdot\frac{\partial\et}{\partial\au}=-\eu\cdot\frac{\partial\et}{\partial\au},
  \ee
  and by eq. (\ref{eq:codazzi0})
  \be
  \frac{\partial\et}{\partial\au}=\frac{A_1}{R_1}\eu,
  \ee
  so finally
  \be
   \frac{\partial\eu}{\partial\au}\cdot\et=-\frac{A_1}{R_1}.
  \ee  
 Again through eqs. (\ref{eq:codazzi1}) and (\ref{eq:codazzi2}) we get 
 \be
 \frac{\partial\eu}{\partial\ad}\cdot\ed=\frac{1}{A_1}\frac{\partial (A_2\ed)}{\partial\au}\cdot\ed-\frac{1}{A_1}\frac{\partial A_1}{\partial\ad}\eu\cdot\ed=\frac{A_2}{A_1}\frac{\partial \ed}{\partial\au}\cdot\ed+\frac{1}{A_1}\frac{\partial A_2}{\partial\au}\ed\cdot\ed=\frac{1}{A_1}\frac{\partial A_2}{\partial\au}
  \ee
  and also, by eq. (\ref{eq:codazzi0})
  \be
  \frac{\partial\eu}{\partial\ad}\cdot\et=
  \frac{\partial(\eu\cdot\et)}{\partial\ad}-\eu\cdot\frac{\partial\et}{\partial\ad}=-\eu\cdot\frac{\partial\et}{\partial\ad}=-\frac{A_2}{R_2}\eu\cdot\ed=0.
  \ee
The derivatives of $\ed$ can be found in the same way and  resuming, we have
\be
\label{eq:codazzi3}
\begin{array}{ccccc}
\dfrac{\partial\eu}{\partial\au}&=&&-\dfrac{1}{A_2}\dfrac{\partial A_1}{\partial\ad}\ed&-\dfrac{A_1}{R_1}\et,\medskip\\
\dfrac{\partial\eu}{\partial\ad}&=&&\dfrac{1}{A_1}\dfrac{\partial A_2}{\partial\au}\ed,\medskip\\
\dfrac{\partial\ed}{\partial\au}&=&\dfrac{1}{A_2}\dfrac{\partial A_1}{\partial\ad}\eu,\medskip\\
\dfrac{\partial\ed}{\partial\ad}&=&-\dfrac{1}{A_1}\dfrac{\partial A_2}{\partial\au}\eu&&-\dfrac{A_2}{R_2}\et,\medskip\\
\dfrac{\partial\et}{\partial\au}&=&\dfrac{A_1}{R_1}\eu,\medskip\\
\dfrac{\partial\et}{\partial\ad}&=&&\dfrac{A_2}{R_2}\ed.
\end{array}
\ee

  Passing now to the 2nd-order derivatives, imposing the equality of mixed derivatives, gives some important differential relations between the Lamé's parameters $A_i$ and the radiuses of curvatures $R_i$. In fact, from the identity
  \be
  \frac{\partial^2\et}{\partial\au\partial\ad}=\frac{\partial^2\et}{\partial\ad\partial\au},
  \ee
  and eqs. (\ref{eq:codazzi3})$_{5,6}$ we get
  \be
  \frac{\partial}{\partial\ad}\left(\dfrac{A_1}{R_1}\eu\right)=\frac{\partial}{\partial\au}\left(\dfrac{A_2}{R_2}\ed\right),
  \ee
  whence
  \be
  \frac{\partial}{\partial\ad}\left(\dfrac{A_1}{R_1}\right)\eu+\dfrac{A_1}{R_1}\frac{\partial\eu}{\partial\ad}=
  \frac{\partial}{\partial\au}\left(\dfrac{A_2}{R_2}\right)\ed+\dfrac{A_2}{R_2}\frac{\partial\ed}{\partial\au}.
  \ee
  Injecting now eqs. (\ref{eq:codazzi3})$_{2,3}$ into the last result and rearranging the terms gives
  \be
  \left[ \frac{\partial}{\partial\ad}\left(\dfrac{A_1}{R_1}\right)-\frac{1}{R_2}\frac{\partial A_1}{\partial\ad}\right]\eu-
  \left[ \frac{\partial}{\partial\au}\left(\dfrac{A_2}{R_2}\right)-\frac{1}{R_1}\frac{\partial A_2}{\partial\au}\right]\ed=0,
  \ee
that to be true needs that the two following conditions be identically satisfied:
\be
\label{eq:codazzi4}
\besp
&\frac{\partial}{\partial\ad}\left(\dfrac{A_1}{R_1}\right)-\frac{1}{R_2}\frac{\partial A_1}{\partial\ad}=0,\\
& \frac{\partial}{\partial\au}\left(\dfrac{A_2}{R_2}\right)-\frac{1}{R_1}\frac{\partial A_2}{\partial\au}=0.
\end{split}
\ee
The above equations are known as the {\it Codazzi conditions}. Let us now consider the other identity
\be
\frac{\partial^2\eu}{\partial\au\partial\ad}=\frac{\partial^2\eu}{\partial\ad\partial\au};
\ee
 still using eq. (\ref{eq:codazzi3}), with some standard passages this identity can be transformed to
 \be
 \left[\frac{\partial}{\partial\au}\left(\frac{1}{A_1}\frac{\partial A_2}{\partial\au}\right)+\frac{\partial}{\partial\ad}\left(\frac{1}{A_2}\frac{\partial A_1}{\partial\ad}\right)+\frac{A_1}{R_1}\frac{A_2}{R_2}\right]\ed+\left[\frac{\partial}{\partial\ad}\left(\frac{A_1}{R_1}\right)-\frac{1}{R_2}\frac{\partial A_1}{\partial\ad}\right]\et=0.
 \ee
  Again, for this equation to be identically satisfied, each of the expressions in square brackets must vanish, which gives two further differential conditions, but only the first one is new, as the second one corresponds to eq. (\ref{eq:codazzi4})$_1$. The new condition is hence
  \be
  \label{eq:codazzi5}
  \frac{\partial}{\partial\au}\left(\frac{1}{A_1}\frac{\partial A_2}{\partial\au}\right)+\frac{\partial}{\partial\ad}\left(\frac{1}{A_2}\frac{\partial A_1}{\partial\ad}\right)+\frac{A_1}{R_1}\frac{A_2}{R_2}=0,
  \ee
  which is known as the {\it Gauss condition}. The last identity
 \be
\frac{\partial^2\ed}{\partial\au\partial\ad}=\frac{\partial^2\ed}{\partial\ad\partial\au}
\ee 
  does not add any further independent condition, as it can be easily checked. The meaning of the {\it Gauss-Codazzi conditions}, eqs. (\ref{eq:codazzi4}) and (\ref{eq:codazzi5}), is that of {\it compatibility conditions}: only when these conditions are satisfied by functions $A_1,A_2,R_1$ and $R_2$, then such functions represent the Lamé's parameters and the principal radiuses of curvature of a surface, i.e. only in this case they define a surface, except for its position in space.

\section{Exercices}
\begin{enumerate}
\item The curve whose polar equation is 
\begin{equation*}
r=a\ \theta,\ \ a\in\mathbb{R},
\end{equation*}
is an {\it Archimede's spiral}. Find its curvature, its length for $\theta\in[0,2\pi)$ and prove that any straight line passing by the origin is divided by the spiral in segments of constant length $2\pi\ a$ (that is why it is used to record disks).

\item The curve whose polar equation is 
\begin{equation*}
r=a\ \mathrm{e}^{b \theta},\ \ a,b\in\mathbb{R},
\end{equation*}
is the {\it logarithmic spiral}. Prove that the origin is an asymptotic point of the curve, find its curvature and the length of the segment in which a straight line by the origin is divided by two consecutive intersections with the spiral. Then prove that the curve is plane and its {\it equiangular property}: $(p(\theta)-o)\cdot\btau(\theta)=const.$

\item The curve whose parametric equation is 
\begin{equation*}
p(\theta)=a(\cos\theta+\theta\sin\theta)\gr{e}_1+a(\sin\theta-\theta\cos\theta)\gr{e}_2
\end{equation*}
with the parameter $\theta$ the angle formed by $p(\theta)-o$ with the $x_1-$axis is the {\it involute of the circle}. Find its curvature and length for $\theta\in[0,2\pi)$ and prove that the geometrical set of the points $p(\theta)+\rho(\theta)\bnu(\theta)$ is exactly the circle of center $o$ and radius $a$ (that is why the involute of the circle is used to profile engrenages).

\item The curve whose parametric equation is 
\begin{equation*}
p(\theta)=a\cos\omega\theta\gr{e}_1+a\sin\omega\theta\gr{e}_2+b\omega\theta\gr{e}_3
\end{equation*}
is a {\it helix} that winds on a circular cylinder of radius $a$. Show that the angle formed by the helix and any generatrix of the cylinder is constant (a property that defines a helix in the general case). Then, find its length for $\theta\in[0,2\pi)$, curvature, torsion and pitch (the distance, on a same generatrix, between two successive intersections with the helix). %Prove then that the osculating circle is a diametral circle of the osculating sphere, find the frame of Frenet-Serret in $\theta=2\pi$ and write the equation of the helix in this frame. 
Prove then the {\it Bertrand's theorem}: a curve is a cylindrical helix if and only if the ratio $c/\vartheta=const.$ %Using this theorem, prove that for a circular helix $\vartheta=const.$ 
Finally, prove that for the above circular helix there are two constants $A$ and $B$ such that
\begin{equation*}
p'\times p''=A\gr{u}(\theta)+B\gr{e}_3,
\end{equation*}
with
\begin{equation*}
\gr{u}=\sin\omega\theta\gr{e}_1-\cos\omega\theta\gr{e}_2;
\end{equation*}
find then $A$ and $B$.

\item For the curve whose cylindrical equation is
\begin{equation*}
\left\{
\begin{split}
&r=1,\\
&z=\sin\theta
\end{split}
\right.
\end{equation*}
find the highest curvature and determine whether or not it is planar.

 \item Prove that a function of the type $x_3=f(x_1,x_2)$, with $f:\Omega\subset\R^2\rightarrow\R$ smooth, defines a surface.
 \item Show that the catenoid is the rotation surface of a catenary, then find its Gaussian curvature.
 \item Show that the pseudo-sphere is the rotation surface of a tractrice and explain why the surface has this name (hint: look for its Gaussian curvature).
% \item Prove that the regularity of a cone is satisfied at each point exception made for the apex and for the points on  straight lines tangent to $\bg(u)$. 
\item Prove that the hyperbolic hyperboloid is a doubly ruled surface.
\item Prove that the hyperbolic paraboloid, whose Cartesian equation is $x_3=x_1x_2$, is a doubly ruled surface.
\item Consider the parameterization
\be
 \f(u,v)=(1-v)\bg(u)+v\bl(u),
\ee
 with
 \be
\bg(u)=(\cos(u-\alpha),\sin(u-\alpha),-1),\ \ \bl(u)=(\cos(u+\alpha),\sin(u+\alpha),1).
 \ee
 Show that:
\begin{itemize}
\item for $\alpha=0$ one gets a cylinder with equation $x_1^2+x_2^2=1$;
\item for $\alpha=\dfrac{\pi}{2}$ one gets a cone with equation $x_1^2+x_2^2=x_3^2$;
\item for $0<\alpha<\dfrac{\pi}{2}$ one gets a hyperbolic hyperboloid with equation 
\be\dfrac{x_1^2+x_2^2}{\cos^2\alpha}-\dfrac{x_3^2}{\cot^2\alpha}=1.\ee
\end{itemize}
\item Calculate the first fundamental form of a sphere of radius $R$, determine the metric on it, the area of a sector of surface between the longitudes $\theta_1$ and $\theta_2$ and the length of the parallel at the latitude $\pi/4$ between these two longitudes.
\item  Prove that the surface defined by
\be
\f(u,v):\Omega=\R\times(-\pi,\pi]\rightarrow\Eu|\ \ \f(u,v)=\left(\frac{\cos v}{\cosh u},\frac{\sin v}{\cosh u},\frac{\sinh u}{\cosh u}\right)
\ee
is a sphere; then show that the image of any straight line on $\Omega$ is a loxodromic line on the sphere. 
\item Calculate the vectors of the natural basis, the first and second fundamental form and the tensors $\g,\X,\B$ for the catenoid.
\item Make the same for the helicoid of parametric equation
\be
 \f(u,v)=\bg(u)+v\bl(u),
 \ee
 with
 \be
  \bg(u)=(0,0,u),\ \ \bl(u)=(\cos u,\sin u,0).
\ee
\item Show that the catenoid and the helicoid are made of hyperbolic points.
%\item Determine the geodesic lines of a circular cylinder.
 \end{enumerate}

\chapter{Cables}
\label{ch:2}
\section{Introduction}
We define as a {\it cable} a special type of continuum body: a curve endowed with a linear mass density. This definition is not sufficient to characterize mechanically a cable (also a curved rod geometrically is a curve): the fundamental relation mechanically defining a cable must be given in terms of internal actions, which is done in the next section.

%Mechanically speaking, a cable has not bending stiffness nor compressive strength 

A cable is called {\it inextensible} if the distance 
\begin{equation}
\Delta s=s_B-s_A
\end{equation}
measured along the cable between two arbitrary points $p_A$ and $p_B$ of the curve, determined on it by two curvilinear abscissae $s_B>s_A$, is constant at any time $t$.
If this does not happen, the cable is {\it extensible}.

\section{Mechanical definition of a cable}
We use the {\it Euler Sections Principle}, valid for any continuum body, to define mechanically a cable. This principle states simply that {\it a body is in equilibrium if and only if any of its parts is in equilibrium}.

For the whole body, this necessitates the equilibrium of all the external forces applied to the body, actions and reactions. But if one isolates, ideally, a part of the body, then only a part of the whole external forces will be directly applied to the isolated portion of the body, so, generally speaking, the equilibrium of the isolated part will be restored only by admitting the existence of some actions,  called the {\it internal actions}, that are transmitted to the isolated part by the remaining parts of the body, throughout the contact zones. Physically speaking, these internal actions are  the resultant and resultant moment of the contact forces, i.e. of the stresses that the different parts of the body exchange at any point of it. The Euler's section in the case of a curve reduces of course to a point, called the {\it separation point}.

This concept is shared by all the continuum bodies, that differ for the definition of the type of internal actions that they can develop at any point of the continuum.

We give then the following mechanical definition:
 \begin{quote}{\it a cable is a  curvilinear unidimensional continuum body whose internal actions that two ideally separated parts of it exchange through the separation point are statically equivalent to a single tension force applied exactly in correspondence of the separation point}.
 \end{quote}

Such a tensile force is currently named {\it tension in the cable} and will be indicated in the following by $\bt(s)$, and its norm, the intensity of the force, by $\theta(s)$.

This is the physical property that   mechanically defines a cable in our model; it is true for extensible or inextensible cables. Let us see now what are the consequences of this definition.

To this purpose, we write the statics balance equations for the part (1) of a cable of length $\ell$,  from $p_0=p(s=0)$, the beginning of the cable, to the point $p(s)$, see Fig. \ref{fig:f2_1}. For the while, we assume that only distributed loads, whose linear intensity is $\gr{f}(s)$, act upon the cable. $\bt(s)$ is the tension that the part (2) of the cable, that between $p(s)$ and $p_1=p(s=\ell)$ apply to the part (1):
\begin{figure}[h]
\begin{center}
\includegraphics[scale=.8]{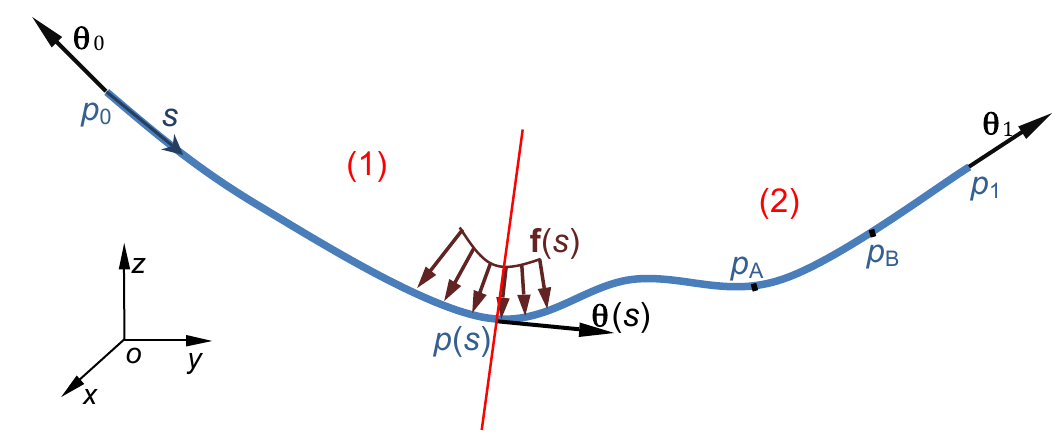}
\caption{General scheme of a cable}
\label{fig:f2_1}
\end{center}
\end{figure}
\begin{itemize}
\item balance of the forces:
\begin{equation}
\bt_0+\int_0^s\gr{f}(s^*)\ ds^*+\bt(s)=\gr{o},
\end{equation}
\item balance of the moments, with respect to a point $o$:
\begin{equation}
(p_0-o)\times{\bt}_0+\int_0^s(p(s^*)-o)\times\gr{f}(s^*)\ ds^*+(p(s)-o)\times\bt(s)=\gr{o}.
\end{equation}
\end{itemize}

Differentiating these equations, we get the {\it local balance equations} for the cable:
\begin{equation}
\label{eq:locequil}
\begin{split}
&\gr{f}(s)+\frac{d\bt(s)}{ds}=\gr{o},\\
&(p(s)-o)\times\gr{f}(s)+\frac{d}{ds}[(p(s)-o)\times\bt(s)]=\gr{o}.
\end{split}
\end{equation}

After differentiating the second term in eq. (\ref{eq:locequil})$_2$ we get
\begin{equation}
\begin{split}
&(p(s)-o)\times\gr{f}(s)+\btau(s)\times\bt(s)+(p(s)-o)\times\frac{d\bt(s)}{ds}=\gr{o}\ \rightarrow\\
&(p(s)-o)\times\left(\gr{f}(s)+\frac{d\bt(s)}{ds}\right)+\btau(s)\times\bt(s)=\gr{o},
\end{split}
\end{equation}
which gives, through the (\ref{eq:locequil})$_1$,
\begin{equation}
\btau(s)\times\bt(s)=\gr{o}\ \ \forall s,
\end{equation}
so that
\begin{equation}
\label{eq:tau1}
\bt(s)=\theta(s)\ \btau(s)\ \ \forall s.
\end{equation}
Then, at the equilibrium the tension is a force which is constantly tangent to the cable. Because the fundamental assumption is that the tension is a tensile force, it is necessarily
\begin{equation}
\label{eq:tau2}
\theta(s)\geq0\ \ \forall s.
\end{equation}
Eqs. (\ref{eq:tau1}) and (\ref{eq:tau2}) defines mechanically a cable. To remark that a cable cannot transmit couples; in fact, the moment of $\bt(s)$ with respect to $p(s)$ vanishes identically $\forall s$, as a consequence of the fact that $\bt(s)$ is applied exactly in $p(s)$. Finally, a cable is a unidimensional continuum body without compression, shear and bending stiffness, but with only tension stiffness.

In a dual approach, we could assume eqs. (\ref{eq:tau1}) and (\ref{eq:tau2}) as the mathematical conditions defining a cable. Then, the local balance of the moments  is always fulfilled because, through eqs. (\ref{eq:locequil})$_1$, (\ref{eq:tau1}) and (\ref{eq:tau2}) we get
\begin{equation}
\begin{split}
&(p(s)-o)\times\gr{f}(s)+\frac{d}{ds}[(p(s)-o)\times\bt(s)]=\gr{o}\ \rightarrow\\
&-(p(s)-o)\times\frac{d}{ds}[\theta(s)\btau(s)]+\btau(s)\times\theta(s)\btau(s)+\\
&(p(s)-o)\times\frac{d}{ds}[\theta(s)\btau(s)]=\gr{o}\ \ \mathrm{identically}.
\end{split}
\end{equation}
Finally, for cables it is the balance of the forces, eq. (\ref{eq:locequil})$_1$, that defines equilibrium. The balance of moments, eq. (\ref{eq:locequil})$_2$, can be used either to define the tension $\bt(s)$, eq. (\ref{eq:tau1}), or, if this last  is assumed {\it a priori}, it is automatically satisfied $\forall s$.

Intrinsically, eq. (\ref{eq:locequil})$_1$ states that the equilibrium of the cable, no matter if it is extensible or not, is possible only thanks to the configuration of this last: in fact, applying the above vector equation to the infinitesimal interval $[s,s+ds]$, equilibrium can be satisfied, if $\gr{f}(s),\ \bt(s)$ and $\bt(s+ds)$ are not collinear, only thanks to a curvature of the cable: the statics of cables is a {\it matter of geometry}. So, if the cable is free, it will find a configuration able to ensure the equilibrium under the action of the applied loads, while if it is wrapped on a surface, the configuration of the cable on the surface will develop contact forces equilibrated with the tension of the cable at the ends.

The balance equations of the whole cable are:
\begin{equation}
\label{eq:globequil}
\begin{split}
&\bt_0+\int_0^\ell\gr{f}(s)\ ds+\bt_1=\gr{o},\\
&(p_0-o)\times{\bt}_0+\int_0^\ell(p(s)-o)\times\gr{f}(s)\ ds+(p_1-o)\times\bt_1=\gr{o}.
\end{split}
\end{equation}
In the previous equations, it is $\bt_0=\bt(s=0)$ and $\bt_1=\bt(s=\ell)$. From eq. (\ref{eq:globequil}), we get immediately that
\begin{equation}
\bt_0=-\theta(s=0)\btau(s=0),\ \ \ \bt_1=\theta(s=\ell)\btau(s=\ell).
\end{equation}

\section{The intrinsic balance equations}
The local force balance equation can be projected onto the intrinsic frame of Frenet-Serret. From the first formula of Frenet-Serret (\ref{eq:fs1}) we get
\begin{equation}
\gr{f}(s)+\frac{d}{ds}[\theta(s)\ \btau(s)]=\gr{f}(s)+\frac{d\theta(s)}{ds} \btau(s)+c(s)\ \theta(s)\ \bnu(s)=\gr{o}.
\end{equation}
If we put, in the Frenet-Serret frame, $\gr{f}=(f_\tau,f_\nu,f_\beta)$, we obtain the three scalar {\it intrinsic balance equations} (Jc. Bernoulli, 1698):
\begin{equation}
\label{eq:intreq}
\begin{split}
&\frac{d\theta}{ds}+f_\tau=0,\\
&c\ \theta+f_\nu=0,\\
&f_\beta=0.
\end{split}
\end{equation}

To remark that $f_\nu<0$. In several problems, however, the equilibrium configuration is not known and constitutes, together with $\theta(s)$, the unknown of the problem.

The intrinsic equations state that the cable finds always a configuration of equilibrium where, pointwise,  the component of the external load on $\bb$ is null; i.e. the loads belong to the osculating plane.

The first of (\ref{eq:intreq}) gives us a general result: whenever $f_\tau=0$, $\theta(s)=$const. The second one also gives us an interesting result: if $f_\nu=0$, then $c\ \theta=0$ and there are two possible cases: the first one is $\theta\neq0\ \forall s\ \Rightarrow c=0\ \forall s$: the equilibrium configuration is a straight line. So, e.g., all the parts of a cable that are unloaded assume as configuration of equilibrium a straight line. The other possibility is $\theta=0\ \forall s\ \Rightarrow\ c$ can take any value, i.e. the equilibrium configuration is undetermined: in the case of null tension, the cable can take any possible configuration, all of them are equilibrated; this is the case, e.g., of a cable simply lying on a horizontal plane, without tension forces applied to its ends.

\section{Forces depending upon a potential}
Be $\gr{f}(s)=\nabla U$; then
\begin{equation}
f_\tau=\gr{f}(s)\cdot\btau(s)=\nabla U\cdot\btau(s)=\nabla U\cdot\frac{dp(s)}{ds}=\frac{\partial U}{\partial x}\frac{dx}{ds}+\frac{\partial U}{\partial y}\frac{dy}{ds}+\frac{\partial U}{\partial z}\frac{dz}{ds}=\frac{dU}{ds}.
\end{equation}
So
\begin{equation}
f_\tau+\frac{d\theta}{ds}=0\ \rightarrow\ \frac{d(\theta+U)}{ds}=0\ \Rightarrow\ \theta+U=\mathrm{const.}
\end{equation}

This means that $\theta$ is maximum where $U$ is minimum; that is why in a cable in equilibrium under the action of its own weight the highest tension is in correspondence of its highest point.

\section{Parallel and coplanar forces}
Be $\gr{f}(s)=f(s)\ \gr{e}$, with $\gr{e}$ a constant unit vector. The balance of forces between $s=0$ and $s$ gives then
\begin{equation}
\theta(s)\btau(s)=\theta(0)\btau(0)-\int_0^sf(s^*)ds^*\ \gr{e}.
\end{equation}
By consequence, $\bt(s)=\theta(s)\btau(s)$ is a linear combination of $\btau(0)$ and $\gr{e}\ \forall s$, i.e. the curve lies in the plane passing through $p(0)$ and containing $\gr{e}$.

This result is valid also for the case of coplanar forces, i.e. of forces of the type $\gr{f}(s)=f_1\gr{e}_1+f_2\gr{e}_2$, with $\gr{e}_1$ and $\gr{e}_2$ two constant unit vectors. In fact:
\begin{equation}
\theta(s)\btau(s)=\theta(0)\btau(0)-\int_0^sf_1(s^*)ds^*\ \gr{e}_1-\int_0^sf_2(s^*)ds^*\ \gr{e}_2.
\end{equation}
In addition, this is true also for the case of concentrated forces, see below.

\section{Concentrated forces}
The local and intrinsic balance equations are valid uniquely if $\gr{f}(s)$ is regular, i.e. if it has not discontinuities, like in the case of concentrated forces.

We can tackle such a problem in the following way: be $\gr{f}_p(s)$ a distribution of forces in the interval $(s-\eps,s+\eps)$; we define the concentrated load associated to $\gr{f}_p(s)$ as the vector
\begin{equation}
\gr{F}=\lim_{\eps\rightarrow0^+}\int_{s-\eps}^{s+\eps}\gr{f}_p(s^*)ds^*.
\end{equation}

In such a way, the concentrated force is treated as a particular distributed load over an interval that tends toward zero. Writing the balance of the forces between $s-\eps$ and $s+\eps$ we get
\begin{equation}
\theta(s+\eps)\btau(s+\eps)-\theta(s-\eps)\btau(s-\eps)+\int_{s-\eps}^{s+\eps}\gr{f}(s^*)ds^*+\int_{s-\eps}^{s+\eps}\gr{f}_p(s^*)ds^*=\gr{o},
\end{equation}
whose limit for $\eps\rightarrow0$ gives
\begin{equation}
\label{eq:raccordo}
\gr{F}=\theta^-\btau^--\theta^+\btau^+,
\end{equation}
where $\theta^-=\lim_{\eps\rightarrow0}\theta(s-\eps)$ etc. This law let us see that the presence of a concentrated load produces, generally speaking, a discontinuity of both $\theta$, the tension, and $\btau$, the direction of the cable.

\begin{figure}[h]
\begin{center}
\includegraphics[scale=1.2]{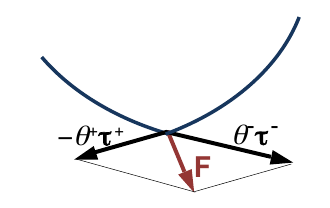}
\caption{Concentrated loads.}
\label{fig:f2_2}
\end{center}
\end{figure}

The above equation is nothing but the rule of parallelogram of the forces, which implies that the three forces are coplanar, see Fig. \ref{fig:f2_2}.

The discontinuity on $\theta(s)$ vanishes if and only if $\gr{F}$ acts along the bissectrice of the angle formed by $\btau^-$ and $-\btau^+$. 

Equation (\ref{eq:raccordo}) must be written in correspondence of each concentrated load, while in all the other parts, the local or intrinsic balance equations govern the problem.

\section{Cables on surfaces}
Let us consider the case of a cable wrapped on a surface, which gives the reaction of contact $\bp$; we assume that $\bp$ is much greater than all the other distributed forces, like the weight of the cable, so that all of them are negligible with respect to $\bp$. In such a situation, the intrinsic balance equations become
\begin{equation}
\label{eq:intrsurf}
\begin{split}
&\frac{d\theta}{ds}+\phi_\tau=0,\\
&c\ \theta+\phi_\nu=0,\\
&\phi_\beta=0.
\end{split}
\end{equation}

In the case of a frictionless contact between the cable and the surface, if this last is regular and of unit normal $\gr{N}$,  it is 
\begin{equation}
\begin{split}
&\bp(s)\times\gr{N}(s)=\gr{o},\\
&\btau(s)\cdot\gr{N}(s)=0,
\end{split}\ 
\Rightarrow\ \ \phi_\tau=0\ \ \forall s.
\end{equation}
Hence, by the (\ref{eq:intrsurf})$_1$, we get
\begin{equation}
\frac{d\theta}{ds}=0,
\end{equation}
i.e. the tension is constant everywhere in the cable, and in particular it is equal to the tensile loads applied to the ends of the cable. Moreover, because $\phi_\tau=0,\phi_\nu<0$, if $\phi(s)$ is the norm of $\bp(s)$, then
\begin{equation}
\bp(s)=-\phi(s)\ \bnu(s)=\pm\phi(s)\gr{N}(s)\ \iff\ \bnu(s)\times\gr{N}(s)=\gr{o}\ \forall s.
\end{equation}
This condition characterizes geodetic lines: the cable {\it wraps the surface along one of its geodetic lines}. 

Let us now suppose that there is friction in the contact between the cable and the surface, and let assume the Coulomb's non-slipping condition:
\begin{equation}
\sigma|\phi_{N}(s)|\geq|\phi_\tau(s)|\ \forall s,
\end{equation}
with $\sigma$ the {\it friction coefficient}. If the cable is wrapped on the surface along a geodetic line, then $\bnu(s)=\pm\gr{N}(s)$, so we get
\begin{equation}
\sigma|\phi_\nu(s)|\geq|\phi_\tau(s)|\ \forall s,
\end{equation}
and by the intrinsic equations (\ref{eq:intrsurf}) we get
\begin{equation}
\sigma \ c\ \theta\geq\left|\frac{d\theta}{ds}\right|\ \rightarrow\ \sigma\ c\geq\left|\frac{1}{\theta}\frac{d\theta}{ds}\right|=\left|\frac{d\log\theta}{ds}\right|.
\end{equation}

For the whole cable we get hence the overall non slipping condition
\begin{equation}
\int_0^\ell\sigma\ c(s)\ ds\geq\int_0^\ell\left|\frac{d\log\theta(s)}{ds}\right|ds\geq\left|\int_0^\ell\frac{d\log\theta(s)}{ds}ds \right|=\left|\log\frac{\theta(\ell)}{\theta(0)}\right|,
\end{equation}
with $\ell$ the winding length of the cable on the surface. If now we assume $\theta(\ell)>\theta(0)$, then we obtain the non-slipping condition that links the tension at $s=\ell$ with that at $s=0$:
\begin{equation}
\label{eq:nonslipp}
\theta(\ell)\leq\theta(0)\ \mathrm{e}^{\int_0^\ell\sigma\ c(s)\ ds}.
\end{equation}
This condition depends upon the friction coefficient, the winding length and the curvature of the cable.

\section{Applications}
\subsection{The catenary}
\begin{figure}[h]
\begin{center}
\includegraphics[scale=.8]{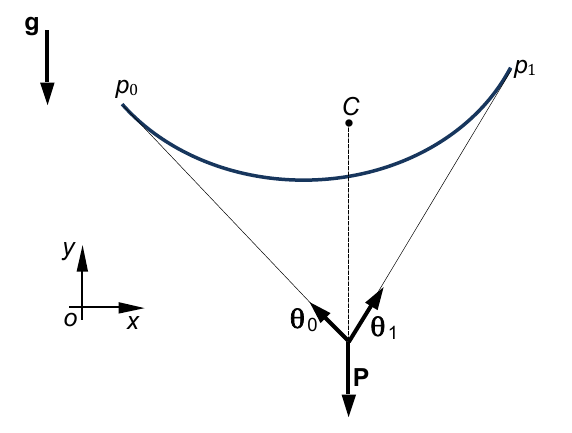}
\caption{The catenary.}
\label{fig:f2_3}
\end{center}
\end{figure}

The {\it catenary} is the equilibrium curve of an inextensible, homogeneous cable that is supported at the ends and that is acted upon uniquely by its own weight\footnote{The problem of the catenary has been very important in the history of mechanics and mathematics, because it is one of the problems at the origin of both the differential calculus and of the calculus of variations. It was proposed by Jc. Bernoulli to scientists in 1690, and besides his solution, he obtained different methods of solution by his brother, Jh. Bernoulli, Leibniz and Huygens.}, see Fig. \ref{fig:f2_3}.

Be $\ell$ the length of the catenary, which is fixed in $p_0=(x_0,y_0)$ and $p_1=(x_1,y_1)$, $|p_0-p_1|<\ell$. The only force is 
\begin{equation}
\label{eq:forzacat}
\gr{f}(s)=-\mu \ g\ \ed,
\end{equation}
with $\mu$ the mass per unit length of the cable; because $\ed$ is a constant vector, the equilibrium curve is contained in a vertical plane, the one passing by $p_0$ and $p_1$. Looking for the equilibrium curve in the form $y=y(x)$ we have
 \begin{equation}
 \begin{split}
 &p-o=x\eu+y(x)\ed,\\
 &\btau=(\eu+y'\ed)\frac{1}{\sqrt{1+y'^2}},\\
 &\bnu=(-y'\eu+\ed)\frac{1}{\sqrt{1+y'^2}},\\
 &c=\frac{y''}{(1+y'^2)^{\frac{3}{2}}}.
 %,\\ &y''>0.
 \end{split}
 \end{equation}
 
 Then,
 \begin{equation}
 \begin{split}
 &f_\tau=\gr{f}\cdot\btau=-\mu\ g\frac{y'}{\sqrt{1+y'^2}},\\
& f_\nu=\gr{f}\cdot\bnu=-\frac{\mu\ g}{\sqrt{1+y'^2}},
 \end{split}
 \end{equation}
 so the intrinsic balance equations are
 \begin{equation}
  \label{eq:intrcaten}
 \begin{split}
 &\frac{d\theta}{ds}=\mu\ g\frac{y'}{\sqrt{1+y'^2}},\\
 &\theta\frac{y''}{(1+y'^2)^{\frac{3}{2}}}=\frac{\mu\ g}{\sqrt{1+y'^2}}.
 \end{split}
 \end{equation}
 
 To obtain $y(x)$ we project the local balance equation onto $\eu$, and because $\eu$ is a constant vector we get
 \begin{equation}
 \frac{d\bt}{ds}\cdot\eu+\gr{f}\cdot\eu=0\ \rightarrow\  \frac{d(\bt\cdot\eu)}{ds}-\mu\ g\ed\cdot\eu=0
 \end{equation}
 which gives  the following first integral of the problem:
 \begin{equation}
 \label{eq:firstintcable}
 \bt\cdot\eu=const:=\theta_0,
 \end{equation}
 where the scalar $\theta_0$ is the horizontal component of the tension in the cable. $\theta_0$ is hence a constant of the problem; in particular, it is the horizontal component of the reactions in $p_0$ and $p_1$ and the tension in the cable in its lowest point. Moreover,
 \begin{equation}
 \label{eq:tensioncable}
 \bt\cdot\eu=\frac{\theta}{\sqrt{1+y'^2}}=\theta_0\ \ \Rightarrow\ \ \theta=\theta_0\sqrt{1+y'^2},
 \end{equation}
that injected into eq. (\ref{eq:intrcaten})$_2$ gives the differential equation of the equilibrium curve:
\begin{equation}
\frac{y''}{\sqrt{1+y'^2}}=\frac{\mu\ g}{\theta_0}.
\end{equation}
 
To solve the above equation, we put
\begin{equation}
y''=\frac{dy'}{dx}
\end{equation}
and write it as
\begin{equation}
 \frac{dy'}{\sqrt{1+y'^2}}=\frac{\mu\ g}{\theta_0}dx,
\end{equation}
 whose solution is
 \begin{equation}
 \label{eq:catprim}
 \mathrm{arcsinh}\ y'=\frac{\mu\ g}{\theta_0}x+c_1\ \rightarrow\ y'=\sinh\left(\frac{\mu\ g}{\theta_0}x+c_1\right),
 \end{equation}
 and integrating again
 \begin{equation}
 \label{eq:catenaria}
 y=\frac{\theta_0}{\mu\ g}\cosh\left(\frac{\mu\ g}{\theta_0}x+c_1\right)+c_2.
 \end{equation}
 This is the equation of the catenary. 
 The three constant $c_1,c_2$ and $\theta_0$ are determined  using the two boundary conditions
 \begin{equation}
 y(x=x_0)=y_0,\ \ y(x=x_1)=y_1,
 \end{equation}
and the condition that the length of the equilibrium curve is $\ell$, because the cable is inextensible:
\begin{equation}
\label{eq:teta0}
\begin{split}
\ell=&\int_{p_0}^{p_1}ds=\int_{x_0}^{x_1}\sqrt{1+y'^2}dx=\int_{x_0}^{x_1}\sqrt{1+\sinh^2\left(\frac{\mu\ g}{\theta_0}x+c_1\right)}dx=\\
&\int_{x_0}^{x_1}\cosh\left(\frac{\mu\ g}{\theta_0}x+c_1\right)dx=
\frac{\theta_0}{\mu\ g}\left[\sinh\left(\frac{\mu\ g}{\theta_0}x_1+c_1\right)-\sinh\left(\frac{\mu\ g}{\theta_0}x_0+c_1\right)\right],
\end{split}
\end{equation}

We can now find the value of the tension $\theta(y)$: injecting the expression (\ref{eq:catprim}) of $y'$ into eq. (\ref{eq:tensioncable}) we get
\begin{equation}
\theta=\theta_0\sqrt{1+\sinh^2\left(\frac{\mu\ g}{\theta_0}x+c_1\right)}=\theta_0\cosh\left(\frac{\mu\ g}{\theta_0}x+c_1\right),
\end{equation}
but for the (\ref{eq:catenaria}), it is
\begin{equation}
\cosh\left(\frac{\mu\ g}{\theta_0}x+c_1\right)=\frac{\mu\ g}{\theta_0}(y-c_2),
\end{equation}
and finally
\begin{equation}
\theta=\mu\ g(y-c_2).
\end{equation}
 
The tension in the cable is hence a linear function of the vertical position of the cable, so it is minimum at the lowest point, where it is equal to $\theta_0$, and maximum for $\max\{y_0,y_1\}$, which could be predicted because $\gr{f}$ depends upon a potential.

As a particular case, we consider a cable where $x_0=0,x_1=\alpha\ell,\ 0<\alpha<1,\ y_0=y_1=0.$ If we put $|p_1-p_0|=x_1-x_0=L$, then $\alpha=L/\ell$. In such a situation, the equations that determine $c_1$ and $c_2$ become
\begin{equation}
\begin{split}
&c_2=-\frac{\theta_0}{\mu g}\cosh c_1,\\
&c_2=-\frac{\theta_0}{\mu g}\cosh\left(\frac{\mu g}{\theta_0}\alpha\ell+ c_1\right),
\end{split}
\ \ \ \ \rightarrow\ \ \ \
\begin{split}
&c_1=-\frac{\mu  g \alpha \ell}{2\theta_0},\\
&c_2=-\frac{\theta_0}{\mu g}\cosh\left(-\frac{\mu  g \alpha \ell}{2\theta_0}\right).
\end{split}
\end{equation}

\begin{figure}[h]
\begin{center}
\includegraphics[scale=.8]{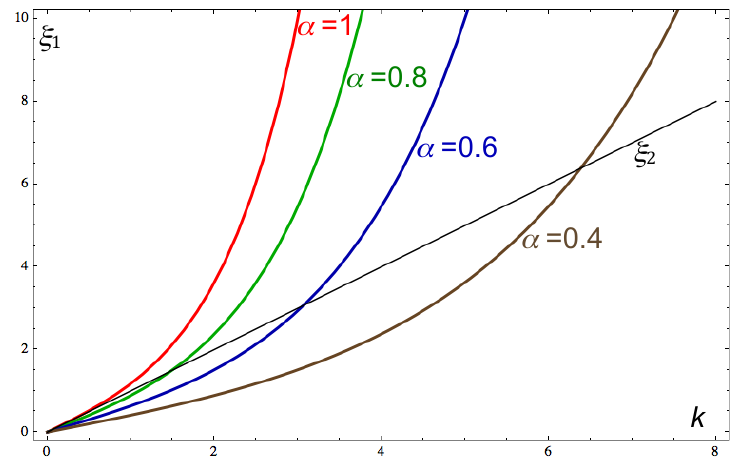}
\caption{The tension in the cable as function of $\alpha$.}
\label{fig:f2_4}
\end{center}
\end{figure}

To determine $\theta_0$, we inject these results in eq. (\ref{eq:teta0}):
\begin{equation}
\begin{split}
&\frac{\theta_0}{\mu\ g}\left[\sinh\left(\frac{\mu g}{\theta_0}\alpha \ell+c_1\right)-\sinh c_1\right]=\ell\ \rightarrow\\
&\sinh\left(\frac{\mu  g\alpha \ell}{2\theta_0}\right)-\sinh\left(-\frac{\mu  g \alpha \ell}{2\theta_0}\right)=\frac{\mu \ g\  \ell}{\theta_0}.
\end{split}
\end{equation}
Because $\sinh$ is an odd function, i.e. $\sinh(-x)=-\sinh x\ \forall x$, we get the equation
\begin{equation}
\label{eq:eqtensio}
\sinh\alpha k=k,\ \ \ k=\frac{\mu g \ell}{2\theta_0}.
\end{equation}
The parameter $k>0$ is half the ratio of the total weight of the cable to the minimal tension $\theta_0$. Putting
\begin{equation}
\begin{split}
&\xi_1=\sinh\alpha k,\\
&\xi_2=k,
\end{split}
\end{equation}
the solutions to eq. (\ref{eq:eqtensio}) are the intersections of $\xi_1(k)$ and $\xi_2(k)$, see Fig. \ref{fig:f2_4}.

We observe that $k$, the abscissa of the intersection point, increases if $\alpha$ decreases, i.e. for a  looser cable; because $\mu g\ell/2=const.$, this implies that $\theta_0$ increases with $\alpha$: the more stretched the cable, the higher $\theta_0$. In particular, because for $\alpha=1$
\begin{equation}
\left.\dfrac{d\sinh k}{dk}\right|_0=1,
\end{equation}
the tangent in $k=0$ to the curve $\xi_1(k)$ is just $\xi_2=k$; no other intersections between $\xi_1$ and $\xi_2$ are possible for $\alpha=1$, because $\xi_1(k)$ is always increasing, so for $\alpha=1$, i.e. for a straight equilibrium configuration of the cable, the only possible solution is $k=0$, which means that $\theta\rightarrow\infty$. Because this is physically impossible, {\it a cable can never take a rectilinear equilibrium configuration under the only action of its own weight}.

The Taylor's expansion of the catenary's equation (\ref{eq:catenaria}) around  a point $\bar{x}$ is
\begin{equation}
\begin{split}
y=&\frac{\theta_0}{\mu\ g}\cosh\left(\frac{\mu\ g}{\theta_0}\bar{x}+c_1\right)+c_2+\sinh\left(\frac{\mu\ g}{\theta_0}\bar{x}+c_1\right)(x-\bar{x})+\\
&\frac{1}{2}\frac{\mu\ g}{\theta_0}\cosh\left(\frac{\mu\ g}{\theta_0}\bar{x}+c_1\right)(x-\bar{x})^2+o(|x-\bar{x}|^3).
\end{split}
\end{equation}
The terms of order $o(|x-\bar{x}|^3)$ decrease with the ratio $\mu\ g/\theta_0$, so, for the same unit mass $\mu$, with the stretch of the cable. Finally, the above equation states that for a stretched cable, the catenary can be well approximated by a parabola.

From the overall balance of the forces on the cable, it is immediate to see that the tangents to the catenary in $p_0$ and $p_1$ cross exactly in correspondence of the vertical drawn from the barycenter $C$ of the curve, see Fig. \ref{fig:f2_3}. The weight $\gr{P}$ of the cable is in fact decomposed by the two tensions $-\bt_0$ and $\bt_1$; this property was discovered by I. G. Pardies in 1673, well before the finding of the catenary's equation.

If the catenary is put upside down, it becomes the equilibrium curve of a body able to transmit only normal compressive forces: the inverse catenary is hence  the equilibrium curve of a masonry or stone or concret arch of constant section submitted uniquely to its own weight. This property was observed by Hooke, that published it under the form of an anagram in 1675\footnote{Translated from Latin, the anagram gives the statement: as hangs the flexible line, so but inverted will stand the rigid arch (J. Heyman, {\it The stone skeleton}, Cambridge University Press, 1995.}, and later confirmed analytically by Jc. Bernoulli in 1704.

In 1740, L. Euler discovered that the catenary is also the {\it catenoid}, i.e. the curve joining two points at different distances from an axis that, when turned about this axis, produces the surface of revolution of least area.

There is a more elegant way to find the equation of the catenary, that using the calculus of variation. In fact, the equilibrium configuration can be found applying the {\it Torricelli's principle}: a body acted upon only by its weight is in a stable configuration of equilibrium when its barycenter occupies the lowest position. 

The position of the barycenter $C$ is
\begin{equation}
y_C=\frac{\int_{p_0}^{p_1}y\ ds}{\int_{p_0}^{p_1} ds},
\end{equation}
and by eq. (\ref{eq:ds}) and because the cable is inextensible, so that $\ell=\int_{p_0}^{p_1} ds=const.$, the problem can be reduced to
\begin{equation}
\min J=\int_{p_0}^{p_1}y\sqrt{1+y'^2}dx
\end{equation}
with the isoperimetric constraint
\begin{equation}
\int_{p_0}^{p_1}\sqrt{1+y'^2}dx=\ell.
\end{equation}

Applying the Lagrange multiplier's method, we minimize the functional 
\begin{equation}
\min J^*=\int_{p_0}^{p_1}f(y,y';\psi)dx;
\end{equation}
with
\begin{equation}
\label{eq:effecate}
f(y,y';\psi)=(y+\psi)\sqrt{1+y'^2}.
\end{equation}
The Euler-Lagrange equation with respect to $y$ (the one relative to $\psi$ gives the isoperimetric constraint),
\begin{equation}
\frac{d}{dx}\frac{\partial f}{\partial y'}-\frac{\partial f}{\partial y}=0,
\end{equation}
because $f$ does not depend explicitly upon $x$, gives the condition\footnote{
This is a classical result in analytical mechanics and can be shown introducing the expression of $\dfrac{\partial f}{\partial y}$ obtained by the Euler-Lagrange equation into $\dfrac{df(y,y')}{dx}$:
\begin{equation*}
\frac{\partial f}{\partial y}=\frac{d}{dx}\frac{\partial f}{\partial y'}\rightarrow\frac{df}{dx}=\frac{\partial f}{\partial y}\frac{dy}{dx}+\frac{\partial f}{\partial y'}\frac{dy'}{dx}=\frac{d}{dx}\frac{\partial f}{\partial y'}
\frac{dy}{dx}+\frac{\partial f}{\partial y'}\frac{dy'}{dx}=\frac{d}{dx}\left(y'\frac{\partial f}{\partial y'}\right)\rightarrow\frac{d}{dx}\left(f-y'\frac{\partial f}{\partial y'}\right)=0.
\end{equation*}
The quantity in the brackets of the last expression is $-H$, the opposite of the integral of Jacobi of the problem. In the end, the constant $\gamma$ is the opposite of $H$.
} 
\begin{equation}
\label{eq:deeffe}
\frac{d}{dx}\left(f-y'\frac{\partial f}{\partial y'}\right)=0\ \Rightarrow\ f-y'\frac{\partial f}{\partial y'}=\gamma,\ \gamma\in\mathbb{R}.
\end{equation}
From eq. (\ref{eq:effecate}) we obtain
\begin{equation}
\frac{\partial f}{\partial y'}= (y+\psi)\frac{y'}{\sqrt{1+y'^2}},
\end{equation}
that injected into eq. (\ref{eq:deeffe}) gives
\begin{equation}
y+\psi=\gamma\sqrt{1+y'^2}\ \rightarrow\ \frac{dy}{\sqrt{(y+\psi)^2-\gamma^2}}=\frac{dx}{\gamma}.
\end{equation}
Putting 
\begin{equation}
y+\psi=\gamma\cosh z\ \rightarrow\ dy=\gamma\sinh z\ dz
\end{equation}
we get
\begin{equation}
\frac{\gamma\sinh z}{\gamma\sqrt{\cosh^2 z-1}}dz=\frac{dx}{\gamma}\ \rightarrow\ dz=\frac{dx}{\gamma}\ \rightarrow\ z=\frac{x}{\gamma}+\beta,\ \beta\in\mathbb{R}.
\end{equation}
Because
\begin{equation}
z=\mathrm{arccosh}\frac{y+\psi}{\gamma},
\end{equation}
we obtain finally
\begin{equation}
y=\gamma\cosh\left(\frac{x}{\gamma}+\beta\right)-\psi.
\end{equation}
This equation coincides with the equation (\ref{eq:catenaria}) of the catenary, with $\gamma=\dfrac{\theta_0}{\mu\ g}, \beta=c_1$ and $\psi=-c_2$.

\subsection{The suspension bridge}
The problem of the suspended bridge (solved by Beeckman en 1615) is similar to that of the catenary, the difference is that the load is uniformly distributed not along the cable, but along its horizontal projection\footnote{This is justified by the fact that the deck of a suspended bridge normally has a weight per unit length $q$ which is far greater than that of the cable, $\lambda $, so that to a first approximation this last  can be neglected.}:
\begin{equation}
\gr{q}(x)=-q\ed,\ \ q>0.
\end{equation}
So, now the load to be used into the intrinsic equations
\begin{equation}
\gr{f}(s)=-f(s)\ed,\ f(s)>0\ \forall s,
\end{equation}
is unknown. It can be deduced as follows:
\begin{equation}
\begin{split}
&\int_{s_1}^{s_2}f(s)\ ds=\int_{x_1}^{x_2}f(s)\sqrt{1+y'^2}\ dx=\\
&\int_{x_1}^{x_2}q\ dx=q(x_2-x_1)\ \rightarrow\ f(s)=\frac{q}{\sqrt{1+y'^2}}.
\end{split}
\end{equation}

Now, we can proceed just like for the catenary: the intrinsic balance equations are
 \begin{equation}
 \label{eq:intrparab}
 \begin{split}
 &\frac{d\theta}{ds}=q\frac{y'}{{1+y'^2}},\\
 &\theta\frac{y''}{(1+y'^2)^{\frac{3}{2}}}=\frac{q}{{1+y'^2}}.
 \end{split}
 \end{equation}
From the first equation, we obtain once more the same first integral (\ref{eq:firstintcable}), having exactly the same mechanical meaning, and by consequence also eq. (\ref{eq:tensioncable}), that injected into eq. (\ref{eq:intrparab})$_2$ gives
\begin{equation}
\label{eq:suspbridge}
y''=\frac{q}{\theta_0}\ \rightarrow\ y=\frac{q}{2\theta_0}x^2+c_1x+c_2.
\end{equation}
The equilibrium curve is hence a parabola; once more, the constants 
$c_1,c_2$ and $\theta_0$ are determined  using the conditions
 \begin{equation}
 \begin{split}
 &y(x=x_0)=y_0,\\
  &y(x=x_1)=y_1,\\
  &\ell=\int_{p_0}^{p_1}ds=\int_{x_0}^{x_1}\sqrt{1+y'^2}dx.
  \end{split}
 \end{equation}
This last, once solved, gives the explicit condition
\begin{equation}
\begin{split}
&\frac{\theta_0}{2q}\left[\left(\frac{q}{\theta_0}x_1+c_1\right)\sqrt{1+\left(\frac{q}{\theta_0}x_1+c_1\right)^2}-\left(\frac{q}{\theta_0}x_0+c_1\right)\sqrt{1+\left(\frac{q}{\theta_0}x_0+c_1\right)^2}+\right.\\
&\left.\log\frac{\dfrac{q}{\theta_0}x_1+\sqrt{1+\left(\dfrac{q}{\theta_0}x_1+c_1\right)^2}}{\dfrac{q}{\theta_0}x_0+\sqrt{1+\left(\dfrac{q}{\theta_0}x_0+c_1\right)^2}}\right]=\ell.
\end{split}
\end{equation}
The tension can still be recovered using eq. (\ref{eq:tensioncable}), that in this case gives
\begin{equation}
\theta(x)=\theta_0\sqrt{1+\left(\dfrac{q}{\theta_0}x+c_1\right)^2}.
\end{equation}

If, like in the case of the catenary, we consider a cable where $x_0=0,x_1=\alpha\ell,$ \-$ 0<\alpha<1,\ y_0=y_1=0,$ then it is easy to check that now
\begin{equation}
\begin{split}
&c_1=-\alpha\ \zeta,\\
&c_2=0,\\
&2\alpha\ \zeta\sqrt{1+\alpha^2\zeta^2}+\log\left(\frac{2\alpha\ \zeta}{\sqrt{1+\alpha^2\zeta^2}}+1\right)=4\zeta,
\end{split}
\end{equation}
where
\begin{equation}
\zeta=\frac{q\ell}{2\theta_0}.
\end{equation}

\subsection{The curve of uniform vertical load}
We consider a cable wrapped on a rigid plane  profile described by the function $y(x)$; the cable is stretched by a tension $\theta$ applied at its ends, so that the cable is everywhere in contact with the profile. The weight of the cable is negligible and the contact between the cable and the profile is frictionless. We want to determine $y(x)$ so that the vertical component of the contact force between the cable and the profile is constant everywhere.

Because there is no friction, the reaction on the cable is of the type
\begin{equation}
\label{eq:contactforce}
\bvp=-\varphi\bnu,\ \varphi>0;
\end{equation}
so, $f_\tau=0,\ f_\nu=-\varphi$ and by the first intrinsic balance equation (\ref{eq:intreq}) we get $\theta=const.$ and of course equal to the tension applied at the ends of the cable. Hence, this problem cannot be equal to the previous one of the suspension bridge, though at a first sight it could seem to be.

The second intrinsic equation gives now
\begin{equation}
\label{eq:intr2parab}
\varphi=c\ \theta,
\end{equation}
while $\bnu$ has components
\begin{equation}
\bnu=\frac{1}{\sqrt{1+y'^2}}(-y',1).
\end{equation}
The condition that determines the shape $y(x)$ of the curve is
\begin{equation}
-\bvp\cdot\ed=p,\ p>0,
\end{equation}
so by eq. (\ref{eq:contactforce})
\begin{equation}
\varphi\bnu\cdot\ed=p\ \rightarrow\ \frac{\varphi}{\sqrt{1+y'^2}}=p,
\end{equation}
that inserted into eq. (\ref{eq:intr2parab}) gives the differential equation
\begin{equation}
\frac{y''}{(1+y'^2)^2}=\frac{p}{\theta}.
\end{equation}

This differential equation has not a solution in closed form, and it should be resolved numerically; nevertheless, if the curve is slowly changing, i.e. if $|y'|\ll1$, then 
\begin{equation}
\label{eq:parabcable}
y''=\frac{p}{\theta}\ \rightarrow\ y=\frac{p}{2\theta}x^2+c_1x+c_2.
\end{equation}
So, in such an approximation, the solution is a parabola. The two constants $c_1$ and $c_2$ can be determined using the boundary conditions: $y(x_0)=y_0,\ y(x_1)=y_1$. The value of the vertical load $p$ can be obtained easily: for a parabola whose equation is $y=a\ x^2+c_1x+c_2$, because in the above approximation $c\simeq y''=2a$, we get comparing with eq. (\ref{eq:parabcable}),
\begin{equation}
p=2a \theta,
\end{equation}
so it depends linearly upon the tension that stretches the cable.

\subsection{The cable coiled on a rough cylinder}
We consider the case of a cable coiled on a rough cylinder, whose radius is $R$; the friction coefficient is $\sigma$ and we assume that the cable is coiled on a helix of pitch $2\pi b$, the helix being a geodetic of the cylinder. The cable is pulled by a tension $\theta_0$ at the end $s=0$; we want to know what is the highest tension $\theta_1$ that can be applied to the other end before slipping of the cable on the cylinder.

The equation of the helix is
\begin{equation}
p(\alpha)-o=R\cos\alpha\eu+R\sin\alpha\ed+b\alpha\et,
\end{equation}
$\alpha$ being the winding angle of the cable on the cylinder. The curvature of the helix is constant and equal to
\begin{equation}
c=\frac{R}{R^2+b^2},
\end{equation}
so that the non-slipping condition (\ref{eq:nonslipp}) becomes
\begin{equation}
\theta_1\leq\theta_0\ \mathrm{e}^{\sigma\int_0^\ell\frac{R}{R^2+b^2}ds}=\theta_0\ \mathrm{e}^{\sigma\frac{R}{R^2+b^2}\ell},
\end{equation}
$\ell$ being the length of the cable in contact with the cylinder. For a circular helix,
\begin{equation}
s=\sqrt{R^2+b^2}\alpha,
\end{equation}
so we get the non-slipping condition
\begin{equation}
\theta_1\leq\theta_0\ \mathrm{e}^{\sigma\frac{R}{\sqrt{R^2+b^2}}\alpha_\ell},
\end{equation}
with  
\begin{equation}
\alpha_\ell=\frac{\ell}{\sqrt{R^2+b^2}}
\end{equation}
the winding angle. The friction force that the cable can exert depends hence upon the exponential of $\alpha_\ell$. For the case of $b\ll R$, we get
\begin{equation}
\theta_1<\sim\theta_0\ \mathrm{e}^{\sigma\alpha_\ell}.
\end{equation}
Just as an exemple, if $\sigma=1/2$ and $\alpha_\ell=2\pi$, we get $\theta_1<\sim23.14\theta_0$: winding a cable on a rough cylinder is a very effective way to anchor it!

\section{Elastic cables}

The results in the previous sections concern inextensible cables; we consider now what happens if the cable is elastic. To fix the ideas, we consider the equation of an elastic cable whose unstretched length is $\ell$ and whose points are determined by the vector function
\begin{equation}
\gr{r}(s)=p(s)-o,\ \ s\in[0,\ell],
\end{equation}
and with end points
\begin{equation}
p_0=p(s=0)=(0,0),\ \ \ p_1=p(s=\ell)=(x_1,y_1).
\end{equation}
Once more the tangent vector is still defined by eq. (\ref{eq:tau}), but now, because the cable is extensible, $|p'(s)|\neq1$:
\begin{equation}
\btau(s)=\frac{p'(s)}{\lambda(s)},
\end{equation}
with 
\begin{equation}
\lambda(s)=|p'(s)|
\end{equation}
the {\it stretch} or {\it elongation} of the elastic cable. If the cable is inextensible, $\lambda(s)=1$, while generally speaking,
\begin{equation}
\label{eq:lambdacable}
\lambda(s)=\sqrt{\left(\frac{dx}{ds}\right)^2+\left(\frac{dy}{ds}\right)^2}.
\end{equation}

The unique balance equation for an elastic cable is still eq. (\ref{eq:locequil})$_1$, as it can be easily recognized:
\begin{equation}
\label{eq:equilcable}
\begin{split}
%&-\bt(s)+\bt(s+ds)+\gr{f}(s)ds=\gr{o}\ \rightarrow\\
%&-\bt(s)+\bt(s)+\frac{d\bt(s)}{ds}ds+\gr{f}(s)ds=\gr{o}\ \rightarrow\\
%&\ 
\bt'(s)+\gr{f}(s)=\gr{o}.
\end{split}
\end{equation}
%Intrinsically, this equation states that the equilibrium of the cable, no matter if it is extensible or not, is possible only thanks to the configuration of this last: in fact, the above vector equation can be satisfied, if $\gr{f}(s),\ \bt(s)$ and $\bt(s+ds)$ are not collinear, only thanks to a curvature of the cable: the statics of cables is a matter of geometry.

About the constitutive law, we restrict our attention to the case of {\it linearly elastic cables}. 
In such a case, the constitutive law, generalizing to elastic cables the classical  Hooke's law, is
\begin{equation}
\label{eq:constlawcable}
\theta(s)=\kappa(\lambda(s)-1)
\end{equation}
The parameter 
\begin{equation}
\kappa=E A
\end{equation}
is the {\it stiffness} of the cable, $E$ being the Young's modulus of the material and $A$ the area of the cross section of the cable.

\subsection{The catenary of an elastic cable}

As an application, we search for the catenary of an elastic cable (problem solved by Routh in 1891);  in this case, we can have also $\ell<\sqrt{x_1^2+y_1^2}$, because the cable can be stretched to join two points whose distance is $>\ell$ \footnote{However, $\ell$ cannot be much less than  $\sqrt{x_1^2+y_1^2}$ for applying the Hooke's law.}.
The unique load is still given by eq. (\ref{eq:forzacat}). Projecting eq. (\ref{eq:equilcable}) onto the two axes and taking into account that the directions of the axes are fixed, gives
\begin{equation}
\label{eq:eqelast}
\begin{split}
&(\bt'(s)+\gr{f}(s))\cdot\gr{e}_1=0\ \rightarrow\ \bt'(s)\cdot\gr{e}_1=0\ \rightarrow\ \frac{d}{ds}(\bt(s)\cdot\gr{e}_1)=0\\
&(\bt'(s)+\gr{f}(s))\cdot\gr{e}_2=0\ \rightarrow\ \bt'(s)\cdot\gr{e}_2=\mu g\ \rightarrow\ \frac{d}{ds}(\bt(s)\cdot\gr{e}_2)=\mu g
\end{split}
\end{equation}

The first of the above equations gives onces more the first integral of the problem of the catenary, eq. (\ref{eq:firstintcable}): the horizontal component of the tension is still a constant, also for an elastic cable.
We can transform eq. (\ref{eq:eqelast})$_1$ as follows:
\begin{equation}
\label{eq:eqelast1}
\begin{split}
\frac{d}{ds}(\bt\cdot\gr{e}_1)=&\frac{d}{ds}(\theta\btau\cdot\gr{e}_1)=\frac{d}{ds}\left(\theta\frac{dx}{ds}\right)=\frac{d}{ds}\left(\frac{\theta dx}{\sqrt{dx^2+dy^2}}\right)=\\
&\frac{d}{ds}\left(\frac{\theta\dfrac{dx}{ds}}{{\sqrt{\left(\dfrac{dx}{ds}\right)^2+\left(\dfrac{dy}{ds}\right)^2}}}\right)=\frac{d}{ds}\left(\frac{\theta\dfrac{dx}{ds}}{\lambda}\right)=0.
\end{split}
\end{equation}
In the same way, eq. (\ref{eq:eqelast})$_2$ can be transformed to
\begin{equation}
\label{eq:eqelast2}
\frac{d}{ds}\left(\frac{\theta\dfrac{dy}{ds}}{\lambda}\right)=\mu g.
\end{equation}
The first integrals of eqs. (\ref{eq:eqelast1}) and (\ref{eq:eqelast2}) are
\begin{equation}
\label{eq:eqelast3}
\begin{split}
&\frac{dx}{ds}=\lambda\frac{\theta_0}{\theta},\\
&\frac{dy}{ds}=\frac{V+\mu g\ s}{\theta}\lambda,
\end{split}
\end{equation}
with $\theta_0$ still the horizontal component of the tension and $V$ the vertical component of the tension at the end $s=0$.
To find a solution, we square and add the two equations above, to obtain
\begin{equation}
\theta^2\left[\left(\dfrac{dx}{ds}\right)^2+\left(\dfrac{dy}{ds}\right)^2\right]=\lambda^2\left[\theta_0^2+(V+\mu g\ s)^2\right],
\end{equation}
ant through eq. (\ref{eq:lambdacable}) finally
\begin{equation}
\label{eq:cabletheta}
\theta=\sqrt{\theta_0^2+(V+\mu g\ s)^2}.
\end{equation}

Inverting the constitutive law (\ref{eq:constlawcable}) we get
\begin{equation}
\lambda=\frac{\theta}{\kappa}+1,
\end{equation}
that injected into eq. (\ref{eq:eqelast3}) gives, through eq. (\ref{eq:cabletheta}),
\begin{equation}
\begin{split}
&\frac{dx}{ds}=\theta_0\left(\frac{1}{\kappa}+\frac{1}{\theta}\right)=\theta_0\left(\frac{1}{\kappa}+\frac{1}{\sqrt{\theta_0^2+(V+\mu g\ s)^2}}\right),\\
&\frac{dy}{ds}=(V+\mu g\ s)\left(\frac{1}{\kappa}+\frac{1}{\theta}\right)=(V+\mu g\ s)\left(\frac{1}{\kappa}+\frac{1}{\sqrt{\theta_0^2+(V+\mu g\ s)^2}}\right),
\end{split}
\end{equation}
These are the differential equations that describe the equilibrium curve; once integrated, they give
\begin{equation}
\begin{split}
&x=\frac{\theta_0s}{\kappa}+\frac{\theta_0}{\mu g}\left(\mathrm{arcsinh}\frac{V+\mu g\ s}{\theta_0}-\mathrm{arcsinh}\frac{V}{\theta_0}\right),\\
&y=\frac{\mu g\ s}{\kappa}\left(\frac{V}{\mu g}+\frac{s}{2}\right)+\frac{\theta_0}{\mu g}\left(\sqrt{1+\left(\frac{V+\mu g\ s}{\theta_0}\right)^2}-\sqrt{1+\left(\frac{V}{\theta_0}\right)^2}\right).
\end{split}
\end{equation}

The two still unknown constants $\theta_0$ and $V$ can be determined imposing the boundary conditions for $s=\ell:\ x(s=\ell)=x_1,\ y(s=\ell)=y_1$, which gives the two conditions
\begin{equation}
\label{eq:condcontcablelast}
\begin{split}
&x_1=\frac{\theta_0\ell}{\kappa}+\frac{\theta_0}{\mu g}\left(\mathrm{arcsinh}\frac{V+\mu g\ell}{\theta_0}-\mathrm{arcsinh}\frac{V}{\theta_0}\right),\\
&y_1=\frac{\mu g\ell}{\kappa}\left(\frac{V}{\mu g}+\frac{\ell}{2}\right)+\frac{\theta_0}{\mu g}\left(\sqrt{1+\left(\frac{V+\mu g\ell}{\theta_0}\right)^2}-\sqrt{1+\left(\frac{V}{\theta_0}\right)^2}\right).
\end{split}
\end{equation}
These two conditions must usually be solved numerically; in the case of $y_1=0$, i.e. with the two end supports at the same level, we get, from eq. (\ref{eq:condcontcablelast})$_2$, the expected condition
\begin{equation}
V=-\frac{1}{2}\mu g\ell,
\end{equation}
while eq. (\ref{eq:condcontcablelast})$_1$ becomes
\begin{equation}
\frac{\mu g\ell}{2\theta_0}=\sinh\left(\frac{\mu gx_1}{2\theta_0}-\frac{\mu g\ell}{2\kappa}\right),
\end{equation}
equation that for $\kappa\rightarrow\infty$ coincides with eq. (\ref{eq:eqtensio}), being $x_1=\alpha\ell$.

\begin{figure}[h]
\begin{center}
\includegraphics[scale=1]{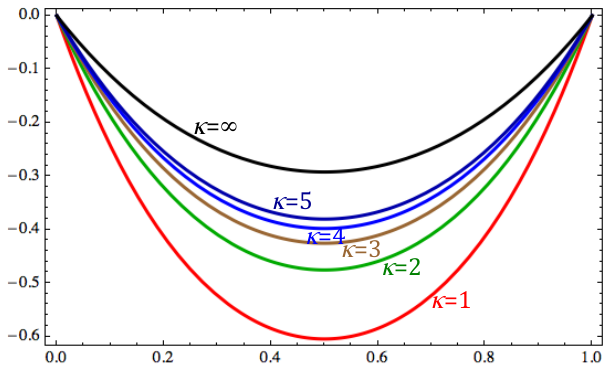}
\caption{Elastic catenaries.}
\label{fig:f2_5}
\end{center}
\end{figure}

In Fig. \ref{fig:f2_5} we show the catenaries of different cables: all the cables are hung between the points $(0,0)$ and $(0,1)$, and the unstretched length of the cables is always $\ell=1.2$, while $\mu g=1$ (all the dimensions are in appropriate units), while the cables differ for the value of $\kappa$. As expected, the more the $\kappa$, the less the depth of the cable; the black curve corresponds to the catenary of an inextensible cable.

\section{Exercices}
\begin{enumerate}
\item An inextensible  cable of length $2\ell$ is fixed at two points, e.g. two pitons, at the same height and at a   distance $2d<2\ell$. The cable is taut by a concentrated force $\gr{f}=f(\cos\alpha,\sin\alpha)$ at mid distance from the ends, see Fig. \ref{fig:f2_6}. Determine, as a function of $\alpha$, the equilibrium configuration and the tension in each one of the two parts of the cable. Such a system is currently used by alpinists to realize a relais during a climbing; could you say which is the minimum length of the rope of the relais to make it safer than a relais using a single piton (consider a vertical force)?

\begin{figure}[h]
\begin{center}
\includegraphics[scale=1]{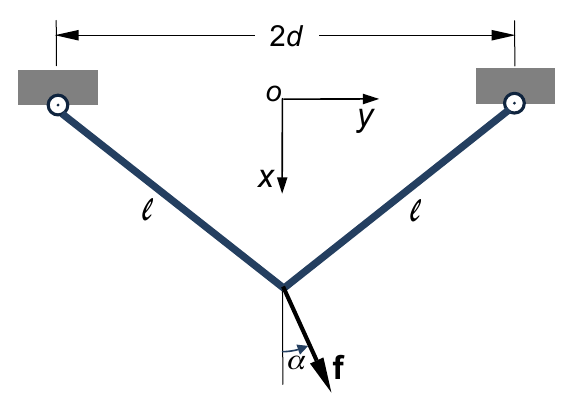}
\caption{Scheme of exercice 1.}
\label{fig:f2_6}
\end{center}
\end{figure}

\item Make the same exercice but this time the cable is linearly elastic with elastic constant $\kappa$. Trace the curve describing the dependence of the force upon the displacement of its point (consider a vertical force).

\item The {\it velaria} is the equilibrium configuration of a cable acted upon uniquely by wind; if the action of the wind can be represented by a uniform load $p$ acting orthogonally to the cable (a characteristic of all the actions of inviscid fluids), show that the velaria is an arc of circle (this problem was solved by Jh. Bernoulli).

\item A variant of the above problem is that of a parallel stream; in such a case, show that the equilibrium configuration of the velaria is actually the catenary.

\item A cable is coiled on a rough cylinder and the friction coefficient is $\sigma$. The cylinder is restrained to rotation by an angular spring whose elastic constant is $\mu$ and the cable must bear, at the free end, a mass $m$, while at the other end it is fixed to the cylinder by a device whose resistance to tension is $\varphi$. If the radius of the cylinder is $R$ and the cable is coiled three times on the cylinder along a helix whose pitch is  $2\pi b\ll R$, which is the greatest mass that the cable can support?

\end{enumerate}

\chapter{Straight rods}
\label{ch:3}

\section{Introduction}
The results of the Saint-Venant Problem greatly simplify the study of beams. In fact, the whole stress state at any point of the beam can be found if the internal actions, $N, T_1, T_2, M_1, M_2$ and $M_3$ are known.

The problem of the study of structures composed by beams is hence reduced to the study of the internal actions. These ones depend only upon the position along the beam axis, say the axis $z$. So the equations concerning $N, T_1$ etc. can be only ordinary differential equations (ODEs), not partial differential equations (PDEs), which simplifies considerably the problem and motivates for the study of beams reduced, ideally, to their axis. 

Such theories, idealizing a beam as a one-dimensional element, are called {\it rod theories} (a rod is considered here to be a beam reduced to its only axis). The objective of the rod theories is hence to provide balance, compatibility and constitutive equations for rods, i.e. for this special type of continuum.

In many practical cases, rods belong to a plane that contains one of their principal axes of inertia of the cross section and are acted upon by loads that belong to such a plane. This is the case of {\it plane rods}: the rods belong, also after the deformation, to their original plane, where the loads act. 

The case of plane rods is much simpler than the general one, because the only possible internal actions reduce to only $N, T_2$ and $M_1$ (that we will indicate, in the following, simply by $N, T$ and $M$, because there is no possibility of ambiguity in the plane case). 

In the remainder of this Chapter, we will focus on  a particularly important case of plane rods, that of {\it straight rods}; nevertheless, it is not difficult to generalize the results to the more general cases of plane or also of three-dimensional rods, following the same approach illustrated below. 

The objective is to write the balance, compatibility and constitutive equations for straight rods, to arrive to a mechanical model for such elements. We will, namely, introduce two classical models of rods, the more general Timoshenko's one and the very classical Euler-Bernoulli rod model.

A last remark: we remain, in this Chapter, in the framework of classical, linear elasticity. We will generalize the linear theory of rods  to arches, i.e. plane curved rods, in Chapter \ref{ch:4}, while  general rods undergoing large displacements are studied in Chapter \ref{ch:5}.

\section{Balance equations}
Let us begin the study of straight rods with the balance equations. The general situation is sketched in Fig. \ref{fig:f27}; loads $p(z)$ and $q(z)$ are the data of the problem. The balance equations can be obtained applying the principle of the sections of Euler to a piece of road between the positions $z$ and $z+dz$. The assumed positive internal actions are those depicted in the figure. The equilibrium of the rod implies that of the segment under scrutiny, submitted to the external loads and to the internal actions transmitted to the segment by the rest of the rod through the end sections. The balance gives hence:
\begin{figure}[h]
	\begin{center}
         \includegraphics[scale=1]{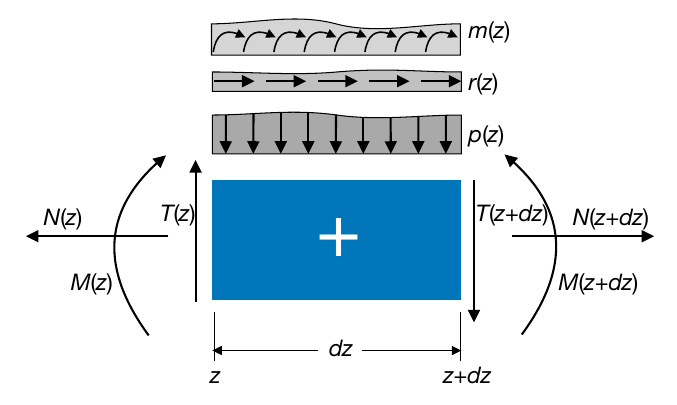}
	\caption{General sketch for the rod's balance equations.}
	\label{fig:f27}
	\end{center}
\end{figure}
\begin{itemize}
\item equilibrium to axial force:
\begin{equation}
N(z+dz)-N(z)+r(z)dz=0;
\end{equation}
\item equilibrium to shear force:
\begin{equation}
T(z+dz)-T(z)+p(z)dz=0;
\end{equation}
\item equilibrium to bending moment (e. g. around the point of abscissa $z$):
\begin{equation}
M(z+dz)-M(z)-T(z+dz)dz-p(z)\frac{dz^2}{2}-m(z)dz=0.
\end{equation}
\end{itemize}

Developing the above expressions gives
\begin{equation}
\begin{split}
&N(z)+\frac{dN(z)}{dz}dz-N(z)+r(z)dz=0,\\
&T(z)+\frac{dT(z)}{dz}dz-T(z)+p(z)dz=0,\\
&M(z)+\frac{dM(z)}{dz}dz-M(z)-T(z)dz-\frac{dT(z)}{dz}dz^2-p(z)\frac{dz^2}{2}-m(z)dz=0,
\end{split}
\end{equation}
and neglecting the terms of order greater than the first we finally obtain the {\it balance equations for straight rods}:
\begin{equation}
\label{eq:balancerod}
\begin{split}
&\frac{dN}{dz}=-r,\\
&\frac{dT}{dz}=-p,\\
&\frac{dM}{dz}=T+m.
\end{split}
\end{equation}
In the special and very common case of $m=0$, we remark that $T$ is the derivative of $M$.
From the two last relations, we get also, by differentiation, 
\begin{equation}
\label{eq:equilM2}
\frac{d^2M}{dz^2}=-p+\frac{dm}{dz},
\end{equation}
a second-order differential equilibrium equation relating directly the bending moment to the loads.

\section{Compatibility equations}
\begin{figure}[h]
	\begin{center}
         \includegraphics[width=.8\textwidth]{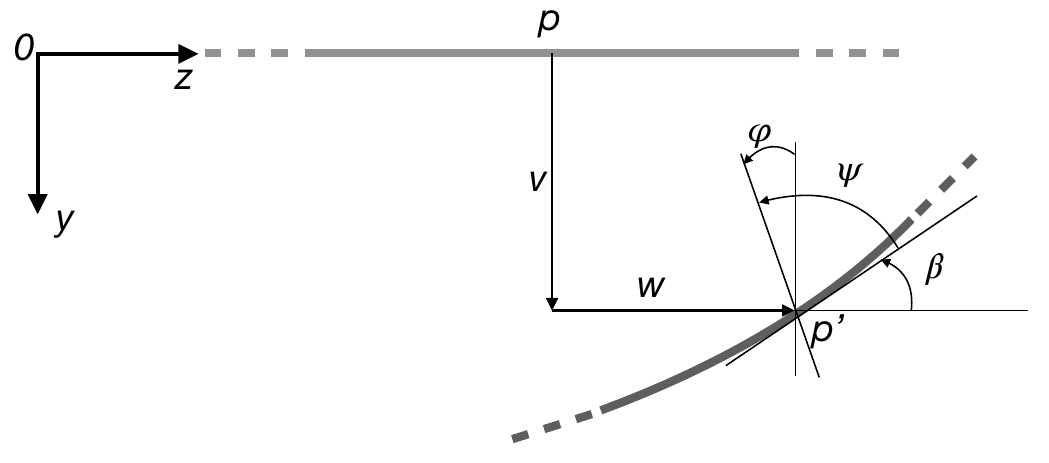}
	\caption{General scheme of the kinematics of a rod.}
	\label{fig:f28}
	\end{center}
\end{figure}

Let us now turn the attention on geometric considerations. In fact, we need a link, the {\it compatibility equations of the rods}, between the displacements of the rod and some internal kinematical quantities defining the deformation of the rod. The general situation is that sketched in Fig. \ref{fig:f28}, where $w$ is the axial displacement, $v$ the {\it deflexion}, i.e. the displacement along $y$, $\beta$ is the local rotation of the axis $z$ and $\varphi$ that of the normal to the undeformed axis $z$.

We introduce first the {\it extension} $\eps$, the internal kinematical descriptor of the stretching of the axis $z$:
\begin{equation}
\label{eq:compat1}
\eps=\frac{dw}{dz}.
\end{equation}
Then, we define the {\it curvature} $\kappa$ of the rod
\begin{equation}
\label{eq:compat2}
\kappa=-\frac{d\varphi}{dz};
\end{equation}
the sign $-$ is due to the fact that in the Saint-Venant Problem the positive bending moments are opposite to the positive concavity, see below.

Finally, we introduce the {\it angular sliding} or {\it shear} $\gamma$ of the rod axis, describing how a segment initially parallel to the rod axis changes in the deformation:
\begin{equation}
\gamma=\beta-\varphi.
\end{equation}
From fig. \ref{fig:f28} we see that 
\begin{equation}
\psi=\frac{\pi}{2}+\varphi-\beta\ \rightarrow\ \gamma=\beta-\varphi=\frac{\pi}{2}-\psi:
\end{equation}
$\gamma$ measures the variation of the angle existing between the axis and a segment orthogonal to it, from its initial value of $\pi/2$ to the final one of $\psi$.
For small perturbations, 
\begin{equation}
\tan\beta=\frac{dv}{dz}\simeq\beta,
\end{equation}
so we have
\begin{equation}
\label{eq:compat3}
\gamma=\frac{dv}{dz}-\varphi.
\end{equation}

We remark that there is a substantial difference between a rod and a classical continuum: in the rod theories, derivatives of angular quantities appear: a rod is a {\it polar continuum}, i.e., unlike classical continuum bodies, it can transmit couples.

\section{Constitutive equations}
We have for the while only six equations, the balance, eq. (\ref{eq:balancerod}), and the compatibility ones, eqs. (\ref{eq:compat1}), (\ref{eq:compat2}) and (\ref{eq:compat3}), for a set of 9 unknowns on the whole: $N, T, M, v, w,\varphi,\eps,\kappa$ and $\gamma$. We need hence three {\it constitutive equations for the rods}; they can be derived using the results of the Saint-Venant Problem. The approach is energetic: we write first the strain energy for a beam, $U_b$, between the two sections $1$ and $2$:
\begin{equation} 
\begin{split}
U_b&=\frac{1}{2}\int_\Omega\bsig\cdot\beps\ dv=\frac{1}{2}\int_\Omega\frac{\sigma_{33}^2}{E}+\frac{1}{\mu}(\sigma_{13}^2+\sigma_{23}^2)\ dv\\
&=\frac{1}{2}\int_1^2\int_S\left(\frac{N}{A}+\frac{M\ y}{J}\right)^2\frac{1}{E}+\frac{1}{\mu}\left(\frac{T\mathcal{S}}{b\ J_1}\right)^2\left(1+\frac{4\tan^2\alpha\ x^2}{b^2}\right)ds\ dz\\
&=\frac{1}{2}\int_1^2\frac{N^2}{EA}+\frac{\chi T^2}{\mu A}+\frac{M^2}{EJ}\ dz.
\end{split}
\end{equation}

Now, thinking at the beam as a rod, we write the  energy $U_r$:
\begin{equation}
U_r=\frac{1}{2}\int_1^2(N\eps+T\gamma+M\kappa)\ dz.
\end{equation}
This result can be obtained applying the Clapeyron's theorem at the segment $dx$ of the rod, considered as charged uniquely by the internal actions. 
Of course, the solid being the same, it must be
\begin{equation}
U_b=U_r\ \ \ \forall N, T, M,
\end{equation}
which gives the conditions
\begin{equation}
N\eps=\frac{N^2}{EA},\ \ T\gamma=\frac{\chi T^2}{\mu A},\ \ M\kappa=\frac{M^2}{EJ},
\end{equation}
and finally the three constitutive equations for the rods:
\begin{equation}
\label{eq:consteq}
\begin{split}
&N=EA\eps,\\
&T=\frac{\mu A}{\chi}\gamma,\\
&M=EJ\kappa.
\end{split}
\end{equation}

We remark that, because of the linearity of the problem, the internal actions are proportional to their corresponding kinematical parameter. We dispose now of all the equations for the rod theories.

\section{The Timoshenko's rod}
\label{sec:timo}
The compatibility equations can be injected into the constitutive equations to get (from now on, for the sake of shortness, we denote by a prime the derivative with respect to $z$):
\begin{equation}
\label{eq:timo1}
\begin{split}
&N=EAw',\\
&T=\frac{\mu A}{\chi}(v'-\varphi),\\
&M=-EJ\varphi'.
\end{split}
\end{equation}

If now we inject these equations into the balance equations, we get
\begin{equation}
\label{eq:timo2}
\begin{split}
&(EAw')'=-r,\\
&\left[\frac{\mu A}{\chi}(v'-\varphi)\right]'=-p,\\
&(-EJ\varphi')'=T+m=\frac{\mu A}{\chi}(v'-\varphi)+m.
\end{split}
\end{equation}
Injecting eq. (\ref{eq:timo2})$_2$ into the differentiated eq. (\ref{eq:timo2})$_3$ gives an equation for $\varphi$ only:
\begin{equation}
\label{eq:timo3}
(EJ\varphi')''=p-m'.
\end{equation}
The above equations are the {\it elastic equilibrium equations} of a straight rod; in fact, they include in the equilibrium equations the constitutive law and the compatibility equations, so they describe the equilibrium of an elastically deformable rod.

Equations (\ref{eq:timo1}) and (\ref{eq:timo2}) define the so-called {\it Timoshenko's rod model}. In this model, any straight segment originally orthogonal to the rod axis remains straight after the deformation, but $not$ necessarily orthogonal to the tangent of the deformed axis, because $\gamma\neq0$, generally speaking, which implies that $v'\simeq\beta\neq\varphi$.

$N$ and $w$ are uncoupled from $T, M, v$ and $\varphi$, but these last are coupled, which complicates the resolution. Anyway, from eq. (\ref{eq:timo1})$_2$ one gets 
\begin{equation}
\label{eq:timo6}
\varphi=v'-\frac{\chi T}{\mu A}\ \rightarrow\ \varphi'=v''-\left(\frac{\chi T}{\mu A}\right)',
\end{equation}
that inserted into eq. (\ref{eq:timo1})$_3$ gives
\begin{equation}
\label{eq:timo5}
v''=-\frac{M}{EJ}+\left(\frac{\chi T}{\mu A}\right)'.
\end{equation}
In the particular, and very frequent, case of a homogeneous rod of uniform section, $EJ$ and $\mu A/\chi$ are independent from $z$; then, from eq. (\ref{eq:timo2})$_2$ we get
\begin{equation}
\varphi'=v''+\frac{\chi p}{\mu A},
\end{equation}
 that inserted into eq. (\ref{eq:timo1})$_{3}$ gives
\begin{equation}
v''=-\frac{M}{EJ}-\frac{\chi }{\mu A}p.
\end{equation}

The above expressions of $v''$ can be used to find the deflection of the rod, $v$, whenever the functions $M(z)$ and $T(z)$ are known, namely from equilibrium equations, e.g. integrating eq. (\ref{eq:equilM2}). In this procedure, we need two {\it geometric} boundary conditions, fixing the value of $v$ or of $\varphi$ at the rod's ends, plus two {\it natural} boundary conditions, i.e. on $M$ or $T$, if the equilibrium problem is solved integrating eq. (\ref{eq:equilM2}). The rotation $\varphi$ can be obtained from eq. (\ref{eq:timo6}) or directly upon integration of eq. (\ref{eq:timo1})$_3$, which needs just one boundary condition, specifying the value of $\varphi$ at a point of the rod (not necessarily at the edges).

If, on the contrary, the functions  $M(z)$ and $T(z)$ cannot be determined upon simple equilibrium considerations, then 
integrating three times eq. (\ref{eq:timo3}) gives an expression for $\varphi$, to be injected into eq. (\ref{eq:timo2})$_3$ to obtain an equation for $v'$:
\begin{equation}
\label{eq:timo4}
v'=\varphi-\frac{\chi}{\mu A}\left((EJ\varphi')'+m\right),
\end{equation}
that integrated gives the deflection $v$. Because we have integrated four times, four boundary conditions are needed to determine the four integration constants. These can be of two types: geometric, fixing the value of $v$ or $\varphi$ at the rod's ends, or natural, imposing a value for $M$ or $T$ at the boundaries. The natural boundary conditions concern $\varphi$ or $\varphi'$; in fact, if at a boundary $z=z_b$ it is $M(z_b)=M_b$, then from eq. (\ref{eq:timo1})$_3$ we get 
\begin{equation}
\varphi'(z_b)=-\frac{M_b}{EJ},
\end{equation}
while if it is $T(z_b)=T_b$ then eq. (\ref{eq:timo2})$_3$ gives the condition
\begin{equation}
(EJ\varphi')'|_{z_b}=-T_b-m(z_b).
\end{equation}

\section{The Euler-Bernoulli rod}
\label{sec:EB}
Since the XVIII$^{th}$ century a simplified model has been proposed by L. Euler and Jacob Bernoulli for the bending case: the basic assumption of the Euler-Bernoulli rod theory is that
\begin{equation}
\beta=\varphi\ \Rightarrow\ \gamma=0.
\end{equation}

Geometrically, this corresponds to  the vanishing of the angular sliding, i.e., finally, of the shear deformation: unlike in the Timoshenko's model, {\it a segment originally orthogonal to the axis remains orthogonal to the deformed axis}. This is the so-called {\it  hypothesis of conservation of the normals}.

For small perturbations, this assumption gives 
\begin{equation}
\label{eq:EB0}
\varphi=\beta\simeq v'\ \rightarrow\ \kappa=-\varphi'=-v''.
\end{equation}
The equations for $N$ and $\eps$ are not affected by this assumption, while for $M$ and $v$ we get
\begin{equation}
\label{eq:eb1}
\begin{split}
&M=-EJv'',\\
&(EJv'')''=p-m'.
\end{split}
\end{equation}
These are the celebrated equations of the {\it Euler-Bernoulli rod model}. The problems for $N, T$ and $M$ are  uncoupled and the bending problem is reduced to a fourth-order differential equation for $v$. For what concerns $T$, the constitutive equation cannot be used, because this should give $T=0$ identically, which would imply $M=$ constant, which is false\footnote{The assumption (\ref{eq:EB0})  renders eqs. (\ref{eq:timo1})$_2$ and (\ref{eq:timo2})$_2$ impossible to be satisfied, and hence meaningless, unless $T=0$ and $p=0$ everywhere. Hence, in principle the Euler-Bernoulli model is correct only for rods submitted to a pure bending state. However, the model is used also when $T\neq 0$; in such a case, $T$ can be retrieved only through equilibrium, once $M$ determined, eq. (\ref{eq:shearEB}). Practically, the use of the Euler-Bernoulli rod model should be restricted to cases where the shear deformation is negligible with respect to the bending one: this happens for {\it slender rods}, i.e. for rods where the {\it slenderness ratio} $\rho/\ell\rightarrow0$, with $\rho=\sqrt{J/A}$  the radius of gyration of the rod's cross-section,  see Ex. \ref{ex:A}.}. The shear distribution can however be get through the equilibrium equation (\ref{eq:balancerod})$_3$ and using eq. (\ref{eq:eb1})$_1$:
\begin{equation}
\label{eq:shearEB}
T=M'-m=(-EJv'')'-m.
\end{equation}

The above equations are valid for straight rods of any cross section, also variable with $z$. For the very common case of constant cross section and material, they simplify to 
\begin{equation}
\label{eq:eb2}
\begin{split}
&M=-EJv'',\\
&T=-EJv'''-m,\\
&EJv^{ \textrm{iv}}=p-m'.
\end{split}
\end{equation}
Also in this case, if the function $M$ can be obtained by equilibrium considerations, then the deflection $v$ is obtained by eq. (\ref{eq:eb1})$_1$,
\begin{equation}
\label{eq:EBv}
v''=-\frac{M}{EJ},
\end{equation}
otherwise eq. (\ref{eq:eb1})$_2$ must be integrated four times. The geometric or natural boundary conditions now concern always $v$ and its derivatives. In fact, the geometric conditions fix the value of $v$ or of $v'$ at one rod's end, while the natural boundary conditions on $M$ and $T$, through eq. (\ref{eq:eb1})$_2$ and (\ref{eq:shearEB}), fix the value of $v''$ or of $v'''$ respectively.

The solution of eqs. (\ref{eq:eb1}) or (\ref{eq:eb2}), or in the general case of the Timoshenko's model, provides the displacement of the rod, determining hence its deformed shape, the so-called $elastica$. %Through the constitutive laws, the corresponding internal actions can also be obtained. The compatibility equations give finally the internal kinematical descriptors of the deformation. The basic variables are hence the components of displacement: $v, w$ and, in the general theory, $\varphi$.

\section{Reduction of the Timoshenko's problem}
It is possible to reduce eq. (\ref{eq:timo3}) of the Timoshenko's rod theory to an equation of the fourth order similar to that of the Euler-Bernoulli model, eq. (\ref{eq:eb1})$_2$. This can be done introducing the {\it auxiliary function} $\eta(z)$ such that
\begin{equation}
\label{eq:auxi1}
\varphi=\eta',
\end{equation}
which inserted into eq. (\ref{eq:timo3}) gives immediately
\begin{equation}
(EJ\eta'')''=p-m',
\end{equation}
formally identical to eq. (\ref{eq:eb1})$_2$, provided that $v$ is replaced by $\eta$. If eq. (\ref{eq:auxi1}) is injected into eq. (\ref{eq:timo1})$_3$, then we get
\begin{equation}
\label{eq:auxi2}
M=-EJ\eta'',
\end{equation}
analogous to eq. (\ref{eq:eb1})$_1$. Moreover, if eq. (\ref{eq:auxi1}) is inserted into eq. (\ref{eq:timo4}) we obtain the link between $v$ and $\eta$:
\begin{equation}
\label{eq:vshear}
v'=\eta'-\frac{\chi}{\mu A}((EJ\eta'')'+m).
\end{equation}
Putting this expression of $v'$ into eq. (\ref{eq:timo1})$_2$ gives
\begin{equation}
\label{eq:auxi3}
T=(-EJ\eta'')'-m=M'-m,
\end{equation}
i.e. the equilibrium equation (\ref{eq:balancerod})$_3$ is automatically satisfied.

For the very common case of a rod of constant stiffness, i.e. such that $EJ$ and $\mu A/\chi$ are constant $\forall z$, and without distributed couples, $m=0$, the above equation gives (the integration constant is inessential, as it can be easily recognized considering that the link imposed by equation (\ref{eq:vshear}) concerns $v'$)
\begin{equation}
v=\eta-\frac{\chi EJ}{\mu A}\eta''.
\end{equation}
Along with eqs. (\ref{eq:auxi1}), (\ref{eq:auxi2}) and (\ref{eq:auxi3}),   this gives the boundary conditions to be satisfied at the rod's ends:
\begin{itemize}
\item clamped edge:
\begin{equation}
v=0,\ \ \varphi=0\ \ \rightarrow\ \ \eta-\frac{\chi EJ}{\mu A}\eta''=0,\ \ \eta'=0;
\end{equation}
\item simply supported edge:
\begin{equation}
v=0,\ \ M=0\ \ \rightarrow\ \ \eta''=0,\ \ \eta=0;
\end{equation}
\item slide edge:
\begin{equation}
\varphi=0,\ \ T=0\ \ \rightarrow\ \ \eta'=0,\ \ \eta'''=0;
\end{equation}
\item free edge:
\begin{equation}
M=0,\ \ T=0\ \ \rightarrow\ \ \eta''=0,\ \ \eta'''=0.
\end{equation}
\end{itemize}

\section{Isostatic and hyperstatic rods}
\label{sec:ipoiper}
The general problem for a rod is: knowing the applied actions $p, q, m$ and the boundary conditions, determine $N, T, M, v, w$ and $\varphi$. Three cases are possible: the rod is {\it hypostatic}, {\it isostatic} or {\it hyperstatic}. 

Generally speaking, a structure is said to be {\it isostatic} if the equilibrium equations can be solved uniquely, i.e. if they are sufficient to determine the distribution of the internal actions and of the reaction forces. In the case of a rod, the equilibrium equations are
\begin{equation}
\label{eq:equiliso}
\begin{split}
&N'=-r,\\
%&T'=-p,\\
&M''=-p+m'.
\end{split}
\end{equation}
$T$ can be obtained  through eq. (\ref{eq:balancerod})$_3$:
\begin{equation}
\label{eq:equilT}
T=M'-m,
\end{equation}
once solved the bending equation. Such equations concern internal forces and need, on the whole, three boundary conditions, one for  eq. (\ref{eq:equiliso})$_1$ and two for  eq. (\ref{eq:equiliso})$_2$. These conditions are necessarily of the {\it natural} type, i.e. they concern the value taken by $N,T$ or $M$ at the rod's ends. So, a rod will be isostatic if and only if it is possible to specify the right number of natural boundary conditions for eqs. (\ref{eq:equiliso}), in particular if it is possible to write:
\begin{itemize}
\item one and only one boundary condition specifying the value of $N$ at one of the rod's ends;
\item two and no more than two boundary conditions specifying the value of $M$ or $T$ at the rod's ends, but at least one of them must concern $M$.
\end{itemize}
In fact, only in such a case it is possible to determine a unique solution to the equilibrium equations and hence to determine uniquely the distribution of the internal actions $N,T$ and $M$ everywhere in the rod.

If the number of natural boundary conditions that can be written is greater than three, then eqs. (\ref{eq:equiliso}) cannot satisfy all of them, generally speaking. Equilibrium is {\it impossible} and the rod is said to be {\it hypostatic}: the constraint conditions are not sufficient to guarantee equilibrium for every possible external loading.

If, on the contrary, the number of natural boundary conditions that can be written is less than three, then the number of independent constraint conditions is too high to allow writing natural boundary conditions. The rod is said to be {\it hyperstatic}, in the sense that it is  {\it statically undetermined}: the equations of statics, that is, the equilibrium equations, are not sufficient to determine the distribution of the internal actions and of the reaction forces. 

In such a case, the compatibility equations and the constitutive law must enter the problem, i.e., the elastic equilibrium equations (\ref{eq:timo2}) or (\ref{eq:eb1})$_2$ should be used. In particular: 
\begin{itemize}
\item for extension, eq. (\ref{eq:timo2})$_1$ is to be used, which needs two  boundary conditions, of the geometric or natural type, specifying respectively the value of $w$ or of $w'$ at the rod's ends; at least one of the two boundary conditions must be concern  $w$ i.e. it must be of the {\it geometric} type;
\item for bending, eqs. (\ref{eq:timo3}) and (\ref{eq:timo4}) are to be used for the Timoshenko's rod, while eq. (\ref{eq:eb1})$_2$ for the Euler-Bernoulli one. In both the cases, four boundary conditions must be specified; they can be of the natural or geometric type, they have been  discussed in Sects. \ref{sec:timo} and \ref{sec:EB}, but at least  one of them must concern $v$, i.e. it must be of the {\it geometric} type. %In addition, the boundary conditions must be {\it independent}, i.e. they must lead to a linear system of four equations whose determinant is not null, otherwise the problem is {\it undetermined} and the equilibrium impossible.
\end{itemize}

For isostatic rods, once the distribution of $M$ determined, the deflection can be calculated through eq. (\ref{eq:EBv}) for the Euler-Bernoulli rod, or through eq. (\ref{eq:timo5}) for the Timoshenko's one; in this case, the rotation $\varphi$ is then determined through eq. (\ref{eq:timo6}), see Sect. \ref{sec:timo}.

We remark that an imposed displacement of a rod's edge {\it stresses an hyperstatic rod, but not an isostatic one}. In fact, an imposed displacement corresponds to a geometric boundary condition, that concerns only hyperstatic rods. In the case of an isostatic one, only natural boundary conditions, i.e. concerning $N,T$ or $M$, are needed to determine the distribution of the internal actions; as a consequence, these last are insensitive to any imposed displacement of rod's edge.

For ending this Section, we remark  that the difference in the resolution of isostatic or hyperstatic rods is in the number and types of boundary conditions to be specified, besides the differential equations, of the first or second order in the first case, of the second or fourth order for the second one. 

It is important to notice that, because equilibrium equations are sufficient to determine the distribution of the internal actions for isostatic rods, such distributions are not affected by the stiffness characteristics of the rod, i.e. $N,T$ and $M$ are independent from the distributions $EA,\mu A/\chi$ and $EJ$, also in the case where these last are not constant throughout the rod. This is not the case for hyperstatic rods (and in general for hyperstatic structures): the distribution of the internal actions, and hence the reaction forces, depend upon the distribution of the stiffnesses. However, for the Euler-Bernoulli model, if the rod has constant properties, $M$ and $T$ are still independent from the stiffnesses, as it can be easily checked; this is true also for the extension behavior.

\section{The torsion equations}
The results found for straight rods can be easily generalized to include an out-of-plane effect, that of torsion $M_T$. The general scheme is sketched in Fig. \ref{fig:f29}:
\begin{figure}[h]
	\begin{center}
         \includegraphics[scale=1]{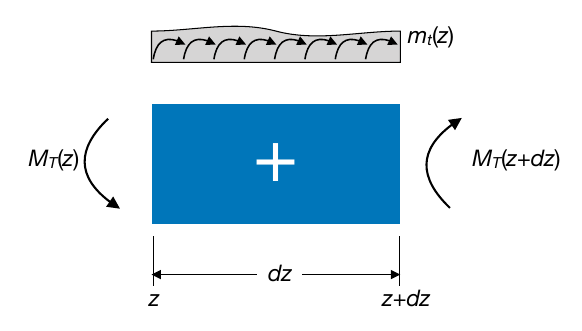}
	\caption{Scheme for the torsion of a rod.}
	\label{fig:f29}
	\end{center}
\end{figure}

\begin{itemize}
\item balance of the torsion:
\begin{equation}
\begin{split}
&M_T(z+dz)-M_T(z)+m_t(z)dz=0\ \rightarrow\\
&M_T(z)+\frac{dM_T(z)}{dz}dz-M_T(z)+m_t(z)dz=0\ \rightarrow\\
&\frac{dM_T}{dz}=-m_t;
\end{split}
\end{equation}
\item compatibility equations: the kinematical descriptor of torsion is the {\it twist angle} $\theta$; it is linked to the internal descriptor $\alpha$, giving the relative rotation of two sections separated by a distance $dz$ by the relation
\begin{equation}
\alpha=\frac{d\theta}{dz};
\end{equation}
\item constitutive law: the strain energy of torsion for a beam is (see Ex. 8 Chapt. 4)
\begin{equation}
U_b=\frac{1}{2}\int_1^2\frac{q\ M_T^2}{\mu J_0}dz,
\end{equation}
and as a rod
\begin{equation}
U_r=\frac{1}{2}\int_1^2M_T\ \alpha\ dz,
\end{equation}
so we get
\begin{equation}
M_T=\frac{\mu\  J_0}{q}\alpha,
\end{equation}
a result already known from the Sant-Venant theory.
\end{itemize}

Finally we have
\begin{equation}
\begin{split}
& M_T=\frac{\mu\  J_0}{q}\theta',\\
&\left(\frac{\mu\  J_0}{q}\theta'\right)'=-m_t,
\end{split}
\end{equation}
These equations for torsion are formally identical to those for extension, eqs. (\ref{eq:timo1})$_1$ and (\ref{eq:timo2})$_1$.

\section{The Mohr's theorems}
\label{sec:mohr}
Let us now consider the case of a bent rod of constant stiffness $EJ$ and without distributed couples; then, in the framework of the Euler-Bernoulli model,
\begin{equation}
M'=T,\ \ \ T'=-p\ \rightarrow\ M''=-p,\ \ v''=-\frac{M}{EJ}.
\end{equation}
These two differential equations are formally identical; so, the {\it elastica} of a rod coincides with the diagram of the bending moment $M^*$ generated by a fictitious load 
\begin{equation}
\label{eq:mohr1}
p^*=\frac{M}{EJ}\ \rightarrow\ v=M^*.
\end{equation}
This is the {\it Theorem of Mohr} (1868); to remark that $p^*$ is the curvature. Deriving eq. (\ref{eq:mohr1})$_2$ gives
\begin{equation}
v'=(M^*)'=T^*,
\end{equation}
i.e. the inclination $\varphi$ of the {\it elastica} is given by the fictitious shear $T^*$ ($\beta\simeq\tan\beta=v'$ for small perturbations). This is the {\it corollary of Mohr}.

The fictitious load $p^*$ is to be applied to a fictitious rod, having the same dimensions of the real rod but whose boundary conditions must in general be changed:
\begin{itemize}
\item for a simply supported rod, the {\it elastica} has $v=0$ and $v'\neq0$ at the edges;  the corresponding edges of the fictitious rod can be found considering that in the correspondence it must be  $M^*=0$ and $T^*\neq0$; hence  the fictitious rod must be simply supported, like the real rod;
\item for a cantilever: at the clamped edge, $v=0$ and $v'=0$: the corresponding edge in the fictitious rod must be a free edge, because in such a way $M^*=0$ and $T^*=0$; at the free edge of the cantilever, $v\neq0$ and $v'\neq0$: in the corresponding edge of the fictitious rod it must be $M^*\neq0$ and $T^*\neq0$, so this edge  must be clamped: a cantilever rod is transformed into a fictitious cantilever where the edges are swapped.
\end{itemize}
Other situations can be studied in a similar way. The use of the Mohr's theorem, and corollary, is normally bounded to isostatic rods. In fact, in such a case the use of the Mohr's technique allows for finding deflections and rotations  using exclusively equilibrium considerations, so without the need of solving differential equations. 

In the case of hyperstatic rods, as we have seen above, the solution of the static problem passes through the determination of the displacements and rotations, so in this case the Mohr's theorem and corollary become useless in this context. Nevertheless,  the corollary of Mohr can be used in some methods for the static resolution of hyperstatic systems of rods.

\section{Hyperstatic systems of rods}

The equations of rods allow, in principle, for studying  any problem of rod structures, regardless of the degree of hyperstaticity. However, in practice they can be used only in simple cases, e.g. for single rods, because very quickly their use becomes too much complicate. 

Actually, this approach is complete: it provides any type of information ($v,w,M$ etc.) everywhere in a rod. So, the question is to know whether or not it can exist an approach which, paying the price of a lower information, can be nevertheless used effectively for more complicated rod structures.

The answer is yes, and the approach is  based upon the Principle of Virtual Displacements (PVD); such a method is sometimes called the {\it force method for solving hyperstatic rod structures}, because the unknown of the method are generalized forces (forces or couples). We introduce it in the following Section, specifying since now that the method, though based upon the PVD, valid for any type of material behavior, is {\it valid only for a linear structural behavior}. This assumption implies actually two distinct and equally important hypotheses: the material is linearly elastic, on one hand, and the perturbations are small, on the other hand.

\subsection{The Principle of Virtual Displacements for rods}
The PVD can be adapted to rods; first of all, we define a {\it state of virtual displacements-deformations for a rod} a state for which $v,w,\varphi$ are
\begin{enumerate}[i.]
\item  regular (continuous and with piecewise continuous derivatives);
\item infinitesimal;
\item independent from time.
\end{enumerate}

Be $N, T, M, p, r$ a {\it field of equilibrated actions}, i. e.
\begin{equation}
\label{eq:eqeq}
\begin{split}
&\frac{dN}{dz}=-r,\\
&\frac{dT}{dz}=-p,\\
&\frac{dM}{dz}=T,
\end{split}
\ \ \ \ \ \mathrm{+\ b.\ c.}
\end{equation}

We can then prove the following
\begin{teo}{(Principle of Virtual Displacements for rods)}: be $\{v^*,w^*,\varphi^*\}$ a field of virtual displacements for a rod of length $\ell$ in equilibrium under the action of external and internal actions; then
\begin{equation}
\int_0^\ell(N\eps^*+T\gamma^*+M\kappa^*)\ dz=\int_0^\ell(p\ v^*+r\ w^*)\ dz.
\end{equation}
\begin{proof}
We remark first that the the left-hand side of the above equation is the internal virtual work, produced by the internal actions for the deformations corresponding to the considered virtual displacements field, while to the right-hand side we have the external virtual work, i.e. that produced by the external applied loads. Hence, once more, the PVD states the equality of the internal and external virtual works.

Because, by hypothesis, $v^*, w^*$ and $\varphi^*$ are sufficiently regular, we can calculate the internal virtual deformations for the rod
\begin{equation}
\eps^*=\frac{dw^*}{dz},\ \ \ \gamma^*=\frac{dv^*}{dz}-\varphi^*,\ \ \ \kappa^*=-\frac{d\varphi^*}{dz}.
\end{equation}
Then, the internal virtual work becomes
\begin{equation}
\int_0^\ell\left[N\frac{dw^*}{dz}+T\left(\frac{dv^*}{dz}-\varphi^*\right)-M\frac{d\varphi^*}{dz}\right]\ dz,
\end{equation}
and integrating by parts we get
\begin{equation}
\left[N\ w^*+T\ v^*-M\ \varphi^*\right]_0^\ell+\int_0^\ell\left[\varphi^*\left(\frac{dM}{dz}-T\right)-w^*\frac{dN}{dz}-v^*\frac{dT}{dz}\right]\ dz.
\end{equation}

The boundary term, the first one in the above equation, vanishes because at the edges it is either an internal action either its dual kinematical descriptor to vanish. Because the actions are equilibrated, eqs. (\ref{eq:eqeq}) are satisfied, so that the term in brackets under the sign of integral vanishes and finally we get
\begin{equation}
\int_0^\ell(N\eps^*+T\gamma^*+M\kappa^*)\ dz=\int_0^\ell(p\ v^*+r\ w^*)\ dz.
\end{equation}
\end{proof}
\end{teo}

We remark that constitutive equations have not been used in the proof of the PVD, so it is valid for any type of material behavior, not only for the elastic rods.

\subsection{The Müller-Breslau equations}
We introduce the method through an example, shown in Fig. \ref{fig:f30}; the structure in object is twice hyperstatic and, thanks to the assumption of linear behavior, using the principle of superposition of the effects, we can think to the structure as the sum of three isostatic structures. This is a key point of the method: {\it the original hyperstatic structures is transformed into the sum of isostatic structures, that can be solved separately using nothing but equilibrium conditions}. In particular, the original structure is decomposed into:
\begin{figure}[h]
	\begin{center}
         \includegraphics[width=\textwidth]{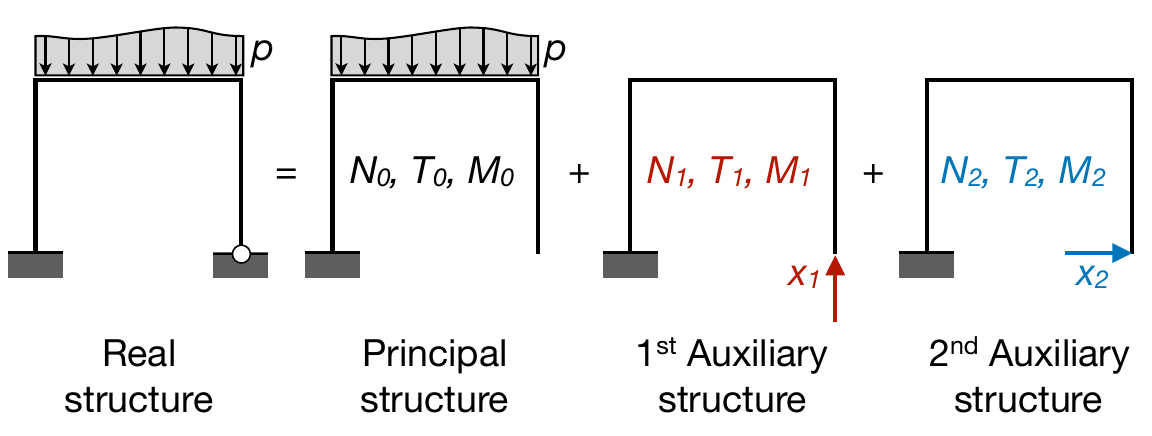}
	\caption{Scheme for the analysis of hyperstatic rod structures.}
	\label{fig:f30}
	\end{center}
\end{figure}
\begin{itemize}
\item a {\it principal structure}: the hyperstatic constraints have been removed (their choice is not unique and anyway arbitrary) and the isostatic structure so obtained is acted upon only by the given, known loads;
\item a number of {\it auxiliary structures} equal to the number of hyperstatic constraints removed (in the example, two); each one of the auxiliary structures is like the principal one, hence isostatic, but  it is {\it loaded uniquely by an unknown generalized force}, a reaction, statically dual of the corresponding removed hyperstatic constraint. 
\end{itemize}

Hence, there is an unknown generalized force for each degree of hyerstaticity, and each one of them is applied to an auxiliary structure; in our example, we have hence two unknowns, $x_1$ and $x_2$, indicated in Fig. \ref{fig:f30}. These unknowns are determined imposing the {\it geometrical condition that their corresponding displacement is equal to the one prescribed in the real structure (usually, it is null)}. We remark hence that in this method the unknowns are forces and the equations, compatibility conditions on the displacements. 

Because each one of the structures decomposing the original hyperstatic one is isostatic, it is possible to determine everywhere the internal actions merely using balance equations. We indicate with 
\begin{itemize}
\item $N_0,T_0,M_0$ the internal actions in the principal structure;
\item $N_i,T_i,M_i$ those in the i$^{th}$ auxiliary structure {\it loaded with} $x_i=1.$
\end{itemize}

Thanks to the assumption of linearity, the actual internal actions in the real, hyperstatic structure, are given by the superposition of the effects:
\begin{equation}
\label{eq:comblin}
\begin{split}
&N=N_0+\sum_{i=1}^n x_iN_i,\\
&T=T_0+\sum_{i=1}^n x_iT_i,\\
&M=M_0+\sum_{i=1}^n x_iM_i,\\
\end{split}
\end{equation}
with $n$ the degree of hyperstaticity. 

To determine the unknowns $x_i$ we use the PVD; to this purpose, we consider as virtual displacements those of the real structure, that are of course surely admissible. As forces, we consider those in each one of the auxiliary structures; because these structures are isostatic, the internal actions, calculated using balance equations, are surely equilibrated with the external loads. As a consequence, we are authorized to use the PVD with such a system of forces and field of virtual (actually, in this case real) displacements. We apply the PVD as much times as the auxiliary structures, i.e. as the degree of hyperstaticity:
\begin{itemize}
\item $1^{st}$ auxiliary structure:
\begin{itemize}
\item virtual work of the external forces $(x_1=1)$: if, in the real structure, the point of application of $x_1$ undergoes an imposed displacement $\bdelta_1$, then such virtual work will be equal to
\begin{equation}
\delta_1=\gr{e}_1\cdot\bdelta,
\end{equation}
where $\gr{e}_1$ is the unit vector oriented like $x_1$; usually, $\bdelta=\gr{o}$, so $\delta_1=0$;
\item virtual work of the internal forces (we indicate with $\Omega$ the whole structure and with $\ell$ a generic curvilinear abscissa along the rods composing the structure):
\begin{equation}
\int_\Omega(N_1\eps+T_1\gamma+M_1\kappa)\ d\ell;
\end{equation}
\end{itemize}
the first equation is hence
\begin{equation}
\label{eq:MB2}
\int_\Omega(N_1\eps+T_1\gamma+M_1\kappa)\ d\ell=\delta_1;
\end{equation}
\item $2^{nd}$ auxiliary structure: proceeding in the same way we obtain
\begin{equation}
\label{eq:MB3}
\int_\Omega(N_2\eps+T_2\gamma+M_2\kappa)\ d\ell=\delta_2.
\end{equation}
\end{itemize}

In the above equations, the internal kinematical descriptors $\eps,\gamma$ and $\kappa$ are those in the real structure. So, using the constitutive equations of elastic rods and the superposition of the effects we get
\begin{equation}
\label{eq:MB1}
\begin{split}
&\eps=\frac{N}{EA}=\frac{N_0+x_1N_1+x_2N_2}{EA},\\
&\gamma=\chi\frac{T}{\mu A}=\chi\frac{T_0+x_1T_1+x_2T_2}{\mu A},\\
&\kappa=\frac{M}{EJ}=\frac{M_0+x_1M_1+x_2M_2}{EJ}.\\
\end{split}
\end{equation}

Replacing the relations above in the two PVD equations, after regrouping the terms we get
\begin{equation}
\begin{split}
&x_1\int_\Omega\left(\frac{N_1^2}{EA}+\chi\frac{T_1^2}{\mu A}+\frac{M_1^2}{EJ}\right)\ d\ell
+x_2\int_\Omega\left(\frac{N_1N_2}{EA}+\chi\frac{T_1T_2}{\mu A}+\frac{M_1M_2}{EJ}\right)\ d\ell\\
&+\int_\Omega\left(\frac{N_1N_0}{EA}+\chi\frac{T_1T_0}{\mu A}+\frac{M_1M_0}{EJ}\right)\ d\ell=\delta_1,\\
&x_1\int_\Omega\left(\frac{N_1N_2}{EA}+\chi\frac{T_1T_2}{\mu A}+\frac{M_1M_2}{EJ}\right)\ d\ell
+x_2\int_\Omega\left(\frac{N_2^2}{EA}+\chi\frac{T_2^2}{\mu A}+\frac{M_2^2}{EJ}\right)\ d\ell\\
&+\int_\Omega\left(\frac{N_2N_0}{EA}+\chi\frac{T_2T_0}{\mu A}+\frac{M_2M_0}{EJ}\right)\ d\ell=\delta_2.
\end{split}
\end{equation}

These equations have the form of a symmetric system of linear algebraic equations; in the general case of $n$ degrees of hyperstaticity, we have a system of $n$ equations with $n$ unknowns $x_i$ that can be synthetically written in the form
\begin{equation}
\label{eq:mb1}
\eta_{ij}x_j=\eta_{i0}+\delta_i,
\end{equation}
with 
\begin{equation}
\label{eq:mb2}
\begin{split}
&\eta_{ij}=\eta_{ji}=\int_\Omega\left(\frac{N_iN_j}{EA}+\chi\frac{T_iT_j}{\mu A}+\frac{M_iM_j}{EJ}\right)\ d\ell,\\
&\eta_{i0}=-\int_\Omega\left(\frac{N_iN_0}{EA}+\chi\frac{T_iT_0}{\mu A}+\frac{M_iM_0}{EJ}\right)\ d\ell.
\end{split}
\end{equation}

The terms on the diagonal, $\eta_{ii}$ are necessarily positive, as it is apparent from the above equations. Equations (\ref{eq:mb1}) are the {\it Müller-Breslau equations} (1886), with the coefficients of the unknowns given by eq. (\ref{eq:mb2}). They provide the classical method for the resolution of hyperstatic systems of elastic rods. In the very frequent case of slender rods, the extension and  shear deformations are much smaller than the bending one, so they can be neglected and the calculation of the coefficients is greatly simplified, as it is reduced to the only bending terms.

It is worth notice that  imposed displacements to a constrained point, like those  produced by ground settlements in a bridge or a building foundation structure, stress an hyperstatic structure. This is not the case for an isostatic one, because in that case the distribution of the internal actions is uniquely determined by the equilibrium equations. As imposed displacements are kinematical conditions, they do not enter the equilibrium equations and, by consequence, do not stress an isostatic structure.

As a last point we remark that once determined the unknowns $x_i$, the real distribution of the internal actions in the structure can be easily calculated  using eq. (\ref{eq:comblin}).

\subsection{The dummy load method}
The method of the forces does not allow to retrieve all the information about the rod, e.g. it does not give the displacements. Anyway, we can calculate the displacement $v$ in a point still using the PVD, by the so-called {\it dummy load method}. To this purpose, we apply the PVD using as displacement field the real one and as forces those in any equilibrated isostatic system acted upon uniquely by  a unit force, the {\it dummy load}, dual of the displacement $v$ to be found. 

The virtual work of the external forces is hence equal to $v$, while that of the internal forces is
\begin{equation}
\int_\Omega\left(\frac{N\ N_d}{EA}+\chi\frac{T\ T_d}{\mu A}+\frac{M\ M_d}{EJ}\right)\ d\ell,
\end{equation}
where $N, T, M$ are the real internal actions; they are already known by a previous calculation, for instance a merely static one if the structure is isostatic or having solved the Müller-Breslau equations if it is hyperstatic. $N_d,T_d,M_d$ are the internal actions produced on the isostatic structure by the dummy load; being the structure isostatic, they can be calculated by simple static conditions, hence they are equilibrated. 

Finally, the PVD gives
\begin{equation}
v=\int_\Omega\left(\frac{N\ N_d}{EA}+\chi\frac{T\ T_d}{\mu A}+\frac{M\ M_d}{EJ}\right)\ d\ell.
\end{equation}

\section{Effects of a temperature change}
A final question concerns the effects of the  temperature changes. We still use the Hooke-Duhamel model (see Ex. 12, Chapt. 3):
\begin{equation}
\label{eq:duhamel}
\beps=\beps_m+\beps_t,
\end{equation}
where $\beps_m$ is the mechanical deformation, given by the Lamé's inverse equations ($E$: Young's modulus, $\nu$: Poisson's ratio),
\begin{equation}
\label{eq:lame2bis}
\beps=\frac{1+\nu}{E}\bsig-\frac{\nu}{E}\tr\bsig\ \gr{I},
\end{equation}
 while the thermal deformation $\beps_t$ is given by 
\begin{equation}
\label{eq:tempeps}
\beps_t=\Delta t\ \alpha\gr{I},
\end{equation}
where $\Delta t$ is the temperature variation with respect to a state where, conventionally, $\beps_t=\gr{o}$, and $\alpha$ is the {\it coefficient of thermal expansion}.

\begin{figure}[b]
	\begin{center}
         \includegraphics[scale=.7]{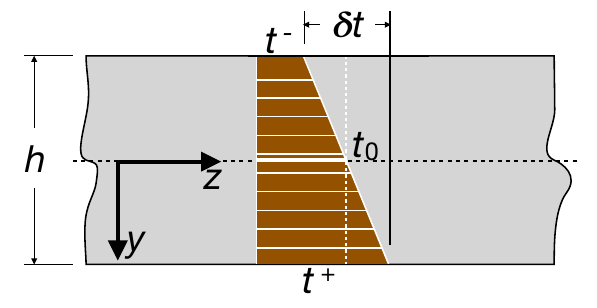}
	\caption{Decomposition of a linear temperature change on a rod.}
	\label{fig:f31}
	\end{center}
\end{figure}

For what concerns $\Delta t$, the usual assumption in the rod theory is that it has a linear variation through the thickness of the rod, see Fig. \ref{fig:f31}, which is rigorously true in a stationary heat flow; if $h$ is the thickness of the section,
\begin{equation}
t(y)=t_0+\frac{\delta t}{h}y, \ \ \ t_0=\frac{t^++t^-}{2},\ \ \ \delta t=t^+-t^-.
\end{equation}
The global temperature change is hence decomposed into a uniform, $t_0$, and an antisymmetric one, $\delta t$.
Now, we need to link $\beps_t$ to the  descriptors $\eps_t,\gamma_t,\kappa_t$ of the {\it thermal deformation}. To this end, we consider a length of rod between the sections $1$ and $2$ and we write the strain energy of the beam and of the rod, that must be equal. Because for a Saint-Venant beam it is $\sigma_{11}=\sigma_{22}=\sigma_{12}=0$, we have 
\begin{equation}
\frac{1}{2}\int_1^2(\sigma_{33}{\eps_{33}}_t+2\sigma_{13}{\eps_{13}}_t+2\sigma_{23}{\eps_{23}}_t)\ dv=\frac{1}{2}\int_1^2(N\eps_t+T\gamma_t+M\kappa_t)\ dz.
\end{equation}

But   ${\eps_{13}}_t={\eps_{23}}_t=0$, because of eq. (\ref{eq:tempeps}); then, using the Saint-Venant Problem results and the linear variation of $t$ through the rod thickness we get, for the left-hand term,
\begin{equation}
\frac{1}{2}\int_1^2\left[\int_S\alpha\left(\frac{N}{A}+\frac{M\ y}{J}\right)\left(t_0+\frac{\delta t}{h}y\right)\ ds\right]dz=\frac{1}{2}\int_1^2\alpha \left(N\ t_0+M\frac{\delta t}{h}\right)dz,
\end{equation}
the other terms are null because the frame axes have their origin in the centroid of the cross section $S$. So, because this equation must be true for all the possible choices of the sections $1$ and $2$, the two integrands must be equal, which gives
\begin{equation}
\label{eq:temp1}
\begin{split}
&\eps_t=\alpha\ t_0,\\
&\gamma_t=0,\\
&\kappa_t=\alpha\frac{\delta t}{h}.
\end{split}
\end{equation}

Shear deformation is not affected by temperature changes; extension is influenced only by uniform and bending only by antisymmetric changes of temperature. According to eq. (\ref{eq:duhamel}), the deformations of the rod are hence given by
\begin{equation}
\label{eq:temp2}
\begin{split}
&\eps=\eps_m+\eps_t,\\
&\gamma=\gamma_m,\\
&\kappa=\kappa_m+\kappa_t,
\end{split}
\end{equation}
where the subscript $m$ indicates the {\it mechanical} part of the deformation, linked to the internal actions by the constitutive equations (\ref{eq:consteq}):
\begin{equation}
N=EA\ \eps_m,\ \ \ T=\frac{\mu A}{\chi}\gamma_m,\ \ \ M=EJ\ \kappa_m,
\end{equation}
while the total deformation is still linked to the displacement components  by the compatibility equations (\ref{eq:compat1}), (\ref{eq:compat2}) and (\ref{eq:compat3})
\begin{equation}
\eps=-w',\ \ \ \gamma=v'-\varphi,\ \ \ \kappa=-\varphi'.
\end{equation}
From the previous equations we get hence
\begin{equation}
\label{eq:thermoel3}
\begin{split}
&N=EA(\eps-\eps_t)=EA(w'-\alpha t_0),\\
&T=\frac{\mu A}{\chi}(v'-\varphi),\\
&M=EJ(\kappa-\kappa_t)=-EJ\left(\varphi'+\alpha\frac{\delta t}{h}\right),
\end{split}
\end{equation}
that inserted into the equilibrium equations (\ref{eq:balancerod}) give finally the {\it thermo-elastic equilibrium equations} for the Thimoshenko's rod model:
\begin{equation}
\begin{split}
&\left(EA(w'-\alpha t_0)\right)'=-r,\\
&\left(\frac{\mu A}{\chi}(v'-\varphi)\right)'=-p,\\
&\left(EJ\left(\varphi'+\alpha \frac{\delta t}{h}\right)\right)'=-\left(\frac{\mu A}{\chi}(v'-\varphi)\right)-m.
\end{split}
\end{equation}
For the Euler-Bernoulli rod, the first equation above does not change, the second one is, as usual, meaningless, while eq. (\ref{eq:thermoel3})$_3$ becomes ($\varphi=v'$),
\begin{equation}
M=EJ(\kappa-\kappa_t)=-EJ\left(v''+\alpha\frac{\delta t}{h}\right),
\end{equation}
that inserted into the equilibrium equation (\ref{eq:equilM2}) finally gives
\begin{equation}
\label{eq:thermoel4}
\left(EJ\left(v''+\alpha\frac{\delta t}{h}\right)\right)''=p-m'.
\end{equation}
In the Müller-Breslau equations, the presence of a temperature field can be accounted for inserting eqs. (\ref{eq:temp1}) and (\ref{eq:temp2}) in the expressions of $\eps$ and $\kappa$ in eqs. (\ref{eq:MB1}):
\begin{equation}
\begin{split}
&\eps=\eps_m+\eps_t=\frac{N}{EA}+\alpha t_0=\frac{N_0+x_1N_1+x_2N_2}{EA}+\alpha t_0,\\
&\gamma=\gamma_m=\chi\frac{T}{\mu A}=\chi\frac{T_0+x_1T_1+x_2T_2}{\mu A},\\
&\kappa=\kappa_m+\kappa_t=\frac{M}{EJ}+\alpha\frac{\delta t}{h}=\frac{M_0+x_1M_1+x_2M_2}{EJ}+\alpha\frac{\delta t}{h}.\\
\end{split}
\end{equation}
Once these expressions inserted into the PVD equations, e.g.  eqs. (\ref{eq:MB2}) and (\ref{eq:MB3}), we get the final form of the Müller-Breslau equations for the thermo-elastic case:
\begin{equation}
\eta_{ij}x_j=\eta_{i0}+\eta_{it}+\delta_i,
\end{equation}
where
\begin{equation}
\eta_{it}=-\int_\Omega \alpha \left(t_0N_i+\frac{\delta t}{h}M_i\right)\ d\ell.
\end{equation}
 
To end this Section, we recall that for an isostatic equation the distribution of the internal actions $N,T$ and $M$ is uniquely determined by the equilibrium equations (\ref{eq:balancerod})$_1$ and (\ref{eq:equilM2}), see Sect. \ref{sec:ipoiper}. Because in such equations the effects of temperature changes, that are deformation effects, do not enter,  {\it temperature changes do not produce any internal action in isostatic structures, but only deformations}. This is not the case of hyperstatic structures, like eqs. (\ref{eq:thermoel3}) to (\ref{eq:thermoel4}) clearly show. Hence, unlike isostatic rods, {\it hyperstatic rods are  stressed by temperature changes}.

\section{Exercises}
{\bf Nota bene:} {\it unless otherwise specified, the exercices below are to be solved using the Euler-Bernoulli rod model.}\smallskip
\begin{enumerate}
\item \label{ex:1} Determine the elastica of a cantilever beam with $EJ=$ const., loaded:
\begin{enumerate}[i.]
\item by a uniform load $p$;
\item by a concentrated force $F$ at the free edge (this is the {\it Galileo's problem});
\item by a couple $M$ at the free edge.
\end{enumerate} 

\item Determine the function $J(z)$ that a homogeneous uniformly loaded cantilever must have to bend along a circular arch of radius $R$.

\item Determine the axial force $N$ and displacement $w$ of a vertical rod with $EA=$ const. clamped at its ends and submitted to its own weight.

\item Determine the variation $h(z)$ of the height of a rectangular cross section of a cantilever loaded by a concentrated force at its free edge in order to have everywhere the same highest stress $\sigma_{33}$ ({\it Galileo's problem of the rod of uniform strength}).

\item What does it change in the previous problem if it is the highest Von Mises equivalent stress to be constant throughout the rod length?

\item Determine the displacement $v$ of the center of a clamped-clamped rod loaded at mid-span by a concentrated force $F$.

\item Study  the structure in the figure and determine the maximum deflection.
%\begin{figure}[h]
	\begin{center}
         \includegraphics[scale=.7]{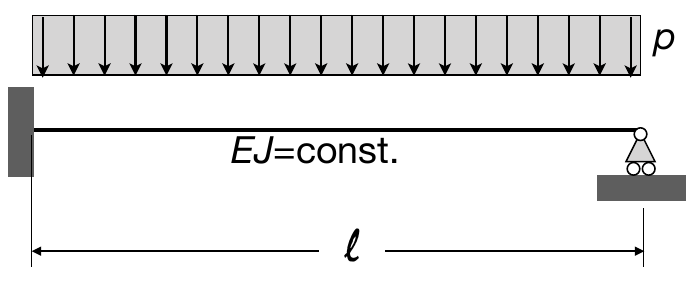}
	\end{center}
%\end{figure}
\vspace{-3mm}

\item \label{ex:A}Study the same problem but now using the Timoshenko's rod model. Show that in this case the solution tends to that of the Euler-Bernoulli model when the slenderness ratio $\rho/\ell\rightarrow0$; how to interpret this result?

\item \label{ex:B} Find the displacement of the free edge of the rod in the figure using the results of the previous exercices.
%\begin{figure}[h]
	\begin{center}
         \includegraphics[scale=.7]{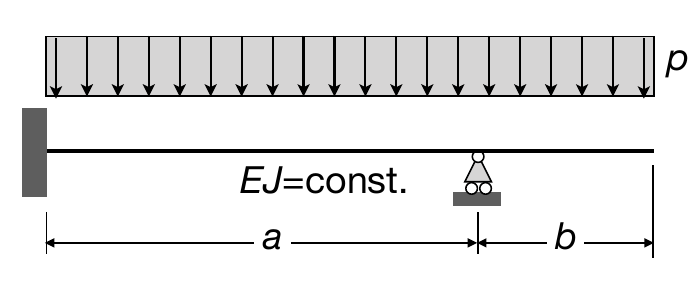}
	\end{center}
%\end{figure}
\vspace{-5mm}

\item Find the deflexion and rotation of the beam of exercise \ref{ex:1} using the Mohr's theorem and corollary.

\item \label{ex:C} A rod clamped at the ends is thermally loaded by:
\begin{enumerate}[i.]
\item a uniform temperature $t_0$;
\item a through the thickness linear variation of the type $t^-=-t,\ \ t^+=+t$.
\end{enumerate}
Study the structure in both the cases, finding the reactions, internal actions and displacements.

\item Consider again the two cases of the previous exercice, but now the rod is simply supported; what changes for the rod?

\item \label{ex:D} Study the structure in the figure, finding also the rotation of the central point.
%\begin{figure}[h]
	\begin{center}
         \includegraphics[scale=.7]{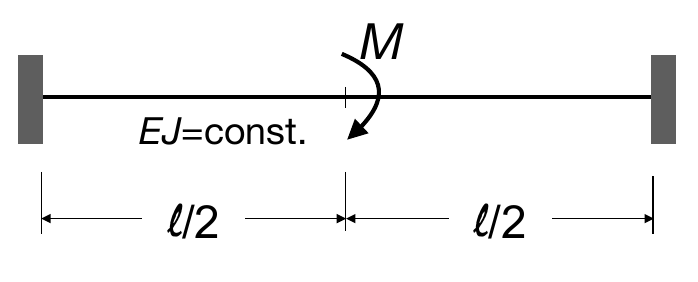}
	\end{center}
%\end{figure}
\vspace{-5mm}

%\item For the structure in the following figure, determine $F$ so as to have a null displacement of point $C$.
%\begin{figure}[h]
%	\begin{center}
%         \includegraphics[scale=.8,angle=1.6]{fig/f35}
%	\end{center}
%\end{figure}

%\item Solve the previous problem with the span $AB$ now loaded by a through the thickness linear variation of the temperature of the type $t^-=-t,\ \ t^+=+t$.

\item \label{eq:E} Study the structure in the following figure.
%\begin{figure}[h]
	\begin{center}
         \includegraphics[scale=.7]{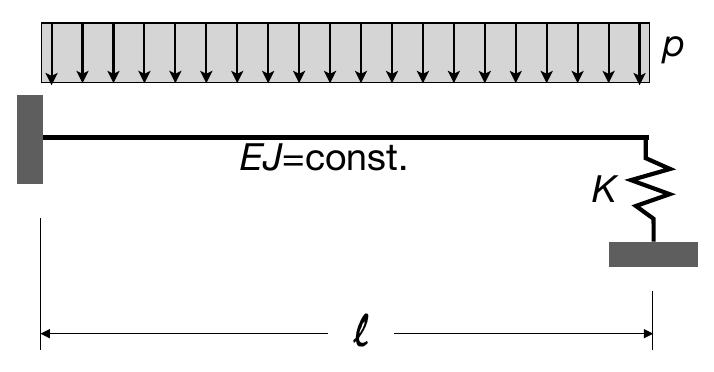}
	\end{center}
%\end{figure}
\vspace{-5mm}

\item Make the same, loading now the rod uniquely by a thermal load of the type $t^-=-t,\ \ t^+=+t$.

\item Determine the support reactions and internal actions of an elastic simply supported rod of length $\ell$ and constant bending stiffness $EJ$ when a point at the abscissa $z=a$ is submitted to an imposed  deflection $\delta$. Why the problem cannot be solved using uniquely equilibrium equations, i.e., why contrarily to what seems at a first sight, this problem is not isostatic?

\item An infinitely long pipe, whose weight per unit length is $p$ and whose constant bending stiffness is $EJ$, lays on a horizontal plane, that can be considered as infinitely rigid. The pipe must be lifted at a certain point, by a crane, of a height equal to $h$. Which is the lifting force that the crane must produce?

\item A rod whose length is $\ell$, weight per unit length $p$ and  constant bending stiffness  $EJ$, lays on a horizontal plane, that can be considered as infinitely rigid. At the left end, the rod is acted upon by a vertical force $F$. For what conditions the equilibrium is ensured? How much the left end of the rod will be lifted up by $F$?

\item A rod whose length is $\ell$, weight per unit length $p$, thickness $h$ and  constant bending stiffness  $EJ$, lays on a horizontal plane, that can be considered as infinitely rigid. The rod is heated on its upper surface to a temperature $t$, while the lower part is at the temperature $-t$. Determine the vertical displacement of the mid point of the rod. Which is the minimum value of $t$ to lift up the rod?

\item Study the case of an elastic rod of length $\ell\rightarrow\infty$ that is submitted to a concentrated load $F$  and that lays on an elastic substrate whose elastic constant is $k$ (this is the case of a {\it rod on a Winkler's soil}). 

\item Study now the case of a rod on an elastic soil, with finite length $\ell$ and submitted to a uniform load $p$. Can you predict the result?

\item Study the case of a pile of length $\ell$  driven into an elastic soil of elastic constant $k$ and submitted, at its top, to an horizontal force $F$. 

\item Solve  exercise \ref{ex:A} using the Müller-Breslau equations.

\item Solve exercise \ref{ex:B} using the dummy load method.

\item Solve exercise \ref{ex:C} using the Müller-Breslau equations and the dummy load method.

\item Solve exercice \ref{ex:D} using the Müller-Breslau equations and the dummy load method

\item Solve exercice \ref{eq:E}  using  the Müller-Breslau equations.

\item Solve the structure in the figure using the Müller-Breslau equations.
%\begin{figure}[h]
	\begin{center}
         \includegraphics[scale=.7]{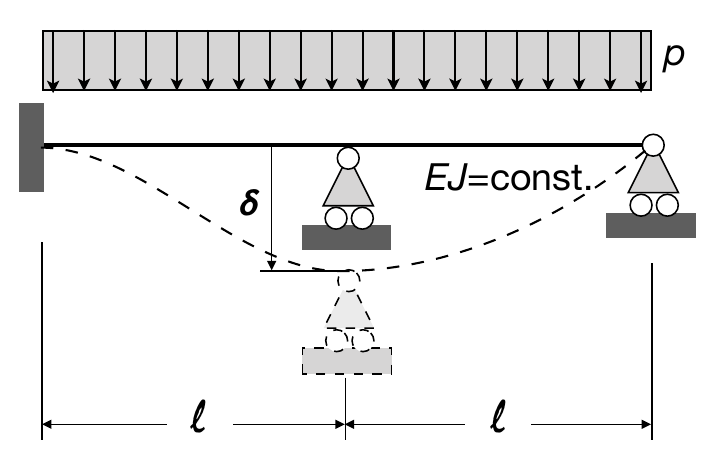}
	\end{center}
%\end{figure}
\vspace{-5mm}

\item Make the same, loading now the rod uniquely by a thermal load of the type $t^-=-t,\ \ t^+=+t$.

\item Solve the structure in the following figure using the Müller-Breslau equations (neglect the axial and shear deformations). What happens when $\lambda\rightarrow\infty$? 
%\begin{figure}[h]
	\begin{center}
         \includegraphics[scale=.7]{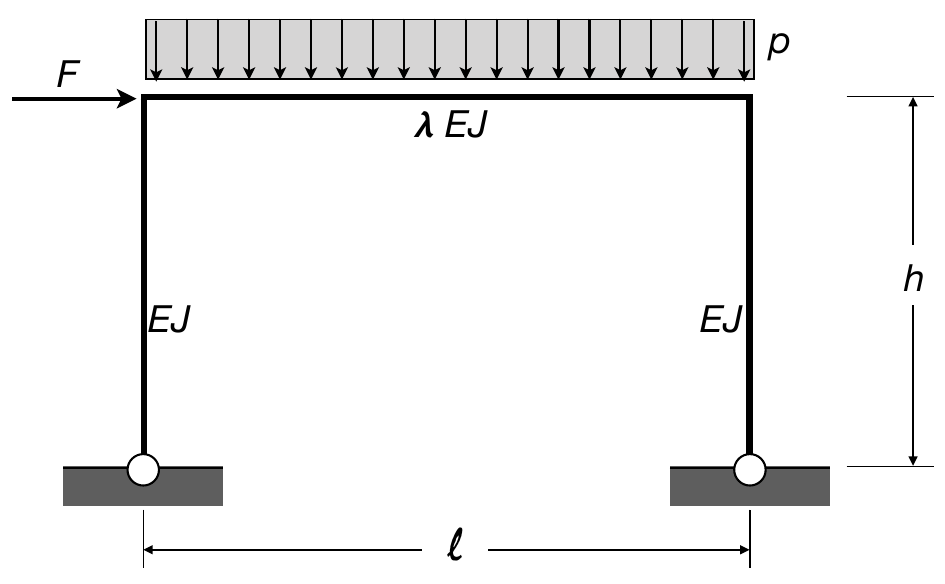}
	\end{center}
%\end{figure}

\end{enumerate}

\chapter{Arches}
\label{ch:4}
\section{Introduction}
\begin{figure}[h]
\begin{center}
\includegraphics[scale=.9]{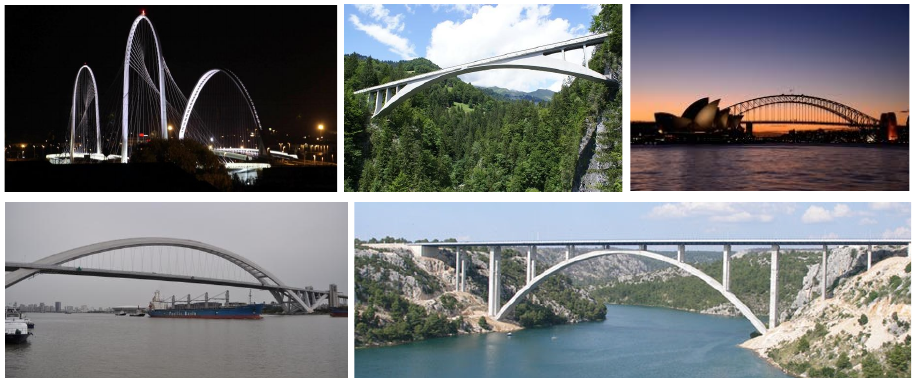}
\caption{Some examples of arch bridges.}
\label{fig:f3_1}
\end{center}
\end{figure}

Arches are plane curved rods. They are currently used in a lot of situations; in civil engineering, the most impressive realizations are some bridges, see Fig. \ref{fig:f3_1}.
Arches are used since long time (probably the most ancient arches in the world are two arches in Italy, at Velia and Volterra, both from the IVth century b.C.), though in the form of arches composed of carved stone voussoirs or masonry bricks. Arches, in fact, have a great advantage with respect to rectilinear rods: while rods are mainly subjected to bending,  internal axial forces are dominating in arches: strain energy is stored in the form of bending energy in rods, while mainly, though not exclusively,  in the form of extension energy in arches. This is rather advantageous, because axial stiffness and strength are normally much greater than the bending ones: arches bear greater loads than rectilinear loads. A simple scheme gives account of this: let us consider a rectilinear, simply supported rod of span $\ell$ submitted to a uniform weight $p$. The highest bending moment, at mid-span, is $M_{max}=\dfrac{p\ell^2}{8}$. For equilibrium, such a couple must be balanced by an internal couple given by internal stresses. This internal couple has a small lever arm, less than the thickness of the cross section. As a consequence, the cross section must have a rather great thickness, in order on one side to increase the internal lever arm, on the other side to decrease stresses and  bring them under the admissible value.

If now we consider an arch hinged at the ends and submitted to the same load, the horizontal reactions give a couple at mid-span that balances the couple given by $p$; in the limit, the bending moment at mid-span can be null, and the section is submitted exclusively to an axial force equal to the horizontal reactions. This is a much more favorable structural situation, that allows to the structure, for the same cross section and material,  to carry more important loads or to cover much longer spans. 

This simple case allows also for understating a basic fact: in a simply supported or clamped arch vertical loads produce also horizontal reactions. If the arch is simply supported at one of the two ends, then these horizontal reactions do not exist and the arch is actually just a curved beam, it works almost exclusively in bending: it is the presence of the horizontal reactions that makes a curved rod an effective arch.

In the following, we study the statics of elastic curved rods and then we apply the theory to the statics of arches. 

\section{Balance equations}
\begin{figure}[h]
\begin{center}
\includegraphics[scale=.9]{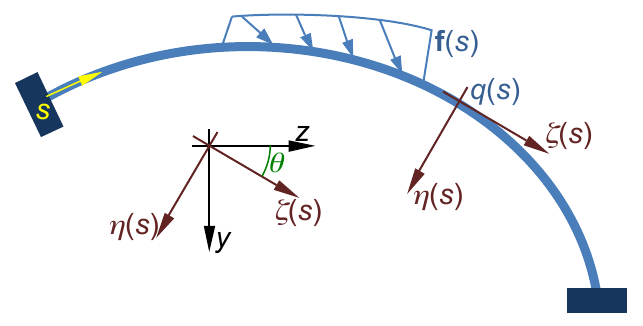}
\caption{General scheme of an arch.}
\label{fig:f3_2}
\end{center}
\end{figure}

\begin{figure}[h]
\begin{center}
\includegraphics[scale=1]{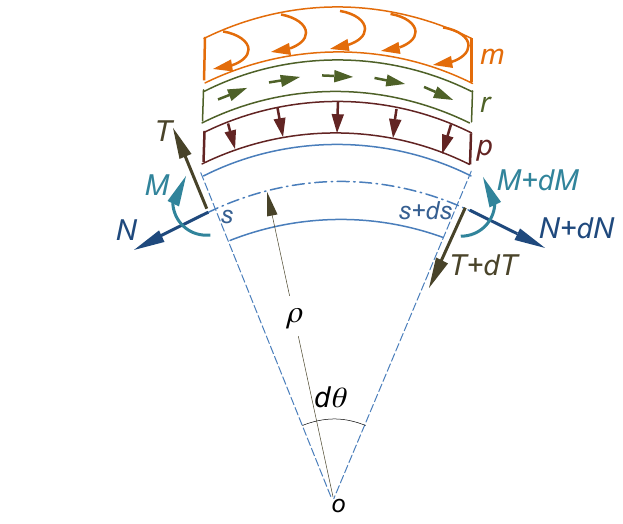}
\caption{General scheme for the balance equations.}
\label{fig:f3_3}
\end{center}
\end{figure}

We consider a curvilinear plane rod like in Fig. \ref{fig:f3_2}; $s$ is a curvilinear abscissa, that we will put, conventionally, equal to zero at the left end of the rod. We then consider a part of the rod between two sections at the abscissae $s$ and $s+ds$ infinitesimally close together, Fig. \ref{fig:f3_3}, so that we can consider as constant the curvature of the rod in the part $ds$, and be $d\theta$ the infinitesimal angle subtended by the two normals to the rod at $s$ and $ds$, that meet together in $o$, the  centre of the local osculating circle.

We remain in the framework of small strain and displacements and write the balance equations in the undeformed configuration. Then, referring to Fig. \ref{fig:f3_3}, it is easy to write the balance equations:
\begin{itemize}
\item horizontal equilibrium:
\begin{equation}
T \sin\frac{d\theta}{2}+N\cos\frac{d\theta}{2}-r\ ds+(T+dT)\sin\frac{d\theta}{2}-(N+dN)\cos\frac{d\theta}{2}=0;
\end{equation}
\item vertical equilibrium:
\begin{equation}
T \cos\frac{d\theta}{2}-N\sin\frac{d\theta}{2}-p\ ds-(T+dT)\cos\frac{d\theta}{2}-(N+dN)\sin\frac{d\theta}{2}=0;
\end{equation}
\item rotation equilibrium (written here with respect to $o$):
\begin{equation}
M+m\ ds+(N+dN)\rho +r\rho\ ds-N\rho-(M+dM)=0,
\end{equation}
\end{itemize}

with $\rho$ the radius of curvature. Because $ds$ is infinitesimal, $d\theta\rightarrow0$ so that $\cos\dfrac{d\theta}{2}\simeq 1$ while $\sin\dfrac{d\theta}{2}\simeq \dfrac{d\theta}{2}$. By consequence, once simplified the balance equations transform to
\begin{equation}
\label{eq:eqarc1}
\begin{split}
&dN=T\ d\theta-r\ ds,\\
&dT=-N\ d\theta-p\ ds,\\
&dM=\rho\ dN+\rho\ r\ ds+m\ ds.
\end{split}
\end{equation}
Injecting eq. (\ref{eq:eqarc1})$_1$ into eq. (\ref{eq:eqarc1})$_3$ and considering that $ds=\rho\ d\theta$, we finally get
\begin{equation}
\label{eq:eqarc2}
\begin{split}
&\frac{dN}{ds}=\frac{T}{\rho}-r,\\
&\frac{dT}{ds}=-\frac{N}{\rho}-p,\\
&\frac{dM}{ds}=T+m.
\end{split}
\end{equation}

Unlike the case of rectilinear rods, now all the balance equations are coupled; to remark that for $\rho\rightarrow\infty$, i.e. for a curvilinear rod that tends to become rectilinear, the above equations tend to those of the straight rods.

Introducing now the symbolic matrix 
\begin{equation}
\gr{D}_1=\left[\begin{array}{ccc}
\dfrac{d}{ds} & -\dfrac{1}{\rho} & 0 \\
\dfrac{1}{\rho} & \dfrac{d}{ds} & 0 \\
0 & -1 & \dfrac{d}{ds},
\end{array}\right]
\end{equation}
along with the vectors of the internal actions, $\gr{S}$, and of the applied loads, $\gr{f}$,
\begin{equation}
\gr{S}=\left\{\begin{array}{c}N \\T \\M\end{array}\right\},\ \ \ \gr{f}=\left\{\begin{array}{c}r \\p \\m\end{array}\right\},
\end{equation}
eq. (\ref{eq:eqarc2}) can be given in matrix form:
\begin{equation}
\label{eq:matrixeq1}
\gr{D}_1\gr{S}+\gr{f}=\gr{o}.
\end{equation}

Implicitly, the above equations have been written in the local frame $\{\zeta(s),\eta(s)\}$, see Fig. \ref{fig:f3_2}. They can be written  in a fixed frame $\{z,y\}$  simply writing that:
\begin{equation}
\label{eq:matrixQ}
\gr{f}=\gr{Q\ f^*}\ \rightarrow\ 
\left\{\begin{array}{c}r \\p \\m\end{array}\right\}=
\left[\begin{array}{ccc}\cos\theta & \sin\theta & 0 \\-\sin\theta & \cos\theta & 0 \\0 & 0 & 1\end{array}\right]
\left\{\begin{array}{c}r^* \\p^* \\m^*\end{array}\right\},
\end{equation}
where the symbol * denotes a vector written in the fixed frame  $\{z,y\}$ and $\gr{Q}$ the rotation tensor operating the change of frame from $\{z,y\}$ to $\{\zeta(s),\eta(s)\}$. Of course, $\gr{Q}=\gr{Q}(s)$, because $\theta=\theta(s)$.

So, finally, the matrix equation (\ref{eq:matrixeq1}) written in the fixed frame $\{z,y\}$ is simply
\begin{equation}
\label{eq:matrixeq2}
\gr{D}_1\gr{S}+\gr{Q f^*}=\gr{o}.
\end{equation}

\section{Compatibility equations}
\begin{figure}[h]
\begin{center}
\includegraphics[scale=1]{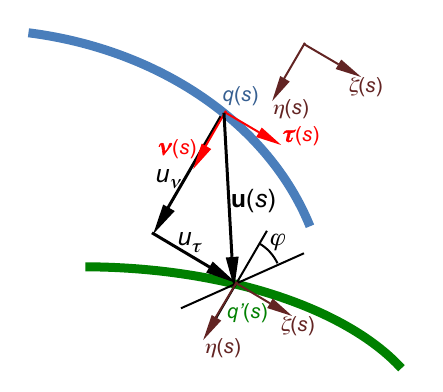}
\caption{General scheme for the compatibility equations.}
\label{fig:f3_4}
\end{center}
\end{figure}

Just as for straight rods, we need to link the displacement of the arch to the kinematical quantities defining its deformation. We write such relations in the local frame $\{\zeta(s),\eta(s)\}$, see Fig. \ref{fig:f3_4}.

A point $q(s)$ of the arch centerline becomes, after deformation, the point $q'(s)$, so the displacement vector is 
\begin{equation}
\label{eq:unuutau}
\gr{u}(s)=q'(s)-q(s)=u_\tau(s)\btau(s)+u_\nu(s)\bnu(s).
\end{equation}
We want to express the link between some {\it internal kinematical quantities} defining the deformation of the arch and the following geometrical quantities, defining the transformation of the cross section of the arch in correspondence of $q(s)$: 
\begin{itemize}
\item $v(s)$, the displacement along $\eta(s)$, i.e. parallel to the principal normal $\bnu(s)$ (it corresponds to the component $u_\nu$ in eq. (\ref{eq:unuutau}) and in Fig. \ref{fig:f3_4});
\item $w(s)$, the displacement along $\zeta(s)$, i.e. parallel to the tangent $\btau(s)$ (it corresponds to the component $u_\tau$ in eq. (\ref{eq:unuutau}) and in Fig. \ref{fig:f3_4});
\item $\varphi(s)$, the rotation of the normal $\bnu(s)$.  
\end{itemize}

The internal kinematical quantities that we will use to describe the deformation of the arch are:
\begin{itemize}
\item $\eps(s)$: the extension of the arch mid-line;
\item $\kappa(s)$: the bending curvature of the arch centerline (not to be confused with the {\it geometric} curvature $c$ of the arch centerline, i.e. the curvature of the undeformed arch);
\item $\gamma(s)$: the shear of the cross section.
\end{itemize}

To define and put in relation these quantities, we need to analyze how a generic point of the cross section is transformed. Remembering that the arch is a plane curve and that the problem is planar, i.e. the arch bends in its plane, it is sufficient to study the transformation of a generic point $\hat{q}(s,\eta)$ of the cross section at the distance $\eta$ from the centerline (so, actually, $q(s)=\hat{q}(s,0)$).

We then introduce the displacement of $\hat{q}(s,\eta)$:
\begin{equation}
\hat{\gr{u}}(s,\eta)=\hat{u}_\tau(s,\eta)\btau(s)+\hat{u}_\nu(s,\eta)\bnu(s),
\end{equation}
and considering the scheme of Fig. \ref{fig:f3_4} we recognize immediately that, thanks to  the assumptions of  small deformations and displacements (namely, $\varphi\rightarrow0\Rightarrow \sin\varphi\simeq\varphi,\ \cos\varphi\simeq1$), we can write
\begin{equation}
\begin{split}
&\hat{u}_\tau(s,\eta)=w(s)-\eta\sin\varphi\simeq w(s)-\eta\varphi,\\
&\hat{u}_\nu(s,\eta)=v(s)-\eta(1-\cos\varphi)\simeq v(s).
\end{split}
\end{equation}

\begin{figure}[h]
\begin{center}
\includegraphics[scale=1]{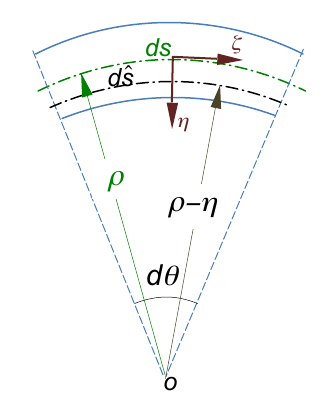}
\caption{Scheme for the relation between $ds$ and $d\hat{s}$.}
\label{fig:f3_5}
\end{center}
\end{figure}

We calculate now the derivatives of $\hat{\gr{u}}(s,\eta)$ with respect to $\eta$ and to the curvilinear abscissa of $\hat{q}(s,\eta)$; this is $\hat{s}\neq s$, due to the geometrical curvature of the arch. Nevertheless, see Fig. \ref{fig:f3_5}, it is easy to find the relation between $s$ and $\hat{s}$:
\begin{equation}
d\theta=\frac{ds}{\rho}=\frac{d\hat{s}}{\rho-\eta}\ \rightarrow\ \frac{ds}{d\hat{s}}=\frac{\rho}{\rho-\eta},
\end{equation}
relation that gives also
\begin{equation}
\label{eq:derivs}
\frac{\partial \cdot}{\partial \hat{s}}=\frac{\partial \cdot}{\partial s}\frac{ds}{d\hat{s}}=\frac{\partial \cdot}{\partial s}\frac{\rho}{\rho-\eta}.
\end{equation}

The derivatives are hence:
\begin{equation}
\begin{split}
&\frac{\partial\hat{\gr{u}}}{\partial\hat{s}}=\frac{\partial v(s)}{\partial\hat{s}}\bnu(s)+v(s)\frac{\partial\bnu(s)}{\partial\hat{s}}+\frac{\partial(w(s)-\eta\varphi(s))}{\partial\hat{s}}\btau(s)+\\
&\hspace{10mm}(w(s)-\eta\varphi(s))\frac{\partial\btau(s)}{\partial\hat{s}},\\
&\frac{\partial\hat{\gr{u}}}{\partial\eta}=-\varphi(s)\btau(s).
\end{split}
\end{equation}

Applying eq. (\ref{eq:derivs}) to $\dfrac{\partial v(s)}{\partial\hat{s}},\dfrac{\partial\btau(s)}{\partial\hat{s}}$ and $\dfrac{\partial\bnu(s)}{\partial\hat{s}}$, using the Frenet-Serret formulae and remembering that  $\vartheta(s)=0\ \forall s$ because the arch is a plane curve, gives finally (we do not write explicitly the dependence on $s$ for the sake of conciseness)
\begin{equation}
\begin{split}
&\frac{\partial\hat{\gr{u}}}{\partial\hat{s}}=\frac{\rho}{\rho-\eta}\left(w'-\frac{v}{\rho}-\eta\varphi'\right)\btau+\frac{\rho}{\rho-\eta}\left(v'+\frac{w}{\rho}-\eta\frac{\varphi}{\rho}\right)\bnu,
\end{split}
\end{equation}
where the  prime denotes differentiation with respect to $s: w'=\dfrac{\partial w}{\partial s}$ etc.

We define now mathematically the kinematical quantities introduced above to describe the deformation of the arch:
\begin{equation}
\begin{split}
&\eps_\eta(s,\eta):=\left(\frac{\partial \hat{\gr{u}}}{\partial\hat{s}}\right)_\tau=\frac{\rho}{\rho-\eta}\left(w'-\frac{v}{\rho}-\eta\varphi'\right),\\
&\kappa(s):=-\varphi',\\
&\gamma_\eta(s,\eta):=\left(\frac{\partial \hat{\gr{u}}}{\partial\hat{s}}\right)_\nu+\left(\frac{\partial \hat{\gr{u}}}{\partial\eta}\right)_\tau=\frac{\rho}{\rho-\eta}\left(v'+\frac{w}{\rho}-\eta\frac{\varphi}{\rho}\right)-\varphi,
\end{split}
\end{equation}
that can be rearranged to give finally
\begin{equation}
\begin{split}
&\eps_\eta(s,\eta)=\frac{\rho}{\rho-\eta}\left(w'-\frac{v}{\rho}+\eta\kappa\right),\\
&\kappa(s)=-\varphi',\\
&\gamma_\eta(s,\eta)=\frac{\rho}{\rho-\eta}\left(v'+\frac{w}{\rho}-\varphi\right).
\end{split}
\end{equation}

For {\it small curvature arches}, i.e. when the radius of curvature $\rho$ is far greater than the thickness of the cross section: $\dfrac{\rho}{\max \eta}\gg1 \ \forall s\Rightarrow \dfrac{\rho}{\rho-\eta}\rightarrow1$, the above equations become
\begin{equation}
\begin{split}
&\eps_\eta(s,\eta)=w'-\frac{v}{\rho}+\eta\kappa,\\
&\kappa(s)=-\varphi',\\
&\gamma(s)=v'+\frac{w}{\rho}-\varphi.
\end{split}
\end{equation}
The term $\eps_\eta(s,\eta)$ is the extension of the fibers at a distance $\eta$ from the centerline; we can write it as
\begin{equation}
\eps_\eta(s,\eta)=\eps(s)+\eta\kappa,
\end{equation}
where
\begin{equation}
\eps(s)=\eps_\eta(s,\eta=0)=w'-\frac{v}{\rho}.
\end{equation}

Finally, for small curvature arches, the case which is by far the most interesting in applications and that we will consider in the following, the kinematical quantities describing its deformation can be reduced to quantities referring to the deformation of the centerline, and are
\begin{equation}
\begin{split}
\label{eq:compatarch}
&\eps(s)=w'-\frac{v}{\rho},\\
&\gamma(s)=v'+\frac{w}{\rho}-\varphi,\\
&\kappa(s)=-\varphi'.
\end{split}
\end{equation}
These are the compatibility equations for small curvature arches. We remark that in the limit case of $\rho
\rightarrow\infty$ the above quantities become
\begin{equation}
\begin{split}
&\eps(s)=w',\\
&\gamma(s)=v'-\varphi,\\
&\kappa(s)=-\varphi',
\end{split}
\end{equation}
i.e. they coincide, as it must be, with those of the straight rods.

If now we introduce the symbolic matrix 
\begin{equation}
\gr{D}_2=\left[\begin{array}{ccc}
\dfrac{d}{ds} & -\dfrac{1}{\rho} & 0 \\
\dfrac{1}{\rho} & \dfrac{d}{ds} & -1 \\
0 & 0 & -\dfrac{d}{ds},
\end{array}\right]
\end{equation}
along with the vectors of the displacement, $\boldsymbol{\xi}$, and of the deformation, $\boldsymbol{\delta}$,
\begin{equation}
\boldsymbol{\xi}=\left\{\begin{array}{c}w \\v \\\varphi\end{array}\right\},\ \ \ \boldsymbol{\delta}=\left\{\begin{array}{c}\eps \\\gamma \\\kappa\end{array}\right\},
\end{equation}
eq. (\ref{eq:compatarch}) can be written in matrix form:
\begin{equation}
\label{eq:matrixeq3}
\gr{D}_2\boldsymbol{\xi}=\boldsymbol{\delta}.
\end{equation}

The same equation can be written in the fixed frame $\{z,y\}$:
\begin{equation}
\label{eq:matrixeq4}
\gr{D}_2\gr{Q}\boldsymbol{\xi^*}=\boldsymbol{\delta},
\end{equation}
where $\boldsymbol{\xi^*}$ is the displacement vector written in the frame $\{z,y\}$ and $\gr{Q}$ is the rotation matrix (\ref{eq:matrixQ}).

\section{Constitutive equations}
In the framework of linear elasticity, we assume as constitutive equations for plane arches the same of the straight rods:
\begin{equation}
\label{eq:constarch1}
\begin{split}
&N=EA\eps,\\
&T=\frac{\mu A}{\chi}\gamma,\\
&M=EJ\kappa,
\end{split}
\end{equation} 
with $E$ the Young's modulus and $\mu$ the shear modulus of the material, $A$ the area of the cross section, $J$ the moment of inertia of the cross section about an horizontal axis passing by the barycenter and $\chi$ the shear factor of the cross section.

Introducing the  matrix
\begin{equation}
\gr{C}=\left[\begin{array}{ccc}EA & 0 & 0 \\0 & \dfrac{\mu A}{\chi} & 0 \\0 & 0 & EJ\end{array}\right],
\end{equation}
we can write eq. (\ref{eq:constarch1}) in matrix form:
\begin{equation}
\label{eq:constarch2}
\gr{C}\boldsymbol{\delta}=\gr{S}.
\end{equation}

\section{The problem of elastic equilibrium for the arches}
We can now put together the constitutive, compatibility and equilibrium equations found above to write the equations of the elastic equilibrium for plane arches. Some simple calculations give
\begin{equation}
\label{eq:archeq1}
\begin{split}
&\left[EA\left(w'-\frac{v}{\rho}\right)\right]'=\frac{\mu A}{\rho\chi}\left(v'+\frac{w}{\rho}-\varphi\right)-r,\\
&\left[\frac{\mu A}{\chi}\left(v'+\frac{w}{\rho}-\varphi\right)\right]'=-\frac{EA}{\rho}\left(w'-\frac{v}{\rho}\right)-p,\\
&\left(-EJ\varphi'\right)'=\frac{\mu A}{\chi}\left(v'+\frac{w}{\rho}-\varphi\right)+m.
\end{split}
\end{equation}
The above equations show that all the equations are coupled; in particular, unlike the case of straight rods, as an effect of the geometry extension and bending are coupled. 

We can put the above equations in a matrix form; to this purpose, we inject successively eqs. (\ref{eq:constarch2}) and (\ref{eq:matrixeq3}) into eq. (\ref{eq:matrixeq1}) to get easily
\begin{equation}
\gr{D}_1\gr{CD}_2\boldsymbol{\xi}+\gr{f}=\gr{o}.
\end{equation}
To write the same equation in the fixed frame $\{z,y\}$, it is sufficient to do the same but with eqs. (\ref{eq:matrixeq2}), (\ref{eq:matrixeq4}) and (\ref{eq:constarch2}), and left-multiply by $\gr{Q}^\top$:
\begin{equation}
\gr{Q}^\top\gr{D}_1\gr{CD}_2\gr{Q}\boldsymbol{\xi^*}+\gr{f^*}=\gr{o}.
\end{equation}

No matter of the form given to the above equations, they remain a system of three second-order coupled linear differential equations, that need 6 boundary conditions, three for each end of the arch. Their solution is, normally, impossible analytically and numerical methods must be used.

\section{Transforming the equations of the arches}
The equations found in the previous Sections can be transformed, so as to obtain equations that can be more easily solved; some assumptions on the geometry or the kinematics of the arch can also be introduced, with the same purpose. We consider first the case of the balance equations (\ref{eq:eqarc2}), then that of the elastic equilibrium equations (\ref{eq:archeq1}).

\subsection{Transforming the balance equations}
\label{sec:trasfbaleq}
\begin{figure}[h]
\begin{center}
\includegraphics[scale=.8]{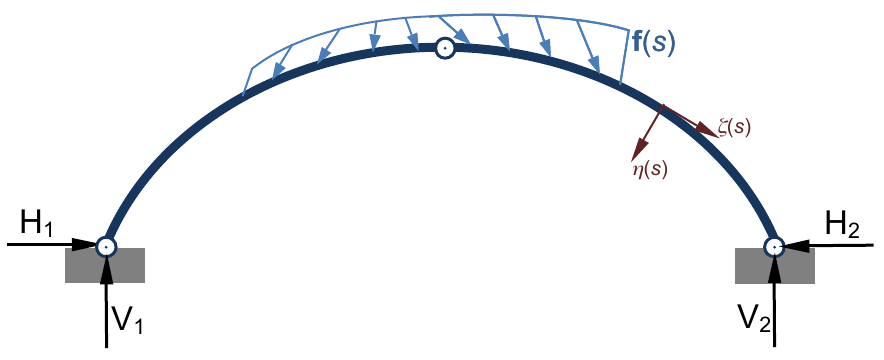}
\caption{Scheme of a three-hinged arch.}
\label{fig:f3_6}
\end{center}
\end{figure}

For an isostatic arch the balance equations are sufficient to determine the internal actions $N, T$ and $M\ \forall s$. A typical example, very used in the applications, especially in bridge constructions, is that of a three-hinged arch, see Fig. \ref{fig:f3_6}.

We can rearrange the balance equations (\ref{eq:eqarc2}) in order to eliminate the shear from eqs. (\ref{eq:eqarc2})$_{1,3}$ so as to obtain a system of two linear differential equations. This can be done as follows: from eq. (\ref{eq:eqarc2})$_3$ we get 
\be
\label{eq:balarch1}
T=\frac{dM}{ds}-m,
\ee
that injected into eq. (\ref{eq:eqarc2})$_1$ gives
\be
\label{eq:balarch2}
\frac{dN}{ds}=\frac{1}{\rho}\frac{dM}{ds}-\frac{m}{\rho}-r.
\ee
Then, we differentiate eq. (\ref{eq:eqarc2})$_3$ to obtain
\be
\frac{d^2M}{ds^2}=\frac{dT}{ds}+\frac{dm}{ds}
\ee
and using eq. (\ref{eq:eqarc2})$_2$ we get
\be
\label{eq:balarch3}
\frac{d^2M}{ds^2}=-\frac{N}{\rho}-p+\frac{dm}{ds}.
\ee

Equations (\ref{eq:balarch2}) and (\ref{eq:balarch3}) constitute a system of two coupled differential equations where the unknowns are the functions $N(s)$ and $M(s)$. In the end, we need 3 boundary conditions for the solution of the equilibrium problem of isostatic arches. The shear force $T(s)$ can be calculated by eq. (\ref{eq:balarch1}) once $M(s)$ known.

Of course, no information about the deformation of the arch is given by the solution of the balance equations; once the problem solved, one can introduce $N,\ T$ and $M$ into the constitutive laws, eqs. (\ref{eq:constarch1}), to obtain $\eps, \ \gamma$ and $\kappa$, that introduced in the compatibility equations (\ref{eq:compatarch}), once integrated, give the components of displacement $w,\ v$ and$\varphi$. Nevertheless, it is normally preferable, at least to find the components of displacement at a specific point, to employ other methods, like the dummy load method.
% it is uncoupled from $M(s)$ and $T(s)$ and can be obtained integrating twice eq. (\ref{eq:balarch7}) and using two boundary conditions, normally the value of $N$ at the two ends of the arch. 

%Once eq. (\ref{eq:balarch7}) solved, $N(s)$ can be replaced into eq. (\ref{eq:balarch5}) to obtain $T(s)$ and into eq. (\ref{eq:balarch3}) to get, after a double integration, $M(s)$. Alternatively, $T(s)$ can be injected into eq. (\ref{eq:eqarc2})$_3$ for obtaining $M(s)$ after only a quadrature.

\subsection{Transforming the elastic equilibrium equations}
First of all, for the sake of conciseness, we call
\begin{equation}
\alpha:=EA,\ \ \beta:=\frac{\mu A}{\chi},\ \ \lambda:=EJ.
\end{equation}
Then, eqs. (\ref{eq:archeq1}) become
\be
\label{eq:archeq2}
\begin{split}
&\left[\alpha\left(w'-\frac{v}{\rho}\right)\right]'=\frac{\beta}{\rho}\left(v'+\frac{w}{\rho}-\varphi\right)-r,\\
&\left[\beta\left(v'+\frac{w}{\rho}-\varphi\right)\right]'=-\frac{\alpha}{\rho}\left(w'-\frac{v}{\rho}\right)-p,\\
&\left(-\lambda\varphi'\right)'=\beta\left(v'+\frac{w}{\rho}-\varphi\right)+m.
\end{split}
\ee

%\be
%\label{eq:archeq3}
%\bs
%&\alpha w''+\alpha'w'-\frac{\beta}{\rho^2}w-\frac{\alpha+\beta}{\rho}v'-\left(\frac{\alpha'}{\rho}-\frac{\alpha\rho'}{\rho^2}\right)v+\frac{\beta}{\rho}\varphi=-r,\\
%&\frac{\alpha+\beta}{\rho}w'+\left(\frac{\beta'}{\rho}-\frac{\beta\rho'}{\rho^2}\right)w+\beta v''+\beta'v'-\frac{\alpha}{\rho^2}v-\beta\varphi'-\beta'\varphi=-p,\\
%&\frac{\beta}{\rho}w+\beta v'+\lambda\varphi''+\lambda'\varphi'-\beta\varphi=-m.
%\end{split}
%\ee

We remark that in the limit case of an arch that tends to be a straight rod, i.e. for $\rho\rightarrow\infty$, the above equations become
\be
\label{eq:archeq3}
\begin{split}
&\left(\alpha w'\right)'=-r,\\
&\left[\beta\left(v'-\varphi\right)\right]'=-p,\\
&\left(-\lambda\varphi'\right)'=\beta\left(v'-\varphi\right)+m,
\end{split}
\ee
that are just the equations of the Timoshenko's rod model.

%It is worth to make the load $p$ appear in the bending equation (\ref{eq:archeq2})$_3$; to this purpose, we derive it to obtain
%\be
%\left(-\lambda\varphi'\right)''=\left[\beta\left(v'+\frac{w}{\rho}-\varphi\right)\right]'+m';
%\ee
%the first term on the right-hand side is just given by eq. (\ref{eq:archeq2})$_2$, so finally we obtain the general equations in the form
%\be
%\label{eq:archeq4}
%\begin{split}
%&\left[\alpha\left(w'-\frac{v}{\rho}\right)\right]'=\frac{\beta}{\rho}\left(v'+\frac{w}{\rho}-\varphi\right)-r,\\
%&\left[\beta\left(v'+\frac{w}{\rho}-\varphi\right)\right]'=-\frac{\alpha}{\rho}\left(w'-\frac{v}{\rho}\right)-p,\\
%&\left(-\lambda\varphi'\right)''=-\frac{\alpha}{\rho}\left(w'-\frac{v}{\rho}\right)-p+m'.
%\end{split}
%\ee

We can obtain an equivalent expression of eqs. (\ref{eq:archeq2}), where $p$ appears directly into the third equation and where the first and third equations contain exclusively the extension and bending stiffnesses, $\alpha$ and $\lambda$. To this purpose, it is sufficient to replace successively eqs. (\ref{eq:constarch1}) and (\ref{eq:compatarch}) into eqs. (\ref{eq:balarch2}) and (\ref{eq:balarch3}) to obtain
\be 
\label{eq:archeq5}
\begin{split}
&\left[\alpha\left(w'-\frac{v}{\rho}\right)\right]'=-\frac{1}{\rho}\left(\lambda\varphi'\right)'-r-\frac{m}{\rho},\\
&\left[\beta\left(v'+\frac{w}{\rho}-\varphi\right)\right]'=-\frac{\alpha}{\rho}\left(w'-\frac{v}{\rho}\right)-p,\\
&\left(\lambda\varphi'\right)''=\frac{\alpha}{\rho}\left(w'-\frac{v}{\rho}\right)+p-m'.
\end{split}
\ee

Unlike the case of the balance equations, it is not possible to uncouple the elastic equilibrium equations. Six boundary conditions complete eqs. (\ref{eq:archeq5}), they specify either the values of $v,\ w$ and $\varphi$ or of their derivatives, at the ends of the arch.

\subsection{The Euler-Bernoulli model for arches}

Let us now generalize the Euler-Bernoulli rod model to arches, assuming that the cross section remains plane and orthogonal to the deformed centerline of the arch, i.e.
\be
\varphi=v'.
\ee
Then, eqs. (\ref{eq:archeq5}) become
\be
\label{eq:archeq6}
\begin{split}
&\left[\alpha\left(w'-\frac{v}{\rho}\right)\right]'=-\frac{1}{\rho}\left(\lambda v''\right)'-r-\frac{m}{\rho},\\
&\left(\lambda v''\right)''=\frac{\alpha}{\rho}\left(w'-\frac{v}{\rho}\right)+p-m'.
\end{split}
\ee
The third equation, where the shear stiffness $\beta$ appears, is now meaningless; the shear $T$ can be recovered, once solved the above equations, which still needs six boundary conditions, using eq. (\ref{eq:balarch1}). 

It is easily checked that for $\rho\rightarrow\infty$ we get the Euler-Bernoulli rod equations. 

\subsection{Arches of constant section}
A particular case is that of an arch with a constant section; in such a case $\alpha'=\beta'=\lambda'=0$, so eqs. (\ref{eq:archeq5}) become
\be
\label{eq:archeq7}
\begin{split}
&\alpha\left(w'-\frac{v}{\rho}\right)'=-\frac{1}{\rho}\lambda\varphi''-r-\frac{m}{\rho},\\
&\beta\left(v'+\frac{w}{\rho}-\varphi\right)'=-\frac{\alpha}{\rho}\left(w'-\frac{v}{\rho}\right)-p,\\
&\lambda\varphi'''=\frac{\alpha}{\rho}\left(w'-\frac{v}{\rho}\right)+p-m'.
\end{split}
\ee
and eqs. (\ref{eq:archeq6})
\be
\label{eq:archeq8}
\begin{split}
&\alpha\left(w'-\frac{v}{\rho}\right)'=-\frac{1}{\rho}\lambda v'''-r-\frac{m}{\rho},\\
&\lambda v^{iv}=\frac{\alpha}{\rho}\left(w'-\frac{v}{\rho}\right)+p-m'.
\end{split}
\ee

\subsection{Circular arches}
A particularly important case of arches is that of circular arches; in such a case, $\rho=const. \ \Rightarrow\ \rho'=0$; actually, $\rho$ is  the radius of the (osculating) circle of which the arch is just a part.

%First of all, the uncoupled balance equation (\ref{eq:balarch7}) becomes
%\be
%\label{eq:circarc1}
%\frac{d^2N}{ds^2}+\frac{1}{\rho^2}N= -\frac{dr}{ds}-\frac{p}{\rho}.
%\ee

It is interesting to see what happens to the equations for the case of constant section: eqs. (\ref{eq:archeq7}) become
\be
\label{eq:circarc2}
\begin{split}
&\alpha\left(w''-\frac{v'}{\rho}\right)=-\frac{1}{\rho}\lambda\varphi''-r-\frac{m}{\rho},\\
&\beta\left(v''+\frac{w'}{\rho}-\varphi'\right)=-\frac{\alpha}{\rho}\left(w'-\frac{v}{\rho}\right)-p,\\
&\lambda\varphi'''=\frac{\alpha}{\rho}\left(w'-\frac{v}{\rho}\right)+p-m',
\end{split}
\ee
while eqs. (\ref{eq:archeq8}) become
\be
\label{eq:circarc3}
\begin{split}
&\alpha\left(w''-\frac{v'}{\rho}\right)=-\frac{1}{\rho}\lambda v'''-r-\frac{m}{\rho},\\
&\lambda v^{iv}=\frac{\alpha}{\rho}\left(w'-\frac{v}{\rho}\right)+p-m'.
\end{split}
\ee

\section{Examples}
\label{sec:exarch}

We give here some examples of the use of the theory developed in this Chapter. All the cases concern circular arches with a constant section. The first two examples have been treated using eqs. (\ref{eq:circarc2}), i.e. the general theory. Then, they have been also treated using eqs. (\ref{eq:circarc3}) of the Euler-Bernoulli model; the results coincide almost perfectly with those reported below, that is why they have not been reported here. The third example concerns an isostatic case, hence the balance equations are sufficient. All the results have been found numerically, using a standard commercial code (Mathematica). All the diagrams in the figures have been normalized, and the deformed shape exaggerated, hence the scale of the diagrams is not real. 

\subsection{Example 1}
\label{sec:exarc1}
\begin{figure}[h]
\begin{center}
\includegraphics[scale=.72]{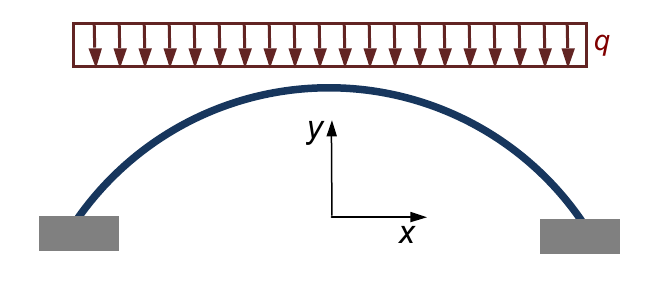}
\caption{Example 1: a uniformly loaded circular arch.}
\label{fig:f3_7_0}
\end{center}
\end{figure}

As a first example, we consider the case of a circular arch of constant section loaded by a uniform load $q=10$ t/m, with $\rho=10$ m, spanning a chord $\ell=17.1$ m, see Fig. \ref{fig:f3_7_0}. The cross section is rectangular, with the base $b=1$ m and the thickness $h=2$ m; the Young's modulus is $E=200$ MPa, the Poisson's ratio $\nu=0.2$, the shear factor $\chi=1.2$. Hence, we get $\alpha=4\times10^7$ kN, $\beta=\alpha/2.88=1.39\times10^8$ kN and $\lambda=\alpha/3=1.33\times10^8$ kN $m^2$. The ends of the arch are clamped. Hence, the appropriate boundary conditions are $w=v=\varphi=0$ at both the ends.

In the following Figures, we report the diagrams, referred to an horizontal axis or to the arch axis, in grey, of $w,\ v,\ \varphi,\ N,\ T, \ M$ and finally of the deformed shape. $N$ is negative everywhere, i.e. the arch is compressed all along $s$, as it must be, while $M$ changes of sign, as a consequence of the clamped edges. 
\begin{figure}[h]
\begin{center}
\includegraphics[width=.48\textwidth]{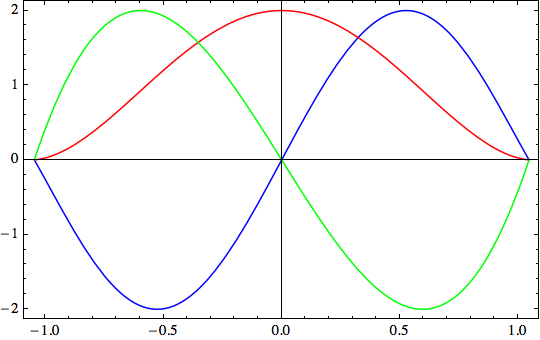}
\includegraphics[width=.48\textwidth]{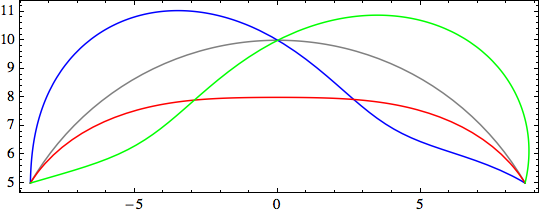}
\caption{Diagrams of $w$, blue, $v$, red, and $\varphi$, green, example 1.}
\label{fig:f3_7}
\end{center}
\end{figure}

\begin{figure}[h]
\begin{center}
\includegraphics[width=.49\textwidth]{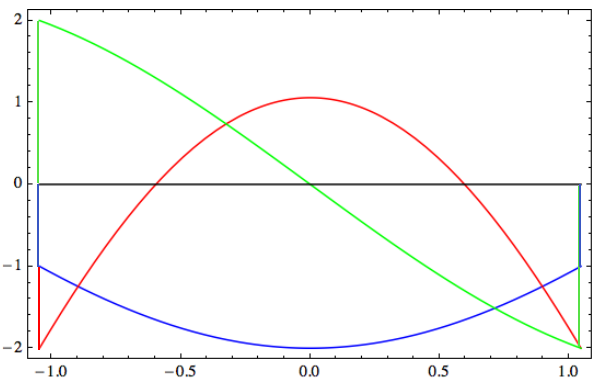}
\includegraphics[width=.49\textwidth]{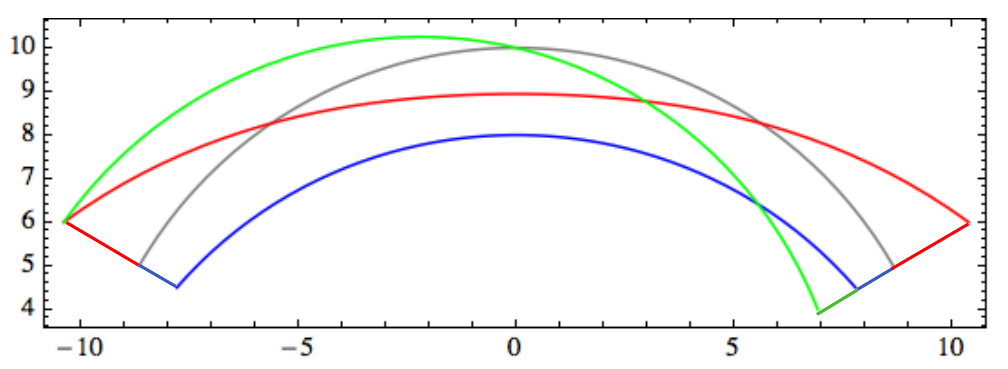}
\caption{Diagrams of $N$, blue, $M$, red, and $T$, green, example 1.}
\label{fig:f3_8}
\end{center}
\end{figure}

\begin{figure}[h]
\begin{center}
\includegraphics[scale=.4]{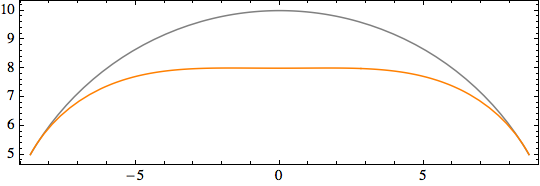}
\caption{Deformed shape of the arch, example 1.}
\label{fig:f3_9}
\end{center}
\end{figure}

\subsection{Example 2}
\begin{figure}[h]
\begin{center}
\includegraphics[scale=.55]{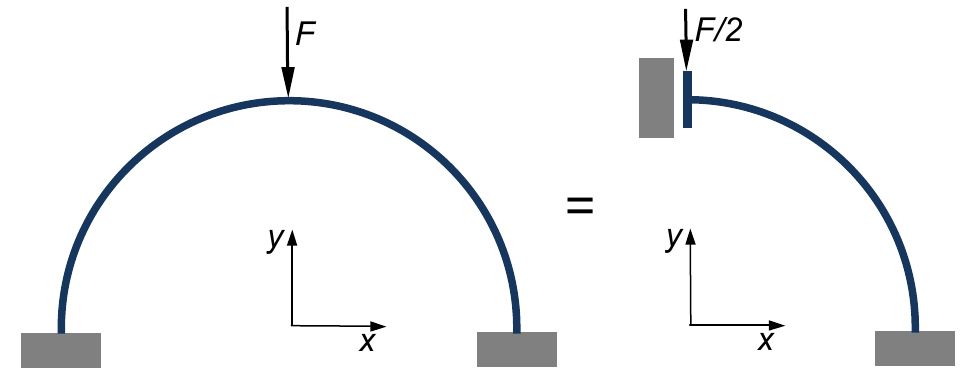}
\caption{Example 2: a circular arch with a concentrated vertical load.}
\label{fig:f3_10}
\end{center}
\end{figure}

A second example is that of a semicircular arch, with constant section, acted upon by a vertical force $F=5$ kN on the top, see Fig. \ref{fig:f3_10}. In this case, $\rho=0.865$ m, $\alpha=1.5\times 10^7$ MPa, $\beta=\alpha/2.88$ and $\lambda=\alpha/12\times10^4$.

Thanks to the symmetry of the problem, we can study just one half of the structure, acted upon by a force equal to $F/2$ and with a slide as a constraint on the top of the arch.  The appropriate boundary conditions are hence: at the left end, $w=\varphi=0$, $T=-F/2\ \Rightarrow\ \beta (v'+{w}/{\rho}-\varphi)=-{F}/{2}$, while at the right end it is $w=v=\varphi=0$.
The following diagrams, referring to only half of the arch, show the results for this example.

\begin{figure}[h]
\begin{center}
\includegraphics[width=.45\textwidth]{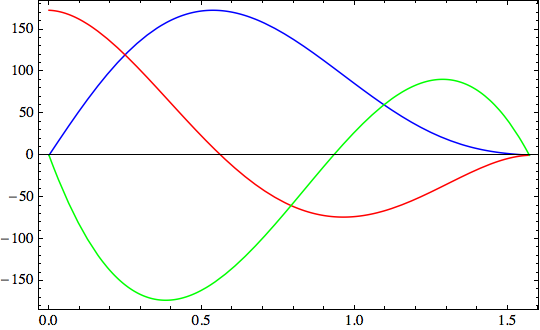}
\includegraphics[width=.45\textwidth]{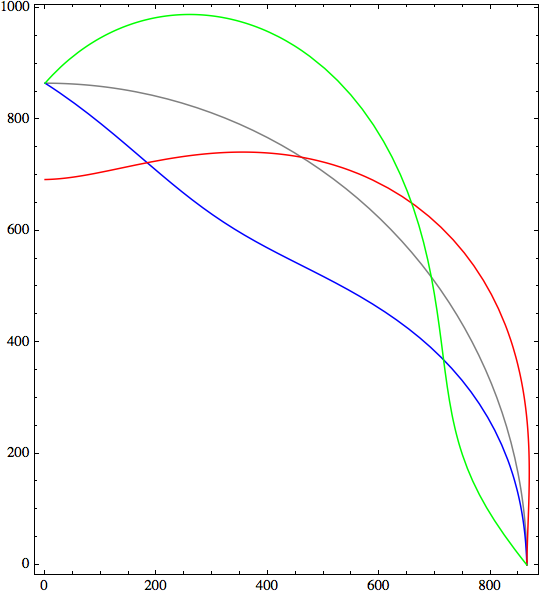}
\caption{Diagrams of $w$, blue, $v$, red, and $\varphi$, green, example 2.}
\label{fig:f3_11}
\end{center}
\end{figure}

\begin{figure}[h]
\begin{center}
\includegraphics[width=.45\textwidth]{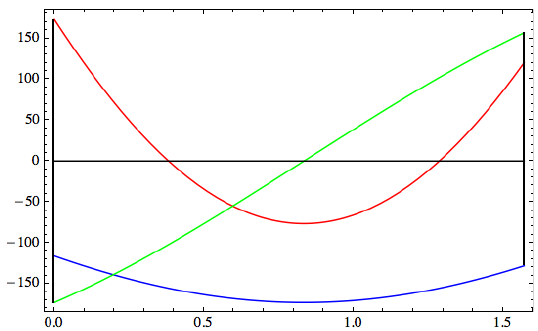}
\includegraphics[width=.45\textwidth]{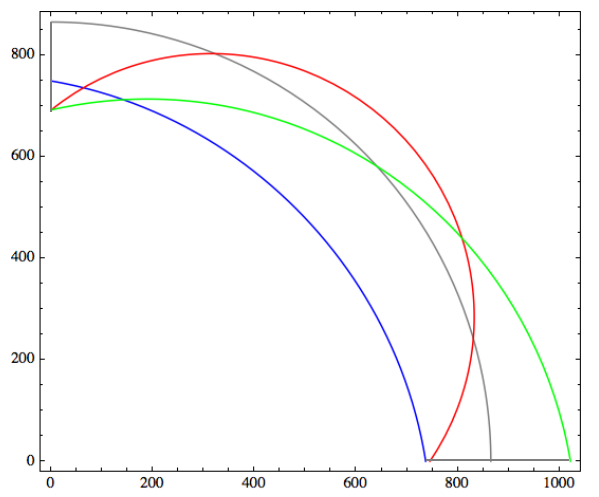}
\caption{Diagrams of $N$, blue, $M$, red, and $T$, green, example 2.}
\label{fig:f3_12}
\end{center}
\end{figure}

\begin{figure}[h]
\begin{center}
\includegraphics[scale=.25]{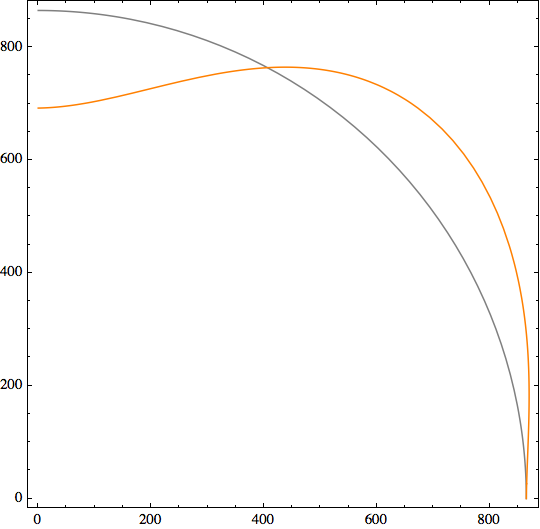}
\caption{Deformed shape of the arch, example 2.}
\label{fig:f3_13}
\end{center}
\end{figure}

\subsection{Example 3}
This last example concerns a three-hinge arch, hence an isostatic case. For this reason, we can  simply consider the equilibrium equations, either in the form (\ref{eq:eqarc2}) or in the form of eqs. (\ref{eq:balarch2}) and (\ref{eq:balarch3}).

\begin{figure}[h]
\begin{center}
\includegraphics[scale=.6]{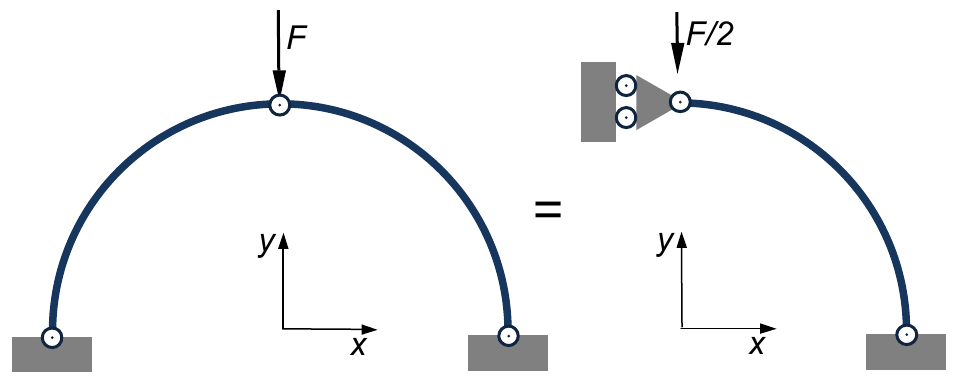}
\caption{Example 3: three-hinged arch.}
\label{fig:f3_14}
\end{center}
\end{figure}

The problem is analogous to that of Example 2 and in particular the data are the same, but now the arch has a hinge also at the top, in correspondence of the force $F$.
Again, we can study just one half of the structure, see Fig. \ref{fig:f3_14}. The boundary conditions now concern exclusively $T$ and $M$. In fact, no information is given about the value of $N$ at the edges of the arch. In particular, the boundary conditions are: at the left edge, $M=0$ and $T=-F/2$, while at the right edge $M=0$. If the system of the two differential equations (\ref{eq:balarch2}) and (\ref{eq:balarch3}) is used, then the condition of the shear force must be written in terms of the derivative of $M$, eq. (\ref{eq:eqarc2})$_3$: $T=-F/2\ \Rightarrow\ M'-m=-F/2$. In this case, we have just solved numerically eqs. (\ref{eq:balarch2}) and (\ref{eq:balarch3}), and then obtained $T$ from (\ref{eq:eqarc2})$_3$. 
The displacement components $w,\ v$ and $\varphi$ have been obtained by the procedure described in Sect. \ref{sec:trasfbaleq}; in particular, the three boundary conditions for integrating the compatibility equations (\ref{eq:compatarch}) are $w=0$ at the left end and $w=v=0$ at the right one.
The results are presented in Figs. \ref{fig:f3_15} to \ref{fig:f3_17}. To remark that the rotation $\varphi$ is almost constant.

%Two remarks about this last example. The first one concerns the use of eq. (\ref{eq:balarch7}): in this example, we can remark that it cannot be used to solve the equilibrium problem. In fact, none of the above boundary conditions, the only ones that can be written, concerns $N$. In addition, now, together with eq. (\ref{eq:balarch3}), we have a system of two differential equations both of the second order: four boundary conditions are needed, in particular, because no information is given about the value of $N$ at the arch ends, one should know the value of $N'$ at one of the edges, but such a value is unknown. Anyway,  (\ref{eq:balarch7}) remains coupled to eq. (\ref{eq:balarch3}) by the boundary conditions.

We remark that in this particular case the balance equations have an analytical solution, that can be easily found   using either 
eqs.  (\ref{eq:eqarc2}) or eqs. (\ref{eq:balarch2}) and (\ref{eq:balarch3}):
\be
\besp
&N=-\frac{F}{2}(\sin\theta+\cos\theta),\\
&T=\frac{F}{2}(\sin\theta-\cos\theta),\\
&M=\frac{F}{2}\rho(1-\sin\theta-\cos\theta).
\end{split}
\ee

\section{Exercices}
\begin{enumerate}
\item Write the equations of the elastic equilibrium of arches using as variable the angle $\theta$ of a polar frame instead of $s$.

\item Write and solve analytically the balance  equations for the circular arch of constant section in  Fig. \ref{fig:f3_17}, right. Try also to find the functions $N, \ T$ and $M$ using basic equilibrium considerations.

\item For the previous exercice, find the horizontal displacement of the loaded node using the dummy load method ($A, J, E, \nu$ and $\chi$ are given).

\item Consider again the Example 1, Sect. \ref{sec:exarc1}; in place of the clamped edges, consider a hinge at the left end and a simple horizontal support at the right one. Using the dummy load method find the displacement of the right edge (neglect the effects of $N$ and $T$).

\item Still for the Example 1 of Sect. \ref{sec:exarc1}, be $A=A_0\cos\theta,\ J=J_0\cos\theta$ the variation of the cross section area and moment of inertia, respectively, with the angle $\theta$, the angle of the polar coordinates in a frame centered in the center of the arch. $A_0$ and $J_0$ are the values for $\theta=0$ and they coincide with those specified in the Example 1. Using $\theta$ as independent variable, write the equations of the elastic equilibrium for the two cases of the general theory and of the Euler-Bernoulli model.

%\item Solve numerically the previous exercice.

\begin{figure}[h]
\begin{center}
\includegraphics[height=.15\textheight]{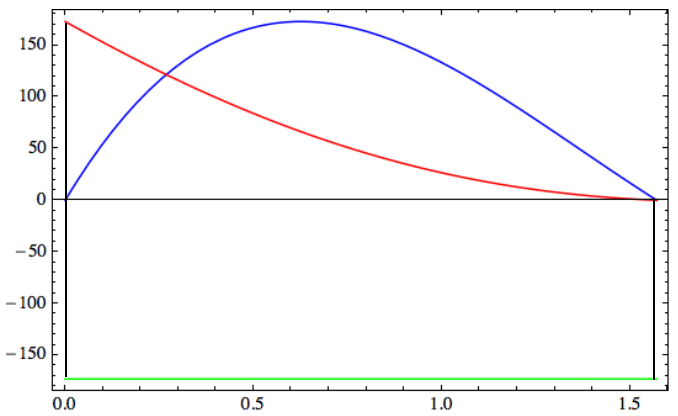}
\includegraphics[height=.15\textheight]{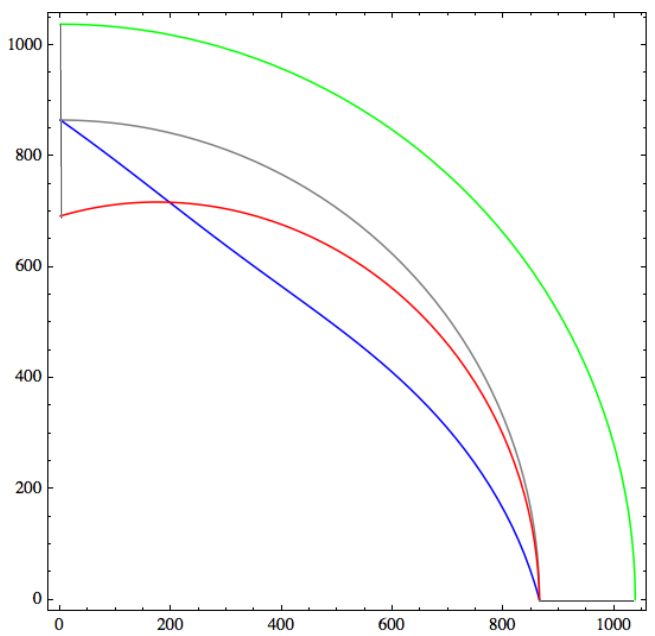}
\caption{Diagrams of $w$, blue, $v$, red, and $\varphi$, green,  example 3.}
\label{fig:f3_15}
\end{center}
\end{figure}

\begin{figure}[h]
\begin{center}
\includegraphics[height=.2\textheight]{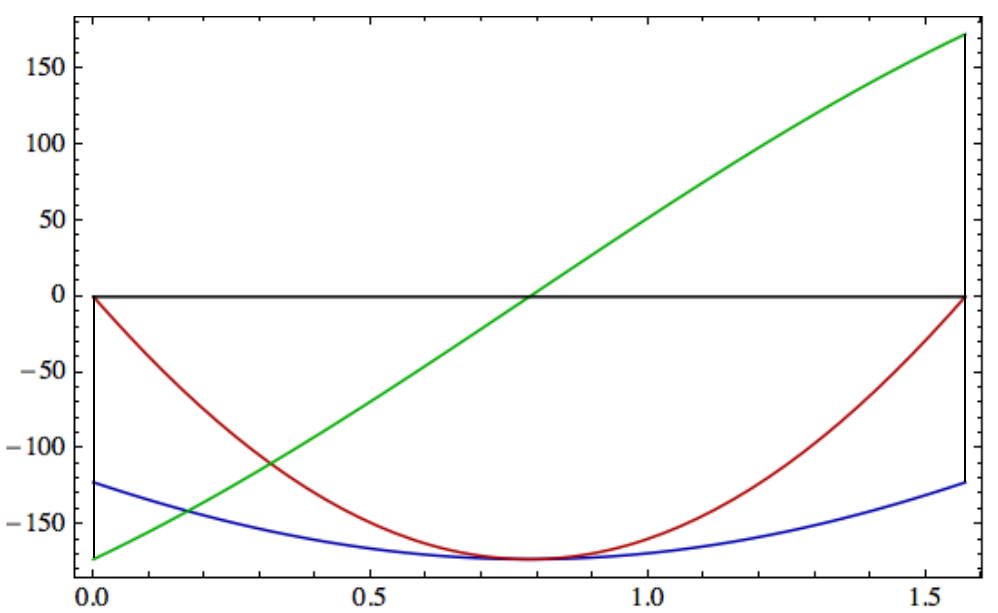}
\includegraphics[height=.2\textheight]{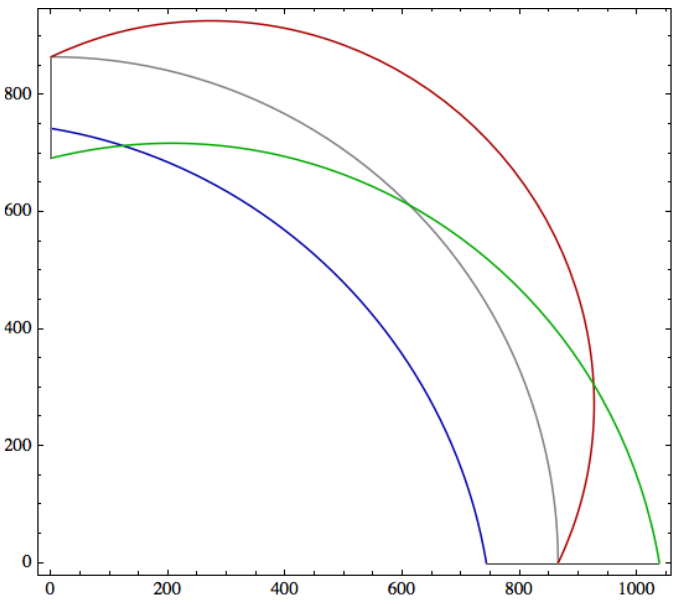}
\caption{Diagrams of $N$, blue, $M$, red, and $T$, green, example 3.}
\label{fig:f3_16}
\end{center}
\end{figure}

\begin{figure}[h]
\begin{center}
\includegraphics[height=.2\textheight]{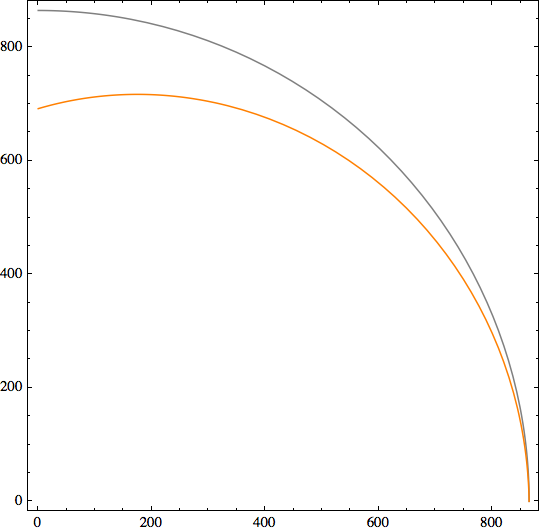}\hspace{20mm}
\includegraphics[height=.2\textheight]{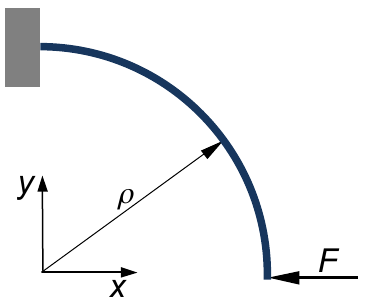}
\caption{Left: deformed shape of the arch, example 3; right: arch of the exercice 2.}
\label{fig:f3_17}
\end{center}
\end{figure}

\end{enumerate}

\chapter{Slender rods}
\label{ch:5}
\section{Directors rods' theory}
We consider a continuum body in the form of a {\it slender rod}, i.e. a  body where one dimension, the length, is much greater than the largest chord of the cross-section, that it is assumed to be constant throughout the rod's length. The rod is not necessarily straight in its undeformed configuration.

We look for a theory particularly adapted to such a continuum body, i.e. taking into account for the geometrical characteristics. In particular, this theory must be able to take into account for the two principal deformation modes: bending and twisting of the rod.

This can be done in the following way: the rod is represented by a line, physically the centerline of the cross-sections, endowed with some directions , i.e. vectors, that may transform independently of the deformation of the line they are attached to.

Such directions attached to the centerline are called {\it directors}; finally, a rod is a material curve that is able to deform independently of the collection of directors assigned to each one of its particles.

The actual configuration of a Cosserat rod is specified by a vector function $ \gr{r}(s,t)=p(s,t)-o$, that gives $\forall s\in[s_1,s_2]$, the curvilinear abscissa, the position of a point $p$ of the curve at the instant $t$ with respect to a fixed frame $\mathcal{R}=\{o;\gr{e}_1,\gr{e}_2,\gr{e}_3\}$, and by the set of {\it directors}
\be
\gr{d}_k(s,t),\ k=1,...,n_d.
\ee 
The most important case is $n_d=2$; it gives the simplest theory able to account for bending, twisting, extension and also for shear and transverse extensions, i.e. for the deformation of the cross section.

We require that, under the change of frame $\gr{r}(s,t)\rightarrow q(t)+\gr{Q}(t)\gr{r}(s,t)$, with $q(t)$ a translation and $\gr{Q}(t)$ a proper rotation $\forall t$, the directors transform as $\gr{d}_k(s,t)\Rightarrow\gr{Q}(t)\gr{d}_k(s,t)$, i.e.  directors are insensitive to translations. In addition, we require the following {\it continuity condition}:
\be
\label{eq:contcond1}
\gr{d}_1\times\gr{d}_2\cdot p'>0\ \forall s,t.
\ee
The physical meaning of this condition will be explained in Sect. \ref{sec:1_2}. Here and in the following of the Chapter, the symbol of prime denotes differentiation with respect to $s$. The continuity condition imposes the vectors $\gr{d}_1,\ \gr{d}_2$ and $p'$ to be independent. We recall that $p'(s)=\r'(s)=\btau(s)\ \forall s$.

\subsection{Reference configuration}
Conventionally, we take the configuration of the rod at $t=0$ as the reference configuration and will indicate any quantity attached to it by the superscript $0$. So, the reference configuration is determined by
\be
\besp
&\gr{r}^0=p^0(s)-o,\\
&\gr{d}_k^0(s),\ k=1,...,n_d.
\end{split}
\ee

So, for the case $n_d=2$, the initial configuration is determined by the three vectors $\gr{r}^0,\ \gr{d}_1^0,\ \gr{d}_2^0$, along with the continuity condition
\be
\label{eq:contcond2}
\gr{d}^0_1\times\gr{d}^0_2\cdot p'^0>0\ \forall s,t.
\ee

Knowing $\gr{r}^0,\ \gr{d}_1^0,\ \gr{d}_2^0$, it is important to dispose of a differential description of the undeformed rod. There is not a unique way to do that, let us see two classical approaches.

\subsubsection{First approach} 
We put $\gr{d}_1^0=\bnu$, the principal normal (if this is not defined, like for straight rods, it can be fixed arbitrarily, e.g. as coincident with one of the principal axes of inertia of the cross section) and $\gr{d}_2^0=\bb$, the binormal. If we call $p'^0=\btau:=\gr{d}_3^0$, then the triad $\{\gr{d}_1^0,\gr{d}_2^0,\gr{d}_3^0\}$ forms a positively oriented  frame that coincides with the Frenet-Serret frame (without loss of generality, we take the directors as unit vectors).

Hence, the variation of the the directors along $s$, in the reference configuration, is given by the Frenet-Serret formulae ($c$ and $\vartheta$ are, respectively, the curvature and the torsion of the undeformed rod):
\be
\besp
&{\gr{d}_1^0}'={\bnu}'=-c\ \gr{d}_3^0-\vartheta\ \gr{d}_2^0,\\
&{\gr{d}_2^0}'={\bb}'=\vartheta\ \gr{d}_1^0,\\
&{\gr{d}_3^0}'={\btau}'=c\ \gr{d}_1^0.
\end{split}
\ee
Of course, such a result is correct only upon the assumption that the directors remain coincident with $\bnu^0,\ \bb^0,\ \btau^0$ to the curve as $s$ varies. This is not necessary: we can choose also a set of directors attached to the curve but turning independently of it.

\subsubsection{Second approach}
We use a unique director $\d_1^0$ orthogonal to the curve:
\be
{p'^0}\cdot\d_1^0=0,
\ee
subtending the angle $\eta$ with $\bnu$:
\be
\d_1^0=\bnu \cos\eta+\bb\sin\eta.
\ee
The quantity $\eta'(s)$ is the twist of $\d_1^0$ with respect to the axis. Because
\be
\d_1^0\cdot\bnu=\cos\eta,
\ee
it is
\be
(\d_1^0\cdot\bnu)'={\d_1^0}'\cdot\bnu+\d_1^0\cdot\bnu'=-\eta'\sin\eta,
\ee
and using the Frenet-Serret formula for $\bnu$ and the expression of $\d_1^0$, we get
\be
%(\d_1^0\cdot\bnu)'-\vartheta_0\sin\varphi=-\varphi'\sin\varphi.
{\d_1^0}'\cdot\bnu=\vartheta\sin\eta-\eta'\sin\eta.
\ee
This equation, for $\eta\neq0$, allows for determining the twist of the rod in terms of the torsion and the derivative of $\d_1^0$.

\subsection{Deformed configuration}
\label{sec:1_2}
The deformed configuration is defined by
\be
\r(s,t)=p(s,t)-o,
\ee
together with a relation linking the new directors to the old ones:
\be
\d_k(s,t)=\d_k(\d_k^0(s);s,t).
\ee
Generally speaking, the $\d_k$s are no more unit vectors nor mutually orthogonal. However, the following assumptions are usually done:
\begin{itemize}
\item $\d_1$ and $\d_2$ preserve their initial length ad remain orthogonal: $\d_1\cdot\d_2=0\ \forall s,t$;
\item the plane whose normal is
\be
\d_3=\d_1\times\d_2
\ee
determines the  section of the rod at $s\ \forall t$. So defined, $\d_3$ is a unit vector orthogonal to $\d_1$ and $\d_2$; the set $\{\d_1,\d_2,\d_3\}$ forms hence a positively oriented orthonormal basis $\forall s,t$;
\item $\d_3$, unlike $\d_3^0$, is not necessarily tangent to the deformed rod: generally speaking
\be
\d_3\times \r'\neq\bo;
\ee
\item  the {\it continuity condition} (\ref{eq:contcond1}) holds:
\be
\label{eq:contcond}
\d_1\times\d_2\cdot p'=\d_3\cdot p'>0\ \ \forall s,t;
\ee
this condition is, for rods, the corresponding of the general 3D condition $\det \mathbf{F}>0$, with {\bf F} the {\it deformation gradient}: it excludes that the  section may be so severely sheared so as to contain the tangent to the axis.
\end{itemize}
To remark, hence, that generally speaking the section not necessarily coincides with the cross section, this last being contained by the plane orthogonal to the tangent $p'$ to the deformed rod.

\begin{figure}[th]
\begin{center}
\includegraphics[scale=.5]{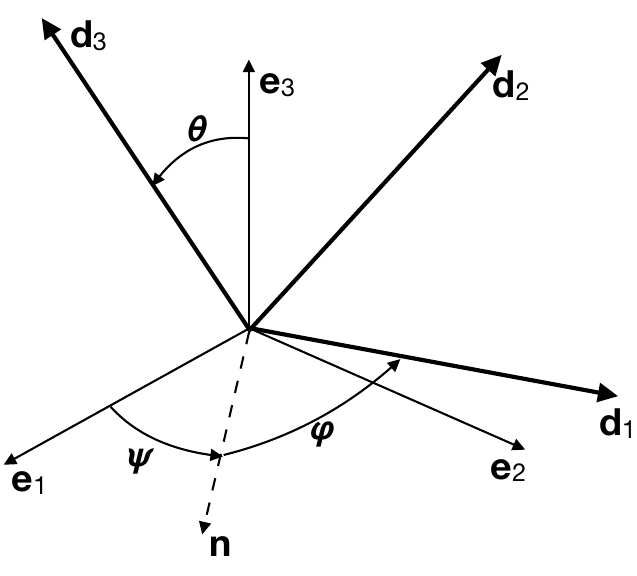}
\caption{Euler's angles.}
\label{fig:f4_1}
\end{center}
\end{figure}

\begin{figure}[th]
\begin{center}
\includegraphics[width=.7\textwidth]{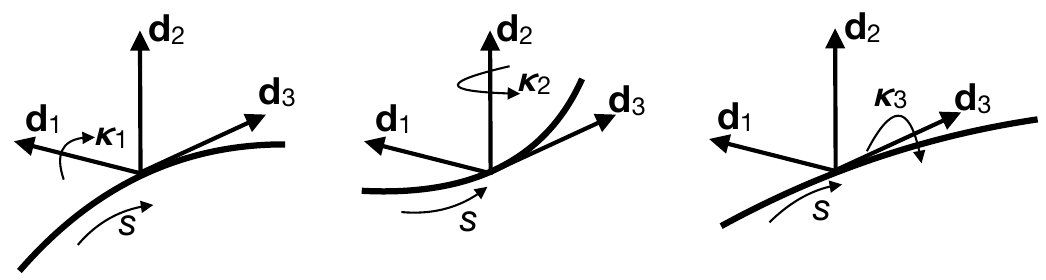}
\caption{Material curvatures.}
\label{fig:f4_2}
\end{center}
\end{figure}

For the kinematical description of the rod's deformed configuration, we determine the triad $\{\d_1,\d_2,\d_3\}$ with respect to a fixed one, e.g. $\{\e_1,\e_2,\e_3\}$, the triad of $\mathcal{R}$. Because both these triads are orthonormal and positively oriented, they are related by a proper rotation tensor $\Q(s)$, i.e. $\forall s\ \exists\ \Q(s),\ \Q^{-1}=\Q^\top,\det\Q=1|$
\be
\label{eq:rod1}
\d_k(s)=\Q(s)\e_k,\ \ \ k=1,2,3.
\ee
The tensor $\Q(s)$ depends upon three parameters; a classical choice is that of the {\it Euler's angles}: the {\it precession} $\psi\in[0,2\pi)$, the {\it nutation} $\theta\in[0,\pi]$ and the {\it proper rotation} $\varphi\in[0,2\pi)$, see  Fig. \ref{fig:f4_1}. In such a case, it can be shown that 
\be
\Q=\left[
\begin{small}
\begin{array}{ccc}
\cos\psi\cos\varphi-\sin\psi\sin\varphi\cos\theta & -\cos\psi\sin\varphi-\sin\psi\cos\varphi\cos\theta & \sin\psi\sin\theta \medskip\\
\sin\psi\cos\varphi+\cos\psi\sin\varphi\cos\theta & -\sin\psi\sin\varphi+\cos\psi\cos\varphi\cos\theta & -\cos\psi\sin\theta \medskip\\
\sin\varphi\sin\theta & \cos\varphi\sin\theta & \cos\theta
\end{array}
\end{small}
\right].
\ee
Let, in the fixed frame (for the sake of shortness, in the following the dependence upon $t$ is omitted but tacitly understood)
\be
\r(s)=x_k(s)\e_k\rightarrow\r'(s)=x_{k}'(s)\e_k=y_k(s)\d_k(s),
\ee
with $x_k$ the components of $\r(s)$ in the fixed frame $\Rep=\{o;\e_1,\e_2,\e_3\}$ and $y_k$ those of $\r'(s)$ in the local basis $\{\d_1,\d_2,\d_3\}$ at $s$. So, we get
\be
x'_{k}(s)\e_k=y_k(s)\Q(s)\e_k=\Q(s)\left(y_k(s)\e_k\right),
\ee
which gives
\be
\besp
x'_{1}=&y_1(\cos\psi\cos\varphi-\sin\psi\sin\varphi\cos\theta)-y_2(\cos\psi\sin\varphi+\sin\psi\cos\varphi\cos\theta)+\\
&y_3\sin\psi\sin\theta,\\
x'_{2}=&y_1(\sin\psi\cos\varphi+\cos\psi\sin\varphi\cos\theta)-y_2(\sin\psi\sin\varphi-\cos\psi\cos\varphi\cos\theta)-\\
&y_3\cos\psi\sin\theta,\\
x'_{3}=&y_1\sin\varphi\sin\theta+y_2\cos\varphi\sin\theta+y_3\cos\theta.
\end{split}
\ee
We introduce now the following {\it local kinematical descriptors} of the rod's deformation:
\be
\besp
\kappa_1:=\d'_{2}\cdot\d_3,\\
\kappa_2:=\d'_{3}\cdot\d_1,\\
\kappa_3:=\d'_{1}\cdot\d_2.
\end{split}
\ee
Geometrically, these quantities represent the local variation of the directors, i.e. of their direction, when passing from a point of the rod to another one infinitely close to it. In particular, see Fig. \ref{fig:f4_2},
\begin{itemize}
\item $\kappa_1$ is a {\it material curvature} associated with $\d_1$;
\item $\kappa_2$ is a {\it material curvature} associated with $\d_2$;
\item $\kappa_3$, the {\it twist}, is a {\it material curvature} associated with $\d_3$.
\end{itemize}
Using eq. (\ref{eq:rod1}) and after differention, we get
\be
\label{eq:curvangEul}
\besp
&\kappa_1=\theta'\cos\varphi+\psi'\sin\theta\sin\varphi,\\
&\kappa_2=-\theta'\sin\varphi+\psi'\sin\theta\cos\varphi,\\
&\kappa_3=\varphi'+\psi'\cos\theta.
\end{split}
\ee
The functions $y_k,\kappa_1,\kappa_2$ and $\kappa_3$ constitute the {\it strains}, i.e. the rod's deformation descriptors in the directors theory. They are almost universally used as measures of strain in a rod, e.g. Clebsch (1862) and Love (1927). Some authors, e.g. Saint Venant, preferred other approaches, namely the  definition of twist given in the second approach described above. 

We can also express the variations $\d'_{k}$ of the directors as functions of the curvatures and of the twist. This can be done if one considers that, being the $\d_k$s unit orthogonal vectors, it is (no summation on the dummy indexes)
\be
\besp
&\d_k\cdot\d'_{k}=0,\\
&\d_k\cdot\d_j=0\Rightarrow\\
&(\d_k\cdot\d_j)'=0\rightarrow\d'_{k}\cdot\d_j=-\d_k\cdot\d'_{j}.
\end{split}
\ee
So, because
\be
\besp
&\kappa_1=\d'_2\cdot\d_3=-\d_2\cdot\d'_3,\\
&\kappa_2=\d'_3\cdot\d_1=-\d_3\cdot\d'_1,\\
&\kappa_3=\d'_1\cdot\d_2=-\d_1\cdot\d'_2,
\end{split}
\ee
then we get
\be
\label{eq:darboux1}
\besp
&\d'_1=\kappa_3\d_2-\kappa_2\d_3,\\
&\d'_2=-\kappa_3\d_1+\kappa_1\d_3,\\
&\d'_3=\kappa_2\d_1-\kappa_1\d_2,
\end{split}
\ee
or in a succinct form
\be
\label{eq:darboux2}
\d'_j=\bom\times\d_j,\ \ \ j=1,2,3,
\ee
with $\bom$ the {\it Darboux vector}:
\be
\label{eq:darboux3}
\bom=\kappa_1\d_1+\kappa_2\d_2+\kappa_3\d_3.
\ee
The Darboux vector $\bom$ can be interpreted as the rotation velocity of the material local basis $\{\d_1,\d_2,\d_3\}$ when the rod is followed at unit speed. So, $\kappa_1,\kappa_2$ and $\kappa_3$ can be seen as the velocity of rotation (or equivalently, the rotation per unit length of $s$) around $\d_1,\d_2$ and $\d_3$ respectively. To remark that $\kappa_1$ and $\kappa_2$ are different from $c$, and $\kappa_3$ from $\vartheta$, as defined in Chapter \ref{ch:1}.

\subsection{Balance equations}
We assume the following fundamental stress principle for rods: across any section ideally dividing the rod into two parts, the action of one part onto the other one is equivalent to a resultant force {\bf F} and a resultant couple {\bf M}\footnote{For this reason, somebody calls such a rod model a  {\it Cosserat rod}, from the brothers Cosserat, who gave the first general theory for bodies able to transmit locally not only  forces but also  couples, 1907.}. Hence, contrarily to classical elastic 3D bodies, a rod is a continuum able to transmit couples, not only forces. Be $\q(s)$ the vector function of the forces distributed along the rod's axis and $\m(s)$ that representing distributed couples. The balance equations for a portion $[a,b]\subset[s_1,s_2]$ of the rod  are:
\be
\besp
&\F_b-\F_a+\int_a^b\q\ ds=\bo,\medskip\\
&\M_b-\M_a+\r_a\times\F_a-\r_b\times\F_b+\int_a^b\m+\r\times\q\ ds=\bo.
\end{split}
\ee
when $a\rightarrow b$, the above equations take their local form:
\be
\label{eq:loceqrod}
\besp
&\F'+\q=\bo,\\
&\M'+\r'\times\F+\m+\r\times\q=\bo.
\end{split}
\ee
The boundary conditions to be associated to eqs. (\ref{eq:loceqrod}) prescribe the value of $\F,\M,\r$ and/or $\r'$ at $s_1$ and $s_2$, the rod's ends. If
\be
\besp
&\F=F_k\ \d_k,\\
&\M=M_k\ \d_k,
\end{split}
\ee
so that
\be
\besp
&\F'=F'_k\ \d_k+F_k\ \d'_k,\\
&\M'=M'_k\ \d_k+M_k\ \d'_k,
\end{split}
\ee
and using eqs. (\ref{eq:darboux1}) we get the following form of the local balance equations:
\be
\label{eq:equilrod1}
\besp
&F'_1+\kappa_2\ F_3-\kappa_3\ F_2+q_1=0,\\
&F'_2+\kappa_3\ F_1-\kappa_1\ F_3+q_2=0,\\
&F'_3+\kappa_1\ F_2-\kappa_2\ F_1+q_3=0,\\
\end{split}
\ee
and
\be
\label{eq:equilrod2}
\besp
&M'_1-\kappa_3\ M_2+\kappa_2\ M_3+m_1+y_2\ F_3-y_3\ F_2+x_2\ q_3-x_3\ q_2=0,\\
&M'_2+\kappa_3\ M_1-\kappa_1\ M_3+m_2+y_3\ F_1-y_1\ F_3+x_3\ q_1-x_1\ q_3=0,\\
&M'_3-\kappa_2\ M_1+\kappa_1\ M_2+m_3+y_1\ F_2-y_2\ F_1+x_1\ q_2-x_2\ q_1=0.
\end{split}
\ee

\subsection{Constitutive law}
We assume an elastic behavior for the rod and we put, generally speaking, 
\be
\besp
&\F=\F(y_k,\kappa_1,\kappa_2,\kappa_3,s),\\
&\M=\M(y_k,\kappa_1,\kappa_2,\kappa_3,s);
\end{split}
\ee
we also assume that such functions are at least of class C$^1$ with respect to all the variables. Concerning these last, they can get any value, except $s$ that belongs to the interval $[s_1,s_2]$, and the continuity condition (\ref{eq:contcond}). Once the constitutive law injected into the balance equations, we obtain a system of six nonlinear differential equations in six unknowns. The problem is hence posed correctly, but the equations can be solved analytically only in a few simple cases.

\section{The Kirchhoff's theory for slender rods}
Kirchhoff (1859) has extended a previous result of Euler (1744) in formulating a theory for slender rods undergoing large displacements but small strains.
 The consequence of the large displacements assumption is that the elastic equilibrium problem is, in general, nonlinear because the balance equations cannot be written in the undeformed, known, configuration, but in the deformed, unknown, one.

The assumptions of the Kirchhoff's theory are:
\begin{enumerate}[i.]
\item the axis of the rod is inextensible;
\item the stress couple $\M$ depends linearly on the curvatures and on the twist;
\item there is no shear between the cross-section and the axis;
\item the cross-section is nondeformable in its plane.
\end{enumerate}
We remark that assumptions iii and iv are the classical assumptions of the Euler-Bernoulli theory of rods. As a consequence, regardless of the deformation, 
\be
\d_3\times\r'=\bo,
\ee
i.e. $\d_3$ remains tangent to the deformed rod's axis. Hence,
\be
\r'=y_j\ \d_j=y_3\ \d_3\Rightarrow y_1=y_2=0\ \ \forall s.
\ee
and 
\be
y_3=|\r'|\ \ \forall s.
\ee
The assumption of inextensibility gives
\be
|\r'|=1\Rightarrow y_3=1\rightarrow\r'=\d_3\ \ \forall s.
\ee
The assumption on the constitutive law implies not only that the components of $\M$ are linear functions of the dual curvature/twist, but also that they are uncoupled:
\be
\label{eq:momrods}
M_1=A\ \kappa_1,\ \ \ M_2=B\ \kappa_2,\ \ \ M_3=C\ \kappa_3,
\ee
with $A,B,C$ constants depending upon the elastic properties of the material and on the geometry of the cross-section. The assumption of non-deformability of the cross-section implies that the directors $\d_1^0$ and $\d_2^0$ of the undeformed shape remain orthogonal and of unit length $\forall s,t$. 

Several interesting solutions have been given for Kirchhoff's rods in the case of initially straight rods; in this case
\be
y_1=y_2=0,\ \ y_3=1,\ \  \kappa_1^0=\kappa_2^0=\kappa_3^0=0.
\ee
The most famous application is that of the {\it elastica of Euler}.

\section{The Euler's elastica}
The problem of the so-called {\it elastica} (Euler, 1744) is that of determining the forms in which an initially straight rod can be held by forces and couples applied to its ends only, when the rod is bent in a principal inertia plane, so that the deformed axis is a plane curve and there is no twist. Hence:
\begin{itemize}
\item $\m=\q=\bo$, for the assumption on the loads;
\item $\kappa_1=\kappa_3=0$, for the assumption on the deformation;
\item $y_1=y_2=0$, for the assumption on the undeformability of the cross-section;
\item $y_3=1$, for the assumption of inextensibility;
\item $M_1=M_3=0$, for the assumption on the deformation.
\end{itemize}
Hence eqs. (\ref{eq:equilrod1}) and (\ref{eq:equilrod2}) become
\be
\label{eq:equilrod3}
\besp
&F'_1+\kappa_2\ F_3=0,\ \ \ F'_2=0,\ \ \ F'_3-\kappa_2\ F_1=0,\\
&F_2=0,\ \ \ M'_2+F_1=0,
\end{split}
\ee
while the last of (\ref{eq:equilrod2}) is now an identity. 

\begin{figure}[th]
\begin{center}
\includegraphics[width=.4\textwidth]{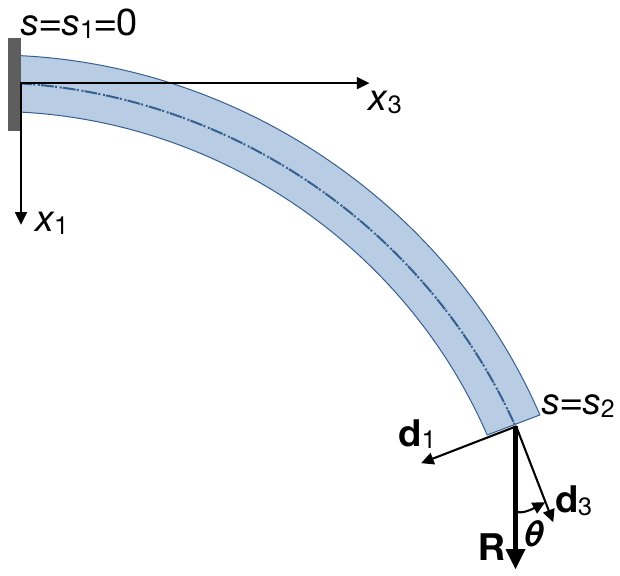}
\caption{Scheme of the Euler's elastica.}
\label{fig:f4_3}
\end{center}
\end{figure}

We consider, for the load, a unique force
\be
\mathbf{R}=R\ \e_1
\ee
applied at $s=s_2$. This gives, see Fig. \ref{fig:f4_3},
\be
F_1(s)=R\ \sin\theta(s),\ \ \ F_3(s)=R\ \cos\theta(s).
\ee
Injecting these relations into eqs. (\ref{eq:equilrod3})$_{1,3}$ gives the same result:
\be
\theta'R\cos\theta+\kappa_2R\cos\theta=0\rightarrow \kappa_2=-\theta',
\ee
while eqs. (\ref{eq:equilrod3})$_5$ and (\ref{eq:momrods}) give
\be
B\ \kappa'_2+R\sin\theta=0\rightarrow \kappa'_2=-\frac{R}{B}\sin\theta,
\ee
and finally we obtain the second order differential equation
\be
\theta''=\frac{R}{B}\sin\theta.
\ee
The resolution of this equation is not easy; to this purpose, we first integrate once to get
\be
\frac{1}{2}(\theta')^2-\frac{R}{B}(\cos\theta_0-\cos\theta)=0,
\ee
with  $\theta_0$ the value of $\theta$ at $s=s_2$. We can rewrite the above equation as 
\be
\frac{d\theta}{ds}=\pm\sqrt{\frac{2R}{B}(\cos\theta_0-\cos\theta)};
\ee
for the case in Fig. \ref{fig:f4_3} the sign minus is to be taken; we then get
\be
\frac{d\theta}{\sqrt{\cos\theta_0-cos\theta)}}=-\sqrt{\frac{2R}{B}}\ ds.
\ee
This equation can be solved only numerically, through an elliptic integral or directly, to obtain $\theta=\theta(s)$.
%and finally
%\be
%\int_0^{\theta_0}\frac{d\theta}{\sqrt{\cos\theta_0-\cos\theta}}=-\sqrt{\frac{2R}{B}}\int_0^Lds=-L\sqrt{\frac{2R}{B}}.
%\ee
%The integral above is an elliptic integral that can be rewritten as
%\be
%\int_0^{\frac{\pi}{2}}\frac{d\phi}{\sqrt{1-k^2\sin^2\phi}}=L\sqrt{\frac{R}{B}},\ \ \ \ \ k=\sin\frac{\theta_0}{2},\ \ k\sin\phi=\sin\frac{\theta}{2}.
%\ee
%This equation can be rewritten as
%\be
%L=\sqrt{\frac{B}{R}}\ \mathcal{E}(k),
%\ee
%with
%\be
%\mathcal{E}(k)=\int_0^\frac{\pi}{2}\frac{d\phi}{\sqrt{1-k^2\sin^2\phi}}
%\ee
%a complete elliptic normal integral of modulus $k$; this modulus is fixed once $\theta_0$ fixed. Hence, $\forall\theta_0$ we can obtain the value of $R$ necessary to get $\theta_0$:
%\be
%R=\frac{B}{L}\ \mathcal{E}^2(k).
%\ee
For any point of the rod it is 
\be
\frac{dx_1}{ds}=\cos\theta,\ \ \ \frac{dx_3}{ds}=\sin\theta;
\ee
hence, with the boundary conditions $x_1=x_3=0$ for $s=0$, we get
\be
x_1=\int\cos\theta(s)\ ds,\ \ \ x_3=\int\sin\theta(s)\ ds,
\ee
which in the end gives, after integration, the parametric equation of the elastica, $(x_1(s),x_3(s))$.

\begin{figure}[th]
\begin{center}
\includegraphics[width=.4\textwidth]{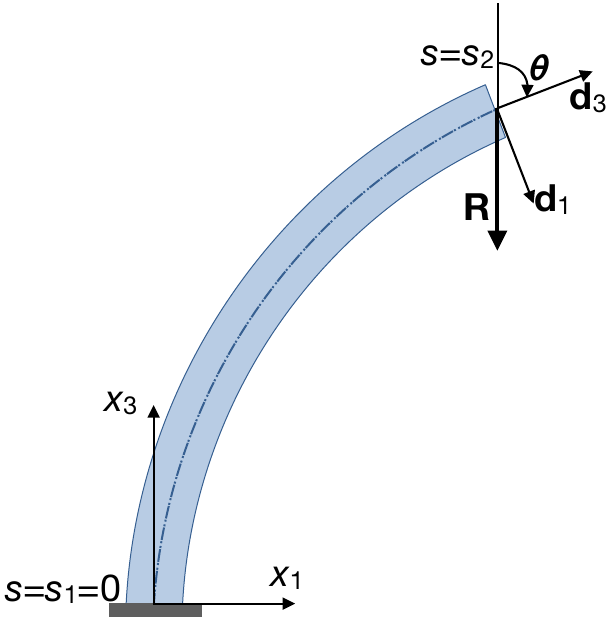}
\caption{Buckling of a rod.}
\label{fig:f4_4}
\end{center}
\end{figure}

We consider now the case of a compressive force $\mathbf{R}=-R\ \et,\ R>0$. We repeat {\it verbatim} the procedure, but now we change the definition of the angle $\theta$, indicated in Fig. \ref{fig:f4_4}, so that
\be
F_1=R\ \sin\theta,\ \ \ F_3=-R\ \cos\theta,
\ee
which gives $\kappa_2=\theta'$ and
\be
B\ \theta''+R\ \sin\theta=0.
\ee
The solution does not change, as it can be checked easily, but now it is
\be
\frac{dx_1}{ds}=\sin\theta,\ \ \ \frac{dx_3}{ds}=\cos\theta,
\ee
so
\be
x_1=\int\sin\theta(s)\ ds,\ \ \ x_3=\int\cos\theta(s)\ ds.
\ee
We can study the possibility of the rod to buckle: when  $R$ increases, a bent state can exist close to the straight one\footnote{This classical approach, called {\it adjacent equilibrium method}, is due to Euler.}, i.e. $\sin\theta\simeq\theta$.
We can hence linearize the equilibrium equation:
\be
B\ \theta''+R\ \theta=0,
\ee
whose solution is
\be
\theta=c_1\cos\left(\sqrt{\frac{R}{B}}\ s\right)+c_2\sin\left(\sqrt{\frac{R}{B}}\ s\right).
\ee
The constants $c_1$ and $c_2$ can be determined through the boundary conditions
\be
\theta(0)=0,\ \ \ \theta'(L)=0;
\ee
with such boundary conditions, the only solution is the trivial one, $\theta(s)=0\ \ \forall s$ (fundamental configuration), unless
\be
R=R_{b}=\frac{\pi^2B}{4L^2};
\ee
this is the {\it Euler buckling load}: when $R\geq R_b$, the rod can buckle. To determine the deflection of point $s=s_2=L$ in the buckled state, we go back to the first integral of the equilibrium equation, that in this case is 
\be
\frac{1}{2}\left(\frac{d\theta}{ds}\right)^2+\frac{R}{B}(\cos\theta_0-\cos\theta)=0.
\ee
Because now $dx_1=ds\ \sin\theta$, we get, in the end,
\be
x_1(L)=%2k\sqrt{\frac{B}{R}}\int_0^\frac{\pi}{2}\sin\phi\ d\phi=2k\sqrt{\frac{B}{R}}=
2\sin\frac{\theta_0}{2}\sqrt{\frac{B}{R}};
\ee
if the value of $\theta_0$ is known, namely after having solved numerically the elastica problem in this case, $x_1(L)$ can be easily calculated. It can be shown, using elliptic integrals, that $x_1(L)$ is effectively approximated by 
\be
x_1(L)\simeq4\left[\frac{2L}{\pi}-\left(\frac{B}{R}\right)^{\frac{1}{2}}\right]^{\frac{1}{2}}\left(\frac{B}{R}\right)^{\frac{1}{4}}.
\ee
It is possible to remove the assumption of axial inextensibility; in this case, the length of the rod just before buckling becomes $L\left(1-\frac{R}{EA}\right)$ and the buckling load changes to
\be
R_{b}=\frac{\pi^2B}{4L^2\left(1-\frac{R}{EA}\right)^2};
\ee
practically, this correction has no importance.

\section{Straight rods with generic end loads}
Let us now consider a more general case, where the rod is acted upon by a force {\bf R} and  couple {\bf M} at one end, balanced by another force and couple at the other end. No distributed loads nor couples are present. So, eqs. (\ref{eq:equilrod1}) and (\ref{eq:equilrod2}) become
 
\be
\label{eq:equilrod3D1}
\besp
&F'_1+\kappa_2\ F_3-\kappa_3\ F_2=0,\\
&F'_2+\kappa_3\ F_1-\kappa_1\ F_3=0,\\
&F'_3+\kappa_1\ F_2-\kappa_2\ F_1=0,\\
\end{split}
\ee
and
\be
\label{eq:equilrod3D2}
\besp
&M'_1-\kappa_3\ M_2+\kappa_2\ M_3+y_2\ F_3-y_3\ F_2=0,\\
&M'_2+\kappa_3\ M_1-\kappa_1\ M_3+y_3\ F_1-y_1\ F_3=0,\\
&M'_3-\kappa_2\ M_1+\kappa_1\ M_2+y_1\ F_2-y_2\ F_1=0.
\end{split}
\ee
Using the Kirchhoff's theory, i.e. putting $y_1=y_2=0,\ y_3=1$, we get
\be
\label{eq:equilrod3D3}
\besp
&M'_1-\kappa_3\ M_2+\kappa_2\ M_3=\ F_2,\\
&M'_2+\kappa_3\ M_1-\kappa_1\ M_3=- F_1,\\
&M'_3-\kappa_2\ M_1+\kappa_1\ M_2=0,
\end{split}
\ee
and introducing the constitutive laws (\ref{eq:momrods}) we obtain
\be
\label{eq:equilrod3D4}
\besp
&A\ \kappa_1'-(B-C)\kappa_2\ \kappa_3=F_2,\\
&B\ \kappa_2'-(C-A)\kappa_1\ \kappa_3=-F_1,\\
&C\ \kappa_3'-(A-B)\kappa_1\ \kappa_2=0.
\end{split}
\ee
These equations are formally identical to the Euler's rigid body motion equations:
\be
\besp
&A\ \dot{p}-(B-C)q\ r=M_1,\\
&B\ \dot{q}-(C-A)r\ p=M_2,\\
&C\ \dot{r}-(A-B)p\ q=M_3,
\end{split}
\ee
where $\bom=(p,q,r)$ is the angular velocity and $A,B,C$ are the principal moments of inertia. There is hence a mechanical analogy between the problem of the elastic rod and that of a rigid body motion.

From eq. (\ref{eq:equilrod3D4}) we can obtain a first integral of the rod problem: injecting the expression of $F_1$ and $F_2$ given by eqs. (\ref{eq:equilrod3D4})$_{1,2}$ into eq. (\ref{eq:equilrod3D1})$_3$ we get
\be
\label{eq:equilrod3D5}
F_3'+A\ \kappa_1\ \kappa_2'+B\ \kappa_2\ \kappa_2'+(A-B)\ \kappa_1\ \kappa_2\ \kappa_3=0,
\ee
and from eq. (\ref{eq:equilrod3D4})$_{3}$ we have
\be
(A-B)\ \kappa_1\ \kappa_2\ \kappa_3=C\ \kappa_3\ \kappa_3',
\ee
that injected into eq. (\ref{eq:equilrod3D5}) gives
\be
\label{eq:equilrod3D6}
\besp
&\frac{d}{ds}\left[F_3+\frac{1}{2}(A\ \kappa_1^2+B\ \kappa_2^2+C\ \kappa_3^2)\right]=0\rightarrow \\
&F_3+\frac{1}{2}(A\ \kappa_1^2+B\ \kappa_2^2+C\ \kappa_3^2)=const.
\end{split}
\ee
Injecting relations (\ref{eq:curvangEul}) into the previous equation gives an equation for the Euler's angles:
\be
\besp
F_3+\frac{1}{2}&[A\left(\theta'\cos\varphi+\psi'\sin\theta\sin\varphi\right)^2+\\
&B\left(-\theta'\sin\varphi+\psi'\sin\theta\cos\varphi\right)^2+\\
&C\left(\varphi'+\psi'\cos\theta\right)^2]=const.
\end{split}
\ee
Another first integral can be obtained observing that $\forall s$ the projection of {\bf M}(s) onto the fixed vector $\et$ is a constant; introducing the Euler's angles,
\be
\besp
\d_1\cdot\et=\sin\varphi\sin\theta,\\
\d_2\cdot\et=\cos\varphi\sin\theta,\\
\d_3\cdot\et=\cos\theta,
\end{split}
\ee
we get
\be
\M\cdot\et=A\ \kappa_1\sin\varphi\sin\theta+B\ \kappa_2\cos\varphi\sin\theta+C\ \kappa_3\cos\theta=const.
\ee
It should be worth to obtain a third first integral, for having then three relations for $\varphi,\theta$ and $\psi$. Generally speaking, this third relation is not known. So, let us assume that $A=B$; this is the case, e.g., of circular or square cross-sections. Then, from eq. (\ref{eq:equilrod3D4})$_{3}$ we get immediately
\be
C\ \kappa_3'=0\Rightarrow\ \kappa_3=const.
\ee
The twist of the rod is necessarily constant; the system (\ref{eq:equilrod3D4}) can now be integrated in closed form:
\be
\besp
&\kappa_1'=\alpha\ \kappa_2+\varphi_2,\\
&\kappa_2'=\beta\ \kappa_1-\varphi_1,
\end{split}
\ee
with 
\be
\alpha=\frac{B-C}{A}\kappa_3=const.,\ \ \ \beta=\frac{C-A}{B}\kappa_3=const.,\ \ \ \varphi_1=\frac{F_1}{B},\ \ \ \varphi_2=\frac{F_2}{A}.
\ee
If $F_1=F_2=0$, i.e. for the case of a rod acted upon only by couples, we get the system of two differential equations
\be
\label{eq:dynrod}
\besp
&\kappa_1'=\alpha\ \kappa_2,\\
&\kappa_2'=\beta\ \kappa_1,
\end{split}
\ee
with, because $A=B$, 
\be
\alpha=-\beta=\left(1-\frac{C}{A}\right)\kappa_3.
\ee
The integral of the above system is
\be
\besp
\kappa_1=c_1\cos(\kappa_3\ s)-c_2\sin(\kappa_3\ s),\\
\kappa_2=c_2\cos(\kappa_3\ s)+c_1\sin(\kappa_3\ s);
\end{split}
\ee
$c_1$ and $c_2$ can be determined using the boundary conditions
\be
\kappa_1(s=0)=\frac{M_1(s=0)}{A},\ \ \ \kappa_2(s=0)=\frac{M_2(s=0)}{B}.
\ee
An interesting interpretation of these results can be given in the framework of the linear stability theory of dynamical systems\footnote{The reader is addressed to the book of Strogatz given in the suggested literature for an insight in the linear stability theory of dynamical systems. It is nice to notice that eq. (\ref{eq:dynrod}) is a particular case of the so-called Romeo and Juliet model, proposed by Strogatz to model the dynamics of a love affair. Rather unexpectedly, there is some analogy between the equilibrium of slender rods and falling in love!}. In this framework, the equilibrated configuration is given by $\kappa_1=\kappa_2=0$ and the Jacobian matrix is 
\be
J=\left[
\begin{array}{cc}
0&\left(1-\frac{C}{A}\right)\kappa_3\medskip\\
-\left(1-\frac{C}{A}\right)\kappa_3&0
\end{array}
\right]
\ee
with two complex conjugate eigenvalues:
\be
\lambda_{1,2}=\pm \mathrm{i}\left|\kappa_3\left(1-\frac{C}{A}\right)\right|.
\ee
Hence, this configuration is an {\it attractor} and in particular it is  a {\it center}: the attractor is stable. Mechanically speaking, this can be interpreted in the following way: starting from an initial perturbed condition $(\kappa_1^0,\kappa_2^0)\neq(0,0)$, the two curvatures along $s$, the independent variable for this problem, change so as to follow a circle in the {\it phase space} $(\kappa_1,\kappa_2)$ with center $(0,0)$ and radius $\sqrt{(\kappa_1^0)^2+(\kappa_2^0)^2}$. In particular, it is possible that for certain values of $s$, the two curvatures take again the value $(\kappa_1^0,\kappa_2^0)$, but generally speaking they change continuously with $s$ and both of them can change of sign. This is in accordance with the analytical solution given above. To remark that, because 
\be
C\simeq\frac{A}{1+\nu},
\ee
it is impossible, if some couple is applied at the rod's ends, i.e. for not null initial conditions, that the rod preserves the configuration of a straight line (here, $\nu$ is the Poisson's ratio).

Another solution of eqs. (\ref{eq:equilrod3D6}) and (\ref{eq:dynrod}) is that of a rod bent in a circular helix. In fact, let  the rod be acted upon by a longitudinal force $R_3$ and a twisting couple $M_3$;  because, by the same definition of helix,
 the inclination of a helix is constant, i.e. $\frac{d\theta}{ds}=0$, then eq. (\ref{eq:equilrod3D4}) gives
 \be
 \label{eq:helix}
 \besp
 &\kappa_1=\psi'\sin\varphi\cos\alpha,\\
 &\kappa_2=\psi'\cos\varphi\cos\alpha,\\
 &\kappa_3=\varphi'+\psi'\sin\alpha.
 \end{split}
 \ee
Because $\kappa_3=const.$ and $A=B$, from eq. (\ref{eq:equilrod3D6})  we get
\be
\kappa_1^2+\kappa_2^2=\psi'^2\cos^2\alpha=const.,
\ee
so, being $\alpha=const., \psi'=const.$ and as a consequence of eq. (\ref{eq:helix}), also $\varphi'=const.$ Finally, the curve has a constant curvature and torsion: it is then a helix traced on a circular cylinder.

\subsection{The Michell-Prandtl bifurcation}
Some peculiar bifurcation phenomena concern slender rods. The {\it Michell-Prandtl} or {\it Prandtl-Timoshenko} bifurcation happens when a rod of length $\ell$, initially straight when unstrained, is subjected to only bending couples at its ends.
\begin{figure}[th]
\begin{center}
\includegraphics[width=.6\textwidth]{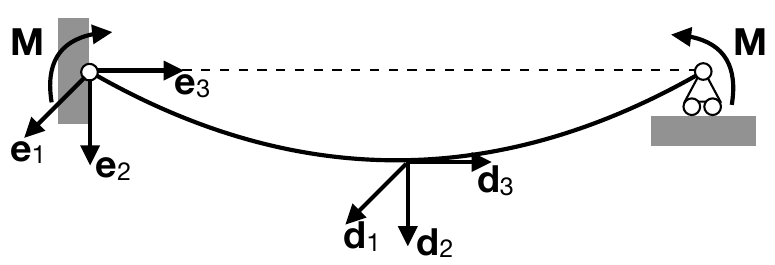}
\caption{Rod acted upon by couples at its ends.}
\label{fig:f4_5}
\end{center}
\end{figure}

To fix the ideas, we consider the case in Fig. \ref{fig:f4_5}, where the couple at the ends is $\M=M\eu$. No other loads are applied to the rod, so that eqs. (\ref{eq:equilrod3D4}) are now 
\be
\label{eq:MichPran1}
\besp
&A\ \kappa_1'-(B-C)\kappa_2\ \kappa_3=0,\\
&B\ \kappa_2'-(C-A)\kappa_1\ \kappa_3=0,\\
&C\ \kappa_3'-(A-B)\kappa_1\ \kappa_2=0.
\end{split}
\ee
These equations are formally identical to those of a {\it Poinsot's motion},  e.g. the equations ruling the motion of a gyroscope. We take care of choosing $\d_1$ and $\d_2$ corresponding with two principal axes of inertia of the cross-section, and initially parallel to the corresponding $\e_i$, see Fig. \ref{fig:f4_5}. The linear, fundamental solution of the above equations is, of course, 
\be
\kappa_1=\frac{M}{A},\ \ \ \kappa_2=0,\ \ \ \kappa_3=0,
\ee
i.e., the rod bends in the plane orthogonal to $\M$.

To  investigate the existence of a bifurcated solution, we use again the {\it Euler's method of adjacent equilibrium}, i.e. we look for an equilibrated configuration close to the fundamental one. To this end, we write the equilibrium equation near the  fundamental configuration, retaining only linear terms. In other words, we linearize the equilibrium equations in the variated configuration. For doing that, we write all the equations as, e.g.,
\be
\kappa_1=\kappa_1^0+\overline{\kappa}_1,
\ee
where $\kappa_1^0$ is the value of $\kappa_1$ in correspondence of the fundamental solution, while $\overline{\kappa}_1$ is the variation with respect to such a solution and $\kappa_1$ is the value in correspondence of the variated solution. We write hence the equilibrium equations for the variations of the different quantities with respect to the fundamental solution. In particular,
\be
\besp
&\kappa_1=\kappa_1^0+\overline{\kappa}_1=\frac{M}{A}+\overline{\kappa}_1\rightarrow \kappa_1'=\overline{\kappa}_1',\\
&\kappa_2=\kappa_2^0+\overline{\kappa}_2=\overline{\kappa}_2\rightarrow \kappa_2'=\overline{\kappa}_2',\\
&\kappa_3=\kappa_3^0+\overline{\kappa}_3=\overline{\kappa_3}\rightarrow\kappa_3'=\overline{\kappa}_3'.
\end{split}
\ee
So, eqs. (\ref{eq:MichPran1}) become
\be
\label{eq:MichPran2}
\besp
&A\ \overline{\kappa}_1'-(B-C)\overline{\kappa}_2\ \overline{\kappa_3}=0,\\
&B\ \overline{\kappa}_2'-(C-A)\left(\frac{M}{A}+\overline{\kappa}_1\right)\ \overline{\kappa_3}=0,\\
&C\ \overline{\kappa}_3'-(A-B)\left(\frac{M}{A}+\overline{\kappa}_1\right)\ \overline{\kappa}_2=0,
\end{split}
\ee
and after linearization we obtain
\be
\label{eq:MichPran3}
\besp
&A\ \overline{\kappa}_1'=0,\\
&B\ \overline{\kappa}_2'-(C-A)\frac{M}{A}\ \overline{\kappa_3}=0,\\
&C\ \overline{\kappa}_3'-(A-B)\frac{M}{A}\ \overline{\kappa}_2=0,
\end{split}
\ee
or
\be
\label{eq:MichPran4}
\besp
&\overline{\kappa}_1'=0,\\
&\overline{\kappa}_2'=\mu_1\ \overline{\kappa_3},\\
&\overline{\kappa}_3'=\mu_2\ \overline{\kappa}_2,
\end{split}
\ee
with
\be
\mu_1=-\frac{A-C}{AB}M,\ \ \ \mu_2=\frac{A-B}{AC}M.
\ee
We need now to write the boundary conditions, $\M=M\eu$ at $s=0,\ell$, in the frame of the directors, $\{\d_1,\d_2,\d_3\}$. Because we are dealing with small variations with respect to the fundamental solution, the rotation transforming  the basis $\{\eu,\ed,\et\}$ into $\{\d_1,\d_2,\d_3\}$ is a small rotation, say $\gamma$. As a consequence, the rotation tensor $\Q(\gamma)$ of this transformation, that generally speaking can be represented in the form
\be
\Q(\gamma)=\I+\sin\gamma\ \W+(1-\cos\gamma)\ \W^2,
\ee
can be linearized: for small $\gamma, \sin\gamma\simeq\gamma,\cos\gamma\simeq1$, so that
\be
\Q(\gamma)\simeq\I+\gamma\ \W,
\ee
and finally
\be
\d_i=\Q(\gamma)\ \e_i\simeq(\I+\gamma\ \W)\ \e_i,\ \ \ i=1,2,3.
\ee
In the above equations, $\W$ is the axial tensor of the rotation axis, $\W=-\W^\top$. To invert this relation, we write
\be
\besp
&(\I+\gamma\ \W)^\top\d_i=(\I+\gamma\ \W)^\top(\I+\gamma\ \W)\e_i\rightarrow\\
&(\I-\gamma\ \W)\d_i=(\I-\gamma\ \W+\gamma\ \W-\gamma^2\ \W^2)\e_i,
\end{split}
\ee
so finally, neglecting higher order terms, 
\be
\e_i=\d_i-\gamma\ \W\ \d_i=\d_i-\bgamma\times\d_i,
\ee
with $\bgamma=(\gamma_1,\gamma_2,\gamma_3)$ the axial vector corresponding to tensor $\gamma\ \W$, representing the infinitesimal rotation of the cross-section. It is then natural to identify the changes in the rotation, per unit $s$, with curvatures:
\be
\kappa_1=\gamma_1',\ \ \ \kappa_2=\gamma_2',\ \ \ \kappa_3=\gamma_3'.
\ee
Hence, the boundary conditions become
\be
\M=M\ \eu=M(\I-\gamma\ \W)\d_1=M(\d_1-\bgamma\times\d_1)\ \ \ \mathrm{at}\ s=0,\ell.
\ee
Along the three directors we have then
\be
\besp
&M_1=\M\cdot\d_1=M,\\
&M_2=\M\cdot\d_2=M(\d_1-\gamma\times\d_1)\cdot\d_2=-M\ \gamma_3,\\
&M_3=\M\cdot\d_3=M(\d_1-\gamma\times\d_1)\cdot\d_3=M\ \gamma_2,
\end{split}
\ee
and hence, putting
\be
\gamma_i=\gamma_i^0+\overline{\gamma}_i,\ \ \ M_i=M_i^0+\overline{M}_i,\ \ \ i=1,2,3,
\ee
we have, because $M_1^0=M,M_2^0=M_3^0=0$ and $\gamma_2^0=\gamma_3^0=0$,
\be
\begin{array}{lll}
M_1^0+\overline{M}_1=M&\rightarrow &\overline{M}_1=0,\medskip\\
M_2^0+\overline{M}_2=-M\ \gamma_3&\rightarrow &\overline{M}_2=-M\ \overline{\gamma}_3,\medskip\\
M_3^0+\overline{M}_3=M\ \gamma_2&\rightarrow& \overline{M}^3=M\ \overline{\gamma}_2.
\end{array}
\ee
Then, because
\be
M_1=A\ \kappa_1\rightarrow M_1^0+\overline{M}_1=A\ \kappa_1^0+A\ \overline{\kappa}_1\rightarrow \overline{M}_1=A\ \overline{\kappa}_1,
\ee
and similarly for $M_2$ and $M_3$, we have, for $s=0,\ell$,
\be
\label{eq:MichPran5}
\begin{array}{lll}
\overline{M}_1=A\ \overline{\kappa}_1=0&\rightarrow&\overline{\kappa}_1=0,\medskip\\
\overline{M}_2=B\ \overline{\kappa}_2=-M\ \overline{\gamma}_3&\rightarrow&\overline{\kappa}_2=-\dfrac{M}{B}\overline{\gamma}_3,\medskip\\
\overline{M}_3=C\ \overline{\kappa_3}&\rightarrow&\overline{\kappa_3}=\dfrac{M}{C}\overline{\gamma}_2.
\end{array}
\ee
Hence, from eqs. (\ref{eq:MichPran4})$_1$ and (\ref{eq:MichPran5})$_1$  we get  that the solution for direction $\d_1$ is
\be
\overline{\kappa}_1=0\Rightarrow \kappa_1=\frac{M}{A}\ \ \ \forall s,
\ee
and, as a consequence,
\be
\gamma_1'=\kappa_1\Rightarrow\gamma_1=\int\kappa_1\ ds=\frac{M}{A}s+\gamma_1(s=0).
\ee
The two remaining equations (\ref{eq:MichPran6})$_{2,3}$ are coupled:
\be
\label{eq:MichPran6}
\besp
&\overline{\kappa}_2'=\mu_1\ \overline{\kappa_3},\\
&\overline{\kappa_3}'=\mu_2\ \overline{\kappa}_2,
\end{split}
\ee
Once more, this system of two differential equations is similar to system (\ref{eq:dynrod}) and it can be interpreted in the same way. In particular, if $\mu_1\mu_2>0$, the solution $\overline{\kappa}_2=\overline{\kappa_3}=0$, i.e. the fundamental solution, is a {\it saddle point}, i.e. it is unstable: the curvatures $\overline{\kappa}_2$ and $\overline{\kappa_3}$ can change of sign along $s$. 
If $\mu_1\mu_2<0$, then the fundamental solution is a {\it center} and $\overline{\kappa}_2$ and $\overline{\kappa_3}$ can change cyclically along $s$. 

We can easily solve eqs. (\ref{eq:MichPran6}) differentiating the first one:
\be
\overline{\kappa}''_2=\mu_1\ \overline{\kappa_3}'\rightarrow \overline{\kappa}''_2+\lambda^2\ \overline{\kappa}_2=0,\ \ \ \mathrm{with}\ \lambda^2=-\mu_1\mu_2.
\ee
The solution is
\be
\kappa_2=\overline{\kappa}_2=c_1\sin(\lambda \ s)+c_2\cos(\lambda\ s),
\ee
which gives
\be
\kappa_3=\overline{\kappa_3}=\frac{\overline{\kappa}_2'}{\mu_1}=\frac{\lambda}{\mu_1}\left[c_1\cos(\lambda \ s)-c_2\sin(\lambda\ s)\right].
\ee

The two other rotations can be found as well:
\be
\besp
&\gamma_2=\overline{\gamma}_2=\overline{\kappa}'_2\rightarrow\gamma_2=\int{\overline{\kappa}_2\ ds=\frac{1}{\lambda}}\left[-c_1\cos(\lambda \ s)+c_2\sin(\lambda\ s)+c_3\right],\\
&\gamma_3=\overline{\gamma}_3=\overline{\kappa_3}'\rightarrow\gamma_3=\int{\overline{\kappa_3}\ ds=\frac{1}{\mu_1}}\left[c_1\sin(\lambda \ s)+c_2\cos(\lambda\ s)+c_4\right].
\end{split}
\ee
The constants $c_1,c_2,c_3,c_4$ can be determined through the boundary conditions
\be
\overline{\kappa}_2(s=0,\ell)=-\frac{M}{B}\overline{\gamma}_3(s=0,\ell),\ \ \ \overline{\kappa_3}(s=0,\ell)=\frac{M}{C}\overline{\gamma}_2(s=0,\ell),
\ee
that give the homogeneous system 
\be
\begin{array}{l}
\left(1+\dfrac{M}{B\ \mu_1}\right) c_2+\dfrac{M}{B\ \mu_1}\ c_4=0,\medskip\\
\left(1+\dfrac{M}{B\ \mu_1}\right)\sin(\lambda\ell)\ c_1+\left(1+\dfrac{M}{B\ \mu_1}\right)\cos(\lambda\ell)\ c_2+\dfrac{M}{B\ \mu_1}\ c_4=0,\medskip\\
\left(\dfrac{\lambda}{\mu_1}+\dfrac{M}{\lambda\ C}\right)c_1-\dfrac{M}{\lambda\ C}\ c_3=0,\medskip\\
\left(\dfrac{\lambda}{\mu_1}+\dfrac{M}{\lambda\ C}\right)\cos(\lambda\ell)\ c_1-\left(\dfrac{\lambda}{\mu_1}+\dfrac{M}{\lambda\ C}\right)\sin(\lambda\ell)\ c_2-\dfrac{M}{\lambda\ C}\ c_3=0.
\end{array}
\ee
This system gives a non-trivial solution (i.e., $c_1=c_2=c_3=c_4=0$, corresponding to the fundamental solution for the rod) only if the determinant of its matrix vanishes, which gives the condition
\be
\frac{2M^2(M+B\ \mu_1)(M\ \mu_1+\lambda^2\ C)}{B^2C^2\lambda^2\mu_1^3}[\cos(\lambda\ell)-1]=0.
\ee
Replacing the expressions of $\mu_1,\mu_2$ and $\lambda$ we get
\be
\frac{2\ A^2B^2}{(A-C)^3(A-B)}[\cos(\lambda\ell)-1]=0,
\ee
which shows that it must be $A\neq C$ (always true, as shown below) and $A\neq B$: the Michell-Prandtl bifurcation cannot happen for rods with such cross-section, e.g. for square or circular sections. 

\begin{figure}[th]
\begin{center}
\includegraphics[width=.4\textwidth]{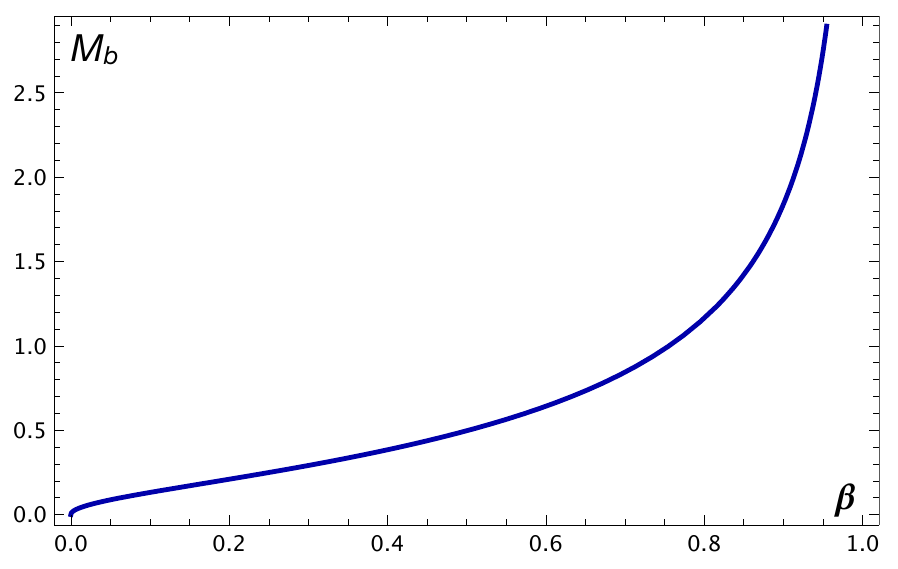}
\caption{The function $M_b(\beta)$, for the case $A=1,\nu=0.3,q=1.5,\ell=10$.}
\label{fig:f4_6}
\end{center}
\end{figure}

Hence, it must be
\be
\cos(\lambda\ell)=1\rightarrow \lambda_n=2 n \frac{\pi}{\ell}\rightarrow-\mu_1\mu_2=4n^2\frac{\pi^2}{\ell^2}\rightarrow\frac{A-C}{A\ B}\frac{A-B}{A\ C}\ M_n^2=4n^2\frac{\pi^2}{\ell^2},
\ee
so finally
\be
M_n=2n\frac{\pi}{\ell}\sqrt{\frac{A\ B}{A-B}\frac{A\ C}{A-C}}.
\ee
The least value, for $n=1$, is the {\it critical Michell-Prandtl bifurcation couple}:
\be
M_{b}=\frac{2\pi}{\ell}\sqrt{\frac{A\ B}{A-B}\frac{A\ C}{A-C}}.
\ee
It is apparent that a bifurcated solution can exist if and only if
\be
(A-B)(A-C)>0\rightarrow A>B\ \mathrm{and}\ A>C\ \mathrm{or}\ A<B\ \mathrm{and}\ A<C.
\ee
Let us better consider this condition: because
\be
A=E\ J_1,\ \ \ B=E\ J_2,\ \ \ C=\frac{G\ J_0}{q},
\ee
with $E$ the Young's modulus, $G$ the shear modulus, $q$ the torsion's factor, $J_1$ and $J_2$ the principal moments of inertia of the cross-section and $J_0$ the polar moment of inertia. As it is, for isotropic bodies,
\be
G=\frac{E}{2(1+\nu)}
\ee
and
\be
J_0=J_1+J_2,
\ee
we have that
\be
C=\frac{A+B}{2(1+\nu)\ q}.
\ee
Because the torsion factor $q\geq1$ ($q=1$ for circular cross-sections, i.e. when A=B, a case that is to be excluded for Michell-Prandtl bifurcation, as seen above), we have that, for ordinary materials, i.e. with a positive Poisson's ratio,
\begin{itemize}
\item if $A>B, \ C<\dfrac{A}{(1+\nu)\ q}\Rightarrow C<A$;
\item if $A<B, \ C>\dfrac{A}{(1+\nu)\ q}\Rightarrow A<(1+\nu)\ C\ q$.
\end{itemize}
Then, the Michell-Prandtl bifurcation can happen in  two possible cases: $A>B$ and $A>C$ or $A<\min\{B;(1+\nu)\ C\ q\}$. The first case is the most interesting for practical applications; it shows that the phenomenon exists only if the rod bends in the plane of the highest inertia. In addition, if we put $\beta=B/A$  we see easily that
\be
\lim_{\beta\rightarrow0}M_b=\lim_{\beta\rightarrow0}\frac{2\pi}{\ell}A\sqrt{\frac{\beta}{1-\beta}\frac{1+\beta}{2(1+\nu)\ q-1-\beta}}=0.
\ee
In other words, $M_b$ decreases for narrow high sections. In Fig. \ref{fig:f4_6}, the function $M_b(\beta)$ is plotted;  it can be noticed that not only $M_b\rightarrow0$  for $\beta\rightarrow0$, but also that $M_b\rightarrow\infty$ for $\beta\rightarrow1$, as seen above ($\beta\rightarrow1$ when $B\rightarrow A$).

\begin{figure}[th]
\begin{center}
\includegraphics[width=.6\textwidth]{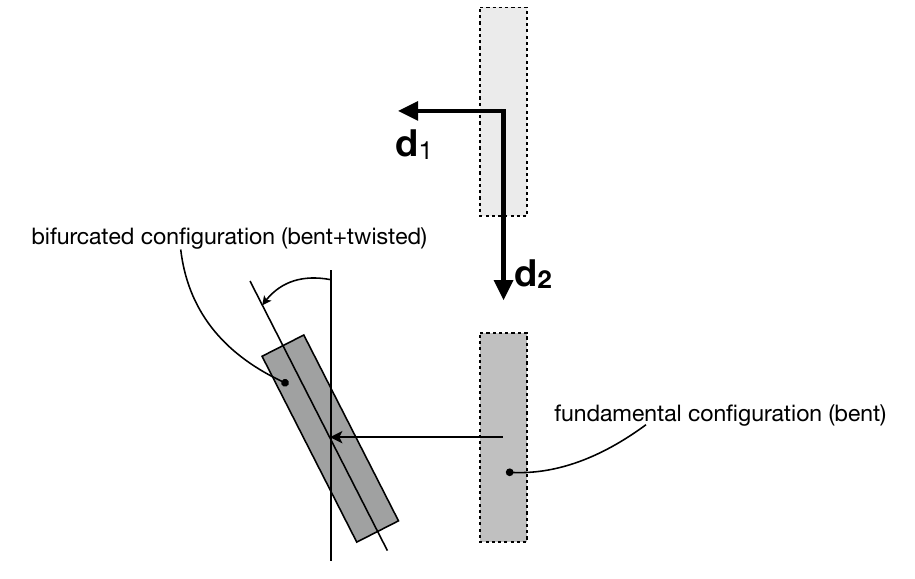}
\caption{Sketch of the deformation in the Michell-Prandtl bifurcation.}
\label{fig:f4_7}
\end{center}
\end{figure}

To end this Section, we remark that the bifurcation configuration corresponds to a torsion plus a bending in the plane of $B$, i.e. the plane $\d_1-\d_3$, see Fig. \ref{fig:f4_7}. Such a situation is particularly dangerous for I-shaped steel struts: such slender rods when bent can undergo a similar phenomenon, if not properly constrained.

\subsection{Helicoidal bifurcation}
A problem similar to the previous one is that of a straight rod acted upon, at its ends, by a twisting couple $\M=M\ \et$. This problem can be solved following the same approach of the previous Section: the equilibrium equations are still eqs. (\ref{eq:MichPran1}) and the same remarks can be done. The fundamental solution now is
\be
\kappa_1=0,\ \ \ \kappa_2=0,\ \ \ \kappa_3=\frac{M}{C}.
\ee 
Once linearized, the variated equilibrium equations are
\be
\besp
&A\ \overline{\kappa}_1'-(B-C)\frac{M}{C}\overline{\kappa}_2=0,\\
&B\ \overline{\kappa}_2'-(C-A)\frac{M}{C}\overline{\kappa}_1=0,\\
&C\ \overline{\kappa_3}'=0,
\end{split}
\ee
while the boundary conditions for $s=0,\ell$ become
\be
 \overline{\kappa}_1=-\dfrac{M}{A}\overline{\gamma}_2,\ \ \
 \overline{\kappa}_2=\dfrac{M}{B}\ \overline{\gamma}_1,\ \ \
\overline{\kappa_3}=0.
\ee
This gives, on one hand,
\be
\overline{\kappa_3}=0\ \ \forall s,\ \ \ \overline{\gamma}_3=const.,
\ee
and, on the other hand,
\be
\besp
&\overline{\kappa}_1'=\beta_1\ \overline{\kappa}_2,\\
&\overline{\kappa}_2'=\beta_2\ \overline{\kappa}_1,
\end{split}
\ee
with
\be
\beta_1=-\frac{C-B}{A\ C}M,\ \ \ \beta_2=\frac{C-A}{B\ C}M.
\ee
The resolution of this system of differential equations is quite similar to the problem of the previous Section and at the end we find a critical value of the twisting couple
\be
M_t=\frac{2\pi}{\ell}\sqrt{\frac{A\ C}{C-A}\frac{B\ C}{C-B}}.
\ee
In this case, the bifurcated solution can exist if and only if
\be
(C-A)(C-B)>0\rightarrow C>A\ \mathrm{and}\ C>B\ \mathrm{or}\ C<A\ \mathrm{and}\ C<B.
\ee
In particular, for $A<C<B$ or $A>C>B$ this type of bifurcation is not possible. This is the case of a ribbon-like rod, i.e. of a rectangular section with $h\gg b$, where $h$ is the height and $b$ the width of the cross-section. In fact, in this case,
\be
A=\frac{b\ h^3}{12}E,\ \ \ B=\frac{b^3\ h}{12}E,\ \ \ C\simeq \frac{b^3\ h}{3}G=\frac{2B}{1+\nu}.
\ee
Hence, because $-1<\nu\leq 1/2$, it is $A>C>B$: for this type of sections, this kind of bifurcation is not possible. At the opposite side, let us consider cross-sections with an inertia ellipse which is a circle, i.e. $A=B$, like the case of circular or square sections. In this case, it is either $C>A=B$ or $C<A=B$, so bifurcation is possible. In particular,
\be
A=B=E\ J,\ \ \ C=2\frac{G\ J}{q}=\frac{A}{(1+\nu)\ q},
\ee
and if $\nu>0$, like for ordinary isotropic materials, being $q>1$ for a non-circular section, it is $C<A$.

The bifurcated configuration is that of a rod with a deflection in both the planes $\d_1-\d_3$ and $\d_2-\d_3$, besides  a constant twist: such a curve is a helix.

%\newpage
%\section{Exercises}
%\begin{enumerate}
%\item A cantilever beam is acted upon by a  force $R$ orthogonal to the axis applied at the free end of the rod. The Young's modulus is $E=20 GPa$, the moment of inertia is $J=10 cm^4$ and the length is $L=5m$. Using the tables for elliptic integrals, calculate the value of $R$ necessary to produce a rotation of $\pi/2$ of the section where the force is applied.

%\item A steel bar is acted upon by two identical, opposite couples at the ends, whose value is $M_1=100kNm$. The moments of inertia of the cross-section are $J_1=150cm^4$ and $J_2=15cm^4$, while the torsion factor is $q=1.5$. Determine the maximal length that the bar can have before bifurcation.

%\end{enumerate}

\chapter{Plates}
% !TEX root = modsol.tex
\label{ch:6}
\section{Problem definition: basic assumptions}

\begin{figure}[th]
\begin{center}
\includegraphics[scale=1.2]{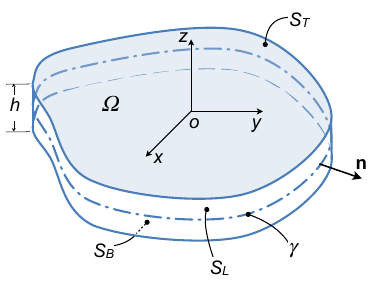}
\caption{General sketch of a plate.}
\label{fig:f5_1}
\end{center}
\end{figure}

A plate, see Fig. \ref{fig:f5_1}, is a solid $\Omega$ bounded by two parallel planes, the upper and lower {\it faces} $S_T$ and $S_B$, and by a {\it lateral surface} $S_L$, orthogonal to the two faces, so that, finally, a plate is just a {\it flat cylinder}. The {\it mid-plane} is the plane at equal distance from the two faces. We introduce the orthonormal Cartesian frame $\{o;x,y,z\}$, with the axes $x$ and $y$ that belong to the mid-plane. Be $h$ the {\it thickness} of the plate; for a solid to be really a plate, it must be $h<<d$, with $d$ the mean chord of the mid-plane. The {\it contour} of the plate is the line $\gamma$ intersection of $S_L$ with the mid-plane. Usually, plates are classified as follows:
\begin{itemize}
\item thin plates: $\dfrac{h}{d}\lesssim\dfrac{1}{10}$;
\item moderately thick plates: $\dfrac{1}{10}\lesssim\dfrac{h}{d}\lesssim\dfrac{1}{5}$;
\item thick plates: $\dfrac{h}{d}\gtrsim\dfrac{1}{5}$.
\end{itemize}
We assume that displacements, rotations and strains are small. As a consequence, the small strain tensor $\beps$ can be taken as an appropriate measure of deformation and the equilibrium equations can be written in the undeformed configuration $\Omega$.

We consider elastic plates, i.e. plates composed by linearly elastic materials whose constitutive law is specified by the Lamé's equations:
\begin{equation}
\label{eq:lamedirect}
\bsig=\frac{E}{1+\nu}\left(\beps+\frac{\nu}{1-2\nu}\tr\beps\ \I\right).
\end{equation}
Finally, the problem so defined is linear. 

Concerning the loading, we consider:
\begin{itemize}
\item distributed body forces $\b(p)$ on $\Omega$;
\item distributed surface tractions $\btau\left(x,y,\pm\frac{h}{2}\right)$ on $S_T$ and $S_B$;
\item distributed line  forces $\hat{\mathbf{F}}$ and couples $\hat{\mathbf{M}}$ on $\gamma$;
\item distributed couples inside $\Omega$ are excluded.
\end{itemize}
About the body forces $\b(p)$, we consider an {\it even distribution} with respect to the mid-plane, i.e.
\be
\b(x,y,z)=\b(x,y,-z).
\ee
For what concerns the surface tractions\footnote{Often, only lateral forces perpendicular to the plate's surfaces are considered and body forces are neglected; we prefer to develop the theory in the most general case and then consider particular cases; this allows to have a deeper insight in the mechanics of the plates.\medskip}, let us denote by $\btau^T=(\tau^T_x,\tau^T_y,\tau^T_z)$ the tractions on the surface $S_T$ and by $\btau^B=(\tau^B_x,\tau^B_y,\tau^B_z)$ those on $S_B$; generally speaking, we admit that $\btau^T\neq\btau^B$. For the sake of convenience, we decompose the surface tractions into a {\it symmetric}, $\btau^s$, and {\it antisymmetric}, $\btau^a$, parts:
\be
\btau^s=\frac{\btau^T+\btau^B}{2},\ \ \ \btau^a=\frac{\btau^T-\btau^B}{2},
\ee
so that
\be
\btau^T=\btau^s+\btau^a,\ \ \ \btau^B=\btau^s-\btau^a.
\ee
The outward unit normals on $S_T$ and $S_B$ are respectively $\n^T=(0,0,1)$ and $\n^B=(0,0,-1)$, so, by the Cauchy's Theorem, on $S_T$ it is\footnote{We use the symbol $T$ to denote any quantity evaluated on $S_T$, i.e. for $z=h/2$, and $B$ for denoting any other quantity evaluated on $S_B$, i.e. for $z=-h/2$.}
\be
\label{eq:tautop}
\btau^T=\bsig^T\n^T\ \rightarrow\ \sigma_{xz}^T=\tau^T_x=\tau_x^a+\tau_x^s,\ \ \sigma_{yz}^T=\tau^T_y=\tau_y^a+\tau_y^s,\ \ \sigma_{zz}^T=\tau^T_z=\tau_z^a+\tau_z^s,
\ee
while on $S_B$
\be
\label{eq:taubot}
\btau^B=\bsig^B\n^B\ \rightarrow\ \sigma_{xz}^B=-\tau^B_x=\tau_x^a-\tau_x^s,\ \ \sigma_{yz}^B=-\tau^B_y=\tau_y^a-\tau_y^s,\ \ \sigma_{zz}^B=-\tau^B_z=\tau_z^a-\tau_z^s.
\ee

\section{The Kirchhoff model for thin plates}
There are different possible approaches to the construction of a plate theory; we will follow here a classical {\it axiomatic approach}, i.e. an approach based upon the {\it a-priori} (i.e. axiomatic) choice of a displacement field. Such choice corresponds to a kinematical model, characterizing the theory.

Putting $\bu=(u,v,w)$ the {\it displacement vector}, we need to specify how the three components  $u=u(x,y,z),v=v(x,y,z),w=w(x,y,z)$ of $\bu$ vary, and in particular how they change across the thickness, i.e. along $z$. This choice is suggested, of course, by the particular geometry of the plate. In particular, being $h/d$ a fundamental parameter defining the plate, kinematical models differ from each other as a function of $h/d$. 

\subsection{The displacement field}
\label{sec:displkirch}
\begin{figure}[t]
\begin{center}
\includegraphics[width=\linewidth]{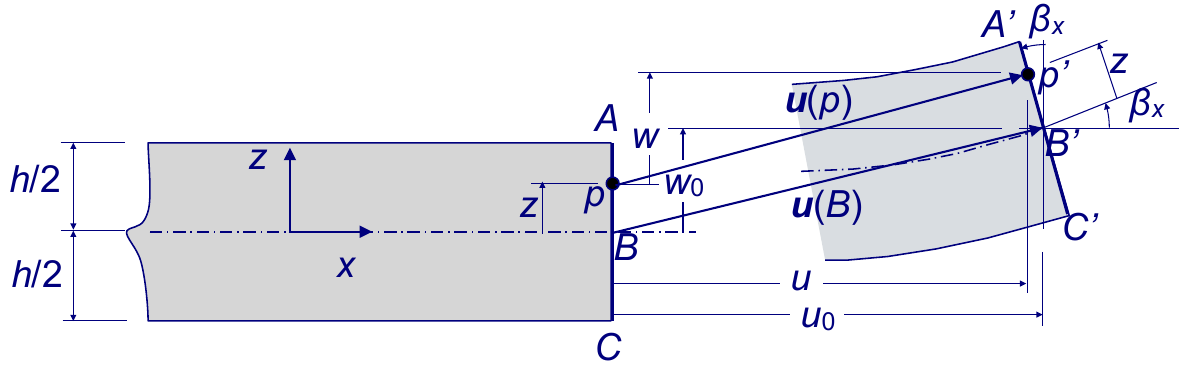}
\caption{Kirchhoff's kinematics in the $\{x,z\}$ plane.}
\label{fig:f5_2}
\end{center}
\end{figure}

In Fig. \ref{fig:f5_2} it is schematically represented a cross-section of the plate, along the plane $x-z$, before and after the deformation.
 Let us consider the segment $AC$, orthogonal to the mid-plane, with $B$ its intersection with the mid-plane and $p$ a point of $AC$ with the distance $z$ from the mid-plane. After the deformation each material point of the plate moves to a new position indicated by a prime: $p$ goes into $p'$, $B$ into $B'$ and so on. We can express the displacement $\bu(p)$ as
 \be
 \bu(p)=p'-p=p'-B'+B'-B+B-p=\bu(B)-z\et+p'-B',
 \ee
 i.e. we can decompose the displacement of any point of $\Omega$ as the sum of the displacement of its projection onto the mid-plane plus a position vector, $B-p$, and a local displacement, $p'-B'$, which is the displacement of $p'$ relative to $B'$. Because $B=(x,y,0)$, $\bu(B)=\bu(x,y,0)$, so we put
 \be
 \bu_0:=\bu(x,y,0)=(u_0(x,y),v_0(x,y),w_0(x,y));
 \ee
 $\bu_0(x,y)$ is a bi-dimensional vector field describing the displacement of the points of the mid-plane. We need now to define the local displacement $p'-B'$. It is exactly at this point that a kinematical model is needed. 
 
 The {\it Kirchhoff model} is based upon three assumptions concerning the deformation of any segment $AC$. According to the {Kirchhoff model}, any material segment orthogonal to the mid-plane remains:
 \begin{itemize}
 \item i. a straight segment;
 \item ii. orthogonal to the deformed mid-surface;
 \item iii. of the same length.
 \end{itemize}
 
\begin{figure}[ht]
\begin{center}
\includegraphics[width=.5\linewidth]{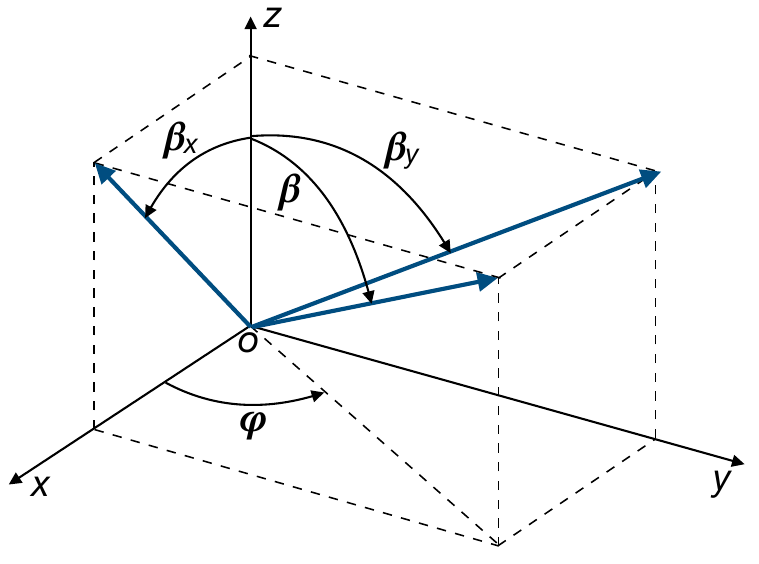}
\caption{Angles scheme.}
\label{fig:f5_3}
\end{center}
\end{figure}

 The third assumption of Kirchhoff states that 
 \be
 |p-B|=|p'-B'|=z;
 \ee
if $\beta$ is the angle that the deformed segment $A'C'$ forms with the axis $z$, then, because of the first Kirchhoff assumption,
\be
p'-B'=(-z\ \sin\beta_x,-z\ \sin\beta_y,z\ \cos\beta),
\ee 
where $\beta_x$ and $\beta_y$ are the angles that $z$ forms with the projections of $A'C'$ onto the planes $x-z$ and $y-z$, respectively. It is an easy task\footnote{With reference to Fig. \ref{fig:f5_3}, it is
\begin{equation*}\besp
&\cos\beta_x=\frac{\cos\beta}{\sqrt{\cos^2\beta+\sin^2\beta\cos^2\phi}},\ \ \cos\beta_y=\frac{\cos\beta}{\sqrt{\cos^2\beta+\sin^2\beta\sin^2\phi}}\ \Rightarrow\\ 
&\cos\beta_x\geq\cos\beta,\ \cos\beta_y\geq\cos\beta\ \Rightarrow\ {\beta_x\leq\beta,\ \beta_y\leq\beta}.
\end{split}\end{equation*}}
 to show that $\beta_x\leq\beta,\beta_y\leq\beta$, Fig. \ref{fig:f5_3}.
The second Kirchhoff assumption lets  now affirm that the angle formed by the line intersection of the plane containing axis $z$ and $A'C'$ with the tangent plane, in $B'$, to the bent mid-surface and by the projection of $A'C'$ onto the plane $x-y$ is exactly $\beta$; the same is valid also for the projection angles $\beta_x$ and $\beta_y$. If now we use the assumption of small rotations, then
\be
\beta\rightarrow0\Rightarrow\beta_x\rightarrow0,\beta_y\rightarrow0\Rightarrow\left\{
\besp
&\beta\simeq\sin\beta\simeq\tan\beta,\ \ \cos\beta\simeq1,\\
&\beta_x\simeq\sin\beta_x\simeq\tan\beta_x=\frac{\partial w_0}{\partial x},\\
&\beta_y\simeq\sin\beta_y\simeq\tan\beta_y=\frac{\partial w_0}{\partial y}.
\end{split}
\right.
\ee
Hence,
\be
p'-B'\simeq\left(-z\frac{\partial w_0}{\partial x},-z\frac{\partial w_0}{\partial y},z\right),
\ee
i.e. within the Kirchhoff model, the local displacement is a linear function of $z$. Finally
\be
\label{eq:kinemkirch}
\bu(p)=\left(
\begin{array}{c}
u_0(x,y)-z\dfrac{\partial w_0(x,y)}{\partial x}\\v_0(x,y)-z\dfrac{\partial w_0(x,y)}{\partial y}\\w_0(x,y)
\end{array}
\right).
\ee
Some remarks about this result:
\begin{itemize}
\item $\bu(p)$ depends linearly upon $z$, so all the problem is reduced to find  the planar vector field $\bu_0(x,y)$ describing the displacement of the points of the mid-plane; the 3D problem has hence been reduced to a 2D one, but this passage, that simplifies remarkably the problem, has heavy consequences, as we will see in the following;
\item the kinematics assumptions of the Kirchhoff model are rather  heavy; namely, the second one eliminates out-of-plane shear deformations, giving on one hand an increased stiffness to the plate, that cannot deform for shear, and, on the other hand, causing difficulties in the computation of the transversal shear stresses;
\item the Kirchhoff assumptions are plausible only for thin plates; when this is not true, they should be removed, especially the first and second ones, passing in this way to higher order theories.
\end{itemize}

\subsection{Strain field}
The strain field $\beps(p)$ can be easily calculated using the displacement field $\bu(p)$:
\be
\besp
\beps(p)=&\frac{\nabla\bu(p)+\nabla\bu^\top(p)}{2}=\\
&\left[\begin{array}{ccc}
\dfrac{\partial u_0}{\partial x}-z\dfrac{\partial^2 w_0}{\partial x^2} & \dfrac{1}{2}\left(\dfrac{\partial u_0}{\partial y}+\dfrac{\partial v_0}{\partial x}\right)-z\dfrac{\partial^2 w_0}{\partial x\partial y} & 0 \medskip\\
\dfrac{1}{2}\left(\dfrac{\partial u_0}{\partial y}+\dfrac{\partial v_0}{\partial x}\right)-z\dfrac{\partial^2 w_0}{\partial x\partial y}  & \dfrac{\partial v_0}{\partial y}-z\dfrac{\partial^2 w_0}{\partial y^2}  & 0\medskip \\
0 & 0 & 0\end{array}\right].
\end{split}
\ee
So, $\varepsilon_{xz}=\varepsilon_{yz}=\varepsilon_{zz}=0$: as a consequence of the Kirchhoff model, $\beps$ is planar. Contrarily to what often said, however, this is {\it not} a plane strain state, because $w$ is not identically null and $\bu$ and $\beps$ depend upon $z$ too. Actually, a plane strain state is typical of long cylinders acted upon by forces that do not vary along the axis of the cylinder; here, the only thing that can be said is that in the Kirchhoff model $\beps$ is planar.

\subsection{Stress field}
\label{sec:stressplate}
About the stress field $\bsig(p)$ a supplementary assumption is made: $\forall p\in\Omega$,
\be
\sigma_{zz}=0.
\ee
This assumption, common also to other plate theories, can be justified as follows:
\begin{itemize}
\item from the inverse Lamé's equation,
\be
\label{eq:inverselame}
\beps=\frac{1+\nu}{E}\bsig-\frac{\nu}{E}\tr\bsig\ \I,
\ee
we get
\be
\eps_{zz}=\frac{\sigma_{zz}-\nu(\sigma_{xx}+\sigma_{yy})}{E},
\ee
and because  in the Kirchhoff model $\eps_{zz}=0$, we get
\be
\sigma_{zz}=\nu(\sigma_{xx}+\sigma_{yy});
\ee
\item for common materials, $0<\nu<1/2$, so $\sigma_{zz}$ is at most of the same order of magnitude of $\sigma_{xx}$ and $\sigma_{yy}$;
\item for the assumptions on the loads, $\sigma_{zz}$ is a smooth function of $z$; so, because $\sigma_{zz}$ is regular and the plate thin, its value cannot increase very much inside the plate with respect to the values that $\sigma_{zz}$ has on the faces of the plate, $S_B$ and $S_T$;
\item because the value of $\sigma_{zz}$ on $S_B$ and $S_T$ is just the value of the contact forces applied to the faces, i.e. $\tau_z$, it has not, in usual situations, a great value compared to the design values of $\sigma_{xx}$ and $\sigma_{yy}$. For this reason, we can neglect $\sigma_{zz}$ and finally consider that it is null.
\end{itemize}
These arguments are merely empirical; of course, they are not valid in some situations, i.e. in correspondence of concentrated loads, e.g. impact forces, or of supporting parts. In such cases the stress state is properly a 3D one and in such zones the behavior is, locally, far from that of a plate.

Because $\eps_{xz}=\eps_{yz}$, from eq. (\ref{eq:lamedirect}) we get also 
\be
\label{eq:sigmatransverse}
\sigma_{xz}=\frac{E}{1+\nu}\eps_{xz}=0,\ \ \sigma_{yz}=\frac{E}{1+\nu}\eps_{yz}=0,
\ee
so finally also the stress field is planar:
\be
\bsig=\left[\begin{array}{ccc}
\sigma_{xx} & \sigma_{xy} & 0 \\
\sigma_{xy} & \sigma_{yy} & 0 \\
0 & 0 & 0
\end{array}\right].
\ee
This is a strange situation, because in elasticity, a plane stress state is incompatible with a plane strain one and vice-versa. However, because $\beps(p)$ depends upon $x,y$ and $z$, this is the same also for the stress field: $\bsig(p)$  is not a plane field.

A dramatic consequence of eq. (\ref{eq:sigmatransverse}) is that the plate cannot be equilibrated under the action of loads that have a component orthogonal to the mid-plane: strictly speaking, the Kirchhoff model concerns plates that are loaded in their plane or by couples on the boundary. This problem will be solved making use directly of the equilibrium equations, cf. Sec. \ref{sec:transvstress}.

We  remark also that eq. (\ref{eq:sigmatransverse}) is a consequence not only of the kinematical model, but also of the constitutive law: for generally oriented anisotropic plates, $\sigma_{xz}$ and $\sigma_{yz}$ are in general not null, also for the Kirchhoff model.

We can now  obtain the constitutive law for a Kirchhoff plate: from eq. (\ref{eq:inverselame}) we get
\be
\besp
&\eps_{xx}=\frac{1}{E}(\sigma_{xx}-\nu\sigma_{yy}),\\
&\eps_{yy}=\frac{1}{E}(\sigma_{yy}-\nu\sigma_{xx}),\\
&\eps_{xy}=\frac{1+\nu}{E}\sigma_{xy},
\end{split}
\ee
that once inverted give
\be
\label{eq:sigmakirch}
\besp
&\sigma_{xx}=\frac{E}{1-\nu^2}(\eps_{xx}+\nu\eps_{yy}),\\
&\sigma_{yy}=\frac{E}{1-\nu^2}(\eps_{yy}+\nu\eps_{xx}),\\
&\sigma_{xy}=\frac{E}{1+\nu}\eps_{xy}=2G\eps_{xy}.
\end{split}
\ee

We can put the last result in matrix form:
\be
\label{eq:loiplates}
\left\{\begin{array}{c}\sigma_{xx} \bigskip\\\sigma_{yy} \bigskip\\\sigma_{xy}\end{array}\right\}=
\left[\begin{array}{ccc}
\dfrac{E}{1-\nu^2} & \dfrac{\nu E}{1-\nu^2} & 0 \medskip\\
\dfrac{\nu E}{1-\nu^2} & \dfrac{E}{1-\nu^2} & 0 \medskip\\
0 & 0 & G
\end{array}\right]
\left\{\begin{array}{c}\eps_{xx} \bigskip\\\eps_{yy} \bigskip\\2\eps_{xy}\end{array}\right\},
\ee
or synthetically 
\be
\bsig=\mathbb{D}\ \beps,
\ee
with $\mathbb{D}$ the matrix in eq. (\ref{eq:loiplates}). It can be checked that $\mathbb{D}$ corresponds to the so-called {\it reduced stiffness matrix}, typical of a plane stress state: finally, though the assumptions of the Kirchhoff model do not coincide with those typical of a plane stress state ($\bsig$ is not a plane field), the plate's constitutive law, eq. (\ref{eq:loiplates}) is just like that of a plane stress state.

\subsection{Internal actions}
\begin{figure}[ht]
\begin{center}
\includegraphics[width=\linewidth]{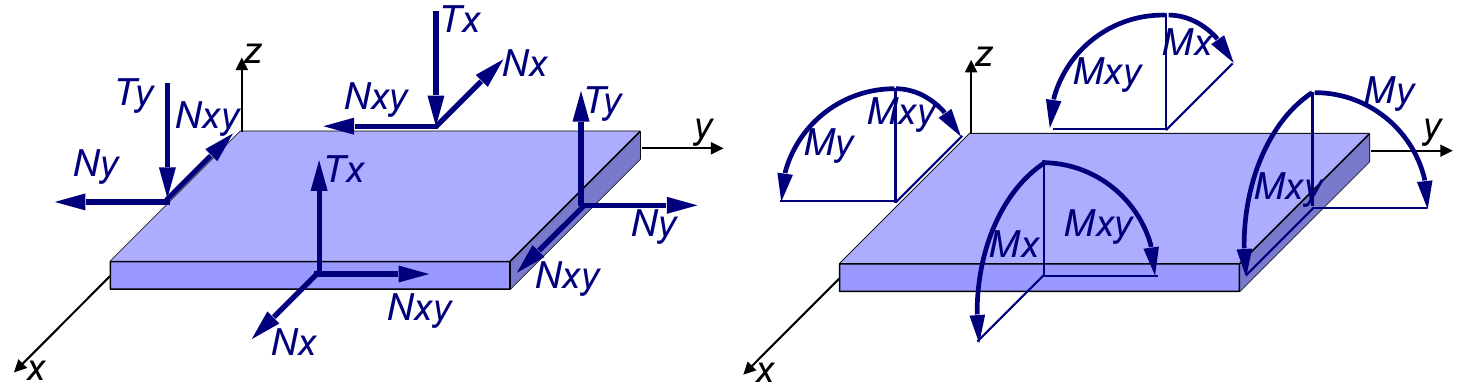}
\caption{Internal actions.}
\label{fig:f5_4}
\end{center}
\end{figure}

The internal actions, forces and couples, are obtained integrating through the thickness the stress field. In particular, we introduce:
\begin{itemize}
\item the {\it extension tensor} $\N$, defined as
\be
\label{eq:defN}
\N=\int_{-\frac{h}{2}}^{+\frac{h}{2}}\bsig\ dz,
\ee
whose components 
\be
\label{eq:compN}
\left\{\begin{array}{c}N_x \bigskip\\N_y \bigskip\\N_{xy}\end{array}\right\}=
\left\{
\begin{split}&\int_{-\frac{h}{2}}^{+\frac{h}{2}}\sigma_{xx}\ dz \medskip\\
&\int_{-\frac{h}{2}}^{+\frac{h}{2}}\sigma_{yy}\ dz \medskip\\
&\int_{-\frac{h}{2}}^{+\frac{h}{2}}\sigma_{xy}\ dz
\end{split}
\right\}
\ee
are represented in Fig. \ref{fig:f5_4};
\item the {\it transverse shear forces}
\be
\label{eq:compT}
T_x=\int_{-\frac{h}{2}}^{+\frac{h}{2}}\sigma_{xz}\ dz,\ \ \ T_y=\int_{-\frac{h}{2}}^{+\frac{h}{2}}\sigma_{yz}\ dz,
\ee
also represented in Fig. \ref{fig:f5_4},
\item the {\it bending tensor} $\M$, defined as
\be
\M=\int_{-\frac{h}{2}}^{+\frac{h}{2}}z\ \bsig\ dz,
\ee
whose components 
\be
\label{eq:compM}
\left\{\begin{array}{c}M_x \bigskip\\M_y \bigskip\\M_{xy}\end{array}\right\}=
\left\{
\begin{split}&\int_{-\frac{h}{2}}^{+\frac{h}{2}}z\ \sigma_{xx}\ dz \medskip\\
&\int_{-\frac{h}{2}}^{+\frac{h}{2}}z\ \sigma_{yy}\ dz \medskip\\
&\int_{-\frac{h}{2}}^{+\frac{h}{2}}z\ \sigma_{xy}\ dz
\end{split}
\right\}
\ee
are still represented in Fig. \ref{fig:f5_4}.
\end{itemize}
Being integrals of a planar symmetric second-rank tensor, both $\N$ and $\M$ are planar symmetric second-rank tensors too; in addition, they are plane fields too: $\N=\N(x,y),\M=\M(x,y)$. To remark that the units of $\N$ and of the shear forces $T_x$ and $T_y$ are a force per unit length, while those of $\M$ are a force.

\subsection{Uncoupling bending and extension}
The expression of $\beps$ can be written in the form
\be
\beps=\beps^0+z\ \bk,
\ee
where
\be
\beps^0=\left\{\begin{array}{c}\beps_{xx}^0\bigskip \\\beps_{yy}^0 \bigskip\\2\beps_{xy}^0\end{array}\right\}=
\left\{\begin{array}{c}\dfrac{\partial u_0}{\partial x} \medskip\\\dfrac{\partial v_0}{\partial y} \medskip\\\dfrac{\partial u_0}{\partial y}+\dfrac{\partial v_0}{\partial x}\end{array}\right\}
\ee
is the {\it in-plane deformation tensor}, describing the strain of the mid-plane, while
\be
\bk=\left\{\begin{array}{c}\kappa_{x} \bigskip\\\kappa_{y} \bigskip\\2\kappa_{xy}\end{array}\right\}=
\left\{\begin{array}{c}-\dfrac{\partial^2 w_0}{\partial x^2} \medskip\\-\dfrac{\partial^2 w_0}{\partial y^2} \medskip\\-2\dfrac{\partial^2 w_0}{\partial x\partial y}\end{array}\right\}
\ee
is the {\it curvatures tensor}, describing the bending of the mid-plane. Using this decomposition of $\beps$, we can write
\be
\label{eq:sigma1}
\bsig=\mathbb{D}\beps=\mathbb{D}\beps^0+z\mathbb{D}\bk
\ee
and because $\mathbb{D},\beps^0$ and $\kappa$ do not depend upon $z$, we get
\be
\label{eq:uncoupledeq}
\besp
&\N=\mathbb{D}\beps^0\int_{-\frac{h}{2}}^{+\frac{h}{2}} dz+\mathbb{D}\kappa\int_{-\frac{h}{2}}^{+\frac{h}{2}}z\ dz=h\ \mathbb{D}\beps^0,\\
&\M=\mathbb{D}\beps^0\int_{-\frac{h}{2}}^{+\frac{h}{2}}z\ dz+\mathbb{D}\kappa\int_{-\frac{h}{2}}^{+\frac{h}{2}}z^2\ dz=\frac{h^3}{12}\mathbb{D}\kappa,\\
\end{split}
\ee
i.e. $\N$ depends only upon in-plane strains and $\M$ only upon curvatures: there is no  coupling between extension and curvatures nor between bending and in-plane deformation. Actually, the absence of any coupling effect is due to two facts: on one hand, geometry (the flatness of the plate) and on the other hand homogeneity (the plate is composed by a unique layer): generally speaking for shells, i.e. "curved plates", and laminates, i.e. plates obtained superposing layers, coupling effects are present.
Finally, extension and bending of the plate can be examined separately, they do not interact. 

A plate that is subjected only to in-plane forces, so that $\bk=\bo\Rightarrow\M=\bo$, is called a (flat) {\it membrane} (curved membranes exist too, e.g. an inflated balloon, they are considered in Chapter \ref{ch:7}). The membrane regime, however, is compatible only with a distribution of the forces  which is point-wise symmetric with respect to the mid-plane, assumption that we have already made.

\subsection{Balance equations}

There are different ways for writing  local balance equations. A standard approach is to write the equilibrium to translation and rotation of a small part of the plate, and then discard higher order terms. We prefer here a more concise approach: starting from the 3D equilibrium equations,
\be
\label{eq:equil3D}
\div\bsig+\b=\bo,
\ee
where $\b(p)$ is the field of body forces (forces per unit volume) on $\Omega$, we obtain, after integration on $z$,
\be
\label{eq:equileqplates}
\besp
&\int_{-\frac{h}{2}}^\frac{h}{2}\left(\frac{\partial \sigma_{xx}}{\partial x}+\frac{\partial \sigma_{xy}}{\partial y}+\frac{\partial \sigma_{xz}}{\partial z}+b_x\right)dz=0,\\
&\int_{-\frac{h}{2}}^\frac{h}{2}\left(\frac{\partial \sigma_{xy}}{\partial x}+\frac{\partial \sigma_{yy}}{\partial y}+\frac{\partial \sigma_{yz}}{\partial z}+b_y\right)dz=0,\\
&\int_{-\frac{h}{2}}^\frac{h}{2}\left(\frac{\partial \sigma_{xz}}{\partial x}+\frac{\partial \sigma_{yz}}{\partial y}+\frac{\partial \sigma_{zz}}{\partial z}+b_z\right)dz=0.
\end{split}
\ee
Because the integration bounds are fixed, we can bring the differential operators outside the integration symbol:
\be
\besp
&\frac{\partial }{\partial x}\int_{-\frac{h}{2}}^\frac{h}{2}\sigma_{xx}\ dz+
\frac{\partial }{\partial y}\int_{-\frac{h}{2}}^\frac{h}{2}\sigma_{xy}\ dz+
\int_{-\frac{h}{2}}^\frac{h}{2}\left(\frac{\partial \sigma_{xz}}{\partial z}+b_x\right)dz=0,\\
&\frac{\partial }{\partial x}\int_{-\frac{h}{2}}^\frac{h}{2}\sigma_{xy}\ dz+
\frac{\partial }{\partial y}\int_{-\frac{h}{2}}^\frac{h}{2}\sigma_{yy}\ dz+
\int_{-\frac{h}{2}}^\frac{h}{2}\left(\frac{\partial \sigma_{yz}}{\partial z}+b_y\right)dz=0,\\
&\frac{\partial }{\partial x}\int_{-\frac{h}{2}}^\frac{h}{2}\sigma_{xz}\ dz+
\frac{\partial }{\partial y}\int_{-\frac{h}{2}}^\frac{h}{2}\sigma_{yz}\ dz+
\int_{-\frac{h}{2}}^\frac{h}{2}\left(\frac{\partial \sigma_{zz}}{\partial z}+b_z\right)dz=0,
\end{split}
\ee
so that, introducing eqs. (\ref{eq:compN}) and (\ref{eq:compT}), we get
\be
\label{eq:membraneeq}
\besp
&\frac{\partial N_x}{\partial x}+\frac{\partial N_{xy}}{\partial y}+f_x=0,\\
&\frac{\partial N_{xy}}{\partial x}+\frac{\partial N_{y}}{\partial y}+f_y=0,
\end{split}
\ee
and
\be
\label{eq:sheareq}
\frac{\partial T_x}{\partial x}+\frac{\partial T_{y}}{\partial y}+f_z=0,
\ee
where we have denoted by $\gr{f}(x,y)=(f_x,f_y,f_z)$ the vector of the {\it total applied loads per unit area}:
%\footnote{\label{note:1}Thanks to the assumption of symmetrical distribution of the actions,  $\btau\left(x,y,{-\frac{h}{2}}\right)=\btau\left(x,y,{\frac{h}{2}}\right)=(\tau_x,\tau_y,\tau_z)$; but 
%$\btau\left(x,y,{\pm\frac{h}{2}}\right)=\bsig\left(x,y,{\pm\frac{h}{2}}\right)\n\left(x,y,{\pm\frac{h}{2}}\right)$
%and $\n\left(x,y,{-\frac{h}{2}}\right)=-\n\left(x,y,{\frac{h}{2}}\right)=(0,0,-1)$, so that 
%$\btau\left(x,y,{-\frac{h}{2}}\right)=(\tau_x,\tau_y,\tau_z)=(-\sigma_{xz},-\sigma_{yz},-\sigma_{zz})_{-\frac{h}{2}}$ and $\btau\left(x,y,{\frac{h}{2}}\right)=(\tau_x,\tau_y,\tau_z)=(\sigma_{xz},\sigma_{yz},\sigma_{zz})_{\frac{h}{2}}\Rightarrow\left[\sigma_{xz}\right]_{-\frac{h}{2}}^\frac{h}{2}=2\tau_x,\ \left[\sigma_{yz}\right]_{-\frac{h}{2}}^\frac{h}{2}=2\tau_y,\ \left[\sigma_{zz}\right]_{-\frac{h}{2}}^\frac{h}{2}=2\tau_z$.}
\be
\besp
\label{eq:resforces}
&f_x(x,y)=\left[\sigma_{xz}\right]_{-\frac{h}{2}}^\frac{h}{2}+\int_{-\frac{h}{2}}^\frac{h}{2}b_xdz=\sigma_{xz}^T-\sigma_{xz}^B+\beta_x=2\tau_x^s+\beta_x,\\
&f_y(x,y)=\left[\sigma_{yz}\right]_{-\frac{h}{2}}^\frac{h}{2}+\int_{-\frac{h}{2}}^\frac{h}{2}b_ydz=\sigma_{yz}^T-\sigma_{yz}^B+\beta_y=2\tau_y^s+\beta_y,\\
&f_z(x,y)=\left[\sigma_{zz}\right]_{-\frac{h}{2}}^\frac{h}{2}+\int_{-\frac{h}{2}}^\frac{h}{2}b_zdz=\sigma_{zz}^T-\sigma_{zz}^B+\beta_z=2\tau_z^s+\beta_z;
\end{split}
\ee
in the above equations, $\bb(x,y)=(\beta_x,\beta_y,\beta_z)$ is the vector of {\it body forces per unit area}:
\be
\bb(x,y)=\int_{-\frac{h}{2}}^\frac{h}{2}\b(x,y,z) dz.
\ee
Equations (\ref{eq:membraneeq}) are the {\it extension, or membrane, equilibrium equations} of the plate, while eq. (\ref{eq:sheareq}) is the {\it transverse shear equilibrium equation}. 

The bending equilibrium can be obtained in a similar way, now integrating eq. (\ref{eq:equil3D}) once multiplied by $z$, the distance of each component $\sigma_{ij}(p)$ from the mid-plane:
\be
\label{eq:bendingeq0}
\besp
&\int_{-\frac{h}{2}}^\frac{h}{2}z\left(\frac{\partial \sigma_{xx}}{\partial x}+z\frac{\partial \sigma_{xy}}{\partial y}+z\frac{\partial \sigma_{xz}}{\partial z}+z\ b_x\right) dz=0,\\
&\int_{-\frac{h}{2}}^\frac{h}{2}\left(z \frac{\partial \sigma_{xy}}{\partial x}+z\frac{\partial \sigma_{yy}}{\partial y}+z\frac{\partial \sigma_{yz}}{\partial z}+z\ b_y\right) dz=0,\\
&\int_{-\frac{h}{2}}^\frac{h}{2}\left(z\frac{\partial \sigma_{xz}}{\partial x}+z\frac{\partial \sigma_{yz}}{\partial y}+z\frac{\partial \sigma_{zz}}{\partial z}+z\ b_z\right) dz=0.
\end{split}
\ee
Because we are interested in the bending of the plate, the third equation of (\ref{eq:bendingeq0}) can be discarded, as it concerns the equilibrium to rotation about the axis $z$.
Observing that
\be
z\frac{\partial \sigma_{\alpha\beta}}{\partial\alpha}=\frac{\partial (z\ \sigma_{\alpha\beta})}{\partial\alpha} \ \ \alpha,\beta\in\{1,2\},
\ee
and that, because $\b(p)$ is an even function, $z\b(p)$ is an odd one so that
\be
\int_{-\frac{h}{2}}^\frac{h}{2}z\ b_x\ dz=\int_{-\frac{h}{2}}^\frac{h}{2}z\ b_y\ dz=0,
\ee
we get
\be
\label{eq:bendingeq}
\besp
&\frac{\partial }{\partial x}\int_{-\frac{h}{2}}^\frac{h}{2}z\sigma_{xx}\ dz+
\frac{\partial }{\partial y}\int_{-\frac{h}{2}}^\frac{h}{2}z\sigma_{xy}\ dz+
\int_{-\frac{h}{2}}^\frac{h}{2}z\frac{\partial \sigma_{xz}}{\partial z}\ dz=0,\\
&\frac{\partial }{\partial x}\int_{-\frac{h}{2}}^\frac{h}{2}z\sigma_{xy}\ dz+
\frac{\partial }{\partial y}\int_{-\frac{h}{2}}^\frac{h}{2}z\sigma_{yy}\ dz+
\int_{-\frac{h}{2}}^\frac{h}{2}z\frac{\partial \sigma_{yz}}{\partial z}\ dz=0.%,\\
%&\frac{\partial }{\partial x}\int_{-\frac{h}{2}}^\frac{h}{2}z\sigma_{xz}\ dz+
%\frac{\partial }{\partial y}\int_{-\frac{h}{2}}^\frac{h}{2}z\sigma_{yz}\ dz+
%\int_{-\frac{h}{2}}^\frac{h}{2}z\frac{\partial \sigma_{zz}}{\partial z}\ dz=0.
\end{split}
\ee
Integration by parts gives, cfr.  eqs.  (\ref{eq:taubot}),  (\ref{eq:tautop}) and (\ref{eq:compT}),
\be
\besp
&\int_{-\frac{h}{2}}^\frac{h}{2}z\frac{\partial \sigma_{xz}}{\partial z}\ dz=[z\sigma_{xz}]_{-\frac{h}{2}}^\frac{h}{2}-\int_{-\frac{h}{2}}^\frac{h}{2}\sigma_{xz}\ dz=\\
&=\frac{h}{2}(\sigma_{xz}^T+\sigma_{xz}^B)-T_x=\frac{h}{2}(\tau_x^T-\tau_x^B)-T_x,=h\tau_x^a-T_x,\\
&\int_{-\frac{h}{2}}^\frac{h}{2}z\frac{\partial \sigma_{yz}}{\partial z}\ dz=[z\sigma_{yz}]_{-\frac{h}{2}}^\frac{h}{2}-\int_{-\frac{h}{2}}^\frac{h}{2}\sigma_{yz}\ dz=\\
&=\frac{h}{2}(\sigma_{yz}^T+\sigma_{yz}^B)-T_y=\frac{h}{2}(\tau_y^T-\tau_y^B)-T_y,=h\tau_y^a-T_y.
\end{split}
\ee
If now we use eq. (\ref{eq:compM}) we obtain
\be
\label{eq:bendshearequil}
\besp
&\frac{\partial M_x}{\partial x}+\frac{\partial M_{xy}}{\partial y}=T_x-h\tau_x^a,\\
&\frac{\partial M_{xy}}{\partial x}+\frac{\partial M_{y}}{\partial y}=T_y-h\tau_y^a,
\end{split}
\ee
and injecting this result into eq. (\ref{eq:sheareq}) we get the unique second-order equation including bending and shear equilibrium:
\be
\label{eq:2ndorderbendeq}
\frac{\partial^2 M_x}{\partial x^2}+2\frac{\partial^2 M_{xy}}{\partial x\partial y}+\frac{\partial^2 M_y}{\partial y^2}+\widehat{f}_z=0
\ee
where
\be
\widehat{f}_z=f_z+h\left(\frac{\partial \tau_x^a}{\partial x}+\frac{\partial \tau_y^a}{\partial y}\right).
\ee
To remark that the for the eight unknowns $N_x,N_y,N_{xy},T_x,T_y,M_x,M_y$ and $M_{xy}$ we dispose of only five equilibrium equations, eqs. (\ref{eq:membraneeq}), (\ref{eq:sheareq}) and (\ref{eq:bendshearequil}): plates are {\it intrinsically hyperstatic bodies}. To solve the equilibrium problem is hence necessary to introduce  a constitutive law, namely the elastic one.

\subsection{Elastic equilibrium equations}
Injecting into eq. (\ref{eq:uncoupledeq})$_1$ the expressions of $\mathbb{D}$ and $\beps^0$ we get
\be
\besp
&N_x=\frac{E\ h}{1-\nu^2}\left(\frac{\partial u_0}{\partial x}+\nu\frac{\partial v_0}{\partial y}\right),\medskip\\
&N_y=\frac{E\ h}{1-\nu^2}\left(\nu\frac{\partial u_0}{\partial x}+\frac{\partial v_0}{\partial y}\right),\medskip\\
&N_{xy}=G\left(\frac{\partial u_0}{\partial y}+\frac{\partial v_0}{\partial x}\right),
\end{split}
\ee
which injected into eq. (\ref{eq:membraneeq}) gives ($G=E/2(1+\nu)$)
\be
\label{eq:2ordermembraneeq}
\besp
&\frac{E\ h}{1+\nu}\left[\frac{1}{1-\nu}\left(\frac{\partial^2 u_0}{\partial x^2}+\nu\frac{\partial^2 v_0}{\partial x\partial y}\right)+\frac{1}{2}\left(\frac{\partial^2 u_0}{\partial y^2}+\frac{\partial^2 v_0}{\partial x\partial y}\right)\right]+f_x=0,\medskip\\
&\frac{E\ h}{1+\nu}\left[\frac{1}{2}\left(\frac{\partial^2 u_0}{\partial x\partial y}+\frac{\partial^2 v_0}{\partial x^2}\right)+\frac{1}{1-\nu}\left(\nu\frac{\partial^2 u_0}{\partial x\partial y}+\frac{\partial^2 v_0}{\partial y^2}\right)\right]+f_y=0.
\end{split}
\ee
These are the {\it elastic membrane equilibrium equations } for isotropic homogeneous plates; they are the 2D equivalent of the rod extension equilibrium equation $EA\ w"+p_z=0$.

Similarly, if now we inject into eq. (\ref{eq:uncoupledeq})$_2$ the expressions of $\mathbb{D}$ and $\bk$, we obtain
\be
\besp
&M_x=-\frac{h^3}{12}\frac{E}{1-\nu^2}\left(\frac{\partial^2 w_0}{\partial x^2}+\nu\frac{\partial^2 w_0}{\partial y^2}\right),\medskip\\
&M_y=-\frac{h^3}{12}\frac{ E}{1-\nu^2}\left(\frac{\partial^2 w_0}{\partial y^2}+\nu\frac{\partial^2 w_0}{\partial x^2}\right),\medskip\\
&M_{xy}=-\frac{h^3}{12}\frac{E}{1+\nu}\frac{\partial^2 w_0}{\partial x\partial y},
\end{split}
\ee
that inserted into eq. (\ref{eq:2ndorderbendeq}) gives
\be
\label{eq:4orderbendeq}
\frac{h^3}{12}\frac{E}{1-\nu^2}\left(\frac{\partial^4 w_0}{\partial x^4}+2\frac{\partial^4 w_0}{\partial x^2\partial y^2}+\frac{\partial^4 w_0}{\partial y^4}\right)=\widehat{f}_z.
\ee
Usually, the symbol $D$ is used to denote the bending stiffness of the plate per unit of length:
\be
D=\frac{h^3}{12}\frac{E}{1-\nu^2},
\ee
so that, once introduced the classical {\it double laplacian} differential operator,
\be
\Delta^2=\frac{\partial^4 \bullet}{\partial x^4}+2\frac{\partial^4 \bullet}{\partial x^2\partial y^2}+\frac{\partial^4 \bullet}{\partial y^4},
\ee
eq. (\ref{eq:4orderbendeq}) can be rewritten in the form
\be
\label{eq:germainlagrange}
D\Delta^2w_0=\widehat{f}_z.
\ee
This is the celebrated {\it Germain-Lagrange equation} (1816); it is the 2D corresponding of the elastic bending equilibrium equation $EJ\ v^{iv}=p_y$ for rods: the Kirchhoff model is, for  plates, the equivalent of the Euler-Bernoulli model for the bending of rods. Actually, because $h^3/12$ is the moment of inertia, about axes $x$ or $y$, of a unit length cross section of the plate, the true difference between the one-dimensional case, rods, and the bi-dimensional one, plates, is in the term $1-\nu^2$, which is a consequence of the reduced stiffness $\mathbb{D}$, hence of the fact that $\bsig(p)$ is planar: plates are intrinsically stiffer than rods.

Equations (\ref{eq:2ordermembraneeq}) and (\ref{eq:germainlagrange}) are the {\it elastic equilibrium equations of plates}, respectively for membrane and for bending.

In the same way, from the equilibrium equations for bending, eq. (\ref{eq:bendshearequil}), we get:
\be
\besp
&T_x=\frac{\partial M_x}{\partial x}+\frac{\partial M_{xy}}{\partial y}+h\tau_x^a=-\frac{h^3}{12}\frac{E}{1-\nu^2}\left[w_{0,xxx}+\nu w_{0,xyy}+(1-\nu)w_{0,xyy}\right]+h\tau_x^a,\\
&T_y=\frac{\partial M_{xy}}{\partial x}+\frac{\partial M_{y}}{\partial y}+h\tau_y^a=-\frac{h^3}{12}\frac{E}{1-\nu^2}\left[(1-\nu)w_{0,xxy}+\nu w_{0,xxy}+w_{0,yyy}\right]+h\tau_y^a,
\end{split}
\ee
i.e.
\be
\label{eq:tagliopiastreelastic}
\besp
&T_x=-D\Delta w_{0,x}+h\tau_x^a,\\
&T_y=-D\Delta w_{0,y}+h\tau_y^a.
\end{split}
\ee

\subsection{Expressions for stresses}
Inverting eqs. (\ref{eq:uncoupledeq}) gives 
\be
\mathbb{D}\beps^0=\frac{\N}{h},\ \ \ \mathbb{D}\bk=\frac{12}{h^3}\M,
\ee
that injected into eq. (\ref{eq:sigma1}) gives
\be
\label{eq:Navierplates}
\bsig=\frac{1}{h}\N+\frac{12}{h^3}\M\ z.
\ee
This formula generalizes to plates the {\it Navier's formula}, giving the normal stress distribution in beams subjected to bending and extension.

\subsection{Transverse shear stresses}
\label{sec:transvstress}

We can now go back to the problem of determining the {\it transverse shear stresses} $\sigma_{xz}$ and $\sigma_{yz}$, that, as we have seen, are incompatible with the Kirchhoff's assumptions. Actually, we can find them  using the equilibrium equations. Let us consider first the component $\sigma_{xz}$: from the first component of eq. (\ref{eq:equil3D}) we get
\be
\besp
\label{eq:equiltransverse1}
&\frac{\partial \sigma_{xz}}{\partial z}=-\frac{\partial \sigma_{xx}}{\partial x}-\frac{\partial \sigma_{xy}}{\partial y}-b_x,\\
\end{split}
\ee
and through the first component of eq. (\ref{eq:Navierplates}) we get
\be
\besp
\label{eq:equiltransverse2}
&\frac{\partial \sigma_{xz}}{\partial z}=-\frac{1}{h}\left(\frac{\partial N_{x}}{\partial x}+\frac{\partial N_{xy}}{\partial y}\right)
-\frac{12}{h^3}\left(\frac{\partial M_{x}}{\partial x}+\frac{\partial M_{xy}}{\partial y}\right)z-b_x,\\
\end{split}
\ee
If now we use the equilibrium equations, i.e. eqs. (\ref{eq:membraneeq}), (\ref{eq:resforces}) and (\ref{eq:bendshearequil}), we get
\be
\besp
&\frac{\partial \sigma_{xz}}{\partial z}=\frac{f_x}{h}-b_x-12\frac{z}{h^2}\left(\frac{T_x}{h}-\tau_x^a\right),\\
\end{split}
\ee
and integrating
\be
\besp
&\sigma_{xz}=f_x\frac{z}{h}-\int_0^zb_xdz-6\frac{z^2}{h^2}\left(\frac{T_x}{h}-\tau_x^a\right)+\phi_x(x,y).
\end{split}
\ee
The unknown function $\phi_x(x,y)$ can be determined through the boundary conditions; %Because we have integrated two first-order equilibrium equations, we only have one boundary condition for each stress equation, that allows for determining just one integration constant, while there are two boundaries, $S_B$ and $S_T$, on which the stresses $\sigma_{xz}$ or $\sigma_{yz}$ must match the value of the applied surface tractions. Because  $\sigma_{xz}$ and $\sigma_{yz}$ are quadratic functions of $z$, this is possible only if the boundary conditions on $S_B$ and $S_T$ are the same, i.e. if the plate is loaded symmetrically with respect to the mid-plane, assumption that we have made.
if we put that, see eq. (\ref{eq:tautop}),
\be
\sigma_{xz}^T=\tau_x^T=\tau_x^s+\tau_x^a,
\ee
we get
\be
\besp
\phi_x(x,y)&=\tau_x^s+\tau_x^a-\frac{f_x}{h}\frac{h}{2}+\int_0^\frac{h}{2}b_xdz+\frac{6}{h^2}\frac{h^2}{4}\left(\frac{T_x}{h}-\tau_x^a\right)\\
&=\tau_x^s+\tau_x^a-\frac{\beta_x}{2}-\tau_x^s+\frac{\beta_x}{2}+\frac{3}{2}\left(\frac{T_x}{h}-\tau_x^a\right)\\
&=\frac{3}{2}\frac{T_x}{h}-\frac{\tau_x^a}{2},
\end{split}
\ee
so finally, 
\be
\besp
&\sigma_{xz}=f_x\frac{z}{h}-\int_0^zb_xdz-6\frac{z^2}{h^2}\left(\frac{T_x}{h}-\tau_x^a\right)+\frac{3}{2}\frac{T_x}{h}-\frac{\tau_x^a}{2}.
\end{split}
\ee
or better
\be
\label{eq:shearxzplates}
\besp
&\sigma_{xz}=\beta_x\frac{z}{h}+2\tau_x^s\frac{z}{h}-\int_0^zb_xdz+\left(6\frac{z^2}{h^2}-\frac{1}{2}\right)\tau_x^a+
6\frac{T_x}{h}\left(\frac{1}{4}-\frac{z^2}{h^2}\right).
\end{split}
\ee
Of course, proceeding in the same way we get a similar expression for $\sigma_{yz}$:
\be
\label{eq:shearyzplates}
\besp
&\sigma_{yz}=\beta_y\frac{z}{h}+2\tau_y^s\frac{z}{h}-\int_0^zb_ydz+\left(6\frac{z^2}{h^2}-\frac{1}{2}\right)\tau_y^a+
6\frac{T_y}{h}\left(\frac{1}{4}-\frac{z^2}{h^2}\right).
\end{split}
\ee
To remark that the two conditions at the surfaces $S_T$ and $S_B$ are matched, see eqs. (\ref{eq:tautop}) and (\ref{eq:taubot}):
\be
\besp
&\sigma_{xz}^T=\sigma_{xz}\left(\frac{h}{2}\right)=\tau_x^a+\tau_x^s=\tau_x^T,\\
&\sigma_{xz}^B=\sigma_{xz}\left(-\frac{h}{2}\right)=\tau_x^a-\tau_x^s=-\tau_x^B,\\
\end{split}
\ee
and similarly for $\sigma_{yz}$.

\subsection{Plates submitted uniquely to orthogonal loads}
The results in the previous Sections have been obtained for the case of a general load, i.e. of a vector of body forces and of surface tractions having all the components different from zero. However, it is interesting to consider the special and very common case of plates loaded by only {\it orthogonal loads}, i.e. such that
\be
b_x=0,\ b_y=0,\ \tau_x^T=0,\  \tau_x^B=0,\ \tau_y^T=0,\ \tau_y^B=0.
\ee
In such a case, $f_x(x,y)=f_y(x,y)=0$ and the equilibrium equations (\ref{eq:bendshearequil})  simplify to
\be
\label{eq:bendshearequilorth}
\besp
&\frac{\partial M_x}{\partial x}+\frac{\partial M_{xy}}{\partial y}=T_x,\\
&\frac{\partial M_{xy}}{\partial x}+\frac{\partial M_{y}}{\partial y}=T_y,
\end{split}
\ee
and eq.  (\ref{eq:2ndorderbendeq}) to
\be
\label{eq:2ndorderbendeqorth}
\frac{\partial^2 M_x}{\partial x^2}+2\frac{\partial^2 M_{xy}}{\partial x\partial y}+\frac{\partial^2 M_y}{\partial y^2}+f_z=0,
\ee
while the elastic equilibrium equations (\ref{eq:germainlagrange}) and (\ref{eq:tagliopiastreelastic}) respectively to
\be
\label{eq:germainlagrangeorth}
D\Delta^2w_0=f_z
\ee
and to
\be
\label{eq:tagliopiastreelasticorth}
\besp
&T_x=-D\Delta w_{0,x},\\
&T_y=-D\Delta w_{0,y}.
\end{split}
\ee
Finally, the expressions of the transverse shear stresses, eqs. (\ref{eq:shearxzplates}) and (\ref{eq:shearyzplates}), become
\be
\label{eq:shearxyzplatesorth}
\besp
&\sigma_{xz}=\frac{6}{h} T_x\left(\frac{1}{4}-\frac{z^2}{h^2}\right),\\
&\sigma_{yz}=\frac{6}{h} T_y\left(\frac{1}{4}-\frac{z^2}{h^2}\right).
\end{split}
\ee
These formulae generalize to plates the classical result of the {\it Jourawski's formula} for the shear stresses in beams. They state that the variation of the transverse shear stresses $\sigma_{xz}$ and $\sigma_{yz}$ through the thickness of the plate is parabolic and that the maximum is get in correspondence of the mid-plane, where they attain 1.5 times their average value:
\be
\sigma_{xz}^{max}=\frac{3}{2}\frac{T_x}{h},\ \ \ \sigma_{yz}^{max}=\frac{3}{2}\frac{T_y}{h}.
\ee

\subsection{Evaluation of the transverse normal stress}
\begin{figure}[th]
\begin{center}
\includegraphics[width=.21\linewidth]{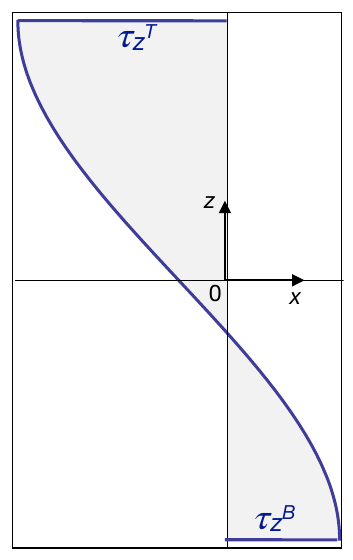}
\caption{Diagram of $\sigma_{zz}$.}
\label{fig:f5_0}
\end{center}
\end{figure}

We can assess now the assumption  $\sigma_{zz}=0$ made and empirically justified in Sect. \ref{sec:stressplate}. Let us consider, for the sake of simplicity, the case of $\b=\gr{o}$ and of only orthogonal lateral loads, i.e. $\btau^T=(0,0,\tau_z^T), \btau^B=(0,0,\tau_z^B)$. From the third component of the equilibrium equation (\ref{eq:equil3D}) and from eqs. (\ref{eq:sheareq}) and (\ref{eq:shearxyzplatesorth}) we get 
\be
\besp
\frac{\partial \sigma_{zz}}{\partial z}=-\frac{\partial \sigma_{zx}}{\partial x}-\frac{\partial \sigma_{zy}}{\partial y}&=
-\frac{6}{h}\left(\frac{1}{4}-\frac{z^2}{h^2}\right)\left(\frac{\partial T_x}{\partial x}+\frac{\partial T_y}{\partial y}\right)\\
&=\frac{12\tau_z^s}{h}\left(\frac{1}{4}-\frac{z^2}{h^2}\right).
\end{split}
\ee
Integrating,
\be
\sigma_{zz}=\frac{12\tau_z^s}{h}z\left(\frac{1}{4}-\frac{1}{3}\frac{z^2}{h^2}\right)+\phi_z(x,y),
\ee
and from the boundary condition on $S_T$,
\be
\sigma_{zz}^T=\sigma_{zz}\left(\frac{h}{2}\right)=\tau_z^T=\tau_z^s+\tau_z^a\ \rightarrow\ \phi_z(x,y)=\tau_z^a,
\ee
so finally
\be
\sigma_{zz}=\tau_z^s\frac{z}{h}\left(3-4\frac{z^2}{h^2}\right)+\tau_z^a.
\ee
The distribution of $\sigma_{zz}$ so found matches the value of the applied tractions on the two surfaces $S_T$ and $S_B$:
\be
\sigma_{zz}^T=\tau_z^a+\tau_z^s=\tau_z^T,\ \ \sigma_{zz}^B=\tau_z^a-\tau_z^s=-\tau_z^B,
\ee
and its variation, cubic, is shown in Fig. \ref{fig:f5_0}; the maximum value is get either on $S_T$ or $S_B$ and it is equal to the applied traction, which is, usually, far below  the value of the in-plane stresses $\sigma_{\alpha\beta},\alpha,\beta\in\{1,2\}$. This considerations better justify the assumption  $\sigma_{zz}=0$  everywhere in the plate. 

\subsection{Boundary conditions}
As already said,  plates are  intrinsically hyperstatic bodies, i.e. the equilibrium equations (\ref{eq:membraneeq}) and (\ref{eq:2ndorderbendeq}) are not sufficient to determine all the static unknowns. We need hence to introduce in the computation not only the balance equations, but also the constitutive law, i.e. we need to make use of the {\it elastic} equilibrium equations (\ref{eq:2ordermembraneeq}) and (\ref{eq:germainlagrange}). These are  partial differential equations of the second and fourth order respectively. The question of how many and which boundary conditions are to be associated to eqs. (\ref{eq:2ordermembraneeq}) and (\ref{eq:germainlagrange}) has needed a long period to be solved and the arguments that lead to the solution are detailed below.
\begin{figure}[t]
\begin{center}
\includegraphics[width=.5\linewidth]{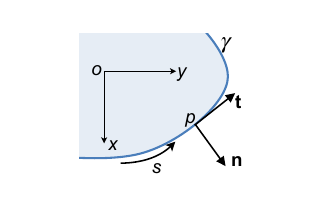}
\includegraphics[width=.45\linewidth]{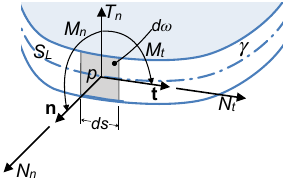}
\caption{Boundary actions.}
\label{fig:f5_5}
\end{center}
\end{figure}
In order to write the boundary conditions, we refer to a generic situation for the boundary of the plate, the lateral surface $S_L$ whose trace in the mid-plane is the curve $\gamma$, see Fig. \ref{fig:f5_5}; at a point $p\in\gamma, \n=(n_x,n_y,0)$ and $\t=(-n_y,n_x,0)$ are the unit vectors of the mid-plane respectively orthogonal and tangent to $\gamma$.

On an infinitesimal surface element $d\omega\in  S_L$ of width $ds$, there are the actions $\N_\n=(N_n,N_{t})$ for extension, $T_n$, for shear, and $\M_\n=(M_n,M_{t})$ for bending, see again Fig. \ref{fig:f5_5}. Because $\N$ and $\M$ are second-rank tensors, we have that
\be
\N_\n=\N\ \n\rightarrow\left\{
\begin{array}{l}
N_n=(\n\otimes\n)\cdot\N=N_xn_x^2+2N_{xy}n_xn_y+N_yn_y^2,\\
N_{t}=(\t\otimes\n)\cdot\N=N_{xy}(n_x^2-n_y^2)+n_xn_y(N_y-N_x),
\end{array}
\right.
\ee
and in the same way
\be
\M_\n=\M\ \n\rightarrow\left\{
\begin{array}{l}
M_n=(\n\otimes\n)\cdot\M=M_xn_x^2+2M_{xy}n_xn_y+M_yn_y^2,\\
M_{t}=(\t\otimes\n)\cdot\M=M_{xy}(n_x^2-n_y^2)+n_xn_y(M_y-M_x).
\end{array}
\right.
\ee
For what concerns $T_n$, it is obtained by simple projection on $\n$ of the shear forces vector $\T=(T_x,T_y)$:
\be
T_n=\T\cdot\n=T_xn_x+T_yn_y.
\ee
In the same way, we can proceed for the displacement vector of the mid-plane, $\bu_0=(u_0,v_0,w_0)$. In particular, we need to project $\bu_0$ on $\n, \t$ and $\e_z$, the unit vector of the axis $z$:
\be
\besp
&u_n=\bu_0\cdot\n=(u_0,v_0,w_0)\cdot(n_x,n_y,0)=u_0n_x+v_0n_y,\\
&u_t=\bu_0\cdot\t=(u_0,v_0,w_0)\cdot(-n_y,n_x,0)=-u_0n_y+v_0n_x,\\
&u_z=\bu_0\cdot\e_z=(u_0,v_0,w_0)\cdot(0,0,1)=w_0.
\end{split}
\ee

We can now write the boundary conditions for any point of the border and let us begin with the boundary conditions for the membrane equations:
\begin{itemize}
\item {\it kinematical} (or {\it geometrical}) conditions:
\be
u_n=\hat{u}_n,\ \ \ u_t=\hat{u}_t;
\ee
\item {\it natural} conditions:
\be
N_n=\hat{F}_n,\ \ \ N_{t}=\hat{F}_t,
\ee
\end{itemize}
where $\hat{u}_n,\hat{u}_t,\hat{F}_n$ and $\hat{F}_t$ are known values.
A list, not exhaustive but comprehending the more important and common cases of boundary conditions for the membrane equations, is given below:
\begin{itemize}
\item supported border: $u_n=0,\ u_t=0$;
\item supported border free to slide along $\t:\ u_n=0,\ N_t=0$;
\item free unloaded border: $N_n=0,\ N_{t}=0$;
\item free loaded border: $N_n=\hat{F}_n,\ N_{t}=\hat{F}_t$.
\end{itemize}

Let us now consider the boundary conditions for the bending equation; now, a point $p\in\gamma$ can have:
\begin{itemize}
\item a displacement along $z$: $w_0$;
\item a rotation about $\t$: $\dfrac{\partial w_0}{\partial \n}$;
\item a rotation about $\n$: $\dfrac{\partial w_0}{\partial \t}$.
\end{itemize}
Statically, the possible actions are:
\begin{itemize}
\item a shear: $T_n$;
\item a bending moment: $M_n$;
\item a twisting moment: $M_t$.
\end{itemize}
The conditions to be written must be specified for each case, let us see the most important ones.
\subsubsection{Simply supported edges}
In this case the boundary conditions are 
\be
\label{eq: bcss}
w_0=0,\ \ \ M_n=0.
\ee
For the case of a border orthogonal to the $x$-axis, i.e. for $\n=\e_x$, 
\be
M_n=M_x=-\frac{h^3}{12}\frac{E}{1-\nu^2}\left(\frac{\partial^2 w_0}{\partial x^2}+\nu\frac{\partial^2 w_0}{\partial y^2}\right),
\ee
so the natural condition becomes
\be
\frac{\partial^2 w_0}{\partial x^2}+\nu\frac{\partial^2 w_0}{\partial y^2}=0.
\ee
Because on such a border $x=const.$ and $w_0=0$, 
\be
\frac{\partial^2 w_0}{\partial y^2}=0,
\ee
and finally the natural boundary condition can be written
\be
\frac{\partial^2 w_0}{\partial x^2}=0.
\ee
For this particular case, originally, Navier proposed the boundary conditions
\be
w_0=0,\ \ \ \Delta w_0=0,
\ee
that actually are equivalent to the above ones. 

\subsubsection{Clamped edges}
For such a case, the boundary conditions are
\be
w_0=0,\ \ \ \dfrac{\partial w_0}{\partial \n}=0.
\ee
To recall that
\be
\dfrac{\partial w_0}{\partial \n}=\nabla w_0\cdot\n=\left(\dfrac{\partial w_0}{\partial x},\dfrac{\partial w_0}{\partial y},\dfrac{\partial w_0}{\partial x}\right)\cdot(n_x,n_y,0)=\dfrac{\partial w_0}{\partial x}n_x+\dfrac{\partial w_0}{\partial y}n_y.
\ee

\subsubsection{Free edges}
According to the approach of Poisson (1829), there are three boundary conditions to be written for a free edge:
\be
T_n=\hat{F}_z,\ \ \ M_n=\hat{M}_n,\ \ \ M_t=\hat{M}_t.
\ee
Actually, this is true also for the two previous cases, where one could write:
\begin{itemize}
\item simply supported edge: \medskip\\$w_0=0\Rightarrow T_n\neq0,\ \ \ \dfrac{\partial w_0}{\partial \t}=0\Rightarrow M_t\neq0,\ \ \ M_n=0\Rightarrow \dfrac{\partial w_0}{\partial \n}\neq0$;
\item clamped edge: \medskip\\$w_0=0\Rightarrow T_n\neq0,\ \ \ \dfrac{\partial w_0}{\partial \t}=0\Rightarrow M_t\neq0,\ \ \ \dfrac{\partial w_0}{\partial \n}=0\Rightarrow M_n\neq0$.
\end{itemize}
However, this approach is {\it false}: in fact, $\dfrac{\partial w_0}{\partial \n}$ and $\dfrac{\partial w_0}{\partial \t}$ are not independent, so the number of boundary conditions cannot be three. The reduction of the number of boundary conditions from 3 to 2, and their correct expression, can be illustrated by a classical procedure due to Kelvin, detailed in the next Section\footnote{The method proposed by Kelvin (1867) to introduce the correct boundary conditions was first established by Kirchhoff through a variational approach (1850).}.

\subsubsection{The Kelvin's reduction of the boundary conditions}
\label{sec:Lordkelvinbc}
\begin{figure}[t]
\begin{center}
\includegraphics[width=.65\linewidth]{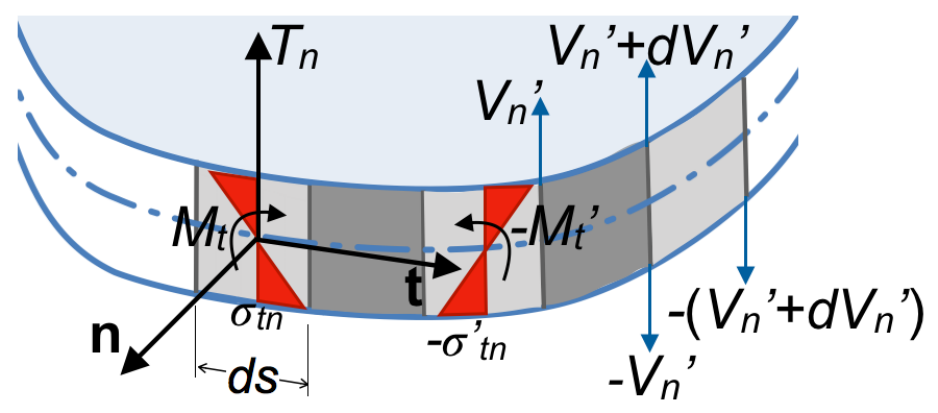}
\caption{Kelvin's reduction of the boundary conditions.}
\label{fig:f5_6}
\end{center}
\end{figure}

Let us imagine of subdividing the plate's edge into a series of parallel stripes of infinitesimal width $ds$, see Fig. \ref{fig:f5_6}. On each stripe acts a shear 
\be
T_n\ ds
\ee
and a twisting moment
\be
M_t\ ds.
\ee
We add, ideally, two equal and opposite couples to each stripe: the first one is the fictitious  couple
\be
-M'_t\ ds,
\ee
result of the shear stress $-\sigma'_{tn}$ and opposite to the real moment $M_t$. The second couple is composed by two vertical, and again fictitious, forces $\pm V'_n$ applied on the edges of the stripe, so as that
\be
V'_n\ ds=M'_t\ ds\Rightarrow V'_n=M'_t.
\ee
On the following stripe, in the sense of $\t$, we add the couple $-(M'_t+dM'_t)ds$ and the forces $\pm(V'_n+dV'_n)ds$, giving a moment equal and opposite to $-(M'_t+dM'_t)ds$, and so on for all the stripes. On the edge shared by two neighbouring stripes, the forces $-V'_n$ and $(V'_n+dV'_n)$ give the net force, per unit length, 
\be
\frac{dV'_n}{ds}=\frac{dM'_t}{ds},
\ee
while the twist moment per unit length is $M_t-M'_t$.
For the Saint Venant's Principle, because the moments $-M'_tds$ and $V'_nds$ are equilibrated, the static and elastic regime of the plate is not affected sufficiently far from the edge (for a distance, say, of the order of the plate's thickness). If now we chose $|M'_t|=|M_t|$, the twisting moment is null on the boundary and it is substituted by a distribution of vertical forces, per unit length, equal to 
\be
\frac{dV_n}{ds}=\frac{dM_t}{ds}.
\ee
These forces, called {\it substitution forces}, are statically equivalent to the twist moment $M_t$. The vertical reaction of a support is hence not $T_n$, but 
\be
T_n^*=T_n+\frac{dM_t}{ds}.
\ee
The force $T_n^*$ is called the {\it Kirchhoff's shear}. Mechanically, this result shows that in the Kirchhoff's model the action of $M_t$ on an edge does not give rise to a deformation, but only to a reaction, in the form of a vertical force.

Finally, the correct boundary conditions are just two and namely:
\begin{itemize}
\item for a free edge:
\be
T_n^*=\hat{F}_z,\ \ \ M_n=\hat{M_n}\ \rightarrow\ w_0\neq0,\ \
\frac{\partial w_0}{\partial \n}\neq0,\ \ \frac{\partial w_0}{\partial \t}\neq0;
\ee
\item for a simply supported edge:
\be
w_0=0,\ \ \ M_n=0\ \rightarrow\ T_n^*\neq 0,
\ \ \frac{\partial w_0}{\partial \n}\neq0,\ \ \frac{\partial w_0}{\partial \t}=0;
\ee
\item for a clamped edge:
\be
w_0=0,\ \ \ \frac{\partial w_0}{\partial \n}=0\ \rightarrow\ T_n^*\neq 0,\ \
M_n\neq0,\ \ \frac{\partial w_0}{\partial \t}=0.
\ee
\end{itemize}
We can also obtain an expression of $T_n^*$ as function of $w_0$: if, e.g., $\n=\e_x$, then
\be
T_n^*=T_x^*=T_x+\frac{\partial M_{xy}}{\partial y}=-\frac{h^3}{12}\frac{E}{1-\nu^2}\left[\frac{\partial^3 w_0}{\partial x^3}+(2-\nu)\frac{\partial^3 w_0}{\partial x\partial y^2}\right];
\ee
in the same way, if $\n=\e_y$, we get
\be
T_n^*=T_y^*=T_y+\frac{\partial M_{xy}}{\partial x}=-\frac{h^3}{12}\frac{E}{1-\nu^2}\left[\frac{\partial^3 w_0}{\partial y^3}+(2-\nu)\frac{\partial^3 w_0}{\partial x^2\partial y}\right].
\ee
Then,
\be
T_n^*=T_x^*n_x+T_y^*n_y.
\ee

\begin{figure}[t]
\begin{center}
\includegraphics[width=.5\linewidth]{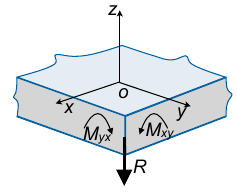}
\caption{The substitution force near a right angle corner.}
\label{fig:f5_7}
\end{center}
\end{figure}

As a last point, we need to consider the case of a discontinuous $M_t$ at some $s=s_0$. In such a case, the substitution force is 
\be
R=\lim_{\varepsilon\rightarrow0}\int_{s_0-\varepsilon}^{s_0+\varepsilon}\frac{dM_t}{ds}\ ds=M_t^+-M_t^-,
\ee
i.e., $R$ is exactly the difference of the twist moment immediately after and before $s_0$. This situation is typical of corners; in particular, for a right angle,
\be
M_t^+=M_t^-\Rightarrow R=-2M_t^-.
\ee
For a corner formed by two orthogonal sides parallel to the axes $x$ and $y$, we get, see Fig. \ref{fig:f5_7}
\be
R=-2|M_{xy}|=-\frac{h^3}{6}\frac{E}{1+\nu}\left|\frac{\partial^2 w_0}{\partial x\partial y}\right|.
\ee
This is a vertical downward reaction that the supports must give near the corners of a rectangular bent plate.

\subsection{Some exact solutions}
\subsubsection{Elliptic plate}

For an elliptic plate loaded by a uniformly distributed load $q$ and clamped on the edge, the deflection is 
\be
w_0(x,y)=\frac{q}{D}\frac{\left(\frac{x^2}{a^2}+\frac{y^2}{b^2}-1\right)^2}{\left(\frac{24}{a^4}+\frac{24}{b^4}+\frac{16}{a^2b^2}\right)},
\ee
where $a$ and $b$ are the semi axes along $x$ and $y$, respectively; the maximum deflection is
\be
w_0^{max}=w_0(0,0)=\frac{q}{D}\frac{1}{\left(\frac{24}{a^4}+\frac{24}{b^4}+\frac{16}{a^2b^2}\right)}.
\ee

\subsubsection{Circular plate}
A clamped, uniformly loaded, circular plate of radius $R$  is a particular case of the previous elliptic plate, where $a=b=R$:
\be
w_0(x,y)=\frac{q}{64\ D}\left(x^2+y^2-R^2\right)^2,
\ee
with a maximum deflection in the plate's center:
\be
w_0^{max}=w(0,0)=\frac{q\ R^4}{64\ D}.
\ee
In polar coordinates it is
\be
w_0(r,\theta)=\frac{q}{64\ D}(R^2-r^2)^2,\ \ 0\leq r\leq R:
\ee
the deflection depends uniquely on the distance $r$ from the plate's center, while it is independent from the angle $\theta$.

\subsubsection{Equilateral triangular plate}
\begin{figure}[t]
\begin{center}
\includegraphics[width=.35\linewidth]{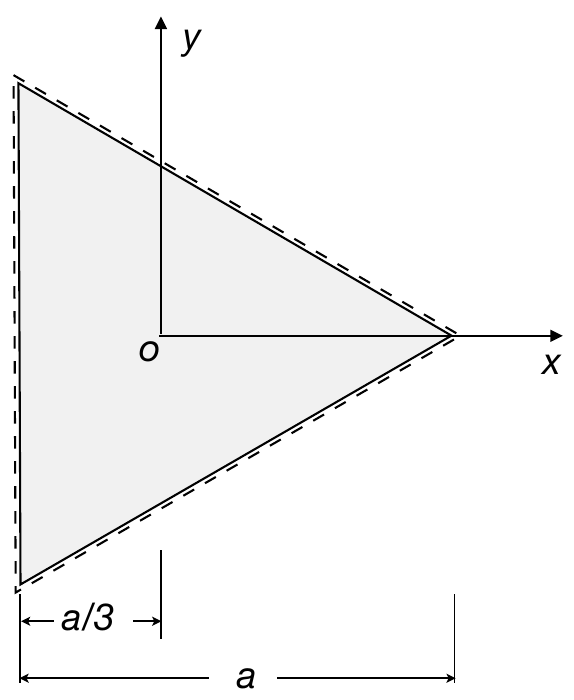}
\caption{Triangular simply supported plate.}
\label{fig:f5_8}
\end{center}
\end{figure}

For  a plate like that in Fig. \ref{fig:f5_8}, uniformly loaded by a load $q$ and simply supported on the edges, it is
\be
w_0(x,y)=\frac{q}{64\ a\ D}\left[x^3-3x^2y-a(x^2+y^2)+\frac{4}{27}a^3\right]\left(\frac{4}{9}a^2-x^2-y^2\right).
\ee

\subsubsection{Anticlastic bending}
\begin{figure}[h]
\begin{center}
\includegraphics[width=.45\linewidth]{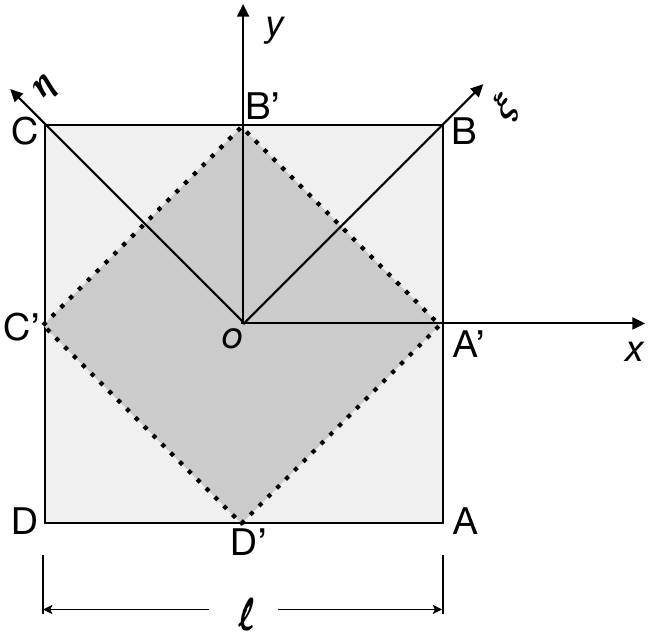}
\caption{Anticlastic bending.}
\label{fig:f5_9}
\end{center}
\end{figure}

\begin{figure}[h]
\begin{center}
\includegraphics[width=.8\linewidth]{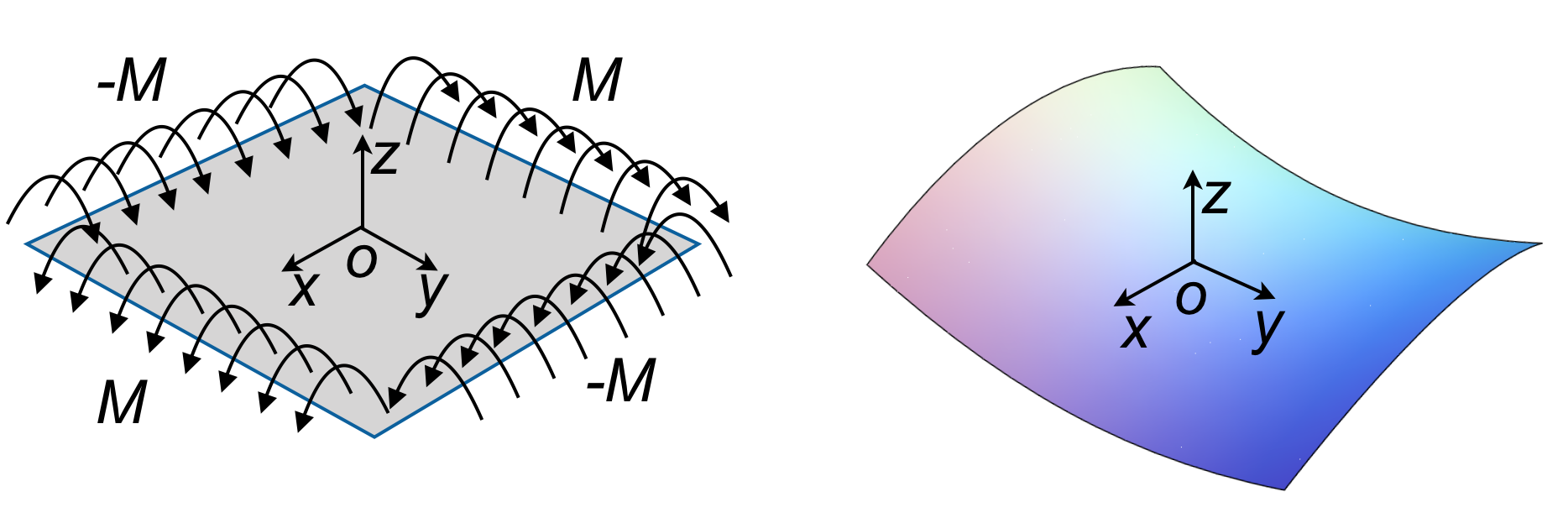}
\caption{Anticlastic bending: actions on the boundary and bent plate.}
\label{fig:f5_10}
\end{center}
\end{figure}

\begin{figure}[h]
\begin{center}
\includegraphics[width=\linewidth]{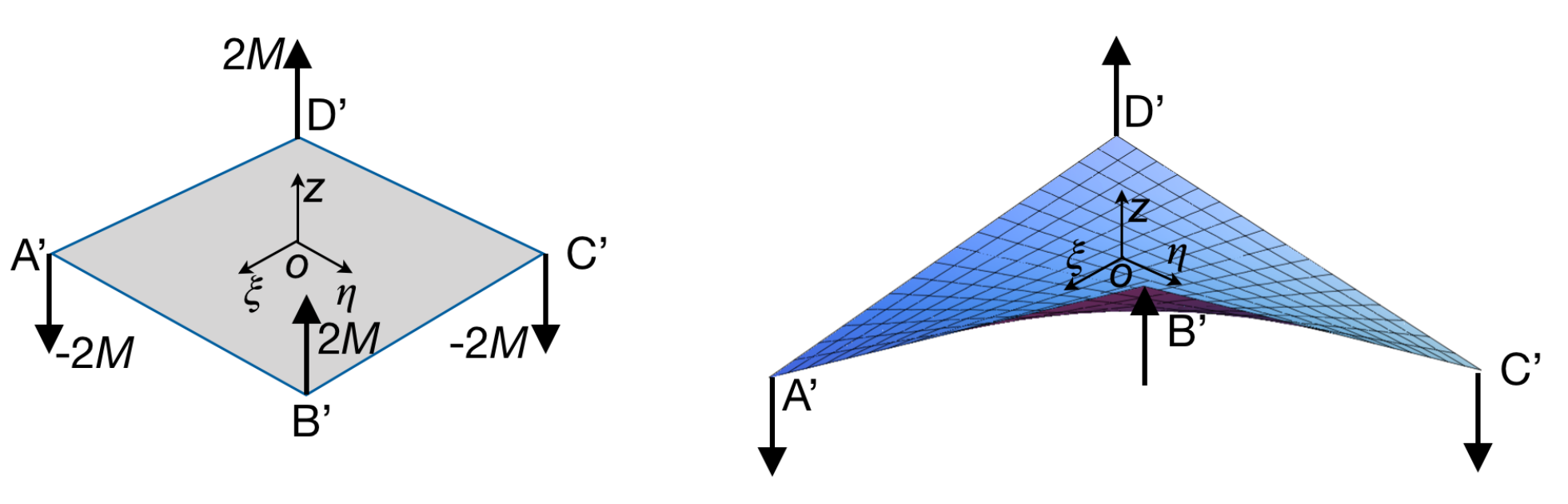}
\caption{Anticlastic bending: actions on the corners and bent plate.}
\label{fig:f5_11}
\end{center}
\end{figure}

Let us consider the square plate ABCD in Fig. \ref{fig:f5_9}; we want to examine to which load and boundary conditions corresponds the solution
\be
\label{eq:anticlastic}
w_0(x,y)=-\frac{M(x^2-y^2)}{2D(1-\nu)}.
\ee
Because
\be
\frac{\partial^2w_0}{\partial x^2}=-\frac{\partial^2w_0}{\partial y^2}=-\frac{M}{D(1-\nu)},
\ee
then
\be
\Delta w_0=0\Rightarrow\Delta^2w_0=0\rightarrow p=0,
\ee
i.e. there is not a distributed load on the plate. We remark that for $y=\pm x$, i.e. along the square's diagonals, $w_0=0$: the diagonals do not move nor deform. Also,
\be
\frac{\partial w_0}{\partial \e}=\nabla w_0\cdot\e=-\frac{M}{D(1-\nu)}(x,-y)\cdot(e_x,e_y).
\ee
So, if $\e$ is parallel to one of the diagonals, $\e=\frac{1}{\sqrt{2}}(1,\pm1)\Rightarrow$
\be
\frac{\partial w_0}{\partial \e}=\nabla w_0\cdot\e=-\frac{M}{\sqrt{2}D(1-\nu)}(x\mp y),
\ee
and on a line parallel to $\e,\ y=\pm x+a$, $a=const$, we get
\be
\frac{\partial w_0}{\partial \e}=\nabla w_0\cdot\e=\pm\frac{M\ a}{\sqrt{2}D(1-\nu)}:
\ee
the slope along the segments parallel to the diagonals is constant: such segments remain straight lines and the final bent surface is a ruled surface (i.e. through any point of the plate pass two orthogonal lines entirely belonging to the deformed surface). Then,
\be
\besp
&M_x=-D\left(\frac{\partial^2 w_0}{\partial x^2}+\nu\frac{\partial^2 w_0}{\partial y^2}\right)=M,\medskip\\
&M_y=-D\left(\frac{\partial^2 w_0}{\partial y^2}+\nu\frac{\partial^2 w_0}{\partial x^2}\right)=-M,\medskip\\
&M_{xy}=-D({1-\nu})\frac{\partial^2 w_0}{\partial x\partial y}=0,
\end{split}
\ee
From eq. (\ref{eq:anticlastic}) we see that lines parallel to axis $x$ deform into a concave parabola, while those parallel to axis $y$ into a convex one. In addition, the edges of the plate are not fixed, as it is easily checked:
\be
\besp
&w_0\left(\pm\frac{\ell}{2},y\right)\neq0, \ \ w_0\left(x,\pm\frac{\ell}{2}\right)\neq0,\\ 
&\left.\frac{\partial w_0}{\partial \n}\right|_{x=\pm\frac{\ell}{2}}=\left.\pm\frac{\partial w_0}{\partial x}\right|_{x=\pm\frac{\ell}{2}}\neq0,\\ &\left.\frac{\partial w_0}{\partial \n}\right|_{y=\pm\frac{\ell}{2}}=\left.\pm\frac{\partial w_0}{\partial y}\right|_{y=\pm\frac{\ell}{2}}\neq0,
\end{split}
\ee
and 
\be
\besp
&T_x^*=T_x+\frac{\partial M_{xy}}{\partial y}=T_x=-D\Delta w_{0,x}=0,\\
&T_y^*=T_y+\frac{\partial M_{xy}}{\partial x}=T_y=-D\Delta w_{0,y}=0.
\end{split}
\ee
So, finally, the only way to deform the plate is to act upon it with distributed bending couples on the edges: $M$ on $x=\pm\frac{\ell}{2}$, $-M$ on $y=\pm\frac{\ell}{2}$, like in Fig. \ref{fig:f5_10}.
Now, let us consider the part A'B'C'D' of the plate shaded in Fig. \ref{fig:f5_9}; $w_0$ is still the same, but now the boundary is different. Nevertheless, because the boundary is parallel to the diagonals, we already know that it remains straight. As a consequence, bending moments are null along these lines, but not necessarily the twisting moments. In fact, because the unit normal to the boundary
of the shaded part is $\n=\frac{1}{\sqrt{2}}(\pm1,\pm1)$, we get
\be
\begin{array}{l}
M_n=M_xn_x^2+2M_{xy}n_xn_y+M_yn_y^2=M(n_x^2-n_y^2)=0,\medskip\\
M_{t}=M_{xy}(n_x^2-n_y^2)+n_xn_y(M_y-M_x)=-2M\ n_xn_y=\pm M,\medskip\\
T_n=T_xn_x+T_yn_y=0.
\end{array}
\ee
So, in A' and C' we have a net vertical force $R=-2M$ while in B' and D' $R=2M$; finally, the plate A'B'C'D' is like in Fig. \ref{fig:f5_11}. For ending, we remark that the same problem can be studied in the frame $\{\xi,\eta\}$, Fig. \ref{fig:f5_9}. Because $x=\xi\cos\theta-\eta\sin\theta,y=\xi\sin\theta+\eta\cos\theta$, we get
\be
w_0(\xi,\eta)=\frac{M\ \xi\ \eta}{D(1-\nu)}.
\ee
This problem is called the {\it anticlastic bending} of plates. It is currently used in laboratory tests, because, on one hand, it is rather easy to be realized, on the other hand, the stress and strain fields are homogeneous, so allowing reliable measures.

\subsubsection{Navier's method for rectangular plates}
Let us consider a rectangular simply supported plate, $0\leq x\leq a$, $0\leq y\leq b$. If the plate is loaded by a distributed load of the type
\be
q(x,y)=q_{mn}\sin\frac{m\pi x}{a}\sin\frac{{n\pi y}}{b},
\ee
then it is easy to check that the solution $w_0(x,y)$ that satisfies both the field equation (\ref{eq:germainlagrange}) and the boundary conditions (\ref{eq: bcss}) is 
\be
w_0(x,y)=w_{mn}\sin\frac{m\pi x}{a}\sin\frac{{n\pi y}}{b}.
\ee
Because
\be
\Delta^2w_0=\pi^4\left(\frac{m^2}{a^2}+\frac{n^2}{b^2}\right)^2w_{mn}\sin\frac{m\pi x}{a}\sin\frac{{n\pi y}}{b},
\ee
we get that $\forall m,n$ it must be
\be
\pi^4\left(\frac{m^2}{a^2}+\frac{n^2}{b^2}\right)^2w_{mn}=\frac{q_{mn}}{D}\Rightarrow w_{mn}=\frac{q_{mn}}{\pi^4\left(\frac{m^2}{a^2}+\frac{n^2}{b^2}\right)^2D}.
\ee
As any load function can be expressed by a double Fourier series,
\be
q(x,y)=\sum_{m=0}^\infty\sum_{n=0}^\infty q_{mn}\sin\frac{m\pi x}{a}\sin\frac{{n\pi y}}{b},
\ee
with
\be
q_{mn}=\frac{4}{ab}\int_0^a\int_0^bq(x,y)\sin\frac{m\pi x}{a}\sin\frac{{n\pi y}}{b}\ dx\ dy,
\ee
a solution for a generic load can be obtained superposing the solutions $w_{mn}$ of each harmonic. In particular, for $q(x,y)=q=const$, we get
\be
q_{mn}=\frac{16\ q}{\pi^2mn},\ \ \ m,n=1,3,5,...
\ee
The double series converges quickly for $w_0$ but slowly for $\M$.% Also, it is worth noting that to represent correctly a load, it is often needed to use  a large number fo terms in the series describing $q(x,y)$.

\subsubsection{Levy's method for rectangular plates}
This method for simply supported plates $0\leq x\leq a$, $-\dfrac{b}{2}\leq y\leq\dfrac{b}{2}$, makes use of simple series: the deflection $w_0$ is expressed as
\be
w_0(x,y)=w_g(x,y)+w_p(x,y),
\ee
where $w_g(x,y)$ is the general integral of the homogeneous equation
\be
\label{eq:intgen}
\Delta^2w_0=0,
\ee
while $w_p(x,y)$ is a particular integral. Generally speaking, 
\be
w_g(x,y)=\sum_{n=1}^\infty Y_n(y)\sin a_nx,\ \ \ a_n=\frac{n\pi}{a}.
\ee
This expression is a solution of (\ref{eq:intgen}) if and only if
\be
Y^{iv}_n(y)-2a_n^2Y''_n(y)+a_n^4Y_n(y)=0\ \ \forall n.
\ee
The general integral of this equation is 
\be
Y_n(y)=A_n\sinh a_ny+B_n\cosh a_ny+C_na_ny\sinh a_ny+D_na_ny\cosh a_ny.
\ee
The particular integral $w_p(x,y)$ depends on the load; for a uniformly distributed load $q$,
\be
w_p(x,y)=\frac{q}{24D}(x^4-2ax^3+a^3x)=\frac{4qa^4}{\pi^5D}\sum_{n=1}^\infty\frac{1}{n^5}\sin a_nx,\ \ \ n=1,3,5,...
\ee
It can be easily checked that $w_g(x,y)$ and $w_p(x,y)$ satisfy the boundary conditions for $x=0$ and $x=a$. In addition, for a uniform load only the second and third terms are not null, because for symmetry it must be $Y_n(y)=Y_n(-y)$. In such a case we get
\be
w_0(x,y)=\frac{qa^4}{D}\sum_{n=1}^\infty\left(\frac{4}{n^5\pi^5}+\xi_n\cosh a_ny+\eta_n\frac{n\pi y}{a}\sinh a_ny\right)\sin\frac{n\pi x}{a},\ \ n=1,3,5,...
\ee
The coefficients $\xi_n$ and $\eta_n$ are determined by the boundary conditions:
\be
w_0(x,\pm\frac{b}{2})=0,\ \ \left.\frac{\partial^2 w_0}{\partial y^2}\right|_{y=\pm\frac{b}{2}}=0,
\ee
that give
\be
\xi_n=-\frac{4+2\varphi_n\tanh\varphi_n}{n^5\pi^5\cosh\varphi_n},\ \ \ \eta_n=\frac{2}{n^5\pi^5\cosh\varphi_n},\ \ \ \varphi_n=\frac{n\pi b}{2a}.
\ee
This series is very quickly convergent, practically a very good estimation for $w_0$ is obtained just with the first term; for $\M$ more terms are needed, but the convergence is still rapid.

\section{The Reissner-Mindlin theory}

The main problem with the Kirchhoff's theory is the absence of transverse shear strains, due to the second kinematical assumption of the model, see Sect. \ref{sec:displkirch}, which engenders the vanishing of the transverse shear stresses\footnote{We have seen that, with the assumption of symmetric loads distribution, these can be calculated using the equilibrium equations}. It seems hence obvious, in order to let appear transverse stresses $\sigma_{xz}$ and $\sigma_{yz}$, to remove the second kinematical assumption of conservation of the orthogonality of the normal segments. This has been done independently, and following two different approaches, by Reissner (1945) and Mindlin (1951). 

Removing the second Kirchhoff's kinematical assumption means that a segment originally orthogonal to the mid-plane will be no more 
orthogonal to the deformed mid-surface. However, we preserve the first and third kinematical assumptions of Kirchhoff: any orthogonal segment remains straight and of the same length.

\begin{figure}[t]
\begin{center}
\includegraphics[width=\linewidth]{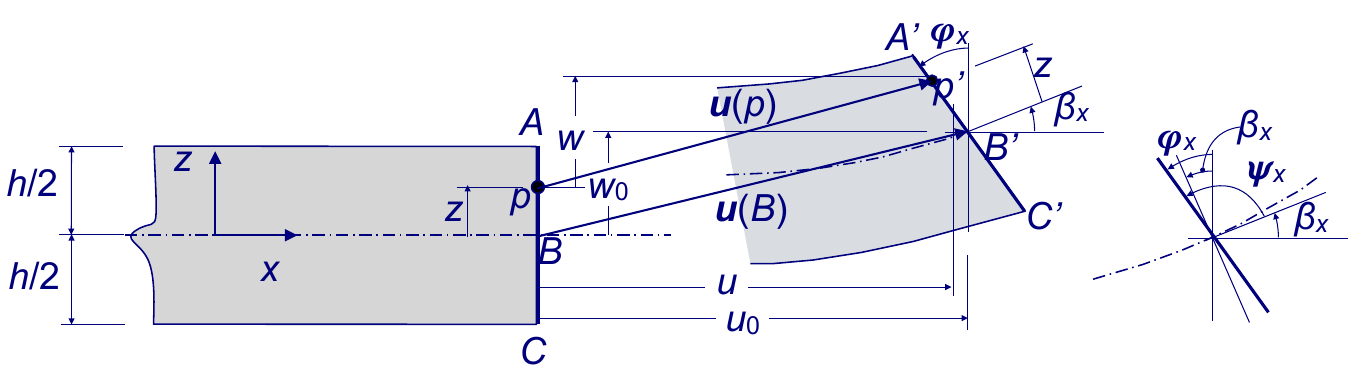}
\caption{Reissner-Mindlin's kinematics in the $\{x,z\}$ plane.}
\label{fig:f5_12}
\end{center}
\end{figure}

Geometrically speaking, this means that in the plane $\{x,z\}$, see Fig. \ref{fig:f5_12}, it will be
\be
\psi_x=\frac{\pi}{2}+\varphi_x-\beta_x.
\ee
The shear strain is hence (recall that for the small strain assumption, $\beta_x\simeq\tan\beta_x=\frac{\partial w_0}{\partial x})$
\be
\gamma_x=\frac{\pi}{2}-\psi_x=\beta_x-\varphi_x\simeq\frac{\partial w_0}{\partial x}-\varphi_x.
\ee
Similarly,
\be
\gamma_y=\frac{\pi}{2}-\psi_y=\beta_y-\varphi_y\simeq\frac{\partial w_0}{\partial y}-\varphi_y,
\ee
and the displacement field is then (the small strain assumption is still used)
\be
\bu(x,y,z)=\left(
\begin{array}{c}
u_0(x,y)-z\varphi_x(x,y)\\
v_0(x,y)-z\varphi_y(x,y)\\
w_0(x,y)
\end{array}
\right)
\ee
so that now the strain tensor becomes
\be
\besp
&\beps(p)=\frac{\nabla\bu(p)+\nabla\bu^\top(p)}{2}=\\
&\left[\begin{array}{ccc}
\frac{\partial u_0}{\partial x}-z\frac{\partial \varphi_x}{\partial x} & \frac{1}{2}\left(\frac{\partial u_0}{\partial y}+\frac{\partial v_0}{\partial x}\right)-\frac{1}{2}z\left(\frac{\partial \varphi_x}{\partial y}+\frac{\partial \varphi_y}{\partial x}\right) & \frac{1}{2}\left(\frac{\partial w_0}{\partial x}-\varphi_x\right)\medskip\\
\frac{1}{2}\left(\frac{\partial u_0}{\partial y}+\frac{\partial v_0}{\partial x}\right)-\frac{1}{2}z\left(\frac{\partial \varphi_x}{\partial y}+\frac{\partial \varphi_y}{\partial x}\right)  & \frac{\partial v_0}{\partial y}-z\frac{\partial \varphi_y}{\partial y}  & \frac{1}{2}\left(\frac{\partial w_0}{\partial y}-\varphi_y\right)\medskip \\
\frac{1}{2}\left(\frac{\partial w_0}{\partial x}-\varphi_x\right) & \frac{1}{2}\left(\frac{\partial w_0}{\partial y}-\varphi_y\right) & 0\end{array}\right].
\end{split}
\ee
Hence, we still can write eq. (\ref{eq:sigma1}), but now the definition of the curvatures changes:
\be
\bk=\left\{
\begin{array}{c}
-\dfrac{\partial \varphi_x}{\partial x}\medskip\\
-\dfrac{\partial \varphi_y}{\partial y}\medskip\\
-\left(\dfrac{\partial \varphi_x}{\partial y}+\dfrac{\partial \varphi_y}{\partial x}\right)
\end{array}
\right\}.
\ee
Also, now $\varepsilon_{xz}$ and $\varepsilon_{yz}$ are different from zero. For the planar part, we still can write eq. (\ref{eq:sigma1}), but now in addition we have also the components of the transverse shear
\be
\left\{
\begin{array}{c}
\sigma_{xz}\medskip\\
\sigma_{yz}
\end{array}
\right\}=
\left[
\begin{array}{cc}
G&0\medskip\\0&G
\end{array}
\right]
\left\{
\begin{array}{c}
2\varepsilon_{xz}\medskip\\
2\varepsilon_{yz}
\end{array}
\right\}.
\ee
Hence, with the Reissner-Mindlin theory, we can obtain the transverse shear stresses directly from the constitutive law. However,   we find here a problem of this theory: $\sigma_{xz}$ and $\sigma_{yz}$ are constant through the thickness, because so are $\varepsilon_{xz}$ and $\varepsilon_{yz}$, as a consequence of the assumption that any orthogonal segment remains straight also in the bent plate. This result is mechanically inconsistent: on one hand, we obtain tangential stresses $\sigma_{zx}=\sigma_{xz}$ and $\sigma_{zy}=\sigma_{yz}$ different from zero on $S_T$ and $S_B$, which is correct only if an equal external tangential load is applied. On the other hand, we know, see Sect. \ref{sec:transvstress}, that the correct variation with $z$ of $\sigma_{xz}$ and $\sigma_{yz}$ is parabolic. 

To solve this inconsistency, one should pass to higher order theories (namely, to the Third Order Shear Deformation Theory of Reddy).
In the framework of the Reissner-Mindlin theory, however, a {\it correction function} $\xi(z)$ is then introduced to have a parabolic variation of $\sigma_{xz}$ and $\sigma_{yz}$ through the thickness:
\be
\xi(z)=\tau\left[1-\left(\frac{z}{h/2}\right)^2\right].
\ee
The constant $\tau$ can be determined in different ways. A first one, is to conserve the value of the shear stress in correspondance of $z=0$, which results in 
\be
\tau=\frac{T}{h},
\ee
i.e. $\tau$ corresponds to the mean value of the shear stress produced by the shear force $T$ in the Reissner-Mindlin model without corrections. If, instead, we conserve the shear force $T$, then
\be
\tau=\frac{3}{2}\frac{T}{h}.
\ee
Finally, the choice commonly done is to put
\be
\tau=\frac{5}{4}\frac{T}{h},
\ee
i.e. a mean value: neither the shear stress nor the shear force are conserved. With such a choice for $\tau$ we get
\be
\besp
&\sigma_{xz}=2G\varepsilon_{xz}(x,y)\xi(z)=\frac{5}{4}\frac{E}{2(1+\nu)}\left[1-\left(\frac{z}{h/2}\right)^2\right]\left(\frac{\partial w_0}{\partial x}-\varphi_x\right),\\
&\sigma_{yz}=2G\varepsilon_{yz}(x,y)\xi(z)=\frac{5}{4}\frac{E}{2(1+\nu)}\left[1-\left(\frac{z}{h/2}\right)^2\right]\left(\frac{\partial w_0}{\partial y}-\varphi_y\right),
\end{split}
\ee
which gives
\be
\besp
& T_x=\int_{-\frac{h}{2}}^{+\frac{h}{2}}\sigma_{xz}\ dz=\frac{5}{6}\frac{E\ h}{2(1+\nu)}\left(\frac{\partial w_0}{\partial x}-\varphi_x\right),\\ 
& T_y=\int_{-\frac{h}{2}}^{+\frac{h}{2}}\sigma_{yz}\ dz=\frac{5}{6}\frac{E\ h}{2(1+\nu)}\left(\frac{\partial w_0}{\partial y}-\varphi_y\right).
\end{split}
\ee
The correction factor is hence $\lambda=\frac{5}{6}$ for the shear forces, while $\lambda=\frac{5}{4}$ for the shear stress.

The expression of $\N$ does not change with respect to the Kirchhoff's theory, while for $\M$ we get
\be
\M=\frac{h^3}{12}\mathbb{D}\bk\Rightarrow\
\left\{
\begin{array}{l}
M_x=-\dfrac{h^3}{12}\dfrac{E}{1-\nu^2}\left(\dfrac{\partial \varphi_x}{\partial x}+\nu\dfrac{\partial \varphi_y}{\partial y}\right),\medskip\\
M_y=-\dfrac{h^3}{12}\dfrac{ E}{1-\nu^2}\left(\dfrac{\partial \varphi_y}{\partial y}+\nu\dfrac{\partial \varphi_x}{\partial x}\right),\medskip\\
M_{xy}=-\dfrac{h^3}{12}\dfrac{E}{1+\nu}\left(\dfrac{\partial \varphi_x}{\partial y}+\dfrac{\partial \varphi_y}{\partial x}\right).
\end{array}
\right.
\ee
The equilibrium equations (\ref{eq:membraneeq}), (\ref{eq:sheareq}) and (\ref{eq:bendshearequil}) do not change, of course; passing to a unique, higher order equation for bending equilibrium, like eq. (\ref{eq:2ndorderbendeq}), is now meaningless, because we cannot obtain an equation of elastic equilibrium depending on a unique unknown, $w_0(x,y)$. In fact,  the elastic equilibrium equations  can be obtained, as usual, injecting the expressions of $T_x, T_y, M_x, M_y$ and $M_{xy}$ into the shear and bending equilibrium equations (the extension equilibrium equations (\ref{eq:2ordermembraneeq})  do not change), to obtain
\be
\besp
&\frac{h^2}{6(1-\nu)}\left[\frac{\partial^2 \varphi_x}{\partial x^2}+(1-\nu)\frac{\partial^2 \varphi_x}{\partial y^2}+\frac{\partial^2 \varphi_y}{\partial x\partial y}\right]+\lambda\left(\frac{\partial w_0}{\partial x}-\varphi_x\right)=0,\\
&\frac{h^2}{6(1-\nu)}\left[(1-\nu)\frac{\partial^2 \varphi_y}{\partial x^2}+\frac{\partial^2 \varphi_y}{\partial y^2}+\frac{\partial^2 \varphi_x}{\partial x\partial y}\right]+\lambda\left(\frac{\partial w_0}{\partial y}-\varphi_y\right)=0,\\
&\lambda\frac{E\ h}{2(1+\nu)}\left(\frac{\partial^2 w_0}{\partial x^2}+\frac{\partial^2 w_0}{\partial y^2}-\frac{\partial \varphi_x}{\partial x}-\frac{\partial \varphi_y}{\partial y}\right)+f_z=0.
\end{split}
\ee
These are three second-order coupled partial differential equations; the unknown functions are $w_0(x,y),\varphi_x(x,y)$ and $\varphi_y(x,y)$. Three boundary conditions are hence needed for each point of the boundary; they must prescribe either
\begin{itemize}
\item kinematical conditions: the value of $w_0,\varphi_x,\varphi_y$ (or, more generally, $\varphi_n$ and $\varphi_t$), or:
\item natural conditions: the value of $T_n, M_n, M_t$.
\end{itemize}
Finally, with the Reissner-Mindlin theory there is no more the problem of the boundary conditions found in the Kirchhoff's theory. Actually, this was a consequence of the kinematical constraint imposed by the second Kirchhoff's assumption so that, once this last removed, the problem disappears.

\section{The Von Karman theory}
The assumption of linear strains, done in the two previous theories, fails to account for the so-called {\it membrane effect}, by which the tensions in a deflected plate help to react the applied {\it lateral loads}, i.e. the surface loads orthogonal to the mid-plane.

In order to take into account for such an effect, we introduce a measure of the strain suitable for geometric non-linearities: the {\it Green-Lagrange strain tensor} $\L$:
\be
\L=\frac{1}{2}(\nabla\bu+\nabla\bu^\top+\nabla\bu^\top\nabla\bu);
\ee
if, for the sake of convenience,  we still indicate $\L$ by $\beps$, then the components $\varepsilon_{ij}$ now are\footnote{For the sake of shortness, we indicate partial derivatives by a comma followed by the differentiation variable, e.g. $\frac{\partial u}{\partial x}= u_{,x}$ etc.}:
\be
\besp
&\eps_{xx}=u_{,x}+\frac{1}{2}\left(u_{,x}^2+v_{,x}^2+w_{,x}^2\right),\\
&\eps_{yy}=v_{,y}+\frac{1}{2}\left(u_{,y}^2+v_{,y}^2+w_{,y}^2\right),\\
&\eps_{zz}=w_{,z}+\frac{1}{2}\left(u_{,z}^2+v_{,z}^2+w_{,z}^2\right),\\
&\eps_{xy}=\frac{1}{2}\left(u_{,y}+v_{,x}+u_{,x}u_{,y}+v_{,x}v_{,y}+w_{,x}w_{,y}\right),\\
&\eps_{xz}=\frac{1}{2}\left(u_{,z}+w_{,x}+u_{,x}u_{,z}+v_{,x}v_{,z}+w_{,x}w_{,z}\right),\\
&\eps_{yz}=\frac{1}{2}\left(v_{,z}+w_{,y}+u_{,y}u_{,z}+v_{,y}v_{,z}+w_{,y}w_{,z}\right).
\end{split}
\ee
In the theory of plates of Von Karman (1910), only some of the quadratic terms are retained, namely:
the quadratic terms in $w_{,x}, w_{,y}$ are retained, the other ones, discarded. Von Karman justifies this choice as follows:
\begin{itemize}
\item $u_{,x}^2$ is negligible with respect to $u_{,x}$; the same is true for $v_{,y}^2$ and $w_{,z}^2$;
\item a similar argument is used also for the terms $v_{,y}v_{,z},w_{,y}w_{,z},u_{,z}u_{,x},w_{,z}w_{,x},u_{,x}u_{,y},v_{,x}v_{,y}$;
\item the terms $v_{,x}^2,u_{,y}^2,w_{,y}^2,u_{,z}^2,v_{,z}^2$ are of the same order of the dropped terms;
\item $w_{,x}$ and $w_{,y}$ are the slopes of cross sections of the deformed mid-plane: they can be large compared to the strain components.
\end{itemize}
So, finally:
\be
\besp
&\eps_{xx}=u_{,x}+\frac{1}{2}w_{,x}^2,\\
&\eps_{yy}=v_{,y}+\frac{1}{2}w_{,y}^2,\\
&\eps_{zz}=w_{,z},\\
&\eps_{xy}=\frac{1}{2}\left(u_{,y}+v_{,x}+w_{,x}w_{,y}\right),\\
&\eps_{xz}=\frac{1}{2}\left(u_{,z}+w_{,x}\right),\\
&\eps_{yz}=\frac{1}{2}\left(v_{,z}+w_{,y}\right).
\end{split}
\ee
We introduce again the vector $\bu_0(x,y)$, displacement of the points of the mid-plane and we need to specify how $u,v,w$ vary through the thickness. To this purpose, the Von Karman's theory makes the same kinematical assumptions of the Kirchhoff's theory, so that eq. (\ref{eq:kinemkirch}) still holds and
\be
\label{eq:defvonkar}
\besp
&\eps_{xx}=e_x-z\ w_{0,xx},\\
&\eps_{yy}=e_y-z\ w_{0,yy},\\
&\gamma_{xy}=2\eps_{xy}=e_{xy}-2z\ w_{0,xy},\\
&\eps_{xz}=\eps_{yz}=\eps_{zz}=0,
\end{split}
\ee
where we have put
\be
\besp
&e_x=u_{0,x}+\frac{1}{2}w_{0,x}^2,\\
&e_y=v_{0,y}+\frac{1}{2}w_{0,y}^2,\\
&e_{xy}=u_{0,y}+v_{0,x}+w_{0,x}w_{0,y}.
\end{split}
\ee
Still like in the Kirchhoff's theory, we assume $\sigma_{zz}=0$, while $\sigma_{xz}=\sigma_{yz}=0$ as a consequence of the constitutive law and of the kinematical assumptions, that lead, as in the Kirchhoff's theory, to $\eps_{xz}=\eps_{yz}=0$. Finally, for $\bsig$ we obtain once more the results of the Kirchhoff's theory, eq. (\ref{eq:sigmakirch}), but now the definition of the strain is different.

Let us now follow a variational approach for finding the equilibrium equations (the same approach can be used also for the classical theory of Kirchhoff\footnote{To this purpose, the reader is addressed to the classical text of Langhaar, see the suggested references. In such a text, a presentation of the the variational approach to the boundary conditions is also given.}).

The strain energy density is
\be
\besp
U=\frac{1}{2}\bsig\cdot\beps&=\frac{1}{2}\frac{E}{1-\nu^2}\left[(\eps_{xx}+\nu\eps_{yy})\eps_{xx}+(\eps_{yy}+\nu\eps_{xx})\eps_{yy}+\frac{1-\nu}{2}\gamma_{xy}^2\right]\\
&=\frac{G}{1-\nu}\left(\eps_{xx}^2+\eps_{yy}^2+2\nu\eps_{xx}\eps_{yy}+\frac{1-\nu}{2}\gamma_{xy}^2\right),
\end{split}
\ee
 by consequence, the total strain energy of the plate is
\be
U_p=\int_\Omega U\ d\omega=\int_S\left(\int_{-\frac{h}{2}}^\frac{h}{2}U\ dz\right)dx \ dy,
\ee
where $S$ denotes the mid-surface. Once the integration over $z$ done, we get
\be
U_p=U_m+U_b,
\ee
where
\begin{itemize}
\item $U_m$: membrane strain energy, linear in $h$:
\be
\besp
U_m=\frac{G\ h}{1-\nu}\int_S&\left[\left(u_{0,x}+\frac{w_{0,x}^2}{2}\right)^2+
\left(v_{0,y}+\frac{w_{0,y}^2}{2}\right)^2\right.\\
&+2\nu\left(u_{0,x}+\frac{w_{0,x}^2}{2}\right)\left(v_{0,y}+\frac{w_{0,y}^2}{2}\right)\\
&\left.+\frac{1-\nu}{2}(u_{0,x}+v_{0,y}+w_{0,x}w_{0,y})^2\right]dx;
\end{split}
\ee
\item $U_b$: bending strain energy, cubic in $h$:
\be
U_b=\frac{G\ h^3}{12(1-\nu)}\int_S\left[w_{0,xx}^2+w_{0,yy}^2+2\nu w_{0,xx}w_{0,yy}+2(1-\nu)w_{0,xy}^2\right]dx\ dy.
\ee
\end{itemize}
The potential energy of a lateral load $q(x,y)$ acting on the plate is
\be
U_q=-\int_Sq(x,y)\ w_0(x,y)\ dx\ dy,
\ee
so finally the total potential energy is the functional
\be
V=U_m+U_b+U_q.
\ee
The principle of the minimum total potential energy is then used: the equilibrium configuration is that giving the least value of $V$. We need hence to write the conditions giving the minimum of the quadratic functional $V$ of the two independent variables $x$ and $y$. These conditions are the Euler-Lagrange equations for $V$: they are the equilibrium equations for the plate. To this purpose, we notice that
\be
V=\int_SF(u_0,u_{0,x},u_{0,y};v_0,v_{0,x},v_{0,y};w_0,w_{0,x},w_{0,y},w_{0,xx},w_{0,xy},w_{0,yy};x,y)dx\ dy,
\ee
i.e. $F$ is a functional of three independent functions, $u_0,v_0$ and $w_0$, so we need to write three Euler-Lagrange equations:
\be
\frac{\partial F}{\partial \xi}-\frac{\partial}{\partial x}\frac{\partial F}{\partial \xi_{,x}}-\frac{\partial}{\partial y}\frac{\partial F}{\partial \xi_{,y}}+\frac{\partial^2}{\partial x^2}\frac{\partial F}{\partial \xi_{,xx}}+\frac{\partial^2}{\partial x\partial y}\frac{\partial F}{\partial \xi_{,xy}}+
\frac{\partial^2}{\partial y^2}\frac{\partial F}{\partial \xi_{,yy}}=0,
\ee
where $\xi$ can be either $u_0,v_0$ or $w_0$. Because
\be
\besp
F&=\frac{G\ h}{1-\nu}\left[\left(u_{0,x}+\frac{w_{0,x}^2}{2}\right)^2+
\left(v_{0,y}+\frac{w_{0,y}^2}{2}\right)^2\right.\\
&\left.+2\nu\left(u_{0,x}+\frac{w_{0,x}^2}{2}\right)\left(v_{0,y}+\frac{w_{0,y}^2}{2}\right)+\frac{1-\nu}{2}(u_{0,x}+v_{0,y}+w_{0,x}w_{0,y})^2\right]\\
&+\frac{G\ h^3}{12(1-\nu)}\left[w_{0,xx}^2+w_{0,yy}^2+2\nu w_{0,xx}w_{0,yy}+2(1-\nu)w_{0,xy}^2\right]-q\ w_0,
\end{split}
\ee
we get finally the three equilibrium equations of the plate (the third one is simplified using the first two equations):
\be
\label{eq:equilvonkarm1}
\besp
&(e_x+\nu e_y)_{,x}+\frac{1-\nu}{2}e_{xy,y}=0,\\
&\frac{1-\nu}{2}e_{xy,x}+(e_y+\nu e_x)_{,y}=0,\\
&\Delta^2w_0=\frac{q}{D}+\frac{12}{h^2}\left[(e_x+\nu e_y)w_{0,xx}+(e_y+\nu e_x)w_{0,yy}+(1-\nu)e_{xy}w_{0,xy}\right].
\end{split}
\ee
The internal actions as defined in eq. (\ref{eq:defN}) become now, through eqs. (\ref{eq:sigmakirch}) and (\ref{eq:defvonkar}),
\be
\besp
&N_x=\frac{E\ h}{1-\nu^2}(e_x+\nu e_y),\\
&N_y=\frac{E\ h}{1-\nu^2}(e_y+\nu e_x),\\
&N_{xy}=G\ h\ e_{xy}=\frac{E\ h}{2(1+\nu)}e_{xy}.
\end{split}
\ee
By consequence, eqs. (\ref{eq:equilvonkarm1})$_{1,2}$ can be written in terms of the components of the membrane internal actions; some simple calculations give
\be
\label{eq:equilvonkarm2}
\besp
&N_{x,x}+N_{xy,y}=0,\\
&N_{xy,x}+N_{y,y}=0.
\end{split}
\ee
We find again eqs. (\ref{eq:membraneeq}) for the case $f_x=f_y=0$, i.e. when the loads parallel to the mid-plane are null (an assumption tacitly done in this case, where the only action is the lateral load $q$). The above two extension equilibrium equations can be solved using the technique of the {\it Airy's stress function} $\chi(x,y)$:
\be
N_x=\chi_{,yy},\ \ \ N_y=\chi_{,xx},\ \ \ N_{xy}=-\chi_{,xy}.
\ee 
If $\N$ is related to $\chi(x,y)$ by the previous relations, than eqs. (\ref{eq:equilvonkarm2}) are automatically satisfied: the problem of determining $\N$ is reduced to the that of finding a unique scalar function, $\chi(x,y)$.
For what concerns eq. (\ref{eq:equilvonkarm1})$_{3}$, it becomes
\be
D\Delta^2w_0=q+N_xw_{0,xx}+N_yw_{0,yy}+2N_{xy}w_{0,xy},
\ee
and introducing $\chi$,
\be
\label{eq:equilvonkarm3}
D\Delta^2w_0=q+\chi_{,yy}w_{0,xx}+\chi_{,xx}w_{0,yy}-2\chi_{,xy}w_{0,xy}.
\ee
A second relation between $w_0(x,y)$ and $\chi(x,y)$ can be easily obtained: from the equations of $N_x,N_y$ and $N_{xy}$ we get
\be
\besp
&h\ e_x=\frac{1}{E}(N_x-\nu N_y)=\frac{1}{E}(\chi_{,yy}-\nu \chi_{,xx}),\\
&h\ e_y=\frac{1}{E}(N_y-\nu N_x)=\frac{1}{E}(\chi_{,xx}-\nu \chi_{,yy}),\\
&h\ e_{xy}=\frac{N_{xy}}{G}=-\frac{\chi_{,xy}}{G},
\end{split}
\ee
while
\be
e_{x,yy}+e_{y,xx}-e_{xy,xy}=w_{0,xy}^2-w_{0,xx}w_{0,yy}.
\ee
Eliminating $e_x,e_y$ and $e_{xy}$ by means of the previous equations gives
\be
\label{eq:equilvonkarm4}
\Delta^2\chi=E\ h(w_{0,xy}^2-w_{0,xx}w_{0,yy}).
\ee
Equations (\ref{eq:equilvonkarm3}) and (\ref{eq:equilvonkarm4}) are the fundamental relations in the Von Karman theory of plates. They reduce the problem to the search of two bi-dimensional functions, $w_0(x,y)$ and $\chi(x,y)$. The use of $\chi$ automatically ensures the in-plane equilibrium and $\N$ is found by derivation, once $\chi$ known.

\newpage
\section{Exercises}
\begin{enumerate}
\item Check that the solution for a circular clamped plate of radius $R$, loaded by a uniform load $q$ is, in polar coordinates,
\begin{equation*}
w(r,\theta)=\frac{q}{64D}(R^2-r^2)^2.
\end{equation*}
Calculate then the deflection in the center and the couples on the boundary.

%\item Find the solution for a circular plate acted upon uniquely by constant normal couples $M_n$ on the boundary.

\item Using the inverse method, study the case of a rectangular plate, whose sides $2a$ and $2b$ are parallel to the axes $x$ and $y$ respectively, with origin at the plate's center, and whose solution is the field of vertical displacements
\begin{equation*}
w=c[(x^2-a^2)^2+(y^2-b^2)^2], \ c<0.
\end{equation*}
In particular, find the conditions on the boundary (displacements and forces) and calculate the deflection and the moments in the plate's center.

\item Study a simply supported rectangular plate of sides $a$ and $b$ subjected to the sinusoidal load
\begin{equation*}
p(x,y)=p_0\sin\frac{\pi\ x}{a}\sin\frac{\pi\ y}{b}.
\end{equation*}
\item Consider a simply supported square plate of side $\ell$, acted upon by a uniform load $p$; compare the results for the  deflection and moments in the plate's center using first the Navier and then the Levy approach. 

\end{enumerate}

\chapter{Membranes}
% !TEX root = modsol.tex
\label{ch:7}
\section{Introduction}
\label{sec:introdchap7}
We call {\it membrane} a solid $\Omega$ that has the form of a surface (i.e., $\Omega$ is delimited by two  outer surfaces, parallel to a  central {\it middle surface} having a distance $h/2$ from each one of the outer surfaces, so that the {\it membrane thickness} $h$ is very small compared to the other dimensions of $\Omega$) and which is able to transmit between its parts {\it exclusively forces tangential to the middle surface}. Hence, {\it  a membrane cannot transmit couples or forces perpendicular to the middle surface% and, as a consequence, bending and shear strain energy are null everywhere in $\Omega$
}.

This definition, geometrical and mechanical, states that:
\begin{enumerate}[i.]
\item a membrane is a {\it two-dimensional solid}, i.e. it can be represented by its middle surface, not necessarily flat nor of a precise form: a membrane can be either a solid whose form is determined by the applied loads and constraints, or a solid whose form is perfectly defined; %in all the cases, its study can be reduced to that of its middle surface;
%\item a membrane does not stock energy in the form of bending energy; this can happen:
%\begin{enumerate}[a.]
\item when the membrane has not a precise form, % it has not flexural stiffness and cannot withstand compressive forces, 
 e.g. a tissue, a cell membrane etc. (all these bodies have a null bending stiffness  and cannot withstand compressive forces), then it is %in this case, only efforts in the tangent plane  can be transmitted by the membrane, which is hence a 
the two-dimensional equivalent of cables and like cables, the equilibrium configuration depends upon the applied loads; the equilibrium problem is in this case {\it  intrinsically nonlinear}, because the equilibrium equations must be written in the deformed, unknown, configuration;
\item they can be studied as membranes also two-dimensional bodies having a precise form and a bending stiffness, like plates and shells, when they are submitted to loads that produce   deformations  only in the tangent plane and with bending strains everywhere null%, so that deformation curvatures are everywhere null and the solid transmits only extension forces
; this is the case, e.g., of plates loaded by in-plane forces, a case already treated in Chap. \ref{ch:6}, but also of domes submitted to their own weight, where the membrane regime is, normally, the prevailing one.
%\end{enumerate}
\end{enumerate}

In two practical cases the regime of membrane is particularly important: 
\begin{itemize}
\item for tissues; in such a case the membrane can transmit only tensile forces and  the impossibility of compressive forces can produce  {\it wrinkling}, a local instability  phenomenon that can be avoided prestressing the membrane;
\item for masonry-like structures, e.g. walls, vaults, domes; this is the opposite case of tissues: the structure can practically transmit only compressive forces, because tensile strength is so small and unreliable that often it can be considered as null, and the presence of tensile internal actions produce cracks, the mechanical equivalent of wrinkles in this case.
\end{itemize}

We consider in the following different problems concerning membranes. 

\section{Linear theory of prestressed plane membranes}
\begin{figure}[th]
\begin{center}
\includegraphics[width=.6\textwidth]{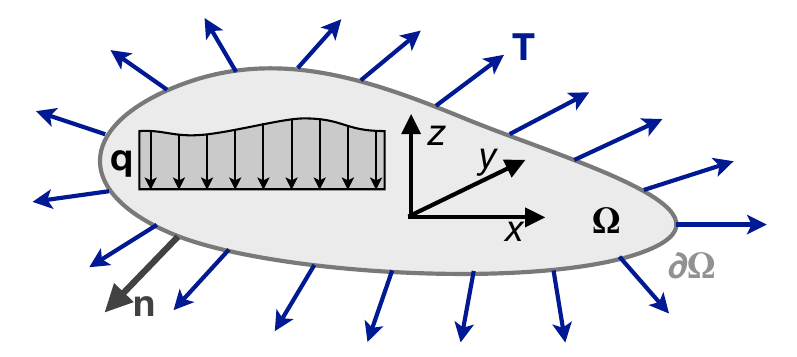}
\caption{General sketch of a prestressed membrane.}
\label{fig:f7_1}
\end{center}
\end{figure}

% dal libro di Villaggio e dagli appunti di dottorato, corso di Villaggio
We consider the case of a flat membrane $\Omega$ that has not flexural stiffness, e.g. a tissue or a drumhead, see Fig. \ref{fig:f7_1}. The membrane belongs to the plane $x-y$, its boundary is $\partial\Omega$ and $\n (p),\ \forall p\in\partial\Omega$, is the outward normal to $\partial\Omega$, $\n=(n_x,n_y,0)$; $\{\ex,\ey,\ez\}$ is the orthonormal basis associated with the frame $\{x,y,z\}$ and $\bu(p)=(u,v,w)$, $p\in\Omega$, is the displacement vector. The membrane is acted upon by :
\begin{itemize}
\item a distributed load $q(p)\bf \ez\ \forall p\in\Omega$;
\item a distribution of tensile forces 
{\bf T}(p)=$T_0\ \n (p)\ \forall p\in\partial\Omega,\ T_0>0$.
\end{itemize}

We first remark that the membrane cannot find its equilibrium in the undeformed configuration, i.e., it cannot remain plane, whenever the lateral load $q(p)\neq0$. In fact, because the only internal actions are the so-called {\it membrane forces}, i.e. forces laying in the tangent plane to the membrane $\forall p\in\Omega$, to equilibrate a lateral load $q(p) \ez$, the membrane cannot remain in the plane $x-y$ and the equilibrium equations must be written in the deformed configuration: the equilibrium problem is intrinsically nonlinear, as already said.

Nonetheless, a linear theory of prestressed plane membranes can be established if we assume that:
\begin{itemize}
\item the vertical displacement $w$ is so small compared to any in-plane dimension of $\Omega$ to let us consider $q(p)\ez$ as unchanged, i.e. independent from the equilibrium configuration $\forall p\in\Omega$ also in the case of following forces;
\item the tension $T_0$ is sufficiently large so as to guarantee a tensile stress state $\forall p\in\Omega$ (i.e., the principal stresses  are everywhere positive): wrinkling does not occur;
\item $T_0$ is much larger than any tension stress that can be produced by the lateral load $q(p)$, so that $T(p)\simeq T_0\ \forall p\in\Omega$: the stress state is constant everywhere in $\Omega$;
\item {\it membrane shear forces}, i.e. shear forces in the middle plane of the membrane, are always negligible with respect to $T_0$.
\end{itemize}

\begin{figure}[th]
\begin{center}
\includegraphics[width=.45\textwidth]{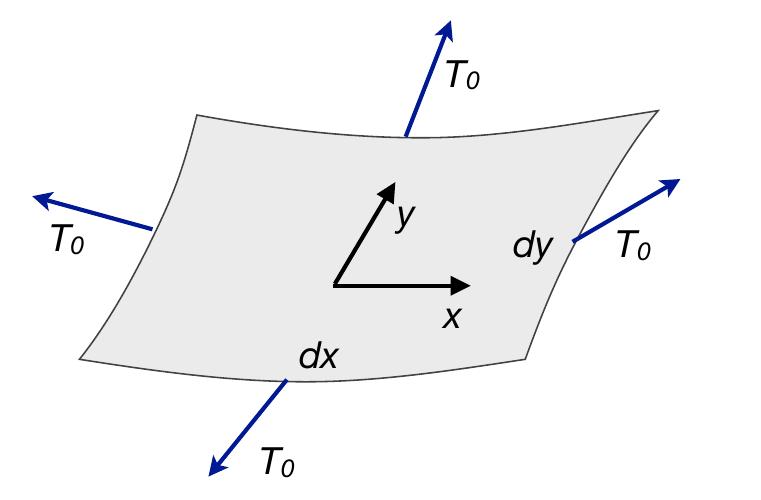}
\includegraphics[width=.45\textwidth]{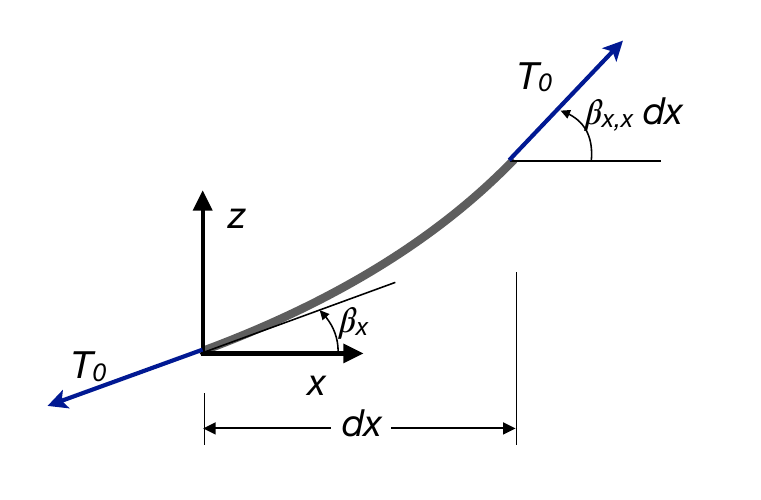}
\caption{Scheme for the equilibrium equation of a prestressed membrane.}
\label{fig:f7_2}
\end{center}
\end{figure}

With these assumptions, let us write the vertical equilibrium equation of a small portion of membrane. Because of the small displacements assumption and for the absence of wrinkles, the deformed surface is regular and with a small slope in any direction. As a consequence, the angles $\beta_x$ and $\beta_y$ that the tangents to the deformed surface form with the axes $x$ and $y$ are small, hence, see Fig. \ref{fig:f7_2}, we get:
\be
\besp
&\beta_x\simeq\sin\beta_x\simeq\tan\beta_x=\frac{\partial w}{\partial x},\\
&\beta_y\simeq\sin\beta_y\simeq\tan\beta_y=\frac{\partial w}{\partial y}.
\end{split}
\ee

So, the equilibrium equation becomes:
\be
\left[T_0\left(\frac{\partial w}{\partial x}+\frac{\partial^2 w}{\partial x^2}dx\right)-T_0\frac{\partial w}{\partial x}\right]dy+\left[T_0\left(\frac{\partial w}{\partial y}+\frac{\partial^2 w}{\partial y^2}dy\right)-T_0\frac{\partial w}{\partial y}\right]dx=-q(x,y)\ dx\ dy,
\ee
giving
\be
\label{eq:prestmemb}
T_0\left(\frac{\partial^2 w}{\partial x^2}+\frac{\partial^2 w}{\partial y^2}\right)=-q(x,y)\ \rightarrow\ \Delta w(x,y)=-\frac{q(x,y)}{T_0}.
\ee

This equation is the two-dimensional corresponding of  the equilibrium equation of taut cables, see e.g. eq. (\ref{eq:suspbridge}). Depending upon the Laplacian of $w$, eq. (\ref{eq:prestmemb}) is a partial differential equation of the elliptic type. 

About the boundary $\partial\Omega$, we need to give some more details: we assume that $\partial\Omega$ is a regular, simple and closed planar curve. So, if $p(t)=(x(t),y(t))$ is a parametric equation of $\partial\Omega,\ 0\leq t\leq\ell$, then we admit that
\be
p(t_1)\neq p(t_2)\ \forall t_1,t_2\in\ ]0,\ell[, \ p(0)=p(\ell),
\ee
i.e. the curve is closed and with no intersecting points. In addition, we assume that $\forall p(t)\in\partial\Omega$, exception made for at most a finite number of points, it is
\be
\dot{x}^2+\dot{y}^2>0.
\ee
This means that $\partial\Omega$ has the tangent defined everywhere but a finite number of points, i.e. polygonal $\Omega$s are allowed.

Finally, the equilibrium problem of a prestressed membrane is in the form of a Dirichlet problem:
\be
\left\{
\besp
&\Delta w=-\frac{q}{T_0}\ \ \mathrm{in}\ \Omega,\\
&w=w_0 \ \ \mathrm{on}\ \partial\Omega.
\end{split}
\right.
\ee

Let us calculate the quantity
\be
\mathcal{E}=\frac{1}{2}\int_\Omega(\nabla w)^2 dx\ dy,
\ee
that represents, to within a scalar factor, the strain energy of the membrane; using integration by parts and the Gauss Theorem, we get successively
\be
\besp
\mathcal{E}&=\frac{1}{2}\int_\Omega(w^2_{,x}+w^2_{,y})\ dx\ dy=\frac{1}{2}\int_\Omega(w_{,x}w)_{,x}+(w_{,y}w)_{,y}\ dx\ dy-\frac{1}{2}\int_\Omega w\ \Delta w\ dx\ dy\\
&=\frac{1}{2}\int_{\partial\Omega}w(w_{,x}n_x+w_{,y}n_y) d\ell-\frac{1}{2}\int_\Omega w\ \Delta w\ dx\ dy\\
&=\frac{1}{2}\int_{\partial\Omega}w\frac{\partial w}{\partial \n}d\ell-\frac{1}{2}\int_\Omega w\ \Delta w\ dx\ dy.
\end{split}
\ee
If we put $w_0=0$ on the boundary, because of eq. (\ref{eq:prestmemb}) we get
\be
\frac{1}{2}\int_\Omega(\nabla w)^2 dx\ dy=\frac{1}{2}\int_\Omega\frac{q\ w}{T_0}\ dx\ dy\ \rightarrow\ \frac{1}{2}\int_\Omega\left[(\nabla w)^2-\frac{q\ w}{T_0}\right]\ dx\ dy=0.
\ee
If $q(x,y)=0$, then the last equation is satisfied $\iff (\nabla w)^2=0\Rightarrow\ w_{,x}=w_{,y}=0\ \forall p\in\Omega\Rightarrow\ w=const.$ everywhere; because $w=0$ on $\partial\Omega$, then $w=0\ \forall p\in\Omega$.

The Laplacian has a maximum principle: if $\Delta w>0$ in $\Omega$, then $w$ cannot have internal maxima. In fact, let us suppose that there is a maximum in $\Omega$; then $w_{,xx}\leq0$ and $w_{,yy}\leq0$ in a neighborhood of the maximum. So, therein, $\Delta w=w_{,xx}+w_{,yy}\leq0$, which contradicts the hypothesis.

In the end, 
\be
w(x,y)\leq\max_{\partial\Omega}w,
\ee
i.e., $w(x,y)$ is everywhere less than the maximum value that $w$ takes on the boundary: all the deformed surface of the membrane is  on one side of the contour plane.

To remark that the principle is not valid for non-quadratic operators, namely for the bi-Laplacian, that holds the bending equilibrium of plates, eq. (\ref{eq:germainlagrange}): for a plate, due to bending, not all the points are necessarily on the same side of the contour plane.

%Another important theoretical result is a consequence of the maximum principle: let us consider the function
%\be
%\varphi=w+\frac{1}{4}\max_\Omega\left|\frac{q}{T_0}\right|(x^2+y^2),
%\ee
%and calculate $\Delta\varphi$:
%\be
%\Delta\varphi=\Delta w+\max_\Omega\left|\frac{q}{T_0}\right|=-\frac{q}{T_0}+\max_\Omega\left|\frac{q}{T_0}\right|%\Rightarrow\Delta\varphi\geq0.
%\ee
%Hence, $\varphi$ satisfies the maximum principle:
%\be
%\varphi\leq\max_{\partial\Omega}w+\frac{1}{4}\max_\Omega\left|\frac{q}{T_0}\right|R^2,\ \ \forall R^2\geq x^2+y^2,\  p=(x,y)\in\partial\Omega.
%\ee
%If now we invert the proceeding, inverting the signs, we can also show that 
%\be
%|w|\leq\max_{\partial\Omega}w+\frac{1}{4}\max_\Omega\left|\frac{q}{T_0}\right|R^2,
%\ee
%i.e., the solution is unique. To remark that we did not use the linearity of the problem to show the uniqueness of the solution.

\section{Curved thin membranes}
\subsection{Internal actions}
\label{sec:internalactionsmembranes}
The  already mentioned {\it membrane forces} are the internal forces laying in the plane tangent to the middle surface; they are the components of the second-rank tensor {\bf N}:
\begin{figure}[th]
\begin{center}
\includegraphics[height=.2\textheight]{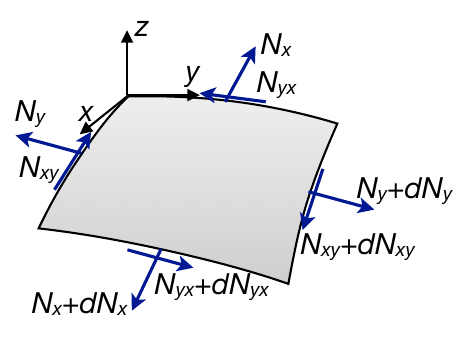}
\includegraphics[height=.2\textheight]{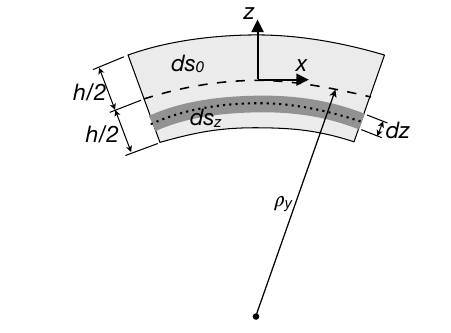}
\caption{Scheme of the internal actions in a curved membrane.}
\label{fig:f7_3}
\end{center}
\end{figure}

\be
\gr{N}=\frac{1}{ds_0}\int_{-\frac{h}{2}}^{\frac{h}{2}}\bsig\ ds_z\ dz,
\ee
and because
\be
ds_z=ds_0\left(1+\frac{z}{\rho}\right),
\ee
we get
\be
\besp
&N_x=\int_{-\frac{h}{2}}^{\frac{h}{2}}\sigma_{xx}\left(1+\frac{z}{\rho_y}\right)\ dz,\ \ \ 
N_y=\int_{-\frac{h}{2}}^{\frac{h}{2}}\sigma_{yy}\left(1+\frac{z}{\rho_x}\right)\ dz,\\
&N_{xy}=\int_{-\frac{h}{2}}^{\frac{h}{2}}\sigma_{xy}\left(1+\frac{z}{\rho_x}\right)\ dz,\ \ 
N_{yx}=\int_{-\frac{h}{2}}^{\frac{h}{2}}\sigma_{yx}\left(1+\frac{z}{\rho_y}\right)\ dz.
\end{split}
\ee
In the above expressions, $\rho_x$ and $\rho_y$ are the curvature radiuses of the middle surface along the directions $x$ and $y$, respectively. It can be noticed that, because in general $\rho_x\neq \rho_y,\ N_{xy}\neq N_{yx}$, though $\sigma_{xy}=\sigma_{yx}$. Nevertheless, if we assume that $h\ll\rho_{min}$, then we obtain, just like for plates,
\be
N_x=\int_{-\frac{h}{2}}^{\frac{h}{2}}\sigma_{xx}\ dz,\ \ \ 
N_y=\int_{-\frac{h}{2}}^{\frac{h}{2}}\sigma_{yy}\ dz,\ \ \
N_{xy}=N_{yx}=\int_{-\frac{h}{2}}^{\frac{h}{2}}\sigma_{xy}\ dz.
\ee
In the following, we will make the above assumption, known as {\it hypothesis of thin membranes}. 

\subsection{Constitutive equation}

Because  $N_x,N_y$ and $N_{xy}$ are different from zero, then $\sigma_{xx},\sigma_{yy},\sigma_{xy}\neq0$; actually, $\sigma_{xz}=\sigma_{yz}=0$ as a consequence of the definition of membrane (absence of transverse shear forces). In addition, we make the assumption that, $\forall p\in\Omega$,
\be
|\sigma_{zz}|\ll\max\{|\sigma_{xx}|,|\sigma_{yy}|,|\sigma_{xy}|\},
\ee 
so that we can put 
\be
\sigma_{zz}=0.
\ee
This assumption can be justified as done  in Sect. \ref{sec:stressplate} for plates. Finally, like for the case of plates, we obtain again a planar stress field:
\be
\bsig=\left[\begin{array}{ccc}
\sigma_{xx} & \sigma_{xy} & 0 \\
\sigma_{xy} & \sigma_{yy} & 0 \\
0 & 0 & 0
\end{array}\right]
\ee
and the constitutive law is still the one find in Sect. \ref{sec:stressplate}:
\be
\label{eq:loimemb}
\left\{\begin{array}{c}\sigma_{xx} \bigskip\\\sigma_{yy} \bigskip\\\sigma_{xy}\end{array}\right\}=
\left[\begin{array}{ccc}
\dfrac{E}{1-\nu^2} & \dfrac{\nu E}{1-\nu^2} & 0 \medskip\\
\dfrac{\nu E}{1-\nu^2} & \dfrac{E}{1-\nu^2} & 0 \medskip\\
0 & 0 & G
\end{array}\right]
\left\{\begin{array}{c}\eps_{xx} \bigskip\\\eps_{yy} \bigskip\\2\eps_{xy}\end{array}\right\},
\ee
or synthetically 
\be
\bsig=\mathbb{D}\ \beps,
\ee
with $\mathbb{D}$ the matrix in eq. (\ref{eq:loimemb}). The components $\eps_{xx},\eps_{yy},\eps_{xy}$ define the deformation of the middle surface. The constitutive law for the membrane can then be obtained easily:
\be
\besp
&N_x=\int_{-\frac{h}{2}}^{\frac{h}{2}}\sigma_{xx}\ dz=\dfrac{E}{1-\nu^2}\int_{-\frac{h}{2}}^{\frac{h}{2}}\eps_{xx}+\nu\eps_{yy}\ dz=\dfrac{E\ h}{1-\nu^2}(\eps_{xx}+\nu\eps_{yy}),\\
&N_y=\int_{-\frac{h}{2}}^{\frac{h}{2}}\sigma_{yy}\ dz=\dfrac{E}{1-\nu^2}\int_{-\frac{h}{2}}^{\frac{h}{2}}\eps_{yy}+\nu\eps_{xx}\ dz=\dfrac{E\ h}{1-\nu^2}(\eps_{yy}+\nu\eps_{xx}),\\
&N_{xy}=N_{yx}=\int_{-\frac{h}{2}}^{\frac{h}{2}}\sigma_{xy}\ dz=2G\int_{-\frac{h}{2}}^{\frac{h}{2}}\eps_{xy}\ dz=2G\ h\ \eps_{xy}=\frac{E\ h}{1+\nu}\eps_{xy}.
\end{split}
\ee

\section{Axisymmetric membranes}
\begin{figure}[th]
\begin{center}
\includegraphics[width=\textwidth]{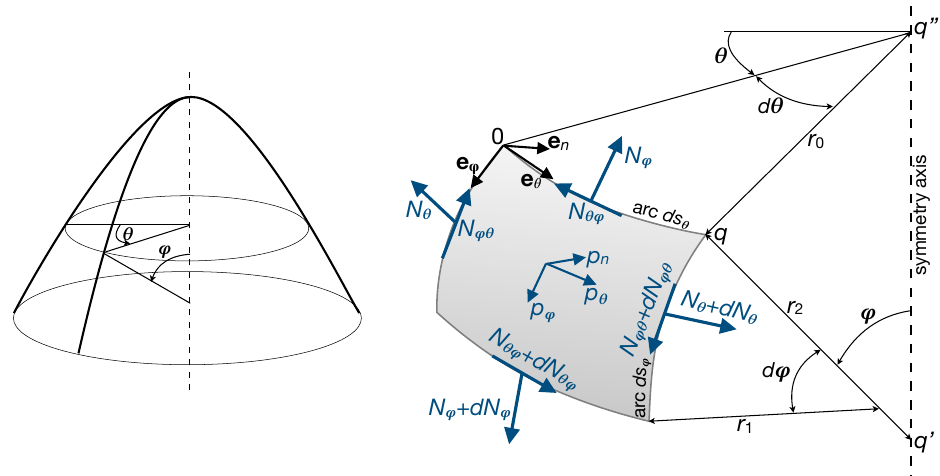}
\caption{Scheme for the equilibrium equations of a  membrane of revolution.}
\label{fig:f7_4}
\end{center}
\end{figure}

Let us now consider a special and rather common case, that of axisymmetric membranes, i.e. in the form of a surface of revolution. The general case is sketched in Fig. \ref{fig:f7_4}.

The elements indicated in Fig. \ref{fig:f7_4} are:
\begin{itemize}
\item $\theta$: longitude angle, measuring the position of a meridian from an arbitrarily fixed direction;
\item $\varphi$: colatitude angle, measuring the angle between the vector $q-q'$ and the symmetry axis;
\item $\{\gr{e}_\theta,\gr{e}_\varphi,\gr{e}_n\}$: positively oriented local basis, with $\gr{e}_\theta$ tangent to the parallel line,$\gr{e}_\varphi$ tangent to the meridian line and $\gr{e}_n$ normal to the membrane surface;
\item $r_1$: curvature radius of the meridian curve;
\item $r_2=|q-q'|$: distance between a point $q$ of the membrane and the symmetry axis, measured along the normal to the membrane at $q$;
\item $r_0=r_2 \sin\varphi$: distance between the point $q$ and the symmetry axis, measured in the plane perpendicular to the symmetry axis;
\item $ds_\theta=r_0d\theta=r_2\sin\varphi\ d\theta$: length of a small arc of parallel;
\item $ds_\varphi=r_1d\varphi$: length of a small arc of meridian;
\item $dA=ds_\theta\ ds_\varphi=r_1r_2\sin\varphi\ d\varphi\ d\theta$: area of a small surface of membrane delimited by two couples of infinitesimally close meridians and parallels;
\item $p_\theta$: component along $\gr{e}_\theta$ of the distributed load {\bf p};
\item $p_\varphi$: component along $\gr{e}_\varphi$ of the distributed load {\bf p};
\item $p_n$: component along $\gr{e}_n$ of the distributed load {\bf p};
\item $N_\varphi$: meridian normal membrane force, per unit length of a parallel arc;
\item $N_\theta$: parallel normal membrane force, per unit length of a meridian arc;
\item $N_{\varphi\theta}$: shear membrane force, per unit length of a meridian arc;
\item $N_{\theta\varphi}$: shear membrane force, per unit length of a parallel arc.
\end{itemize}
In the hypothesis  of  thin membrane, $N_{\theta\varphi}=N_{\varphi\theta}$. 
We write the equilibrium equations for the three directions: $\gr{e}_\varphi$, i.e. tangent to a meridian, $\gr{e}_\theta$, i.e. tangent to a parallel, and $\gr{e}_n$, i.e. normal to the surface. The  forces acting upon the element of surface $dA$  are the membrane forces and the applied forces {\bf p}=$(p_\theta,p_\varphi,p_n)$.

\subsection{Equilibrium along a meridian}
We need to consider all of the force components acting in the direction $\gr{e}_\varphi$, tangent to a meridian line. The resultant of the membrane forces $N_\varphi$ on the upper and lower sides of the surface element $dA$ in Fig. \ref{fig:f7_4}  is\footnote{In the following, we neglect all the infinitesimal terms $o(d\theta d\varphi)$.}
\be
\besp
&(N_\varphi+dN_\varphi)\left(r_0+\dfrac{dr_0}{d\varphi}d\varphi\right)d\theta-N_\varphi r_0d\theta\simeq\left(N_\varphi\frac{dr_0}{d\varphi}d\varphi+dN_\varphi r_0\right)d\theta=\\
&\left(N_\varphi\frac{dr_0}{d\varphi}+\frac{\partial N_\varphi}{\partial\varphi} r_0\right)d\varphi\ d\theta=\frac{\partial}{\partial\varphi}(N_\varphi r_0)d\varphi\ d\theta.
\end{split}
\ee

The resultant of the forces $N_\theta$ on the lateral sides of $dA$ is
\be
\besp
&(N_\theta +   N_\theta+\frac{\partial N_\theta}{\partial \theta}d\theta)\sin\frac{d\theta}{2}  ds_\varphi\simeq(2N_\theta+\frac{\partial N_\theta}{\partial \theta}d\theta)\frac{d\theta}{2}ds_\varphi=\\
&\left(N_\theta\  d\theta+\frac{1}{2}\frac{\partial N_\theta}{\partial \theta}d\theta^2 \right)ds_\varphi\simeq N_\theta\ ds_\varphi\ d\theta;
\end{split}
\ee
this force is horizontal and its component on the tangent to the meridian is
\be
N_\theta\ \cos\varphi\ ds_\varphi\ d\theta=N_\theta\ \cos\varphi\  r_1\ d\varphi\ d\theta.
\ee
\begin{figure}[th]
\begin{center}
\includegraphics[width=.8\textwidth]{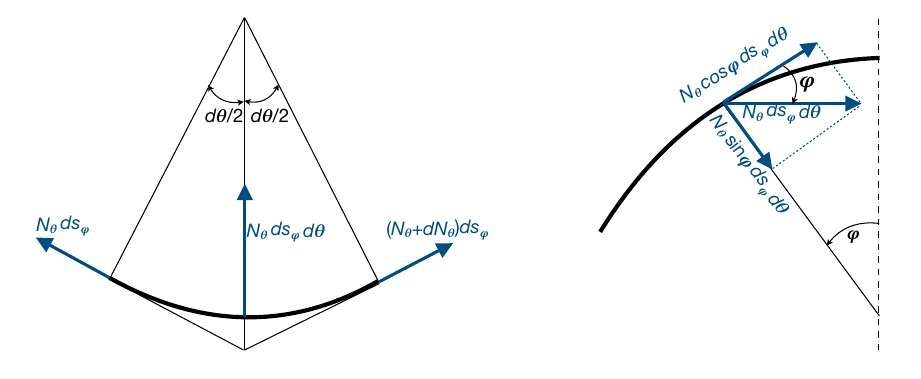}
\caption{Scheme for the forces $N_\theta$.}
\label{fig:f7_5}
\end{center}
\end{figure}

The resultant of the forces $N_{\varphi\theta}$ is
\be
-N_{\varphi\theta}\ ds_\varphi+\left(N_{\varphi\theta}+\frac{\partial N_{\varphi\theta}}{\partial\theta}d\theta\right)ds_\varphi=\frac{\partial N_{\varphi\theta}}{\partial\theta}r_1\ d\varphi\ d\theta,
\ee
while the forces applied to $dA$ are
\be
p_\varphi dA=p_\varphi r_0r_1\ d\varphi\ d\theta.
\ee
Finally, the equilibrium equation of the forces in the direction of the meridian is
\be
\frac{\partial}{\partial\varphi}(N_\varphi r_0)d\varphi\ d\theta-N_\theta\ \cos\varphi\ r_1d\varphi\ d\theta+\frac{\partial N_{\varphi\theta}}{\partial\theta}r_1\ d\varphi\ d\theta+p_\varphi r_0r_1\ d\varphi\ d\theta=0,
\ee
or, because $r_0=r_2 \sin\varphi$,
\be
\frac{\partial}{\partial\varphi}(r_2 N_\varphi \sin\varphi)-r_1\ N_\theta\cos\varphi+r_1\frac{\partial N_{\varphi\theta}}{\partial\theta}=-p_\varphi r_1r_2\sin\varphi.
\ee

\subsection{Equilibrium along a parallel}
\begin{figure}[th]
\begin{center}
\includegraphics[width=\textwidth]{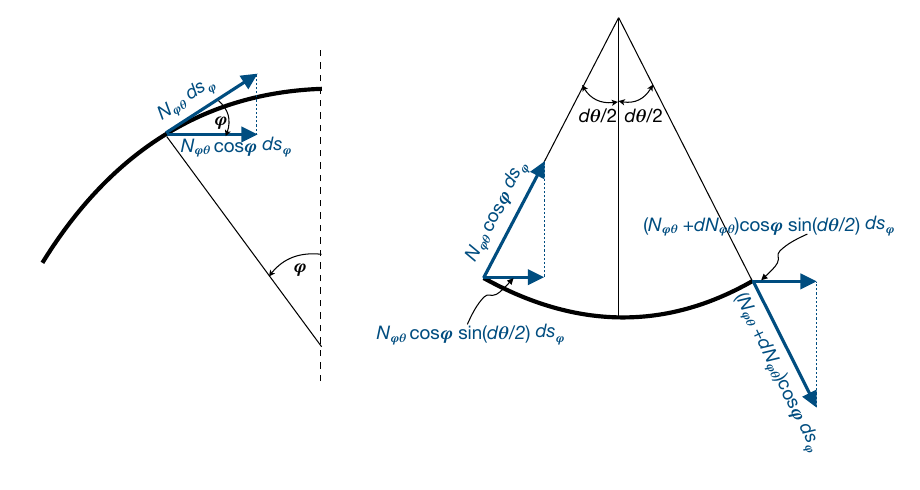}
\caption{Scheme for the forces $N_{\varphi\theta}$.}
\label{fig:f7_6}
\end{center}
\end{figure}

In the direction $\gr{e}_\theta$, along a parallel line, the resultant of the forces $N_\theta$ is, see Fig. \ref{fig:f7_5},
\be
\left(N_\theta+\frac{\partial N_\theta}{\partial \theta}d\theta-N_\theta \right)\cos\frac{d\theta}{2}  ds_\varphi\simeq
\frac{\partial N_\theta}{\partial \theta}ds_\varphi\ d\theta=\frac{\partial N_\theta}{\partial \theta}r_1\ d\varphi\ d\theta,
\ee
while for forces $N_{\theta\varphi}$ it is
\be
\besp
&\left(N_{\theta\varphi}+\frac{\partial N_{\theta\varphi}}{\partial \varphi}d\varphi \right)\left(r_0+\frac{dr_0}{d\varphi}d\varphi\right)d\theta
-N_{\theta\varphi}r_0\ d\theta\simeq\\
&\frac{\partial N_{\theta\varphi}}{\partial \varphi}d\varphi\ r_0\ d\theta+N_{\theta\varphi}\frac{dr_0}{d\varphi}d\varphi\ d\theta=\frac{\partial (r_0N_{\theta\varphi})}{\partial \varphi}d\varphi\ d\theta.
\end{split}
\ee
For the resultant of the forces $N_{\varphi\theta}$ we get, see Fig. \ref{fig:f7_6},
\be
\besp
&\left(N_{\varphi\theta}+N_{\varphi\theta}+\frac{\partial N_{\varphi\theta}}{\partial\theta}d\theta\right)\cos\varphi\sin\frac{d\theta}{2}ds_\varphi\simeq\\
&\left(N_{\varphi\theta}+N_{\varphi\theta}+\frac{\partial N_{\varphi\theta}}{\partial\theta}d\theta\right)\cos\varphi\frac{d\theta}{2}ds_\varphi\simeq
 N_{\varphi\theta}\cos\varphi\ d\theta\  ds_\varphi=N_{\varphi\theta}r_1\cos\varphi\ d\varphi\ d\theta,
\end{split}
\ee
while  the component of the applied loads along a parallel is
\be
p_\theta dA=p_\theta r_0 r_1 d\varphi\ d\theta.
\ee
Finally, the equilibrium equation in the direction of a parallel is
\be
\frac{\partial N_\theta}{\partial \theta}r_1\ d\varphi\ d\theta+\frac{\partial (r_0N_{\theta\varphi})}{\partial \varphi}d\varphi\ d\theta+N_{\varphi\theta}r_1\cos\varphi\ d\varphi\ d\theta+p_\theta r_0 r_1 d\varphi\ d\theta=0,
\ee
i.e., because $r_0=r_2\sin\varphi$, %and for the hypothesis of thin membrane, $N_{\varphi\theta}=N_{\theta\varphi}$,
\be
r_1\frac{\partial N_\theta}{\partial \theta}+\frac{\partial (r_2\sin\varphi N_{\theta\varphi})}{\partial \varphi}+N_{\varphi\theta}r_1\cos\varphi=-p_\theta r_1r_2\sin\varphi.
\ee

\subsection{Equilibrium along the normal to the surface}
\begin{figure}[th]
\begin{center}
\includegraphics[width=.4\textwidth]{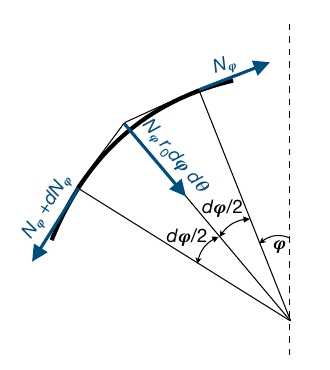}
\caption{Scheme for the forces $N_{\varphi}$.}
\label{fig:f7_7}
\end{center}
\end{figure}

In the direction $\gr{e}_n$, orthogonal to the membrane's surface, the resultant of the forces $N_\varphi$ is, see Fig. \ref{fig:f7_7},
\be
\besp
&\left[ r_0\ N_\varphi+\left(N_\varphi+\frac{\partial N_\varphi}{\partial\varphi}d\varphi\right)\left(r_0+\frac{dr_0}{d\varphi}d\varphi\right)\right]\sin\frac{d\varphi}{2}d\theta\simeq\\
&\left[ r_0\ N_\varphi+\left(N_\varphi+\frac{\partial N_\varphi}{\partial\varphi}d\varphi\right)\left(r_0+\frac{dr_0}{d\varphi}d\varphi\right)\right]\frac{d\varphi}{2}d\theta\simeq N_\varphi r_0d\varphi\ d\theta.
\end{split}
\ee
The component of the resultant of the forces $N_\theta$ along $\gr{e}_\n$ is, see Fig. \ref{fig:f7_5},
\be
N_\theta\ r_1\sin\varphi\ d\varphi\ d\theta,
\ee
while that of the forces $N_{\varphi\theta}$ and $N_{\theta\varphi}$ is null; finally, the component of the applied forces is
\be
p_ndA=p_nr_0\ r_1\ d\varphi\ d\theta.
\ee
The equilibrium equation in the direction of the normal to the membrane's surface is hence
\be
N_\varphi r_0d\varphi\ d\theta+N_\theta\ r_1\sin\varphi\ d\varphi\ d\theta+p_nr_0\ r_1\ d\varphi\ d\theta=0,
\ee
i.e., because $r_0=r_2\sin\varphi$,
\be
N_\varphi r_2\sin\varphi+N_\theta\ r_1\sin\varphi=-p_nr_0\ r_1\ \rightarrow\ \frac{N_\varphi}{r_1}+\frac{N_\theta}{r_2}=-p_n.
\ee

\subsection{The system of equilibrium equations}
The  equilibrium equations are three coupled first-order partial differential equations:
\be
\label{eq:equilmemb1}
\besp
&\frac{\partial}{\partial\varphi}(r_2 N_\varphi \sin\varphi)-r_1\cos\varphi\ N_\theta+r_1\frac{\partial N_{\varphi\theta}}{\partial\theta}=-p_\varphi r_1r_2\sin\varphi,\\
&r_1\frac{\partial N_\theta}{\partial \theta}+\frac{\partial (r_2\sin\varphi N_{\theta\varphi})}{\partial \varphi}+r_1\cos\varphi\ N_{\varphi\theta}=-p_\theta r_1r_2\sin\varphi,\\
&\frac{N_\varphi}{r_1}+\frac{N_\theta}{r_2}=-p_n.
\end{split}
\ee
The four  unknown  functions are $N_\varphi,N_\theta,N_{\varphi\theta}$ and $N_{\theta\varphi}$; in the hypothesis of thin membrane, they reduce to three ($N_{\varphi\theta}=N_{\theta\varphi}$). Hence, in this framework, the equilibrium equations are sufficient to solve the problem: {\it a thin membrane is intrinsically isostatic}.

\subsection{Resolution procedure}
\label{sec:resomemb}
\begin{figure}[th]
\begin{center}
\includegraphics[width=.3\textwidth]{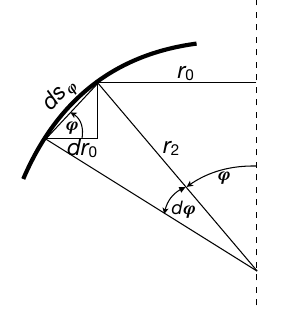}
\caption{Scheme for the geometrical quantities.}
\label{fig:f7_8}
\end{center}
\end{figure}

\begin{figure}[th]
\begin{center}
\includegraphics[height=.4\textwidth]{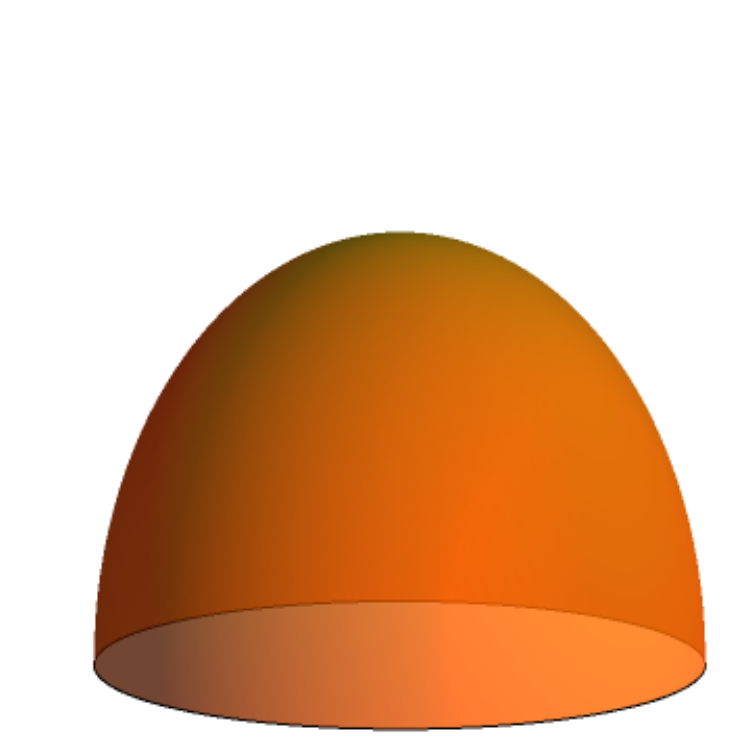}
\includegraphics[height=.3\textwidth]{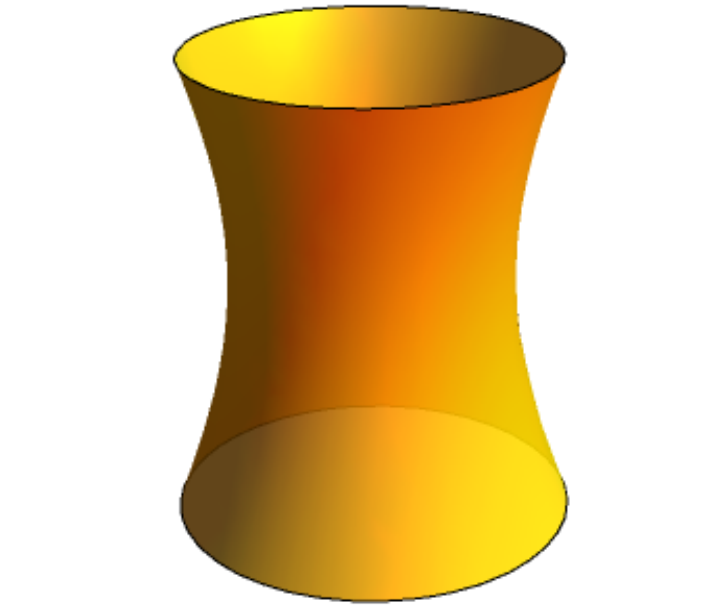}
\caption{Ellipsoidal-like (left) and hyperbolic-like (right) surfaces of revolution.}
\label{fig:f7_9}
\end{center}
\end{figure}

From eq. (\ref{eq:equilmemb1})$_3$ we get
\be
\label{eq:nthetanphi}
N_\theta=-p_nr_2-N_\varphi\frac{r_2}{r_1},
\ee
that injected in eqs. (\ref{eq:equilmemb1})$_{1,2}$ gives
\be
\besp
&\frac{\partial}{\partial\varphi}(r_2\sin\varphi\ N_\varphi)+r_1\frac{\partial N_{\varphi\theta}}{\partial\theta}+r_2\cos\varphi\ N_\varphi=-r_1r_2(p_\varphi\sin\varphi+p_n\cos\varphi),\\
&\frac{\partial}{\partial\varphi}(r_2\sin\varphi\ N_{\varphi\theta})+r_1\cos\varphi\ N_{\varphi\theta}-r_2\frac{\partial N_{\varphi}}{\partial\theta}=-r_1r_2\left(p_\theta\sin\varphi-\frac{\partial p_n}{\partial\theta}\right).
\end{split}
\ee
We introduce now the following change of variables:
\be
\label{eq:fonctUV}
U=r_2\sin^2\varphi\ N_{\varphi},\ \ V=r_2^2\sin^2\varphi\ N_{\varphi\theta},
\ee
that used in the previous equations give
\be
\besp
&\frac{\partial}{\partial\varphi}\left(\frac{U}{\sin\varphi}\right)+r_1\frac{\partial}{\partial\theta}\left(\frac{V}{r_2^2\sin^2\varphi}\right)+\frac{U}{\sin^2\varphi}\cos\varphi=-r_1r_2(p_\varphi\sin\varphi+p_n\cos\varphi),\\
&\frac{\partial}{\partial\varphi}\left(\frac{V}{r_2\sin\varphi}\right)+r_1\cos\varphi \frac{V}{r_2^2\sin^2\varphi}-r_2
\frac{\partial}{\partial\theta}\left(\frac{U}{r_2\sin^2\varphi}\right)=-r_1r_2\left(p_\theta\sin\varphi-\frac{\partial p_n}{\partial\theta}\right),
\end{split}
\ee
and once the derivates done
\be
\besp
\label{eq:equilmemb2}
&\frac{\partial U}{\partial\varphi}\frac{1}{\sin\varphi}-U\frac{\cos\varphi}{\sin^2\varphi}+\frac{r_1}{r_2^2\sin^2\varphi}\frac{\partial V}{\partial\theta} +U\frac{\cos\varphi}{\sin^2\varphi}=-r_1r_2(p_\varphi\sin\varphi+p_n\cos\varphi),\\
&\frac{1}{r_2\sin\varphi}\frac{\partial V}{\partial\varphi}+V\frac{-\frac{d}{d\varphi}(r_2\sin\varphi)}{r_2^2\sin^2\varphi}+
 V\frac{r_1\cos\varphi}{r_2^2\sin^2\varphi}-
\frac{1}{\sin^2\varphi}\frac{\partial U}{\partial\theta}=-r_1r_2\left(p_\theta\sin\varphi-\frac{\partial p_n}{\partial\theta}\right).
\end{split}
\ee
Now, considering that, on the one hand,
\be
dr_0=d(r_2\sin\varphi)
\ee
and that, on the other hand, see Fig. \ref{fig:f7_8},

\be
dr_0=ds_\varphi\cos\varphi=r_1\cos\varphi d\varphi,
\ee
we get
\be
\label{eq:diffmembr}
r_1\cos\varphi=\frac{d}{d\varphi}(r_2\sin\varphi);
\ee
injecting this result into eq. (\ref{eq:equilmemb2})$_2$, after reordering we get:
\be
\besp
\label{eq:equilmemb3}
&\frac{\partial V}{\partial\theta} +\frac{r_2^2}{r_1}\sin\varphi\frac{\partial U}{\partial\varphi}=-r_2^3\sin^2\varphi(p_\varphi\sin\varphi+p_n\cos\varphi),\\
&\frac{\partial V}{\partial\varphi} -\frac{r_2}{\sin\varphi}\frac{\partial U}{\partial\theta}=-r_1r_2^2\sin\varphi\left(p_\theta\sin\varphi-\frac{\partial p_n}{\partial\theta}\right).
\end{split}
\ee
Differentiating the first equation with respect to $\varphi$, the second one with respect to $\theta$ and subtracting we get an equation depending uniquely on $U$
\be
\besp
&\frac{\partial}{\partial\varphi}\left(\frac{r_2^2}{r_1}\sin\varphi\frac{\partial U}{\partial\varphi}\right)+\frac{r_2}{\sin\varphi}\frac{\partial^2U}{\partial\theta^2}=\\
&\frac{\partial}{\partial\varphi}\left(r_2^3\sin^2\varphi(p_n\cos\varphi-p_\varphi\sin\varphi)\right)+r_1r_2^2\sin\varphi\left(\sin\varphi\frac{\partial p_\theta}{\partial\theta}-\frac{\partial^2 p_n}{\partial\theta^2}\right).
\end{split}
\ee
After differentiating and reordering the terms we get finally
\be
\label{eq:equilmemb4}
r_2^2\sin^2 \varphi\ \frac{\partial^2U}{\partial\varphi^2}+r_1r_2\frac{\partial^2U}{\partial\theta^2}+\alpha(\varphi,\theta)\frac{\partial U}{\partial\varphi}=p(\varphi,\theta),
\ee
where\footnote{To obtain the expression of $\alpha(\varphi,\theta)$  we need to consider that
\begin{equation*}
\frac{\partial}{\partial\varphi}\left(\frac{r_2^2}{r_1}\sin\varphi\right)=\frac{1}{r_1^2}\left(r_1r_2\frac{\partial(r_2\sin\varphi)}{\partial\varphi}+r_1r_2\sin\varphi\frac{\partial r_2}{\partial\varphi}-\frac{\partial r_1}{\partial\varphi}r_2^2\sin\varphi\right)=r_2\cos\varphi+\frac{r_2}{r_1}\sin\varphi\frac{\partial r_2}{\partial\varphi}-\frac{r_2^2}{r_1^2}\sin\varphi\frac{\partial r_1}{\partial\varphi},
\end{equation*} 
the last result being obtained thanks to eq. (\ref{eq:diffmembr}).
}
\be
\besp
&\alpha(\varphi,\theta)=r_1r_2\sin\varphi\cos\varphi+r_2\sin^2\varphi\ \frac{\partial r_2}{\partial\varphi}-\frac{r_2^2}{r_1}\sin^2\varphi\frac{\partial r_1}{\partial\varphi},\\
&p(\varphi,\theta)=r_1\sin\varphi\left[\frac{\partial}{\partial\varphi}\left(r_2^3\sin^2\varphi(p_n\cos\varphi-p_\varphi\sin\varphi)\right)+r_1r_2^2\sin\varphi\left(\sin\varphi\frac{\partial p_\theta}{\partial\theta}-\frac{\partial^2 p_n}{\partial\theta^2}\right) \right].
\end{split}
\ee
are two known functions of the geometry and of the loads.
Once the function $U$ obtained as solution of eq. (\ref{eq:equilmemb4}), we obtain $N_\varphi$ from eq. (\ref{eq:fonctUV})$_1$ and then $N_\theta$ from eq. (\ref{eq:nthetanphi}). To obtain $N_{\varphi\theta}$, we differentiate eqs. (\ref{eq:equilmemb3})  with respect to $\theta$ and $\varphi$ respectively and sum up them to obtain
\be
\label{eq:equilmemb5}
\besp
\Delta V=&r_2\frac{\partial^2U}{\partial\varphi\partial\theta}\left(\frac{1}{\sin\varphi}-\frac{r_2}{r_1}\sin\varphi\right)-r_2^3\sin^2\varphi\left(\sin\varphi\frac{\partial p_\varphi}{\partial\theta}+\cos\varphi\frac{\partial p_n}{\partial\theta}\right)-\\
&r_1r_2^2\sin\varphi\left(\sin\varphi\frac{\partial p_\theta}{\partial\theta}-\frac{\partial^2p_n}{\partial\theta^2}\right).
\end{split}
\ee
Once $V$ obtained as solution of the previous equation, it can be injected into eq. (\ref{eq:fonctUV})$_2$ to obtain $N_{\varphi\theta}$.
It is interesting to notice that, while the partial differential equation (\ref{eq:equilmemb5}) is of the elliptic type, because the laplacian $\Delta V=\partial^2 V/\partial \varphi^2+\partial^2 V/\partial \theta^2$ is the typical elliptic operator, eq. (\ref{eq:equilmemb4}) is elliptic if $r_1r_2>0$, while it is hyperbolic if $r_1r_2<0$. The first case is typical of ellipsoidal-like surfaces, while the second one of the hyperbolic hyperboloid-like ones, Fig. \ref{fig:f7_9}.

\subsection{Axisymmetric load}
\label{sec:axisymload}
In case of axisymmetric load, eqs. (\ref{eq:equilmemb3}) can be solved separately, because they are uncoupled: in fact, because nothing depends upon $\theta$, all the derivatives with respect to this variable are null and hence eqs. (\ref{eq:equilmemb1}) reduce to
\be
\label{eq:equilmemb6}
\besp
&\frac{\partial}{\partial\varphi}(r_2 N_\varphi \sin\varphi)-r_1\ N_\theta\cos\varphi=-p_\varphi r_1r_2\sin\varphi,\\
&\frac{\partial (r_2\sin\varphi N_{\theta\varphi})}{\partial \varphi}+r_1\cos\varphi\ N_{\varphi\theta}=-p_\theta r_1r_2\sin\varphi,\\
&\frac{N_\varphi}{r_1}+\frac{N_\theta}{r_2}=-p_n,
\end{split}
\ee
while eqs. (\ref{eq:equilmemb3}) become
\be
\besp
\label{eq:equilmemb7}
&\frac{\partial U}{\partial\varphi}=-r_1r_2\sin\varphi(p_\varphi\sin\varphi+p_n\cos\varphi),\\
&\frac{\partial V}{\partial\varphi}=-r_1r_2^2\sin^2\varphi\ p_\theta.
\end{split}
\ee

Again, after the function $U$ found, i.e. once eq. (\ref{eq:equilmemb7})$_1$ solved, we get from eq. (\ref{eq:fonctUV})$_1$
\be
\label{eq:axisymNphi}
N_{\varphi}=\frac{U}{r_2\sin^2\varphi},
\ee
and, from eq. (\ref{eq:nthetanphi}),
\be
\label{eq:axisymNtheta}
N_\theta=-p_nr_2-\frac{r_2}{r_1}N_\varphi=-p_nr_2-\frac{U}{r_1\sin^2\varphi}.
\ee
For what concerns $N_{\varphi\theta}$, it depends only upon $V$, eq. (\ref{eq:fonctUV})$_2$, hence eq. (\ref{eq:equilmemb7})$_2$ describes the tangential internal forces. If $p_\theta=0$, i.e. if the tangential forces along the parallels are null, then $V=0\Rightarrow N_{\varphi\theta}=0$: the shear forces are everywhere null in the membrane.
\begin{figure}[th]
\begin{center}
\includegraphics[height=.3\textwidth]{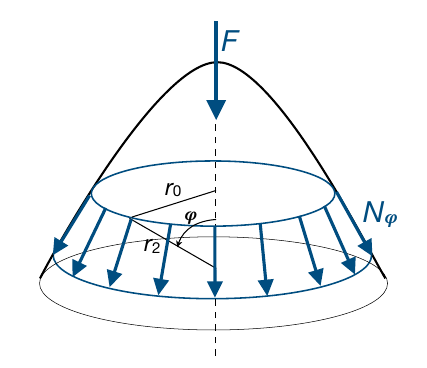}
\caption{Scheme for the equilibrium in case of axisymmetric loads.}
\label{fig:f7_10}
\end{center}
\end{figure}

\subsubsection{Global equilibrium}
\label{sec:globeqmemb}
The above results can be found with no need to integrate the equilibrium equations, that can be stated and solved directly. In fact, for the global equilibrium, in the symmetry axis direction, of the part of  membrane between $\varphi=0$ and $\varphi$, it must be, see Fig. \ref{fig:f7_10}, 
\be
\label{eq:gloeqmemb}
2\pi r_0\sin\varphi\ N_\varphi+F=0\ \rightarrow\ N_\varphi=-\frac{F}{2\pi r_0\sin\varphi}=-\frac{F}{2\pi r_2\sin^2\varphi},
\ee
where $F$ is the resultant of the loads for the part of membrane under consideration, and then
\be
N_\theta=-p_nr_2-\frac{r_2}{r_1}N_\varphi=-p_nr_2+\frac{F}{2\pi r_1\sin^2\varphi}.
\ee

\subsubsection{Strains and displacements}
\begin{figure}[th]
\begin{center}
\includegraphics[height=.3\textwidth]{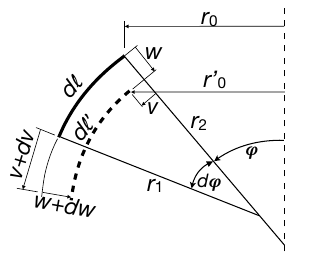}
\caption{Scheme of the deformation of a meridian line.}
\label{fig:f7_11}
\end{center}
\end{figure}

The inverse Lamé's equations\footnote{To remark that, because we have a plane stress state, the Lamé's inverse equations are the same of the 3D case.}
\be
\beps=\frac{1+\nu}{E}\bsig-\frac{\nu}{E}\tr\bsig\ \gr{I},
\ee
can be used to recover the strain field. Considering that
\be
\sigma_\theta=\frac{N_\theta}{h},\ \ \sigma_\varphi=\frac{N_\varphi}{h},\ \ \sigma_{\varphi\theta}=\frac{N_{\varphi\theta}}{h},
\ee
we get
\be
\label{eq:membeps}
\besp
&\eps_\theta=\frac{1+\nu}{E}\sigma_\theta-\frac{\nu}{E}(\sigma_\varphi+\sigma_\theta)=\frac{1}{E}(\sigma_\theta-\nu\ \sigma_\varphi)=\frac{1}{h\ E}(N_\theta-\nu\ N_\varphi),\\
&\eps_\varphi=\frac{1+\nu}{E}\sigma_\varphi-\frac{\nu}{E}(\sigma_\varphi+\sigma_\theta)=\frac{1}{E}(\sigma_\varphi-\nu\ \sigma_\theta)=\frac{1}{h\ E}(N_\varphi-\nu\ N_\theta),\\
&\eps_{\varphi\theta}=\frac{1+\nu}{E}\sigma_{\varphi\theta}=\frac{1+\nu}{h\ E}N_{\varphi\theta}.
\end{split}
\ee
We pass now to determine the displacement field. To this purpose, we need to link the above strain field to the displacement vector $\gr{u}=u \gr{e}_\theta+v \gr{e}_\varphi+w \gr{e}_n$; we already know that, for the symmetry of the problem, $u=0$. Then, let us consider an infinitesimal arc of meridian $d\ell=r_1\ d\varphi$, see Fig. \ref{fig:f7_11}. 
The length of $d\ell$ decreases for the radial displacement $w$ and increases for the tangential displacement $v$ so that the final length will be
\be
\besp
d\ell'&=d\ell+(v+dv)-v-(w+w+dw) \sin\frac{d\varphi}{2}\simeq d\ell+dv-w\ d\varphi-dw\ \frac{d\varphi}{2}\\
&=d\ell+\left(\frac{dv}{d\varphi}-w\right)d\varphi-\frac{1}{2}\frac{dw}{d\varphi}d\varphi^2\simeq d\ell+\left(\frac{dv}{d\varphi}-w\right)d\varphi,
\end{split}
\ee
so that finally
\be
\label{eq:membepsphi}
\eps_\varphi=\frac{d\ell'-d\ell}{d\ell}=\frac{1}{r_1}\left(\frac{dv}{d\varphi}-w\right).
\ee
In the direction of a parallel, the initial radius $r_0$ becomes
\be
r'_0=r_0+v\ \cos\varphi-w\ \sin\varphi,
\ee
so that
\be
\label{eq:membepstheta}
\eps_\theta=\frac{2\pi r'_0-2\pi r_0}{2\pi r_0}=\frac{1}{r_0}(v\ \cos\varphi-w\ \sin\varphi)=\frac{1}{r_2}\left(\frac{v}{\tan\varphi}-w\right).
\ee
From eqs. (\ref{eq:membepsphi}) and (\ref{eq:membepstheta}) we get
\be
\frac{dv}{d\varphi}-\frac{v}{\tan\varphi}=r_1\eps_\varphi-r_2\eps_\theta,
\ee
and using eqs. (\ref{eq:membeps})
\be
\label{eq:membeqdiff}
\frac{dv}{d\varphi}-\frac{v}{\tan\varphi}=f(\varphi),
\ee
with
\be
\label{eq:membeffe}
f(\varphi)=\frac{1}{h\ E}\left[N_\varphi(r_1+\nu\ r_2)-N_\theta(r_2+\nu\ r_1)\right].
\ee
The general solution of eq. (\ref{eq:membeqdiff}) is
\be
\label{eq:membdisp1}
v=\sin\varphi\left(\int\frac{f(\varphi)}{\sin\varphi}d\varphi+c\right),
\ee
with $c$ a constant to be determined upon the boundary conditions; finally, from eqs. (\ref{eq:membeps})$_1$ and  (\ref{eq:membepstheta}) we get
\be
\label{eq:membdisp2}
w=\frac{v}{\tan\varphi}-r_2\eps_\theta=\frac{v}{\tan\varphi}-\frac{r_2}{h\ E}(N_\theta-\nu\ N_\varphi).
\ee

\section{Exemples of axisymmetric membranes}
We consider in this Section some cases of axisymmetric membranes that are particularly interesting for applications.
\begin{figure}[th]
\begin{center}
\includegraphics[height=.3\textwidth]{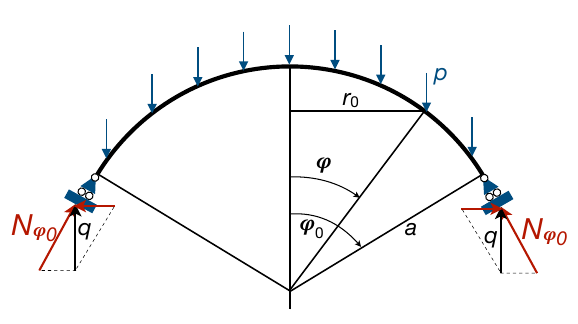}
\caption{Scheme of a spherical dome.}
\label{fig:f7_12}
\end{center}
\end{figure}
\subsection{Spherical domes}
\begin{figure}[th]
\begin{center}
\includegraphics[height=.5\textwidth]{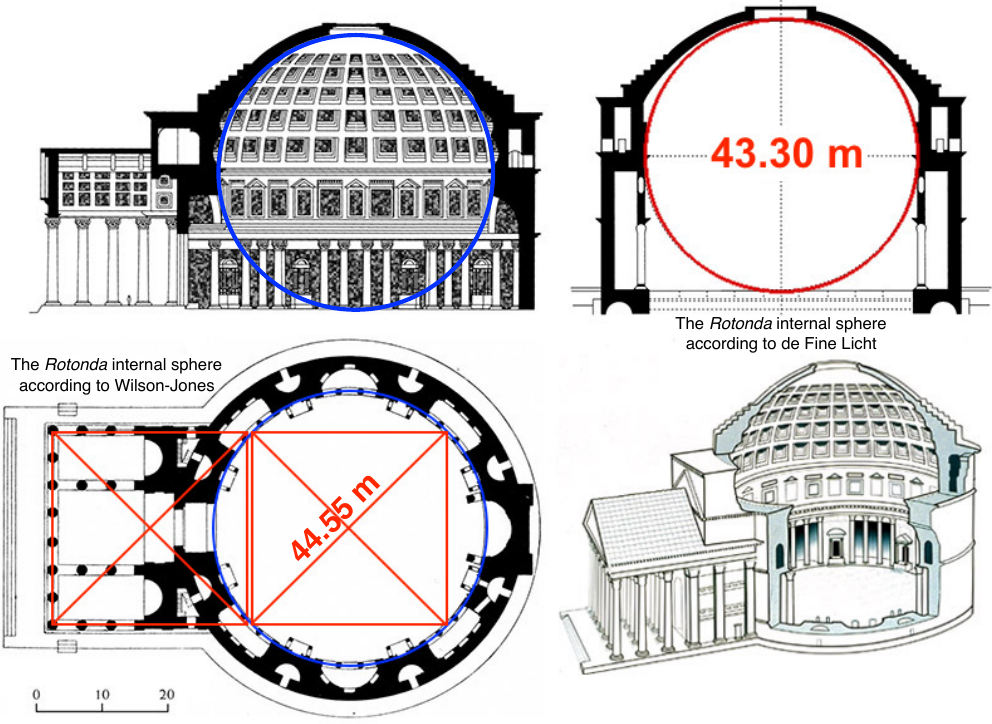}
\caption{Scheme of the Pantheon in Rome.}
\label{fig:f7_13}
\end{center}
\end{figure}

Let us first consider the case of a spherical dome of radius $a$ submitted to its own weight, whose value per unit area is $p$, see Fig. \ref{fig:f7_12}. This problem is historically rather important, because  several are the monumental domes all around the world, the most celebrated one being the dome of the Pantheon\footnote{The dome of the Pantheon, whose thickness is not constant, is still today the largest dome in the world in unreinforced concrete. Its inner surface, sculpted by a coffering, is a perfect hemisphere with a radius of 44.55 m (150 Roman feet). The lower part of the sphere, if existing, should exactly touch the floor of the building, Fig. \ref{fig:f7_13}. The Pantheon was built, probably by the great Nabatean architect Apollodorus of Damascus, under the Emperor Adrian, around 120 a. C.} in Rome, see Fig. \ref{fig:f7_13}. 
In this case, $r_1=r_2=a$ while $\gr{p}=(p_\theta,p_\varphi,p_n)=(0,p\sin\varphi,p\cos\varphi)$.
Hence, eq. (\ref{eq:equilmemb7})$_1$ becomes
\be
\frac{\partial U}{\partial\varphi}=-a^2p\sin\varphi,
\ee
whose solution is 
\be
U=a^2p\cos\varphi+c_1,
\ee
so that, by eq. (\ref{eq:axisymNphi}),
\be
\label{eq:sphericNphi}
N_\varphi=\frac{c+a\ p \cos\varphi}{\sin^2\varphi},
\ee
with $c=c_1/a$ to be determined upon the boundary conditions. We consider two different cases.
\subsubsection{Conical support}
\label{sec:sphericaldomeconic}
This is the case  represented in Fig. \ref{fig:f7_12}: the dome is simply supported at its base and the support has the shape of a conical surface orthogonal to the meridians. In this case, the reaction of the support is tangent to the meridians, i.e. it is directed like, but opposite to, $N_\varphi$.

The boundary condition is hence
\be
\label{eq:bc1}
N_\varphi(\varphi_0)=-\frac{q}{\sin\varphi_0},
\ee
with $q$ the total weight of the dome per unit length:
\be
q=\frac{\int_0^{\varphi_0}2\pi ar_0(\varphi)p\ d\varphi}{2\pi r_0(\varphi_0)}=\frac{p\ a\int_0^{\varphi_0}\sin\varphi\ d\varphi}{\sin\varphi_0}=p\ a\frac{1-\cos\varphi_0}{\sin\varphi_0}.
\ee
Replacing this value in eq. (\ref{eq:bc1}) gives $c=-p\ a$ and finally
\be
\label{eq:spherdome1}
N_\varphi=p\ a\frac{\cos\varphi-1}{\sin^2\varphi}=-\frac{p\ a}{1+\cos\varphi},
\ee
which gives also, eq. (\ref{eq:nthetanphi}),
\be
\label{eq:spherdome2}
N_\theta=-p\ a \cos\varphi-N_\varphi=p\ a\left(\frac{1}{1+\cos\varphi}-\cos\varphi\right).
\ee
It can be noticed that the same result can be obtained easily also by the approach described in Sect. \ref{sec:globeqmemb}; in fact, the weight of the dome between $\varphi=0$ and $\varphi$ is 
\be
F=\int_0^{\varphi}2\pi ar_0(\varphi)p\ d\varphi=\int_0^{\varphi}2\pi p\ a^2 \sin\varphi\ d\varphi=2\pi p\ a^2(1-\cos\varphi),
\ee
so that, using eq. (\ref{eq:gloeqmemb}),
\be
N_\varphi=-\frac{p\ a}{1+\cos\varphi},
\ee
as before.

\begin{figure}[th]
\begin{center}
\includegraphics[width=.4\textwidth]{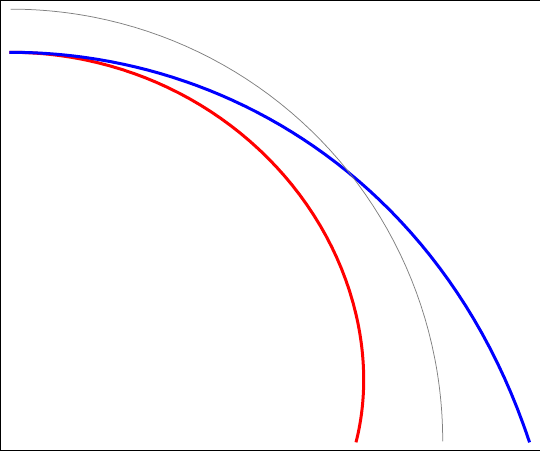}
\caption{Polar diagrams of $N_\varphi$, red, and $N_\theta$, blue, for a spherical shell submitted to its own weight (in grey: the profile of the dome; the radial distance between this curve and the curve of a diagram measures the value of the quantity, positive when outward).}
\label{fig:f7_12g}
\end{center}
\end{figure}

We remark that at the dome's top, $\varphi=0$, we get $N_\theta=N_\varphi=-p\ a/2$, while at the dome's base, for a hemispherical dome, i.e. for $\varphi_0=\pi/2$, it is $N_\varphi=-p\ a$ and $N_\theta=p\ a$. It interesting to note that $N_\varphi<0\ \forall\varphi$, i.e. the meridian lines are completely in compression, while $N_\theta<0\ \iff$
\be
\besp
&\frac{1}{1+\cos\varphi}-\cos\varphi<0\ \rightarrow\ \cos^2\varphi+\cos\varphi-1>0\ \rightarrow\\
& \cos\varphi<\frac{\sqrt{5}-1}{2}\ \rightarrow\ \varphi<\sim51.82^\circ.
\end{split}
\ee
So, the parallels are in compression only in the upper part of the dome, while they are in tension in the lower one\footnote{In the Pantheon's dome  there are, in fact, 14 major cracks in the meridian direction, from the springing of the dome up to about  $57^\circ$; they are produced exactly by the tension, in the direction of the parallels, that cannot be absorbed by unreinforced concrete (see F. Masi, I. Stefanou \& P. Vannucci: {\it A study about the origin of the cracks in the Pantheon’s dome}. Engineering Failure Analysis, 92, 587-596, 2018). Similar problems exist also in the brick's dome of Saint Peter, in Rome and they gave rise to the famous study of the {\it three mathematicians} in the XVIIIth century (see E. Benvenuto: {\it An Introduction to the History of Structural Mechanics - Part II}, Springer, 1991).\medskip}.
The polar diagrams of $N_\varphi$ and $N_\theta$ are plotted in Fig. \ref{fig:f7_12g}.

\subsubsection{Ring beam}
\label{sec:ringbeam}
\begin{figure}[th]
\begin{center}
\includegraphics[width=.8\textwidth]{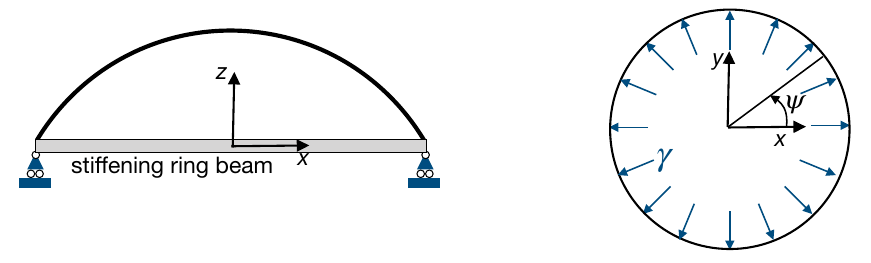}
\caption{Scheme of a spherical dome with a ring beam.}
\label{fig:f7_14}
\end{center}
\end{figure}

We consider now the presence of a ring beam, stiffening the border of the dome and simply supported, see Fig. \ref{fig:f7_14}.
In this case, the boundary condition is
\be
N_\varphi(\varphi_0)\sin\varphi_0=-q,
\ee
so finally we find the same result as in the previous case. However, the traction in the parallel at the springing of the dome, i.e.
\be
N_\theta(\varphi_0)=p\ a\left(\frac{1}{1+\cos\varphi_0}-\cos\varphi_0\right),
\ee
is {\it not} the tension force in the ring itself. In fact, this last is submitted to the radial load
\be
\gamma=-N_\varphi(\varphi_0)\cos\varphi_0,
\ee
see Fig. \ref{fig:f7_14}, that produces the tensile force
\be
\besp
t&=\frac{1}{2}\int_0^\pi\gamma r_0(\varphi_0)\sin\psi\ d\psi=\gamma\ r_0(\varphi_0)
=-N_\varphi(\varphi_0)r_0(\varphi_0)\cos\varphi_0\\&=\frac{p\ a^2}{1+\cos\varphi_0}\sin\varphi_0\cos\varphi_0\neq N_\theta(\varphi_0).
\end{split}
\ee
The difference in the two tensile forces is absorbed by the membrane through a local bending regime, that cannot be described by a membrane theory\footnote{This is a typical mechanical inconsistency that exists in the equilibrium theory of curved membranes; we will find in the following examples other similar inconsistencies, all of them giving rise to local bending states that extinguish rapidly in a small zone close to the membrane's edge, see the textbooks of Timoshenko and Woinowsky-Krieger or of Novozhilov in the references.}. 

In fact, for $\varphi=\varphi_0$, the radial displacement of the membrane, here taken as positive if outward, is, see eq. (\ref{eq:dispwmembspher}) in the next Section, 
\be
w(\varphi_0)=-\frac{p\ a^2}{h\ E}\left(\cos\varphi_0-\frac{1+\nu}{1+\cos\varphi_0}\right),
\ee
while that in the ring beam is
\be
w_b=\frac{t\ a}{E_b\ A_b}=\frac{p\ a^3}{E_b\ A_b}\frac{\sin\varphi_0\cos\varphi_0}{1+\cos\varphi_0}.
\ee
In the above equations, $E$ and $\nu$ are the Young's modulus and Poisson's ratio of the material composing the membrane while $E_b$ and $A_b$ are, respectively, the Young's modulus of the ring beam and the area of its cross-section. 

\begin{figure}[th]
\begin{center}
\includegraphics[width=.6\textwidth]{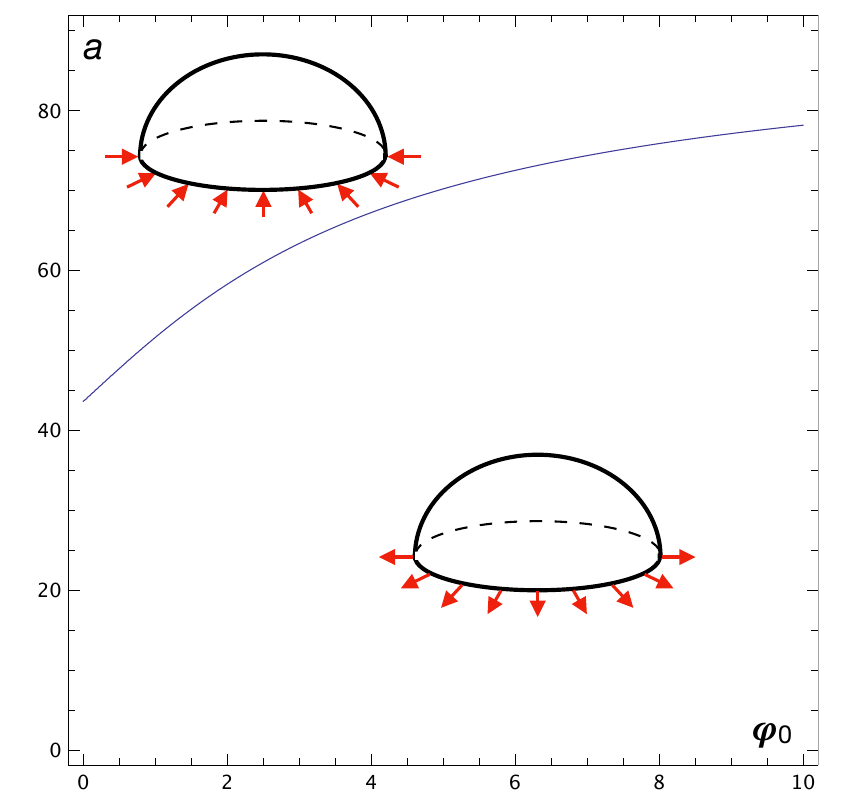}
\caption{Actions of a ring beam on the membrane ($E=E_b, A_b=1,h=0.5,\nu=0.25$).}
\label{fig:f7_14bis}
\end{center}
\end{figure}

When the two displacements are equal, no edge effect is given by the ring beam to the membrane: the regime is everywhere a pure membrane regime. This happens when
\be
w_b-w(\varphi_0)=0\ \rightarrow\ \frac{p\ a^3}{E_b\ A_b}\frac{\sin\varphi_0\cos\varphi_0}{1+\cos\varphi_0}+\frac{p\ a^2}{h\ E}\left(\cos\varphi_0-\frac{1+\nu}{1+\cos\varphi_0}\right)=0.
\ee
For given materials, the solution of this equation depends only upon the geometry of the surface, i.e. upon $a$ and $\varphi_0$. Such a solution  describes a curve in the plane $a-\varphi_0$; in particular, it does not depend on the load $p$. For the points on the curve, $w_b=w(\varphi_0)$ and the regime is of membrane everywhere; for the points above the curve, $w_b<w(\varphi_0)$, so the ring beam applies to the membrane inward radial forces, while  the opposite happens for the points below the curve, see  Fig. \ref{fig:f7_14bis}.

\subsubsection{Deformations and displacements}
\begin{figure}[th]
\begin{center}
\includegraphics[width=.4\textwidth]{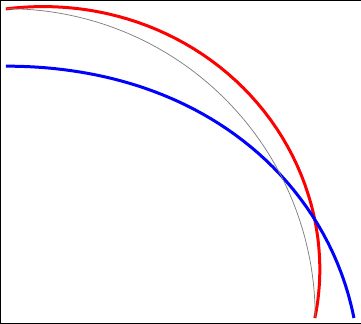}
\caption{Polar diagrams of $v$, red, and $-w$, blue, for a spherical shell submitted to its own weight (in grey: the profile of the dome; the radial distance between this curve and the curve of a diagram measures the value of the quantity, positive when outward).}
\label{fig:f7_15g}
\end{center}
\end{figure}

From eqs. (\ref{eq:membeps}), (\ref{eq:spherdome1}) and (\ref{eq:spherdome2}) we get
\be
\besp
&\eps_\varphi=\frac{p\ a}{h\ E}\frac{\nu\cos^2\varphi+\nu\cos\varphi-\nu-1}{1+\cos\varphi},\\
&\eps_\theta=-\frac{p\ a}{h\ E}\frac{\cos^2\varphi+\cos\varphi-\nu-1}{1+\cos\varphi},
\end{split}
\ee
and from eq. (\ref{eq:membeffe})
\be
f(\varphi)=\frac{p\ a^2(1+\nu)}{h\ E}\frac{\cos^2\varphi+\cos\varphi-2}{1+\cos\varphi},
\ee
that injected into eq. (\ref{eq:membdisp1}) gives
\be
\besp
v&=\frac{p\ a^2(1+\nu)}{h\ E}\sin\varphi\left(\int\frac{\cos^2\varphi+\cos\varphi-2}{(1+\cos\varphi)\sin\varphi}+c\right)\\
&=\frac{p\ a^2(1+\nu)}{h\ E}\sin\varphi\left(\log\frac{1+\cos\varphi}{2}-\frac{1}{1+\cos\varphi}+c\right).
\end{split}
\ee
If we impose a null displacement on the boundary, i.e. $v(\varphi=\varphi_0)=0$, then we get
\be
c=\frac{1}{1+\cos\varphi_0}-\log\frac{1+\cos\varphi_0}{2},
\ee
so finally
\be
v=\frac{p\ a^2(1+\nu)}{h\ E}\sin\varphi\left(\log\frac{1+\cos\varphi}{1+\cos\varphi_0}+\frac{\cos\varphi-\cos\varphi_0}{(1+\cos\varphi)(1+\cos\varphi_0)}\right).
\ee
The displacement $w$ can then be obtained through eq. (\ref{eq:membdisp2}):
\be
\label{eq:dispwmembspher}
\besp
w=&\frac{p\ a^2(1+\nu)}{h\ E}\sin\varphi\left[\cos\varphi\left(\log\frac{1+\cos\varphi}{1+\cos\varphi_0}+\frac{\cos\varphi-\cos\varphi_0}{(1+\cos\varphi)(1+\cos\varphi_0)}\right)\right.\\
&+\left.\frac{\cos\varphi}{1+\nu}-\frac{1}{1+\cos\varphi}\right].
\end{split}
\ee
The polar diagrams of $v$ and $-w$ are plotted in Fig. \ref{fig:f7_15g}\footnote{To remark that the calculation of the radial displacement $w$ has been done according to the basis $\{\gr{e}_\theta,\gr{e}_\varphi,\gr{e}_n\}$, see Fig. \ref{fig:f7_4}. Hence, an inward  displacement is  positive and an outward one is negative. In order to have a graphical representation of the true displacements of the dome, the opposite of $w$ is plotted in Fig.  \ref{fig:f7_15g}.}

\subsubsection{Spherical dome with \textit{oculus}}
\begin{figure}[th]
\begin{center}
\includegraphics[width=.4\textwidth]{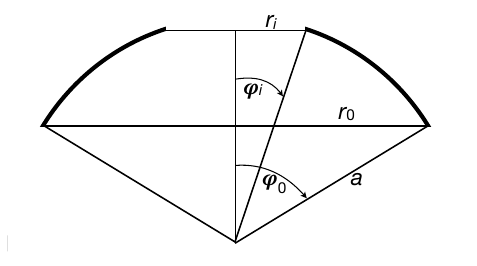}
\caption{Scheme of a spherical dome with an {\it oculus}.}
\label{fig:f7_15}
\end{center}
\end{figure}

Let us now consider the case of a spherical dome with an {\it oculus}, i.e. a hole, on the top, see Fig. \ref{fig:f7_15}, of radius $r_i$. Now, $\varphi_i\leq\varphi\leq\varphi_0$ and $N_\varphi(\varphi_i)=0$ and using this boundary condition gives $c=-p\ a\cos\varphi_i$ in eq. (\ref{eq:sphericNphi}) so that
\be
N_\varphi=p\ a\frac{\cos\varphi-\cos\varphi_i}{\sin^2\varphi},
\ee
and 
\be
N_\theta=p\ a\left(\frac{\cos\varphi_i-\cos\varphi}{\sin^2\varphi}-\cos\varphi\right).
\ee
The edge of the oculus is compressed\footnote{The Romans already knew this fact: the Pantheon's dome, like others built by the Romans, had an oculus whose edge is reinforced by an arch made by bricks and, probably, by a bronze ring.}:
\be
N_\theta(\varphi_i)=-p\ a\cos\varphi_i.
\ee

\subsubsection{Spherical dome with lantern}
\begin{figure}[th]
\begin{center}
\includegraphics[width=.8\textwidth]{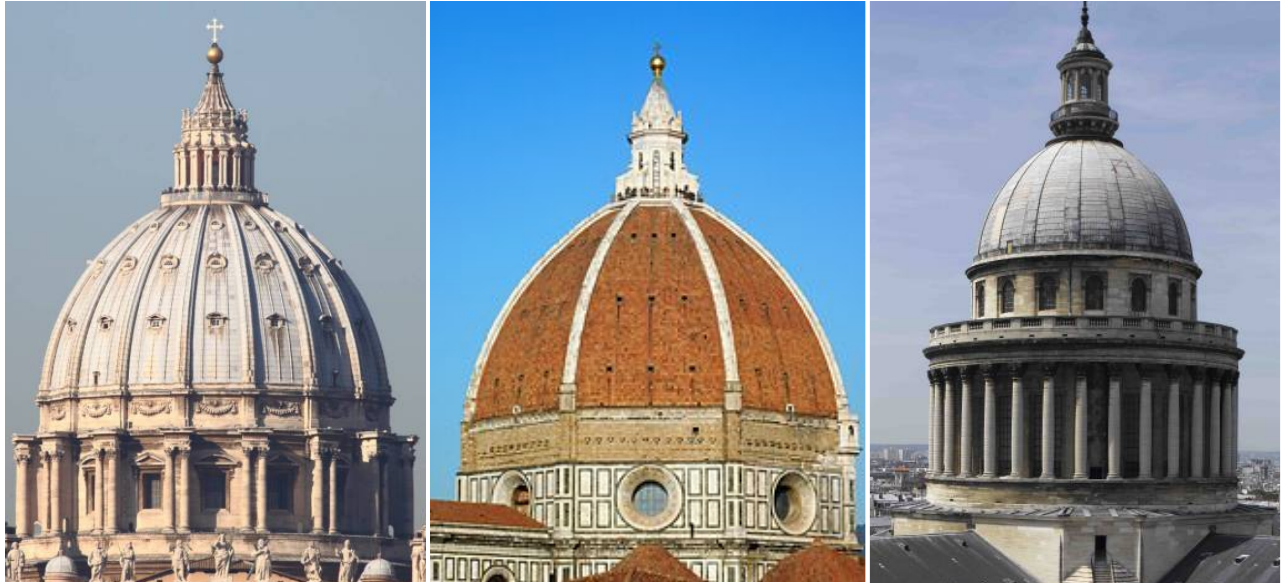}
\caption{Some famous examples of historical domes with lantern.}
\label{fig:f7_16}
\end{center}
\end{figure}

This is the case of the most part of historical domes, see e.g. Fig \ref{fig:f7_16}. The previous case is modified imposing that now
\be
N_\varphi(\varphi_i)=-\frac{q_L}{\sin\varphi_i},
\ee
where $q_L$ is the weight per unit length of the lantern. This gives now
\be
c=-q_L\sin\varphi_i-p\ a\cos\varphi_i,
\ee
so that
\be
\besp
&N_\varphi=\frac{p\ a(\cos\varphi-\cos\varphi_i)-q_L\sin\varphi_i}{\sin^2\varphi},\\
&N_\theta=p\ a\left(\frac{\cos\varphi_i-\cos\varphi}{\sin^2\varphi}-\cos\varphi\right)+q_L\frac{\sin\varphi_i}{\sin^2\varphi}.
\end{split}
\ee
In particular,
\be
N_\theta(\varphi_i)=-p\ a\cos\varphi_i+\frac{q_L}{\sin\varphi_i},
\ee
that can be positive or negative, i.e. the base of the lantern can be either in compression or in tension. 

\begin{figure}[th]
\begin{center}
\includegraphics[width=.8\textwidth]{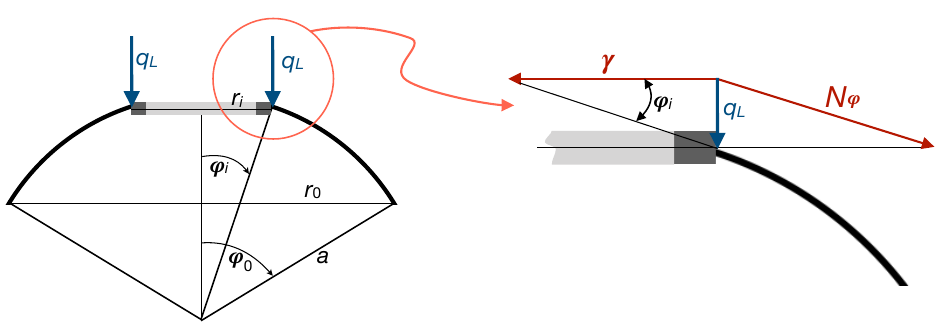}
\caption{Scheme of a spherical dome with lantern.}
\label{fig:f7_17}
\end{center}
\end{figure}

\begin{figure}[th]
\begin{center}
\includegraphics[width=.4\textwidth]{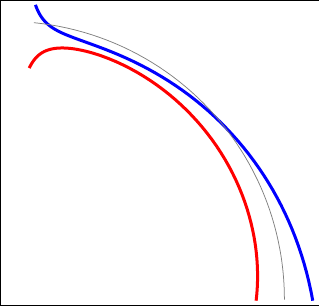}
\caption{Polar diagrams of $N_\varphi$, red, and $N_\theta$, blue, for a spherical shell with lantern (in grey: the profile of the dome; the radial distance between this curve and the curve of a diagram measures the value of the quantity, positive when outward).}
\label{fig:f7_17g}
\end{center}
\end{figure}

Because the membrane cannot take on transverse shear forces, a reinforcing ring is necessary at the base of the lantern, see Fig. \ref{fig:f7_17}; this is submitted to the radial compressive load 
\be
\gamma=\frac{q_L}{\tan\varphi_i}
\ee
that produces in the ring the traction 
\be
t=\gamma r_i=q_L a\cos\varphi_i.
\ee
As in the case of the dome supported by a ring beam, also in this case such a force produces in the ring deformations that are not equal to those in the adjacent part of the dome: this produces local bending that is rapidly dissipated in the dome.
In Fig. \ref{fig:f7_17g} we show the polar diagrams of $N_\varphi$ and $N_\theta$ for the case of a spherical dome with a lantern (the diagrams are referred to a case where the total weight of the lantern is equal to $p a^2/10$ and the radius of the lantern is equal to $a/10$).

\begin{figure}[th]
\begin{center}
\includegraphics[width=.6\textwidth]{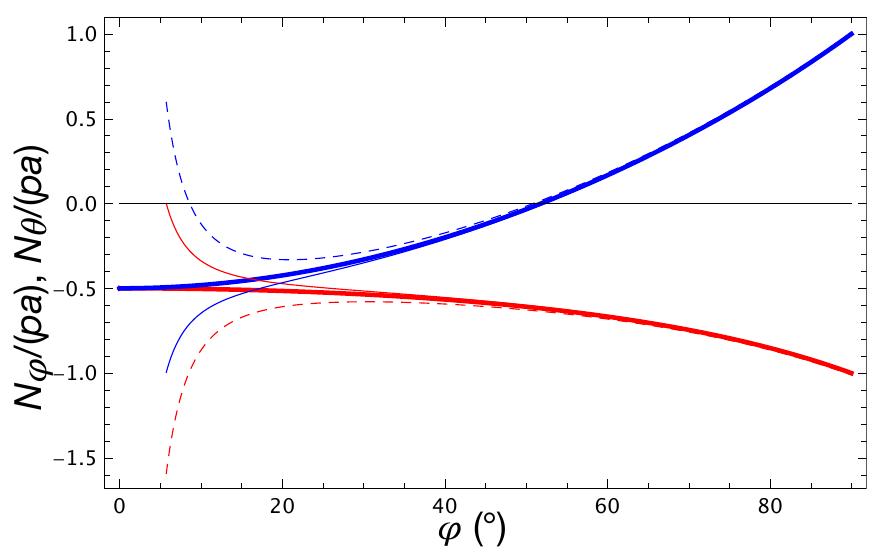}
\caption{Cartesian diagrams of $N_\varphi$, red, and $N_\theta$, blue, for an entire spherical shell, thick lines, with an {\it oculus}, thin lines, and with a lantern, dashed lines.}
\label{fig:f7_18}
\end{center}
\end{figure}

In Fig. \ref{fig:f7_18} we compare the diagrams of $N_\varphi$ and $N_\theta$ for three different cases of spherical domes: entire dome, dome with {\it  oculus} and dome with lantern. The diagrams are referred to a case where the total weight of the lantern is equal to $p a^2/10$ and the radius of the lantern or of the oculus is equal to $a/10$.

\subsubsection{Wind on a spherical dome}
\begin{figure}[th]
\begin{center}
\includegraphics[width=\textwidth]{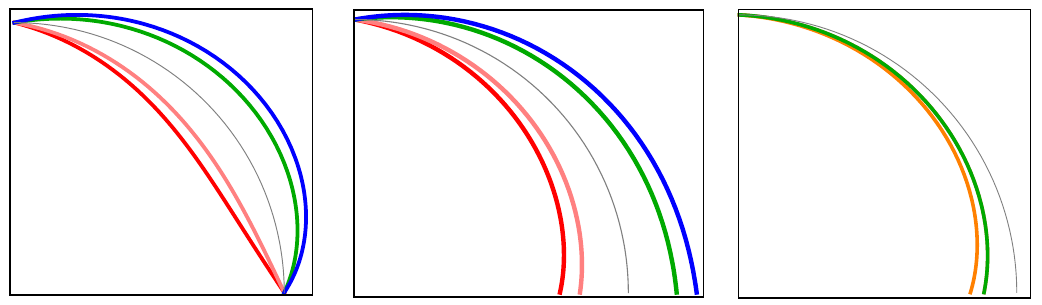}
\caption{Polar diagrams of $N_\varphi$, left, $N_\theta/4$, center, $N_{\varphi\theta}/4$, right, plotted for $\theta=0^\circ$, red, $\theta=45^\circ$, pink, $\theta=90^\circ$, orange, $\theta=135^\circ$, green, and $\theta=180^\circ$, blue (in grey: the profile of the dome; the radial distance between this curve and the curve of a diagram measures the value of the quantity, positive when outward).}
\label{fig:f7_19}
\end{center}
\end{figure}

Let us now consider the action of the wind on a spherical dome; if we consider air as a perfect, i.e. inviscid, fluid, the  wind's load   resume to only a load normal to the dome's surface, i.e. $p_\theta=p_\varphi=0$, while  $p_n$  can be modeled as
\be
p_n=p\sin\varphi\cos\theta.
\ee
In such a case, eqs. (\ref{eq:equilmemb1}) become
\be
\label{eq:equilmembwind1}
\besp
&\frac{\partial}{\partial\varphi}(N_\varphi \sin\varphi)-\cos\varphi\ N_\theta+\frac{\partial N_{\varphi\theta}}{\partial\theta}=0,\\
&\frac{\partial N_\theta}{\partial \theta}+\frac{\partial (\sin\varphi N_{\theta\varphi})}{\partial \varphi}+\cos\varphi\ N_{\varphi\theta}=0,\\
&{N_\varphi}+{N_\theta}=-p\ a\sin\varphi\cos\theta.
\end{split}
\ee
If from the third of the above equations we make $N_\theta$ explicit and inject its expression into the first two equations, after some standard passages we obtain two equations for $N_\varphi$ and $N_{\varphi\theta}$:
\be
\label{eq:membwindload1}
\besp
&\sin\varphi\frac{\partial N_\varphi}{\partial\varphi}+\frac{\partial N_{\varphi\theta}}{\partial\theta}+2\cos\varphi N_\varphi+p\ a\sin\varphi\cos\varphi\cos\theta=0,\\
&\sin\varphi\frac{\partial N_{\varphi\theta}}{\partial\varphi}-\frac{\partial N_{\varphi}}{\partial\theta}+2\cos\varphi N_{\varphi\theta}+p\ a\sin\varphi\sin\theta=0.
\end{split}
\ee
To solve the above partial differential equations, we look for a solution of the form\footnote{This choice is motivated by the symmetry that the solution should have as a consequence of the symmetrical distribution of the wind load $p_n$.}
\be
N_\varphi=S_\varphi(\varphi)\cos\theta,\ \ N_{\varphi\theta}=S_{\varphi\theta}(\varphi)\sin\theta;
\ee
injecting these two expressions into eqs. (\ref{eq:membwindload1}) we get a system of two coupled ordinary differential equations:
\be
\besp
&\sin\varphi\frac{\partial S_\varphi}{\partial\varphi}+2\cos\varphi S_\varphi+S_{\varphi\theta}+p\ a \sin\varphi\cos\varphi=0,\\
&\sin\varphi\frac{\partial S_{\varphi\theta}}{\partial\varphi}+2\cos\varphi S_{\varphi\theta}+S_{\varphi}+p\ a \sin\varphi=0.
\end{split}
\ee
To uncouple and solve the above equations, we first add them together and then we subtract the second one from the first one, to obtain
\be
\label{eq:membwindload2}
\besp
&\sin\varphi\frac{\partial S_1}{\partial\varphi}+(1+2\cos\varphi) S_1+p\ a \sin\varphi(1+\cos\varphi)=0,\\
&\sin\varphi\frac{\partial S_2}{\partial\varphi}-(1-2\cos\varphi) S_2-p\ a \sin\varphi(1-\cos\varphi)=0,
\end{split}
\ee
where
\be
S_1=S_\varphi+S_{\varphi\theta},\ \ S_2=S_\varphi-S_{\varphi\theta}\ \rightarrow\ S_\varphi=\frac{S_1+S_2}{2},\ \ S_{\varphi\theta}=\frac{S_1-S_2}{2}.
\ee
The solution of eqs. (\ref{eq:membwindload2}) is
\be
\besp
&S_1=\frac{1+\cos\varphi}{\sin^3\varphi}\left[c_1+\frac{1}{3}p\ a\cos\varphi\left(3-\cos^2\varphi\right)\right],\\
&S_2=\frac{1-\cos\varphi}{\sin^3\varphi}\left[c_2-\frac{1}{3}p\ a\cos\varphi\left(3-\cos^2\varphi\right)\right],\\
\end{split}
\ee
that finally gives
\be
\besp
&N_\varphi=\frac{\cos\theta}{\sin^3\varphi}\left(\frac{c_1+c_2}{2}+\frac{c_1-c_2}{2}\cos\varphi+\frac{1}{3}p\ a\cos^2\varphi(3-\cos^2\varphi)\right),\\
&N_{\varphi\theta}=\frac{\sin\theta}{\sin^3\varphi}\left(\frac{c_1-c_2}{2}+\frac{c_1+c_2}{2}\cos\varphi+\frac{1}{3}p\ a\cos\varphi(3-\cos^2\varphi)\right),
\end{split}
\ee
and, through eq. (\ref{eq:equilmembwind1})$_3$,
\be
N_\theta=-\frac{\cos\theta}{\sin^3\varphi}\left(\frac{c_1+c_2}{2}+\frac{c_1-c_2}{2}\cos\varphi+\frac{1}{3}p\ a(3\sin^2\varphi+2\cos^4\varphi)\right).
\ee
The two constants $c_1$ and $c_2$ can be determined through the boundary conditions. In the case of a hemispherical dome ($\varphi_0=\pi/2$), we get that on the border
\be
N_\varphi(\varphi_0)=\frac{c_1+c_2}{2}\cos\theta,\ \ \ N_{\varphi\theta}(\varphi_0)=\frac{c_1-c_2}{2}\sin\theta.
\ee
We then consider that the moment of the wind load about the line at $\theta=\pi/2$ of the dome's base, i.e. with respect to the  diameter of the dome's base orthogonal to the direction of the wind, is null\footnote{This is an immediate consequence of the fact that $p_s$ is a radial load.}. Because such a moment is equal and opposite to the moment of the internal forces at the dome's base, also these internal forces have a null moment:
\be
\int_0^{2\pi}N_\varphi(\varphi_0)a^2\cos\theta\ d\theta=0.
\ee
The second boundary condition imposes that the component in the direction of the wind of the resultant of the internal forces at the dome's base must be equal and opposite to the total horizontal wind action:
\be
\int_0^{2\pi}N_{\varphi\theta}(\varphi_0)a\sin\theta\ d\theta=-\int_0^{2\pi}\int_0^{\frac{\pi}{2}}p_na^2\sin^2\varphi\cos\theta\ d\theta\ d\varphi.
\ee
The two boundary conditions give hence the equations
\be
\besp
&\frac{\pi}{2}(c_1+c_2)a^2=0,\\
&\frac{\pi}{2} a(c_1-c_2)=-\frac{2}{3}\pi p\ a^2,
\end{split}\ \ \ \rightarrow\ \ c_1=-\frac{2}{3} p\ a,\ \ c_2=\frac{2}{3} p\ a,
\ee
and finally
\be
\besp
&N_\varphi=-\frac{p\ a}{3}\frac{\cos\varphi\cos\theta}{\sin^3\varphi}(2-3\cos\varphi+\cos^3\varphi),\\
&N_\theta=\frac{p\ a}{3}\frac{\cos\theta}{\sin^3\varphi}(2\cos\varphi-3\sin^2\varphi-2\cos^4\varphi),\\
&N_{\varphi\theta}=-\frac{p\ a}{3}\frac{\sin\theta}{\sin^3\varphi}(2-3\cos\varphi+\cos^3\varphi).
\end{split}
\ee
The polar diagrams of $N_\varphi,N_\theta$ and $N_{\varphi\theta}$ are plotted in Fig. \ref{fig:f7_19}.

%As in the case of the dome supported by a ring beam, also in this case such a force produces in the ring deformations that are not equal to those in the adjacent part of the dome: this results in local bending that is rapidly dissipated in the dome.

\subsection{Elliptical domes}
\begin{figure}[th]
\begin{center}
\includegraphics[width=.4\textwidth]{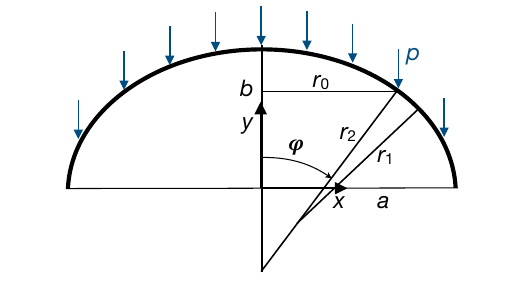}
\caption{Scheme of an elliptical dome submitted to its own weight.}
\label{fig:f7_20}
\end{center}
\end{figure}

We consider the case of an axisymmetric dome whose meridians have the shape of an ellipse of equation
\be
\frac{x^2}{a^2}+\frac{y^2}{b^2}=1,
\ee
like in Fig. \ref{fig:f7_20}. For such a case,
\be
r_1=\frac{a^2b^2}{(a^2\sin^2\varphi+b^2\cos^2\varphi)^{\frac{3}{2}}},\ \ \ r_2=\frac{a^2}{(a^2\sin^2\varphi+b^2\cos^2\varphi)^{\frac{1}{2}}}.
\ee
We consider two load cases: a vault submitted to its own weight and a pressure tank.

\subsubsection{Elliptical vault under its own weight}
The load vector is $\gr{p}=(p_\theta,p_\varphi,p_n)=(0,p\sin\varphi,p\cos\varphi)$, so eq. (\ref{eq:equilmemb7})$_1$ becomes
\be
\label{eq:membellipt1}
\frac{\partial U}{\partial \varphi}=-\frac{a^4b^2}{(a^2\sin^2\varphi+b^2\cos^2\varphi)^2}p\ \sin\varphi.
\ee
This equation can be solved by a change of variable: we put
\be
\label{eq:membellipt2}
\cos\psi=\frac{c}{a}\cos\varphi,
\ee
with
\be
c^2=a^2-b^2.
\ee
Hence, 
\be
\frac{\partial U}{\partial \varphi}=\frac{\partial U}{\partial \psi}\frac{\partial \psi}{\partial \varphi},
\ee
and through eqs. (\ref{eq:membellipt1}) and (\ref{eq:membellipt2}) we get the differential equation
\be
\frac{\partial U}{\partial \psi}=-\frac{p\ a\ b^2}{c\ \sin^3\psi},
\ee
whose solution is
\be
U=-p\frac{a^2b^2}{2}\left(\frac{1}{2a\ c}\log\frac{a-c\ \cos\varphi}{a+c\ \cos\varphi}-\frac{\cos\varphi}{a^2\sin^2\varphi+b^2\cos^2\varphi}+c_1\right).
\ee
To determine the constant $c_1$, we consider that, because of eq. (\ref{eq:axisymNphi}), it must be $U(\varphi=0)=0$ (otherwise, $N_\varphi$ should not be finite at the dome's top). This gives
\be
c_1=\frac{1}{b^2}-\frac{1}{2a\ c}\log\frac{a-c}{a+c},
\ee
and finally, from eqs. (\ref{eq:axisymNphi}) and (\ref{eq:axisymNtheta}) we get 
\be
\besp
&N_\varphi=-p\frac{b^2}{2\sin^2\varphi}\sqrt{a^2\sin^2\varphi+b^2\cos^2\varphi}\ \times\\
&\hspace{10mm}\left[\frac{1}{b^2}-\frac{\cos\varphi}{a^2\sin^2\varphi+b^2\cos^2\varphi}+\frac{1}{2a\ c}\log\left(\frac{a-c\ \cos\varphi}{a+c\ \cos\varphi}\frac{a+c}{a-c}\right)\right],\\
&N_\theta=-p\frac{a^2\cos\varphi}{\sqrt{a^2\sin^2\varphi+b^2\cos^2\varphi}}+\frac{p}{2\sin^2\varphi}(a^2\sin^2\varphi+b^2\cos^2\varphi)^{\frac{3}{2}}\ \times\\
&\hspace{10mm}\left[\frac{1}{b^2}-\frac{\cos\varphi}{a^2\sin^2\varphi+b^2\cos^2\varphi}+\frac{1}{2a\ c}\log\left(\frac{a-c\ \cos\varphi}{a+c\ \cos\varphi}\frac{a+c}{a-c}\right)\right].
\end{split}
\ee
At the dome's top, $\varphi=0$, it is
\be
N_\varphi=N_\theta=-p\frac{a^2}{2b}.
\ee
To remark that for $a=b$ the above formulae coincide with those already found for the spherical dome, cfr. Sect. \ref{sec:sphericaldomeconic}.

\begin{figure}[th]
\begin{center}
\includegraphics[width=.5\textwidth]{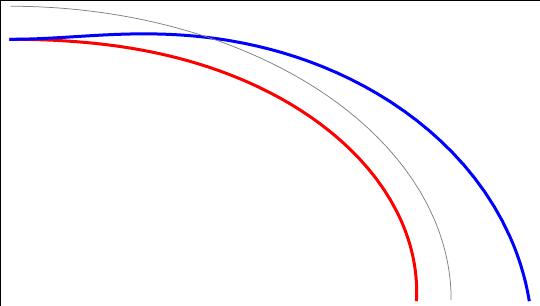}
\caption{Polar diagrams of $N_\varphi$, red, and $N_\theta$, blue, for an elliptical shell (in grey: the profile of the dome; the radial distance between this curve and the curve of a diagram measures the value of the quantity, positive when outward).}
\label{fig:f7_21}
\end{center}
\end{figure}

The diagrams of $N_\varphi$ and $N_\theta$ for the case $a=1.5 b$ are plotted in Fig. \ref{fig:f7_21}.

\subsubsection{Elliptical pressure tank}
Axisymmetric elliptical domes are often used to form the edges of  many cylindrical pressure tanks. In this case the load vector is 
$\gr{p}=(p_\theta,p_\varphi,p_n)=(0,0,-p)$. The procedure followed in the previous case can still be used, but it is easier to follow the procedure of Sect. \ref{sec:globeqmemb}. In fact, the resultant of the pressure load is
\be
F=-\pi p r_0^2,
\ee
so from eq. (\ref{eq:gloeqmemb}) we get
\be
N_\varphi=\frac{p\ r_0}{2\sin\varphi}=\frac{1}{2}\frac{p\ a^2}{\sqrt{a^2\sin^2\varphi+b^2\cos^2\varphi}}.
\ee
and from eq. (\ref{eq:axisymNtheta})
\be
N_\theta=\frac{p\ a^2}{\sqrt{a^2\sin^2\varphi+b^2\cos^2\varphi}}\left(1-\frac{a^2\sin^2\varphi+b^2\cos^2\varphi}{2b^2}\right).
\ee
We see hence that, while $N_\varphi>0\ \forall\varphi$, i.e. the meridians are always in tension, $N_\theta$ can change its sign. In particular, at the dome's top
\be
N_\varphi(\varphi=0)=N_\theta(\varphi=0)=\frac{p\ a^2}{2b},
\ee
while at the equator
\be
N_\varphi\left(\varphi=\frac{\pi}{2}\right)=\frac{p\ a}{2},\ \ N_\theta\left(\varphi=\frac{\pi}{2}\right)=p\ a\left(1-\frac{a^2}{2b^2}\right),
\ee
so at the equator $N_\theta<0$ (compression) if $a>b\sqrt{2}$. Finally,  a spherical tank corresponds to the special case $a=b$.

The diagrams of $N_\varphi$ and $N_\theta$ for the case $a=2.5 b$ are plotted in Fig. \ref{fig:f7_21}.

\begin{figure}[th]
\begin{center}
\includegraphics[width=.5\textwidth]{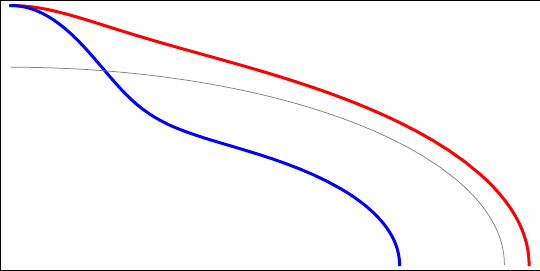}
\caption{Polar diagrams of $N_\varphi$, red, and $N_\theta$, blue, for an elliptical pressure tank (in grey: the profile of the dome; the radial distance between this curve and the curve of a diagram measures the value of the quantity, positive when outward).}
\label{fig:f7_22}
\end{center}
\end{figure}

\subsection{Conical vaults}

\begin{figure}[th]
\begin{center}
\includegraphics[width=.4\textwidth]{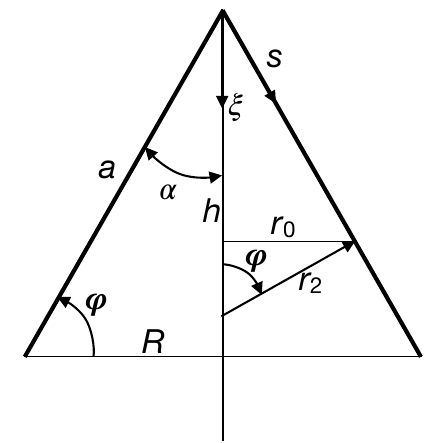}
\caption{Scheme of a conical vault.}
\label{fig:f7_23}
\end{center}
\end{figure}

We consider a conical vault with the meridians inclined of an angle $\alpha$ on the cone's axis, like in Fig. \ref{fig:f7_23}; in this case, $r_1\rightarrow\infty$ and $\varphi=\pi/2-\alpha$ cannot be used as a variable because it is a constant. Hence, a curvilinear coordinate must be chosen to determine a position on a chosen meridian. This one can be, e.g., the coordinate $\xi$, measuring along the axis of the cone the distance from the  tip, or the distance $s$, still from the cone's tip but measured directly along a meridian, see Fig. \ref{fig:f7_23}. 
Of course, 
\be
\xi=s \cos\alpha,
\ee
and
\be
r_2=\xi\frac{\tan\alpha}{\cos\alpha}=s\tan\alpha,\ \ r_0=r_2\cos\alpha=\xi\tan\alpha=s\sin\alpha.
\ee
$R$ is the radius of the cone's base, $h$ the cone's height and $a$ the length of a meridian:
\be
a=\sqrt{R^2+h^2}.
\ee
If we consider eq. (\ref{eq:axisymNtheta}), we see that, because $r_1\rightarrow\infty$,
\be
\label{eq:Nthetacone}
N_\theta=-p_nr_2=-p_n\ \xi\frac{\tan\alpha}{\cos\alpha}=-p_n\ s\tan\alpha;
\ee
so, in this case, $N_\theta$ and $N_s$ are uncoupled, in particular, $N_\theta$ does not depend upon $N_s$ but uniquely upon the load and the distance from the tip.
\begin{figure}[ht]
\begin{center}
\includegraphics[width=.7\textwidth]{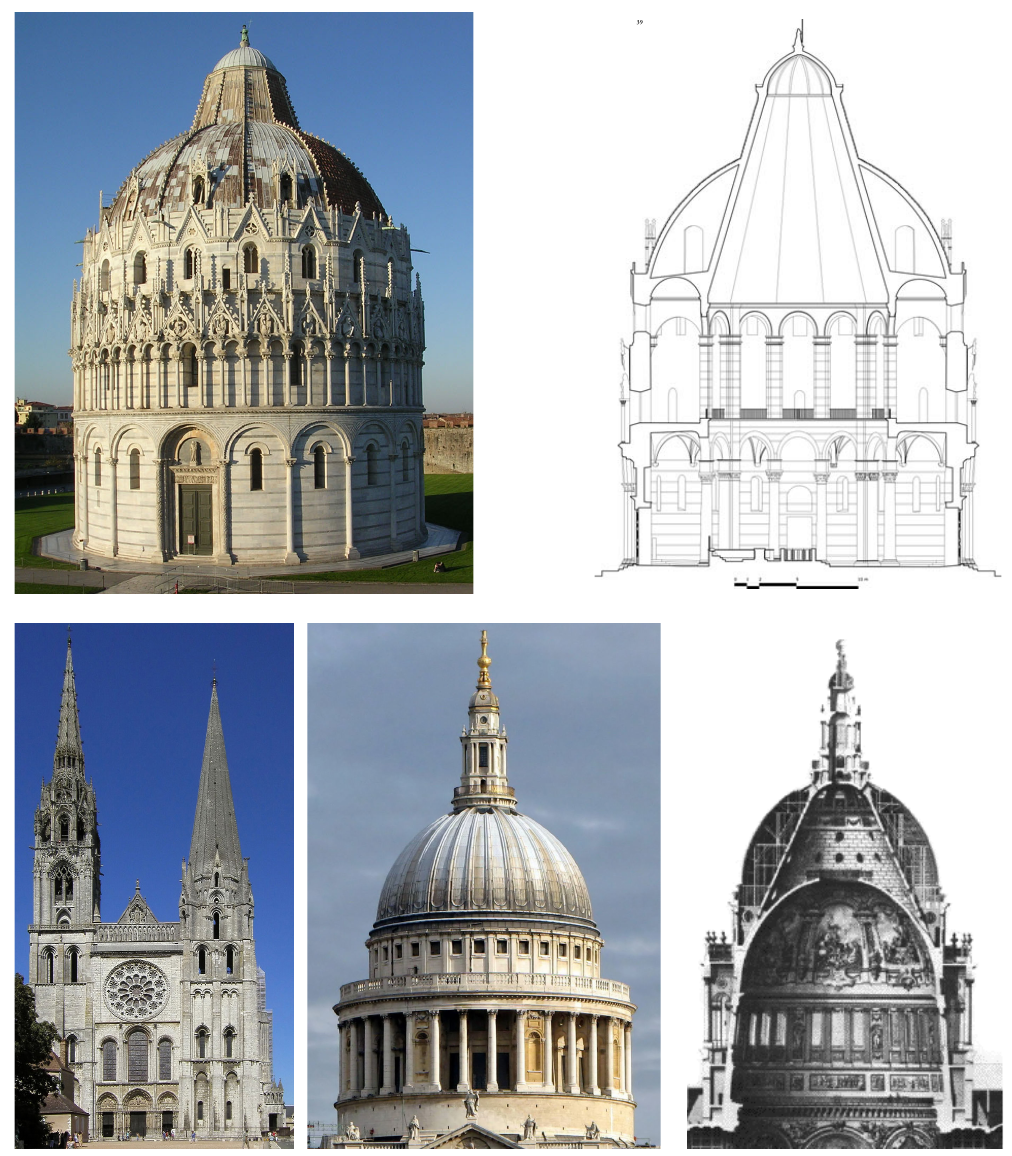}
\caption{Three historical conical vaults; the Baptistery of Pisa (from 1152) and a cross-section of its structure (from the thesis of G. Verdiani, 2003); the Cathedral of Chartres (from 1194), the dome of Saint Paul's Cathedral (from 1675) and a cross-section of its structure.}
\label{fig:f7_24}
\end{center}
\end{figure}

Historically, some remarkable structures of the Middle Ages have the form of a conical vault, they are shown in Fig. \ref{fig:f7_24}: the roof of the South Tower of the Chartres Cathedral, the internal supporting structures of the dome of the Baptistery of Pisa and of the dome of Saint Paul's Cathedral, in London.

We consider in the following Sections different load cases for conical vaults.

\subsubsection{Vertical tip load}
\label{sec:verttipload}
We consider the presence of a vertical concentrated force $P$ applied on the tip of the cone. From eq. (\ref{eq:gloeqmemb}) we get (of course, $F=P$)\footnote{Because $\varphi$ cannot be  used anymore as a variable, we denote by $N_s$ the membrane force along the meridians; in the same way, the index $\varphi$ is replaced everywhere by $s$.}:
\be
N_s=-\frac{P}{2\pi r_0\sin\varphi}=-\frac{P}{2\pi r_0\cos\alpha}=-\frac{P}{2\pi \xi\sin\alpha}=-\frac{P}{\pi s\sin2\alpha}.
\ee
while, because $\gr{p}=(p_s,p_\theta,p_n)=(0,0,0)$, it is $N_\theta=0$, see eq. (\ref{eq:Nthetacone}). This is an interesting result, because, along with the general result $N_{s\theta}$=0 for axisymmetric membranes acted upon by an axisymmetric load, it clearly indicates that the membrane works like if it was constituted by an array of struts, the meridians, working independently. The singularity of $N_s$ at the tip is of course due to the model, that considers a concentrated load.
\subsubsection{Conical vault under its own weight}
In this case, $\gr{p}=(p_s,p_\theta,p_n)=(p\cos\alpha,0,p\sin\alpha)$
\be
\besp
F(\xi)&=\int_0^\xi2\pi p\ r_0(\xi^*)\frac{1}{\cos\alpha}d\xi^*=2\pi p\ \frac{\tan\alpha}{\cos\alpha}\int_0^\xi\xi^*d\xi^*\\
&=\pi p\ \xi^2\frac{\tan\alpha}{\cos\alpha}=\pi p\ s^2\tan\alpha\cos\alpha.
\end{split}
\ee
So, still from eqs. (\ref{eq:gloeqmemb})  and (\ref{eq:Nthetacone}), we obtain
\be
\besp
&N_s=-\pi p\ \xi^2\frac{\tan\alpha}{\cos\alpha}\frac{1}{2\pi\ \xi\tan\alpha\sin\varphi}=-\frac{p\ \xi}{2\cos^2\alpha}=-\frac{p\ s}{2\cos\alpha},\\
&N_\theta=-p\ \xi\tan^2\alpha=-p\ s\frac{\sin^2\alpha}{\cos\alpha}.
\end{split}
\ee
So, both $N_s$ and $N_\theta$ are compression forces and their intensities increase linearly with the distance from the tip.\\

\subsubsection{Conical water reservoir}
\begin{figure}[th]
\begin{center}
\includegraphics[width=.4\textwidth]{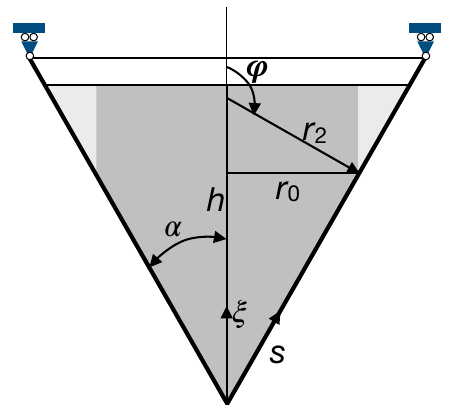}
\caption{Scheme of a conical water reservoir.}
\label{fig:f7_25}
\end{center}
\end{figure}

\begin{figure}[th]
\begin{center}
\includegraphics[width=.2\textwidth]{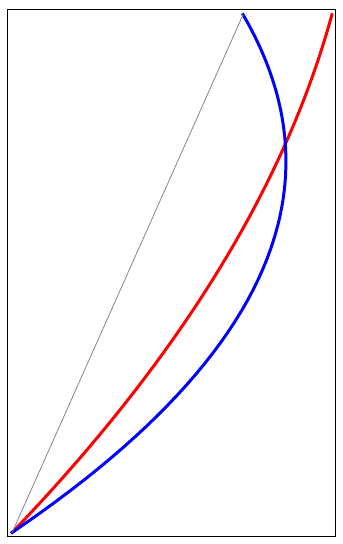}
\caption{Diagrams of $N_s$, red, and $N_\theta$, blue, for a suspended conical reservoir (in grey: the profile of the reservoir; the horizontal distance between this curve and the curve of a diagram measures the value of the quantity, positive when outward).}
\label{fig:f7_26}
\end{center}
\end{figure}

We refer to a conical reservoir like in Fig. \ref{fig:f7_25}, filled with water (or anther liquid) of density $\rho$ and we consider it hanged at the upper part. We have now $\varphi=\pi/2+\alpha$ and the load is $\gr{p}=(p_s,p_\theta,p_n)=(0,0,-\rho g(h-\xi))$. The force $F$ for a position determined by a given $\xi$, or $s$, is the weight of the mass of water shadowed in Fig. \ref{fig:f7_25}:
\be
\besp
F&=\int_0^s 2\pi r_0 p_n(s^*)\sin\alpha\ ds^*=-\int_0^s 2\pi  \rho g\ {s^*}^2\sin^2\alpha\ ds^*\\
&=-\pi\rho\ g\ s^2\sin^2\alpha\left(h-\frac{2}{3}s\cos\alpha\right)=-\pi\rho\ g\ \xi^2\tan^2\alpha\left(h-\frac{2}{3}\xi\right).
\end{split}
\ee
Hence, again from eqs.  (\ref{eq:gloeqmemb})  and (\ref{eq:Nthetacone}), we obtain
\be
\besp
&N_s=\rho\ g\ \xi\frac{\tan\alpha}{2\cos\alpha}\left(h-\frac{2}{3}\xi\right)=\frac{1}{2}\rho\ g\ s\ \tan\alpha\left(h-\frac{2}{3}s\cos\alpha\right),\\
&N_\theta=\rho\ g\ \xi\frac{\tan\alpha}{\cos\alpha}(h-\xi)=\rho\ g\ s\tan\alpha(h-s\cos\alpha).
\end{split}
\ee
$N_s$ and $N_\theta$ are hence quadratic functions of $\xi$ or $s$, their variation is plotted in Fig. \ref{fig:f7_26} for the case $\alpha=30^\circ$. It easy to see that
\be
\besp
&N_s^{max}=N_s\left(\xi=\frac{3}{4}h\right)=\frac{3}{16}\rho\ g\ h^2\frac{\tan\alpha}{\cos\alpha},\\
&N_\theta^{max}=N_\theta\left(\xi=\frac{h}{2}\right)=\frac{1}{4}\rho\ g\ h^2\frac{\tan\alpha}{\cos\alpha}.
\end{split}
\ee
If a stiffening ring is added to the base of the cone, like in the case of the spherical  dome in Sect. \ref{sec:ringbeam}, this is submitted to the radial load
\be
\gamma=-N_s(\xi=h)\sin\alpha=-\frac{1}{6}\rho\ g\ h^2\tan^2\alpha,
\ee
which gives in the ring the axial compression force
\be
t=\gamma R=-\frac{1}{6}\rho\ g\ h^3\tan^3\alpha.
\ee
Because at the same level $N_\theta=0$, see Fig. \ref{fig:f7_26}, the deformation in the ring and the adjacent part of the cone is different; once more, this gives rise to local bending strains and stresses that quickly  decrease with the distance from the border.

\begin{figure}[th]
\begin{center}
\includegraphics[width=.2\textwidth]{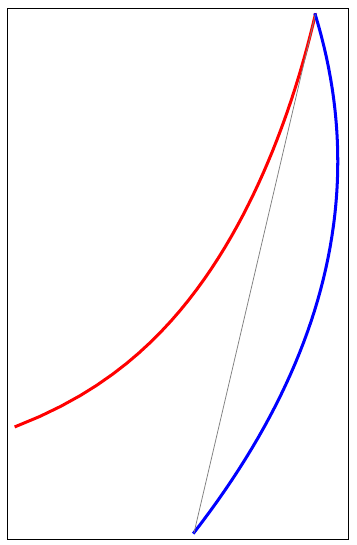}
\caption{Diagrams of $N_s$, red, and $N_\theta$, blue, for a conical reservoir supported at the tip (in grey: the profile of the reservoir; the horizontal distance between this curve and the curve of a diagram measures the value of the quantity, positive when outward).}
\label{fig:f7_27}
\end{center}
\end{figure}

The above results can be used easily to study the case where the cone is supported at its base (for instance, in a water tower); in such a case,
\be
F=\pi\rho\ g\ \tan^2\alpha\left[\frac{h^3}{3}-\xi^2\left(h-\frac{2}{3}\xi\right)\right]
\ee
and, by consequence
\be
N_s=-\frac{\rho\ g}{2\xi} \frac{\tan\alpha}{\cos\alpha}\left[\frac{h^3}{3}-\xi^2\left(h-\frac{2}{3}\xi\right)\right].
\ee
Also in this case, $N_s\rightarrow\infty$ for $s\rightarrow 0$, as a consequence of the assumption of concentrated reaction at the tip. For what concerns $N_\theta$, this does not change with respect to the case of suspended reservoir, because it depends uniquely upon $p_n$, which is the same in the two cases. In Fig. \ref{fig:f7_27} are plotted the diagrams of $N_s$ and $N_\theta$ for this case, $\alpha=30^\circ$ (the curve representing $N_s$ has been cut off near the tip).

\subsubsection{Wind load on a conical vault}
We assume the load to be  $\gr{p}=(p_s,p_\theta,p_n)=(0,0,\frac{1}{2}p\cos\alpha\cos\theta)$; because, from the same definition of curvature radius,
\be
r_1=\frac{ds}{d\varphi},
\ee
eqs. (\ref{eq:equilmemb1}) become 
\be
\label{eq:equilcone}
\besp
&\frac{\partial}{\partial s}(r_2  \sin\varphi N_s)-\cos\varphi\ N_\theta+\frac{\partial N_{s\theta}}{\partial\theta}=-p_s r_2\sin\varphi,\\
&\frac{\partial N_\theta}{\partial \theta}+\frac{\partial (r_2\sin\varphi N_{s\theta})}{\partial s}+\cos\varphi\ N_{s\theta}=-p_\theta r_2\sin\varphi,\\
&N_\theta=-p_nr_2,
\end{split}
\ee
and applied to the present case they give
\be
\label{eq:conewind}
\besp
&\frac{\partial}{\partial s}(s\sin\alpha N_s)-\sin\alpha\ N_\theta+\frac{\partial N_{s\theta}}{\partial\theta}=0,\\
&\frac{\partial N_\theta}{\partial \theta}+\frac{\partial (s\sin\alpha N_{s\theta})}{\partial s}+\sin\alpha\ N_{s\theta}=0,\\
&N_\theta=-\frac{1}{2}p\ s\sin\alpha\cos\theta.
\end{split}
\ee
Injecting the third equation into the second one gives the equation
\be
s\frac{\partial N_{s\theta}}{\partial s}+2N_{s\theta}=-\frac{1}{2}p\ s\sin\theta,
\ee
whose solution is
\be
N_{s\theta}=-\frac{1}{6}p\ s\sin\theta+\frac{f_{s\theta}(\theta)}{s^2}.
\ee
To have  a finite value of $N_{s\theta}$ at the tip, $s=0$,  it must be $f_{s\theta}(\theta)$=0, so finally
\be
N_{s\theta}=-\frac{1}{6}p\ s\sin\theta.
\ee
Injecting this result and eq. (\ref{eq:conewind})$_3$ into eq. (\ref{eq:conewind})$_1$ gives the equation
\be
s\sin\alpha\frac{\partial N_s}{\partial s}+\sin\alpha N_s=\frac{1}{6}p\ s\ \cos\theta(1-3\sin^2\alpha),
\ee
whose solution is
\be
N_s=\frac{1}{12}p\ s\frac{\cos\theta}{\sin\alpha}(1-3\sin^2\alpha)+\frac{f_s(\theta)}{s};
\ee
also in this case, to have a finite value of $N_s$ at the tip, it must be $f_s(\theta)=0$; in the end,
\be
N_s=\frac{1}{12}p\ s\frac{\cos\theta}{\sin\alpha}(1-3\sin^2\alpha).
\ee
To remark that $N_s, N_\theta$ and $N_{s\theta}$ vary linearly with $s$.

\subsubsection{A concentrated force at the tip}
Let au now consider the case of a (non vertical) concentrated force at the tip $\gr{F}=(F_x,F_y,F_z)$. For the equilibrium of the cone,
it must be
\be
\label{eq:forcesconetip}
\besp
&F_x=\int_0^{2\pi} N_s\ s\sin\alpha\cos\alpha\ d\theta,\\
&F_y=\int_0^{2\pi} (N_s\sin\alpha\cos\theta-N_{s\theta}\sin\theta)s\sin\alpha\ d\theta,\\
&F_z=\int_0^{2\pi} (N_s\sin\alpha\sin\theta+N_{s\theta}\cos\theta)s\sin\alpha\ d\theta.
\end{split}
\ee
In this case, $\gr{p}=(p_s,p_\theta,p_n)=(0,0,0)$, so $N_\theta=0$, like in the case of Sect. \ref{sec:verttipload}. 
Eq. (\ref{eq:conewind})$_2$ becomes
\be
\frac{\partial (s\sin\alpha N_{s\theta})}{\partial s}+\sin\alpha\ N_{s\theta}=0\ \rightarrow\ s\frac{\partial  N_{s\theta}}{\partial s}+2 N_{s\theta}=0,
\ee
whose solution is
\be
\label{eq:membconetipNstheta}
N_{s\theta}=\frac{f_1(\theta)}{s^2}.
\ee
Inserting this result into eq. (\ref{eq:conewind})$_1$ gives 
\be
\frac{\partial}{\partial s}(s\sin\alpha N_s)+\frac{1}{s^2}\frac{df_1(\theta)}{d\theta}=0\ \rightarrow\ s\frac{\partial N_s}{\partial s}+N_s+\frac{1}{s^2\sin\alpha}\frac{df_1(\theta)}{d\theta}=0,
\ee
whose solution is
\be
\label{eq:membconetipNs}
N_s=\frac{1}{s^2\sin\alpha}\frac{df_1(\theta)}{d\theta}+\frac{f_2(\theta)}{s}.
\ee
We put the two unknown functions 
\be
f_1(\theta)=A\cos\theta,\ \ \ f_2(\theta)=B\cos\theta,\ \ \ A,B\in\mathbb{R},
\ee
and we determine the two constants $A$ and $B$ inserting eqs. (\ref{eq:membconetipNstheta}) and (\ref{eq:membconetipNs}) into eqs. (\ref{eq:forcesconetip}); to be specific, we consider the case $\gr{F}=(0,F_y,0)$, so we get the conditions
\be
\besp
&\int_0^{2\pi} \left(-\frac{A\sin\theta}{s\sin\alpha}+B\cos\theta\right)\sin\alpha\cos\alpha\ d\theta=0,\\
&\int_0^{2\pi} \left[\left(-\frac{A\sin\theta}{s\sin\alpha}+B\cos\theta\right)\sin\alpha-\frac{1}{s}A\sin\theta\right]\sin\alpha\cos\theta\ d\theta=F_y\,\\
&\int_0^{2\pi} \left[\left(-\frac{A\sin\theta}{s\sin\alpha}+B\cos\theta\right)\sin\alpha\sin\theta+\frac{1}{s}A\cos^2\theta\right]\sin\alpha\ d\theta=0,
\end{split}
\ee
that give
\be
F_y=\pi B\sin^2\alpha\ \rightarrow\ B=\frac{F_y}{\pi\sin^2\alpha}\ \rightarrow\ A=0.
\ee
Finally, $N_\theta=N_{s\theta}=0$ while
\be
N_s=\frac{F_y}{\pi\ s\ \sin^2\alpha}\cos\theta.
\ee
We have already seen in Sect. \ref{sec:verttipload} that a vertical (upward) force $F_x$ produces the internal action
\be
N_s=\frac{F_x}{\pi s\sin2\alpha}.
\ee
If $F_y=F_x\tan\alpha$, the tip force is aligned with a meridian, i.e. it is tangent to the cone's surface,  and, by the superposition of the effects, we get
\be
N_s=\frac{F_y}{2\pi s\sin\alpha\cos\alpha}(1+2\cos\theta)=\frac{F_y}{2\pi \xi\sin\alpha}(1+2\cos\theta).
\ee
Once more, the paradox that the membrane forces tend towards infinity when $s\rightarrow0$ is due to the model of concentrated force.

\subsection{Pipes}
A pipe of radius $r$ is a degenerated conic vault, where $\varphi=\pi/2\-\rightarrow\ \alpha=0$, $r_1\rightarrow\infty$ and $r_2=r$. Eqs. (\ref{eq:equilcone}) become hence
\be
\label{eq:equilpipe}
\besp
&r\frac{\partial N_s}{\partial s}+\frac{\partial N_{s\theta}}{\partial\theta}=-p_s r,\\
&\frac{\partial N_\theta}{\partial \theta}+r\frac{\partial N_{s\theta}}{\partial s}=-p_\theta r,\\
&N_\theta=-p_nr.
\end{split}
\ee
A particularly important case is that of a pipe submitted to an internal pressure $p\ \rightarrow\ \gr{p}=(p_s,p_\theta,p_n)=(0,0,-p),\ p>0$. Then, eq. (\ref{eq:equilpipe})$_3$ gives immediately\footnote{This result is the famous formule of Mariotte, that can  be easily  found  using the global equilibrium approach, cfr. Sect. \ref{sec:globeqmemb}. Mariotte found this result $(1686)$ studying the system of pipes for the fountains of the garden of the {\it Château de Versailles}.}
\be
N_\theta=p\ r.
\ee
Because the load is axisymmetric and $p_\theta=0,  N_{s\theta}=0$, cfr. Sect. \ref{sec:axisymload}. Hence, from eq. (\ref{eq:equilpipe})$_1$ we get
\be
\label{eq:fspipes}
r\frac{\partial N_{s}}{\partial s}=0\ \rightarrow\ N_{s}=f(\theta).
\ee
If we consider an infinitely long pipe, nothing depends upon $s\ \rightarrow\ N_{s}=const$, so $N_s$ has everywhere the value that it takes in a given section (in particular, if it is null somewhere, it is null everywhere).

\subsection{Hyperbolic hyperboloid}
\begin{figure}[th]
\begin{center}
\includegraphics[height=.3\textwidth]{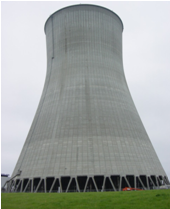}
\includegraphics[height=.3\textwidth]{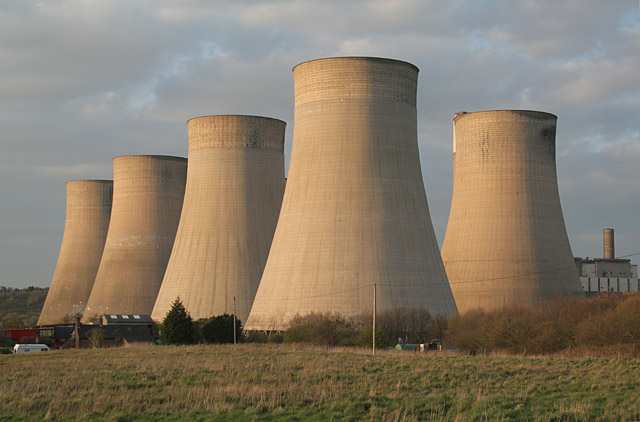}
\includegraphics[height=.3\textwidth]{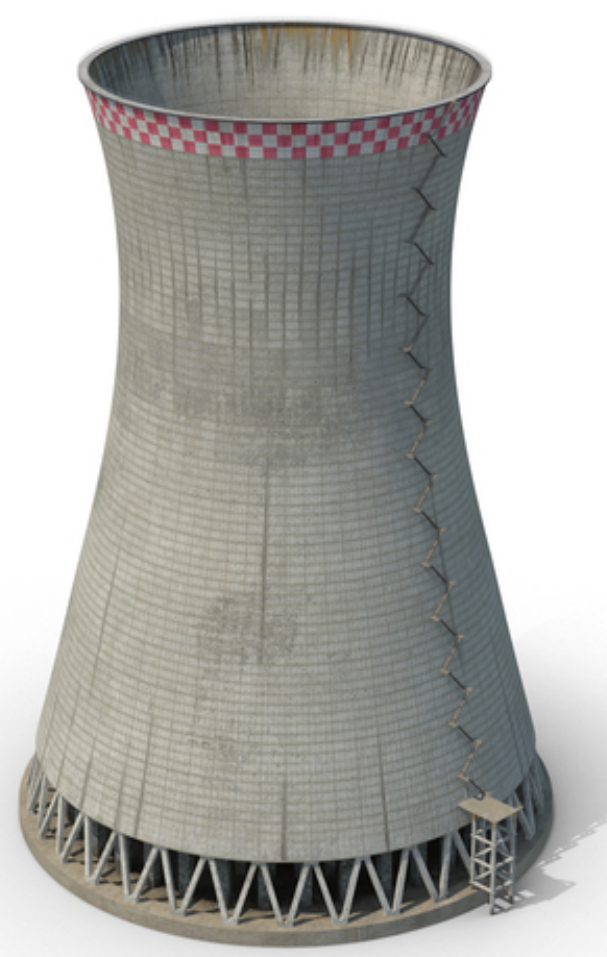}
\caption{Cooling towers in the form of hyperbolic hyperboloids.}
\label{fig:f7_28}
\end{center}
\end{figure}

Hyperbolic hyperboloids are quadric ruled surfaces, often used to shape some structures like cooling towers, Fig. \ref{fig:f7_28}.
The Cartesian equation of an axisymmetric hyperbolic hyperboloid of axis $z$ is
\be
\frac{x^2}{a^2}+\frac{y^2}{a^2}-\frac{z^2}{b^2}=1.
\ee
However, for our purposes the best way to tackle the problem is to use a parametric form of the equation of the surface; there are different ways to parameterize the surface of a hyperbolic hyperboloid, the well suited for our study is the following one
\be
\left\{
\besp
&x=\alpha\cosh u\ \cos\theta,\\
&y=\alpha\cosh u\ \sin\theta,\\
&z=\beta\sinh u+z_0,
\end{split}
\right.
\ \ \ u\in[u_1,u_2].
\ee
The three constants $\alpha,\beta$ and $z_0$ are adjusted to fit the surface to a given case; for  a hyperboloid with $0\leq z\leq h$ and with a minimum radius $R_0$, a radius $R_1$ for $z=0$ and a radius $R_2$ for $z=h$, it is
\begin{figure}[th]
\begin{center}
\includegraphics[width=.64\textwidth]{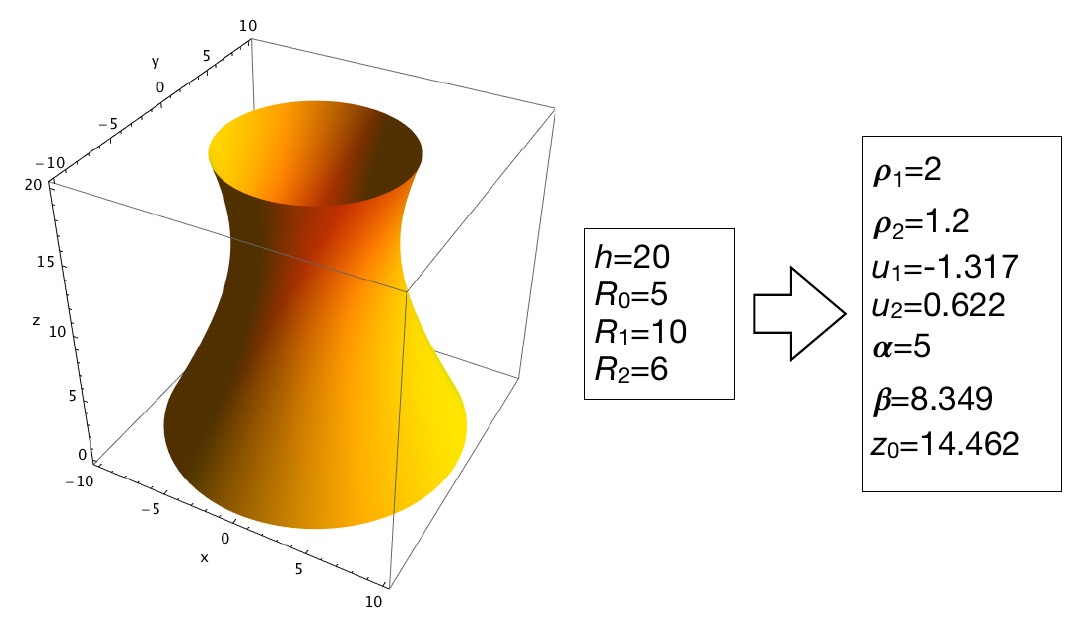}
\includegraphics[width=.34\textwidth]{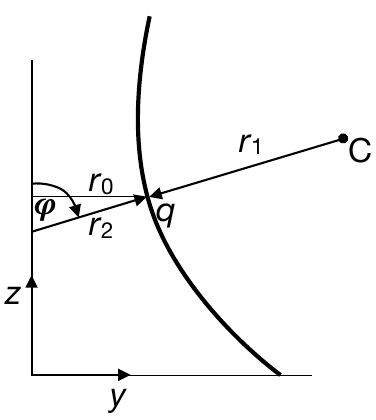}
\caption{Geometry of a hyperboloid; point C denotes the  osculating circle center relative to point $q$.}
\label{fig:f7_29}
\end{center}
\end{figure}
\be
\besp
&\alpha=R_0,\ \ \beta=\frac{h}{\sinh u_2-\sinh u_1},\ \ z_0=-\beta\sinh u_1,\\ 
&u_1=\log\left(\rho_1-\sqrt{\rho_1^2-1}\right),\ \ u_2=\log\left(\rho_2+\sqrt{\rho_2^2-1}\right), \\
\end{split}
\ee
where $\rho_1=R_1/R_0,\rho_2=R_2/R_0$.
In Fig. \ref{fig:f7_29} we show the geometry of a hyperboloid plotted using the above parametric equation.
With a little effort, one can calculate
\be
\besp
&r_1=-\frac{1}{\alpha\beta}(\alpha^2\sinh^2u+\beta^2\cosh^2u)^{\frac{3}{2}},\\
&r_2=\frac{\alpha}{\beta}(\alpha^2\sinh^2u+\beta^2\cosh^2u)^{\frac{1}{2}},
\end{split}
\ee
so\footnote{To remark the fact that $r_1<0$, cfr. Sect. \ref{sec:resomemb} and Fig. \ref{fig:f7_29}.}
\be
r_1=-\frac{r_2}{\alpha^2}(\alpha^2\sinh^2u+\beta^2\cosh^2u),
\ee
and 
\be
\label{eq:sincostanhyperb}
\besp
&\sin\varphi=\frac{\beta\cosh u}{\sqrt{\alpha^2\sinh^2u+\beta^2\cosh^2u}},\\
&\cos\varphi=-\frac{\alpha\sinh u}{\sqrt{\alpha^2\sinh^2u+\beta^2\cosh^2u}},\\
&\tan\varphi=-\frac{\beta}{\alpha\tanh u}.
\end{split}
\ee
We consider here the case of a cooling tower submitted uniquely to its own weight. Then,
\be
\besp
\gr{p}&=(p_\varphi,p_\theta,p_n)=(p\sin\varphi,0,p\cos\varphi)\\
&=\frac{p}{\sqrt{\alpha^2\sinh^2u+\beta^2\cosh^2u}}(\beta\cosh u,0,\alpha\sinh u).
\end{split}
\ee
Replacing the above quantities in eq. (\ref{eq:equilmemb7})$_1$ gives the equation
\be
\frac{\partial U}{\partial \varphi}=\frac{p}{\beta^2}(\alpha^2\sinh^2u+\beta^2\cosh^2u)^2.
\ee
Moreover,
\be
\frac{\partial U}{\partial u}=\frac{\partial U}{\partial \varphi}\frac{\partial \varphi}{\partial u},
\ee
and, through eq. (\ref{eq:sincostanhyperb})$_3$,
\be
\frac{\partial \varphi}{\partial u}=\frac{\partial}{\partial u}\arctan\left(-\frac{\beta}{\alpha\tanh u}\right)=\frac{\alpha\beta}{\alpha^2\sinh^2u+\beta^2\cosh^2u}.
\ee
Hence, we get the equation
\be
\frac{\partial U}{\partial u}=p\frac{\alpha}{\beta}(\alpha^2\sinh^2u+\beta^2\cosh^2u),
\ee
whose solution is
\be
U=p\frac{\alpha}{\beta}\left(\frac{\alpha^2+\beta^2}{4}\sinh 2u+\frac{\beta^2-\alpha^2}{2}u+c\right),
\ee
and hence, through eq. (\ref{eq:axisymNphi}),
\be
N_\varphi=p\frac{\beta^2\cosh^2u}{(\alpha^2\sinh^2u+\beta^2\cosh^2u)^\frac{3}{2}}\left(\frac{\alpha^2+\beta^2}{4}\sinh 2u+\frac{\beta^2-\alpha^2}{2}u+c\right).
\ee
The constant $c$ can be determined imposing that on the upper edge the meridian force is null: $N_\varphi(u=u_2)=0$; this gives
\be
c=-\left(\frac{\alpha^2+\beta^2}{4}\sinh 2u_2+\frac{\beta^2-\alpha^2}{2}u_2\right),
\ee
and finally
\be
\besp
N_\varphi=&p\frac{\beta^2\cosh^2u}{(\alpha^2\sinh^2u+\beta^2\cosh^2u)^\frac{3}{2}}\ \times\\
&\left(\frac{\alpha^2+\beta^2}{4}(\sinh 2u-\sinh2u_2)+\frac{\beta^2-\alpha^2}{2}(u-u_2)\right).
\end{split}
\ee
$N_\theta$ can be calculated once more using eq. (\ref{eq:axisymNtheta}):
\be
\besp
N_\theta=&p\ \alpha^2\left[\frac{\sinh u}{\beta}\frac{\beta^2\cosh^2u}{(\alpha^2\sinh^2u+\beta^2\cosh^2u)^\frac{5}{2}}\right.\ \times\\
&\left.\left(\frac{\alpha^2+\beta^2}{4}(\sinh 2u-\sinh2u_2)+\frac{\beta^2-\alpha^2}{2}(u-u_2)\right)\right].
\end{split} 
\ee
We plot in Fig. \ref{fig:f7_30} the diagrams of $N_\varphi$ and $N_\theta$ for the case of the hyperboloid represented in Fig. \ref{fig:f7_29}.

\begin{figure}[th]
\begin{center}
\includegraphics[height=.4\textwidth]{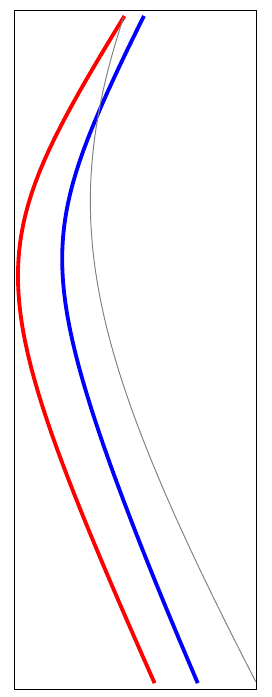}
\caption{Diagrams of $N_s$, red, and $N_\theta$, blue, for  the hyperboloid in Fig. \ref{fig:f7_29} (in grey: the profile of the hyperboloid; the horizontal distance between this curve and the curve of a diagram measures the value of the quantity, positive when outward).}
\label{fig:f7_30}
\end{center}
\end{figure}
We can see that, while $N_\varphi<0\ \forall u$, i.e., the meridians are always in compression, the parallels are in tension in the upper part of the hyperboloid. This solution implicitly implies that the base of the hyperboloid is simply supported in the same way of the spherical dome of Fig. \ref{fig:f7_12}. This is the reason of the supports that are visible in Fig. \ref{fig:f7_28}.

\section{Vaults}
Some kinds of vaults can be treated with the theory of axisymmetric membranes; however, there is a substantial difference with respect to the above cases, it concerns the boundary conditions because now the membrane has two supplementary boundaries; we will see that this limits the application of the theory and determines the need to pass to the shell theory, i.e. of structural surfaces having a non null bending stiffness. In order to use the results of the theory of axisymmetric membranes, we limit our study to the vaults having the cross section in the form of an arc of a circle of radius $r$.

\subsection{Barrel vaults}
\begin{figure}[th]
\begin{center}
\includegraphics[height=.25\textwidth]{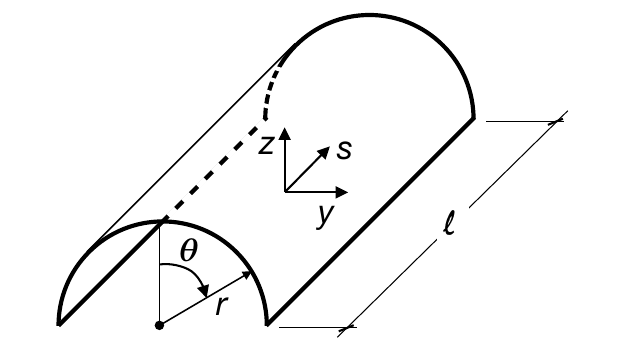}
\caption{Scheme of a barrel vault.}
\label{fig:f7_31}
\end{center}
\end{figure}

A barrel vault is a vault whose cross section is half a circle, see Fig. \ref{fig:f7_31}. We consider the vault loaded by its own weight: $\gr{p}=(p_s,p_\theta,p_n)=(0,p\sin\theta,p\cos\theta)$.
\be
\label{eq:equilvoute}
\besp
&r\frac{\partial N_s}{\partial s}+\frac{\partial N_{s\theta}}{\partial\theta}=0,\\
&\frac{\partial N_\theta}{\partial \theta}+r\frac{\partial N_{s\theta}}{\partial s}=-p\ r\sin\theta,\\
&N_\theta=-p\ r\cos\theta.
\end{split}
\ee
Injecting the third equation in the second one gives
\be
\frac{\partial N_{s\theta}}{\partial s}=-2p\sin\theta,
\ee
whose solution is 
\be
N_{s\theta}=-2p\ s\sin\theta+f(\theta),
\ee
and replacing this in the first equation above we get
\be
r\frac{\partial N_s}{\partial s}-2p\ s\cos\theta+f'(\theta)=0,
\ee
which gives
\be
N_s=\frac{p\ s^2}{r}\cos\theta-\frac{s}{r}f'(\theta)+g(\theta).
\ee
To remark that the effect of the function $g(\theta)$, that account for the boundary conditions, is that of propagating a local effect (that on the boundary) unaltered along a generatrix of the vault. This fact, typical of surfaces where one of the two curvatures is null (this happens also in the conical vaults, see e.g. eq. (\ref{eq:membconetipNstheta}), and pipes, eq. (\ref{eq:fspipes})), is a consequence of the membrane model and it is produced by the equilibrium equations; in an elastic model, where the constitutive law enter the equilibrium equations, this should not happen.

\begin{figure}[th]
\begin{center}
\includegraphics[height=.25\textwidth]{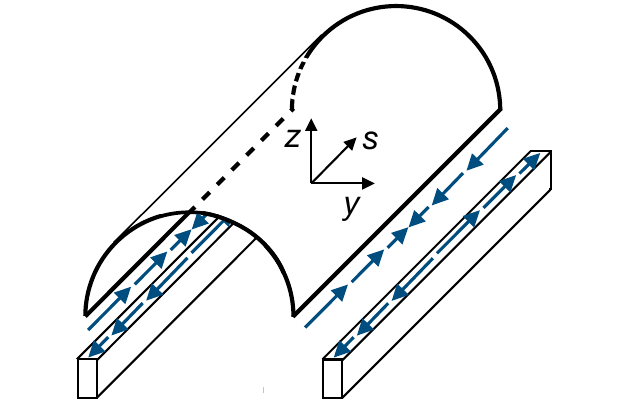}
\caption{Scheme of a barrel vault with supporting beams.}
\label{fig:f7_32}
\end{center}
\end{figure}

\begin{figure}[th]
\begin{center}
\includegraphics[height=.25\textwidth]{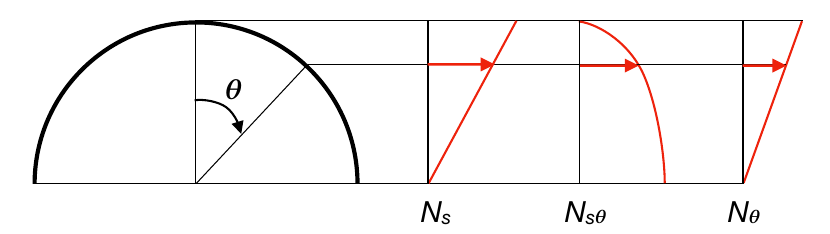}
\caption{Distribution fo the membrane forces for a generic cross section of the vault.}
\label{fig:f7_33}
\end{center}
\end{figure}

It is also worth noticing  that $N_\theta$ is unaffected by the boundary conditions and $N_\theta(\pm\pi/2)=0$: there is no need to have a support for the vault all along the two straight edges. Actually, this solution corresponds to the case of a vault that is supported at the two curved edges and that is bent by its own weight. If we take the reference frame as in Fig. \ref{fig:f7_31}, then $N_{s\theta}(0)=0$  by symmetry, which implies that $f(\theta)=0$, so that
\be
N_{s\theta}=-2p\ s\sin\theta.
\ee
If we impose that at the two curved edges it is $N_s(\pm\ell/2)=0$, i.e. if we assume that  globally the vault corresponds to a simply supported beam, then 
\be
\label{eq:Nsvault}
g(\theta)=-\frac{p\ \ell^2}{4r}\cos\theta\ \rightarrow\ N_s=-\frac{p\ \cos\theta}{4r}(\ell^2-4s^2).
\ee
We remark also that on the two straight edges, $N_s(\pm\pi/2)=N_\theta(\pm\pi/2)=0$, but the shear force is not null: 
\be
\label{eq:shearbarrel}
N_{s\theta}=\mp2p\ s.
\ee
This result is inconsistent with the boundary condition of free straight edges. Hence, to make the above solution valid, the shear (\ref{eq:shearbarrel}) must be applied to the straight edges. This can be done using  beams supporting the straight edges, like in Fig. \ref{fig:f7_32}.
Such beams are submitted to an axial  tension force $t$ 
\be
t=-\int_s^\frac{\ell}{2}N_{s\theta}\left(\theta=\frac{\pi}{2}\right)ds^*=\int_s^\frac{\ell}{2}2p\ s^*ds^*=\frac{p}{4}(\ell^2-4s^2).
\ee
So, from eq. (\ref{eq:Nsvault}), we get also that
\be
N_s=-\frac{t}{r}\cos\theta,
\ee
from which we see that  $N_s\leq0$ everywhere: the vault is compressed axially and the resultant of $N_s$ is equal to $2t$. In correspondence of a generic cross section, the diagrams of $N_s,N_{s\theta}$ and $N_\theta$ are shown in Fig. \ref{fig:f7_33}. At the two curved edges, 
\be
N_{s\theta}\left(s=\pm\frac{\ell}{2}\right)=\mp p\ \ell\sin\theta,
\ee
and also this force must be provided through supporting arches or diaphragms. The solution found in this Section, hence, does {\it not} correspond to that of a classical barrel vault supported at the two straight edges. 

In fact, for such a case we should satisfy the boundary conditions
\be
N_\theta\left(\theta=\pm\frac{\pi}{2}\right)=-p\ r,\ \ N_s\left(s=\pm\frac{\ell}{2}\right)=N_{s\theta}\left(s=\pm\frac{\ell}{2}\right)=0,
\ee
that give
\be
\besp
&\mp p\ell\sin\theta+f(\theta)=0,\\
&\frac{p\ell^2}{4r}\cos\theta\mp\frac{\ell}{2r}f'(\theta)+g(\theta)=0,
\end{split}
\ee
and also the conditions
\be
N_s\left(\theta=\pm\frac{\pi}{2}\right)=N_{s\theta}\left(\theta=\pm\frac{\pi}{2}\right)=0,
\ee
that give
\be
f(\theta)=\pm p\ell,\ \ g(\theta)=0.
\ee
This result does not coincide with the values found above for $f(\theta)$ and $g(\theta)$, which confirms that the solution found in this Section cannot represent the case of a classical barrel vault supported at the straight edges. 

Actually, such a problem is not determined because there are too much boundary conditions to be satisfied: those on the two straight edges. This can be seen also if one considers that it is not possible that $N_\theta$ satisfy at $\theta=\pm \pi/2$ boundary conditions other than $N_\theta=0$. Finally, such a problem is intrinsically hyperstatic and cannot be solved by the membrane theory, based uniquely upon equilibrium equations, i.e. intrinsically isostatic.

\subsection{Cross vaults}
Cross vaults are vaults composed by intersecting vaults; in the simplest scheme, two identical barrel vaults cross at a right angle, so their intersection happens along two diagonal elliptical arches. If these two arches are not reinforced, then the vault is called a {\it groin vault}. Groin vaults have been widely used, especially for vaults on a square plan; normally, they are realized in bricks and normally used to cover small spans (some exceptions are the vaults of some great Roman {\it thermae}).

\begin{figure}[th]
\begin{center}
\includegraphics[height=.28\textwidth]{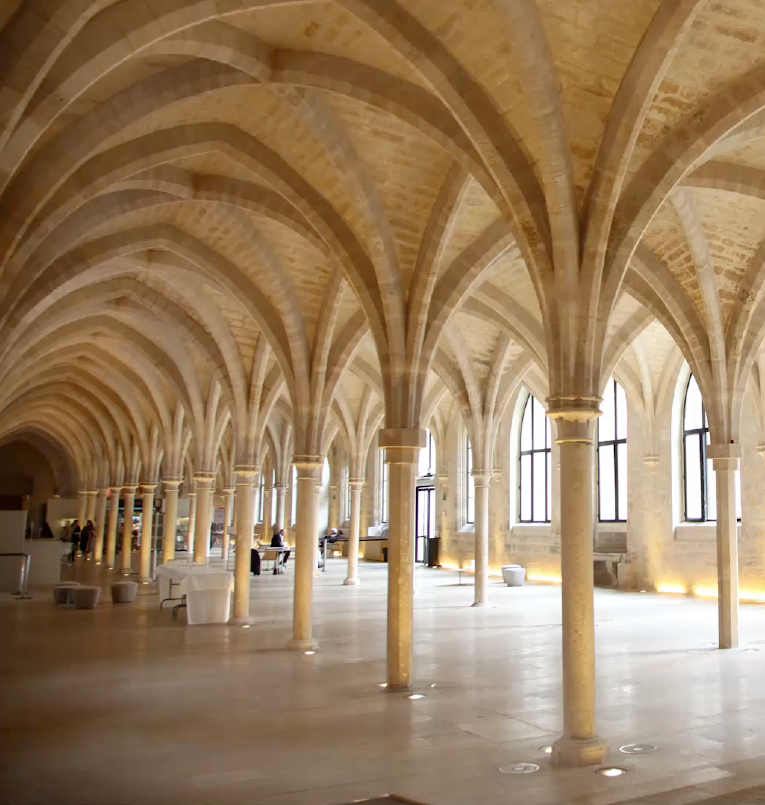}
\includegraphics[height=.28\textwidth]{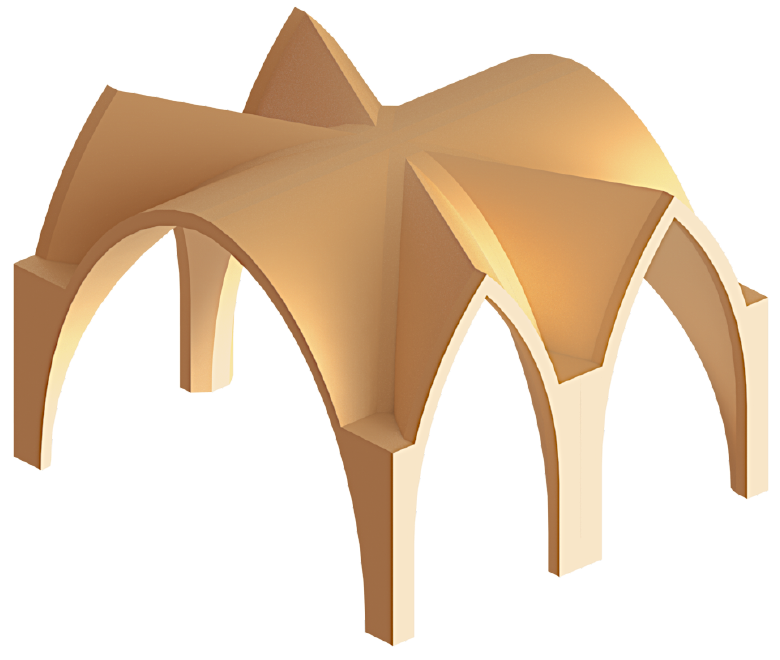}
\includegraphics[height=.28\textwidth]{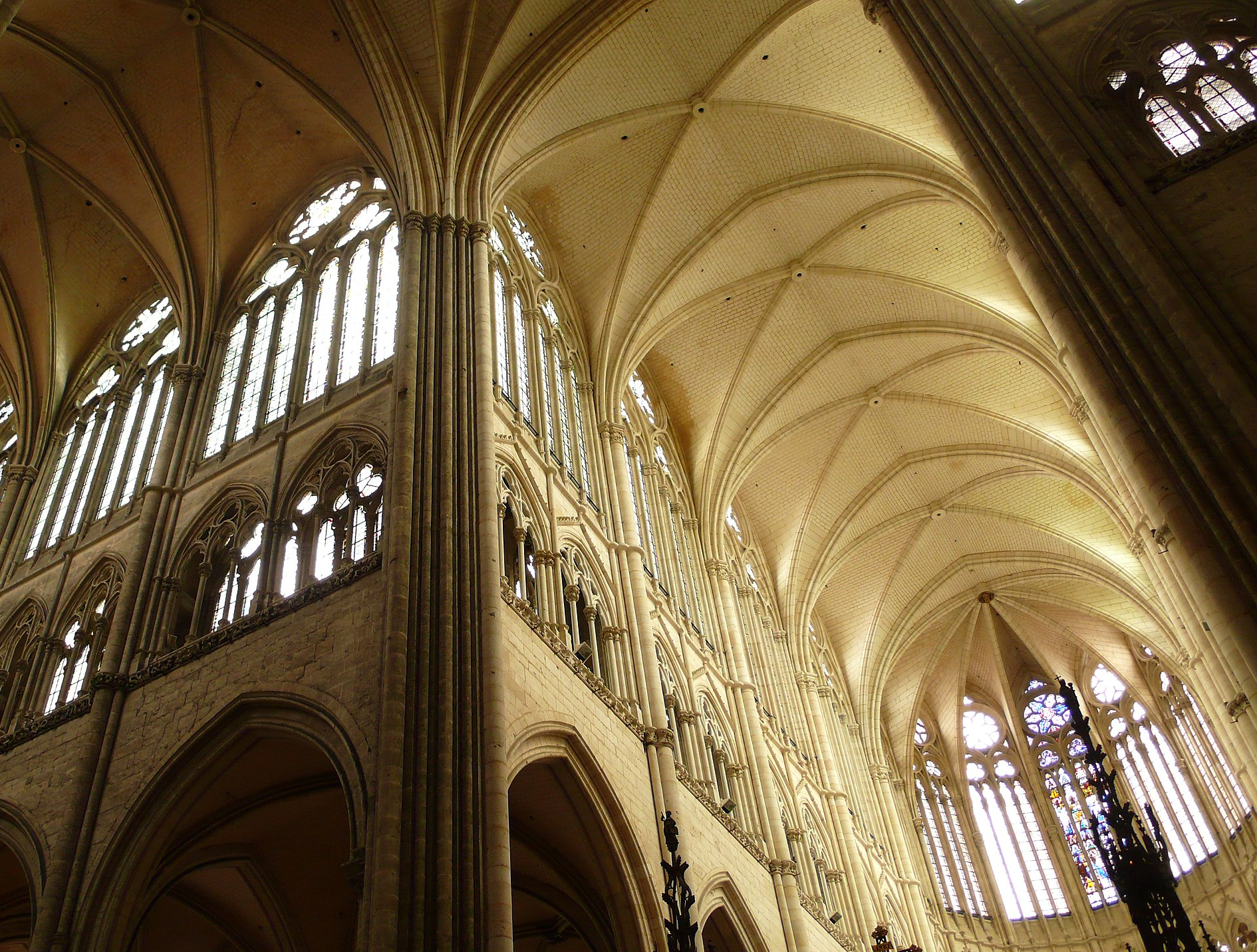}
\caption{Different types of rib vaults. From the left: quadripartite vault of the {\it Collège des Bernardins}, Paris, scheme of a sexpartite vault, {\it barlong} vaults of the Cathedral of Amiens.}
\label{fig:f7_34}
\end{center}
\end{figure}

\begin{figure}[b]
\begin{center}
\includegraphics[width=.5\textwidth]{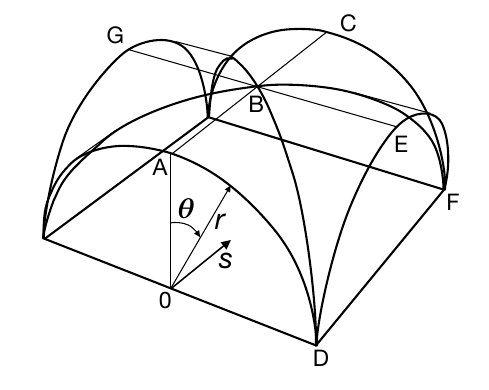}
\caption{Scheme for the calculation of a quadripartite rib vault.}
\label{fig:f7_35}
\end{center}
\end{figure}

\begin{figure}[h]
\begin{center}
\includegraphics[width=.65\textwidth]{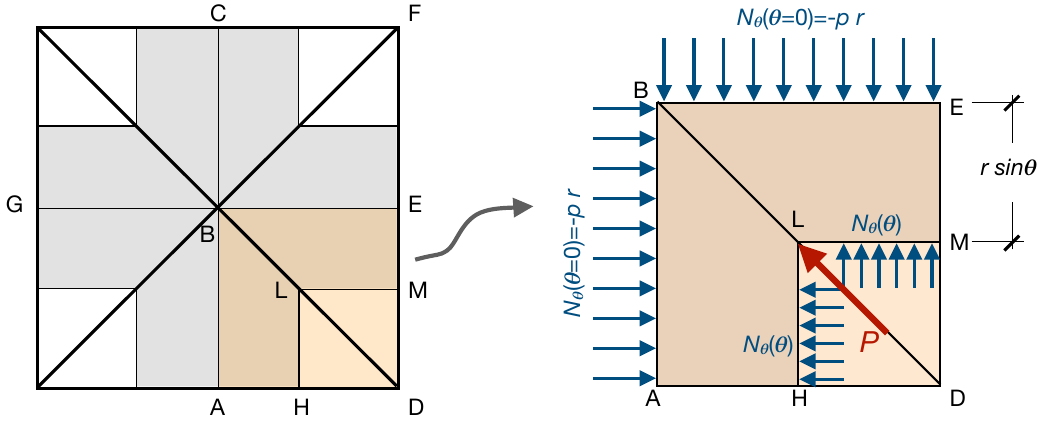}
\includegraphics[width=.33\textwidth]{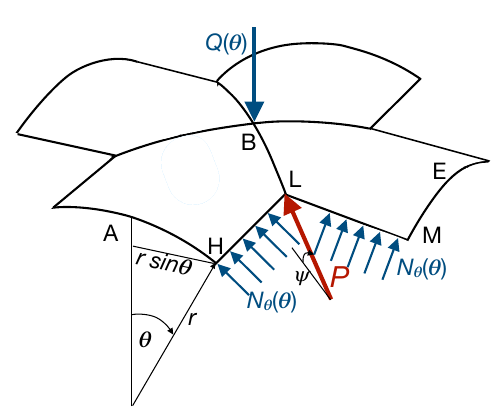}
\caption{Scheme for the calculation of the membrane forces in a quadripartite rib vault.}
\label{fig:f7_36}
\end{center}
\end{figure}

\begin{figure}[h]
\begin{center}
\includegraphics[width=.5\textwidth]{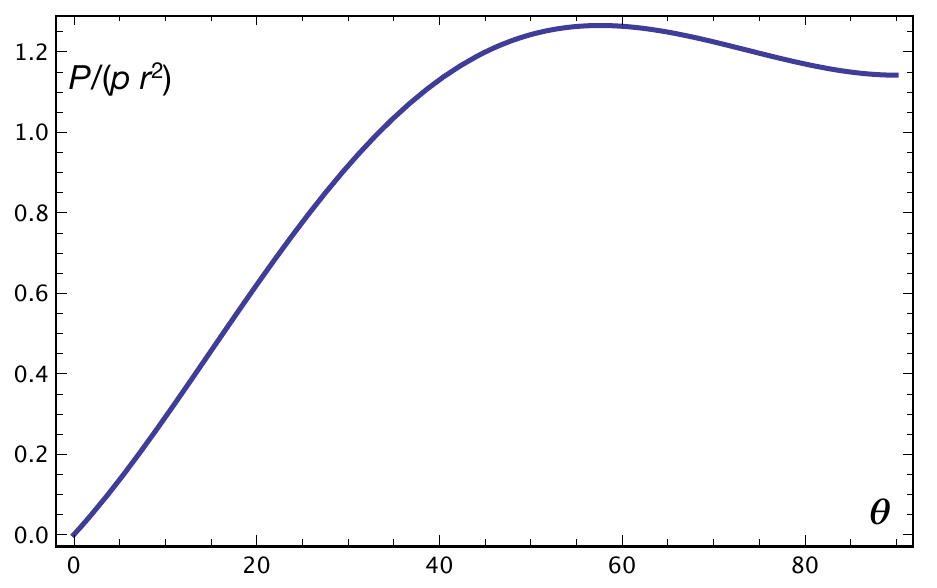}
\caption{Diagram of the function $P(\theta)$.}
\label{fig:f7_37}
\end{center}
\end{figure}

\begin{figure}[th]
\begin{center}
\includegraphics[width=.5\textwidth]{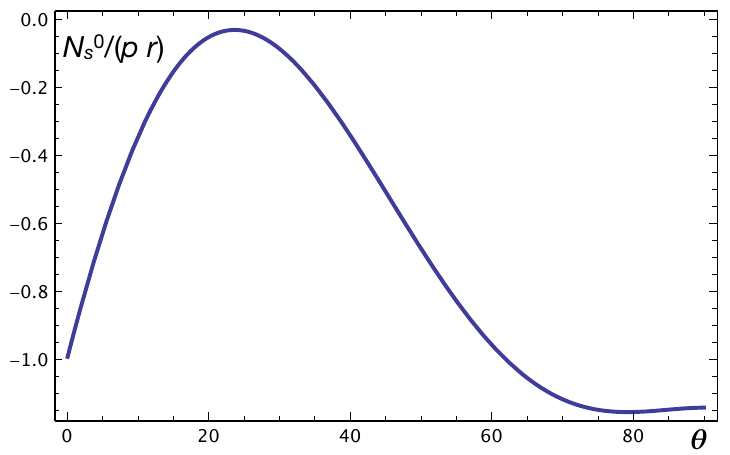}
\caption{Diagram of $N_s^0(\theta)$.}
\label{fig:f7_39}
\end{center}
\end{figure}

For larger spans to be covered, the diagonal arches are reinforced by ribs and in this case we talk of {\it rib vaults}; the rib vault is the main architectural element of Gothic architecture and it can take different dimensions and geometries: quadripartite, {\it barlong}, sexpartite etc. see Fig. \ref{fig:f7_34}.
A cross vault is a typical structure that stands thanks to its geometry; we study in this Section the equilibrium of a quadripartite rib vault composed by two intersecting identical barrel vaults of radius $r$, see Fig. \ref{fig:f7_35}, and submitted to its own weight\footnote{This part is essentially inspired by the work of J. Heyman, see references. It is worth noting, however, that in this Section we present just the {\it membrane solution} for a rib vault while, if the no-tension behavior of the material is considered, the solution is different and cannot be found with the classical membrane theory, cfr. again the book of Heyman}.
In such a case, the equilibrium equations for each one of the four parts of the vaults are still eqs. (\ref{eq:equilvoute}), whose solution is again the one found in the previous Section:
\be
\label{eq:crossvault0}
\besp
&N_s=\frac{p\ s^2}{r}\cos\theta-\frac{s}{r}f'(\theta)+g(\theta),\\
&N_{s\theta}=-2p\ s\sin\theta+f(\theta),\\
&N_\theta=-p\ r\cos\theta.
\end{split}
\ee
To determine the unknown functions $f(\theta)$ and $g(\theta)$, let us consider one of the eight shells composing the vault in Fig. \ref{fig:f7_35}, e.g. the shell ABD; we impose 
\be
N_{s\theta}(s=0)=0\ \forall\theta\ \rightarrow\ f(\theta)=0.
\ee
Then, to determine $g(\theta)$ we assume the solution to be symmetric with respect to the longitudinal and transversal directions: the distribution of the membrane forces in shells ABD and EBD is the same; this fact implies that $N_{s\theta}=0$ also on the edge EFD.
At the top of the vault, lines AC and EG, $\theta=0$, it is
\be
N_\theta=-p\ r\cos\theta=-p\ r,
\ee
i.e. at the top the shells are compressed. If we imagine a horizontal section of the vault like in Fig. \ref{fig:f7_36}, then the vertical equilibrium can be given only thanks to forces $N_\theta$, because along lines HL and LM the shear forces $N_{s\theta}$ are horizontal. 
The total vertical component of $N_\theta$ on the 8 lines like HL or LM is
\be
N_\theta^{vert}=8N_\theta r\sin\theta(1-\sin\theta)=-8p\ r^2\sin\theta\cos\theta(1-\sin\theta),
\ee
while the total weight of the vault above the horizontal section (the part shadowed in grey in Fig. \ref{fig:f7_36}) is 
\be
\label{eq:pesovolta}
Q(\theta)=8p\ r^2\int_0^\theta1-\sin\theta^*d\theta^*=8p\ r^2(\theta+\cos\theta-1).
\ee
Then, if we consider the vertical equilibrium of this part of the vault, we get
\be
Q(\theta)+N_\theta^{vert}=8p\ r^2(\theta+\cos\theta-1-\sin\theta\cos\theta(1-\sin\theta))\neq0.
\ee
Hence, for the equilibrium to be possible, there is the need of a device that takes on this difference of the vertical forces. Such a device can be only the rib at the intersection of the shells (along the diagonal arches). We then suppose  that the rib is compressed by a force $P$ inclined of $\psi$ on the horizontal; hence
\be
\tan\theta=\sqrt{2}\tan\psi\ \rightarrow\ \sin\psi=\frac{\tan\psi}{\sqrt{1+\tan^2\psi}}=\frac{\tan\theta}{\sqrt{2+\tan^2\theta}}.
\ee
The vertical equilibrium is satisfied if, see Fig. \ref{fig:f7_36},
\be
\label{eq:crossvaultP}
\besp
&4P\sin\psi=Q(\theta)+N_\theta^{vert}\ \rightarrow\ P=\frac{Q(\theta)+N_\theta^{vert}}{4\sin\psi}\ \rightarrow\\
&P(\theta)=2p\ r^2(\theta+\cos\theta-1-\sin\theta\cos\theta(1-\sin\theta))\frac{\sqrt{2+\tan^2\theta}}{\tan\theta}.
\end{split}
\ee
The diagram of the function $P(\theta)$ is plotted in Fig. \ref{fig:f7_37}; numerically, one can calculate that
\be
P_{max}=P(\theta=57.61^\circ)=1.265\ p\ r^2.
\ee
To determine the force $N_s^0$ along the edges like  AD or EF, we make the horizontal equilibrium of  the rib in the part LMDH, see Fig. \ref{fig:f7_38}:
\begin{figure}[th]
\begin{center}
\includegraphics[width=.4\textwidth]{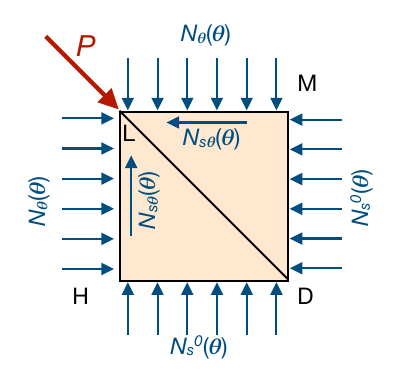}
\caption{Scheme for the calculation of actions $N_s^0$ (the actions in blue are those applied to the vault, $P$, in red, is the compression in the rib).}
\label{fig:f7_38}
\end{center}
\end{figure}
\be
\label{eq:crossvault1}
P\cos\psi+\sqrt{2}\int_0^{r(1-\sin\theta)}N_{s\theta}-N_\theta\cos\theta\ ds+\sqrt{2}\int_\theta^\frac{\pi}{2}N_s^0r\ d\theta=0.
\ee
Then, on one hand
\be
\cos\psi=\frac{1}{\sqrt{1+\tan^2\psi}}=\frac{\sqrt{2}}{\sqrt{2+\tan^2\theta}},
\ee
and, on the other hand,
\be
\int_0^{r(1-\sin\theta)}N_{s\theta}-N_\theta\cos\theta\ ds=p\int_0^{r(1-\sin\theta)}r\cos^2\theta-2s\ \sin\theta\ ds=p\ r^2(1-\sin\theta)^2.
\ee
So, injecting these results into eq. (\ref{eq:crossvault1}) we get
\be
\frac{P}{\sqrt{2+\tan^2\theta}}+p\ r^2(1-\sin\theta)^2+r\int_\theta^\frac{\pi}{2}N_s^0\ d\theta=0,
\ee
and hence
\be
N_s^0(\theta)=\frac{d}{d\theta}\left(\frac{P}{r\sqrt{2+\tan^2\theta}}+p\ r(1-\sin\theta)^2\right).
\ee
Injecting now the expression of $P$, eq. (\ref{eq:crossvaultP}), and differentiating we obtain finally
\be
N_s^0(\theta)=2p\ r\left(\frac{(1-\sin\theta)(1+3\sin^2\theta)}{\tan\theta}-\frac{\theta+\cos\theta-1}{\sin^2\theta}\right).
\ee
Of course, this is actually the expression of function $g(\theta)$ in eq. (\ref{eq:crossvault0})$_1$. The function $N_s^0(\theta)$ is plotted in Fig. \ref{fig:f7_39}. As apparent, $N_s^0$ is a compression and we can remark that
\be
-2\int_0^\frac{\pi}{2}r\ N_s^0\ d\theta=2p\ r^2,
\ee
just as it must be, in order to equilibrate the forces on AC or GE. 

\begin{figure}[th]
\begin{center}
\includegraphics[width=.4\textwidth]{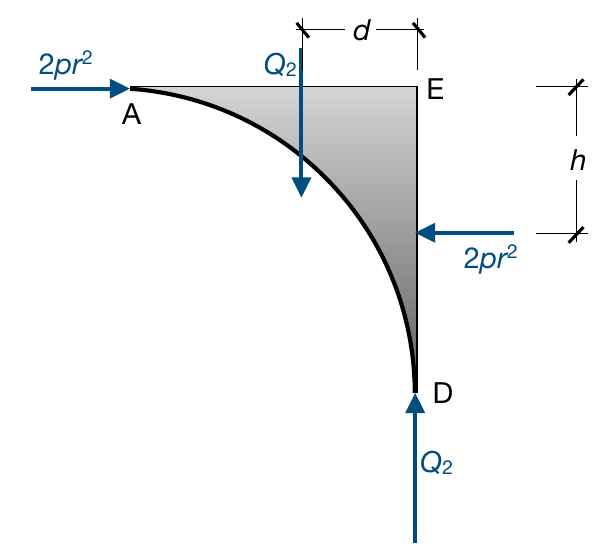}
\caption{Scheme for the calculation of the thrust point.}
\label{fig:f7_40}
\end{center}
\end{figure}

The above results show that on the curved edges like DEF, the cross vault is submitted to horizontal compressions whose resultant is exactly $2p\ r^2$. Such a force must be equilibrated in some way. The device that was introduced by the {\it master builders} of the  Gothic architecture was the flying buttress. To evaluate where the thrust point is located, i.e. at which position on the height of the cross-vault to put the flying buttress, we can make the equilibrium to rotation about E of the shell ACFD, i.e. of half the vault, see Fig. \ref{fig:f7_40}:
\be
Q_2\ d=2p\ r^2\ h\ \rightarrow\ h=\frac{Q_2\ d}{2p\ r^2},
\ee
where $Q_2$ is the total weight of half the vault, cfr. eq. (\ref{eq:pesovolta}):
\be
Q_2=\frac{1}{2}Q\left(\frac{\pi}{2}\right)=4p\ r^2\left(\frac{\pi}{2}-1\right)\simeq2.283p\ r^2.
\ee
For what concerns the distance $d$, this can be calculated with a little effort to be equal to $d\simeq0.468\ r$, so finally $h\simeq 0.534\ r$. 

\section{Equilibrium of membranes of any shape}
\begin{figure}[th]
\begin{center}
\includegraphics[width=.8\textwidth]{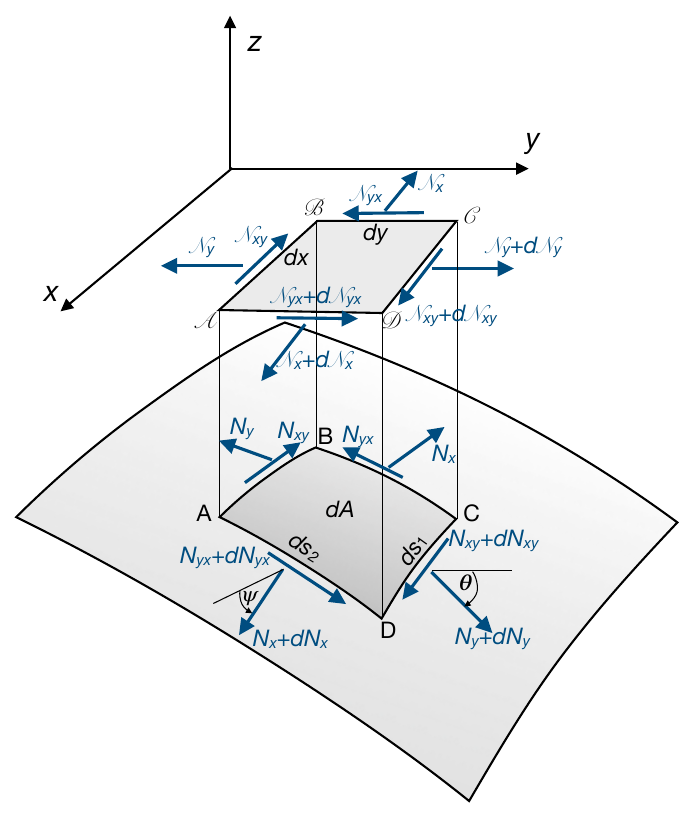}
\caption{Scheme of a generic membrane.}
\label{fig:f7_41}
\end{center}
\end{figure}

We consider in this Section the case of a membrane having a generic form, i.e. not axisymmetric. To this purpose, we refer to the scheme in Fig. \ref{fig:f7_41}.
We consider a small part ABCD of the membrane, whose area is $dA$ and whose projection onto the plane $x-y$ is the rectangle $\mathcal{ABCD}$ with sides $dx$ and $dy$; in addition:
\be
\besp
&\tan\psi=\frac{\partial z}{\partial x},\ \ ds_1\cos\psi=dx,\ \ \cos\psi=\frac{1}{\sqrt{1+\tan^2\psi}}=\frac{1}{\sqrt{1+\left(\frac{\partial z}{\partial x}\right)^2}},\\
&\tan\theta=\frac{\partial z}{\partial y},\ \ ds_2\cos\theta=dy,\ \ \cos\theta=\frac{1}{\sqrt{1+\tan^2\theta}}=\frac{1}{\sqrt{1+\left(\frac{\partial z}{\partial y}\right)^2}},
\end{split}
\ee
so that
\be
\label{eq:anymembgeom}
\besp
&N_xds_2\cos\psi=N_x\frac{\cos\psi}{\cos\theta}dy=\mn_xdy,\ \rightarrow\ \mn_x=N_x\frac{\cos\psi}{\cos\theta},\\
&N_yds_1\cos\theta=N_y\frac{\cos\theta}{\cos\psi}dx=\mn_ydx,\ \rightarrow\ \mn_y=N_y\frac{\cos\theta}{\cos\psi},\\
&N_{yx}ds_2\cos\theta=\mn_{yx}dy,\ \rightarrow\ \mn_{yx}=N_{yx},\\
&N_{xy}ds_1\cos\psi=\mn_{xy}dx,\ \rightarrow\ \mn_{xy}=N_{xy},\\
&N_{xy}=N_{yx}\ \Rightarrow\ \mn_{xy}=\mn_{yx}.
\end{split}
\ee
If $\gr{p}=(p_x,p_y,p_z)$ is the load applied to the membrane, then the equivalent $\boldsymbol{\mpe}=(\mpe_x,\mpe_y,\mpe_z)$ on the projected part $\mathcal{ABCD}$ is defined by the equivalence
\be
\mpe_xdx\ dy=p_xdA,\ \ \mpe_ydx\ dy=p_ydA,\ \ \mpe_zdx\ dy=p_zdA.
\ee
We can now write the equilibrium of the projected part $\mathcal{ABCD}$; in the direction $x$
\be
-\mn_xdy+(\mn_x+d\mn_x)dy-\mn_{xy}dx(\mn_{xy}+d\mn_{xy})dx+\mpe_xdx\ dy=0,
\ee
that gives the partial differential equation
\be
\label{eq:anymemb1}
\frac{\partial\mn_x}{\partial x}+\frac{\partial\mn_{xy}}{\partial y}+\mpe_x=0.
\ee
In a similar way we obtain, for the direction $y$,
\be
\label{eq:anymemb2}
\frac{\partial\mn_{xy}}{\partial x}+\frac{\partial\mn_{y}}{\partial y}+\mpe_y=0.
\ee
We introduce now the {\it Pucher's stress function} (1934), which is the function $\varphi(x,y)$ such that
\be
\label{eq:pucher}
\besp
&\mn_x=\frac{\partial^2\varphi}{\partial y^2}-\int\mpe_xdx,\\
&\mn_y=\frac{\partial^2\varphi}{\partial x^2}-\int\mpe_ydy,\\
&\mn_{xy}=-\frac{\partial^2\varphi}{\partial x\partial y}.
\end{split}
\ee
It is immediate to check that injecting these expressions into eqs. (\ref{eq:anymemb1}) and (\ref{eq:anymemb2}) we obtain two identities. 

A third equilibrium equation must still be written, the one in the direction of $z$; the vertical components of the membrane forces can be obtained as follows
\be
\besp
&N_xds_2\sin\psi=\mn_x\frac{\cos\theta}{\cos\psi}ds_2\sin\psi=\mn_x\tan\psi dy=\mn_x\frac{\partial z}{\partial x}dy,\\
&N_yds_1\sin\theta=\mn_y\frac{\cos\psi}{\cos\theta}ds_1\sin\theta=\mn_y\tan\theta dx=\mn_y\frac{\partial z}{\partial y}dx,\\
&N_{xy}ds_1\sin\psi=N_{xy}ds_1\cos\psi\tan\psi=N_{xy}dx\frac{\partial z}{\partial x}=\mn_{xy}\frac{\partial z}{\partial x}dx,\\
&N_{yx}ds_2\sin\theta=N_{yx}ds_2\cos\theta\tan\theta=N_{yx}dy\frac{\partial z}{\partial y}=\mn_{yx}\frac{\partial z}{\partial y}dy.
\end{split}
\ee
The equilibrium along $z$ is hence
\be
\besp
&-\mn_x\frac{\partial z}{\partial x}dy+\left(\mn_x\frac{\partial z}{\partial x}+\frac{\partial}{\partial x}\left(\mn_x\frac{\partial z}{\partial x}\right)dx\right)dy-\\
&-\mn_y\frac{\partial z}{\partial y}dx+\left(\mn_y\frac{\partial z}{\partial y}+\frac{\partial}{\partial y}\left(\mn_y\frac{\partial z}{\partial y}\right)dy\right)dx-\\
&-\mn_{xy}\frac{\partial z}{\partial x}dx+\left(\mn_{xy}\frac{\partial z}{\partial x}+\frac{\partial}{\partial y}\left(\mn_{xy}\frac{\partial z}{\partial x}\right)dy\right)dx-\\
&-\mn_{yx}\frac{\partial z}{\partial y}dy+\left(\mn_{yx}\frac{\partial z}{\partial y}+\frac{\partial}{\partial x}\left(\mn_{yx}\frac{\partial z}{\partial y}\right)dx\right)dy+\mpe_zdx\ dy=0,
\end{split}
\ee
and after simplifying we obtain the partial differential equation
\be
\frac{\partial}{\partial x}\left(\mn_x\frac{\partial z}{\partial x}\right)+\frac{\partial}{\partial y}\left(\mn_y\frac{\partial z}{\partial y}\right)+\frac{\partial}{\partial y}\left(\mn_{xy}\frac{\partial z}{\partial x}\right)+\frac{\partial}{\partial x}\left(\mn_{yx}\frac{\partial z}{\partial y}\right)+\mpe_z=0.
\ee
After differentiating and once introduced the values of $\partial \mn_x/\partial x$ and of $\partial \mn_y/\partial y$ that can be deduced from eqs. (\ref{eq:anymemb1}) and (\ref{eq:anymemb2}), the last equation becomes
\be
\label{eq:anymemb3}
\mn_x\frac{\partial^2 z}{\partial x^2}+\mn_y\frac{\partial^2 z}{\partial y^2}+2\mn_{xy}\frac{\partial^2 z}{\partial x\partial y}=-\mpe_z+\mpe_x\frac{\partial z}{\partial x}+\mpe_y\frac{\partial z}{\partial y}.
\ee
Introducing the Pucher's stress function $\varphi(x,y)$ we get the linear second-order partial differential equation
\be
\label{eq:anymemb4}
a(x,y)\frac{\partial^2\varphi}{\partial x^2}-2b(x,y)\frac{\partial^2\varphi}{\partial x\partial y}+c(x,y)\frac{\partial^2\varphi}{\partial y^2}=q(x,y),
\ee
where
\be
\besp
&a(x,y)=\frac{\partial^2 z}{\partial y^2},\ \ 
b(x,y)=\frac{\partial^2 z}{\partial x\partial y},\ \ 
c(x,y)=\frac{\partial^2 z}{\partial x^2},\\
&q(x,y)=-\mpe_z+\frac{\partial}{\partial x}\left(\frac{\partial z}{\partial x}\int\mpe_xdx\right)+\frac{\partial}{\partial y}\left(\frac{\partial z}{\partial y}\int\mpe_ydy\right)
\end{split}
\ee
are known functions depending upon the membrane's form and load. When $\partial^2z/\partial y^2$ and $\partial^2z/\partial x^2$ have the same sign, i.e.  if the radiuses of curvature in the two orthogonal directions $x$ and $y$ are on the same side of the membrane's surface, then eq. (\ref{eq:anymemb4}) is of the elliptic type, otherwise hyperbolic. 

The general problem of the equilibrium of a membrane of any shape is hence reduced to the search for a scalar function of two variables, $\varphi(x,y)$; once this found, the internal actions are recovered by differentiation, through eqs. (\ref{eq:pucher}). It is worth noticing, on one hand, that the problem is still intrinsically isostatic (the determination of the distribution of the internal actions needs just the equilibrium equations, there is no need to introduce the elastic behavior of the membrane to find them), and, on the other hand, that the use of the stress function automatically ensures that the distribution of the internal actions, once calculated through eqs. (\ref{eq:pucher}), is equilibrated. 

We consider in the next Sections some applications of the above results.

\subsection{Hyperbolic paraboloid}
\begin{figure}[th]
\begin{center}
\includegraphics[width=.4\textwidth]{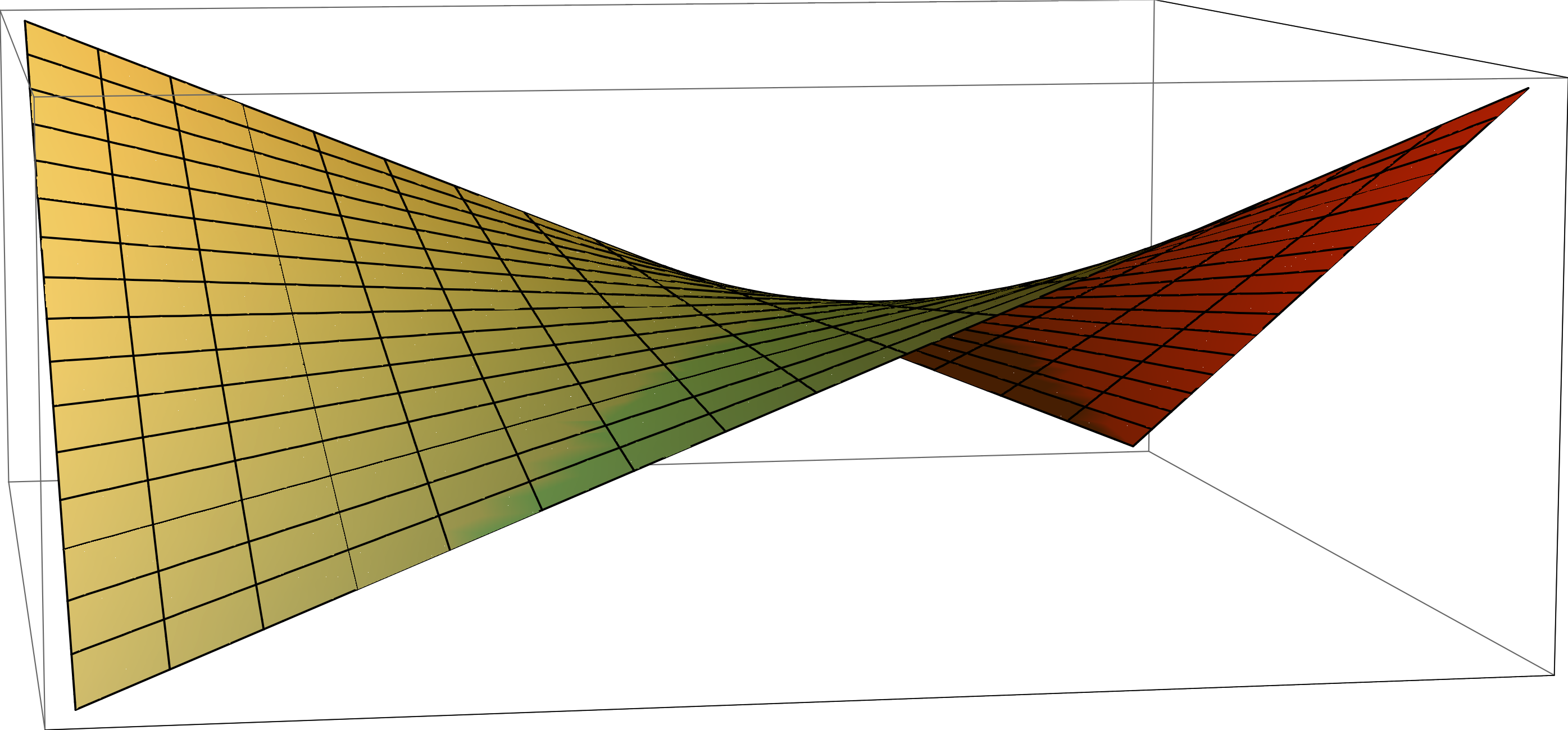}
\caption{Scheme of a hyperbolic paraboloid; the two families of straight lines belonging to the surface are visible in the plot.}
\label{fig:f7_42}
\end{center}
\end{figure}

Let us first consider the case of a membrane in the form of a hyperbolic paraboloid whose Cartesian equation is
\be
\label{eq:anymembhyperolparab1}
z=h\frac{x}{a}\frac{y}{b},
\ee
see Fig. \ref{fig:f7_42}. In such a case,
\be
\label{eq:anymembhyperolparab2}
\besp
&\frac{\partial z}{\partial x}=\frac{h\ y}{a\ b},\ \ \frac{\partial z}{\partial y}=\frac{h\ x}{a\ b},\\
&a(x,y)=\frac{\partial^2z}{\partial y^2}=0,\ \ b(x,y)=\frac{\partial^2z}{\partial x\partial y}=\frac{h}{a\ b},\ \ c(x,y)=\frac{\partial^2z}{\partial x^2}=0,\\ 
&q(x,y)=\mpe_z-\frac{h}{ab}(\mpe_xy+\mpe_yx),
\end{split}
\ee
i.e. the principal curvatures are null: as well known, this is a ruled surface. 
Equation (\ref{eq:anymemb4}) becomes
\be
2\frac{h}{a\ b}\frac{\partial^2\varphi}{\partial x\partial y}=q(x,y).
\ee
For the case, very frequent, of vertical loads, $\mpe_x=\mpe_y=0$, and if, in addition, $\mpe_z=\mpe= const.$, then we have the equation
\be
2\frac{h}{a\ b}\frac{\partial^2\varphi}{\partial x\partial y}=\mpe,
\ee
whose solution is
\be
\varphi(x,y)=\frac{a\ b}{2h}\mpe x\ y+f_1(x)+f_2(y).
\ee
Injecting this result into eqs. (\ref{eq:pucher}) gives
\be
\mn_x=\frac{\partial^2f_2}{\partial y^2},\ \ \mn_y=\frac{\partial^2f_1}{\partial x^2},\ \ \mn_{xy}=N_{xy}=-\frac{a\ b\ \mpe}{2h},
\ee
and hence, through eqs. (\ref{eq:anymembgeom}) and (\ref{eq:anymembhyperolparab2}),
\be
\label{eq:hyperbparabNxNy}
\besp
&N_x=\mn_x\frac{\cos\theta}{\cos\psi}=\mn_x\sqrt{\frac{1+\left(\frac{\partial z}{\partial x}\right)^2}{1+\left(\frac{\partial z}{\partial y}\right)^2}}=\frac{\partial^2f_2}{\partial y^2}\sqrt{\frac{c^2+y^2}{c^2+x^2}},\\
&N_y=\mn_y\frac{\cos\psi}{\cos\theta}=\mn_y\sqrt{\frac{1+\left(\frac{\partial z}{\partial y}\right)^2}{1+\left(\frac{\partial z}{\partial x}\right)^2}}=\frac{\partial^2f_1}{\partial x^2}\sqrt{\frac{c^2+x^2}{c^2+y^2}},
\end{split}
\ee
where for the sake of shortness we have posed
\be
c^2=\frac{a^2b^2}{h^2}.
\ee
The two unknown functions $f_1(x)$ and $f_2(y)$ are determined upon the boundary conditions; if, e.g., we fix $\mn_x$, i.e. $N_x$ for $x=0$, this amounts to impose that
\be
\frac{\partial^2f_2}{\partial y^2}=\mn_x=N_x\frac{1}{\sqrt{1+\frac{y^2}{c^2}}}.
\ee
In the same way, fixing $\mn_y$ for $y=0$ implies that
\be
\frac{\partial^2f_1}{\partial x^2}=\mn_y=N_y\frac{1}{\sqrt{1+\frac{x^2}{c^2}}}.
\ee
It is worth noticing that, because $f_2$ depends uniquely  upon $y$, its value, i.e. that of $\mn_x$, is constant on the lines $x=const.$; for the same reason, $f_1$, and hence $\mn_y$, is constant along the lines $y=const.$ Such lines are the characteristic straight lines of the ruled surface.  As a consequence if, e.g., the edges cannot bear the forces $\mn_x$ and $\mn_y$, these are null everywhere. 

It is also interesting to notice that the distribution of the forces $\mn_{xy}=N_{xy}$ does not depend upon the boundary conditions; in other words, these last can concern uniquely $\mn_x$ and $\mn_y$.

\begin{figure}[ht]
\begin{center}
\includegraphics[height=.6\textheight]{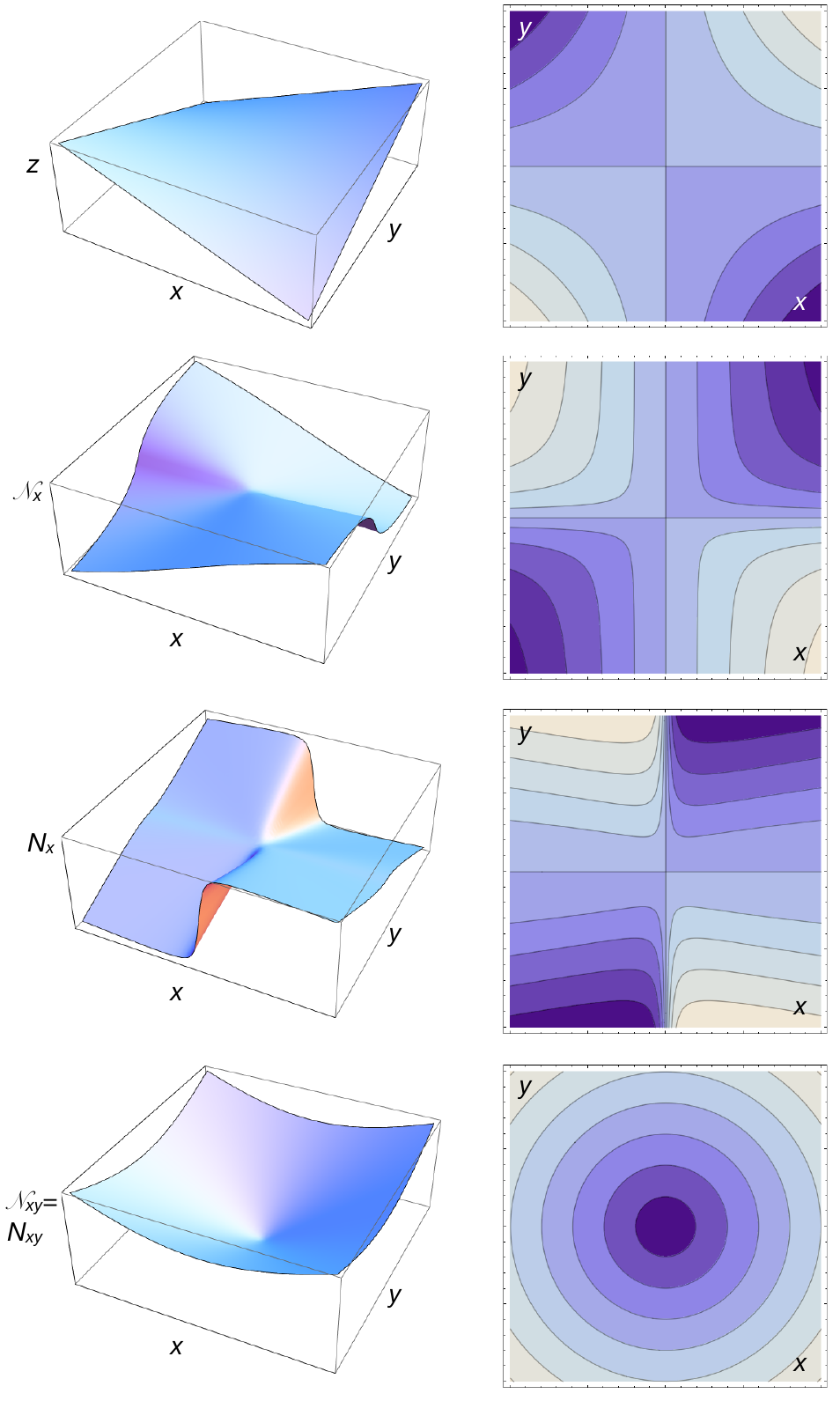}
\caption{Case of a hyperbolic paraboloid with $a=b=1,h=10,p_z=1$, shown at the top of the figure: 3D plots and contourplots of  $\mn_x,N_x,N_{xy}$.}
\label{fig:f7_43}
\end{center}
\end{figure}

A more interesting case is that where not $\mpe_z$ but the load $p_z$ is constant, $p_z=p$; in that case, 
\be
\mpe_z=-p\frac{h}{a\ b}\sqrt{x^2+y^2+c^2}.
\ee
Equation (\ref{eq:anymemb4}) becomes
\be
\frac{\partial^2\varphi}{\partial x\partial y}=-\frac{p}{2}\sqrt{x^2+y^2+c^2},
\ee
and hence, through eq. (\ref{eq:pucher})$_3$,
\be
\mn_{xy}=N_{xy}=\frac{p}{2}\sqrt{x^2+y^2+c^2}.
\ee
The expressions for $\mn_x$ and $\mn_y$ can be obtained integrating and differentiating, still in accordance with eqs. (\ref{eq:pucher}):
\be
\besp
&\mn_x=-\frac{p\ y}{2}\log\frac{x+\sqrt{x^2+y^2+c^2}}{\sqrt{y^2+c^2}},\\
&\mn_y=-\frac{p\ x}{2}\log\frac{y+\sqrt{x^2+y^2+c^2}}{\sqrt{x^2+c^2}}.
\end{split}
\ee
Finally, the expressions of $N_x$ and $N_y$ can be obtained by eqs. (\ref{eq:hyperbparabNxNy}):
\be
\besp
&N_x=-\frac{p\ y}{2}\sqrt{\frac{c^2+y^2}{c^2+x^2}}\log\frac{x+\sqrt{x^2+y^2+c^2}}{\sqrt{y^2+c^2}},\\
&N_y=-\frac{p\ x}{2}\sqrt{\frac{c^2+x^2}{c^2+y^2}}\log\frac{y+\sqrt{x^2+y^2+c^2}}{\sqrt{x^2+c^2}}.
\end{split}
\ee
The above result corresponds to the case where $N_x=N_y=0$ for $x=0$, $y=0$. The variation of the actions $N_x$ and $N_{xy}$ are plotted in Fig. \ref{fig:f7_43} (the plot of $N_y$ is equal to that of $N_x$, but turned of $\pi/2$).
\newpage
\pagebreak
\subsection{Paraboloid of revolution}
\begin{figure}[th]
\begin{center}
\includegraphics[width=.7\textwidth]{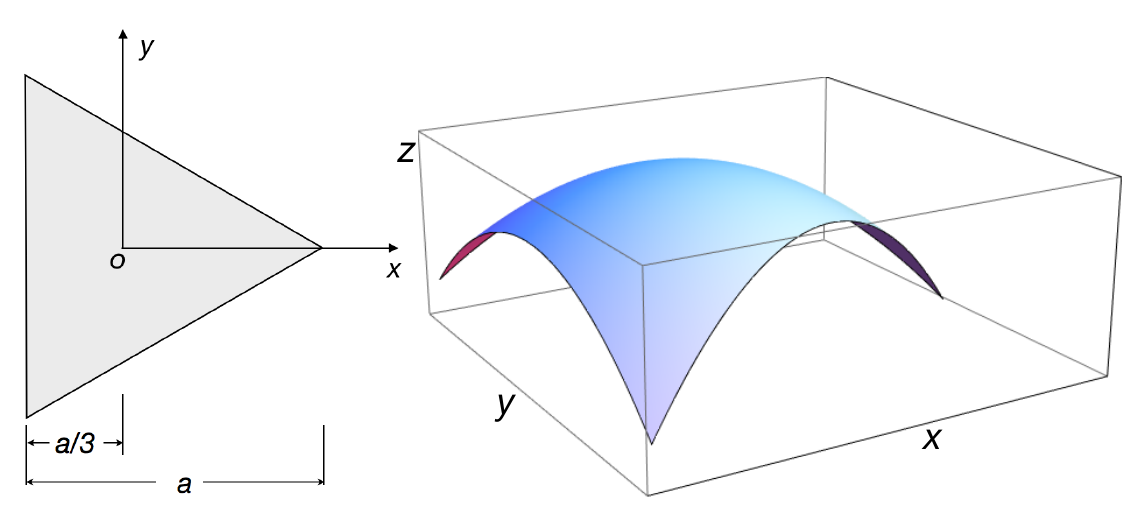}
\caption{Scheme of a paraboloid of revolution on a triangular plan.}
\label{fig:f7_44}
\end{center}
\end{figure}

\begin{figure}[th]
\begin{center}
\includegraphics[width=.6\textwidth]{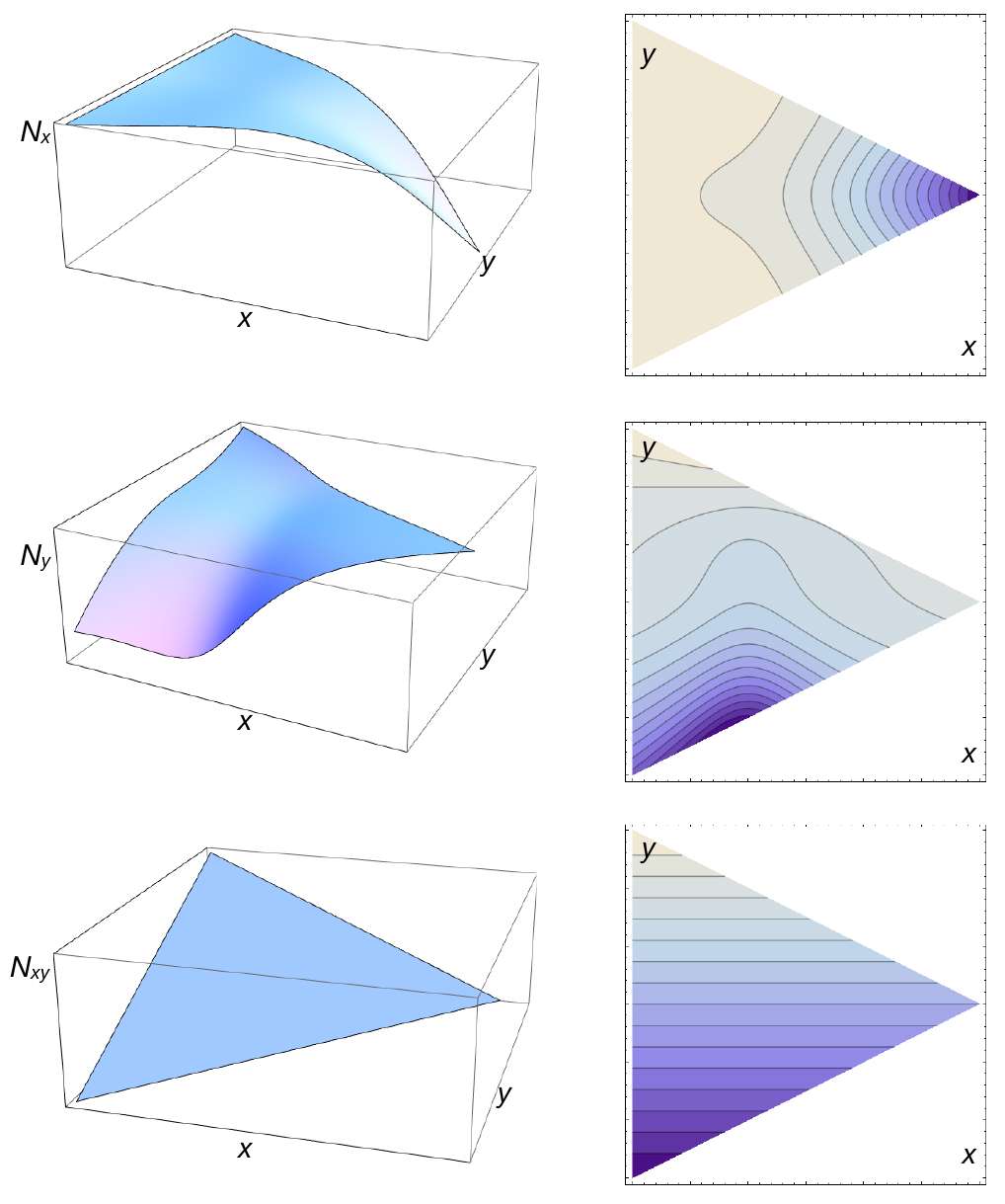}
\caption{Case of a  paraboloid of revolution with $a=3,h=1,\mpe_z=1$: 3D plots and contourplots of  $N_x,N_y,N_{xy}$.}
\label{fig:f7_45}
\end{center}
\end{figure}

We consider the case of a paraboloid of revolution of equation 
\be
z=\frac{x^2+y^2}{h},
\ee
still submitted to a vertical load $\mpe_z=\mpe$ per unit of projected horizontal area. In this case
\be
a(x,y)=\frac{\partial^2 z}{\partial y^2}=\frac{2}{h},\ \  b(x,y)=\frac{\partial^2 z}{\partial x\partial y}=0,\ \ c(x,y)=\frac{\partial^2 z}{\partial x^2}=\frac{2}{h},\ \ q(x,y)=\mpe,
\ee
and eq. (\ref{eq:anymemb4}) becomes
\be
\label{eq:parabrevol1}
\Delta\varphi=-\frac{\mpe h}{2},
\ee
a typical elliptical problem. To make an example, let us consider the case of membrane over a triangular plan like in Fig. \ref{fig:f7_44}.
We apply an inverse method: we fix the stress function
\be
\varphi=-\frac{1}{8}\mpe h\left(x^2+y^2+\frac{x}{a}(3y^2-x^2)\right),
\ee
and verify that it satisfies the equations. Actually, it is immediate to check that this function satisfy eq. (\ref{eq:parabrevol1}) and that on each one of the three sides
\be
\varphi=-\frac{1}{54}\mpe a^2h.
\ee
Hence, the derivatives of $\varphi$ in the direction of the border are null and by consequence the forces orthogonal to the border  are null too, because of eqs. (\ref{eq:pucher}). Hence,
\be
\besp
&\mn_x=\frac{\partial^2\varphi}{\partial y^2}=-\frac{1}{4}\mpe h\left(1+\frac{3}{a}x\right),\\
&\mn_y=\frac{\partial^2\varphi}{\partial x^2}=-\frac{1}{4}\mpe h\left(1-\frac{3}{a}x\right),\\
&\mn_{xy}=N_{xy}=-\frac{\partial^2\varphi}{\partial x\partial y}=\frac{3}{4}\mpe h\frac{y}{a}.
\end{split}
\ee
By consequence, through eqs. (\ref{eq:anymembgeom}), we get
\be
\besp
&N_x=\mn_x\frac{\cos\theta}{\cos\psi}=\mn_x\sqrt{\frac{1+\left(\frac{\partial z}{\partial x}\right)^2}{1+\left(\frac{\partial z}{\partial y}\right)^2}}=-\frac{1}{4}\mpe h\left(1+\frac{3x}{a}\right)\sqrt{\frac{h^2+4x^2}{h^2+4y^2}},\\
&N_y=\mn_y\frac{\cos\psi}{\cos\theta}=\mn_y\sqrt{\frac{1+\left(\frac{\partial z}{\partial y}\right)^2}{1+\left(\frac{\partial z}{\partial x}\right)^2}}=-\frac{1}{4}\mpe h\left(1-\frac{3y}{a}\right)\sqrt{\frac{h^2+4y^2}{h^2+4x^2}}.
\end{split}
\ee
The variation of the actions $N_x$, $N_y$ and $N_{xy}$  for a paraboloid with $a=3,h=1,\mpe_z=1$ are plotted in Fig. \ref{fig:f7_45}.

\subsection{Elliptic paraboloid}
\begin{figure}[th]
\begin{center}
\includegraphics[width=.6\textwidth]{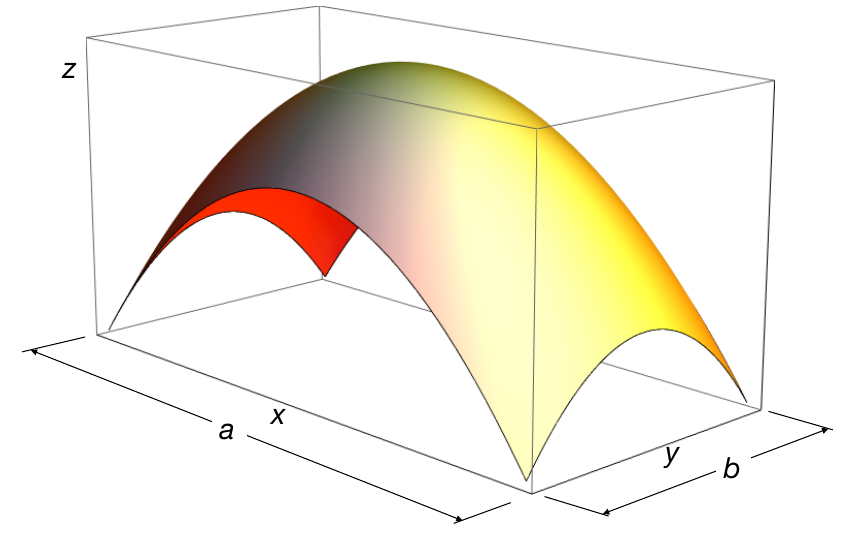}
\caption{Elliptic paraboloid with $a=2,b=1,h_1=2,h_2=1$.}
\label{fig:f7_46}
\end{center}
\end{figure}

As last case, we consider an elliptic paraboloid, on the rectangle $x\in\left[-\frac{a}{2},\frac{a}{2}\right],y\in\left[-\frac{b}{2},\frac{b}{2}\right]$, of equation
\be
z=\frac{x^2}{h_1}+\frac{y^2}{h_2},
\ee
see Fig. \ref{fig:f7_46}, submitted to the vertical force, per unit of horizontal surface, $\mpe_z=\mpe$. 
In this case, 
\be
a(x,y)=\frac{\partial^2 z}{\partial y^2}=\frac{2}{h_2},\ \  b(x,y)=\frac{\partial^2 z}{\partial x\partial y}=0,\ \ c(x,y)=\frac{\partial^2 z}{\partial x^2}=\frac{2}{h_1},\ \ q(x,y)=\mpe,
\ee
and eq. (\ref{eq:anymemb4}) becomes
\be
\label{eq:equilparabellipt}
\frac{1}{h_2}\frac{\partial^2\varphi}{\partial x^2}+\frac{1}{h_1}\frac{\partial^2\varphi}{\partial y^2}=-\frac{\mpe}{2}.
\ee
As boundary conditions, we impose that 
\be
\besp
&N_x\left(x=\pm\frac{a}{2}\right)=0\ \rightarrow\ \frac{\partial^2\varphi}{\partial y^2}\left(x=\pm\frac{a}{2}\right)=0,\\
&N_y\left(y=\pm\frac{b}{2}\right)=0\ \rightarrow\ \frac{\partial^2\varphi}{\partial x^2}\left(y=\pm\frac{b}{2}\right)=0,
\end{split}
\ee
i.e., on the boundaries the normal forces are null (this is the case of free edges or of supports realized by diaphragms unable to sustain  thrusts orthogonal to their plane). 
 To satisfy these conditions, $\varphi(x,y)$ must be linear with respect to $x$ or $y$ along the respective edges. Because the second derivatives are proportional to the membrane forces, without affecting the generality, we can pose 
 \be
 \varphi\left(\pm\frac{a}{2},y\right)=\varphi\left(x,\pm\frac{b}{2}\right)=0.
 \ee
 We then construct $\varphi(x,y)$ as
 \be
 \varphi(x,y)=\varphi_0(x,y)+\varphi_1(x,y),
 \ee
 with
 \be
 \varphi_0(x,y)=\mpe\frac{h_1}{4}\left(\frac{b^2}{4}-y^2\right),
 \ee
a particular solution of (\ref{eq:equilparabellipt}) and
\be
\varphi_1(x,y)=\sum_{n=1}^\infty f_n(x)\sin\frac{n\pi}{2}\cos\frac{n\pi y}{b}
\ee
a form of the integral of the associated homogeneous equation. Both $\varphi_0$ and $\varphi_1$ are null for  $y=\pm b/2$.
If $\varphi_1$ is injected into the associated homogeneous equation
\be
\label{eq:equilparabellipthomog}
\frac{1}{h_2}\frac{\partial^2\varphi}{\partial x^2}+\frac{1}{h_1}\frac{\partial^2\varphi}{\partial y^2}=0,
\ee
this is satisfied for each term if $f_n(x)$ is determined as solution of the equation
\be
\frac{1}{h_2}\frac{d^2f_n}{d x^2}-\frac{1}{h_1}\frac{n^2\pi^2}{b^2}f_n=0,
\ee
which gives
\be
f_n(x)=A_n\cosh\frac{n\pi x}{c}+B_n\sinh\frac{n\pi x}{c},
\ee
with 
\be
c=b\sqrt{\frac{h_1}{h_2}},
\ee
 \begin{figure}[t]
\begin{center}
\includegraphics[width=.6\textwidth]{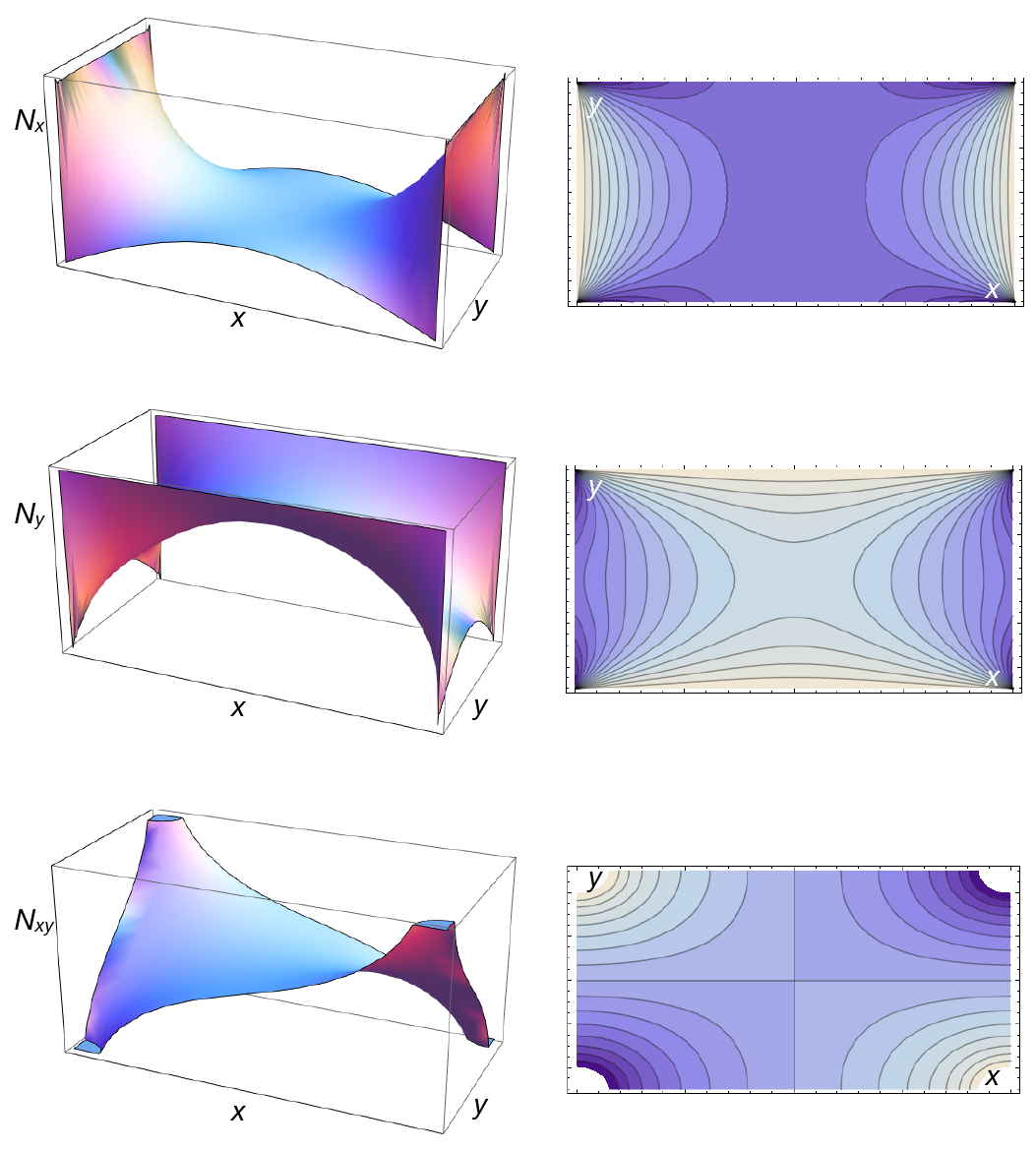}
\caption{Case of an elliptic  paraboloid with  $a=2,b=1,h_1=2,h_2=1,\mpe_z=1$: 3D plots and contourplots of  $N_x,N_y,N_{xy}$.}
\label{fig:f7_47}
\end{center}
\end{figure}
Because $f_n(x)$ must be symmetric with respect to $x=0$, $B_n=0$ and
\be
\varphi_1(x,y)=\sum_{n=1}^\infty A_n\cosh\frac{n\pi x}{c}\sin\frac{n\pi}{2}\cos\frac{n\pi y}{b}
\ee
We still need to satisfy the boundary conditions on $x=\pm a/2$; to this purpose, we rewrite $\varphi_0$ in the equivalent form
\be
\varphi_0(x,y)=\frac{\mpe h_1}{4}\frac{8b^2}{\pi^3}\sum_{n=1}^\infty\frac{1}{n^3}\sin\frac{n\pi}{2}\cos\frac{n\pi y}{b},
\ee
So now imposing that $\varphi=\varphi_0+\varphi_1=0$ for $x=\pm a/2$ and $\forall y\in[-b/2,b/2]$, we get the condition
\be
A_n=-\frac{1}{\cosh\frac{n\pi a}{2c}}\frac{\mpe h_1}{4}\frac{8b^2}{\pi^3}\frac{1}{n^3},
\ee
and finally
\be
\varphi(x,y)=\frac{\mpe h_1}{4}\left(\frac{b^2}{4}-y^2-\frac{8b^2}{\pi^3}\sum_{n=1}^\infty\frac{1}{n^3}\sin\frac{n\pi}{2}\cos\frac{n\pi y}{b}\frac{\cosh\frac{n\pi x}{c}}{\cosh\frac{n\pi a}{2c}}\right).
\ee
The membrane forces can now be calculated following the usual procedure:
\be
\besp
&N_x=-\frac{\mpe h_2}{2}\sqrt{\frac{h_1^2+4x^2}{h_2^2+4y^2}}\left(1-\frac{4}{\pi}\sum_{n=1}^\infty\frac{1}{n}\sin\frac{n\pi}{2}\cos\frac{n\pi y}{b}\frac{\cosh\frac{n\pi x}{c}}{\cosh\frac{n\pi a}{2c}}\right),\\
&N_y=-\frac{2\mpe h_1}{\pi}\sqrt{\frac{h_2^2+4y^2}{h_1^2+4x^2}}\sum_{n=1}^\infty\frac{1}{n}\sin\frac{n\pi}{2}\cos\frac{n\pi y}{b}\frac{\cosh\frac{n\pi x}{c}}{\cosh\frac{n\pi a}{2c}},\\
&N_{xy}=-\frac{2\mpe}{\pi}\sqrt{h_1h_2}\sum_{n=1}^\infty\frac{1}{n}\sin\frac{n\pi}{2}\sin\frac{n\pi y}{b}\frac{\sinh\frac{n\pi x}{c}}{\cosh\frac{n\pi a}{2c}}.
\end{split}
\ee
 The diagrams of $N_x,N_y$ and $N_{xy}$ are plotted in Fig. \ref{fig:f7_47}; it can be noticed that $N_{xy}\rightarrow\pm\infty$ in the corners, as a consequence of the point support.
 
\begin{figure}[th]
\begin{center}
\includegraphics[width=.4\textwidth]{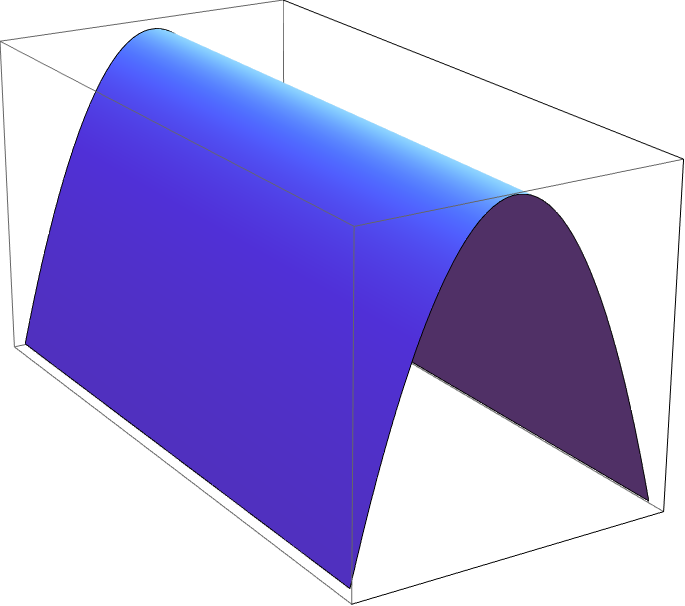}
\caption{Parabolic vault obtained as a limit case of an elliptic  paraboloid with  $h_1/h_2=10^6$.}
\label{fig:f7_48}
\end{center}
\end{figure}

 It is interesting to remark that if the ratio $h_1/h_2\rightarrow\infty$ or to $0$, then the shell tends towards cylindrical vault with parabolic cross-section, see e.g. the case in Fig. \ref{fig:f7_48}.

% Guardare il Belluzzi, vol 3, n° 629 per teoremi interessanti e differenze finite per membrane tese al bordo e capitolo 27 per membrane

\chapter{Shells}
% !TEX root = modsol.tex
\label{ch:8}
\section{Introduction}
\begin{figure}[th]
\begin{center}
\includegraphics[width=.4\textwidth]{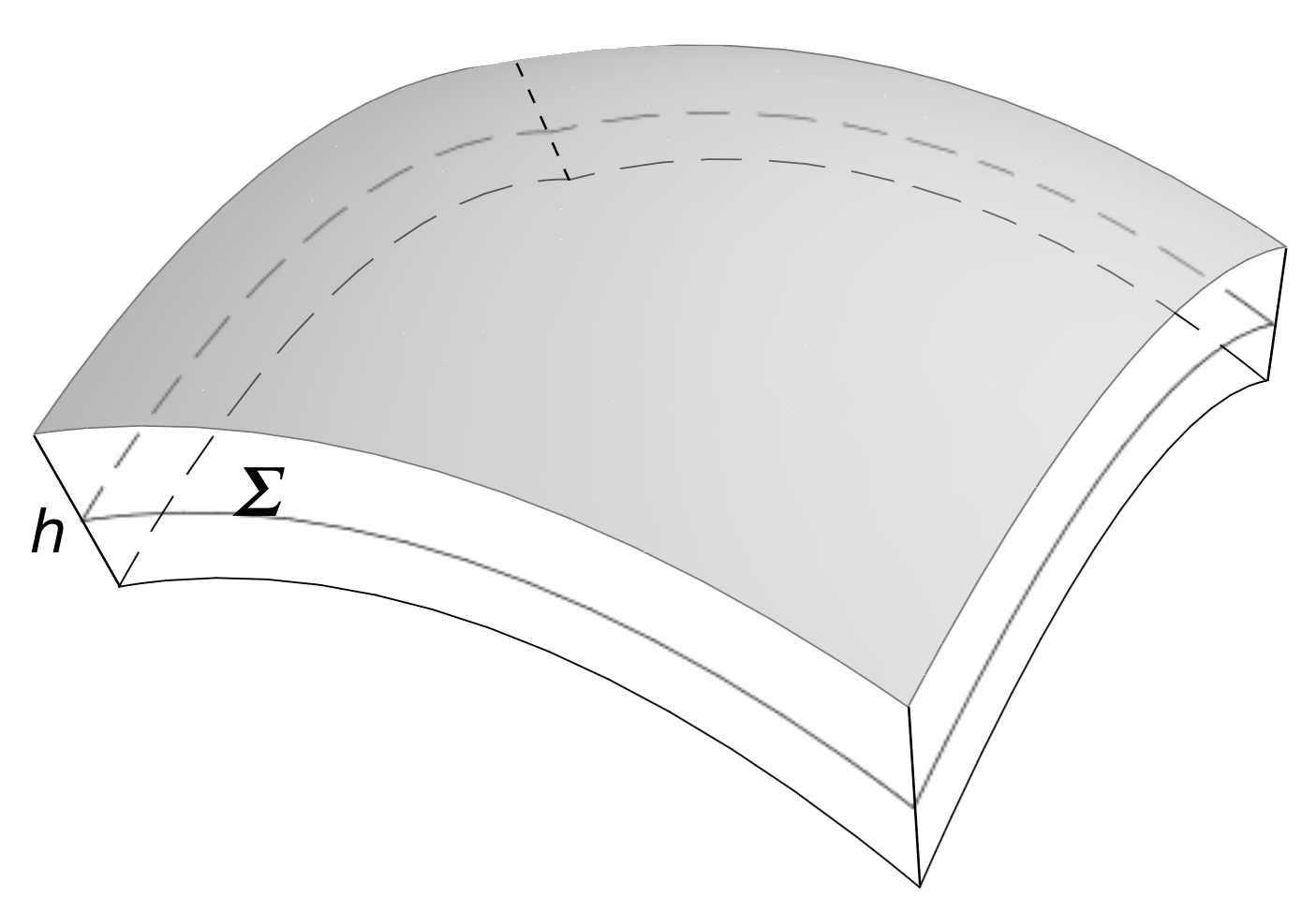}
\caption{Scheme of a shell.}
\label{fig:8_1}
\end{center}
\end{figure}

We call {\it shell} a solid $\Omega$ that is bounded by two curved surfaces whose distance is very small compared to the other dimensions, Fig. \ref{fig:8_1}. The set of points of $\Omega$ that lie at equal distance from the two bounding surfaces forms the {\it middle surface} $\Sigma$. The reader can compare this geometrical definition with  that of membranes, cf. Sect. \ref{sec:introdchap7} of Chap. \ref{ch:7}: there is a noticeable geometric difference: $\Omega$ is bounded by curved surfaces, and hence also $\Sigma$ is not planar, while flat membranes can exist. 

There is another fundamental difference between membranes and shells, mechanical this one: unlike membranes, {\it shells have a bending stiffness}. This is  the true mechanical difference between membranes and shells\footnote{In many texts, this difference is not clearly stressed and authors sometimes reduce the case of membranes to that of thin shells; we will see that this can be misleading, thin shells exist and are actually the topic of this chapter.}, which has three main consequences:
\begin{enumerate}[i.]
\item unlike membranes, shells   {\it have always a precise form}\footnote{We refer, of course, to the case where displacements are small, which is the situation considered here, so that the reference shape $\Omega$ remains, practically, unaltered under the action of the applied loads.} (e.g. inflatable shells do not exist);
\item a shell can stock the elastic energy in the double form of bending and stretching energies;
\item  shells are {\it intrinsically hyperstatic bodies}, while membranes are intrinsically isostatic bodies; hence, the only equilibrium equations are not sufficient to solve the equilibrium problem of shells, a constitutive law must be specified.
\end{enumerate}

Any segment perpendicular to the middle surface crosses the two bounding surface at two points whose distance, measured along the same segment, is the {\it thickness h} of the shell. The thickness can vary point-wise, though in the most part of applications it is contant all over $\Omega$. 
The boundaries of a shell\footnote{It is not excluded the case of a shell without boundaries, like an egg; in this case, the shell is called {\it complete}.} are formed by ruled surfaces orthogonal to $\Sigma$, the straight lines belonging to these ruled surfaces being segments orthogonal to $\Sigma$ (in other words, these surfaces are orthogonal to $\Sigma$).  
 
 Usually, shells are subdivided into two classes: {\it thin} and {\it thick shells}. A shell is considered to be thin if 
 \be
 \frac{h}{R}<\sim\frac{1}{20},
 \ee
 thick otherwise; in the above equation, $R$ is the radius of curvature of the shell\footnote{About $R$, more precisions are given below.}. The class of thin shells is particularly important for applications, because, as a matter of fact, in a large number of cases it is 
 \be
 \frac{1}{1000}<\sim\frac{h}{R}<\sim\frac{1}{50}.
 \ee
 This is mainly due to the fact that, thanks to their geometry, shells can support large loads even when they are very thin\footnote{This is also the case of curved membranes.}.
 
 In this text, we will focus on the classical theory of {\it thin elastic shells}: thin shells of isotropic, linearly elastic materials  are considered. In addition, we assume that the displacements $\bu$ caused by the applied loads are small compared to $h:|\bu|
\ll h$, and that  strains are small: $|\nabla\bu|\ll1$.

Historically, the first attempt of a theory of elastic shells is due to Aron (1874). However, his theory was not completely correct, and successively Love (1888, 1927) corrected some errors, giving a theory that, in the end, is the generalization to curved surfaces of the Kirchhoff's theory of plates, i.e. based upon the well known three assumptions on the kinematics of a material segment orthogonal to $\Sigma$, cf. Sect. \ref{sec:displkirch}. The classical theory of thin elastic shells is also known as the {\it Love's shell theory}; despite its success, there were some inconsistencies in the Love's theory: some small terms were rejected, while others, of the same order of magnitude, conserved in the equations.

After these initial works, the Russian school has largely contributed to the development of the theory of shells, with the works of Galerkin (1934, 1935), who succeeded in obtaining the equations of thick shells from the general theory of elasticity, Lur'e (1937, 1940), who derived the correct equations for thin shells following the approach of Galerkin, Novozhilov\footnote{The theory presented in this text is essentially inspired by the work of Novozhilov.} (1941, 1943, 1946), who simplified the general results of Lur'e, Gol'denweizer (1939, 1940), Mushtari (1948), Vlasov (1949).

 \section{Kirchhoff-Love kinematics of a shell}
  We assume for the shell the same hypotheses introduced by Kirchhoff for the plates, Sect. \ref{sec:displkirch}:  any material straight segment originally orthogonal to $\Sigma$ remains 
  \bi
\i i. a straight segment,
\i ii. orthogonal to $\Sigma'$,
\i iii. of the same length $h$,
  \ei
 with $\Sigma'$ the middle deformed surface. These assumptions, along with those of small displacements and strains, define the {\it linear shell theory of Kirchhoff-Love}.
 
 Let $\r=\r(\au,\ad)$ the function defining the middle surface $\Sigma$ through the curvilinear coordinates $\au,\ad$, that we assume to be  {\it principal coordinates}, i.e.  coordinates taken along the {\it lines of principal curvature of $\Sigma$}, cf. Sect. \ref{sec:linescurvat} and \ref{sec:gausscodazzi}. 
 
\begin{figure}[h]
\begin{center}
\includegraphics[width=.7\textwidth]{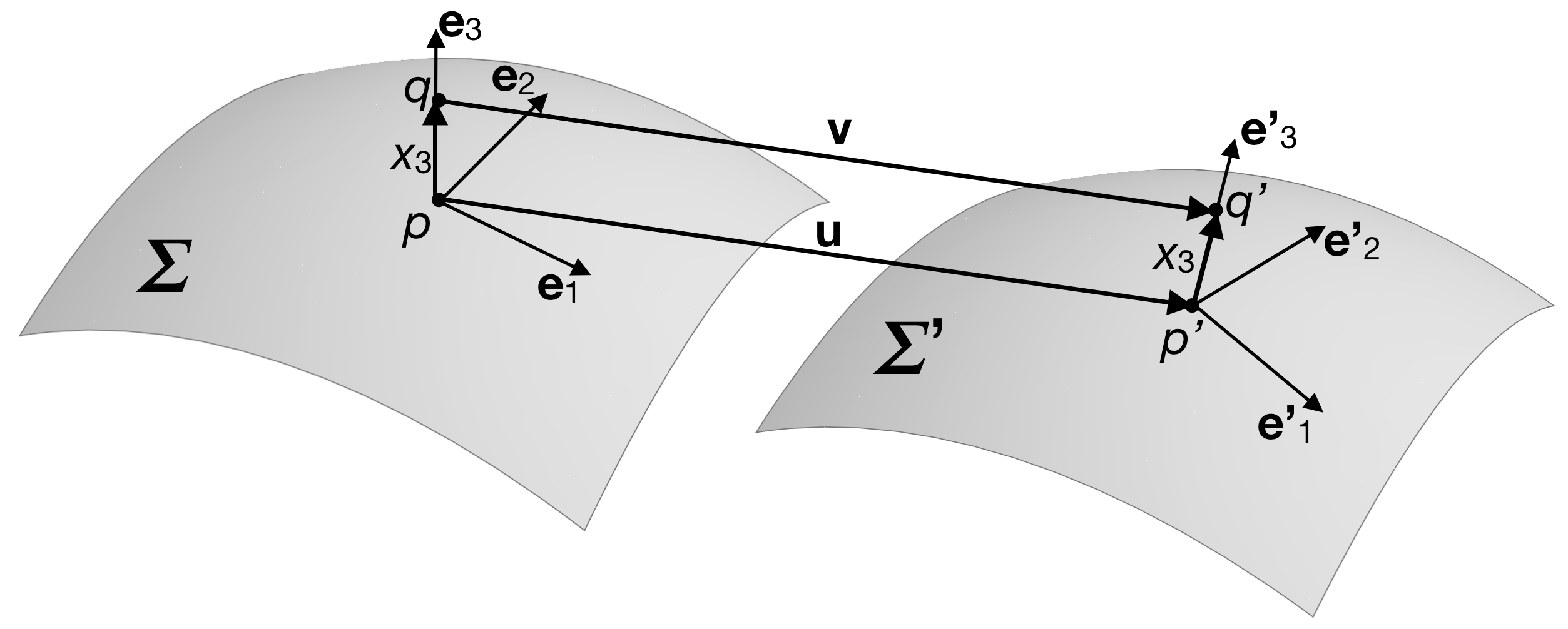}
\caption{Scheme of the displacements.}
\label{fig:8_2}
\end{center}
\end{figure}

 Let $p$ be a point of $\Sigma$ that is transformed into $p'\in\Sigma'$ by the deformation and let us denote by $\bu$ the {\it displacement vector of the points of $\Sigma$}:
 \be
 \bu(\au,\ad)=u_i(\au,\ad)\e_i:=p'-p.
 \ee
 Be $q$  a point of the shell on the normal segment through $p$, at the distance $x_3$ from $p$, that is transformed into the point $q'$ by the deformation and denote by $\bv$ its displacement: $\bv:=q'-q$. Then, by the Kirchhoff-Love assumptions, see Fig. \ref{fig:8_2},
 \be
 \label{eq:shell0}
 \bv=\bu+x_3(\et'-\et).
 \ee
 If we denote by $\r'$ the function defining $\Sigma'$ then
\be
\r'=\r+\bu=\r(\au,\ad)+u_i(\au,\ad)\e_i.
\ee
Differentiating,
\be
\frac{\partial\r'}{\partial\ai}=\frac{\partial\r}{\partial\ai}+\frac{\partial(u_i\e_i)}{\partial\ai},
\ee
 and using eq. (\ref{eq:triedre}) and (\ref{eq:codazzi3}) we get, after some  passages,
 \be
 \label{eq:shell1}
 \besp
& \frac{\partial\r'}{\partial\au}=A_1((1+\eps_1)\eu+\omega_1\ed-\theta\et),\\
 & \frac{\partial\r'}{\partial\ad}=A_2(\omega_2\eu+(1+\eps_2)\ed-\psi\et),
\end{split}
\ee
 where
 \be
 \label{eq:epsomegathetapsi}
 \besp
 &\eps_1=\frac{1}{A_1}\frac{\partial u_1}{\partial\au}+\frac{1}{A_1A_2}\frac{\partial A_1}{\partial\ad}u_2+\frac{u_3}{R_1},\\
 &\eps_2=\frac{1}{A_2}\frac{\partial u_2}{\partial\ad}+\frac{1}{A_1A_2}\frac{\partial A_2}{\partial\au}u_1+\frac{u_3}{R_2},\\
 &\omega_1=\frac{1}{A_1}\frac{\partial u_2}{\partial\au}-\frac{1}{A_1A_2}\frac{\partial A_1}{\partial\ad}u_1,\\
 &\omega_2=\frac{1}{A_2}\frac{\partial u_1}{\partial\ad}-\frac{1}{A_1A_2}\frac{\partial A_2}{\partial\au}u_2,\\
 &\theta=-\frac{1}{A_1}\frac{\partial u_3}{\partial\au}+\frac{u_1}{R_1},\\
 &\psi=-\frac{1}{A_2}\frac{\partial u_3}{\partial\ad}+\frac{u_2}{R_2}.
 \end{split}
 \ee
 By eq. (\ref{eq:triedre}) we have also that 
 \be
\e'_i=\frac{\partial \r'}{\partial\ai}\frac{1}{A'_i},\ \ \ i=1,2,
 \ee
 where, cf. eq. (\ref{eq:paramlamé}),
 \be
 A'_i=\sqrt{\frac{\partial\r'}{\partial\ai}\cdot\frac{\partial\r'}{\partial\ai}},\ \ \ i=1,2.
 \ee
 If in the last equations  we inject eq. (\ref{eq:shell1}), neglecting nonlinear terms\footnote{This is necessary for and coherent with a linear theory of shells. Namely, the products of the terms $\eps_1,\eps_2,\omega_1,\omega_2,\theta,\psi$ are all to be considered nonlinear quantities, hence to be neglected in a first order, linear theory of shells.} we get successively
 \be
 \label{eq:shell2}
 A'_i\simeq A_i(1+\eps_i),\ \ \ i=1,2,
 \ee
and 
\be
\label{eq:shell4}
\besp
&\e'_1=\frac{\partial \r'}{\partial\au}\frac{1}{A'_1}\simeq\eu+\omega_1\ed-\theta\et,\\
&\e'_2=\frac{\partial \r'}{\partial\ad}\frac{1}{A'_2}\simeq\omega_2\eu+\ed-\psi\et.
\end{split}
\ee
 By consequence, still neglecting nonlinear terms, it is also
 \be
 \label{eq:shell3}
 \et'=\eu'\times\ed'\simeq\theta\eu+\psi\ed+\et,
 \ee
 that injected into eq. (\ref{eq:shell0}) gives the expression of the displacement field $\bv$ for the shell:
 \be
 \bv=\bu+x_3(\theta\eu+\psi\ed),
 \ee
 or, by components,
 \be
 \label{eq:shell5}
 \besp
 &v_1=u_1+x_3\theta,\\
 &v_2=u_2+x_3\psi,\\
 &v_3=u_3.
 \end{split}
 \ee
 The hypotheses of Kirchhoff-Love have given for shells the same type of displacement field, i.e. with a linear variation through the thickness. 
 
 We can give a geometrical interpretation to quantities $\omega_1,\omega_2,\theta$ and $\psi$. In fact, %by eq. (\ref{eq:shell2}) we can interpret the $\eps_i$s as a sort of strain of the Lamé's parameters $A_i$, while 
  the norms of the cross products
\be
\besp
&\eu\times\eu'=\theta\ed+\omega_1\et,\\
&\ed'\times\ed=\psi\eu+\omega_2\et,
\end{split}
\ee
  are the sine of the angles by which $\eu$ and $\ed$ rotate by the deformation. Because these angles are small quantities, they can be identified with their sine, so finally the rotation of $\eu$ is of $\theta$ about $\ed$ and of $\omega_1$ about $\et$, while that of $\ed$ is of $\psi$ about $\eu$ and of $\omega_2$ about $\et$. 
 Moreover, $\theta$ and $\psi$ are the projections of $\et'$ on $\eu$ and $\ed$, eq. (\ref{eq:shell3}) and hence, by the smallness of the deformation, they approximate the angles by which $\et$ rotates about $\ed$ and $\eu$, respectively. By the Kirchhoff-Love second assumption, these are also the angles by which the tangents to the lines $\au$ and $\ad$ of $\Sigma$ rotate during the deformation about $\ed$ and $\eu$ respectively.

 \section{The deformation of a shell}
 By eq. (\ref{eq:distanzelinee}) the distances along the lines $\ai$, before and after the deformation, are respectively given by 
\be
\besp
&ds_i=A_id\ai,\\
&ds'_i=A'_id\ai,
\end{split}
\ \ \ \ i=1,2.
\ee
 The {\it strain of a material linear  element $\ai$ of $\Sigma$} is hence, by eq. (\ref{eq:shell2}),
 \be
 \frac{ds'_i-ds_i}{ds_i}=\frac{A'_i}{A_i}-1=\eps_i,\ \ \ i=1,2,
 \ee
which gives the mechanical meaning of the quantities $\eps_1$ and $\eps_2$, whose expression is given in eqs. (\ref{eq:epsomegathetapsi})$_{1,2}$.

If now we define  the {\it shear strain, $\omega$, of the lines $\au$ and $\ad$ of $\Sigma$} as the cosine of the angle formed by $\au$ and $\ad$ (mutually orthogonal before the deformation), then, by eq. (\ref{eq:shell4})
\be
\omega:=\eu'\cdot\ed'=\omega_1+\omega_2+\psi\theta,
\ee
 and   still neglecting nonlinear terms and by eq. (\ref{eq:epsomegathetapsi}),
 \be
 \label{eq:shell8}
 \omega\simeq\omega_1+\omega_2=\frac{A_1}{A_2}\frac{\partial}{\partial\ad}\left(\frac{u_1}{A_1}\right)+\frac{A_2}{A_1}\frac{\partial}{\partial\au}\left(\frac{u_2}{A_2}\right).
 \ee
 To remark the similarity of this result with the expression of $\gamma=2\eps_{12}=u_{1,2}+u_{2,1}$ of the linear theory of strain. 
 
 We need now to find the components of strain elsewhere from the middle surface $\Sigma$. To this end, we consider any other surface $\Sigma^z$, inside the shell's thickness, parallel to the middle surface at a distance $x_3=z$ from it. When the normal to $\Sigma$ passes along the lines $\au,\ad$, the normal to $\Sigma$ describes on $\Sigma^z$ two lines $\au^z,\ad^z$. The vectors $\eu^z$ and $\ed^z$ tangent to the lines $\au^z$ and $\ad^z$ are parallel, by construction, to vectors $\eu,\ed$, respectively, while $\et^z=\eu^z\times\ed^z$ is parallel to $\et$. So, the net of lines $\au^z,\ad^z$ is orthogonal and actually they are lines of principal curvature on $\Sigma^z$. 
 
\begin{figure}[h]
\begin{center}
\includegraphics[width=.4\textwidth]{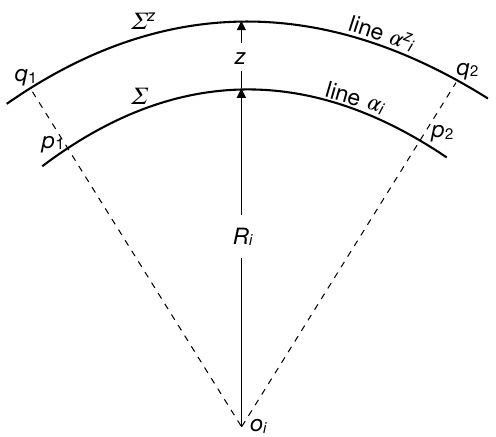}
\caption{Scheme for the strains.}
\label{fig:8_3}
\end{center}
\end{figure}

 Be $p_1,p_2$ two adjacent points on the line $\ai\in\Sigma$ and $q_1,q_2$ the corresponding points in $\Sigma^z$, Fig. \ref{fig:8_3}. 
If $R_i$ is the (principal) radius of curvature along the line $\ai$, then 
\be
\label{eq:shell6}
R_i^z=R_i+z
\ee
 is that of the line $\ai^z$ and if 
 \be
 ds_i=A_id\ai
 \ee 
 is the length of the arc $p_1p_2$, then, by similarity,
 \be
 ds_i^z=A_i\left(1+\frac{z}{R_i}\right) d\ai
 \ee
 is that of the arc $q_1q_2$. Because by construction the value of the coordinates $\ai$ are the same for the points of $\Sigma$ and $\Sigma^z$ on the same normal, 
 \be
 \label{eq:shell19}
 ds_i^z=A_i^zd\ai,
 \ee
and by comparison we get that  the Lamé's parameters for $\Sigma^z$ are 
 \be
 \label{eq:shell7}
 A_i^z=A_i\left(1+\frac{z}{R_i}\right).
 \ee
Putting eqs. (\ref{eq:shell6}) and  (\ref{eq:shell7}) into eq. (\ref{eq:epsomegathetapsi})$_1$ we get the expression of the strain $\eps_1^z$ on $\Sigma^z$:
\be
\eps_1^z=\frac{1}{A_1\left(1+\frac{z}{R_1}\right)}\frac{\partial u_1^z}{\partial\au}+\frac{1}{A_1A_2\left(1+\frac{z}{R_1}\right)\left(1+\frac{z}{R_2}\right)}\frac{\partial\left(A_1\left(1+\frac{z}{R_1}\right)\right)}{\partial\ad}u_2^z+\frac{u_3^z}{R_1+z},
\ee
with $u_i^z$ the components of the displacement vector for the points of $\Sigma^z$.
But
\be
\frac{\partial\left(A_1\left(1+\frac{z}{R_1}\right)\right)}{\partial\ad}=\frac{\partial A_1}{\partial\ad}+z\frac{\partial}{\partial\ad}\left(\frac{A_1}{R_1}\right)=\left(1+\frac{z}{R_2}\right)\frac{\partial A_1}{\partial\ad},
\ee
the last passage obtained by the first condition of Codazzi, eq. (\ref{eq:codazzi4})$_1$. Injecting this result into the expression of $\eps_1^z$ gives
\be
\eps_1^z=\frac{1}{1+\frac{z}{R_1}}\left(\frac{1}{A_1}\frac{\partial u_1^z}{\partial\au}+\frac{1}{A_1A_2}\frac{\partial A_1}{\partial\ad}u_2^z+\frac{u_3^z}{R_1}\right).
\ee
Now, by eq. (\ref{eq:shell5}), for  the components $u_i^z$ of the displacement vector of the points of $\Sigma^z$, we have also
\be
\label{eq:shell9}
\besp
&u_1^z=u_1+z\theta,\\
&u_2^z=u_2+z\psi,\\
&u_3^z=u_3.
\end{split}
\ee
Replacing these relations into the last expression of $\eps_1^z$  and remembering eq. (\ref{eq:epsomegathetapsi})$_1$ we obtain
\be
\label{eq:shell11}
\eps_1^z=\frac{1}{1+\frac{z}{R_1}}(\eps_1+z\kappa_1),
\ee
with
\be
\label{eq:kappa1}
\kappa_1=\frac{1}{A_1}\frac{\partial\theta}{\partial\au}+\frac{1}{A_1A_2}\frac{\partial A_1}{\partial\ad}\psi.
\ee
Proceeding in the same way for $\eps_2^z$, we obtain
\be
\label{eq:shell12}
\eps_2^z=\frac{1}{1+\frac{z}{R_2}}(\eps_2+z\kappa_2),
\ee
with
\be
\label{eq:kappa2}
\kappa_2=\frac{1}{A_2}\frac{\partial\psi}{\partial\ad}+\frac{1}{A_1A_2}\frac{\partial A_2}{\partial\au}\theta.
\ee

We make the same for the shear strain $\omega$: putting into eq. (\ref{eq:shell8}) the expressions for the $A_i^z$s, eq. (\ref{eq:shell7}), and using the Codazzi's equations, (\ref{eq:codazzi4}), we get after some easy passages
\be
\label{eq:shell15}
\omega^z=\frac{1}{1+\frac{z}{R_1}}\left(\frac{1}{A_1}\frac{\partial u_2^z}{\partial\au}-\frac{1}{A_1A_2}\frac{\partial A_1}{\partial\ad}u_1^z\right)+
\frac{1}{1+\frac{z}{R_2}}\left(\frac{1}{A_2}\frac{\partial u_1^z}{\partial\ad}-\frac{1}{A_1A_2}\frac{\partial A_2}{\partial\au}u_2^z\right).
\ee
If now eq. (\ref{eq:shell9}) is used for the $u_i^z$s, then
\be
\label{eq:shell23}
\omega^z=\frac{1}{1+\frac{z}{R_1}}(\omega_1+z\tau_1)+\frac{1}{1+\frac{z}{R_2}}(\omega_2+z\tau_2),
\ee
with $\omega_1,\omega_2$ given by eq. (\ref{eq:epsomegathetapsi})$_{3,4}$ while
\be
\besp
&\tau_1=\frac{1}{A_1}\frac{\partial \psi}{\partial\au}-\frac{1}{A_1A_2}\frac{\partial A_1}{\partial\ad}\theta,\\
&\tau_2=\frac{1}{A_2}\frac{\partial \theta}{\partial\ad}-\frac{1}{A_1A_2}\frac{\partial A_2}{\partial\au}\psi. 
\end{split}
\ee
The above expression can be further reduced, considering on one hand that $\omega_1+\omega_2=\omega$, eq. (\ref{eq:shell8}), and, on the other hand, the identity
\be
\label{eq:shell10}
\tau_1+\frac{\omega_2}{R_1}=\tau_2+\frac{\omega_1}{R_2},
\ee
that can be proved using the expressions for $\tau_1,\tau_2,\omega_1,\omega_2,\psi,\theta$ and the Codazzi's conditions.
So, putting
\be
\tau:=\tau_1+\frac{\omega_2}{R_1}=\tau_2+\frac{\omega_1}{R_2}
\ee
and reducing the expression of $\omega^z$ to a common denominator we get
\be
\label{eq:shell22}
\omega^z=\frac{1}{(1+\frac{z}{R_1})(1+\frac{z}{R_2})}\left(\omega+2\tau z+\frac{z^2}{R_1R_2}(\tau_1R_1+\tau_2R_2)\right).
\ee
Moreover, by eq. (\ref{eq:shell10}),
\be
\tau_1R_1+\tau_2R_2=(R_1+R_2)\tau-\omega,
\ee
and finally
\be
\label{eq:shell13}
\omega^z=\frac{1}{(1+\frac{z}{R_1})(1+\frac{z}{R_2})}\left[\left(1-\frac{z^2}{R_1R_2}\right)\omega+2\left(1+\left(\frac{1}{R_1}+\frac{1}{R_2}\right)\frac{z}{2}\right)\tau z\right].
\ee
Resuming: eqs. (\ref{eq:shell11}), (\ref{eq:shell12}) and (\ref{eq:shell13}) are the extension and shear strains of points of the shell lying on a surface parallel to the middle surface and at a distance $z$ from this one. By these equations, the study of the deformation of a shell is reduced to that of the deformation of the middle surface, through  six  functions,  $\eps_1,\eps_2,\omega,\kappa_1,\kappa_2$ and $\tau$, that operate the transfer from $\Sigma^z$ to $\Sigma$:
\be
\label{eq:shell18}
\besp
 &\eps_1=\frac{1}{A_1}\frac{\partial u_1}{\partial\au}+\frac{1}{A_1A_2}\frac{\partial A_1}{\partial\ad}u_2+\frac{u_3}{R_1},\\
 &\eps_2=\frac{1}{A_2}\frac{\partial u_2}{\partial\ad}+\frac{1}{A_1A_2}\frac{\partial A_2}{\partial\au}u_1+\frac{u_3}{R_2},\\
 &\omega=\frac{A_1}{A_2}\frac{\partial}{\partial\ad}\left(\frac{u_1}{A_1}\right)+\frac{A_2}{A_1}\frac{\partial}{\partial\au}\left(\frac{u_2}{A_2}\right),\\
 &\kappa_1=-\frac{1}{A_1}\frac{\partial}{\partial\au}\left(\frac{1}{A_1}\frac{\partial u_3}{\partial\au}-\frac{u_1}{R_1}\right)-\frac{1}{A_1A_2}\frac{\partial A_1}{\partial\ad}\left(\frac{1}{A_2}\frac{\partial u_3}{\partial\ad}-\frac{u_2}{R_2}\right),\\
  &\kappa_2=-\frac{1}{A_2}\frac{\partial}{\partial\ad}\left(\frac{1}{A_2}\frac{\partial u_3}{\partial\ad}-\frac{u_2}{R_2}\right)-\frac{1}{A_1A_2}\frac{\partial A_1}{\partial\ad}\left(\frac{1}{A_1}\frac{\partial u_3}{\partial\au}-\frac{u_1}{R_1}\right),\\
  &\tau=-\frac{1}{A_1A_2}\left(\frac{\partial^2u_3}{\partial\au\partial\ad}-\frac{1}{A_1}\frac{\partial A_1}{\partial\ad}\frac{\partial u_3}{\partial\au}-\frac{1}{A_2}\frac{\partial A_2}{\partial\au}\frac{\partial u_3}{\partial\ad}\right)\\
&\hspace{7.2mm} +\frac{1}{R_1}\left(\frac{1}{A_2}\frac{\partial u_1}{\partial\ad}-\frac{1}{A_1A_2}\frac{\partial A_1}{\partial\ad}u_1\right)
 +  \frac{1}{R_2}\left(\frac{1}{A_1}\frac{\partial u_2}{\partial\au}-\frac{1}{A_1A_2}\frac{\partial A_2}{\partial\au}u_2\right).
\end{split}
\ee
While the geometrical meaning of $\eps_1,\eps_2$ and $\omega$ has already been specified, we consider now which is that of $\kappa_1,\kappa_2,\tau$. To this end, consider the change $d\eu'$ that the unit vector $\eu'$, tangent to the line $\au$ on the deformed middle surface $\Sigma'$, undergoes for the increment $ds_1'=A_1'd\au$ along $\au$:
\be
d\eu'=\frac{\partial\eu'}{\partial s_1'}ds_1',
\ee
and hence
\be
\frac{d\eu'}{ds_1'}=\frac{1}{A_1'}\frac{\partial\eu'}{\partial\au}.
\ee
The projection of this quantity on $\et'$, the normal to $\Sigma'$, is by definition the curvature of the line $\au$ on $\Sigma'$, i.e.\footnote{The sign $-$ is due to the fact the the positive direction of the normal is in the direction of the centres of negative curvature, cf. above.}
\be
\frac{1}{R_1'}=-\frac{1}{A_1'}\frac{\partial\eu'}{\partial\au}\cdot\et'.
\ee
If now we consider the change $d\eu'$ for an increment $ds_2'=A_2'd\ad$ along the line $\ad$ and projecting it onto $\et'$, operating in the same way gives
\be
\tau_{\ad}'=-\frac{1}{A_2'}\frac{\partial\eu'}{\partial\ad}\cdot\et'.
\ee
In a similar manner, we get also
\be
\frac{1}{R_2'}=-\frac{1}{A_2'}\frac{\partial\ed'}{\partial\ad}\cdot\et',\ \ \tau_{\au}'=-\frac{1}{A_1'}\frac{\partial\ed'}{\partial\au}\cdot\et'
\ee
for the vector $\ed'$. $R_1'$ and $R_2'$ are respectively the radius of curvature of the lines $\au$ and $\ad$ on $\Sigma'$, while $\tau_{\ad}'$ and $\tau_{\au}'$ represent the twist (angle of rotation)  of $\Sigma'$ respectively about $\ad$ and $\au$.
Using eqs. (\ref{eq:codazzi3}),  (\ref{eq:shell4}) and (\ref{eq:shell3}) we can get also
\be
\besp
\frac{1}{R_1'}=&-\frac{1}{A_1'}\left(\frac{\partial\eu}{\partial\au}+\omega_1\frac{\partial\ed}{\partial\au}-\theta\frac{\partial\et}{\partial\au}+\ed\frac{\partial\omega_1}{\partial\au}-\et\frac{\partial\theta}{\partial\au}\right)\cdot(\theta\eu+\psi\ed+\et)\\
=&-\frac{1}{A_1'}\left[\left(\omega_1\frac{1}{A_2}\frac{\partial A_1}{\partial\ad}-\theta\frac{A_1}{R_1}\right)\eu+\left(-\frac{1}{A_2}\frac{\partial A_1}{\partial\ad}+\frac{\partial\omega_1}{\partial\au}\right)\ed\right.\\
&\left.-\left(\frac{A_1}{R_1}+\frac{\partial\theta}{\partial\au}\right)\et\right]\cdot(\theta\eu+\psi\ed+\et).
\end{split}
\ee
Performing the scalar product and neglecting, as usual, higher order terms, we find
\be
\frac{1}{R_1'}=\frac{1}{R_1}(1-\eps_1)+\frac{1}{A_1}\frac{\partial\theta}{\partial\au}+\frac{1}{A_1A_2}\frac{\partial A_1}{\partial\ad}\psi=\kappa_1+\frac{1}{R_1}(1-\eps_1),
\ee
so that the change in curvature of the line $\au$ is
\be
%\delta\kappa_1=
\frac{1}{R_1'}-\frac{1}{R_1}=\kappa_1-\frac{\eps_1}{R_1}.
\ee
Operating in the same way for $R_2'$, we can get also
\be
%\delta\kappa_2=
\frac{1}{R_2'}-\frac{1}{R_2}=\kappa_2-\frac{\eps_2}{R_2},
\ee
while for $\tau_{\au}'$ and $\tau_{\ad}'$ we obtain
\be
\tau_{\au}'=\tau_{\ad}'=\tau_1+\frac{\omega_2}{R_1}=\tau_2+\frac{\omega_1}{R_2}=\tau.
\ee
So, we see that $\kappa_1,\kappa_2$ and $\tau$ characterize the change of curvatures (i.e. the curvatures produced by the deformation) and the twist of $\Sigma$ in the deformation. Finally, the six parameters $\eps_1,\eps_2,\omega,\kappa_1,\kappa_2$ and $\tau$ characterize the deformation of $\Sigma$. This number of six is not accidental: by the theory of surfaces we know that the dimension and shape of a surface are known if the first and second fundamental form are known, and each one of these forms is determined by three parameters.

\section{Gol'denweizer compatibility conditions}

Using eq. (\ref{eq:shell1}), let us calculate the mixed second derivatives of $\r'(\au,\ad)$, the vector that defines the deformed surface $\Sigma'$:
\be
\besp
\frac{\partial^2\r'}{\partial\ad\partial\au}&=\left(\frac{\partial A_1(1+\eps_1)}{\partial\ad}-\omega_1\frac{\partial A_2}{\partial\au}\right)\eu+\left( (1+\eps_1)\frac{\partial A_2}{\partial\au}+\frac{\partial (A_1\omega_1)}{\partial\ad}-\frac{A_1A_2}{R_2}\theta\right)\ed\\
&-\left(\frac{\partial (A_1\theta)}{\partial\ad}+\frac{A_1A_2}{R_2}\omega_1\right)\et,\\
\frac{\partial^2\r'}{\partial\au\partial\ad}&=
\left( (1+\eps_2)\frac{\partial A_1}{\partial\ad}+\frac{\partial (A_2\omega_2)}{\partial\au}-\frac{A_1A_2}{R_1}\psi\right)\eu+
\left(\frac{\partial A_2(1+\eps_2)}{\partial\au}-\omega_2\frac{\partial A_1}{\partial\ad}\right)\ed\\
&-\left(\frac{\partial (A_2\psi)}{\partial\au}+\frac{A_1A_2}{R_1}\omega_2\right)\et.
\end{split}
\ee
Because $\dfrac{\partial^2\r'}{\partial\ad\partial\au}=\dfrac{\partial^2\r'}{\partial\au\partial\ad}$, we get the following three equalities
\be
\label{eq:shell14}
\besp
&\frac{\partial(A_1\eps_1)}{\partial\ad}-\frac{\partial(A_2\omega)}{\partial\au}-\eps_2\frac{\partial A_1}{\partial\ad}=-A_2\left(\frac{\partial\omega_1}{\partial\au}+\frac{A_1}{R_1}\right)\psi,\\
&\frac{\partial(A_2\eps_2)}{\partial\au}-\frac{\partial(A_1\omega)}{\partial\ad}-\eps_1\frac{\partial A_2}{\partial\au}=-A_1\left(\frac{\partial\omega_2}{\partial\ad}+\frac{A_2}{R_2}\right)\theta,\\
&\frac{\partial(A_1\theta)}{\partial\ad}-\frac{\partial(A_2\psi)}{\partial\au}=A_1A_2\left(\frac{\omega_2}{R_1}-\frac{\omega_1}{R_2}\right).
\end{split}
\ee
Now we compute $\dfrac{\partial\et'}{\partial\ad\partial\au}$ and $\dfrac{\partial\et'}{\partial\au\partial\ad}$, using eqs. (\ref{eq:shell1}), (\ref{eq:kappa1}) and (\ref{eq:kappa2}), to get
\be
\besp
\frac{\partial\et'}{\partial\ad\partial\au}&=\left[\frac{\partial}{\partial\ad}\left(\frac{A_1}{R_1}\right)+\frac{\partial(A_1\kappa_1)}{\partial\ad}-\tau_1\frac{\partial A_2}{\partial\au}\right]\eu\\
&+\left(\frac{1}{R_1}\frac{\partial A_2}{\partial\au}+\kappa_1\frac{\partial A_2}{\partial\au}+\frac{\partial(A_1\tau_1)}{\partial\ad}-\theta\frac{A_1A_2}{R_1R_2}\right)\ed\\
&-\left[\frac{A_1A_2}{R_2}\tau_1+\frac{\partial}{\partial\ad}\left(\frac{A_1}{R_1}\theta\right)\right]\et,\\
\frac{\partial\et'}{\partial\au\partial\ad}&=
\left(\frac{1}{R_2}\frac{\partial A_1}{\partial\ad}+\kappa_2\frac{\partial A_1}{\partial\ad}+\frac{\partial(A_2\tau_2)}{\partial\au}-\psi\frac{A_1A_2}{R_1R_2}\right)\eu\\
&+\left[\frac{\partial}{\partial\au}\left(\frac{A_2}{R_2}\right)+\frac{\partial(A_2\kappa_2)}{\partial\au}-\tau_2\frac{\partial A_1}{\partial\ad}\right]\ed\\
&-\left[\frac{A_1A_2}{R_1}\tau_2+\frac{\partial}{\partial\au}\left(\frac{A_2}{R_2}\psi\right)\right]\et.
\end{split}
\ee
Then, the equality  $\dfrac{\partial\et'}{\partial\ad\partial\au}=\dfrac{\partial\et'}{\partial\au\partial\ad}$ gives 
\be
\besp
&\frac{\partial(A_1\kappa_1)}{\partial\ad}-\kappa_2\frac{\partial A_1}{\partial\ad}-\frac{\partial(A_2\tau)}{\partial\au}-\tau\frac{\partial A_2}{\partial\au}+\frac{\omega}{R_1}\frac{\partial A_2}{\partial\au}=-\frac{A_2}{R_2}\left(\frac{\partial\omega_1}{\partial\au}+\frac{A_1}{R_1}\psi\right),\\
&\frac{\partial(A_2\kappa_2)}{\partial\au}-\kappa_1\frac{\partial A_2}{\partial\au}-\frac{\partial(A_1\tau)}{\partial\ad}-\tau\frac{\partial A_1}{\partial\ad}+\frac{\omega}{R_2}\frac{\partial A_1}{\partial\ad}=-\frac{A_1}{R_1}\left(\frac{\partial\omega_2}{\partial\ad}+\frac{A_2}{R_2}\theta\right),\\
&\frac{A_1A_2}{R_2}\tau_1+\frac{\partial}{\partial\ad}\left(\frac{A_1}{R_1}\theta\right)=
\frac{A_1A_2}{R_1}\tau_2+\frac{\partial}{\partial\au}\left(\frac{A_2}{R_2}\psi\right).
\end{split}
\ee
The right-hand side of the first two of the last equations are proportional to the right-hand sides of eqs. (\ref{eq:shell14})$_{1,2}$, which gives
\be
\label{eq:shell16}
\besp
&\frac{\partial(A_1\kappa_1)}{\partial\ad}-\kappa_2\frac{\partial A_1}{\partial\ad}-\frac{\partial(A_2\tau)}{\partial\au}-\tau\frac{\partial A_2}{\partial\au}+\frac{\omega}{R_1}\frac{\partial A_2}{\partial\au}=\frac{1}{R_2}\left(\frac{\partial(A_1\eps_1)}{\partial\ad}-\frac{\partial(A_2\omega)}{\partial\au}-\eps_2\frac{\partial A_1}{\partial\ad}\right),\\
&\frac{\partial(A_2\kappa_2)}{\partial\au}-\kappa_1\frac{\partial A_2}{\partial\au}-\frac{\partial(A_1\tau)}{\partial\ad}-\tau\frac{\partial A_1}{\partial\ad}+\frac{\omega}{R_2}\frac{\partial A_1}{\partial\ad}=\frac{1}{R_1}\left(\frac{\partial(A_2\eps_2)}{\partial\au}-\frac{\partial(A_1\omega)}{\partial\ad}-\eps_1\frac{\partial A_2}{\partial\au}\right),
\end{split}
\ee
while the third equation is actually an identity, which can be checked using eqs. (\ref{eq:codazzi4}) and (\ref{eq:shell15}).
Finally, proceeding in the same way for the second derivatives of $\eu'$ and $\ed'$, one can get six relations analogous to the last ones. After some transformations, some of these become identities, others are equal to eqs. (\ref{eq:shell16}), while one is a new condition on the deformation of $\Sigma$:
\be
\label{eq:shell17}
\besp
\frac{\kappa_1}{R_2}+\frac{\kappa_2}{R_1}+\frac{1}{A_1A_2}\left[\frac{\partial}{\partial\au}\frac{1}{A_1}\left(A_2\frac{\partial\eps_2}{\partial\au}+\frac{\partial A_2}{\partial\au}(\eps_2-\eps_1)-\frac{1}{2}A_1\frac{\partial\omega}{\partial\ad}-\frac{\partial A_1}{\partial\ad}\omega\right)\right.&\\
\left.+\frac{\partial}{\partial\ad}\frac{1}{A_2}
\left(A_1\frac{\partial\eps_1}{\partial\ad}+\frac{\partial A_1}{\partial\ad}(\eps_1-\eps_2)-\frac{1}{2}A_2\frac{\partial\omega}{\partial\au}-\frac{\partial A_2}{\partial\au}\omega\right)\right]=0.&
\end{split}
\ee
Equations (\ref{eq:shell16}) and (\ref{eq:shell17}) are the {\it Gol'denweizer conditions}. They link the six parameters $\eps_1,\eps_2,\omega,\kappa_1,\kappa_2$ and $\tau$, that characterize the deformation of $\Sigma$ and have been derived in the same manner of the Gauss-Codazzi conditions, Sect. \ref{sec:gausscodazzi}. In some sense, they are the Gauss-Codazzi compatibility conditions for  the deformation of $\Sigma$. The Gol'denweizer conditions must be satisfied by any displacement function $\bu=(u_1,u_2,u_3)$ of $\Sigma$, i.e., they become identities if the quantities $\eps_1,\eps_2,\omega,\kappa_1,\kappa_2$ and $\tau$ are expressed through eqs. (\ref{eq:shell18}) and the Gauss-Codazzi conditions are used; this check, not difficult, but long, is left to the reader as an exercice. 

The role of the Gol'denweizer conditions  is, for shells,  the same of the Saint Venant-Beltrami compatibility conditions for the deformation of continuum bodies: they are the continuity  conditions for the deformation of $\Sigma$ and ensure the compatibility of the deformation: only six functions $\eps_1,\eps_2,\omega,\kappa_1,\kappa_2$ and $\tau$ that satisfies the Gol'denweizer conditions are physical deformation functions, i.e. corresponding to a displacement vector function $\bu=(u_1,u_2,u_3)$.  

\section{Internal actions}
\begin{figure}[h]
\begin{center}
\includegraphics[height=.25\textheight]{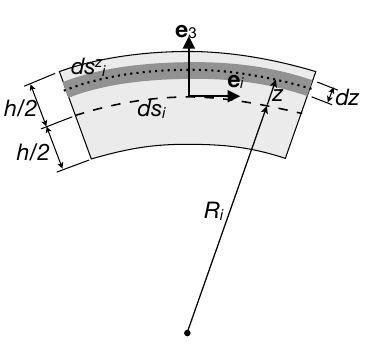}
\caption{Scheme for the stresses.}
\label{fig:8_4}
\end{center}
\end{figure}

\begin{figure}[h]
\begin{center}
\includegraphics[height=.2\textheight]{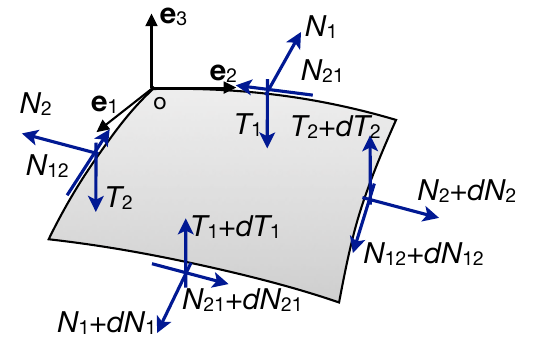}\ \
\includegraphics[height=.195\textheight]{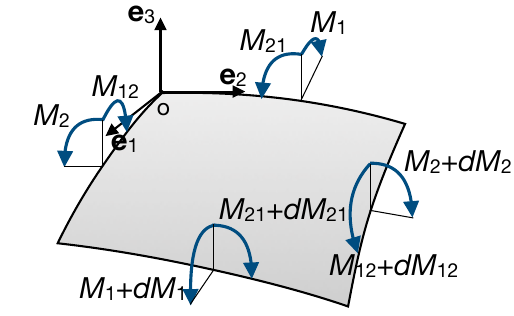}
\caption{Internal forces, left, and  moments, right.}
\label{fig:8_5}
\end{center}
\end{figure}

For a small normal section along the line $\ai$ orthogonal to $\e_j$, i.e. to line $\alpha_j$, see Fig. \ref{fig:8_4}, the resultant of the stress $\sigma_{ij}$   is\footnote{Recall that the length of the arc $\ai$ on $\Sigma^z$ is $ds_i^z=A_i\left(1+\frac{z}{R_i}\right)d\ai$, cf. eqs. (\ref{eq:shell19}) and (\ref{eq:shell7}).}
\be
\textbf{S}_j=S_{kj}\e_k=A_id\alpha_i\int_{-\frac{h}{2}}^{\frac{h}{2}}\sigma_{kj}\left(1+\frac{z}{R_i}\right)dz\ \e_k.
\ee
Because the length of the arc $\ai$ on $\Sigma$ is $ds_i=A_id\ai$, the resultant  $\textbf{s}_j$ per unit length is 
\be
\textbf{s}_j=\frac{\textbf{S}_j}{A_id\alpha_i},
\ee
 whose components are, see Fig. \ref{fig:8_5},
\be
\label{eq:shell25}
\besp
&N_j:=\textbf{s}_j\cdot\e_j=\int_{-\frac{h}{2}}^{\frac{h}{2}}\sigma_{jj}\left(1+\frac{z}{R_i}\right)dz,\\
&N_{ij}:=\textbf{s}_j\cdot\e_i=\int_{-\frac{h}{2}}^{\frac{h}{2}}\sigma_{ij}\left(1+\frac{z}{R_i}\right)dz,\\
&T_j:=\textbf{s}_j\cdot\e_3=\int_{-\frac{h}{2}}^{\frac{h}{2}}\sigma_{3j}\left(1+\frac{z}{R_i}\right)dz.
\end{split}
\ee
$N_j$ and $N_{ij}$ are the {\it membrane forces}, already found in Sect. \ref{sec:internalactionsmembranes}, while the $T_j$s are the {\it shear forces}, that do not exist for membranes. 

Multiplying the stresses $\sigma_{ij}$ by $z$ and proceeding in the same way, we obtain the components of the moments $\textbf{M}_j$ per unit length\footnote{Like for plates, the component of the moment along the normal to $\Sigma$ is meaningless.}:
\be
\label{eq:shell26}
\besp
&M_j:=\int_{-\frac{h}{2}}^{\frac{h}{2}}z\sigma_{jj}\left(1+\frac{z}{R_i}\right)dz,\\
&M_{ij}:=\int_{-\frac{h}{2}}^{\frac{h}{2}}z\sigma_{ij}\left(1+\frac{z}{R_i}\right)dz,
\end{split}
\ee
that are still represented in Fig. \ref{fig:8_5}; the $M_j$s are the {\it bending moments}, while  the $M_{ij}$s are the {\it twisting moments}. Like the shear forces, also bending and twisting moments do not exist in membranes and differentiate the mechanical response of shells with respect to that of membranes. 

To remark that the Kirchhoff-Love's hypotheses, namely the second one,  lead to the mechanical paradox $\sigma_{3j}=0$, already found in the case of the plates, see Sect. \ref{sec:stressplate}, which implies that the shear forces $T_j$ are identically null. Just like in the case of plates, this paradox is ignored and shear forces $T_j$, necessary for equilibrium whenever external loads parallel to $\et$ exist, are introduced as above and used in the equilibrium equations.

\section{Equilibrium equations for shells}
Let us write the the equilibrium equations for  a shell element of dimensions $d\au,d\ad$ like in Fig. \ref{fig:8_5}. If the load $\q$ for unit area of $\Sigma$, comprehending the body forces and the surface loads applied to the shell surfaces, has components, with respect to the local basis $\{\eu,\ed,\et\},\q=(q_1,q_2,q_3)$, then balance of the linear momentum gives
\be
\frac{\partial\textbf{S}_j}{\partial\alpha_j}d\alpha_j+\q A_1A_2d\au d\ad=\bo,
\ee
i.e.
\be
\besp
\frac{\partial}{\partial\au}\left[\left(N_1\eu+N_{21}\ed+T_1\et\right)A_2d\ad\right]d\au+
\frac{\partial}{\partial\ad}\left[\left(N_{12}\eu+N_2\ed+T_2\et\right)A_1d\au\right]d\ad&\\
+\q A_1A_2d\au d\ad=\bo,&
\end{split}
\ee
that gives, using eq. (\ref{eq:codazzi3}), the three equilibrium equations, one for each direction,
\be
\label{eq:shell20}
\besp
&\frac{1}{A_1A_2}\left(\frac{\partial (A_2N_1)}{\partial\au}+\frac{\partial(A_1N_{12})}{\partial\ad}+N_{21}\frac{\partial A_1}{\partial\ad}-N_2\frac{\partial A_2}{\partial\au}\right)+\frac{T_1}{R_1}+q_1=0,\\
&\frac{1}{A_1A_2}\left(\frac{\partial (A_2N_{21})}{\partial\au}+\frac{\partial(A_1N_2)}{\partial\ad}+N_{12}\frac{\partial A_2}{\partial\au}-N_1\frac{\partial A_1}{\partial\ad}\right)+\frac{T_2}{R_2}+q_2=0,\\
&\frac{1}{A_1A_2}\left(\frac{\partial (A_2T_1)}{\partial\au}+\frac{\partial(A_1T_2)}{\partial\ad}\right)-\frac{N_1}{R_1}-\frac{N_2}{R_2}+q_3=0.
\end{split}
\ee

Similarly, the balance of the angular momentum, made about point $o$, Fig. (\ref{fig:8_5}), reads
\be
\frac{\partial \M_j}{\partial\alpha_j}d\alpha_j+(p-o)\times\q A_1A_2d\au d\ad+(p-o)\times\left(\textbf{S}_j+\frac{\partial\textbf{S}_j}{\partial\alpha_j}d\alpha_j\right)-(p-o)\times\textbf{S}_j=\bo.
\ee
Replacing the expressions of the components of the $\textbf{S}_j$s, using once more eqs. (\ref{eq:codazzi3}), 
developing the derivatives and neglecting the terms of order greater than the second one, we get the three equations
\be
\label{eq:shell21}
\besp
&\frac{1}{A_1A_2}\left(\frac{\partial (A_2M_1)}{\partial\au}+\frac{\partial(A_1M_{12})}{\partial\ad}+M_{21}\frac{\partial A_1}{\partial\ad}-M_2\frac{\partial A_2}{\partial\au}\right)-T_1=0,\\
&\frac{1}{A_1A_2}\left(\frac{\partial (A_2M_{21})}{\partial\au}+\frac{\partial(A_1M_2)}{\partial\ad}+M_{12}\frac{\partial A_2}{\partial\au}-M_1\frac{\partial A_1}{\partial\ad}\right)-T_2=0,\\
&N_{21}-N_{12}+\frac{M_{21}}{R_1}-\frac{M_{12}}{R_2}=0.
\end{split}
\ee
However, if in the third equation the expressions of $N_{ij}$s and $M_{ij}$s are used, we get easily 
\be
\int_{-\frac{h}{2}}^{\frac{h}{2}}\left(1+\frac{z}{R_1}\right)\left(1+\frac{z}{R_2}\right)(\sigma_{12}-\sigma_{21})dz=0,
\ee
which is an identity, because of the symmetry of the stress tensor $\bsig$ (and independently from the shell theory). 

Equations (\ref{eq:shell20}) and (\ref{eq:shell21}) are the {\it equilibrium equations for shells}. They are six equations (one of them is an identity) for ten unknowns: $N_1,N_2,N_{12},N_{21},T_1,T_2,M_1,M_2,M_{12}$ and $M_{21}$: like plates, but unlike thin membranes, shells are {\it intrinsically hyperstatic bodies}. A constitutive law must then be introduced, namely the elastic one, to solve the static problem of shells.

The equilibrium equations (\ref{eq:shell20}) and (\ref{eq:shell21}) can be transformed into a form similar to the compatibility equations: first, introduce the quantities
\be
\label{eq:shell32}
S:=N_{21}-\frac{M_{12}}{R_2}=N_{12}-\frac{M_{21}}{R_1},\ \ \ H:=\frac{M_{12}+M_{21}}{2};
\ee
then, use eqs. (\ref{eq:shell21})$_{2,3}$ to eliminate $T_1$ and $T_2$ from eqs. (\ref{eq:shell20}) and use the Codazzi conditions to differentiate, to finally find
\be
\label{eq:shell39}
\besp
\frac{\partial(A_2N_1)}{\partial\au}+\frac{\partial (A_1S)}{\partial\ad}+\frac{\partial A_1}{\partial\ad}S-\frac{\partial A_2}{\partial\au}N_2+\frac{1}{R_1}\left(\frac{\partial(A_2M_1)}{\partial\au}-\frac{\partial A_2}{\partial\au}M_2+2\frac{\partial(A_1H)}{\partial\ad}\right.&\\
\left.+2\frac{R_1}{R_2}\frac{\partial A_1}{\partial\ad}H\right)=-A_1A_2q_1,&\\
\frac{\partial(A_2S)}{\partial\au}+\frac{\partial (A_1N_2)}{\partial\ad}+\frac{\partial A_2}{\partial\au}S-\frac{\partial A_1}{\partial\ad}N_1+\frac{1}{R_2}\left(\frac{\partial(A_1M_2)}{\partial\ad}-\frac{\partial A_1}{\partial\ad}M_1+2\frac{\partial(A_2H)}{\partial\au}\right.&\\
\left.+2\frac{R_2}{R_1}\frac{\partial A_2}{\partial\au}H\right)=-A_1A_2q_2,&\\
\frac{N_1}{R_1}+\frac{N_2}{R_2}-\frac{1}{A_1A_2}\left[\frac{\partial}{\partial\au}\frac{1}{A_1}
\left(\frac{\partial(A_2M_1)}{\partial\au}+\frac{\partial(A_1H)}{\partial\ad}+\frac{\partial A_1}{\partial\ad}H-\frac{\partial A_2}{\partial\au}M_2\right)\right.&\\
+\left.\frac{\partial}{\partial\ad}\frac{1}{A_2}
\left(\frac{\partial(A_2H)}{\partial\au}+\frac{\partial(A_1M_2)}{\partial\ad}+\frac{\partial A_2}{\partial\au}H-\frac{\partial A_1}{\partial\ad}M_1\right)\right]=q_3.&
\end{split}
\ee

\section{The strain energy of a shell}
The strain energy of an elastic isotropic  stressed shell is
\be
V=\frac{1}{2}\int_\Omega\bsig\cdot\beps\ dv=\frac{1}{2}\int_\Sigma\left(\int_{-\frac{h}{2}}^{\frac{h}{2}}\bsig\cdot\beps\  dz\right)d\Sigma^z,
\ee
with 
\be
d\Sigma^z =A_1A_2\left(1+\frac{z}{R_1}\right)\left(1+\frac{z}{R_2}\right)d\au d\ad,
\ee
 a small part of the  surface $\Sigma^z$ at the distance $z$ from the middle surface $\Sigma$. The Lamé's equations give the link between $\bsig$ and $\beps$:
\be
\bsig=\frac{E}{1+\nu}\left(\beps+\frac{\nu}{1-2\nu}\tr\beps\ \I\right),
\ee
with $E$ the Young's modulus and $\nu$ the Poisson's ratio of the material. For the Kirchhoff-Love assumptions, $\eps_{13}=\eps_{23}=0\Rightarrow\sigma_{13}=\sigma_{23}=0$ too. As already done for plates, and for the same reasons, we further assume 
\be
\sigma_{33}=0\Rightarrow\eps_{33}=-\frac{\nu}{1-\nu}(\eps_{11}+\eps_{22}),
\ee
so that
\be
\label{eq:shell30}
\besp
&\sigma_{11}=\frac{E}{1-\nu^2}(\eps_{11}+\nu\eps_{22}),\\
&\sigma_{22}=\frac{E}{1-\nu^2}(\eps_{22}+\nu\eps_{11}),\\
&\sigma_{12}=\frac{E}{1+\nu}\eps_{12}.
\end{split}
\ee
Hence,
\be
\besp
V=&\frac{1}{2}\int_\Sigma\left[\int_{-\frac{h}{2}}^{\frac{h}{2}}(\sigma_{11}\eps_{11}+\sigma_{22}\eps_{22}+2\sigma_{12}\eps_{12})\left(1+\frac{z}{R_1}\right)\left(1+\frac{z}{R_2}\right)dz\right]A_1A_2d\au d\ad\\
=&\frac{1}{2}\int_\Sigma\left[\int_{-\frac{h}{2}}^{\frac{h}{2}}\left(\frac{E}{1-\nu^2}(\eps_{11}+\nu\eps_{22})\eps_{11}+
\frac{E}{1-\nu^2}(\eps_{22}+\nu\eps_{11})\eps_{22}\right.\right.\\
&\left.\left.+2\frac{E}{1+\nu}\eps_{12}^2\right)\left(1+\frac{z}{R_1}\right)\left(1+\frac{z}{R_2}\right)dz\right]A_1A_2d\au d\ad.
\end{split}
\ee
For a point of the shell on $\Sigma^z, \eps_{11}=\eps_1^z,\eps_{22}=\eps_2^z$ and $\eps_{12}=\dfrac{\omega^z}{2}$, which gives
\be
\besp
V=\frac{E}{2(1-\nu^2)}\int_\Sigma\left[\int_{-\frac{h}{2}}^{\frac{h}{2}}
\left((\eps_1^z)^2+2\nu\eps_1^z\eps_2^z+(\eps_2^z)^2+
(1-\nu)(\omega^z)^2\right)\times\right.&\\
\left.\times\left(1+\frac{z}{R_1}\right)\left(1+\frac{z}{R_2}\right)dz\right]A_1A_2d\au d\ad.&
\end{split}
\ee
Using  eqs. (\ref{eq:shell11}), (\ref{eq:shell12}) and (\ref{eq:shell22}), $\eps_1^z,\eps_2^z$ and $\omega^z$ can be expressed as functions of the deformation parameters of $\Sigma$, i.e. $\eps_1,\eps_2,\omega,\kappa_1,\kappa_2,\tau$. In order to simplify, we develop first $\eps_1^z,\eps_2^z$ and $\omega^z$ in a power series of $z$ up to the second order\footnote{This is legitimated by the hypothesis that the shell is thin, hence $z/R_i, i=1,2$ changes few in the thickness of the shell.}:
\be
\besp
&\eps_1^z\simeq\eps_1+z\left(\kappa_1-\frac{\eps_1}{R_1}\right)-\frac{z^2}{R_1}\left(\kappa_1-\frac{\eps_1}{R_1}\right),\\
&\eps_2^z\simeq\eps_2+z\left(\kappa_2-\frac{\eps_2}{R_2}\right)-\frac{z^2}{R_2}\left(\kappa_2-\frac{\eps_2}{R_2}\right),\\
&\omega^z\simeq\omega+2z\left[\tau-\left(\frac{1}{R_1}+\frac{1}{R_2}\right)\frac{\omega}{2}\right]-
z^2\left(\frac{1}{R_1}+\frac{1}{R_2}\right)\left(\tau-\frac{R_1^2+R_2^2}{R_1R_2}\frac{\omega}{R_1R_2}\right).
\end{split}
\ee
Once these quantities injected into the last equation, the integration with respect to $z$ can be done. Then, in order to be consistent with the approximations done up to for a linear theory of shells, higher order terms must be neglected\footnote{This operation is quite delicate and not at all trivial; this discussion is omitted here, but the reader can find the details in the treatise of Novozhilov, see the bibliography, p. 45.}, which finally leads to the expression of the strain energy of the shell:
\be
\label{eq:shell27}
\besp
V&=\frac{E\ h}{2(1-\nu^2)}\int_\Sigma\left[(\eps_1+\eps_2)^2-2(1-\nu)\left(\eps_1\eps_2-\frac{\omega^2}{4}\right)\right]A_1A_2d\au d\ad\\
&+\frac{E\ h^3}{24(1-\nu^2)}\int_\Sigma\left[(\kappa_1+\kappa_2)^2-2(1-\nu)\left(\kappa_1\kappa_2-\tau^2\right)\right]A_1A_2d\au d\ad.
\end{split}
\ee
It is apparent here that the strain energy $V$ is composed by two parts: the first integral accounts for the part of $V$ depending upon the stretching and shearing of $\Sigma$, i.e. it corresponds to the {\it extension strain energy} of the shell, while the second integral corresponds to the part of $V$ depending upon the bending and twisting of $\Sigma$, i.e. it corresponds to the {\it flexural strain energy} of the shell. 

Finally, if eqs. (\ref{eq:shell18}) are used to express $\eps_1,\eps_2,\omega,\kappa_1,\kappa_2$ and $\tau$, then $V$ can be given as function of the displacement components $u_1,u_2,u_3$ of the points of $\Sigma$.

\section{The constitutive law of elastic shells}
We use the previously found expression of the strain energy $V$ to obtain the constitutive law for elastic shells, i.e. the relations between the internal actions $N_1,N_2,N_{12},N_{21},T_1,T_2$, $M_1,M_2,M_{12}, M_{21}$ and the deformations $\eps_1,\eps_2,\omega,\kappa_1,\kappa_2,\tau$. To this purpose, let us write the variation $\delta V$ of $V$ due to a variation $\delta\beps$ of its deformation:
\be
\label{eq:shell24}
\besp
\delta V=\int_\Omega\bsig\cdot\delta\beps\ dv=&\int_\Sigma\left[\int_{-\frac{h}{2}}^{\frac{h}{2}}\left(\sigma_{11}\delta\eps_1^z+\sigma_{22}\delta\eps_2^z+\sigma_{12}\delta\omega^z\right)\times\right.\\
&\left.\times\left(1+\frac{z}{R_1}\right)\left(1+\frac{z}{R_2}\right)dz\right]A_1A_2d\au d\ad.
\end{split}
\ee
The variations $\delta\eps_1^z,\delta\eps_2^z$ and $\delta\omega^z$ can be calculated through eqs. (\ref{eq:shell11}), (\ref{eq:shell12}) and (\ref{eq:shell23}):
\be
\besp
&\delta\eps_1^z=\frac{1}{1+\frac{z}{R_1}}(\delta\eps_1+z\delta\kappa_1),\\
&\delta\eps_2^z=\frac{1}{1+\frac{z}{R_2}}(\delta\eps_2+z\delta\kappa_2),\\
&\delta\omega^z=\frac{1}{1+\frac{z}{R_1}}(\delta\omega_1+z\delta\tau_1)+\frac{1}{1+\frac{z}{R_2}}(\delta\omega_2+z\delta\tau_2).
\end{split}
\ee
Replacing these expressions into eq. (\ref{eq:shell24}) and using equations (\ref{eq:shell25}) and (\ref{eq:shell26}), we get
\be
\label{eq:shell28}
\delta V=\int_\Sigma(N_1\delta\eps_1+N_2\delta\eps_2+S\delta\omega+M_1\delta\kappa_1+M_2\delta\kappa_2+2H\delta\tau)A_1A_2d\au d\ad.
\ee
Let us write now the variation $\delta V$ using eq. (\ref{eq:shell27}):
\be
\label{eq:shell29}
\besp
\delta V&=\frac{E\ h}{2(1-\nu^2)}\int_\Sigma\left[2(\eps_1+\eps_2)(\delta\eps_1+\delta\eps_2)\hspace{-1mm}-\hspace{-1mm}2(1-\nu)\hspace{-1mm}\left(\eps_1\delta\eps_2+\eps_2\delta\eps_1-\frac{\omega\delta\omega}{2}\right)\right]\hspace{-1mm}A_1A_2d\au d\ad\\
&+\frac{E\ h^3}{24(1-\nu^2)}\int_\Sigma\left[(\kappa_1+\kappa_2)(\delta\kappa_1+\delta\kappa_2)\hspace{-1mm}-\hspace{-1mm}2(1-\nu)\left(\kappa_1\delta\kappa_2+\kappa_2\delta\kappa_1-2\tau\delta\tau\right)\right]\hspace{-1mm}A_1A_2d\au d\ad\\
&=\frac{E\ h}{1-\nu^2}\int_\Sigma\left((\eps_1+\nu\eps_2)\delta\eps_1+(\eps_2+\nu\eps_1)\delta\eps_2+\frac{1-\nu}{2}\omega\delta\omega\right)A_1A_2d\au d\ad\\
&+\frac{E\ h^3}{12(1-\nu^2)}\int_\Sigma\left((\kappa_1+\nu\kappa_2)\delta\kappa_1+(\kappa_2+\nu\kappa_1)\delta\kappa_2+2(1-\nu)\tau\delta\tau\right)A_1A_2d\au d\ad.
\end{split}
\ee
Comparing eqs. (\ref{eq:shell28}) and (\ref{eq:shell29}) we see that, for the arbitrariness of the variations $\delta\eps_i,\delta\kappa_i,\delta\omega$ and $\delta\tau$,  it must be
\be
\label{eq:shell33}
\begin{array}{lll}
N_1=\dfrac{E\ h}{1-\nu^2}(\eps_1+\nu\eps_2),&\ \ \ \ \ &N_2=\dfrac{E\ h}{1-\nu^2}(\eps_2+\nu\eps_1),\medskip\\
M_1=\dfrac{E\ h^3}{12(1-\nu^2)}(\kappa_1+\nu\kappa_2),&\ \ \ \ \ &M_2=\dfrac{E\ h^3}{12(1-\nu^2)}(\kappa_2+\nu\kappa_1),\medskip\\
S=\dfrac{E\ h}{2(1+\nu)}\omega,&\ \ \ \ \ &H=\dfrac{E\ h^3}{12(1+\nu)}\tau.
\end{array}
\ee
These equations are the relations between the internal actions and the strains, i.e. they are the {\it constitutive law for shells}. It should be remarked that they give the link between two sets of six quantities: $N_1,N_2,S,M_1,M_2,H$, on one side, with $\eps_1,\eps_2,\omega,\kappa_1\kappa_2,\tau$, on the other side.
However, shear forces $T_1$ and $T_2$ are not included in these equations. This is because of the Kirchhoff-Love assumptions, for which $\eps_{13}=\eps_{23}=0$, that, together with the constitutive law, i.e. the  Lamé's equations, give $\sigma_{13}=\sigma_{23}=0$. $T_1$ and $T_2$ can be obtained by eqs. (\ref{eq:shell21})$_{1,2}$ once $M_{12}$ and $M_{21}$ known. To this purpose, it is necessary to develop  the last two equations:
\be
\besp
&N_{21}-\frac{M_{12}}{R_2}=\dfrac{E\ h}{2(1+\nu)}\omega,\\
&N_{12}-\frac{M_{21}}{R_1}=\dfrac{E\ h}{2(1+\nu)}\omega,\\
&\frac{M_{12}+M_{21}}{2}=\dfrac{E\ h^3}{12(1+\nu)}\tau.
\end{split}
\ee
We see hence that there are three equations for four unknowns. 

The indeterminacy is removed after proving that\footnote{The proof is not immediate and is omitted here, we address the reader to p. 48-50 of the book of Novozhilov.}, within the approximations used for the linear theory, it is actually $M_{12}=M_{21}=H$, so that finally the expressions of forces and moments in terms of the deformation of $\Sigma$ are
\be
\label{eq:shell31}
\begin{array}{lll}
N_1=\dfrac{E\ h}{1-\nu^2}(\eps_1+\nu\eps_2),&\ \ \ \ \ &N_2=\dfrac{E\ h}{1-\nu^2}(\eps_2+\nu\eps_1),\medskip\\
N_{12}=\dfrac{E\ h}{2(1+\nu)}\left(\omega+\dfrac{h^2}{6R_1}\tau\right),&\ \ \ \ \ &N_{21}=\dfrac{E\ h}{2(1+\nu)}\left(\omega+\dfrac{h^2}{6R_2}\tau\right),\medskip\\
M_1=\dfrac{E\ h^3}{12(1-\nu^2)}(\kappa_1+\nu\kappa_2),&\ \ \ \ \ &M_2=\dfrac{E\ h^3}{12(1-\nu^2)}(\kappa_2+\nu\kappa_1),\medskip\\
M_{12}=\dfrac{E\ h^3}{12(1+\nu)}\tau,&\ \ \ \ \ &M_{21}=\dfrac{E\ h^3}{12(1+\nu)}\tau.
\end{array}
\ee

There is another way to obtain these relations: injecting successively eqs. (\ref{eq:shell30}), (\ref{eq:shell11}), (\ref{eq:shell12}), (\ref{eq:shell23}) into eqs. (\ref{eq:shell25}) and (\ref{eq:shell26}). However, this leads to a different form of the expressions for $N_{12}$ and $N_{21}$,
\be
N_{12}=N_{21}=\dfrac{E\ h}{2(1+\nu)}\omega,
\ee
while the other quantities do not change. This expression for $N_{12}$ and $N_{21}$ was that given by Love; its advantage is to be simpler than the one in eq. (\ref{eq:shell31}) and to be analogous to the corresponding formulae for plates. However, though largely used, these last formulae brings to some mechanical  inconsistencies, one of them is that the Maxwell-Betti reciprocal theorem does not hold. Also, if the last equation is used into eq. (\ref{eq:shell21})$_3$, then it is no longer identically satisfied. We have seen that eq. (\ref{eq:shell21})$_3$ is an identity thanks to the symmetry of $\bsig$. So, finally, the last equation contradicts also the basic property of symmetry of the stress tensor. 

These contradictions should not be surprising: they all arise from the kinematical model, on one hand, and on the approximations made, on the other hand, i.e. on the choice of the terms to be discarded in the order of approximation of the theory. To this purpose, it should be noticed that opinions diverge and different approximations exist, where different terms are retained or neglected.  

\section{Boundary conditions}
Let us now consider the case of a boundary coinciding with the line of principal curvature $\ai=const.$ Along this line, which is a line $\alpha_j$ for $\Sigma$, by the integration of the stresses over the thickness, act the forces $N_i,N_{ji}, T_i$ and the moments $M_i$ and $M_{ji}$. It seems hence that there are five boundary conditions  for each line $\ai$, but this is false, the right number is four: the twisting moment $M_{ji}$ can be replaced by a function of the tangential and shear forces. 

\begin{figure}[h]
\begin{center}
\includegraphics[height=.25\textheight]{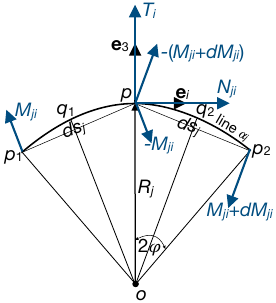}
\caption{Scheme for the boundary conditions.}
\label{fig:8_7}
\end{center}
\end{figure}

To prove this\footnote{The proof is based upon an argument analogous to the one used by Lord Kelvin to state the right number of boundary conditions for the classical theory of plates, cf. Sect. \ref{sec:Lordkelvinbc}.}, let us consider a segment of the boundary on the line $\ai$ around the point $p$, see Fig. \ref{fig:8_7}, and ideally subdivide the arc into small segments of equal length, say $|p-p_1|=|p-p_2|\simeq ds_j=A_jd\alpha_j$, and of course also $|q_1-q_2|\simeq ds_j$, with the points $q_1,q_2$ halfway the previous segments.

Be $M_{ji}$ the twisting moment at $q_1$ and $M_{ji}+dM_{ji}=M_{ji}+\dfrac{\partial M_{ji}}{\partial s_j}ds_j$ that at $q_2$; then, replace the twisting moment over the arc $p_1p$ by a couple of forces\footnote{Recall that bending and twisting moments are actually moments per unit length, i.e. they have the dimensions of a force.} $M_{ji}$, applied at the ends of the arc. Such fictitious forces are directed like the vector $q_1-o$, where $o$ is the center of curvature of the line $\alpha_j$. We make the same type of substitution also for the arc $pp_2$.
Now, projecting the forces at $p_1$ onto $\e_i$ we get
\be
\left(2M_{ji}+\frac{\partial M_{ji}}{\partial s_j}ds_j\right)\varphi,
\ee
with the angle $\varphi$ such that, see Fig. \ref{fig:8_7},
\be
2\varphi=\frac{ds_j}{R_j},
\ee
so finally, neglecting higher order quantities, we get a force per unit length, along $\e_i$, equal to
\be
\frac{M_{ji}}{R_j}.
\ee
 Moreover, projecting onto  $\et$ gives the shear force, per unit length,
 \be
 \frac{\partial M_{ji}}{\partial s_j}.
 \ee
According to the Saint-Venant Principle, this statical substitution made on the boundary $\ai=const.$, i.e. on the line $\alpha_j$, does not affect the real stress state at a sufficient distance, say of the order of $h$, in the shell. Finally, on such a boundary, the four quantities 
\be
N_i,\ N_{ji}+\frac{M_{ji}}{R_j},\ T_i+\frac{1}{A_j}\frac{\partial M_{ji}}{\partial \alpha_j},\ M_{i},
\ee
completely determine the stress state at the shell's edge. As a consequence, the number of boundary conditions at each edge\footnote{The more general case of a generic boundary is more complicate and not dealt with here; the reader can find it in the treatise of Love, see the bibliography.} corresponding to a line of principal curvature is four. We remark also that, by eq. (\ref{eq:shell32}), it is
\be
N_{ji}+\frac{M_{ji}}{R_j}=N_{ji}-\frac{M_{ij}}{R_j}+\frac{M_{ij}+M_{ji}}{R_j}=S+2\frac{H}{R_j},
\ee
while, by eq. (\ref{eq:shell21}) we get
\be
T_i+\frac{1}{A_j}\frac{\partial M_{ji}}{\partial \alpha_j}=\frac{1}{A_1A_2}\left(2\frac{\partial (A_iH)}{\partial\alpha_j}+\frac{\partial (A_jM_i)}{\partial\alpha_i}-\frac{\partial A_j}{\partial\alpha_i}M_j\right),
\ee
so, like for equilibrium equations, $T_{12},T_{21},M_{12}$ and $M_{21}$ do not enter separately the boundary conditions, but only through $H$ and $S$. 

Of course, geometrical boundary conditions are possible too, prescribing the value of some components of displacement, and various combinations are possible, here is some few examples:
\begin{itemize}
\item free edge: 
\be
N_i=0,\ N_{ji}+\frac{M_{ji}}{R_j}=0,\ T_i+\frac{1}{A_j}\frac{\partial M_{ji}}{\partial\alpha_j}=0,\ M_i=0;
\ee
\item hinged edge with fixed support:
\be
M_i=0,\ u_1=0,\ u_2=0,\ u_3=0;
\ee
\item hinged edge with support free to move in the normal direction:
\be
M_i=0,\ T_i+\frac{1}{A_j}\frac{\partial M_{ji}}{\partial\alpha_j}=0, \ u_1=0,\ u_2=0;
\ee
\item clamped edge:
\be
u_1=0,\ u_2=0,\ u_3=0,\ \psi=-\frac{1}{A_i}\frac{\partial u_3}{\partial\ai}+\frac{u_2}{R_i}=0.
\ee
\end{itemize}
Many other combinations are possible, but in all the cases, like for the equilibrium equations and unlike plates, it is not possible to uncouple the boundary conditions for the membrane and flexural problem.

\section{Compatibility conditions in terms of forces and moments}
Using the constitutive equations (\ref{eq:shell33}) we can express the strain of $\Sigma$ in terms of the forces and moments:
\be
\label{eq:shell34}
\begin{array}{lll}
\eps_1=\dfrac{N_1-\nu N_2}{E\ h},&
\eps_2=\dfrac{N_2-\nu N_1}{E\ h},&
\omega=\dfrac{2(1+\nu)}{E\ h}S,\medskip\\
\kappa_1=\dfrac{12}{E\ h^3}(M_1-\nu M_2),&
\kappa_2=\dfrac{12}{E\ h^3}(M_2-\nu M_1),&
\tau=\dfrac{12(1+\nu)}{E\ h^3}H.
\end{array}
\ee
Replacing these equations into eqs. (\ref{eq:shell16}) and (\ref{eq:shell17}) we get
\be
\label{eq:shell35}
\besp
&\frac{\partial A_2(M_2-\nu M_1)}{\partial\au}-(1+\nu)\left(\frac{\partial(A_1H)}{\partial\ad}+\frac{\partial A_1}{\partial \ad}H\right)-\frac{\partial A_2}{\partial\au}(M_1-\nu M_2)-\\
&\frac{h^2}{12R_1}\left(\frac{\partial A_2(N_2-\nu N_1)}{\partial\au}-\frac{\partial A_2}{\partial\au}(N_1-\nu N_2)-\right.\\
&\left.2(1+\nu)\frac{\partial (A_1S)}{\partial\ad}-2(1+\nu)\frac{R_1}{R_2}\frac{\partial A_1}{\partial\ad}S\right)=0,
\end{split}
\ee
\be
\label{eq:shell36}
\besp
&\frac{\partial A_1(M_1-\nu M_2)}{\partial\au}-(1+\nu)\left(\frac{\partial(A_2H)}{\partial\au}+\frac{\partial A_2}{\partial \au}H\right)-\frac{\partial A_1}{\partial\ad}(M_2-\nu M_1)-\\
&\frac{h^2}{12R_2}\left(\frac{\partial A_1(N_1-\nu N_2)}{\partial\ad}-\frac{\partial A_1}{\partial\ad}(N_2-\nu N_1)-\right.\\
&\left.2(1+\nu)\frac{\partial (A_2S)}{\partial\au}-2(1+\nu)\frac{R_2}{R_1}\frac{\partial A_2}{\partial\au}S\right)=0,
\end{split}
\ee
and
\be
\label{eq:shell37}
\besp
&\frac{M_2-\nu M_1}{R_1}+\frac{M_1-\nu M_2}{R_2}+\frac{h^2}{12}\frac{1}{A_1A_2}\left\{\frac{\partial}{\partial\au}\frac{1}{A_1}\left[\frac{\partial A_2(N_2-\nu N_1)}{\partial\au}-\right.\right.\\
&\left.(1+\nu)\left(\frac{\partial(A_1S)}{\partial\ad}+\frac{\partial A_1}{\partial\ad}S\right)-\frac{\partial A_2}{\partial\au}(N_1-\nu N_2)\right]+\\
&\frac{\partial}{\partial\ad}\frac{1}{A_2}
\left[\frac{\partial A_1(N_1-\nu N_2)}{\partial\ad}-(1+\nu)\left(\frac{\partial(A_2S)}{\partial\au}+\frac{\partial A_2}{\partial\au}S\right)-\right.\\
&\left.\left.\frac{\partial A_1}{\partial\ad}(N_2-\nu N_1)\right]\right\}=0.
\end{split}
\ee
These three equations are the {\it compatibility conditions in terms of forces and moments}. Their meaning is the following one: assume that the forces and moments are known; then, substituted into eqs. (\ref{eq:shell34}) and replacing in theses ones the expressions of the strains of $\Sigma$, i.e. $\eps_1,\eps_2,\omega,\kappa_1,\kappa_2,\tau$, in terms of the components of the components $u_1,u_2,u_3$ of the displacement vector $\bu$, eq. (\ref{eq:shell18}), we obtain six partial differential equations (PDEs) for the determination of $u_1,u_2,u_3$. However, such equations could be inconsistent, i.e. it could happen that there not exist three functions $u_1,u_2,u_3$ satisfying all the six equations. Actually, the compatibility conditions in terms of forces and moments are necessary and sufficient conditions for the existence of three independent functions $u_1,u_2,u_3$ satisfying the six PDEs determining the displacements. Actually, eqs. (\ref{eq:shell35}), (\ref{eq:shell36}) and (\ref{eq:shell37}) are the equivalent, for shells, of the Beltrami-Michell compatibility equations for elasticity. 

The above compatibility equations can be transformed using the equilibrium equations of shells, eqs. (\ref{eq:shell20}) and (\ref{eq:shell21}),  to obtain
\be
\label{eq:shell38}
\besp
(1+\nu)T_1&=\frac{1}{A_1}\frac{\partial M}{\partial\au}-\frac{h^2}{12m_1}\left\{\frac{1}{R_1A_1}\frac{\partial T}{\partial\au}-\frac{1+\nu}{A_2}\left(\frac{1}{R_1}\frac{\partial S}{\partial\ad}+\frac{2S}{R_2A_1}\frac{\partial A_1}{\partial\ad}\right)\right.\\
&+\left.\frac{1}{R_1}\left[\frac{1}{R_1A_1}\frac{\partial M}{\partial\au}+\frac{1+\nu}{A_2}\left(\frac{1}{R_1}\frac{\partial H}{\partial\ad}+\frac{2H}{R_2A_1}\frac{\partial A_1}{\partial\ad}\right)\right]+\frac{1+\nu}{R_1}q_1\right\},\\
&\\
(1+\nu)T_2&=\frac{1}{A_2}\frac{\partial M}{\partial\ad}-\frac{h^2}{12m_2}\left\{\frac{1}{R_2A_2}\frac{\partial T}{\partial\ad}-\frac{1+\nu}{A_1}\left(\frac{1}{R_2}\frac{\partial S}{\partial\au}+\frac{2S}{R_1A_2}\frac{\partial A_2}{\partial\au}\right)\right.\\
&+\left.\frac{1}{R_2}\left[\frac{1}{R_2A_2}\frac{\partial M}{\partial\ad}+\frac{1+\nu}{A_1}\left(\frac{1}{R_2}\frac{\partial H}{\partial\au}+\frac{2H}{R_1A_2}\frac{\partial A_2}{\partial\au}\right)\right]+\frac{1+\nu}{R_2}q_2\right\},\\
&\\
&\hspace{-17mm}\frac{M_1-\nu M_2}{R_2}+\frac{M_2-\nu M_1}{R_1}+\frac{h^2}{12}\left\{\Delta T+\frac{1+\nu}{A_1A_2}\left[\frac{\partial}{\partial\au}\left(\frac{A_2}{R_1}N_1\right)\hspace{-1mm}+\hspace{-1mm}\frac{\partial}{\partial\ad}\left(\frac{A_1}{R_2}N_2\right)\right]\right.\hspace{-1mm}+\\
&\hspace{-2mm}\frac{1+\nu}{A_1A_2}\left[\frac{\partial}{\partial\au}\left(\frac{1}{R_1}\frac{\partial H}{\partial\ad}+\frac{2H}{R_2A_1}\frac{\partial A_1}{\partial\ad}\right)+\frac{\partial}{\partial\ad}\left(\frac{1}{R_2}\frac{\partial H}{\partial\au}+\frac{2H}{R_1A_2}\frac{\partial A_2}{\partial\au}\right)\right]+\\
&\hspace{65mm}\left.\frac{1+\nu}{A_1A_2}\left[\frac{\partial(A_2q_1}{\partial\au}+\frac{\partial(A_1q_2}{\partial\ad}\right]\right\}=0,
\end{split}
\ee
where
\be
T:=T_1+T_2,\ \ \ M:=M_1+M_2,\ \ \ m_1=1+\frac{h^2}{12R_1^2},\ \ \ m_2=1+\frac{h^2}{12R_2^2},
\ee
while the symbol $\Delta$ denotes the Laplacian operator in curvilinear coordinates:
\be
\label{eq:laplcurv}
\Delta\cdot=\frac{1}{A_1A_2}\left\{\frac{\partial}{\partial\au}\left[\frac{A_2}{A_1}\frac{\partial\cdot}{\partial\au}\right]+\frac{\partial}{\partial\ad}\left[\frac{A_1}{A_2}\frac{\partial\cdot}{\partial\ad}\right]\right\}.
\ee

\section{The two solution methods in the theory of shells}
Two general solution methods exist in elasticity:
\begin{enumerate}[i.]
\item replacing in the equilibrium equations  the stresses $\sigma_{ij}$ by their expressions in terms of the displacements, to obtain a system of three PDEs, the so-called {\it Lamé's equations};
\item replacing in the Saint Venant-Beltrami compatibility equations the strains $\eps_{ij}$ by their expressions in terms of the stresses $\sigma_{ij}$, to obtain the six PDEs of Beltrami-Michell that, together with the equilibrium equations, permit the solution of the problem directly in terms of the $\sigma_{ij}$.
\end{enumerate}
In a similar manner, the problems of shells can be solved by two methods: in terms of displacements of $\Sigma$ or in terms of internal actions (forces and moments). In the first case, we replace in the equilibrium equations (\ref{eq:shell20}) and (\ref{eq:shell21}) the expressions of the forces and moments in terms of strains of $\Sigma$, i.e. the constitutive law of shells, eqs. (\ref{eq:shell31}), so obtaining a system of three PDEs for the unknowns $u_1,u_2,u_3$, which is equivalent to a unique 8th order PDE.

In the second method, the compatibility equations in terms of forces and moments, i.e. (\ref{eq:shell35}), (\ref{eq:shell36}) and (\ref{eq:shell37}) or eq. (\ref{eq:shell38}), are supplemented to the equilibrium equations (\ref{eq:shell39})  to get six PDEs for the determination of the unknowns $N_1,N_2,S,M_1,M_2,H$, that once more is equivalent to a unique 8th order PDE.

In all the cases, the equations are extremely difficult to be solved, also in the simplest cases, despite the fact that, thanks to the basic assumption of the theory, equations are linear. That is why different simplified shell theories have been proposed, we will see two among them.

\section{The Donnell's simplified theory}
  As said in the introduction, the first theory of shells was proposed by Aron,   who erroneously disregarded the terms depending upon the tangential displacements $u_1$ and $u_2$ in the expressions of curvature and twist, defining these ones as:
  \be
  \besp
 &\kappa_1=-\frac{1}{A_1}\frac{\partial}{\partial\au}\left(\frac{1}{A_1}\frac{\partial u_3}{\partial\au}\right)-\frac{1}{A_1A_2^2}\frac{\partial u_3}{\partial\ad},\\
  &\kappa_2=-\frac{1}{A_2}\frac{\partial}{\partial\ad}\left(\frac{1}{A_2}\frac{\partial u_3}{\partial\ad}\right)-\frac{1}{A_1^2A_2}\frac{\partial u_3}{\partial\au},\\
  &\tau=-\frac{1}{A_1A_2}\left(\frac{\partial^2u_3}{\partial\au\partial\ad}-\frac{1}{A_1}\frac{\partial A_1}{\partial\ad}\frac{\partial u_3}{\partial\au}-\frac{1}{A_2}\frac{\partial A_2}{\partial\au}\frac{\partial u_3}{\partial\ad}\right).
  \end{split}
  \ee
 Though in principle, as we have seen and Love demonstrated, these relations are false,  the approximations so obtained can be considered as fairly good whenever the membrane regime is the dominant one, i.e. whenever the stresses produced by $N_1,N_2,N_{12},N_{21}$ are predominant over those due to $M_1,M_2,M_{12},M_{21}$. This is often the case, especially when the shell is submitted to distributed loads. 
 
 Donnell\footnote{Almost simultaneously and independently, similar results were obtained also by Mushtari and Vlasov.} has shown that neglecting the contribution of $u_1$ and $u_2$ to $\kappa_1,\kappa_2$ and $\tau$, leads to other simplifications in the theory of elastic shells. In particular, it can be shown that  in this approximation equilibrium equations are
  \be
  \label{eq:donnell1}
\besp
&\frac{1}{A_1A_2}\left(\frac{\partial (A_2N_1)}{\partial\au}+\frac{\partial(A_1N_{12})}{\partial\ad}+N_{21}\frac{\partial A_1}{\partial\ad}-N_2\frac{\partial A_2}{\partial\au}\right)+q_1=0,\\
&\frac{1}{A_1A_2}\left(\frac{\partial (A_2N_{21})}{\partial\au}+\frac{\partial(A_1N_2)}{\partial\ad}+N_{12}\frac{\partial A_2}{\partial\au}-N_1\frac{\partial A_1}{\partial\ad}\right)+q_2=0,\\
&\frac{E\ h^3}{12(1-\nu^2)}\Delta\Delta u_3+\frac{N_1}{R_1}+\frac{N_2}{R_2}=q_3,
\end{split}
\ee
while for the shear forces it is
\be
\label{eq:donnell2}
\besp
&T_1=-\frac{E\ h^3}{12(1-\nu^2)}\frac{1}{A_1}\frac{\partial\Delta u_3}{\partial\au},\\
&T_2=-\frac{E\ h^3}{12(1-\nu^2)}\frac{1}{A_2}\frac{\partial\Delta u_3}{\partial\ad}.
\end{split}
\ee
To remark the similarity of eq. (\ref{eq:donnell1})$_3$ and of eqs. (\ref{eq:donnell2}) with, respectively, eqs.  (\ref{eq:germainlagrangeorth}) and (\ref{eq:tagliopiastreelasticorth}) found for the plates. In particular, the equations are the same if $R_1$ and $R_2$ tend to infinity, i.e. if the shell is flat, i.e. a plate.

  \section{The theory of shallow shells}
  Be $z=f(x,y)$ the (explicit) function defining the middle surface $\Sigma$ of a shell in a reference frame $\R=\{o;x,y,z\}$. 
  Then, we say that a shell is {\it shallow} if 
  \be
  \left(\frac{\partial z}{\partial x}\right)^2\ll1, \ \ \  \left(\frac{\partial z}{\partial y}\right)^2\ll1.
  \ee
  Without giving the details, it can be shown that in this case the (approximate) equilibrium equations of the shell reduce to only two PDEs, of two unknown functions $w(x,y)$ (representing the displacement of the points of $\Sigma$ along the $z-$axis) and $\phi(x,y)$, linked to the membrane forces:
  \be
  \besp
  &\Delta\Delta\phi+E\ h\left(\frac{\partial^2z}{\partial x^2}\frac{\partial^2w}{\partial y^2}-2\frac{\partial^2z}{\partial x\partial y}\frac{\partial^2w}{\partial x\partial y}+\frac{\partial^2z}{\partial y^2}\frac{\partial^2w}{\partial x^2}\right)=Q\\
  &\frac{E\ h^3}{12(1-\nu^2)}\Delta\Delta w-\left(\frac{\partial^2z}{\partial x^2}\frac{\partial^2\phi}{\partial y^2}-2\frac{\partial^2z}{\partial x\partial y}\frac{\partial^2\phi}{\partial x\partial y}+\frac{\partial^2z}{\partial y^2}\frac{\partial^2\phi}{\partial x^2}\right),
  \end{split}
  \ee
  where
  \be
Q:=  (1+\nu)\left(\frac{\partial^2Q_x}{\partial y^2}+\frac{\partial^2Q_y}{\partial x^2}\right)-\nu\Delta(Q_x+Q_y),
  \ee
  with $\Delta$ the customary Laplacian operator in Cartesian coordinates and
  \be
  Q_x:=\int q_xdx,\ \ \ Q_y=\int q_ydy,
  \ee
being $q_x,q_y,q_z$ the projections of the loading on $\Sigma$ onto the three Cartesian axes.
  The membrane forces are linked to the function $\phi(x,y)$ by the relations
  \be
  N_1(=N_x)=\frac{\partial^2\phi}{\partial y^2},\ \ \ S(=N_{xy}\simeq N_{yx})=-\frac{\partial^2\phi}{\partial x\partial y},\ \ \ N_2(=N_y)=\frac{\partial^2\phi}{\partial x^2}.
  \ee
  Moreover, if $u,v,w$ are the  components of the displacement vector in the frame $\{o;x,y,z\}$, then
  \be
  \besp
 & \eps_1(=\eps_{xx})=\frac{\partial u}{\partial x}-\frac{\partial^2 z}{\partial x^2}w,\ \ \  \eps_2(=\eps_{yy})=\frac{\partial v}{\partial y}-\frac{\partial^2 z}{\partial y^2}w,\\ 
 &\omega(=2\eps_{xy})=\frac{\partial u}{\partial y}+\frac{\partial v}{\partial x}-2\frac{\partial^2z}{\partial x\partial y}w,\\
 &\kappa_1(=\kappa_x)=-\frac{\partial^2w}{\partial x^2},\ \ \ \kappa_2(=\kappa_y)=-\frac{\partial^2w}{\partial y^2},\ \ \ \tau(=\kappa_{xy})=-\frac{\partial^2w}{\partial x\partial y}.
  \end{split}
  \ee
  Once more, one can easily see how these formulae are similar to those defining the strain of a plate. The constitutive equations (\ref{eq:shell33}), connecting forces and strains, are still valid.
  
  \begin{center}
*\hspace{2mm}   *\hspace{2mm} *
  \end{center}
  
  To end this Chapter, which is just an introduction tot he theory of shells, it is worth to notice the existence of a certain number of (approximated) solutions for different interesting cases:
  \begin{itemize}
  \item surfaces of revolution;
  \item cylindrical vaults;
  \item tubes;
  \item ruled surfaces.
  \end{itemize}
 The reader is addressed to the fundamental text of Timoshenko (see the bibliography) for more details on the matter.

\cleardoublepage
\phantomsection
\addcontentsline{toc}{chapter}{Suggested texts}%{\bibname}
\chapter*{Suggested texts}
\begin{enumerate}
\item A. E. H. Love: \textit{A treatise on the mathematical theory of elasticity}. Fourth edition. Dover, 1944.
\item S. P. Timoshenko: {\it Strength of materials - Parts I \& II}. Second edition. Van Nostrand, 1945.
\item I. S. Sokolnikoff: \textit{Mathematical theory of elasticity}. McGraw-Hill, 1946.
\item S. P. Timoshenko, J. N. Goodier: \textit{Theory of elasticity}. Second edition. McGraw-Hill, 1951.
\item S. P. Timoshenko, S. Woinowsky-Krieger:{\it Theory of plates and shells}. McGraw-Hill, 1959.
\item V. V. Novozhilov: {\it Thin shell theory}. Noordhoff, 1959.
\item H. L. Langhaar: \textit{Energy methods in applied mechanics}. Wiley, 1962.
%\item P. Germain, P. Muller: \textit{Introduction à la mécanique des milieux continus}. Masson, 1980.
\item J. Heyman: {\it Equilibrium of shell structures}. Clarendon Press, 1977.
\item M. E. Gurtin: \textit{An introduction to continuum mechanics}. Academic Press, 1981.
\item F. Hartmann: \textit{The mathematical foundation of structural mechanics}. Springer, 1985.
\item J. R. Barber: \textit{Elasticity}. Kluwer Academic Publishers, 1992.
%\item S. H. Strogatz: {\it Nonlinear dynamics and chaos}, Addison-Weslaey, 1994.
\item P. Villaggio: \textit{Mathematical models for elastic structures}, Cambridge University Press, 1997.
\item P. Podio-Guidugli: \textit{A primer in elasticity}. Journal of Elasticity, v. 58: 1-104, 2000.
%\item W. S. Slaughter: \textit{The linearized theory of elasticity}. Birkhäuser, 2002.
\item B. Audoly, Y. Pomeau: \textit{Elasticity and geometry}, Oxford University Press, 2010.
\item J.-J. Marigo: \textit{Mécanique des milieux continus I}. Ecole Polytechnique, 2014.\\ \href{https://cel.archives-ouvertes.fr/cel-01023392}{https://cel.archives-ouvertes.fr/cel-01023392}
\item P. M. Mariano, L. Galano: \textit{Fundamentals of the Mechanics of Solids}, Birkhä\-user, 2016.
\item \label{ref:geodiff}P. Vannucci: \textit{Tensor algebra and analysis for engineers - With application to differential geometry of curves and surfaces}, World Scientific, 2023. \href{https://doi.org/10.1142/13099}{https://doi.org/10.1142/13099}.
\end{enumerate}

\end{document}